\documentclass{aa}  

\usepackage{graphicx}
\usepackage{txfonts}
\begin{document}

   \title{Questioning the spatial origin of complex organic molecules\\ in 
   young protostars with the CALYPSO survey\thanks{Based on observations 
   carried out with the IRAM Plateau de Bure Interferometer. IRAM is supported 
   by INSU/CNRS (France), MPG (Germany), and IGN (Spain).}$^{\rm ,}$\thanks{The 
   CALYPSO calibrated visibility tables and maps are publicly available at 
   http://www.iram-institute.org/EN/content-page-317-7-158-240-317-0.html}}

   \author{A. Belloche\inst{1}
          \and A.~J. Maury\inst{2,3}
          \and S. Maret\inst{4}
          \and S. Anderl\inst{4}
          \and A. Bacmann\inst{4}
          \and Ph. Andr\'e\inst{2}
          \and S. Bontemps\inst{5}
          \and S. Cabrit\inst{6}
          \and\\ C. Codella\inst{7}
          \and M. Gaudel\inst{2}
          \and F. Gueth\inst{8}
          \and C. Lef\`evre\inst{8}
          \and B. Lefloch\inst{4}
          \and L. Podio\inst{7}
          \and L. Testi\inst{9}
          }

   \institute{Max-Planck-Institut f\"ur Radioastronomie, Auf dem H\"ugel 69,
              53121 Bonn, Germany\\
              \email{belloche@mpifr-bonn.mpg.de}
         \and
             Laboratoire d’Astrophysique (AIM), CEA, CNRS, Universit\'e 
             Paris-Saclay, Universit\'e Paris 
             Diderot, Sorbonne Paris Cit\'e, 91191 Gif-sur-Yvette, France
         \and
             Harvard-Smithsonian Center for Astrophysics, Cambridge, MA 02138, 
             USA
         \and
             Univ. Grenoble Alpes, CNRS, IPAG, 38000 Grenoble, France
         \and
             Laboratoire d'Astrophysique de Bordeaux, Univ. Bordeaux, CNRS, 
             B18N, all\'ee Geoffroy Saint-Hilaire, 33615 Pessac, France
         \and
            PSL Research University, Sorbonne Universit\'e, Observatoire de
            Paris, CNRS, LERMA, 75014 Paris, France
         \and 
            INAF, Osservatorio Astrofisico di Arcetri, Largo E. Fermi 5, 50125 
            Firenze, Italy
         \and
            Institut de Radioastronomie Millim\'etrique (IRAM), 38406 
            Saint-Martin-d'H\`eres, France
         \and
            ESO, Karl Schwarzschild Stra{\ss}e 2, 85748 Garching bei M\"unchen, 
            Germany
             }

   \date{Received 19 December, 2019; accepted 30 January, 2020}

 
  \abstract
   {Complex organic molecules (COMs) have been detected in a few Class 0 
   protostars but their origin is not well understood. While the usual 
   picture of a hot corino explains their presence as resulting from the 
   heating of the inner envelope by the nascent protostar, shocks in the 
   outflow, a disk wind, the presence of a flared disk, or the interaction 
   region between envelope and disk at the centrifugal barrier have also been 
   claimed to enhance the abundance of COMs.}
   {Going beyond studies of individual objects, we want to investigate the 
   origin of COMs in young protostars on a statistical basis.}
   {We use the CALYPSO survey performed with the Plateau de Bure 
   Interferometer of the Institut de Radioastronomie Millim\'etrique (IRAM) to 
   search for COMs at high angular resolution in a sample of 26 solar-type 
   protostars, including 22 Class~0 and four Class~I objects. We derive the 
   column densities of the detected molecules under the local thermodynamic 
   equilibrium approximation and search for correlations between their 
   abundances and with various source properties.}
   {Methanol is detected in 12 sources and tentatively in one source, which
   represents half of the sample. Eight sources (30\%) have detections of at 
   least three COMs. We find a strong chemical differentiation in multiple
   systems with five systems having one component with at least three COMs 
   detected but the other component devoid of COM emission. All sources with a 
   luminosity higher than 4~$L_\odot$ have at least one detected COM whereas 
   no COM emission is detected in sources with internal luminosity lower than 
   2~$L_\odot$, likely because of a lack of sensitivity. The internal 
   luminosity is found to be the source parameter impacting the most the COM 
   chemical composition of the sources, while there is no obvious correlation 
   between the detection of COM emission and that of a disk-like structure.
   A canonical hot-corino 
   origin may explain the COM emission in four sources, an accretion-shock 
   origin in two or possibly three sources, and an outflow origin in three 
   sources. The CALYPSO sources with COM detections can be classified into 
   three groups on the basis of the abundances of oxygen-bearing molecules, 
   cyanides, and CHO-bearing molecules. These chemical groups correlate 
   neither with the COM origin scenarii, nor with the evolutionary status of 
   the sources if we take the ratio of envelope mass to internal luminosity as 
   an evolutionary tracer. We find strong correlations between molecules that 
   are a priori not related chemically (for instance methanol and methyl 
   cyanide), implying that the existence of a correlation does not imply a 
   chemical link.}
   {The CALYPSO survey has revealed a chemical differentiation in multiple
   systems that is markedly different from the case of the prototypical binary 
   IRAS16293-2422. This raises the question whether all low-mass protostars go 
   through 
   a phase showing COM emission. A larger sample of young protostars and a 
   more accurate determination of their internal luminosity will be necessary 
   to make further progress. Searching for correlations between the COM 
   emission and the jet/outflow properties of the sources may also be 
   promising.}


   \maketitle
%

\section{Introduction}
\label{s:intro}

Class~0 protostars are key objects along the evolutionary path that leads to 
the formation of Sun-like stars. They represent the earliest stage of 
the main accretion phase when a stellar embryo has just been formed but most 
of the mass is still stored in the collapsing protostellar envelope 
\citep[][]{Andre93,Andre00}. They retain memory of the initial physical and 
chemical conditions of star formation that prevailed during the prestellar 
phase. The collapse of their envelope also initiates the formation of 
circumstellar disks 
\citep[][]{Maury19} which will turn into protoplanetary disks and set the 
initial conditions for planet formation in the subsequent stages of star 
formation. While the formation of Sun-like stars is broadly understood, 
several issues remain. In particular, how young stars have got rid of most of 
the angular momentum initially stored in their protostellar envelopes remains 
unanswered. This has been formulated as the angular momentum problem of star 
formation \citep[][]{Bodenheimer95}. 

The Continuum And Lines in Young ProtoStellar Objects 
(CALYPSO\footnote{See \url{http://irfu.cea.fr/Projets/Calypso/}.}) IRAM Large 
Program has been set up to tackle this angular momentum problem. CALYPSO is a 
survey of 16 nearby ($d < 500$~pc) Class 0 protostellar systems carried out at 
high angular 
resolution ($\sim$0.5\arcsec) with the IRAM Plateau de Bure interferometer
(PdBI, now called Northern Extended Millimeter Array, NOEMA). This 
interferometric survey is complemented with observations with the IRAM 30~m 
single-dish telescope that provide short-spacing information. The observations 
were performed at three frequencies (94, 219, and 231~GHz) with both narrow- 
and broad-band spectrometers. In addition to the continuum emission used to 
probe circumstellar disks and protostellar multiplicity \citep[][]{Maury19}, 
the frequencies were selected in order to include in particular tracers of 
envelope rotation \citep[][]{Maret14,Gaudel20}, disk rotation 
\citep[][]{Maret20}, jets and outflows 
\citep[][]{Codella14,Santangelo15,Podio16, Lefevre17}, 
and snow lines \citep[][]{Anderl16}. Here, we take 
advantage of the large frequency coverage of the survey ($\sim$11~GHz in 
total) to probe the chemical composition of the targets, focusing on complex 
organic molecules (COMs), which are molecules with at least six atoms 
according to the definition adopted in astrochemistry \citep[][]{Herbst09}.
\citet{Maury14} reported the detection of resolved emission of numerous 
COMs toward NGC1333-IRAS2A on the basis of CALYPSO and a CALYPSO study 
specific to glycolaldehyde was published in \citet{DeSimone17}.

COMs have been detected for more than two decades in several Class 0 
protostars such as IRAS~16293-2422, NGC~1333-IRAS4A, and NGC~1333~IRAS2A 
\citep[see, e.g.,][]{vanDishoeck95,Cazaux03,Bottinelli04,Bottinelli07}. This 
COM emission was found to be compact, confined to regions with temperatures
higher than $\sim$100~K. It is usually interpreted as resulting from the 
sublimation of the ice mantles of dust grains in the hot, inner parts of the 
envelope heated by the stellar embryo. This sublimation process is thought to 
either release COMs previously formed in the solid phase directly into the gas 
phase or trigger a hot gas-phase chemistry that subsequently forms COMs. These 
regions with compact COM emission were called hot corinos by 
\citet{Ceccarelli04}, in analogy to hot cores, their counterparts around 
young high-mass stars \citep[e.g.,][]{Walmsley92,vanDishoeck98,Kurtz00}.

Unsaturated carbon chain molecules, some of them being COMs, were detected
a decade ago in the Class 0 protostar L1527, a source that was not known to 
harbor a hot corino. This led to the definition of a new class of protostars, 
the so-called Warm Carbon Chain Chemistry (WCCC) sources 
\citep[see, e.g., the review of][]{Sakai13}. While unsaturated carbon chain
molecules are known to form in the early stages of the prestellar phase,
before carbon atoms get locked up into CO, the presence of these molecules in 
the warm ($>$25~K) region of the protostellar envelope was interpreted as 
resulting from the desorption of CH$_4$ from the grain surfaces and subsequent 
gas-phase chemistry involving C$^+$. Sakai et al. proposed that hot corinos and 
WCCC sources are distinct classes of protostars, maybe related to different 
ice composition of the gain mantles resulting from different conditions during 
the prestellar phase. However, the detection of methanol at small scales in 
L1527 and the recent detection of a hot corino in L483, a candidate WCCC 
source, call for a revision of this interpretation \citep[][]{Sakai14b,Oya17}.

Saturated COMs have also been detected at even earlier stages, in prestellar 
cores \citep[e.g.,][]{Bacmann12}, albeit with lower abundances compared to hot 
corinos. A fraction of the COMs detected in hot corinos may thus have been 
formed during the prestellar phase. However,
the origin of the COM emission in hot corinos (and hot cores) is not
fully established. The sublimation of the ice mantles of dust grains could also
occur through shocks generated by jets or direct UV irradiation by the 
protostar on the walls of the cavity excavated by these jets. Enhancements of 
COM emission have indeed been reported in shocked regions of 
outflows for more than two decades 
\citep[][]{Avery96,Jorgensen04,Arce08,Sugimura11,Lefloch17}. Numerical
simulations have shown that an enhancement of COM abundances in irradiated 
cavity walls of outflows may be relevant and should strongly depend on the
protostellar luminosity \citep[][]{Drozdovskaya15}. However, this process does 
not seem to dominate the COM emission detected along the outflow cavity of the
high-mass protostar IRAS 20216+4104, which was rather interpreted as produced by
shocks \citep[][]{Palau17}.

The COM emission could also be enhanced by accretion shocks at the centrifugal
barrier in the interaction region between the disk and the collapsing 
envelope. Such an interpretation has been proposed by \citet{Csengeri18} to 
explain the presence of two hot spots detected in COM emission at small 
offsets from the high-mass protostar G328.2551-0.5321. However, the methanol 
emission detected with the Atacama Large Millimeter/submillimeter Array (ALMA) 
toward the Class 0 protostar B335 was found to be 
more extended than the centrifugal barrier \citep[][]{Imai19}. In the evolved 
Class 0 protostar L1527, \citet{Sakai14a} reported a change of chemistry with 
an enhancement of SO abundance at the centrifugal barrier but, due to a lack 
of sensitivity, it is at the moment unclear how COMs like methanol behave at 
this scale in this source \citep[][]{Sakai14b}.

High-angular resolution observations with ALMA have recently revealed that the 
COM emission in the Class 0 protostar HH212, previously interpreted as a
bona-fide hot corino in lower-angular-resolution observations
\citep[][]{Codella16a}, actually originates near the centrifugal barrier from a 
rotating ring in the warm atmosphere above and below the disk detected around 
this source \citep[][]{Lee17}. The rotational temperatures derived from 
the COM emission in HH212 are on the order of 150~K \citep[][]{Lee19}. 
\citet{Lee17} argue that
the COMs were formed in the disk rather than in the rapidly-infalling inner
envelope and that their release in the disk atmosphere may result from the
irradiation of the flared disk by the protostar. A disk wind may also play a
role in their formation or their release into the gas phase 
\citep[][]{Leurini16,Lee19}.

This introduction shows that no consensus exists about the origin of the
COM emission in Class 0 protostars. Do all protostars show COM emission? Is 
this emission consistent with the usual picture of a hot corino or do disks, 
outflows, or disk winds dominate the COM emission? When does this emission 
appear? Can it be used as an evolutionary tracer? Does it depend on the 
envelope mass? Is the COM chemical composition of Class 0 protostars universal? 
In order to start addressing these questions on a statistical basis, we 
take advantage of the unprecedentedly large sample of young embedded 
protostars and the sub-arcsecond angular resolution provided by the CALYPSO 
survey to investigate the origin of small-scale COM emission in Class 0 
protostars, as well as a few Class I protostars that happened to be in the 
covered fields. We use the CALYPSO data set to search for COMs in the targeted 
sources and derive their COM chemical composition.
We provide a short description of the 
observations in Sect.~\ref{s:obs}. The chemical composition of the sources 
derived from the observations is presented in Sect.~\ref{s:results_chemcomp}. 
An analysis of the correlations between the chemical abundances of the 
detected COMs and the correlations between these abundances and various source 
properties is given in Sect.~\ref{s:correlations}. These results are discussed 
in Sect.~\ref{s:discussion} and our conclusions are stated in 
Sect.~\ref{s:conclusions}.

\section{Observations}
\label{s:obs}

The observations of the CALYPSO survey were performed with PdBI at wavelengths 
of 1.29, 1.37, and 3.18~mm. The interferometric data were obtained by combining 
observations carried out with two distinct PdBI array configurations (A 
and C), providing baselines ranging from 16~m to 760~m and sub-arcsecond 
resolution. General information 
about the observations (dates of observation, conditions for each track) and 
the overall data calibration strategy have been published in \citet{Maury19}. 
The phase self-calibration corrections derived from the continuum emission 
gain curves were also applied to the line visibility data presented here 
\citep[when applicable, that is for all sources with a 1.3~mm PdBI integrated 
flux density higher than 100~mJy in Table 4 of][except L1448-N]{Maury19}.

\begin{table*}[!ht]
 \begin{center}
 \caption{
 Properties of the CALYPSO sources analyzed in this work.
}
 \label{t:sources}
 \vspace*{-1.2ex}
 \begin{tabular}{lcccccccccc}
 \hline\hline
 \multicolumn{1}{c}{Source} & \multicolumn{2}{c}{Coordinates (J2000)\tablefootmark{a}} & \multicolumn{1}{c}{$d$\tablefootmark{b}} & \multicolumn{1}{c}{\hspace*{-2ex} Ref.\tablefootmark{c}} & \multicolumn{1}{c}{$M_{\rm env}$\tablefootmark{d}} & \multicolumn{1}{c}{\hspace*{-2ex} Ref.\tablefootmark{e}} & \multicolumn{1}{c}{\hspace*{-1ex} $L_{\rm int}$\tablefootmark{f}} & \multicolumn{1}{c}{\hspace*{-1ex} $S_{\rm 1mm}$\tablefootmark{g}} & \multicolumn{1}{c}{$S_{\rm 3mm}$\tablefootmark{g}} & \multicolumn{1}{c}{\hspace*{-1ex} $F_{\rm 1mm}$\tablefootmark{h}} \\ 
  & \multicolumn{1}{c}{$\alpha$ (hh:mm:ss)} & \multicolumn{1}{c}{$\delta$ (dd:mm:ss)} & \multicolumn{1}{c}{(pc)} & & \multicolumn{1}{c}{($M_\odot$)} & & \multicolumn{1}{c}{($L_\odot$)} & \multicolumn{2}{c}{(mJy/beam)} & \multicolumn{1}{c}{(mJy)} \\ 
  \multicolumn{1}{c}{(1)} & \multicolumn{1}{c}{(2)} & \multicolumn{1}{c}{(3)} & \multicolumn{1}{c}{(4)} & \multicolumn{1}{c}{(5)} & \multicolumn{1}{c}{(6)} & \multicolumn{1}{c}{(7)} & \multicolumn{1}{c}{(8)} & \multicolumn{1}{c}{(9)} & \multicolumn{1}{c}{(10)} & \multicolumn{1}{c}{(11)} \\ 
 \hline
 L1448-2A & 03:25:22.405 & 30:45:13.26 & 293 & \hspace*{-2ex} 1,2 & 1.2 & \hspace*{-2ex} 1 & \hspace*{-1ex} 4.7 (0.5) & \hspace*{-1ex} 23 (2) & 5.8 (1) & \hspace*{-1ex} 38 \\ 
 L1448-2Ab & 03:25:22.355 & 30:45:13.16 & 293 & \hspace*{-2ex} 1,2 & 0.6 & \hspace*{-2ex} 1 & \hspace*{-1ex} $<$ 4.7 (--) & \hspace*{-1ex} 11 (1) & 4.5 (2) & \hspace*{-1ex} 26.8 \\ 
 L1448-NA & 03:25:36.498 & 30:45:21.85 & 293 & \hspace*{-2ex} 1,2 & 0.8 & \hspace*{-2ex} 2 & \hspace*{-1ex} 6.4 (0.6) & \hspace*{-1ex} 46 (4) & 6.7 (0.3) & \hspace*{-1ex} -- \\ 
 L1448-NB1 & 03:25:36.378 & 30:45:14.77 & 293 & \hspace*{-2ex} 1,2 & 3.3 & \hspace*{-2ex} 1 & \hspace*{-1ex} $<$ 3.9 (--) & \hspace*{-1ex} 146 (6) & 69 (2) & \hspace*{-1ex} 155 \\ 
 L1448-NB2 & 03:25:36.315 & 30:45:15.15 & 293 & \hspace*{-2ex} 1,2 & 1.6 & \hspace*{-2ex} 1 & \hspace*{-1ex} 3.9 (--) & \hspace*{-1ex} 69 (3) & <25 (5) & \hspace*{-1ex} 136 \\ 
 L1448-C & 03:25:38.875 & 30:44:05.33 & 293 & \hspace*{-2ex} 1,2 & 1.9 & \hspace*{-2ex} 1 & \hspace*{-1ex} 11 (1) & \hspace*{-1ex} 123 (5) & 19 (1) & \hspace*{-1ex} -- \\ 
 L1448-CS & 03:25:39.132 & 30:43:58.04 & 293 & \hspace*{-2ex} 1,2 & 0.16 & \hspace*{-2ex} 1 & \hspace*{-1ex} 3.6 (--) & \hspace*{-1ex} 8 (2) & 1.6 (0.1) & \hspace*{-1ex} -- \\ 
 IRAS2A1 & 03:28:55.570 & 31:14:37.07 & 293 & \hspace*{-2ex} 1 & 7.9 & \hspace*{-2ex} 1 & \hspace*{-1ex} 47 (5) & \hspace*{-1ex} 132 (5) & 20 (1) & \hspace*{-1ex} -- \\ 
 SVS13B & 03:29:03.078 & 31:15:51.74 & 293 & \hspace*{-2ex} 1 & 2.8 & \hspace*{-2ex} 1 & \hspace*{-1ex} 3.1 (1.6) & \hspace*{-1ex} 127 (7) & 22 (1) & \hspace*{-1ex} -- \\ 
 SVS13A & 03:29:03.756 & 31:16:03.80 & 293 & \hspace*{-2ex} 1 & 0.8 & \hspace*{-2ex} 2 & \hspace*{-1ex} 44 (5) & \hspace*{-1ex} 120 (7) & 21 (1) & \hspace*{-1ex} -- \\ 
 IRAS4A1 & 03:29:10.537 & 31:13:30.98 & 293 & \hspace*{-2ex} 1 & 9.9 & \hspace*{-2ex} 1 & \hspace*{-1ex} $<$ 4.7 (--) & \hspace*{-1ex} 481 (10) & 148 (6) & \hspace*{-1ex} -- \\ 
 IRAS4A2 & 03:29:10.432 & 31:13:32.12 & 293 & \hspace*{-2ex} 1 & 2.3 & \hspace*{-2ex} 1 & \hspace*{-1ex} 4.7 (0.5) & \hspace*{-1ex} 186 (8) & <34 (10) & \hspace*{-1ex} -- \\ 
 IRAS4B & 03:29:12.016 & 31:13:08.02 & 293 & \hspace*{-2ex} 1 & 3.3 & \hspace*{-2ex} 1 & \hspace*{-1ex} 2.3 (0.3) & \hspace*{-1ex} 278 (6) & 75 (3) & \hspace*{-1ex} -- \\ 
 IRAS4B2 & 03:29:12.841 & 31:13:06.84 & 293 & \hspace*{-2ex} 1 & 1.4 & \hspace*{-2ex} 1 & \hspace*{-1ex} $<$ 0.16 (--) & \hspace*{-1ex} 114 (4) & 31 (1) & \hspace*{-1ex} 273 \\ 
 IRAM04191 & 04:21:56.899 & 15:29:46.11 & 140 & \hspace*{-2ex} 3,2 & 0.5 & \hspace*{-2ex} 1 & \hspace*{-1ex} 0.05 (0.01) & \hspace*{-1ex} 4.7 (0.8) & 0.31 (0.09) & \hspace*{-1ex} 5.3 \\ 
 L1521F & 04:28:38.941 & 26:51:35.14 & 140 & \hspace*{-2ex} 3,2 & 0.7 & \hspace*{-2ex} 1 & \hspace*{-1ex} 0.035 (0.010) & \hspace*{-1ex} 1.6 (0.2) & 0.27 (0.05) & \hspace*{-1ex} -- \\ 
 L1527 & 04:39:53.875 & 26:03:09.66 & 140 & \hspace*{-2ex} 3,2 & 1.2 & \hspace*{-2ex} 1 & \hspace*{-1ex} 0.9 (0.1) & \hspace*{-1ex} 129 (8) & 23 (1) & \hspace*{-1ex} 179 \\ 
 SerpM-S68N & 18:29:48.091 & 01:16:43.41 & 436 & \hspace*{-2ex} 4,5 & 11 & \hspace*{-2ex} 1 & \hspace*{-1ex} 11 (2) & \hspace*{-1ex} 35 (3) & 5.3 (0.5) & \hspace*{-1ex} -- \\ 
 SerpM-S68Nb & 18:29:48.707 & 01:16:55.53 & 436 & \hspace*{-2ex} 4,5 & -- & \hspace*{-2ex} -- & \hspace*{-1ex} 1.8 (0.2) & \hspace*{-1ex} --\tablefootmark{i} & 2.9 (0.4) & \hspace*{-1ex} -- \\ 
 SerpM-SMM4a & 18:29:56.716 & 01:13:15.65 & 436 & \hspace*{-2ex} 4,5 & 6.7 & \hspace*{-2ex} 1 & \hspace*{-1ex} 2.2 (0.2) & \hspace*{-1ex} 184 (11) & 48 (2) & \hspace*{-1ex} -- \\ 
 SerpM-SMM4b & 18:29:56.525 & 01:13:11.58 & 436 & \hspace*{-2ex} 4,5 & 1.0 & \hspace*{-2ex} 1 & -- & \hspace*{-1ex} 27 (4) & 9 (1) & \hspace*{-1ex} -- \\ 
 SerpS-MM18a & 18:30:04.118 & $-$02:03:02.55 & 350 & \hspace*{-2ex} 6 & 4.5 & \hspace*{-2ex} 1 & \hspace*{-1ex} 13 (4) & \hspace*{-1ex} 148 (9) & 20 (1) & \hspace*{-1ex} -- \\ 
 SerpS-MM18b & 18:30:03.541 & $-$02:03:08.33 & 350 & \hspace*{-2ex} 6 & 0.9 & \hspace*{-2ex} 1 & \hspace*{-1ex} 16 (4) & \hspace*{-1ex} 62 (4) & 7.8 (0.8) & \hspace*{-1ex} -- \\ 
 SerpS-MM22 & 18:30:12.310 & $-$02:06:53.56 & 350 & \hspace*{-2ex} 6 & 0.9 & \hspace*{-2ex} 1 & \hspace*{-1ex} 0.36 (0.18) & \hspace*{-1ex} 20 (2) & 2.8 (0.7) & \hspace*{-1ex} -- \\ 
 L1157 & 20:39:06.269 & 68:02:15.70 & 352 & \hspace*{-2ex} 2 & 3.0 & \hspace*{-2ex} 1 & \hspace*{-1ex} 4.0 (0.4) & \hspace*{-1ex} 117 (9) & 18 (1) & \hspace*{-1ex} -- \\ 
 GF9-2 & 20:51:29.823 & 60:18:38.44 & 474 & \hspace*{-2ex} 2,7 & 2.8 & \hspace*{-2ex} 1 & \hspace*{-1ex} 1.7 (--) & \hspace*{-1ex} 9.9 (1) & 1.6 (0.4) & \hspace*{-1ex} -- \\ 
 \hline
 \end{tabular}
 \end{center}
 \vspace*{-2.5ex}
 \tablefoot{
 \tablefoottext{a}{Equatorial coordinates of the continuum peak measured by \citet{Maury19}.}
 \tablefoottext{b}{Distance to the source.}
 \tablefoottext{c}{References for the distance: 1: \citet{OrtizLeon18a}; 2: \citet{Zucker19}; 3: \citet{Loinard07}; 4: \citet{OrtizLeon17}; 5: \citet{OrtizLeon18b}; 6: Palmeirim et al., in prep.; 7: C. Zucker, priv. comm..}
 \tablefoottext{d}{Envelope mass. For the binaries with no mass estimate of their individual envelopes, the value corresponds to the fraction of the total envelope mass in proportion of the peak flux densities at 1.3~mm given in Col. 9. The masses have been rescaled to the distances given in Col. 4.}
 \tablefoottext{e}{References for the envelope mass: 1: \citet{Maury19} and references therein; 2: A. Maury, priv. comm..}
 \tablefoottext{f}{Internal luminosity with the uncertainty in parentheses when available. The luminosity has been estimated by Ladjelate et al., in prep., from \textit{Herschel} Gould Belt survey data \citep[http://gouldbelt-herschel.cea.fr/archives,][]{Andre10}, except for GF9-2 for which we use \citet{Wiesemeyer97}. The luminosity has been rescaled to the distance given in Col.~4.}
 \tablefoottext{g}{Peak flux densities at 1.3~mm and 3~mm measured with PdBI by \citet{Maury19} with the uncertainties in parentheses when available.}
 \tablefoottext{h}{Integrated flux density at 1.3~mm over the source size given in Table~\ref{t:sizes}, when the latter is larger than the beam.}
 \tablefoottext{i}{This source is located outside the primary beam of the CALYPSO survey at 1.3~mm but is detected at 3~mm.}
 }
 \end{table*}

Observations of the molecular line emission analyzed in the present paper
were obtained under the form of three spectral setups (setup S1 around 231~GHz,
setup S2 around 219~GHz and setup S3 around 94~GHz) containing each two 
1.8-GHz-wide spectral windows covered with the WideX backends with a spectral
resolution of 1.95 MHz. The frequency ranges covered by the survey are the 
following: 229.242--231.033~GHz and 231.0422--232.8338~GHz in setup S1, 
216.8700--218.6550~GHz and 218.6720--220.4550~GHz in setup S2, and 
91.8560--93.6728~GHz and 93.6753--95.5460~GHz in setup S3. The sources 
IRAM04191, NGC1333-IRAS2A, L1448-NB, L1448-2A, L1448-C, and L1521F were 
already observed with configuration A in setup S1 by the beginning of our 
CALYPSO program 
\citep[see, e.g., the pilot observations presented in][]{Maury10}, hence 
configuration-A WideX data around 231~GHz were not obtained for these sources.

The continuum built from line-free channels was subtracted directly from the 
spectral visibility data sets, for each of the 30 continuum sources detected 
in the 16 fields targeted by the CALYPSO survey 
\citep[see Table 3 in][]{Maury19}. Details about this procedure are 
available on the CALYPSO data release webpage\footnote{See
\url{http://www.iram.fr/ILPA/LP010/calypso-readme.txt}.}.
Here we focus our analysis on 26 of these sources, ignoring SerpM-S68Nc, 
L1448-NW, and SVS13C that are located well outside the primary beam at 1.3~mm, 
and VLA3 the nature of which is unknown.
These 26 sources are listed in Table~\ref{t:sources} along with some of their 
properties. For readability reasons,
we use in Table~\ref{t:sources} and in the rest of the article the short names 
IRAS2A, IRAS4A, and IRAS4B for the IRAS sources located in the NGC~1333 
molecular cloud.

We built spectral maps, covering the six frequency windows, for each source.
The maps were obtained using a robust weighting scheme, resulting in
synthesized beam sizes and rms noise levels reported in Table~\ref{t:beam_rms}.

\section{Results}
\label{s:results_chemcomp}

\subsection{Basic spectral features}
\label{ss:basic}

\begin{figure*}
 \centerline{\resizebox{1.0\hsize}{!}{\includegraphics[angle=0]{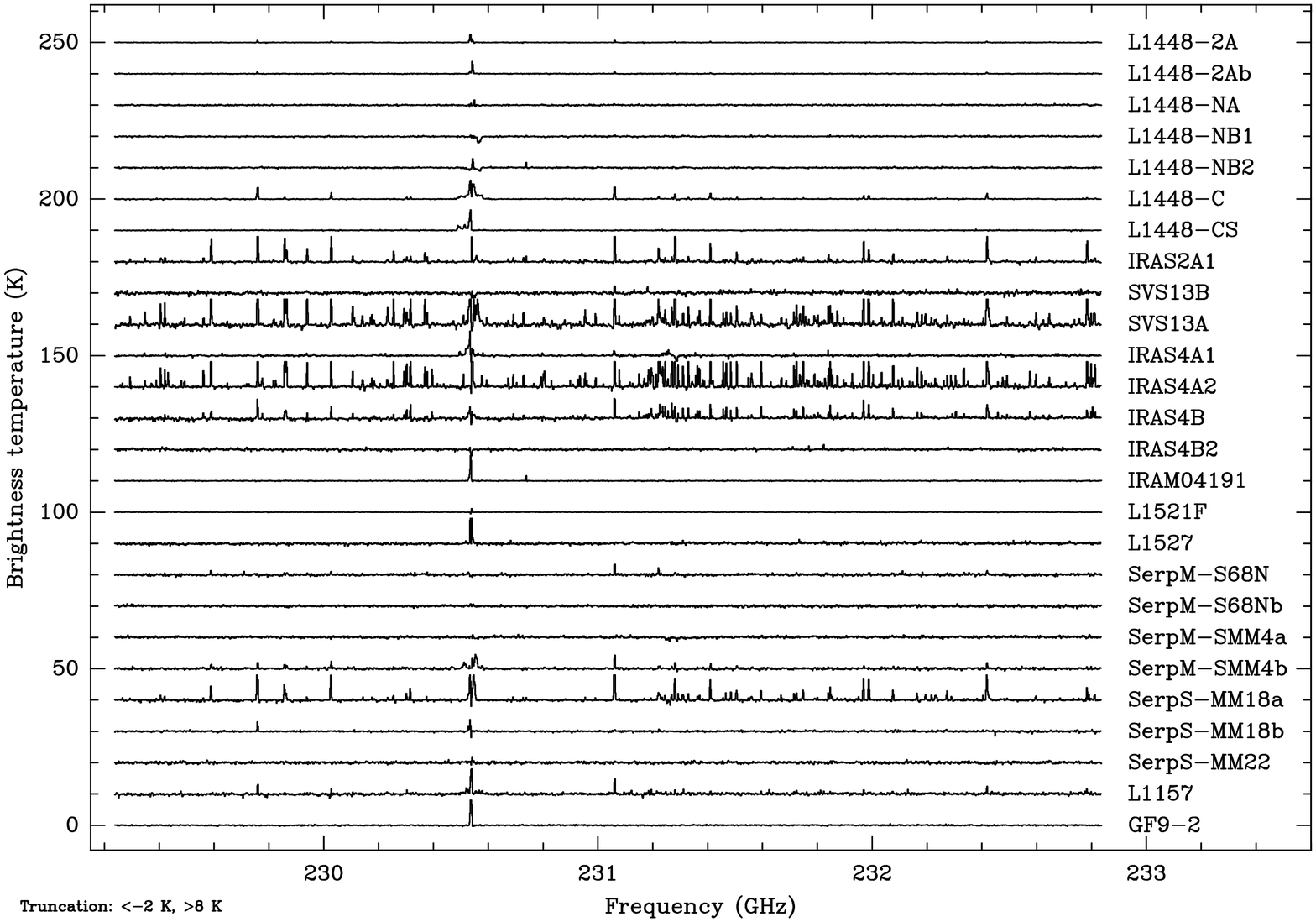}}}
 \caption{Continuum-subtracted WideX spectra at 1.3~mm toward the continuum 
emission peaks of the CALYPSO sources. For display purposes, the spectra were 
truncated to the range [-2, 8] K and shifted by multiples of 10~K vertically. 
In this figure, the spectra are not corrected for primary-beam attenuation.}
 \label{f:spectraS1}
\end{figure*}

The continuum-subtracted WideX spectra obtained in setups S1, S2, and S3 
toward the main and some of the secondary continuum emission peaks found by 
\citet{Maury19} in the CALYPSO sample are shown in Figs.~\ref{f:spectraS1}, 
\ref{f:spectraS2}, and \ref{f:spectraS3}, respectively. Only few lines 
are detected in the 3~mm spectra, but several sources show many spectral lines 
in the 1.4~mm and 1.3~mm bands. Qualitatively, the sources with a high density
of detected spectral lines are IRAS2A1, IRAS4A2, 
IRAS4B, L1448-C, SVS13A, and SerpS-MM18a. The first three sources were 
already known to harbor a hot corino before the CALYPSO survey started 
\citep[][]{Bottinelli04, Bottinelli07}. \citet{Maret04} reported a jump of 
H$_2$CO abundance by three orders of magnitude in the region above 100 K in 
the envelope of L1448-C, suggesting the presence of a hot corino in this 
source as well. The presence of a hot corino in SVS13A was reported 
based on the Astrochemical Surveys At IRAM (ASAI) Large Program 
\citep[][]{Codella16b} and on CALYPSO data \citep[][]{Lefevre17}.

\subsection{Channel count maps and line counts}
\label{ss:counts}

\begin{figure*}
 \centerline{\resizebox{0.24\hsize}{!}{\includegraphics[angle=0]{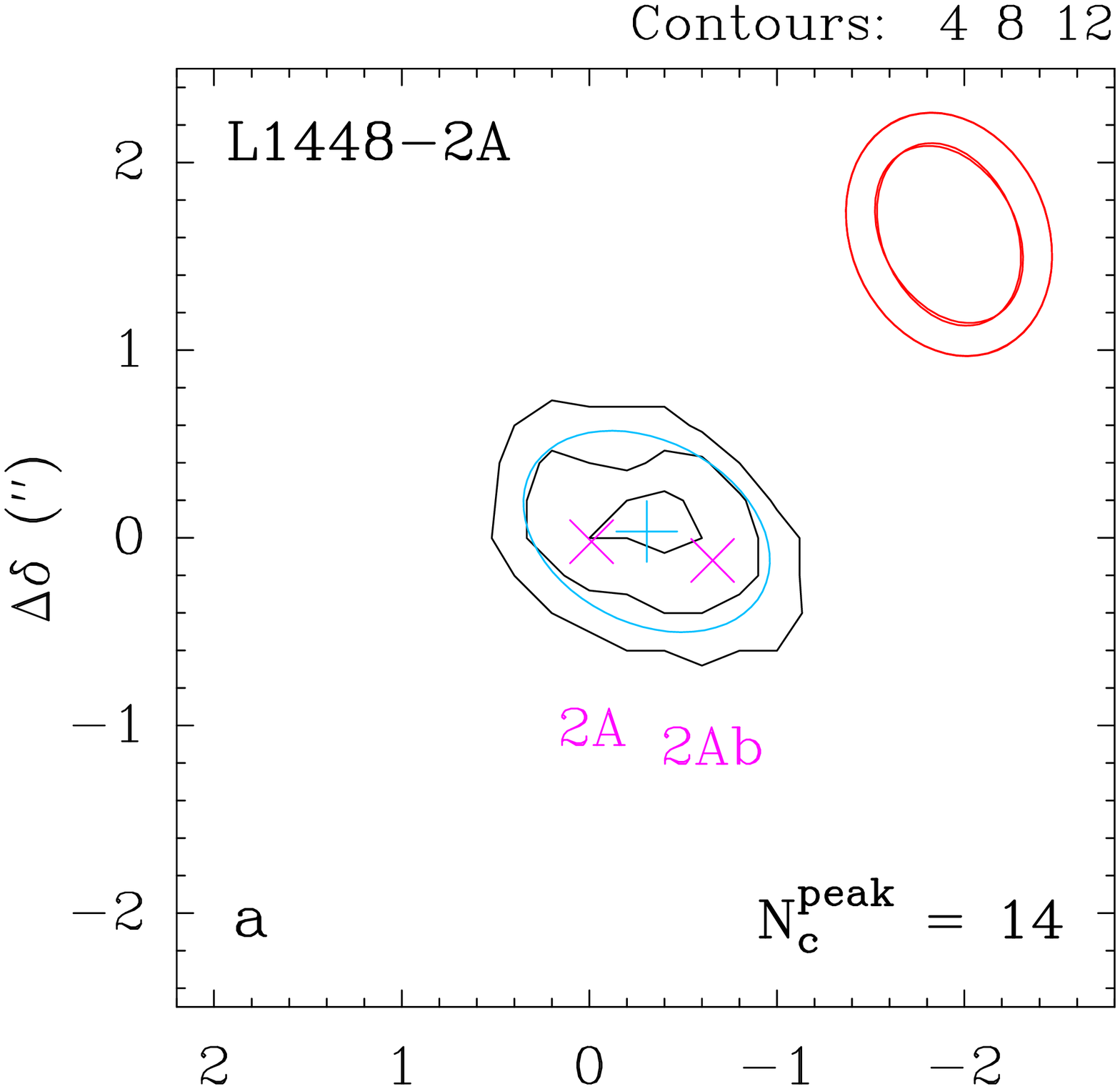}}\resizebox{0.24\hsize}{!}{\includegraphics[angle=0]{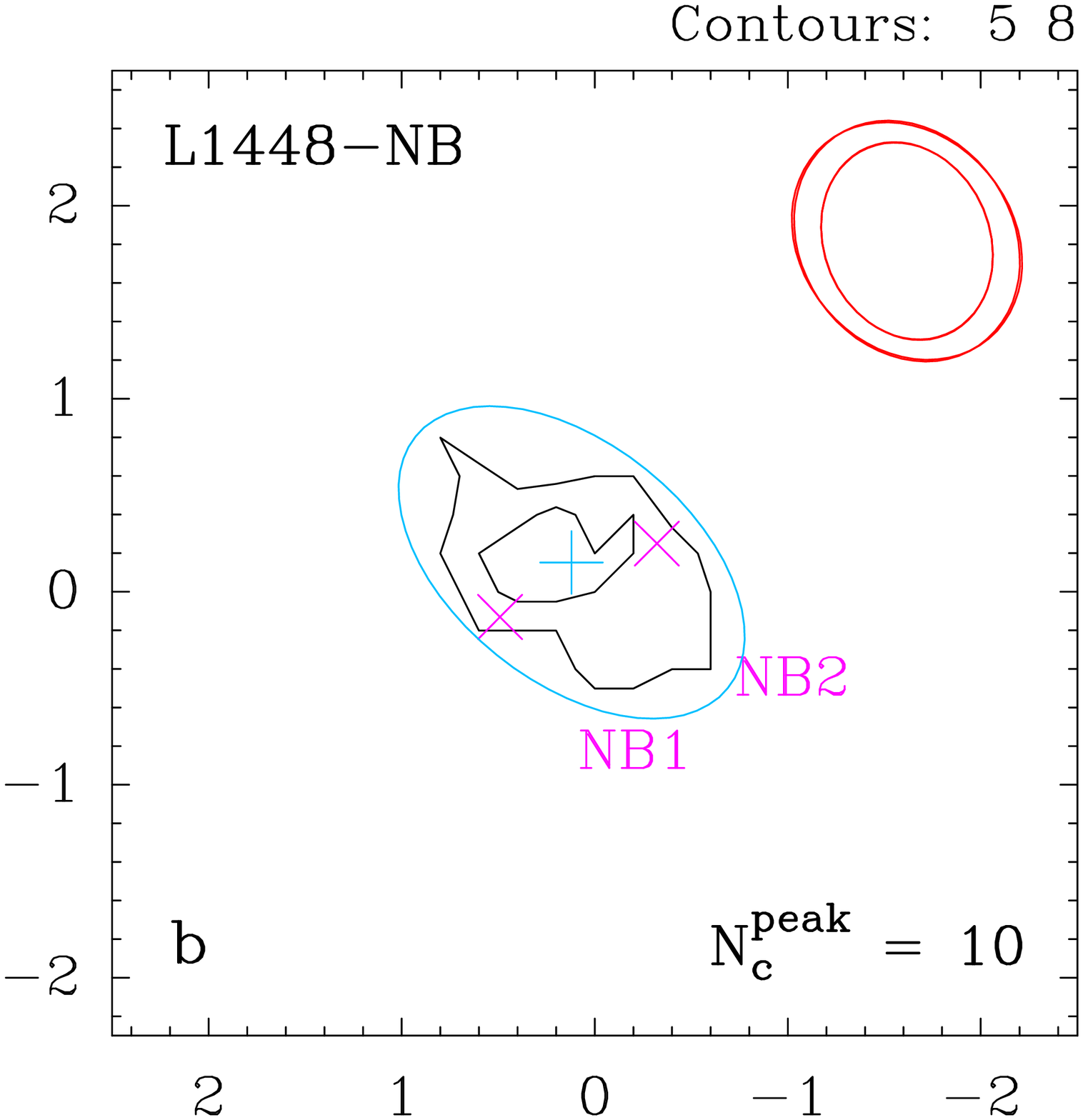}}\resizebox{0.24\hsize}{!}{\includegraphics[angle=0]{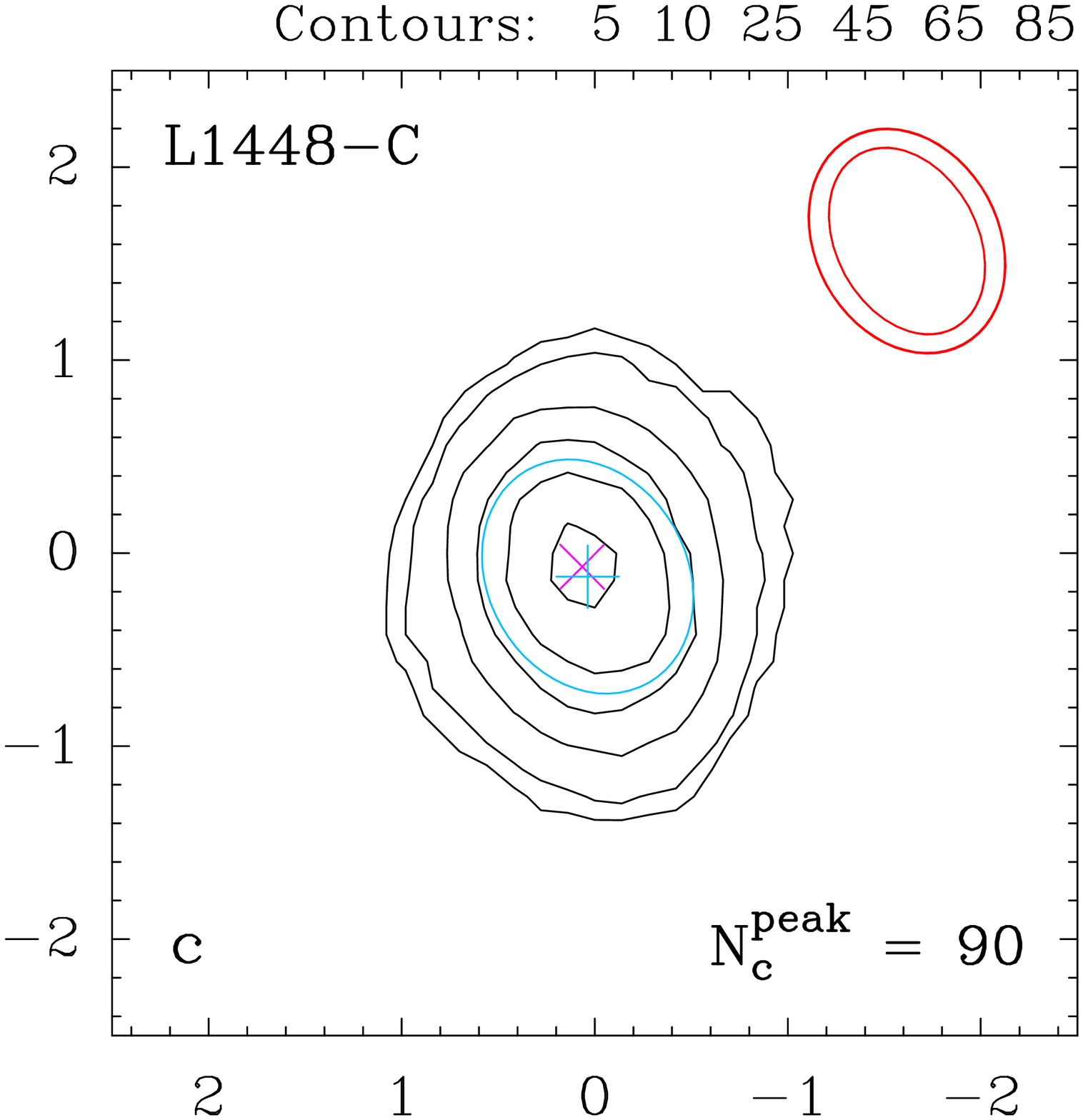}}\resizebox{0.24\hsize}{!}{\includegraphics[angle=0]{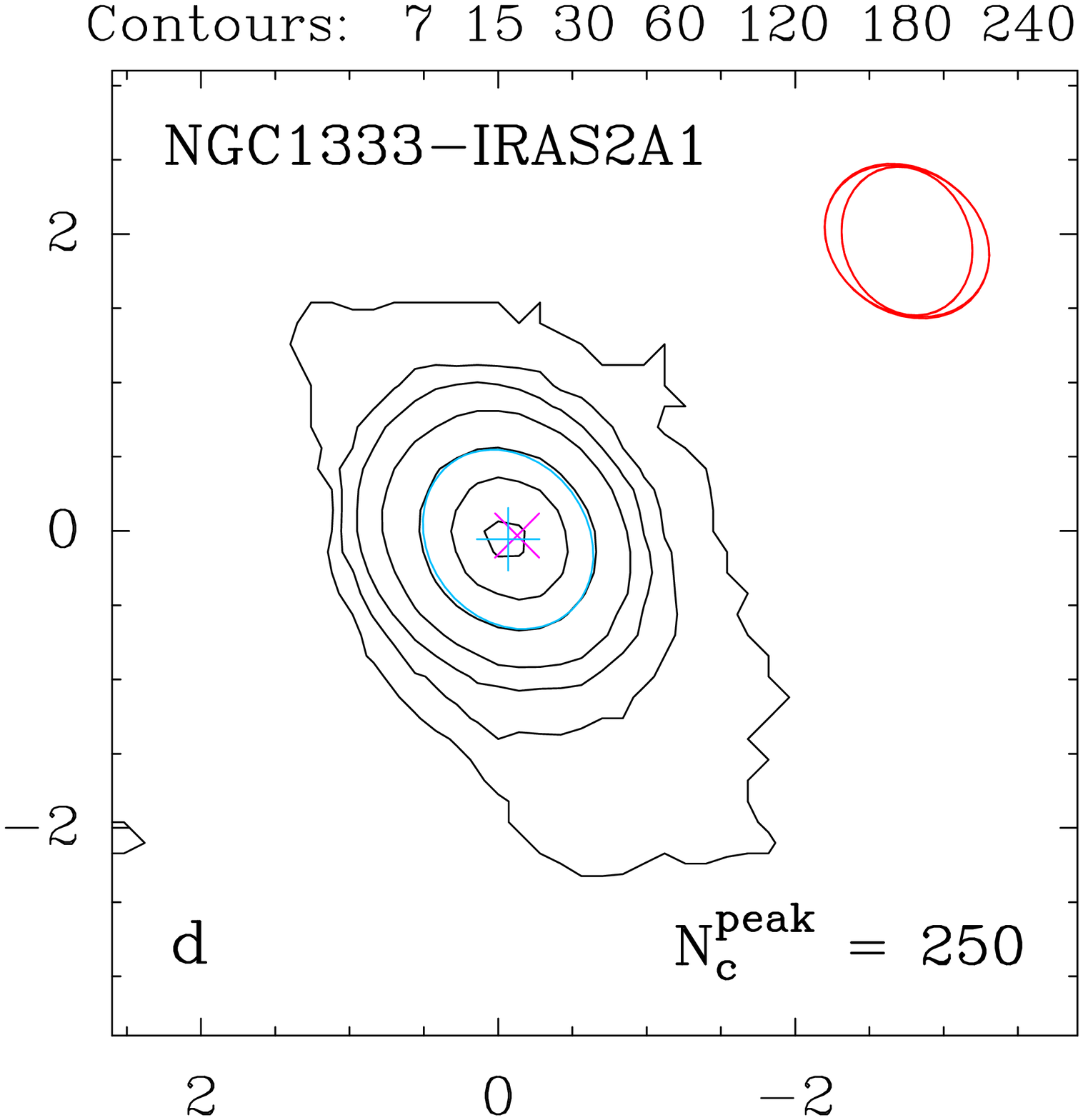}}}
 \centerline{\resizebox{0.24\hsize}{!}{\includegraphics[angle=0]{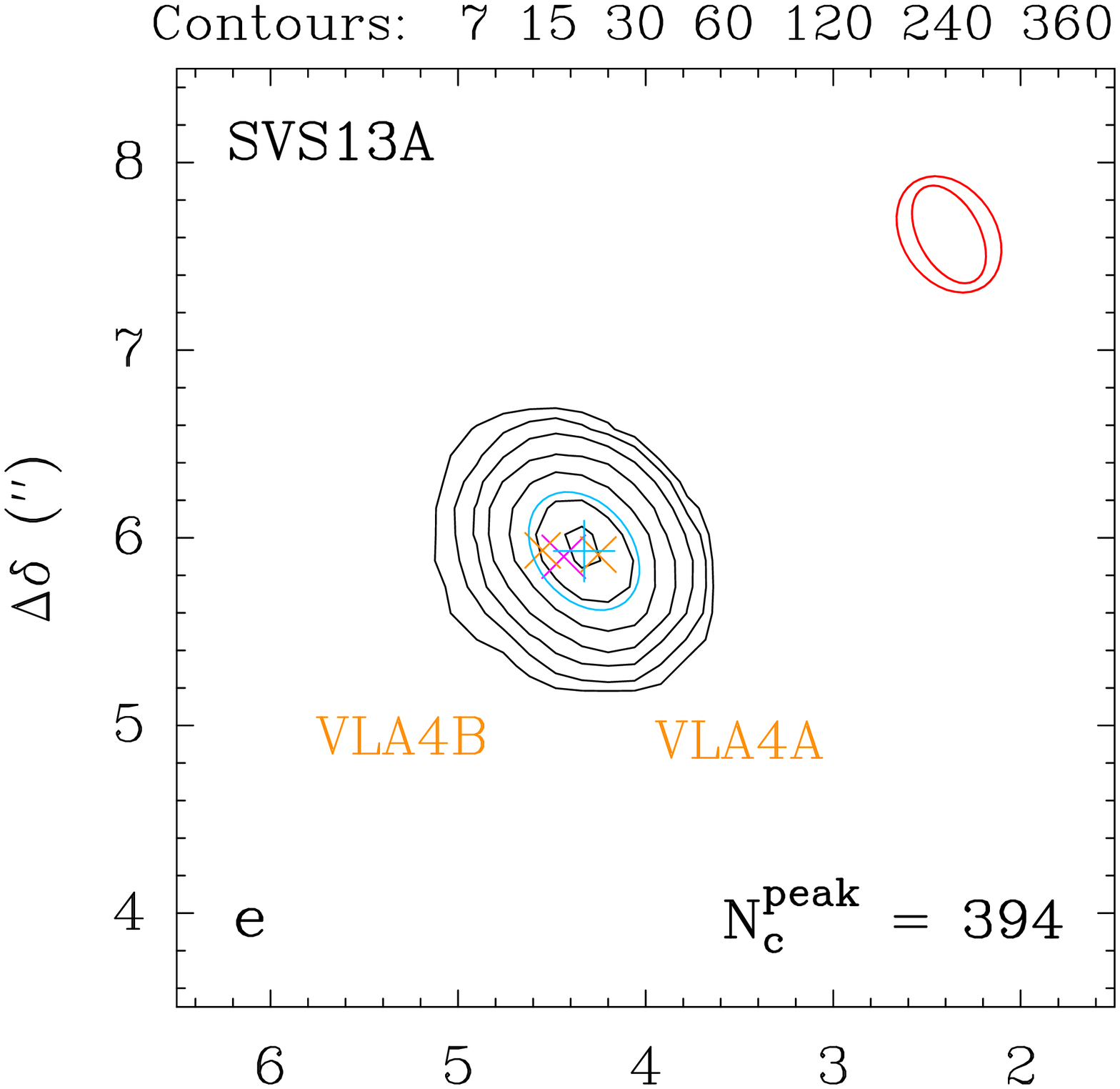}}\resizebox{0.24\hsize}{!}{\includegraphics[angle=0]{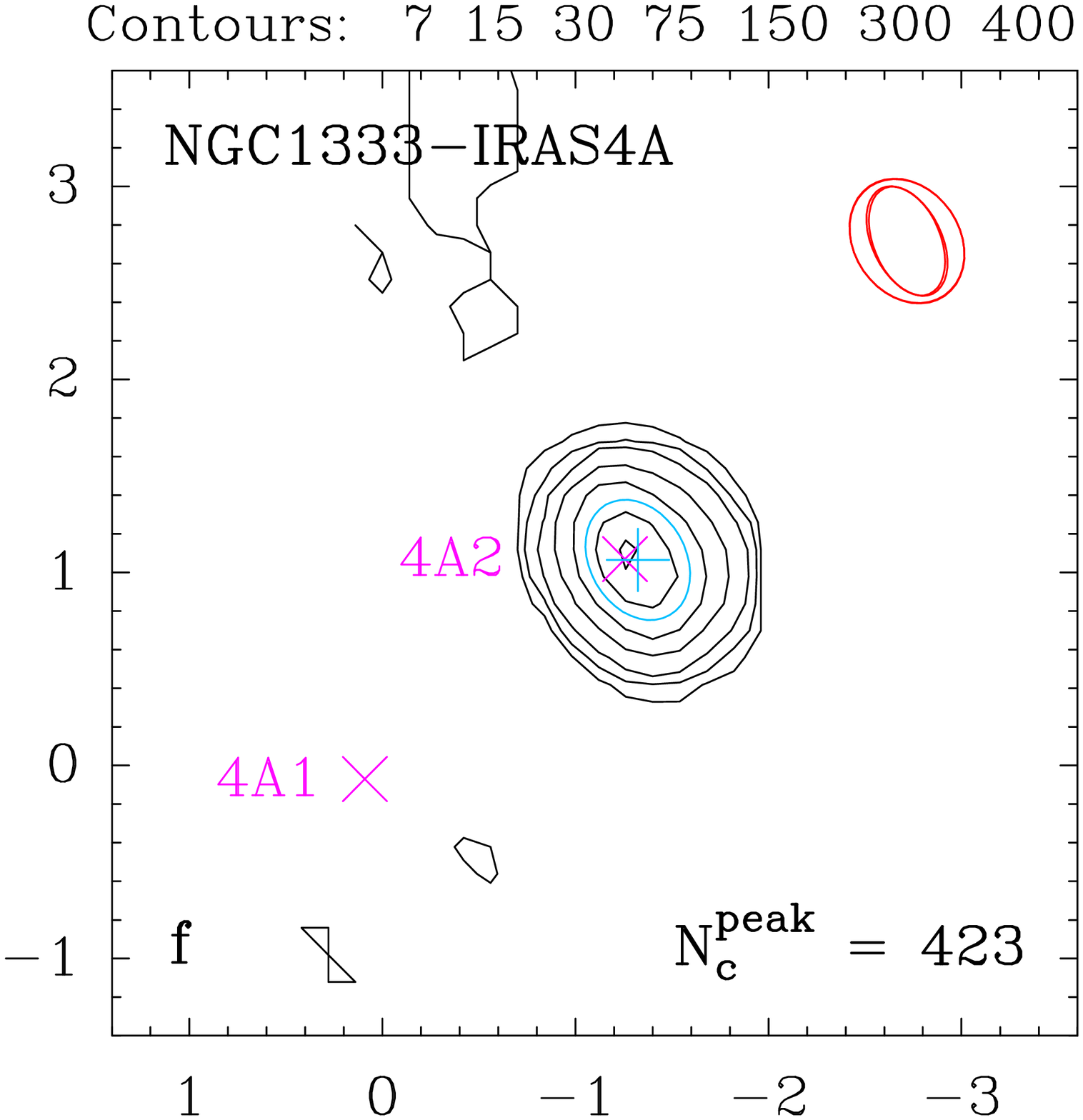}}\resizebox{0.24\hsize}{!}{\includegraphics[angle=0]{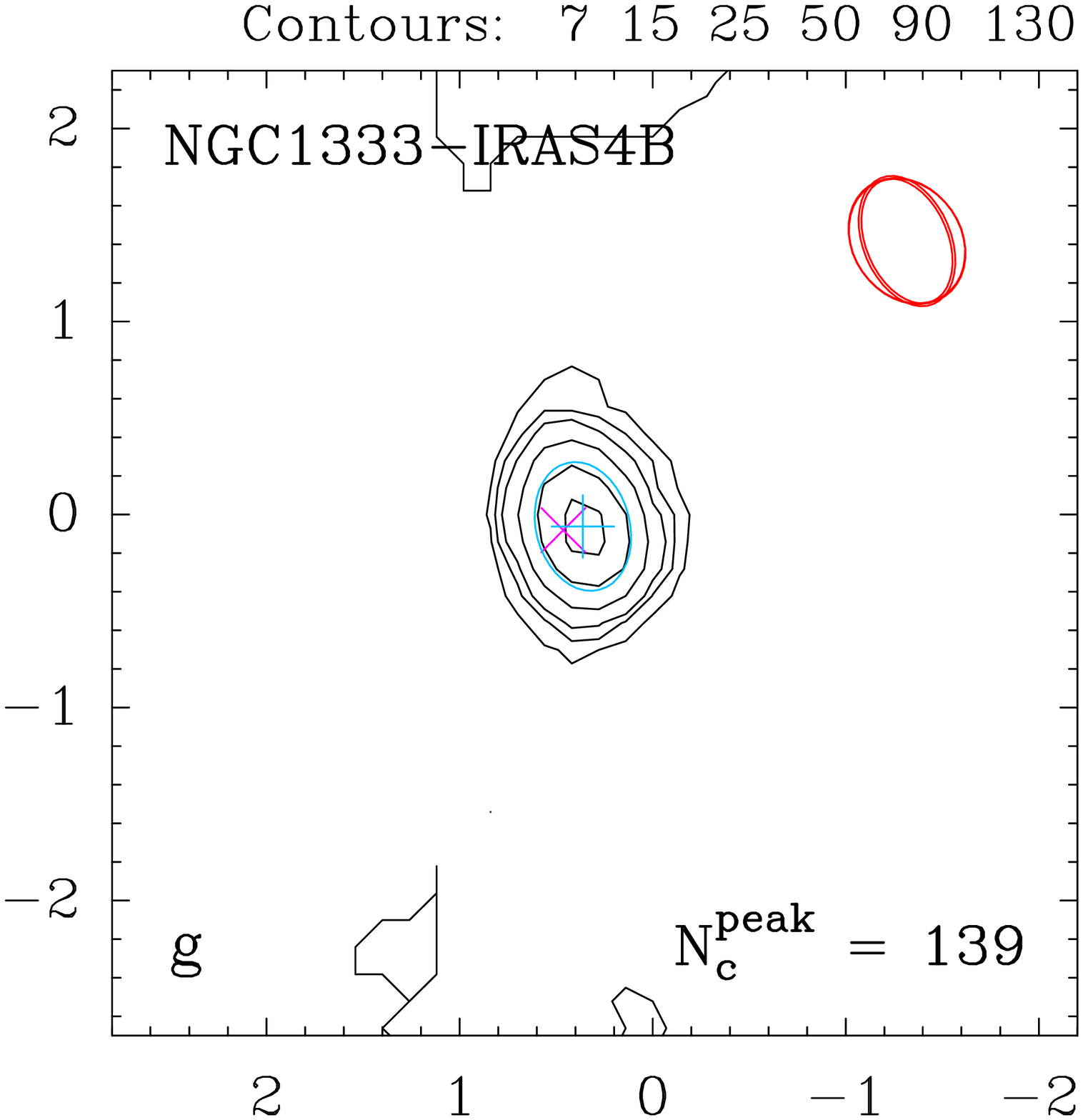}}\resizebox{0.24\hsize}{!}{\includegraphics[angle=0]{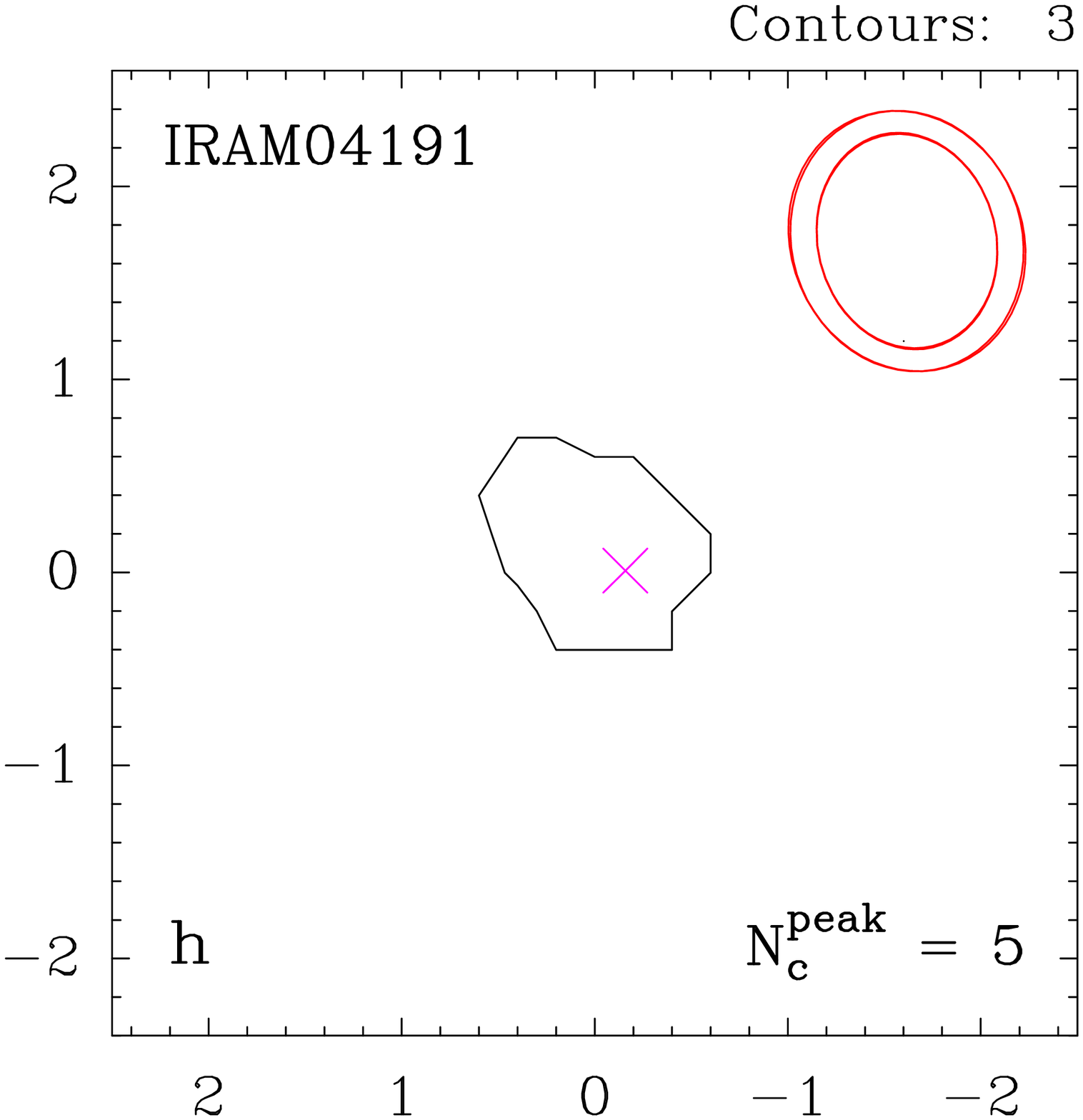}}}
 \centerline{\resizebox{0.24\hsize}{!}{\includegraphics[angle=0]{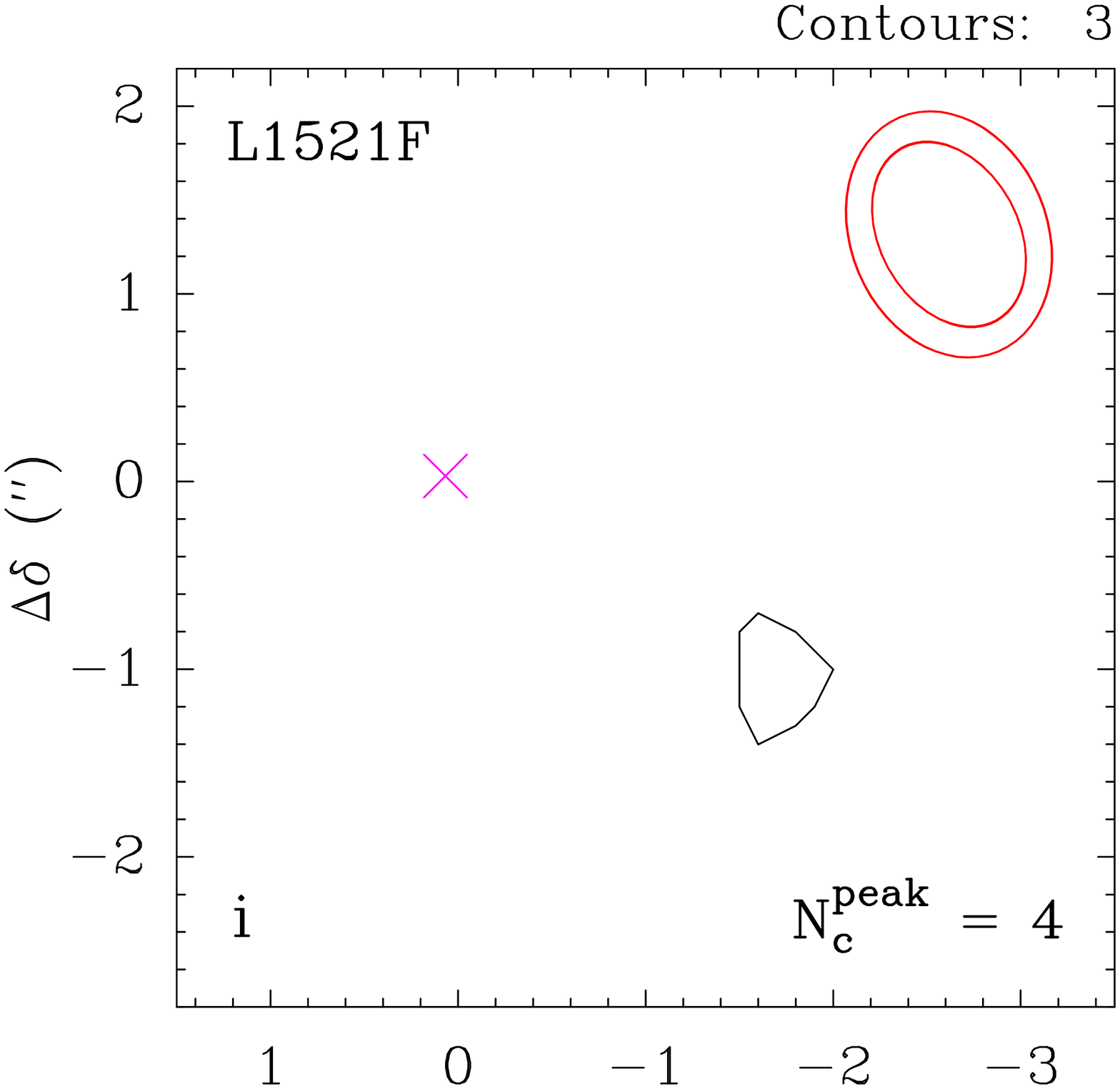}}\resizebox{0.24\hsize}{!}{\includegraphics[angle=0]{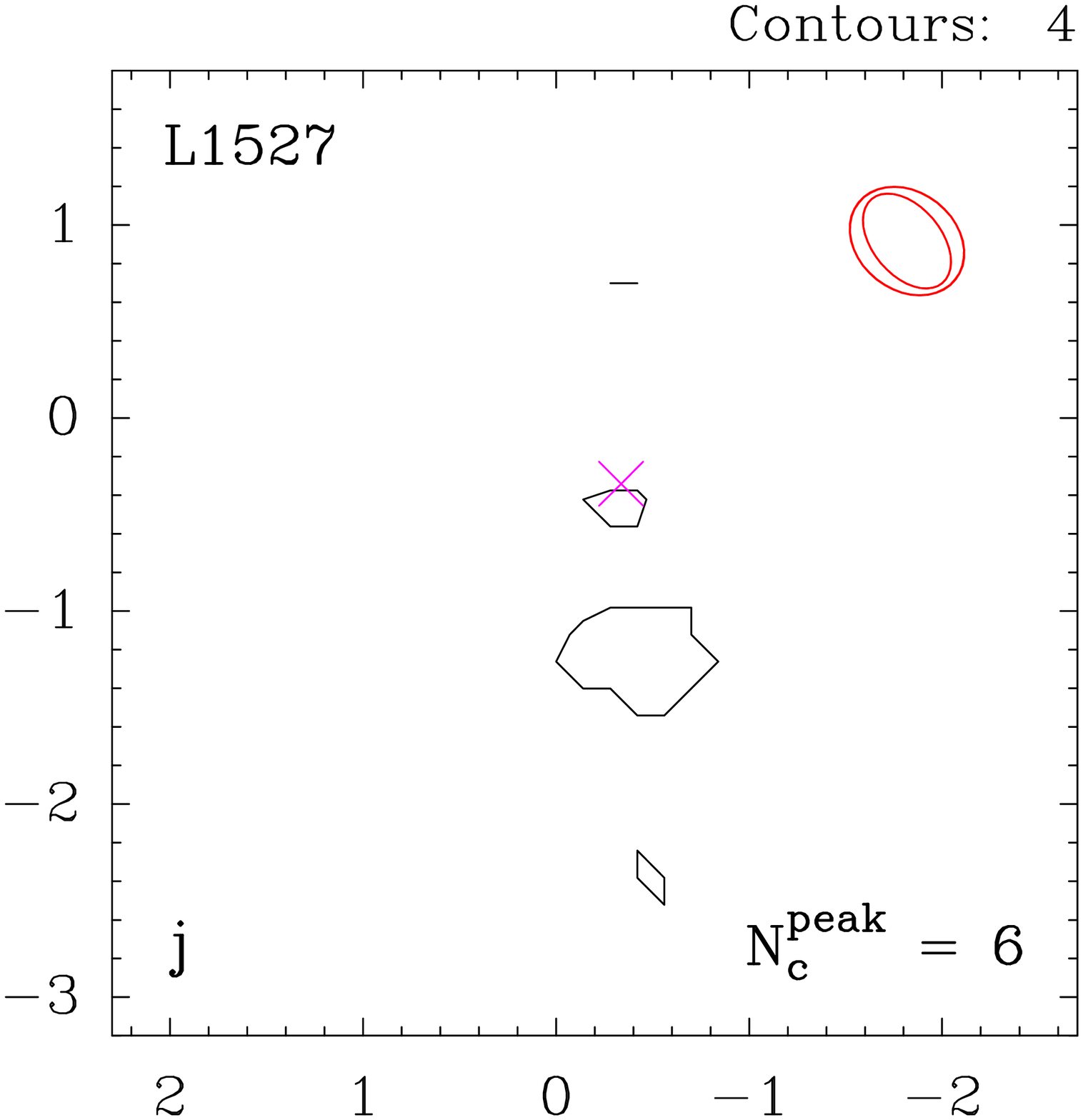}}\resizebox{0.24\hsize}{!}{\includegraphics[angle=0]{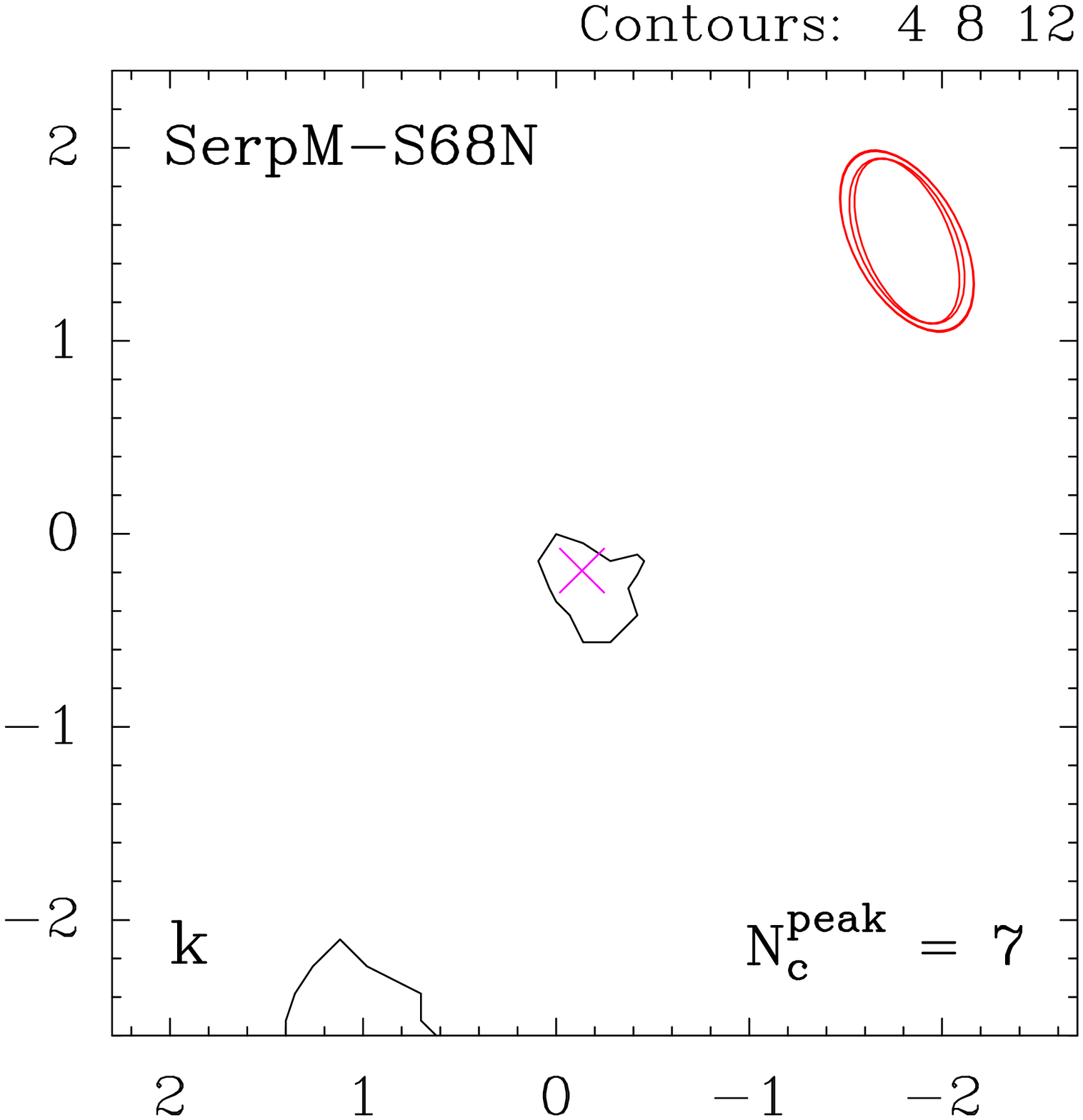}}\resizebox{0.24\hsize}{!}{\includegraphics[angle=0]{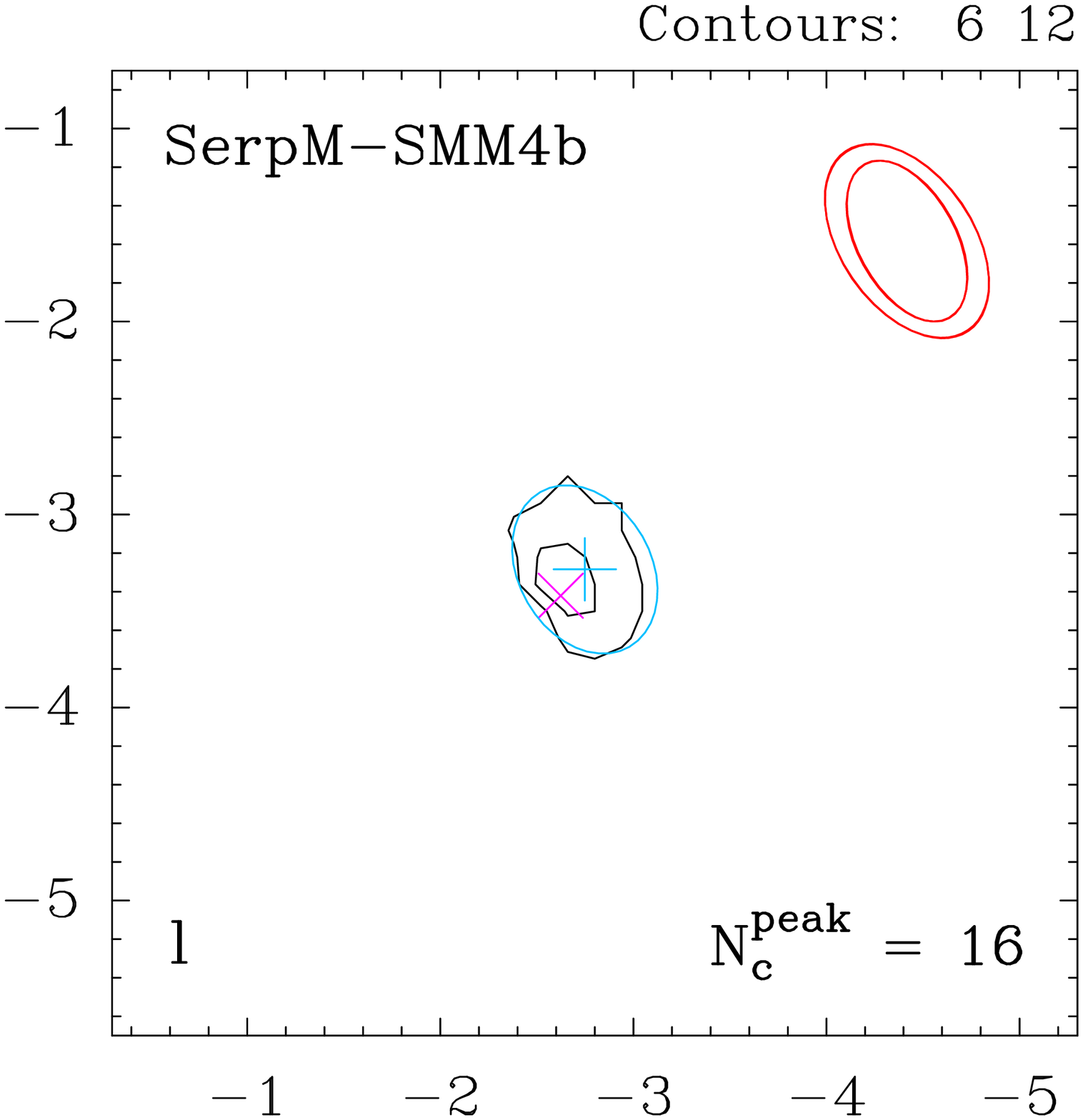}}}
 \centerline{\resizebox{0.24\hsize}{!}{\includegraphics[angle=0]{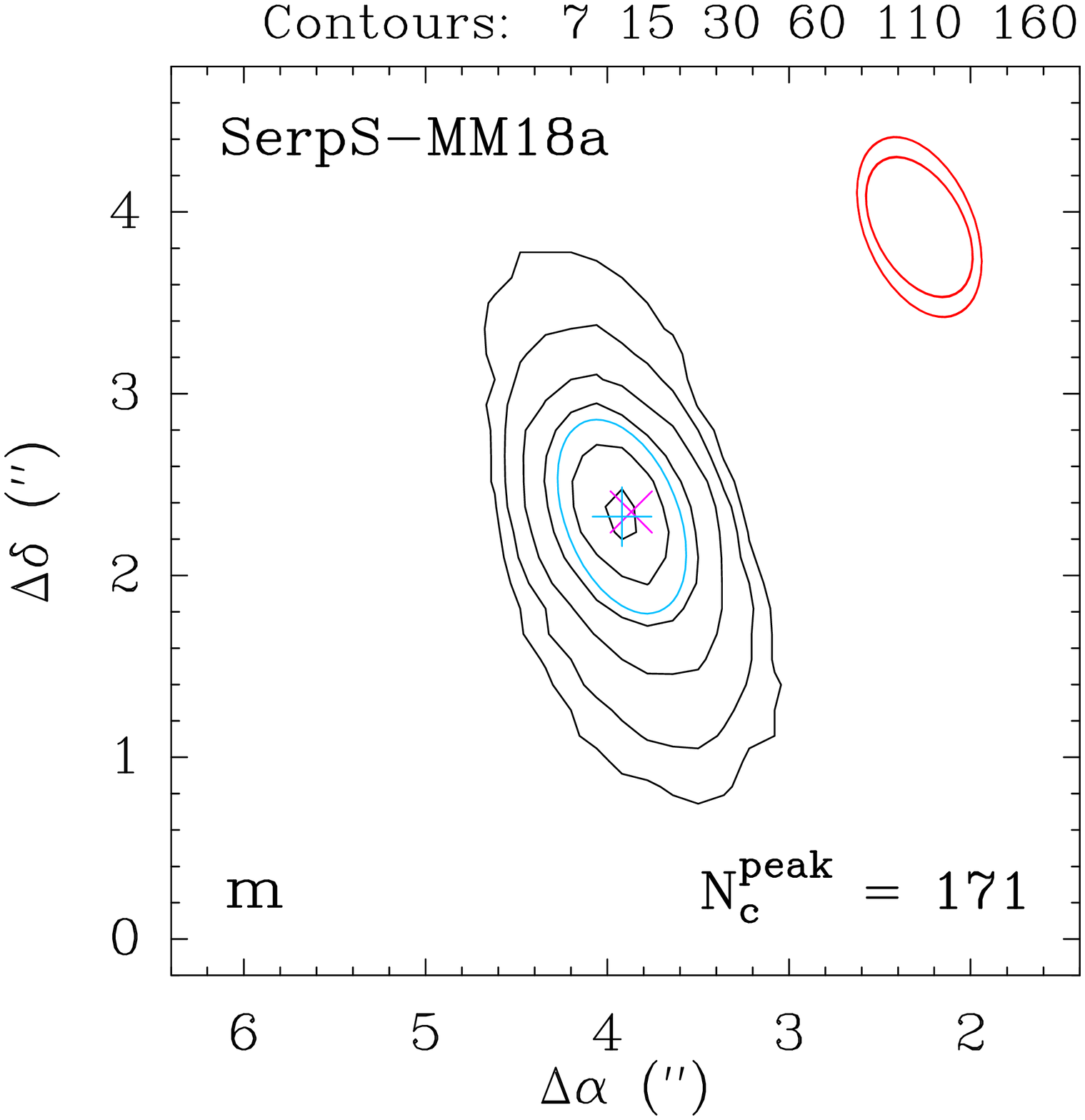}}\resizebox{0.24\hsize}{!}{\includegraphics[angle=0]{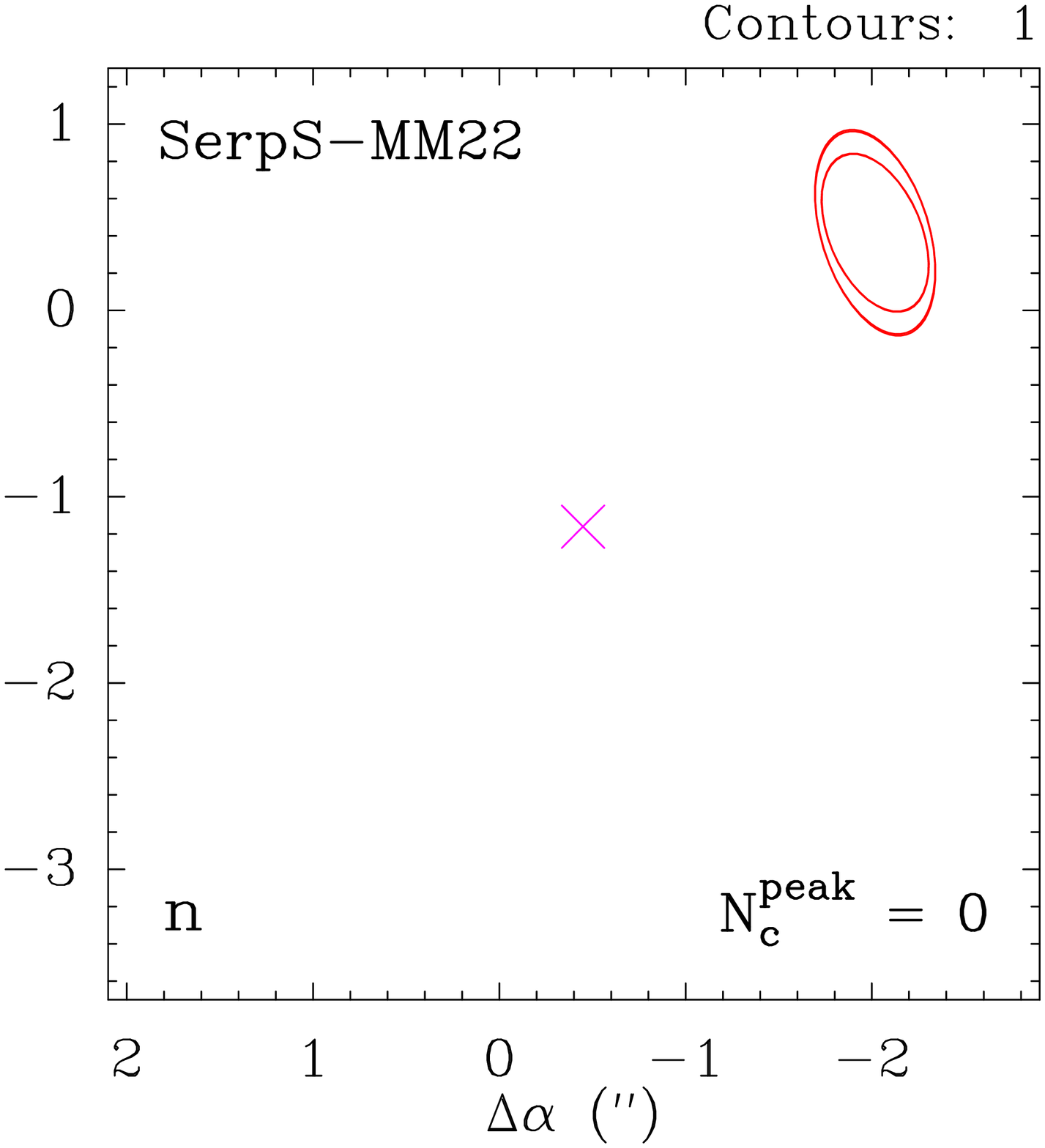}}\resizebox{0.24\hsize}{!}{\includegraphics[angle=0]{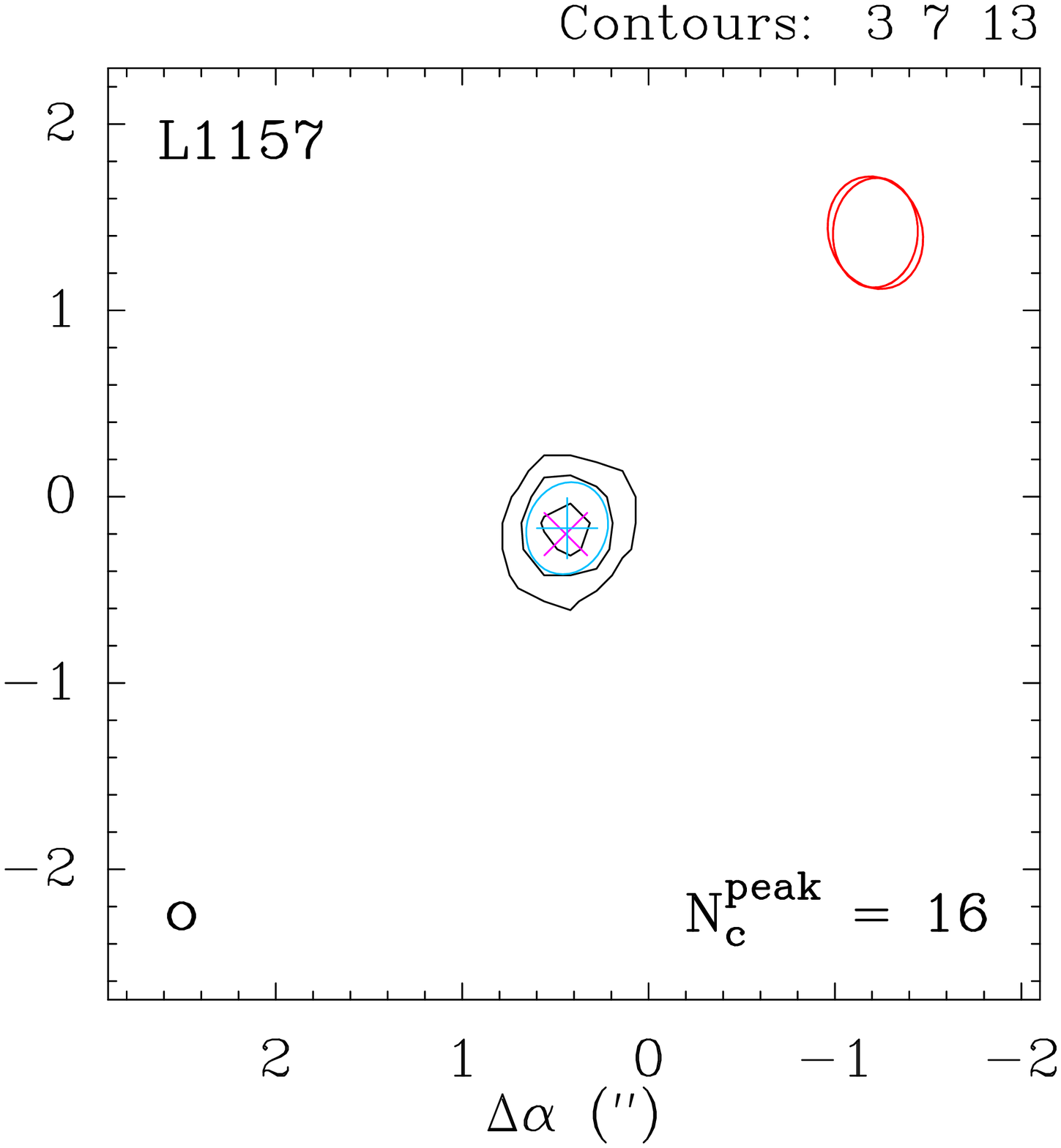}}\resizebox{0.24\hsize}{!}{\includegraphics[angle=0]{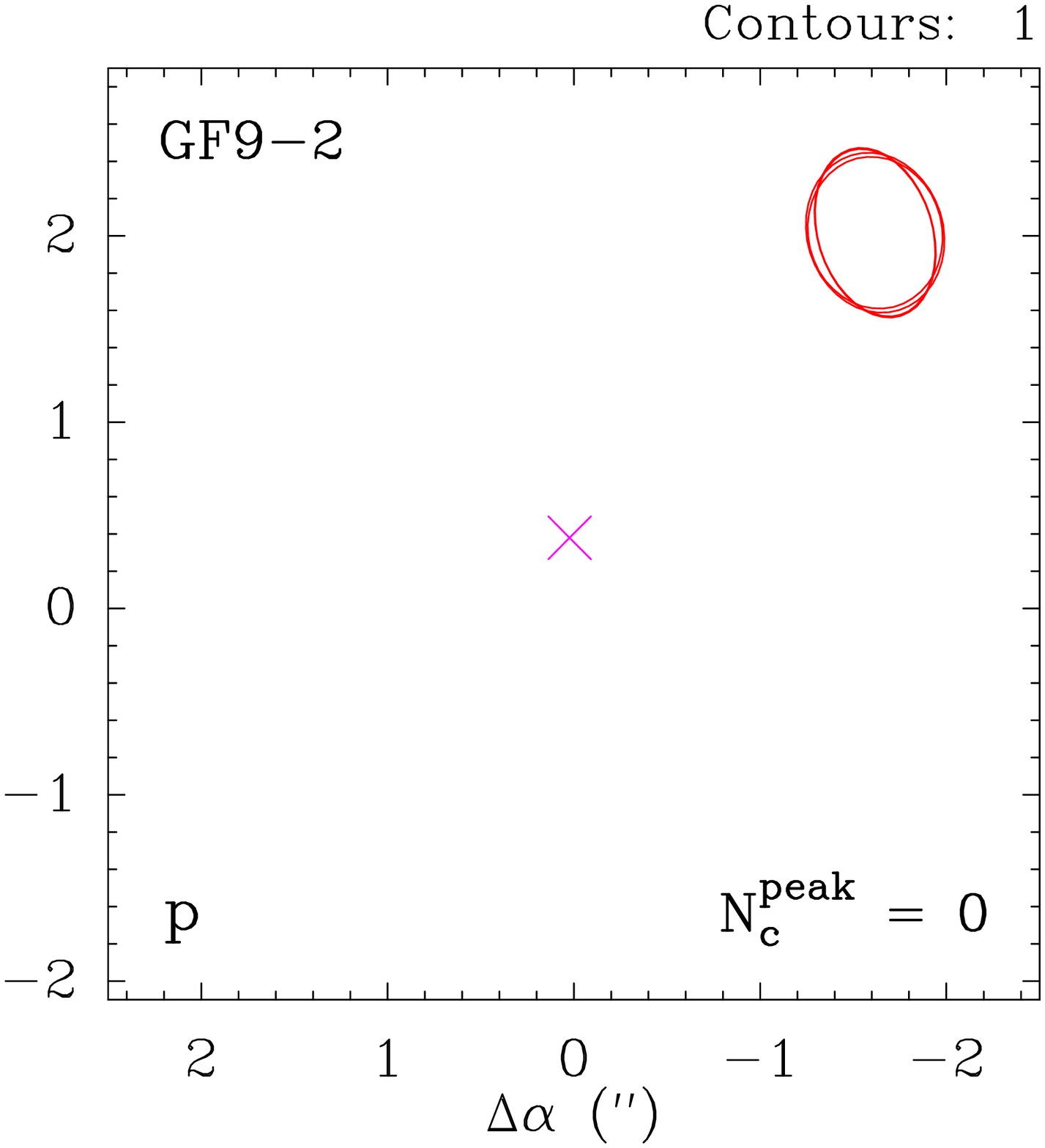}}}
 \caption{Maps of number of channels with signal detected above 6 times the 
rms noise level in the continuum-removed WideX spectra at 1.3~mm and 1.4~mm, 
excluding CO~2--1, $^{13}$CO~2--1, C$^{18}$O 2--1, SiO~5--4, SO~5$_6$--4$_5$, 
OCS 18--17, and OCS 19--18. In each panel, the red ellipses show the 
synthesized beam sizes at 219~GHz and 231~GHz. The pink crosses indicate the 
positions of the continuum emission peaks derived by \citet{Maury19}, and the 
orange crosses in panel e the positions of the binary components VLA4A and 
VLA4B of SVS13A determined by 
\citet{Lefevre17}. The peak count is given in the bottom right corner. For 
sources with a peak count higher than 9, the blue cross and ellipse show the 
position of the peak and the width ($FWHM$) of a Gaussian fit to the map, 
respectively. The contour levels are indicated above each panel. The 
coordinates at the origin are listed in Table~\ref{t:maporigins}. Maps over a 
larger field of view can be found in Fig.~\ref{f:channelcounts_large}.}
 \label{f:channelcounts}
\end{figure*}

We constructed maps of channel counts in order to search in a systematic way 
for hot-corino-like emission in each source of the sample. In the 
continuum-subtracted spectrum of 
each position in the field of view of a source, we counted the channels that 
have a flux density higher than six times the rms noise 
level. The noise level was derived as the median of the dispersions of the 
intensity distributions of all channel maps. This value was obtained with the 
task ``go noise'' of the GREG 
software\footnote{\label{fn:gildas}See http://www.iram.fr/IRAMFR/GILDAS.}.
Because the S3 data cubes have a worse angular resolution, the counting was 
performed on the S2 and S1 data cubes only. In order to avoid counting 
contributions from transitions of diatomic or triatomic molecules that may be 
dominated by emission produced by outflows and/or molecular jets, we excluded 
the following frequency ranges when counting the channels:
230.400--230.680~GHz (CO~2--1), 220.340--220.460~GHz ($^{13}$CO~2--1), 
219.550--219.570~GHz (C$^{18}$O 2--1), 217.040--217.205~GHz (SiO~5--4), 
219.890--220.045~GHz (SO~5$_6$--4$_5$), 218.890--218.920~GHz (OCS 18--17), and 
231.045--231.075~GHz (OCS 19--18). The excluded frequency ranges cover 0.8~GHz 
in total, meaning that the channel counting was performed over 6.4~GHz only. 
The resulting maps are shown in Fig.~\ref{f:channelcounts} and the values of 
the strongest peaks are listed in Table~\ref{t:counts}. Maps over a larger
field of view are displayed in Fig.~\ref{f:channelcounts_large}. Given that 
this article is later focused on COMs, we would like to emphasize that 
molecules such as DCN, c-C$_3$H$_2$, and H$_2$CO may contribute, depending on
the source, with a handful of channels to the maps shown in 
Figs.~\ref{f:channelcounts} and \ref{f:channelcounts_large}. 

\begin{table}
 \begin{center}
 \caption{Statistics of line detections in line-rich CALYPSO sources.}
 \label{t:counts}
 \vspace*{0.0ex}
 \begin{tabular}{lrcrr}
 \hline\hline
 \multicolumn{1}{c}{Source} & \multicolumn{1}{c}{$N_{\rm c}^{\rm peak}$\tablefootmark{a}} & \multicolumn{1}{c}{$N_{\rm l}^{\rm peak}$\tablefootmark{b}} &  \multicolumn{1}{c}{$\frac{N_{\rm c}^{\rm peak}}{N_{\rm l}^{\rm peak}}$\tablefootmark{c}} & \multicolumn{1}{c}{$n_{\rm l}^{\rm peak}$\tablefootmark{d}} \\ 
  & & & & \multicolumn{1}{c}{\small (GHz$^{-1}$)} \\ 
 \hline
 L1448-C         &  90 &  30 & 3.0 &  5 \\  
 IRAS2A          & 250 &  99 & 2.5 & 15 \\  
 SVS13A          & 394 & 142 & 2.8 & 22 \\  
 IRAS4A2         & 423 & 216 & 2.0 & 34 \\  
 IRAS4B          & 139 &  89 & 1.6 & 14 \\  
 SerpS-MM18a     & 171 &  70 & 2.4 & 11 \\  
\hline
 \end{tabular}
 \end{center}
 \vspace*{-2.5ex}
 \tablefoot{
 \tablefoottext{a}{Peak number of channels with a continuum-subtracted flux 
density above $6\sigma$ in setups S1 and S2, excluding channels covering 
CO~2--1, $^{13}$CO~2--1, C$^{18}$O 2--1, SiO~5--4, SO~5$_6$--4$_5$, OCS 18--17, 
and OCS 19--18. This value corresponds to the peak of the contour map shown in 
Fig.~\ref{f:channelcounts}. The total bandwidth is 6.4~GHz.}
 \tablefoottext{b}{Number of spectral lines with a peak flux density 
above $6\sigma$ in setups S1 and S2 toward the continuum peak, excluding 
the transitions listed above.}
 \tablefoottext{c}{Average number of channels with a flux density higher than
$6\sigma$ per detected line with a peak flux density above the same threshold.}
 \tablefoottext{d}{Spectral density of lines detected above $6\sigma$ toward 
the continuum peak.}
 }
 \end{table}

The inspection of Fig.~\ref{f:channelcounts} reveals six sources with a clear 
peak of channel counts associated with one of the continuum peaks, and a value 
at the peak, $N_{\rm c}^{\rm peak}$, higher than 20: 
SerpS-MM18a ($N_{\rm c}^{\rm peak} = 171$), L1448-C (90), IRAS2A1 (250), 
IRAS4A2 (423), IRAS4B (139), and SVS13A (394). All other CALYPSO sources have 
maximum channel counts below 20. We also counted by eye 
the number of spectral lines detected with a peak flux density above $6\sigma$ 
toward the continuum peak positions of the six line-rich sources, excluding 
the transitions listed in the previous paragraph. These line counts translate 
into spectral line densities between 5 lines per GHz for L1448-C and 34 lines 
per GHz for IRAS4A2 (Table~\ref{t:counts}). From the channel and line 
counts listed in Table~\ref{t:counts}, we deduce that the lines detected with 
a peak flux density above 6$\sigma$ have flux densities above this threshold 
over 2--3 channels on average ($\sim$5--8~km~s$^{-1}$). The maps of channel 
counts displayed in 
Fig.~\ref{f:channelcounts} can thus be roughly converted into maps of spectral 
line counts by dividing them with this average number of channel counts per 
line (fourth column of Table~\ref{t:counts}).

\subsection{Associations between channel count peaks and continuum sources, and
relation to outflows}

\begin{table*}[!ht]
 \begin{center}
 \caption{
 Coordinates of channel count peaks and angular distance to the nearest continuum peak, for the sources with a maximum channel count higher than 9.
}
 \label{t:lcountoffset}
 \vspace*{-1.2ex}
 \begin{tabular}{lcccccccccccc}
 \hline\hline
 \multicolumn{1}{c}{Source\tablefootmark{a}} & \multicolumn{2}{c}{Channel count peak\tablefootmark{b}} & \multicolumn{2}{c}{Cont. peak\tablefootmark{c}} & \multicolumn{1}{c}{$\Delta\alpha$\tablefootmark{d}} & \multicolumn{1}{c}{$\Delta\delta$\tablefootmark{d}} & \multicolumn{1}{c}{$D_{\rm c}$\tablefootmark{e}} & \multicolumn{1}{c}{$PA_{\rm c}$\tablefootmark{f}} & \multicolumn{1}{c}{$d_{\rm b}$\tablefootmark{g}} & \multicolumn{1}{c}{$\frac{D_{\rm c}}{d_{\rm b}}$} & \multicolumn{2}{c}{$PA_{\rm j}$\tablefootmark{h}} \\ 
  & \multicolumn{1}{c}{\hspace*{-1ex} $\alpha$ (hh:mm:ss)} & \multicolumn{1}{c}{\hspace*{-2ex} $\delta$ (dd:mm:ss)} & \multicolumn{1}{c}{$\alpha$ (ss)} & \multicolumn{1}{c}{$\delta$ (ss)} & \multicolumn{1}{c}{$\arcsec$} & \multicolumn{1}{c}{$\arcsec$} & \multicolumn{1}{c}{$\arcsec$} & \multicolumn{1}{c}{$^\circ$} & \multicolumn{1}{c}{$\arcsec$} &  & \multicolumn{1}{c}{$^\circ$} & \multicolumn{1}{c}{$^\circ$} \\ 
 \multicolumn{1}{c}{(1)} & \multicolumn{1}{c}{(2)} & \multicolumn{1}{c}{(3)} & \multicolumn{1}{c}{(4)} & \multicolumn{1}{c}{(5)} & \multicolumn{1}{c}{(6)} & \multicolumn{1}{c}{(7)} & \multicolumn{1}{c}{(8)} & \multicolumn{1}{c}{(9)} & \multicolumn{1}{c}{(10)} & \multicolumn{1}{c}{(11)} & \multicolumn{1}{c}{(12)} & \multicolumn{1}{c}{(13)} \\ 
 \hline
 L1448-2A & \hspace*{-1ex} 03:25:22.382 & \hspace*{-2ex} $$30:45:13.31 & 22.405 & 13.26 & $ -0.29$ & $  0.05$ &  0.30 & $-79$ &  0.90 & 0.33 & $ 135$ & $ -45$ \\ 
 L1448-NB2 & \hspace*{-1ex} 03:25:36.349 & \hspace*{-2ex} $$30:45:15.05 & 36.315 & 15.15 & $  0.44$ & $ -0.10$ &  0.45 & $102$ &  1.00 & 0.45 & -- & -- \\ 
 L1448-C & \hspace*{-1ex} 03:25:38.873 & \hspace*{-2ex} $$30:44:05.28 & 38.875 & 05.33 & $ -0.03$ & $ -0.05$ &  0.06 & $-151$ &  1.12 & 0.05 & $ -17$ & $ 163$ \\ 
 IRAS2A1 & \hspace*{-1ex} 03:28:55.575 & \hspace*{-2ex} $$31:14:37.04 & 55.570 & 37.07 & $  0.06$ & $ -0.03$ &  0.07 & $113$ &  0.94 & 0.07 & $ 205$ & $  25$ \\ 
 SVS13A & \hspace*{-1ex} 03:29:03.748 & \hspace*{-2ex} $$31:16:03.83 & 03.756 & 03.80 & $ -0.11$ & $  0.03$ &  0.11 & $-75$ &  0.43 & 0.26 & $ 155$ & $ -25$ \\ 
 IRAS4A2 & \hspace*{-1ex} 03:29:10.427 & \hspace*{-2ex} $$31:13:32.12 & 10.432 & 32.12 & $ -0.07$ & $ -0.00$ &  0.07 & $-94$ &  0.51 & 0.13 & $ 182$ & $   0$ \\ 
 IRAS4B & \hspace*{-1ex} 03:29:12.008 & \hspace*{-2ex} $$31:13:08.04 & 12.016 & 08.02 & $ -0.10$ & $  0.02$ &  0.10 & $-79$ &  0.52 & 0.20 & $ 167$ & $ -13$ \\ 
 SerpM-SMM4b & \hspace*{-1ex} 18:29:56.517 & \hspace*{-2ex} $$01:13:11.72 & 56.525 & 11.58 & $ -0.13$ & $  0.14$ &  0.18 & $-43$ &  0.64 & 0.29 & $  10$ & $ 145$ \\ 
 SerpS-MM18a & \hspace*{-1ex} 18:30:04.121 & \hspace*{-2ex} $-$02:03:02.58 & 04.118 & 02.55 & $  0.05$ & $ -0.03$ &  0.06 & $116$ &  0.55 & 0.11 & $ 188$ & $   8$ \\ 
 L1157 & \hspace*{-1ex} 20:39:06.268 & \hspace*{-2ex} $$68:02:15.73 & 06.269 & 15.70 & $ -0.01$ & $  0.03$ &  0.03 & $-12$ &  0.59 & 0.06 & $ 163$ & $ -17$ \\ 
 \hline
 \end{tabular}
 \end{center}
 \vspace*{-2.5ex}
 \tablefoot{
 \tablefoottext{a}{Continuum source nearest to channel count peak.}
 \tablefoottext{b}{J2000 equatorial coordinates of channel count peak obtained from Gaussian fit in Fig.~\ref{f:channelcounts}.}
 \tablefoottext{c}{J2000 equatorial coordinates of nearest continuum peak from \citet{Maury19}. The hours, minutes, degrees, and arcminutes are not displayed. They are the same as in Cols. 2 and 3.}
 \tablefoottext{d}{Equatorial offsets of channel count peak with respect to nearest continuum peak.}
 \tablefoottext{e}{Angular distance between channel count peak and nearest continuum peak.}
 \tablefoottext{f}{Position angle of vector going from nearest continuum peak to channel count peak, counted east from north.}
 \tablefoottext{g}{Beam diameter ($HPBW$) along $PA_{\rm c}$. This is an average value for setups S1 and S2.}
 \tablefoottext{h}{Position angle of the blueshifted and redshifted lobes of the outflow/jet from the CALYPSO survey (Podio et al., in prep.), except for L1448-2A for which we take the tentative detection of \citet{Kwon19}.}
 }
 \end{table*}

The positions of the channel count peaks higher than 9 were derived from 
Gaussian fits to the maps shown in Fig.~\ref{f:channelcounts}. The results are
listed in Cols.~2 and 3 of Table~\ref{t:lcountoffset}. This table also gives 
the name (Col.~1) and coordinates (Cols.~4 and 5) of the nearest continuum 
source. As will be seen below (Sect.~\ref{ss:chemcomp}), COMs (at least 
methanol) are detected toward all sources listed in Table~\ref{t:lcountoffset} 
except for L1448-NB2. The lines contributing to the channel count toward 
L1448-NB2 are from H$_2$CO and c-C$_3$H$_2$.

The angular separation between the channel count peak and its nearest 
continuum source, and its ratio to the width ($HPBW$) of the beam along the 
same position angle are listed in Cols. 8 and 11, respectively, of 
Table~\ref{t:lcountoffset}. For most 
sources, the positions of the channel count peak and its nearest continuum 
source agree to better than one fifth of the beam width ($HPBW$). There are 
four exceptions: L1448-2A, L1448-NB2, SVS13A, and SerpM-SMM4b. For both 
L1448-2A and L1448-N, the contours of channel counts in 
Figs.~\ref{f:channelcounts}a and b enclose a close binary, and the peak is 
located at roughly equal distance from each binary component. In both cases, 
the binary separation is roughly one beam therefore we cannot exclude that a 
map of channel counts at higher angular resolution would reveal two peaks 
coinciding with the binary components. In the case of SerpM-SMM4b, the contour 
map of channel counts is asymmetric and its actual peak is located half-way 
between the peak derived from the Gaussian fit and the continuum peak, that is 
less than one fifth of the beam from the continuum peak (panel l of 
Fig.~\ref{f:channelcounts}). The offset may thus not be significant.

The only source for which there seems to be a significant offset between the
continuum peak and the channel count peak is SVS13A. The separation is 
0.11$\arcsec$, which corresponds to one fourth of the beam. 
However, SVS13A is a tight binary, with the components VLA4A and VLA4B located 
west and east of the CALYPSO continuum peak, respectively 
\citep[][see orange crosses in Fig.~\ref{f:channelcounts}e]{Lefevre17}. The 
channel count peak is located
at an angular distance of 0.08$\arcsec$ from VLA4A with a position angle of 
77$^\circ$, which is nearly perpendicular to the outflow axis. This offset is
less than one fifth of the beam. VLA4A thus seems to be associated with the 
peak of molecular richness in SVS13A, as already noted by \citet{Lefevre17}, 
and there is no obvious link between the channel count peak and the outflow.
Higher angular resolution ALMA data on SVS13A are currently being analyzed to 
clarify the distribution of COMs in this protostellar system (C. Lef\`evre,
priv. comm.).

\begin{figure}
 \centerline{\resizebox{1.0\hsize}{!}{\includegraphics[angle=0]{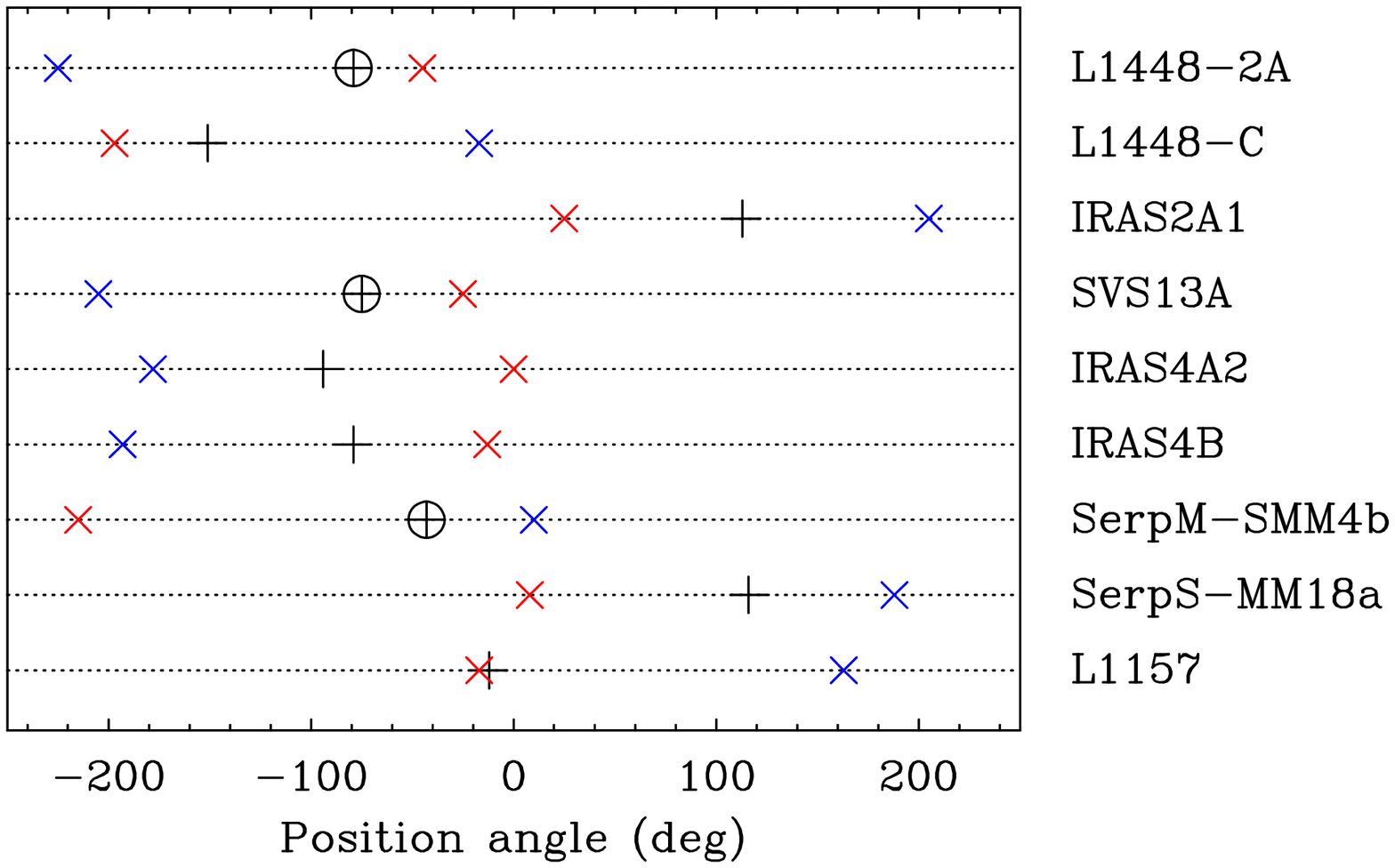}}}
 \caption{Comparison of the position angle $PA_{\rm c}$ of the vector going 
from the continuum peak to the channel count peak (black plus symbol) and the 
position angles of the blueshifted (blue cross) and redshifted (red cross) 
outflow lobes for the sources listed in Table~\ref{t:lcountoffset}, except for
L1448-NB2 which lacks outflow information. The black 
plus symbol is enclosed by a circle when the angular separation between the 
continuum peak and the channel count peak is larger than one fifth of the 
beam. For the other ones, $PA_{\rm c}$ may not be reliable.}
 \label{f:lcountpa}
\end{figure}

Figure~\ref{f:lcountpa} displays the position angle of the offset between the
channel count peak and its nearest continuum peak, along with the position 
angles of the blueshifted and redshifted lobes of the outflow for the 
sources listed in Table~\ref{t:lcountoffset}, except for L1448-NB2 which lacks
outflow information. There is no systematic alignment
of the channel count peaks with the direction of the outflows. For the
three sources with an offset larger than one fifth of the beam, that is the
ones with a measured offset a priori more reliable, the channel count peak 
tends to be close
to the outflow direction ($\sim$ 30--50$^\circ$). However, as discussed above, 
the position angle of the offset for each of these sources may not be 
meaningful. For L1448-2A, this is due to the presence of the close binary that 
is barely resolved in the channel count map. The second contour of 
Fig.~\ref{f:channelcounts}a actually suggests a channel count peak associated 
with each component. For SerpM-SMM4b, the asymmetry of the channel count 
contours leads to an overestimate of the offset. In the case of SVS13A, the 
channel counts peak very close to VLA4A, one of the tight binary components. 
If we take this component as reference, the offset becomes roughly aligned 
with the direction perpendicular to the outflow axis. On the basis of 
Fig.~\ref{f:lcountpa}, there is thus no obvious link between the channel 
count peaks and the outflows for the sample of sources with a maximum
channel count higher than 9.

\subsection{Sizes of the COM emission}
\label{ss:sizes}

For each source, we selected the spectral lines that have a peak flux density 
above $6\sigma$ in the spectrum of the continuum peak that has the highest 
number of detected lines among all continuum peaks in the field of view. We 
determined the size 
of the emission of each selected spectral line by fitting a two-dimensional
Gaussian to the emission of the peak channel in the $uv$ plane. The deconvolved 
sizes derived in this way are displayed in Figs.~\ref{f:sizes_l1448-2a} to 
\ref{f:sizes_gf9-2}. The uncertainties are significant, but the high number
of detected lines in some sources allows to estimate a 
representative size of their COM spectral line emission. The COM emission of 
the six line-rich CALYPSO sources has a size ($HPBW$) ranging between 
0.3$\arcsec$ for SVS13A and 0.5$\arcsec$ for L1448-C and SerpS-MM18a 
(Table~\ref{t:sizes}). The uncertainty is likely on the order of 0.1$\arcsec$. 
For the sources listed in Table~\ref{t:sizes} but not
shown in Figs.~\ref{f:sizes_l1448-2a} to \ref{f:sizes_gf9-2}, we assumed
by default the same source size as (one of) the other source(s) present in the 
field of view.

\begin{table}
 \begin{center}
 \caption{Representative or assumed size of spectral line emission in CALYPSO 
sources.}
 \label{t:sizes}
 \vspace*{0.0ex}
 \begin{tabular}{llrc}
 \hline\hline
 \multicolumn{1}{c}{Source} & \multicolumn{2}{c}{Size\tablefootmark{a}} & \multicolumn{1}{c}{Filter\tablefootmark{b}}  \\ 
  & \multicolumn{1}{c}{\small ($\arcsec$)} & \multicolumn{1}{c}{\small (au)} & \multicolumn{1}{c}{\small ($\arcsec$)} \\ 
 \hline
L1448-2A      & 1.0  & 290 & 0.8 \\
L1448-2Ab     & 1.0\tablefootmark{c}  & 290 & -- \\
L1448-NA      & 1.0  & 290 & 1.2 \\
L1448-NB1     & 1.0\tablefootmark{c}  & 290 & -- \\
L1448-NB2     & 1.0\tablefootmark{c}  & 290 & -- \\
L1448-C       & 0.5  & 150 & 0.8 \\
L1448-CS      & 0.5\tablefootmark{c}  & 150 & -- \\
IRAS2A1       & 0.35 & 100 & 0.8 \\
SVS13B        & 0.5  & 150 & 0.8 \\
SVS13A        & 0.3  & 90 & 0.8 \\
IRAS4A1       & 0.35\tablefootmark{c} & 100 & -- \\
IRAS4A2       & 0.35 & 100 & 0.6 \\
IRAS4B        & 0.4  & 120 & 0.8 \\
IRAS4B2       & 0.4\tablefootmark{c}  & 120 & -- \\
IRAM04191     & 2.0  & 280 & 1.0 \\
L1521F        & 0.5  & 70 & 1.0 \\
L1527         & 1.0  & 140 & 0.8 \\
SerpM-S68N    & 0.5  & 220 & 0.8 \\
SerpM-S68Nb   & 0.5\tablefootmark{c}  & 220 & -- \\
SerpM-SMM4a   & 0.5\tablefootmark{c}  & 220 & -- \\
SerpM-SMM4b   & 0.5  & 220 & 0.8 \\
SerpS-MM18a   & 0.5  & 180 & 0.8 \\
SerpS-MM18b   & 0.5\tablefootmark{c}  & 180 & -- \\
SerpS-MM22    & 1.0  & 350 & 0.8 \\
L1157         & 0.25 & 90 & 0.8 \\
GF9-2         & 1.0  & 470 & 0.8 \\
\hline
 \end{tabular}
 \end{center}
 \vspace*{-2.5ex}
 \tablefoot{
 \tablefoottext{a}{Representative size ($HPBW$) of the spectral line emission. 
The value corresponds to the level of the dashed line in 
Figs.~\ref{f:sizes_l1448-2a} to \ref{f:sizes_gf9-2} for the sources shown in 
these figures.}
\tablefoottext{b}{Only the lines that have a peak offset from the continuum 
peak by less than this value were taken into account to estimate the emission 
size.}
\tablefoottext{c}{Assumed value.} }
 \end{table}

\subsection{Elongations of the COM emission}
\label{ss:elongations}

\begin{figure}
 \centerline{\resizebox{1.0\hsize}{!}{\includegraphics[angle=0]{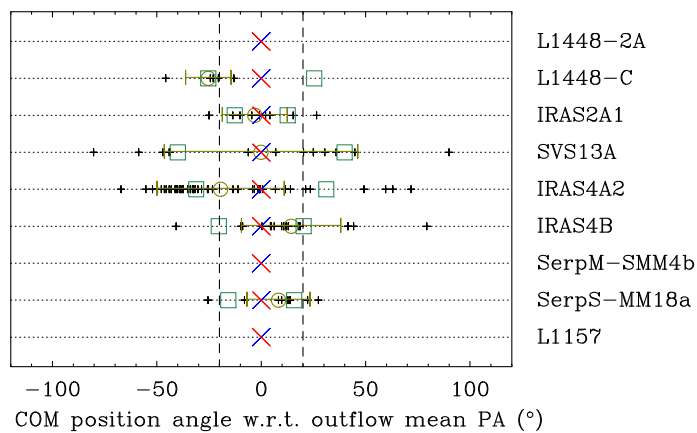}}}
 \caption{Position angles of the COM emission elongation (black crosses) with 
respect to the mean position angle of the outflow lobes (bicolor cross) for 
the sources listed in Table~\ref{t:lcountoffset}. Only the transitions with an 
error on the position angle smaller than 10$^\circ$ were selected. The 
unidentified transitions were not used and the following molecules 
with less than five atoms were 
ignored: CO, $^{13}$CO, C$^{18}$O, SO, SO$_2$, H$_2$CO, H$_2$$^{13}$CO, D$_2$CO, 
OCS, O$^{13}$CS, SiO, DCN, C$^{13}$S, and HNCO. L1448-2A, SerpM-SMM4b, and 
L1157 have no transitions fulfilling the selection criteria. For each source, 
the olive circle and the two green squares indicate the angles 
$\Delta PA_{\rm mean}$ and $\pm$$|\Delta PA|_{\rm mean}$ with
$\Delta PA_{\rm mean}$ and $|\Delta PA|_{\rm mean}$ the mean deviation and mean 
absolute deviation, respectively, of the COM position angles with respect to 
the mean position angle of the outflow. The dispersion around the mean 
deviation is also displayed. The vertical dashed lines mark a deviation of 
$\pm$20$^\circ$ from the mean position angle of the outflow. Only 
IRAS2A1 and SerpS-MM18a fall within these limits with a small dispersion.}
 \label{f:uvfitpa}
\end{figure}

We used the two-dimensional Gaussian fits performed in Sect.~\ref{ss:sizes} to
search for a correlation between the position angle of the ellipses fitted to
the COM emission and the position angles of the outflows. Here we included also
molecules with five atoms (CH$_2$CO, HC$_3$N, c-C$_3$H$_2$, NH$_2$CN, and 
\textit{t}-HCOOH). Figure~\ref{f:uvfitpa} displays the fit results for the 
1.3~mm and 1.4~mm transitions that have an error on the fitted position angle 
smaller than 10$^\circ$. The molecules with less than five atoms are ignored 
for this investigation, as well as the unidentified transitions. The mean 
absolute deviation of the fitted COM position angles with respect to the 
mean position angle of the outflow lobes is less than 20$^\circ$ for two out of 
six sources, IRAS2A1 and SerpS-MM18a. This is also true for the mean algebraic 
(signed) deviations which are $-3^\circ$ and $8^\circ$, respectively, both with a 
small dispersion of $\sim$$15^\circ$. SVS13A and IRAS4A2 have a mean algebraic 
deviation smaller than 20$^\circ$ as well, but with large dispersions 
($\sim$$50^\circ$ and $\sim$$30^\circ$, respectively) so we cannot conclude that
the COM emission is preferentially elongated along the direction of their 
outflows. However, most 
(80\%) transitions fitted for IRAS4B fall within $\pm$20$^{\circ}$ of the mean 
position angle of the outflow and the mean deviation is $14^\circ \pm 24^\circ$ 
so we can consider that IRAS4B behaves like IRAS2A1 and SerpS-MM18a as well. 
It is thus tempting to conclude that the COM emission is preferentially 
elongated along the direction of the outflow in IRAS2A1, IRAS4B, and 
SerpS-MM18a.

\subsection{Spectral line modeling}
\label{ss:modeling}

\longtab{
\begin{longtable}{llrr@{\mbox{ }}l}
 \caption{\label{t:popfit}
 Rotational temperatures derived from population diagrams of selected (complex) organic molecules toward CALYPSO sources.
} \\ 
 \hline\hline
 \multicolumn{1}{c}{Molecule} & \multicolumn{1}{c}{States\tablefootmark{a}} & \multicolumn{1}{c}{$N_{\rm pop}$\tablefootmark{b}} & \multicolumn{2}{c}{$T_{\rm fit}$\tablefootmark{c}} \\ 
  & & & \multicolumn{2}{c}{(K)} \\ 
 \hline
 \endfirsthead
 \caption{continued.} \\ 
 \hline\hline
 \multicolumn{1}{c}{Molecule} & \multicolumn{1}{c}{States\tablefootmark{a}} & \multicolumn{1}{c}{$N_{\rm pop}$\tablefootmark{b}} & \multicolumn{2}{c}{$T_{\rm fit}$\tablefootmark{c}} \\ 
  & & & \multicolumn{2}{c}{(K)} \\ 
 \hline
 \endhead
 \hline
 \endfoot
\multicolumn{5}{c}{L1448-2A} \\ 
CH$_3$OH & $\varv=0$ & 8 &   160 & (108) \\ 
\hline 
\multicolumn{5}{c}{L1448-2Ab} \\ 
CH$_3$OH & $\varv=0$ & 10 &   970 & (3508) \\ 
HNCO & $\varv=0$ & 2 & \multicolumn{2}{c}{--} \\ 
\hline 
\multicolumn{5}{c}{L1448-C} \\ 
CH$_3$OH & $\varv=0$, $\varv_{\rm t}=1$ & 13 & \textbf{  107} & (11) \\ 
CH$_3$OCH$_3$ & $\varv=0$ & 3 & \textbf{   75} & ( 5) \\ 
CH$_3$OCHO & $\varv=0$ & 7 &    39 & (27) \\ 
CH$_3$CHO & $\varv=0$ & 13 & \textbf{   77} & (21) \\ 
NH$_2$CHO & $\varv=0$ & 2 & \multicolumn{2}{c}{--} \\ 
CH$_3$CN & $\varv=0$ & 4 & \textbf{  105} & (23) \\ 
HNCO & $\varv=0$ & 3 & \textbf{   59} & ( 3) \\ 
\hline 
\multicolumn{5}{c}{IRAS2A1} \\ 
CH$_3$OH & $\varv=0$, $\varv_{\rm t}=1$ & 24 & \textbf{  229} & (21) \\ 
C$_2$H$_5$OH & $\varv=0$ & 17 &   703 & (890) \\ 
CH$_3$OCH$_3$ & $\varv=0$ & 6 & \textbf{  113} & ( 9) \\ 
CH$_3$OCHO & $\varv=0$, $\varv_{\rm t}=1$ & 19 & \textbf{  128} & (16) \\ 
CH$_3$CHO & $\varv=0$, $\varv=1$ & 17 & \textbf{  161} & (36) \\ 
NH$_2$CHO & $\varv=0$, $\varv_{12}=1$ & 4 &   404 & (163) \\ 
CH$_3$CN & $\varv=0$, $\varv_8=1$ & 5 & \textbf{  161} & (17) \\ 
(CH$_2$OH)$_2$ & $\varv=0$ & 20 &   173 & (86) \\ 
HNCO & $\varv=0$ & 5 & \textbf{  254} & (22) \\ 
NH$_2$CN & $\varv=0$ & 7 &   200 & (108) \\ 
\hline 
\multicolumn{5}{c}{SVS13A} \\ 
CH$_3$OH & $\varv=0$, $\varv_{\rm t}=1$ & 23 &   285 & (82) \\ 
C$_2$H$_5$OH & $\varv=0$ & 26 &   274 & (115) \\ 
CH$_3$OCH$_3$ & $\varv=0$, $\varv_{11}=1$ & 9 & \textbf{   94} & ( 7) \\ 
CH$_3$OCHO & $\varv=0$, $\varv_{\rm t}=1$, $\varv_{\rm t}=2$ & 28 & \textbf{  109} & (12) \\ 
CH$_3$CHO & $\varv=0$, $\varv=1$ & 26 & \textbf{  113} & (14) \\ 
NH$_2$CHO & $\varv=0$, $\varv_{12}=1$ & 4 &   351 & (117) \\ 
CH$_3$CN & $\varv=0$, $\varv_8=1$ & 6 & \textbf{  222} & (49) \\ 
C$_2$H$_5$CN & $\varv=0$ & 6 &    33 & (16) \\ 
(CH$_2$OH)$_2$ & $\varv=0$ & 24 & \textbf{   98} & (19) \\ 
CH$_2$(OH)CHO & $\varv=0$ & 7 & \textbf{  106} & (24) \\ 
HNCO & $\varv=0$ & 5 & \textbf{  233} & (17) \\ 
NH$_2$CN & $\varv=0$ & 8 &   317 & (136) \\ 
\hline 
\multicolumn{5}{c}{IRAS4A2} \\ 
CH$_3$OH & $\varv=0$, $\varv_{\rm t}=1$ & 24 & \textbf{  307} & (56) \\ 
C$_2$H$_5$OH & $\varv=0$ & 25 & \textbf{  127} & (33) \\ 
CH$_3$OCH$_3$ & $\varv=0$, $\varv_{11}=1$, $\varv_{15}=1$ & 10 & \textbf{  134} & ( 6) \\ 
CH$_3$OCHO & $\varv=0$, $\varv_{\rm t}=1$, $\varv_{\rm t}=2$ & 34 & \textbf{  150} & (11) \\ 
CH$_3$CHO & $\varv=0$, $\varv=1$, $\varv=2$ & 54 & \textbf{  220} & (18) \\ 
NH$_2$CHO & $\varv=0$, $\varv_{12}=1$ & 4 &   536 & (239) \\ 
CH$_3$CN & $\varv=0$ & 5 & \textbf{  142} & (13) \\ 
C$_2$H$_5$CN & $\varv=0$ & 6 & \multicolumn{2}{c}{--} \\ 
(CH$_2$OH)$_2$ & $\varv=0$ & 28 &   126 & (47) \\ 
CH$_2$(OH)CHO & $\varv=0$ & 15 &   196 & (67) \\ 
HNCO & $\varv=0$ & 5 & \textbf{  280} & (56) \\ 
NH$_2$CN & $\varv=0$ & 10 &   280 & (171) \\ 
\hline 
\multicolumn{5}{c}{IRAS4B} \\ 
CH$_3$OH & $\varv=0$, $\varv_{\rm t}=1$ & 18 & \textbf{  305} & (52) \\ 
C$_2$H$_5$OH & $\varv=0$ & 12 &   238 & (123) \\ 
CH$_3$OCH$_3$ & $\varv=0$ & 3 &    95 & (28) \\ 
CH$_3$OCHO & $\varv=0$, $\varv_{\rm t}=1$ & 14 &   328 & (110) \\ 
CH$_3$CHO & $\varv=0$, $\varv=1$, $\varv=2$ & 46 & \textbf{  318} & (53) \\ 
NH$_2$CHO & $\varv=0$ & 2 & \multicolumn{2}{c}{--} \\ 
CH$_3$CN & $\varv=0$ & 3 & \multicolumn{2}{c}{--} \\ 
C$_2$H$_5$CN & $\varv=0$ & 4 &    14 & ( 6) \\ 
CH$_2$(OH)CHO & $\varv=0$ & 8 &   180 & (74) \\ 
HNCO & $\varv=0$ & 6 & \textbf{  379} & (62) \\ 
\hline 
\multicolumn{5}{c}{SerpM-S68N} \\ 
CH$_3$OH & $\varv=0$, $\varv_{\rm t}=1$ & 9 & \textbf{  210} & (37) \\ 
CH$_3$OCHO & $\varv=0$ & 7 &    10 & ( 3) \\ 
CH$_3$CN & $\varv=0$ & 3 & \multicolumn{2}{c}{--} \\ 
\hline 
\multicolumn{5}{c}{SerpM-SMM4b} \\ 
CH$_3$OH & $\varv=0$, $\varv_{\rm t}=1$ & 16 & \textbf{  223} & (46) \\ 
HNCO & $\varv=0$ & 3 &   383 & (267) \\ 
\hline 
\multicolumn{5}{c}{SerpS-MM18a} \\ 
CH$_3$OH & $\varv=0$, $\varv_{\rm t}=1$ & 18 & \textbf{  154} & (22) \\ 
C$_2$H$_5$OH & $\varv=0$ & 9 &   238 & (202) \\ 
CH$_3$OCH$_3$ & $\varv=0$ & 4 & \textbf{  106} & (15) \\ 
CH$_3$OCHO & $\varv=0$, $\varv_{\rm t}=1$ & 9 &   171 & (65) \\ 
CH$_3$CHO & $\varv=0$ & 12 &    78 & (41) \\ 
NH$_2$CHO & $\varv=0$ & 2 &    80 & (41) \\ 
CH$_3$CN & $\varv=0$, $\varv_8=1$ & 6 & \textbf{  244} & (28) \\ 
C$_2$H$_5$CN & $\varv=0$ & 6 &    33 & (12) \\ 
HNCO & $\varv=0$ & 5 & \textbf{  178} & (25) \\ 
\hline 
\multicolumn{5}{c}{SerpS-MM18b} \\ 
CH$_3$OH & $\varv=0$ & 5 & \textbf{   69} & (10) \\ 
CH$_3$CN & $\varv=0$ & 2 &    17 & (18) \\ 
HNCO & $\varv=0$ & 2 &    12 & (28) \\ 
\hline 
\multicolumn{5}{c}{L1157} \\ 
CH$_3$OH & $\varv=0$ & 11 & \textbf{  199} & (50) \\ 
CH$_3$OCHO & $\varv=0$ & 2 &     8 & ( 4) \\ 
CH$_3$CN & $\varv=0$ & 3 &   274 & (536) \\ 
\hline 
 \end{longtable}
 \tablefoot{
 \tablefoottext{a}{Vibrational or torsional states that were taken into account to fit the population diagram.}
 \tablefoottext{b}{Number of detected lines displayed in the population diagram and used to fit the rotational temperature.}
 \tablefoottext{c}{The standard deviation of the fit is given in parentheses. These uncertainties are purely statistical and should be viewed with caution. They may be underestimated. Rotational temperatures with a signal-to-noise ratio higher than 3.5 are indicated in bold face. Negative temperatures or temperatures above 1500~K, likely the result of high uncertainties or line contamination by unidentified species, are not shown and are replaced with a dash.}
 }
 }

We modeled the spectral line emission detected toward the continuum peaks of
the CALYPSO sources assuming local thermodynamic equilibrium (LTE) with the 
Weeds package \citep[][]{Maret11} of the CLASS software (see 
footnote~\ref{fn:gildas}). Weeds takes into account the finite angular
resolution of the telescope and the opacity of the transitions. We corrected
the spectra for primary beam attenuation. We followed
the same modeling strategy as applied for Sgr~B2 by \citet{Belloche16}. For 
each continuum peak position, each species is modeled separately and its 
synthetic spectrum is computed over the same frequency ranges as covered by 
setups S1 to S3. Each species is modeled with five free parameters: source 
size, column density, rotational temperature, linewidth, and velocity offset 
with respect to the systemic velocity of the source. For most 
species, we assumed the (Gaussian) source sizes listed in Table~\ref{t:sizes}.

After a first iteration of modeling all identified molecules, we constructed
population diagrams of the detected complex organic molecules following the
method described in Sect. 3 of \citet{Belloche16}. The diagrams
are corrected for optical depth and contamination from species included in the 
full model that have overlapping transitions. These diagrams are presented in 
Figs.~\ref{f:popdiag_l1448-2a_p1_ch3oh}--\ref{f:popdiag_l1157_p1_ch3cn}.
For each molecule, the rotational temperature is derived from a linear fit to 
the corrected population diagram in linear-logarithmic space. The derived 
temperatures are listed in Table~\ref{t:popfit} 
for all sources and positions 
that are detected in several transitions of, at least, methanol. They are also 
plotted in a synthetic way for each source and position in 
Figs.~\ref{f:trot_l1448-2a_p1}--\ref{f:trot_l1157_p1}. Appendix~\ref{a:trot}
explains in detail how the temperatures for the radiative transfer modeling
were then chosen to derive the column densities.

Most lines are barely resolved spectrally and were modeled assuming linewidths
in the range 3--6 km~s$^{-1}$. With the source size, rotational temperature,
and linewidths set as described above, and the velocity offset derived 
directly from the spectra, the only remaining free parameter for each molecule 
is its column density. This parameter was adjusted by eye until the synthetic
spectrum matched the observed spectrum reasonably well. For each molecule, we 
tried as far as possible not to exceed the peak temperature of any of the 
detected lines (or the upper limit of any of the undetected transitions). The 
parameters resulting from these fits are listed in 
Table~\ref{t:coldens} for a set of twelve (complex) organic molecules. 
Table~\ref{t:coldens} also indicates the number of detected lines per molecule,
as well as the detection status of each species (detected, tentatively 
detected, or not detected, depending on the number of detected lines and/or the
strength of these lines). A species is considered as only tentatively 
detected when its lines are either all weak (below $\sim$3$\sigma$) or it has 
too few lines just above $\sim$3$\sigma$.
Column density upper limits at the $\sim$3$\sigma$ 
level are given for the non-detections.

\longtab{
\begin{longtable}{lcrcccccc}
 \caption{\label{t:coldens}
 Parameters of our best-fit LTE model (or upper limit) of selected organic molecules toward CALYPSO continuum sources.
} \\ 
 \hline\hline
 \multicolumn{1}{c}{Molecule} & \multicolumn{1}{c}{Status\tablefootmark{a}} & \multicolumn{1}{c}{$N_{\rm det}$\tablefootmark{b}} & \multicolumn{1}{c}{Size\tablefootmark{c}} & \multicolumn{1}{c}{$T_{\mathrm{rot}}$\tablefootmark{d}} & \multicolumn{1}{c}{$N$\tablefootmark{e}} & \multicolumn{1}{c}{$C$\tablefootmark{f}} & \multicolumn{1}{c}{$\Delta V$\tablefootmark{g}} & \multicolumn{1}{c}{$V_{\mathrm{off}}$\tablefootmark{h}} \\ 
  & & & \multicolumn{1}{c}{$''$} & \multicolumn{1}{c}{K} & \multicolumn{1}{c}{cm$^{-2}$} & & \multicolumn{1}{c}{km~s$^{-1}$} & \multicolumn{1}{c}{km~s$^{-1}$}\\ 
 \hline
 \endfirsthead
 \caption{continued.} \\ 
 \hline\hline
 \multicolumn{1}{c}{Molecule} & \multicolumn{1}{c}{Status\tablefootmark{a}} & \multicolumn{1}{c}{$N_{\rm det}$\tablefootmark{b}} & \multicolumn{1}{c}{Size\tablefootmark{c}} & \multicolumn{1}{c}{$T_{\mathrm{rot}}$\tablefootmark{d}} & \multicolumn{1}{c}{$N$\tablefootmark{e}} & \multicolumn{1}{c}{$C$\tablefootmark{f}} & \multicolumn{1}{c}{$\Delta V$\tablefootmark{g}} & \multicolumn{1}{c}{$V_{\mathrm{off}}$\tablefootmark{h}} \\ 
 \hline
 \endhead
 \hline
 \endfoot
 \multicolumn{9}{c}{L1448-2A} 
 \smallskip \\ 
 CH$_3$OH & d & 6 & 1.00 &  150 &  5.0 (15) & 1.0 & 4.0 & 0.0 \\ 
 C$_2$H$_5$OH & n & 0 & 1.00 &  150 & $<$  1.6 (15) & 1.32 & 4.0 & 0.0 \\ 
 CH$_3$OCH$_3$ & n & 0 & 1.00 &  150 & $<$  9.0 (14) & 1.0 & 4.0 & 0.0 \\ 
 CH$_3$OCHO & n & 0 & 1.00 &  150 & $<$  6.0 (14) & 1.19 & 4.0 & 0.0 \\ 
 CH$_3$CHO & n & 0 & 1.00 &  150 & $<$  2.5 (14) & 1.02 & 4.0 & 0.0 \\ 
 NH$_2$CHO & n & 0 & 1.00 &  150 & $<$  6.4 (13) & 1.07 & 4.0 & 0.0 \\ 
 CH$_3$CN & t & 0 & 1.00 &  150 &  2.2 (14) & 1.0 & 4.0 & 0.0 \\ 
 C$_2$H$_5$CN & n & 0 & 1.00 &  150 & $<$  1.8 (14) & 1.38 & 4.0 & 0.0 \\ 
 (CH$_2$OH)$_2$ & n & 0 & 1.00 &  150 & $<$  8.1 (14) & 1.61 & 4.0 & 0.0 \\ 
 CH$_2$(OH)CHO & n & 0 & 1.00 &  150 & $<$  3.6 (14) & 1.04 & 4.0 & 0.0 \\ 
 HNCO & n & 0 & 1.00 &  150 & $<$  9.1 (13) & 1.01 & 4.0 & 0.0 \\ 
 NH$_2$CN & n & 0 & 1.00 &  150 & $<$  5.1 (13) & 1.70 & 4.0 & 0.0 \\ 
\hline 
 \multicolumn{9}{c}{L1448-2Ab} 
 \smallskip \\ 
 CH$_3$OH & d & 7 & 1.00 &  150 &  6.0 (15) & 1.0 & 4.0 & 0.3 \\ 
 C$_2$H$_5$OH & n & 0 & 1.00 &  150 & $<$  2.0 (15) & 1.32 & 4.0 & 0.0 \\ 
 CH$_3$OCH$_3$ & n & 0 & 1.00 &  150 & $<$  1.2 (15) & 1.0 & 4.0 & 0.0 \\ 
 CH$_3$OCHO & n & 0 & 1.00 &  150 & $<$  7.1 (14) & 1.19 & 4.0 & 0.0 \\ 
 CH$_3$CHO & n & 0 & 1.00 &  150 & $<$  3.1 (14) & 1.02 & 4.0 & 0.0 \\ 
 NH$_2$CHO & n & 0 & 1.00 &  150 & $<$  5.3 (13) & 1.07 & 4.0 & 0.0 \\ 
 CH$_3$CN & n & 0 & 1.00 &  150 & $<$  2.0 (14) & 1.0 & 4.0 & 0.0 \\ 
 C$_2$H$_5$CN & n & 0 & 1.00 &  150 & $<$  2.1 (14) & 1.38 & 4.0 & 0.0 \\ 
 (CH$_2$OH)$_2$ & n & 0 & 1.00 &  150 & $<$  9.7 (14) & 1.61 & 4.0 & 0.0 \\ 
 CH$_2$(OH)CHO & n & 0 & 1.00 &  150 & $<$  5.2 (14) & 1.04 & 4.0 & 0.0 \\ 
 HNCO & d & 1 & 1.00 &  150 &  1.2 (14) & 1.01 & 4.0 & 0.7 \\ 
 NH$_2$CN & n & 0 & 1.00 &  150 & $<$  4.3 (13) & 1.70 & 4.0 & 0.0 \\ 
\hline 
 \multicolumn{9}{c}{L1448-NA} 
 \smallskip \\ 
 CH$_3$OH & t & 0 & 1.00 &  100 &  1.8 (15) & 1.0 & 4.0 & 0.0 \\ 
 C$_2$H$_5$OH & n & 0 & 1.00 &  100 & $<$  3.3 (15) & 1.09 & 4.0 & 0.0 \\ 
 CH$_3$OCH$_3$ & n & 0 & 1.00 &  100 & $<$  2.0 (15) & 1.0 & 4.0 & 0.0 \\ 
 CH$_3$OCHO & t & 0 & 1.00 &  100 &  1.2 (15) & 1.04 & 5.0 & 1.0 \\ 
 CH$_3$CHO & n & 0 & 1.00 &  100 & $<$  6.0 (14) & 1.00 & 4.0 & 0.0 \\ 
 NH$_2$CHO & n & 0 & 1.00 &  100 & $<$  8.2 (13) & 1.02 & 4.0 & 0.0 \\ 
 CH$_3$CN & n & 0 & 1.00 &  100 & $<$  1.5 (14) & 1.0 & 4.0 & 0.0 \\ 
 C$_2$H$_5$CN & n & 0 & 1.00 &  100 & $<$  2.2 (14) & 1.11 & 4.0 & 0.0 \\ 
 (CH$_2$OH)$_2$ & n & 0 & 1.00 &  100 & $<$  1.1 (15) & 1.17 & 4.0 & 0.0 \\ 
 CH$_2$(OH)CHO & n & 0 & 1.00 &  100 & $<$  7.0 (14) & 1.00 & 4.0 & 0.0 \\ 
 HNCO & n & 0 & 1.00 &  100 & $<$  1.0 (14) & 1.00 & 4.0 & 0.0 \\ 
 NH$_2$CN & n & 0 & 1.00 &  100 & $<$  5.3 (13) & 1.33 & 4.0 & 0.0 \\ 
\hline 
 \multicolumn{9}{c}{L1448-NB1} 
 \smallskip \\ 
 CH$_3$OH & n & 0 & 1.00 &  100 & $<$  1.8 (15) & 1.0 & 4.0 & 0.0 \\ 
 C$_2$H$_5$OH & n & 0 & 1.00 &  100 & $<$  3.3 (15) & 1.09 & 4.0 & 0.0 \\ 
 CH$_3$OCH$_3$ & n & 0 & 1.00 &  100 & $<$  1.5 (15) & 1.0 & 4.0 & 0.0 \\ 
 CH$_3$OCHO & n & 0 & 1.00 &  100 & $<$  5.2 (14) & 1.04 & 4.0 & 0.0 \\ 
 CH$_3$CHO & n & 0 & 1.00 &  100 & $<$  5.0 (14) & 1.00 & 4.0 & 0.0 \\ 
 NH$_2$CHO & n & 0 & 1.00 &  100 & $<$  6.1 (13) & 1.02 & 4.0 & 0.0 \\ 
 CH$_3$CN & n & 0 & 1.00 &  100 & $<$  1.3 (14) & 1.0 & 4.0 & 0.0 \\ 
 C$_2$H$_5$CN & n & 0 & 1.00 &  100 & $<$  1.9 (14) & 1.11 & 4.0 & 0.0 \\ 
 (CH$_2$OH)$_2$ & n & 0 & 1.00 &  100 & $<$  9.4 (14) & 1.17 & 4.0 & 0.0 \\ 
 CH$_2$(OH)CHO & n & 0 & 1.00 &  100 & $<$  7.0 (14) & 1.00 & 4.0 & 0.0 \\ 
 HNCO & n & 0 & 1.00 &  100 & $<$  1.0 (14) & 1.00 & 4.0 & 0.0 \\ 
 NH$_2$CN & n & 0 & 1.00 &  100 & $<$  3.3 (13) & 1.33 & 4.0 & 0.0 \\ 
\hline 
 \multicolumn{9}{c}{L1448-NB2} 
 \smallskip \\ 
 CH$_3$OH & n & 0 & 1.00 &  100 & $<$  1.5 (15) & 1.0 & 4.0 & 0.0 \\ 
 C$_2$H$_5$OH & n & 0 & 1.00 &  100 & $<$  3.3 (15) & 1.09 & 4.0 & 0.0 \\ 
 CH$_3$OCH$_3$ & n & 0 & 1.00 &  100 & $<$  1.2 (15) & 1.0 & 4.0 & 0.0 \\ 
 CH$_3$OCHO & n & 0 & 1.00 &  100 & $<$  7.3 (14) & 1.04 & 4.0 & 0.0 \\ 
 CH$_3$CHO & n & 0 & 1.00 &  100 & $<$  4.0 (14) & 1.00 & 4.0 & 0.0 \\ 
 NH$_2$CHO & n & 0 & 1.00 &  100 & $<$  6.1 (13) & 1.02 & 4.0 & 0.0 \\ 
 CH$_3$CN & n & 0 & 1.00 &  100 & $<$  1.2 (14) & 1.0 & 4.0 & 0.0 \\ 
 C$_2$H$_5$CN & n & 0 & 1.00 &  100 & $<$  2.2 (14) & 1.11 & 4.0 & 0.0 \\ 
 (CH$_2$OH)$_2$ & n & 0 & 1.00 &  100 & $<$  1.1 (15) & 1.17 & 4.0 & 0.0 \\ 
 CH$_2$(OH)CHO & n & 0 & 1.00 &  100 & $<$  5.0 (14) & 1.00 & 4.0 & 0.0 \\ 
 HNCO & n & 0 & 1.00 &  100 & $<$  8.0 (13) & 1.00 & 4.0 & 0.0 \\ 
 NH$_2$CN & n & 0 & 1.00 &  100 & $<$  3.3 (13) & 1.33 & 4.0 & 0.0 \\ 
\hline 
 \multicolumn{9}{c}{L1448-C} 
 \smallskip \\ 
 CH$_3$OH & d & 13 & 0.50 &  100 &  8.0 (16) & 1.0 & 5.0 & 0.0 \\ 
 C$_2$H$_5$OH & n & 0 & 0.50 &  100 & $<$  4.4 (15) & 1.09 & 5.0 & 0.0 \\ 
 CH$_3$OCH$_3$ & d & 1 & 0.50 &  100 &  1.0 (16) & 1.0 & 5.0 & 0.0 \\ 
 CH$_3$OCHO & d & 4 & 0.50 &  100 &  5.2 (15) & 1.04 & 5.0 & 0.0 \\ 
 CH$_3$CHO & d & 11 & 0.50 &  100 &  1.9 (15) & 1.00 & 5.0 & -0.5 \\ 
 NH$_2$CHO & d & 2 & 0.50 &  100 &  1.6 (14) & 1.02 & 5.0 & -0.3 \\ 
 CH$_3$CN & d & 3 & 0.50 &  100 &  2.5 (15) & 1.0 & 5.0 & 0.0 \\ 
 C$_2$H$_5$CN & n & 0 & 0.50 &  100 & $<$  3.9 (14) & 1.11 & 5.0 & 0.0 \\ 
 (CH$_2$OH)$_2$ & n & 0 & 0.50 &  100 & $<$  1.8 (15) & 1.17 & 5.0 & 0.0 \\ 
 CH$_2$(OH)CHO & n & 0 & 0.50 &  100 & $<$  1.0 (15) & 1.00 & 5.0 & 0.0 \\ 
 HNCO & d & 2 & 0.50 &  100 &  8.0 (14) & 1.00 & 5.0 & -0.2 \\ 
 NH$_2$CN & n & 0 & 0.50 &  100 & $<$  8.0 (13) & 1.33 & 5.0 & -0.2 \\ 
\hline 
 \multicolumn{9}{c}{L1448-CS} 
 \smallskip \\ 
 CH$_3$OH & n & 0 & 0.50 &  100 & $<$  4.0 (15) & 1.0 & 5.0 & 0.0 \\ 
 C$_2$H$_5$OH & n & 0 & 0.50 &  100 & $<$  5.5 (15) & 1.09 & 5.0 & 0.0 \\ 
 CH$_3$OCH$_3$ & t & 0 & 0.50 &  100 &  2.6 (15) & 1.0 & 5.0 & 0.0 \\ 
 CH$_3$OCHO & n & 0 & 0.50 &  100 & $<$  1.6 (15) & 1.04 & 5.0 & 0.0 \\ 
 CH$_3$CHO & n & 0 & 0.50 &  100 & $<$  1.1 (15) & 1.00 & 5.0 & 0.0 \\ 
 NH$_2$CHO & n & 0 & 0.50 &  100 & $<$  2.0 (14) & 1.02 & 5.0 & 0.0 \\ 
 CH$_3$CN & n & 0 & 0.50 &  100 & $<$  3.5 (14) & 1.0 & 5.0 & 0.0 \\ 
 C$_2$H$_5$CN & n & 0 & 0.50 &  100 & $<$  5.6 (14) & 1.11 & 5.0 & 0.0 \\ 
 (CH$_2$OH)$_2$ & n & 0 & 0.50 &  100 & $<$  2.3 (15) & 1.17 & 5.0 & 0.0 \\ 
 CH$_2$(OH)CHO & n & 0 & 0.50 &  100 & $<$  1.2 (15) & 1.00 & 5.0 & 0.0 \\ 
 HNCO & n & 0 & 0.50 &  100 & $<$  2.5 (14) & 1.00 & 5.0 & 0.0 \\ 
 NH$_2$CN & n & 0 & 0.50 &  100 & $<$  9.3 (13) & 1.33 & 5.0 & 0.0 \\ 
\hline 
 \multicolumn{9}{c}{IRAS2A1} 
 \smallskip \\ 
 CH$_3$OH & d & 21 & 0.35 &  250 &  1.8 (18) & 1.0 & 4.0 & -0.5 \\ 
 C$_2$H$_5$OH & d & 9 & 0.35 &  150 &  4.0 (16) & 1.32 & 4.0 & -0.5 \\ 
 CH$_3$OCH$_3$ & d & 4 & 0.35 &  100 &  6.3 (16) & 1.0 & 4.0 & -0.5 \\ 
 CH$_3$OCHO & d & 11 & 0.35 &  120 &  7.1 (16) & 1.09 & 4.0 & -0.5 \\ 
 CH$_3$CHO & d & 13 & 0.35 &  150 &  1.0 (16) & 1.02 & 4.0 & -0.5 \\ 
 NH$_2$CHO & d & 3 & 0.35 &  250 &  8.0 (15) & 1.33 & 4.0 & -0.4 \\ 
 CH$_3$CN & d & 4 & 0.35 &  170 &  2.5 (16) & 1.0 & 4.0 & -0.3 \\ 
 C$_2$H$_5$CN & n & 0 & 0.35 &  150 & $<$  1.4 (15) & 1.38 & 4.0 & 0.0 \\ 
 (CH$_2$OH)$_2$ & d & 13 & 0.35 &  150 &  1.8 (16) & 1.61 & 4.0 & -0.5 \\ 
 CH$_2$(OH)CHO & t & 0 & 0.35 &  150 &  4.2 (15) & 1.04 & 4.0 & -0.5 \\ 
 HNCO & d & 5 & 0.35 &  250 &  2.7 (16) & 1.07 & 4.5 & 0.0 \\ 
 NH$_2$CN & d & 7 & 0.35 &  150 &  8.5 (14) & 1.70 & 4.0 & 0.0 \\ 
\hline 
 \multicolumn{9}{c}{SVS13B} 
 \smallskip \\ 
 CH$_3$OH & n & 0 & 0.50 &  150 & $<$  8.0 (15) & 1.0 & 4.0 & 0.0 \\ 
 C$_2$H$_5$OH & n & 0 & 0.50 &  150 & $<$  1.5 (16) & 1.32 & 4.0 & 0.0 \\ 
 CH$_3$OCH$_3$ & n & 0 & 0.50 &  150 & $<$  8.0 (15) & 1.0 & 4.0 & 0.0 \\ 
 CH$_3$OCHO & n & 0 & 0.50 &  150 & $<$  3.6 (15) & 1.19 & 4.0 & 0.0 \\ 
 CH$_3$CHO & n & 0 & 0.50 &  150 & $<$  2.3 (15) & 1.02 & 4.0 & 0.0 \\ 
 NH$_2$CHO & n & 0 & 0.50 &  150 & $<$  3.4 (14) & 1.07 & 4.0 & 0.0 \\ 
 CH$_3$CN & n & 0 & 0.50 &  150 & $<$  6.0 (14) & 1.0 & 4.0 & 0.0 \\ 
 C$_2$H$_5$CN & n & 0 & 0.50 &  150 & $<$  9.0 (14) & 1.38 & 4.0 & 0.0 \\ 
 (CH$_2$OH)$_2$ & n & 0 & 0.50 &  150 & $<$  4.0 (15) & 1.61 & 3.0 & 0.0 \\ 
 CH$_2$(OH)CHO & n & 0 & 0.50 &  150 & $<$  4.7 (15) & 1.04 & 4.0 & 0.0 \\ 
 HNCO & n & 0 & 0.50 &  150 & $<$  5.0 (14) & 1.01 & 4.0 & 0.0 \\ 
 NH$_2$CN & n & 0 & 0.50 &  150 & $<$  2.0 (14) & 1.70 & 4.0 & 0.0 \\ 
\hline 
 \multicolumn{9}{c}{SVS13A} 
 \smallskip \\ 
 CH$_3$OH & d & 22 & 0.30 &  220 &  2.0 (18) & 1.0 & 4.0 & -0.5 \\ 
 C$_2$H$_5$OH & d & 16 & 0.30 &  220 &  1.5 (17) & 1.86 & 4.0 & -0.5 \\ 
 CH$_3$OCH$_3$ & d & 7 & 0.30 &  100 &  2.0 (17) & 1.0 & 4.0 & -0.1 \\ 
 CH$_3$OCHO & d & 21 & 0.30 &  100 &  2.1 (17) & 1.04 & 4.0 & -0.3 \\ 
 CH$_3$CHO & d & 14 & 0.30 &  100 &  1.6 (16) & 1.00 & 4.0 & -0.3 \\ 
 NH$_2$CHO & d & 4 & 0.30 &  300 &  1.2 (16) & 1.54 & 4.0 & -0.4 \\ 
 CH$_3$CN & d & 5 & 0.30 &  180 &  4.0 (16) & 1.0 & 4.0 & 0.0 \\ 
 C$_2$H$_5$CN & d & 2 & 0.30 &  180 &  4.1 (15) & 1.63 & 4.0 & -0.3 \\ 
 (CH$_2$OH)$_2$ & d & 11 & 0.30 &  100 &  1.3 (16) & 1.17 & 3.0 & -0.3 \\ 
 CH$_2$(OH)CHO & d & 2 & 0.30 &  100 &  6.0 (15) & 1.00 & 4.0 & -0.3 \\ 
 HNCO & d & 5 & 0.30 &  220 &  3.1 (16) & 1.04 & 4.0 & -0.3 \\ 
 NH$_2$CN & d & 8 & 0.30 &  220 &  2.2 (15) & 2.46 & 4.0 & -0.3 \\ 
\hline 
 \multicolumn{9}{c}{IRAS4A1} 
 \smallskip \\ 
 CH$_3$OH & n & 0 & 0.35 &  150 & $<$  1.0 (16) & 1.0 & 4.0 & 0.5 \\ 
 C$_2$H$_5$OH & n & 0 & 0.35 &  150 & $<$  1.6 (16) & 1.32 & 4.0 & 0.5 \\ 
 CH$_3$OCH$_3$ & n & 0 & 0.35 &  150 & $<$  8.0 (15) & 1.0 & 4.0 & 0.5 \\ 
 CH$_3$OCHO & n & 0 & 0.35 &  150 & $<$  7.1 (15) & 1.19 & 4.0 & 0.5 \\ 
 CH$_3$CHO & n & 0 & 0.35 &  150 & $<$  2.0 (15) & 1.02 & 4.0 & 0.5 \\ 
 NH$_2$CHO & n & 0 & 0.35 &  150 & $<$  4.3 (14) & 1.07 & 4.0 & 0.5 \\ 
 CH$_3$CN & n & 0 & 0.35 &  150 & $<$  3.0 (15) & 1.0 & 4.0 & 0.5 \\ 
 C$_2$H$_5$CN & n & 0 & 0.35 &  150 & $<$  1.2 (15) & 1.38 & 4.0 & 0.5 \\ 
 (CH$_2$OH)$_2$ & n & 0 & 0.35 &  150 & $<$  8.1 (15) & 1.61 & 4.0 & 0.5 \\ 
 CH$_2$(OH)CHO & n & 0 & 0.35 &  150 & $<$  3.1 (15) & 1.04 & 4.0 & 0.5 \\ 
 HNCO & n & 0 & 0.35 &  150 & $<$  9.1 (14) & 1.01 & 4.0 & 0.5 \\ 
 NH$_2$CN & n & 0 & 0.35 &  150 & $<$  3.4 (14) & 1.70 & 4.0 & 0.5 \\ 
\hline 
 \multicolumn{9}{c}{IRAS4A2} 
 \smallskip \\ 
 CH$_3$OH & d & 21 & 0.35 &  250 &  6.0 (17) & 1.0 & 3.5 & -0.3 \\ 
 C$_2$H$_5$OH & d & 14 & 0.35 &  150 &  5.3 (16) & 1.32 & 3.5 & 0.0 \\ 
 CH$_3$OCH$_3$ & d & 5 & 0.35 &  150 &  6.0 (16) & 1.0 & 3.5 & -0.1 \\ 
 CH$_3$OCHO & d & 20 & 0.35 &  150 &  8.9 (16) & 1.19 & 3.5 & 0.0 \\ 
 CH$_3$CHO & d & 33 & 0.35 &  200 &  2.7 (16) & 1.08 & 2.5 & 0.0 \\ 
 NH$_2$CHO & d & 4 & 0.35 &  300 &  7.7 (15) & 1.54 & 3.5 & 0.0 \\ 
 CH$_3$CN & d & 3 & 0.35 &  150 &  1.1 (16) & 1.0 & 3.5 & 0.0 \\ 
 C$_2$H$_5$CN & d & 3 & 0.35 &  150 &  2.3 (15) & 1.38 & 3.5 & 0.0 \\ 
 (CH$_2$OH)$_2$ & d & 18 & 0.35 &  150 &  1.3 (16) & 1.61 & 3.5 & 0.0 \\ 
 CH$_2$(OH)CHO & d & 10 & 0.35 &  150 &  9.4 (15) & 1.04 & 3.5 & 0.0 \\ 
 HNCO & d & 5 & 0.35 &  250 &  1.1 (16) & 1.07 & 4.0 & 0.0 \\ 
 NH$_2$CN & d & 7 & 0.35 &  150 &  8.5 (14) & 1.70 & 4.5 & 0.0 \\ 
\hline 
 \multicolumn{9}{c}{IRAS4B} 
 \smallskip \\ 
 CH$_3$OH & d & 14 & 0.40 &  300 &  1.5 (17) & 1.0 & 4.0 & 0.0 \\ 
 C$_2$H$_5$OH & d & 3 & 0.40 &  150 &  1.5 (16) & 1.32 & 4.0 & 0.0 \\ 
 CH$_3$OCH$_3$ & d & 1 & 0.40 &  150 &  2.0 (16) & 1.0 & 4.0 & 0.0 \\ 
 CH$_3$OCHO & d & 6 & 0.40 &  200 &  5.4 (16) & 1.46 & 4.0 & 0.0 \\ 
 CH$_3$CHO & d & 37 & 0.40 &  200 &  1.3 (16) & 1.08 & 3.0 & 0.0 \\ 
 NH$_2$CHO & d & 1 & 0.40 &  150 &  3.7 (14) & 1.07 & 3.5 & 0.0 \\ 
 CH$_3$CN & d & 3 & 0.40 &  150 &  2.5 (15) & 1.0 & 5.0 & 0.0 \\ 
 C$_2$H$_5$CN & d & 1 & 0.40 &  150 &  6.9 (14) & 1.38 & 4.0 & 0.5 \\ 
 (CH$_2$OH)$_2$ & n & 0 & 0.40 &  150 & $<$  4.8 (15) & 1.61 & 4.0 & 0.0 \\ 
 CH$_2$(OH)CHO & d & 1 & 0.40 &  150 &  3.1 (15) & 1.04 & 4.0 & 0.0 \\ 
 HNCO & d & 5 & 0.40 &  300 &  5.5 (15) & 1.14 & 4.0 & 0.0 \\ 
 NH$_2$CN & n & 0 & 0.40 &  150 & $<$  3.4 (14) & 1.70 & 4.0 & 0.0 \\ 
\hline 
 \multicolumn{9}{c}{IRAS4B2} 
 \smallskip \\ 
 CH$_3$OH & n & 0 & 0.40 &  150 & $<$  2.0 (16) & 1.0 & 4.0 & 0.0 \\ 
 C$_2$H$_5$OH & n & 0 & 0.40 &  150 & $<$  2.9 (16) & 1.32 & 4.0 & 0.0 \\ 
 CH$_3$OCH$_3$ & n & 0 & 0.40 &  150 & $<$  1.0 (16) & 1.0 & 4.0 & 0.0 \\ 
 CH$_3$OCHO & n & 0 & 0.40 &  150 & $<$  7.1 (15) & 1.19 & 4.0 & 0.0 \\ 
 CH$_3$CHO & n & 0 & 0.40 &  150 & $<$  3.1 (15) & 1.02 & 4.0 & 0.0 \\ 
 NH$_2$CHO & n & 0 & 0.40 &  150 & $<$  7.5 (14) & 1.07 & 4.0 & 0.0 \\ 
 CH$_3$CN & n & 0 & 0.40 &  150 & $<$  2.2 (15) & 1.0 & 4.0 & 0.0 \\ 
 C$_2$H$_5$CN & n & 0 & 0.40 &  150 & $<$  2.1 (15) & 1.38 & 4.0 & 0.0 \\ 
 (CH$_2$OH)$_2$ & n & 0 & 0.40 &  150 & $<$  1.1 (16) & 1.61 & 4.0 & 0.0 \\ 
 CH$_2$(OH)CHO & n & 0 & 0.40 &  150 & $<$  8.3 (15) & 1.04 & 4.0 & 0.0 \\ 
 HNCO & n & 0 & 0.40 &  150 & $<$  1.3 (15) & 1.01 & 4.0 & 0.0 \\ 
 NH$_2$CN & n & 0 & 0.40 &  150 & $<$  5.1 (14) & 1.70 & 4.0 & 0.0 \\ 
\hline 
 \multicolumn{9}{c}{IRAM04191} 
 \smallskip \\ 
 CH$_3$OH & n & 0 & 0.50 &  100 & $<$  2.5 (15) & 1.0 & 5.0 & 0.5 \\ 
 C$_2$H$_5$OH & n & 0 & 0.50 &  100 & $<$  4.9 (15) & 1.09 & 5.0 & 0.5 \\ 
 CH$_3$OCH$_3$ & n & 0 & 0.50 &  100 & $<$  1.5 (15) & 1.0 & 5.0 & 0.5 \\ 
 CH$_3$OCHO & n & 0 & 0.50 &  100 & $<$  7.3 (14) & 1.04 & 5.0 & 0.5 \\ 
 CH$_3$CHO & n & 0 & 0.50 &  100 & $<$  5.0 (14) & 1.00 & 5.0 & 0.5 \\ 
 NH$_2$CHO & n & 0 & 0.50 &  100 & $<$  1.0 (14) & 1.02 & 5.0 & 0.5 \\ 
 CH$_3$CN & n & 0 & 0.50 &  100 & $<$  2.0 (14) & 1.0 & 5.0 & 0.5 \\ 
 C$_2$H$_5$CN & n & 0 & 0.50 &  100 & $<$  3.3 (14) & 1.11 & 5.0 & 0.5 \\ 
 (CH$_2$OH)$_2$ & n & 0 & 0.50 &  100 & $<$  1.8 (15) & 1.17 & 5.0 & 0.5 \\ 
 CH$_2$(OH)CHO & n & 0 & 0.50 &  100 & $<$  7.0 (14) & 1.00 & 5.0 & 0.5 \\ 
 HNCO & n & 0 & 0.50 &  100 & $<$  1.0 (14) & 1.00 & 5.0 & 0.5 \\ 
 NH$_2$CN & n & 0 & 0.50 &  100 & $<$  6.0 (13) & 1.33 & 5.0 & 0.5 \\ 
\hline 
 \multicolumn{9}{c}{L1521F} 
 \smallskip \\ 
 CH$_3$OH & n & 0 & 0.50 &  100 & $<$  1.3 (15) & 1.0 & 4.0 & 0.0 \\ 
 C$_2$H$_5$OH & n & 0 & 0.50 &  100 & $<$  1.7 (15) & 1.09 & 4.0 & 0.0 \\ 
 CH$_3$OCH$_3$ & n & 0 & 0.50 &  100 & $<$  9.0 (14) & 1.0 & 4.0 & 0.0 \\ 
 CH$_3$OCHO & n & 0 & 0.50 &  100 & $<$  7.3 (14) & 1.04 & 4.0 & 0.0 \\ 
 CH$_3$CHO & n & 0 & 0.50 &  100 & $<$  3.0 (14) & 1.00 & 4.0 & 0.0 \\ 
 NH$_2$CHO & n & 0 & 0.50 &  100 & $<$  6.1 (13) & 1.02 & 4.0 & 0.0 \\ 
 CH$_3$CN & n & 0 & 0.50 &  100 & $<$  1.9 (14) & 1.0 & 4.0 & 0.0 \\ 
 C$_2$H$_5$CN & n & 0 & 0.50 &  100 & $<$  2.2 (14) & 1.11 & 4.0 & 0.0 \\ 
 (CH$_2$OH)$_2$ & n & 0 & 0.50 &  100 & $<$  8.2 (14) & 1.17 & 4.0 & 0.0 \\ 
 CH$_2$(OH)CHO & n & 0 & 0.50 &  100 & $<$  4.0 (14) & 1.00 & 4.0 & 0.0 \\ 
 HNCO & n & 0 & 0.50 &  100 & $<$  1.0 (14) & 1.00 & 4.0 & 0.0 \\ 
 NH$_2$CN & n & 0 & 0.50 &  100 & $<$  3.3 (13) & 1.33 & 4.0 & 0.0 \\ 
\hline 
 \multicolumn{9}{c}{L1527} 
 \smallskip \\ 
 CH$_3$OH & n & 0 & 1.00 &  100 & $<$  2.0 (15) & 1.0 & 5.0 & 0.0 \\ 
 C$_2$H$_5$OH & n & 0 & 1.00 &  100 & $<$  3.8 (15) & 1.09 & 5.0 & 0.0 \\ 
 CH$_3$OCH$_3$ & n & 0 & 1.00 &  100 & $<$  2.3 (15) & 1.0 & 5.0 & 0.0 \\ 
 CH$_3$OCHO & n & 0 & 1.00 &  100 & $<$  8.3 (14) & 1.04 & 5.0 & 0.0 \\ 
 CH$_3$CHO & n & 0 & 1.00 &  100 & $<$  7.0 (14) & 1.00 & 5.0 & 0.0 \\ 
 NH$_2$CHO & n & 0 & 1.00 &  100 & $<$  1.0 (14) & 1.02 & 5.0 & 0.0 \\ 
 CH$_3$CN & n & 0 & 1.00 &  100 & $<$  1.0 (14) & 1.0 & 5.0 & 0.0 \\ 
 C$_2$H$_5$CN & n & 0 & 1.00 &  100 & $<$  2.8 (14) & 1.11 & 5.0 & 0.0 \\ 
 (CH$_2$OH)$_2$ & n & 0 & 1.00 &  100 & $<$  1.2 (15) & 1.17 & 5.0 & 0.0 \\ 
 CH$_2$(OH)CHO & n & 0 & 1.00 &  100 & $<$  6.0 (14) & 1.00 & 5.0 & 0.0 \\ 
 HNCO & n & 0 & 1.00 &  100 & $<$  1.4 (14) & 1.00 & 5.0 & 0.0 \\ 
 NH$_2$CN & n & 0 & 1.00 &  100 & $<$  4.7 (13) & 1.33 & 5.0 & 0.0 \\ 
\hline 
 \multicolumn{9}{c}{SerpM-S68N} 
 \smallskip \\ 
 CH$_3$OH & d & 7 & 0.50 &  200 &  3.0 (16) & 1.0 & 4.0 & 0.0 \\ 
 C$_2$H$_5$OH & n & 0 & 0.50 &  150 & $<$  1.1 (16) & 1.32 & 4.0 & 0.0 \\ 
 CH$_3$OCH$_3$ & t & 0 & 0.50 &  150 &  4.0 (15) & 1.0 & 4.0 & -1.0 \\ 
 CH$_3$OCHO & d & 4 & 0.50 &  150 &  8.3 (15) & 1.19 & 4.0 & 0.0 \\ 
 CH$_3$CHO & n & 0 & 0.50 &  150 & $<$  1.3 (15) & 1.02 & 4.0 & 0.0 \\ 
 NH$_2$CHO & n & 0 & 0.50 &  150 & $<$  3.7 (14) & 1.07 & 4.0 & -0.3 \\ 
 CH$_3$CN & d & 1 & 0.50 &  150 &  8.0 (14) & 1.0 & 4.0 & -0.3 \\ 
 C$_2$H$_5$CN & n & 0 & 0.50 &  150 & $<$  8.3 (14) & 1.38 & 4.0 & -0.3 \\ 
 (CH$_2$OH)$_2$ & n & 0 & 0.50 &  150 & $<$  4.0 (15) & 1.61 & 4.0 & -0.3 \\ 
 CH$_2$(OH)CHO & n & 0 & 0.50 &  150 & $<$  3.1 (15) & 1.04 & 4.0 & -0.3 \\ 
 HNCO & t & 0 & 0.50 &  150 &  3.0 (14) & 1.01 & 4.0 & -0.3 \\ 
 NH$_2$CN & n & 0 & 0.50 &  150 & $<$  2.2 (14) & 1.70 & 4.0 & -0.3 \\ 
\hline 
 \multicolumn{9}{c}{SerpM-S68Nb} 
 \smallskip \\ 
 CH$_3$OH & n & 0 & 0.50 &  150 & $<$  2.5 (16) & 1.0 & 3.0 & 0.0 \\ 
 C$_2$H$_5$OH & n & 0 & 0.50 &  150 & $<$  4.0 (16) & 1.32 & 3.0 & 0.0 \\ 
 CH$_3$OCH$_3$ & n & 0 & 0.50 &  150 & $<$  1.5 (16) & 1.0 & 3.0 & 0.0 \\ 
 CH$_3$OCHO & n & 0 & 0.50 &  150 & $<$  8.3 (15) & 1.19 & 3.0 & 0.0 \\ 
 CH$_3$CHO & n & 0 & 0.50 &  150 & $<$  4.1 (15) & 1.02 & 3.0 & 0.0 \\ 
 NH$_2$CHO & n & 0 & 0.50 &  150 & $<$  1.1 (15) & 1.07 & 3.0 & 0.0 \\ 
 CH$_3$CN & n & 0 & 0.50 &  150 & $<$  7.5 (14) & 1.0 & 3.0 & 0.0 \\ 
 C$_2$H$_5$CN & n & 0 & 0.50 &  150 & $<$  3.4 (15) & 1.38 & 3.0 & 0.0 \\ 
 (CH$_2$OH)$_2$ & n & 0 & 0.50 &  150 & $<$  1.3 (16) & 1.61 & 3.0 & 0.0 \\ 
 CH$_2$(OH)CHO & n & 0 & 0.50 &  150 & $<$  8.3 (15) & 1.04 & 3.0 & 0.0 \\ 
 HNCO & n & 0 & 0.50 &  150 & $<$  1.0 (15) & 1.01 & 3.0 & 0.0 \\ 
 NH$_2$CN & n & 0 & 0.50 &  150 & $<$  6.0 (14) & 1.70 & 3.0 & 0.0 \\ 
\hline 
 \multicolumn{9}{c}{SerpM-SMM4a} 
 \smallskip \\ 
 CH$_3$OH & n & 0 & 0.50 &  150 & $<$  8.0 (15) & 1.0 & 3.0 & 0.0 \\ 
 C$_2$H$_5$OH & n & 0 & 0.50 &  150 & $<$  1.3 (16) & 1.32 & 3.0 & 0.0 \\ 
 CH$_3$OCH$_3$ & n & 0 & 0.50 &  150 & $<$  4.5 (15) & 1.0 & 3.0 & 0.0 \\ 
 CH$_3$OCHO & n & 0 & 0.50 &  150 & $<$  3.6 (15) & 1.19 & 3.0 & 0.0 \\ 
 CH$_3$CHO & n & 0 & 0.50 &  150 & $<$  1.7 (15) & 1.02 & 3.0 & 0.0 \\ 
 NH$_2$CHO & n & 0 & 0.50 &  150 & $<$  3.2 (14) & 1.07 & 3.0 & 0.0 \\ 
 CH$_3$CN & n & 0 & 0.50 &  150 & $<$  7.0 (14) & 1.0 & 3.0 & 0.0 \\ 
 C$_2$H$_5$CN & n & 0 & 0.50 &  150 & $<$  9.7 (14) & 1.38 & 3.0 & 0.0 \\ 
 (CH$_2$OH)$_2$ & n & 0 & 0.50 &  150 & $<$  4.8 (15) & 1.61 & 3.0 & 0.0 \\ 
 CH$_2$(OH)CHO & n & 0 & 0.50 &  150 & $<$  3.1 (15) & 1.04 & 3.0 & 0.0 \\ 
 HNCO & n & 0 & 0.50 &  150 & $<$  5.0 (14) & 1.01 & 3.0 & 0.0 \\ 
 NH$_2$CN & n & 0 & 0.50 &  150 & $<$  2.5 (14) & 1.70 & 3.0 & 0.0 \\ 
\hline 
 \multicolumn{9}{c}{SerpM-SMM4b} 
 \smallskip \\ 
 CH$_3$OH & d & 10 & 0.50 &  250 &  5.0 (16) & 1.0 & 3.0 & -0.5 \\ 
 C$_2$H$_5$OH & n & 0 & 0.50 &  250 & $<$  3.3 (16) & 2.18 & 3.0 & 0.0 \\ 
 CH$_3$OCH$_3$ & n & 0 & 0.50 &  250 & $<$  1.3 (16) & 1.0 & 3.0 & 0.0 \\ 
 CH$_3$OCHO & n & 0 & 0.50 &  250 & $<$  1.1 (16) & 1.84 & 3.0 & 0.0 \\ 
 CH$_3$CHO & n & 0 & 0.50 &  250 & $<$  3.5 (15) & 1.17 & 3.0 & 0.0 \\ 
 NH$_2$CHO & n & 0 & 0.50 &  250 & $<$  8.0 (14) & 1.33 & 3.0 & 0.0 \\ 
 CH$_3$CN & n & 0 & 0.50 &  250 & $<$  1.5 (15) & 1.0 & 3.0 & 0.0 \\ 
 C$_2$H$_5$CN & n & 0 & 0.50 &  250 & $<$  2.5 (15) & 2.50 & 3.0 & 0.0 \\ 
 (CH$_2$OH)$_2$ & n & 0 & 0.50 &  250 & $<$  1.5 (16) & 3.65 & 3.0 & 0.0 \\ 
 CH$_2$(OH)CHO & n & 0 & 0.50 &  250 & $<$  7.7 (15) & 1.29 & 3.0 & 0.0 \\ 
 HNCO & d & 1 & 0.50 &  250 &  9.6 (14) & 1.07 & 3.0 & 0.0 \\ 
 NH$_2$CN & n & 0 & 0.50 &  250 & $<$  5.8 (14) & 2.89 & 3.0 & 0.0 \\ 
\hline 
 \multicolumn{9}{c}{SerpS-MM18a} 
 \smallskip \\ 
 CH$_3$OH & d & 17 & 0.50 &  150 &  2.2 (17) & 1.0 & 4.0 & 0.5 \\ 
 C$_2$H$_5$OH & d & 4 & 0.50 &  150 &  1.7 (16) & 1.32 & 4.0 & 0.5 \\ 
 CH$_3$OCH$_3$ & d & 4 & 0.50 &  110 &  2.6 (16) & 1.0 & 4.0 & 0.5 \\ 
 CH$_3$OCHO & d & 7 & 0.50 &  150 &  3.0 (16) & 1.19 & 4.0 & 0.5 \\ 
 CH$_3$CHO & d & 15 & 0.50 &  150 &  6.1 (15) & 1.02 & 4.0 & 0.5 \\ 
 NH$_2$CHO & d & 2 & 0.50 &  150 &  1.2 (15) & 1.07 & 4.0 & 0.9 \\ 
 CH$_3$CN & d & 5 & 0.50 &  200 &  1.3 (16) & 1.0 & 4.0 & 0.9 \\ 
 C$_2$H$_5$CN & d & 5 & 0.50 &  150 &  1.9 (15) & 1.38 & 4.0 & 0.5 \\ 
 (CH$_2$OH)$_2$ & n & 0 & 0.50 &  150 & $<$  4.0 (15) & 1.61 & 4.0 & 0.9 \\ 
 CH$_2$(OH)CHO & n & 0 & 0.50 &  150 & $<$  2.6 (15) & 1.04 & 4.0 & 0.9 \\ 
 HNCO & d & 4 & 0.50 &  150 &  7.1 (15) & 1.01 & 6.0 & 0.9 \\ 
 NH$_2$CN & n & 0 & 0.50 &  150 & $<$  3.4 (14) & 1.70 & 4.0 & 0.9 \\ 
\hline 
 \multicolumn{9}{c}{SerpS-MM18b} 
 \smallskip \\ 
 CH$_3$OH & d & 4 & 0.50 &  120 &  1.7 (16) & 1.0 & 4.0 & 0.0 \\ 
 C$_2$H$_5$OH & n & 0 & 0.50 &  120 & $<$  8.2 (15) & 1.17 & 4.0 & 0.0 \\ 
 CH$_3$OCH$_3$ & n & 0 & 0.50 &  120 & $<$  3.5 (15) & 1.0 & 4.0 & 0.0 \\ 
 CH$_3$OCHO & n & 0 & 0.50 &  120 & $<$  3.3 (15) & 1.09 & 4.0 & 0.0 \\ 
 CH$_3$CHO & n & 0 & 0.50 &  120 & $<$  1.0 (15) & 1.01 & 4.0 & 0.0 \\ 
 NH$_2$CHO & n & 0 & 0.50 &  120 & $<$  3.1 (14) & 1.03 & 4.0 & 0.0 \\ 
 CH$_3$CN & d & 1 & 0.50 &  120 &  4.7 (14) & 1.0 & 4.0 & 0.0 \\ 
 C$_2$H$_5$CN & n & 0 & 0.50 &  120 & $<$  7.2 (14) & 1.20 & 4.0 & 0.0 \\ 
 (CH$_2$OH)$_2$ & n & 0 & 0.50 &  120 & $<$  3.9 (15) & 1.31 & 4.0 & 0.0 \\ 
 CH$_2$(OH)CHO & n & 0 & 0.50 &  120 & $<$  2.5 (15) & 1.01 & 4.0 & 0.0 \\ 
 HNCO & d & 1 & 0.50 &  120 &  5.0 (14) & 1.00 & 4.0 & 0.0 \\ 
 NH$_2$CN & n & 0 & 0.50 &  120 & $<$  1.9 (14) & 1.46 & 4.0 & 0.0 \\ 
\hline 
 \multicolumn{9}{c}{SerpS-MM22} 
 \smallskip \\ 
 CH$_3$OH & n & 0 & 1.00 &  150 & $<$  3.0 (15) & 1.0 & 3.0 & -0.9 \\ 
 C$_2$H$_5$OH & n & 0 & 1.00 &  150 & $<$  5.3 (15) & 1.32 & 3.0 & -0.9 \\ 
 CH$_3$OCH$_3$ & n & 0 & 1.00 &  150 & $<$  2.5 (15) & 1.0 & 3.0 & -0.9 \\ 
 CH$_3$OCHO & n & 0 & 1.00 &  150 & $<$  1.8 (15) & 1.19 & 3.0 & -0.9 \\ 
 CH$_3$CHO & n & 0 & 1.00 &  150 & $<$  7.7 (14) & 1.02 & 3.0 & -0.9 \\ 
 NH$_2$CHO & n & 0 & 1.00 &  150 & $<$  1.4 (14) & 1.07 & 3.0 & -0.9 \\ 
 CH$_3$CN & n & 0 & 1.00 &  150 & $<$  1.8 (14) & 1.0 & 3.0 & -0.9 \\ 
 C$_2$H$_5$CN & n & 0 & 1.00 &  150 & $<$  3.5 (14) & 1.38 & 3.0 & -0.9 \\ 
 (CH$_2$OH)$_2$ & n & 0 & 1.00 &  150 & $<$  2.4 (15) & 1.61 & 3.0 & -0.9 \\ 
 CH$_2$(OH)CHO & n & 0 & 1.00 &  150 & $<$  1.4 (15) & 1.04 & 3.0 & -0.9 \\ 
 HNCO & n & 0 & 1.00 &  150 & $<$  2.0 (14) & 1.01 & 3.0 & -0.9 \\ 
 NH$_2$CN & n & 0 & 1.00 &  150 & $<$  8.5 (13) & 1.70 & 3.0 & -0.9 \\ 
\hline 
 \multicolumn{9}{c}{L1157} 
 \smallskip \\ 
 CH$_3$OH & d & 10 & 0.25 &  200 &  1.3 (17) & 1.0 & 4.0 & 0.0 \\ 
 C$_2$H$_5$OH & n & 0 & 0.25 &  150 & $<$  3.3 (16) & 1.32 & 4.0 & 0.0 \\ 
 CH$_3$OCH$_3$ & n & 0 & 0.25 &  150 & $<$  1.5 (16) & 1.0 & 4.0 & 0.0 \\ 
 CH$_3$OCHO & d & 1 & 0.25 &  150 &  9.5 (15) & 1.19 & 4.0 & 0.8 \\ 
 CH$_3$CHO & n & 0 & 0.25 &  150 & $<$  5.1 (15) & 1.02 & 4.0 & 0.0 \\ 
 NH$_2$CHO & n & 0 & 0.25 &  150 & $<$  7.5 (14) & 1.07 & 4.0 & 0.0 \\ 
 CH$_3$CN & d & 1 & 0.25 &  150 &  4.6 (15) & 1.0 & 4.0 & 0.0 \\ 
 C$_2$H$_5$CN & n & 0 & 0.25 &  150 & $<$  2.8 (15) & 1.38 & 4.0 & 0.0 \\ 
 (CH$_2$OH)$_2$ & n & 0 & 0.25 &  150 & $<$  1.3 (16) & 1.61 & 4.0 & 0.0 \\ 
 CH$_2$(OH)CHO & n & 0 & 0.25 &  150 & $<$  7.3 (15) & 1.04 & 4.0 & 0.0 \\ 
 HNCO & n & 0 & 0.25 &  150 & $<$  1.0 (15) & 1.01 & 4.0 & 0.0 \\ 
 NH$_2$CN & n & 0 & 0.25 &  150 & $<$  5.1 (14) & 1.70 & 4.0 & 0.0 \\ 
\hline 
 \multicolumn{9}{c}{GF9-2} 
 \smallskip \\ 
 CH$_3$OH & n & 0 & 0.50 &  100 & $<$  1.5 (15) & 1.0 & 5.0 & 0.0 \\ 
 C$_2$H$_5$OH & n & 0 & 0.50 &  100 & $<$  3.8 (15) & 1.09 & 5.0 & 0.0 \\ 
 CH$_3$OCH$_3$ & n & 0 & 0.50 &  100 & $<$  1.0 (15) & 1.0 & 5.0 & 0.0 \\ 
 CH$_3$OCHO & n & 0 & 0.50 &  100 & $<$  5.2 (14) & 1.04 & 5.0 & 0.0 \\ 
 CH$_3$CHO & n & 0 & 0.50 &  100 & $<$  4.0 (14) & 1.00 & 5.0 & 0.0 \\ 
 NH$_2$CHO & n & 0 & 0.50 &  100 & $<$  7.1 (13) & 1.02 & 5.0 & 0.0 \\ 
 CH$_3$CN & n & 0 & 0.50 &  100 & $<$  2.5 (14) & 1.0 & 5.0 & 0.0 \\ 
 C$_2$H$_5$CN & n & 0 & 0.50 &  100 & $<$  2.2 (14) & 1.11 & 5.0 & 0.0 \\ 
 (CH$_2$OH)$_2$ & n & 0 & 0.50 &  100 & $<$  9.4 (14) & 1.17 & 5.0 & 0.0 \\ 
 CH$_2$(OH)CHO & n & 0 & 0.50 &  100 & $<$  4.0 (14) & 1.00 & 5.0 & 0.0 \\ 
 HNCO & n & 0 & 0.50 &  100 & $<$  1.0 (14) & 1.00 & 5.0 & 0.0 \\ 
 NH$_2$CN & n & 0 & 0.50 &  100 & $<$  5.3 (13) & 1.33 & 5.0 & 0.0 \\ 
\hline 
 \end{longtable}
 \tablefoot{
 \tablefoottext{a}{d: detection, t: tentative detection, n: nondetection.}
 \tablefoottext{b}{Number of detected lines. One line of a given species may mean a group of transitions of that species that are blended together.}
 \tablefoottext{c}{Source diameter (\textit{FWHM}).}
 \tablefoottext{d}{Rotational temperature.}
 \tablefoottext{e}{Total column density of the molecule. $X$ ($Y$) means $X \times 10^Y$. The upper limits represent the $3\sigma$ level.}
 \tablefoottext{f}{Correction factor that was applied to the column density to account for the contribution of vibrationally or torsionally excited states, in the cases where this contribution was not included in the partition function of the spectroscopic predictions. This factor was estimated in the harmonic approximation.}
 \tablefoottext{g}{Linewidth (\textit{FWHM}).}
 \tablefoottext{h}{Velocity offset with respect to the assumed systemic velocity given in Table~\ref{t:maporigins}.}
 }
 }

\subsection{Chemical composition}
\label{ss:chemcomp}

\begin{figure*}[!ht]
 \centerline{\resizebox{1.0\hsize}{!}{\includegraphics[angle=270]{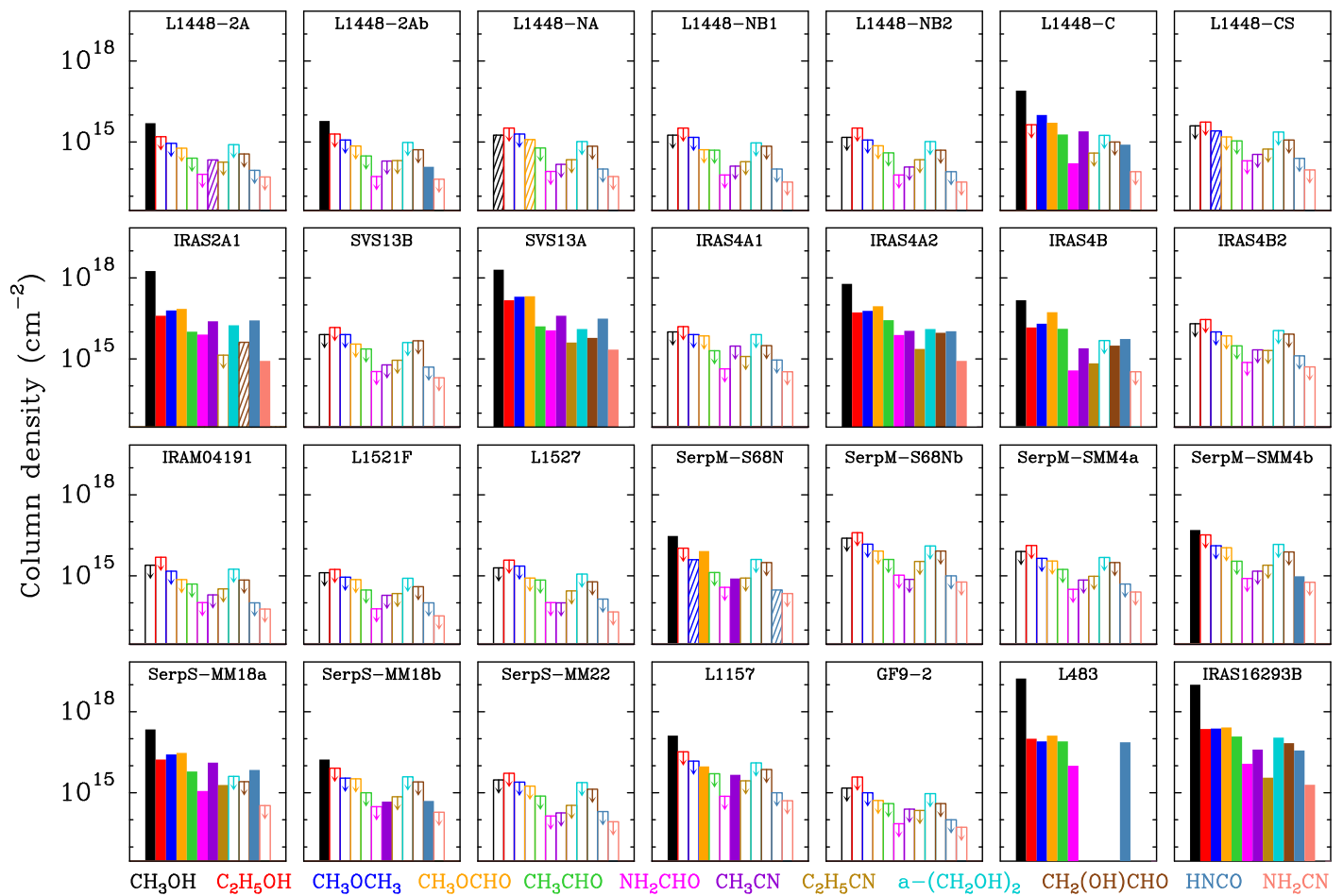}}}
 \caption{Column densities of (complex) organic molecules toward CALYPSO 
sources. The color coding of the molecules is indicated at the bottom. Secure 
detections are shown as plain bars, tentative detections as hatched bars, and 
upper limits as empty bars with a downward arrow. The name of each source is
indicated in each panel. The two rightmost panels in the bottom row show
the column densities of two other Class~0 protostars collected from the 
literature (see Sect.~\ref{ss:chemcomp} for references).}
 \label{f:chemcomp}
\end{figure*}

The chemical composition of all continuum peaks is displayed in terms of column 
densities in Fig.~\ref{f:chemcomp} for a selection of ten COMs and two simpler 
organic 
molecules, HNCO and NH$_2$CN. Twelve of the 26 analyzed CALYPSO sources are 
detected in methanol, plus one (L1448-NA) tentatively. Among these 12 sources,
nine have at least two COMs detected (CH$_3$OH and CH$_3$CN).
We also show in Fig.~\ref{f:chemcomp} the column densities of these 
molecules toward two other Class~0 protostars reported from ALMA 
interferometric measurements in the literature. We compiled the ALMA results of 
\citet{Jacobsen19} and \citet{Oya17} for L483 (distance 200~pc), and the ALMA 
results of \citet{Jacobsen19}, \citet{Calcutt18}, \citet{Jorgensen16}, 
\citet{Ligterink17}, and \citet{Coutens18} for IRAS16293B (distance 120~pc). 
The column density of methanol differs by up to at least four orders of 
magnitude, with L483 and IRAS16293B at the level of 10$^{19}$~cm$^{-2}$, and 
upper limits as low as 10$^{15}$~cm$^{-2}$ for a number of sources where no COMs
are detected (e.g., L1521F, GF9-2). The CALYPSO sources with the highest
column densities of methanol are IRAS2A1, SVS13A, and IRAS4A2
($\sim$$10^{18}$~cm$^{-2}$), followed by L1448C, IRAS4B, SerpM-SMM4b, SerpS-MM18a,
and L1157 ($\sim$$10^{17}$~cm$^{-2}$), and L1448-2A, L1448-2Ab, SerpM-S68N, and
SerpS-MM18b  ($\sim$$10^{16}$~cm$^{-2}$).

Because the size of the molecular emission varies from source to source and 
their H$_2$ column densities at these small scales may vary too, we 
renormalized the column densities of the CALYPSO sources to obtain a quantity 
that may better reflect the abundances of the molecules with respect to H$_2$. 
We used the ratio of the continuum flux density to the measured or assumed 
solid angle of the COM emission as a proxy for the H$_2$ column density at the
scale of the COM emission. We normalized the molecular column densities with 
this proxy. By doing this, we neglected possible differences in 
terms of temperature and dust properties between the sources. We computed the 
normalization using the continuum emission at 1.3~mm or 3~mm. At 1.3~mm 
\citep[from setups S1 and S2, see][]{Maury19}, we used either the peak flux 
density of the continuum emission if the solid angle of the COM emission is 
smaller than the beam, or the continuum flux density integrated over the COM 
emission size if it is larger than the beam. At 3~mm, we used the peak flux
density of the continuum emission.

Figure~\ref{f:chemcompsolang1mmcont} shows the column densities renormalized 
with the 1.3~mm continuum emission as described in the previous paragraph. 
Among the CALYPSO sources detected in methanol, two stand out with a high
``abundance'' of methanol
(\hbox{$\sim$4 $\times 10^{4}$~cm$^{-2}$~sr~mJy$^{-1}$}):
IRAS2A1 and SVS13A. On the contrary, the following sources have an
``abundance'' of methanol lower by more than one order of magnitude 
($\sim$10$^{3}$~cm$^{-2}$~sr~mJy$^{-1}$): L1448-NA (tentative), IRAS4B, 
SerpS-MM18b, and L1157. Seven sources lie in between: L1448-2A, L1448-2Ab, 
L1448-C, IRAS4A2, SerpM-S68N, SerpM-SMM4b, and SerpS-MM18a. Among the sources 
not detected in methanol, several stand out with a low upper limit of methanol 
``abundance'' ($< 3 \times 10^{2}$~cm$^{-2}$~sr~mJy$^{-1}$): L1448-NB1,
L1448-NB2, IRAS4A1, IRAS4B2, L1527, and SerpM-SMM4a. The remaining sources
with a 1.3~mm continuum detection have an upper limit in the range
$\sim$$3 \times 10^{2}$--10$^{4}$~cm$^{-2}$~sr~mJy$^{-1}$ (L1448-CS, SVS13B,
IRAM04191, L1521F, SeprS-MM22, GF9-2). We remind the reader, however, that
these ``abundance'' upper limits are sensitive to the emission size assumed to
derive the upper limits of the methanol column density. The sources remain
grouped in the same way when we consider their column densities renormalized
with the 3~mm continuum emission (see Fig.~\ref{f:chemcompsolang3mmcont}). This
suggests that the 1.3~mm and 3~mm continuum emissions trace to first order the
same dust (and gas) reservoir. SerpM-S68Nb is plotted in
Fig.~\ref{f:chemcompsolang3mmcont}: it falls into the category of sources with a
high upper limit of methanol ``abundance''.

\begin{sidewaystable}[!h]
 \begin{center}
 \caption{\label{t:chemcomp}
 Chemical composition relative to methanol of nine CALYPSO sources and two additional Class 0 protostars, and model results of \citet{Bergner17}.
}
 \begin{tabular}{lccccccccccc}
 \hline\hline
 \multicolumn{1}{c}{Source} & \multicolumn{11}{c}{Column density relative to CH$_3$OH} \\ 
  & \multicolumn{1}{c}{C$_2$H$_5$OH} & \multicolumn{1}{c}{CH$_3$OCH$_3$} & \multicolumn{1}{c}{CH$_3$OCHO} & \multicolumn{1}{c}{CH$_3$CHO} & \multicolumn{1}{c}{NH$_2$CHO} & \multicolumn{1}{c}{CH$_3$CN} & \multicolumn{1}{c}{C$_2$H$_5$CN} & \multicolumn{1}{c}{(CH$_2$OH)$_2$} & \multicolumn{1}{c}{CH$_2$(OH)CHO} & \multicolumn{1}{c}{HNCO} & \multicolumn{1}{c}{NH$_2$CN} \\ 
 \hline
IRAS2A1 &  2.2 ($-2$) &  3.5 ($-2$) &  3.9 ($-2$) &  5.7 ($-3$) &  4.4 ($-3$) &  1.4 ($-2$) & $<$  7.7 ($-4$) &  9.8 ($-3$) &  2.3 ($-3$)$^\star$ &  1.5 ($-2$) &  4.7 ($-4$)\\ 
L483 &  5.9 ($-3$) &  4.7 ($-3$) &  7.6 ($-3$) &  4.7 ($-3$) &  5.9 ($-4$) & n.a.  & n.a.  & n.a.  & n.a.  &  4.4 ($-3$) & n.a. \\ 
IRAS16293B &  2.3 ($-2$) &  2.4 ($-2$) &  2.6 ($-2$) &  1.2 ($-2$) &  1.2 ($-3$) &  4.0 ($-3$) &  3.6 ($-4$) &  1.1 ($-2$) &  6.8 ($-3$) &  3.7 ($-3$) &  2.0 ($-4$)\\ 
\textbf{Group 1} & \textbf{ 1.7 ($-2$)} & \textbf{ 2.1 ($-2$)} & \textbf{ 2.4 ($-2$)} & \textbf{ 7.5 ($-3$)} & \textbf{ 2.1 ($-3$)} & \textbf{ 8.9 ($-3$)} & \textbf{ 3.6 ($-4$)} & \textbf{ 1.0 ($-2$)} & \textbf{ 4.6 ($-3$)} & \textbf{ 7.6 ($-3$)} & \textbf{ 3.4 ($-4$)}\smallskip \\ 
\hline 
SVS13A &  7.4 ($-2$) &  1.0 ($-1$) &  1.0 ($-1$) &  8.0 ($-3$) &  5.8 ($-3$) &  2.0 ($-2$) &  2.0 ($-3$) &  6.4 ($-3$) &  3.0 ($-3$) &  1.6 ($-2$) &  1.1 ($-3$)\\ 
IRAS4A2 &  8.8 ($-2$) &  1.0 ($-1$) &  1.5 ($-1$) &  4.5 ($-2$) &  1.3 ($-2$) &  1.8 ($-2$) &  3.9 ($-3$) &  2.1 ($-2$) &  1.6 ($-2$) &  1.8 ($-2$) &  1.4 ($-3$)\\ 
IRAS4B &  9.7 ($-2$) &  1.3 ($-1$) &  3.6 ($-1$) &  8.6 ($-2$) &  2.5 ($-3$) &  1.7 ($-2$) &  4.6 ($-3$) & $<$  3.2 ($-2$) &  2.1 ($-2$) &  3.6 ($-2$) & $<$  2.3 ($-3$)\\ 
SerpM-S68N & $<$  3.5 ($-1$) &  1.3 ($-1$)$^\star$ &  2.8 ($-1$) & $<$  4.4 ($-2$) & $<$  1.2 ($-2$) &  2.7 ($-2$) & $<$  2.8 ($-2$) & $<$  1.3 ($-1$) & $<$  1.0 ($-1$) &  1.0 ($-2$)$^\star$ & $<$  7.4 ($-3$)\\ 
\textbf{Group 2} & \textbf{ 8.6 ($-2$)} & \textbf{ 1.2 ($-1$)} & \textbf{ 2.2 ($-1$)} & \textbf{ 4.6 ($-2$)} & \textbf{ 7.0 ($-3$)} & \textbf{ 2.0 ($-2$)} & \textbf{ 3.5 ($-3$)} & \textbf{ 1.4 ($-2$)} & \textbf{ 1.3 ($-2$)} & \textbf{ 2.0 ($-2$)} & \textbf{ 1.3 ($-3$)}\smallskip \\ 
\hline 
L1448-2A & $<$  3.2 ($-1$) & $<$  1.8 ($-1$) & $<$  1.2 ($-1$) & $<$  5.1 ($-2$) & $<$  1.3 ($-2$) &  4.4 ($-2$)$^\star$ & $<$  3.6 ($-2$) & $<$  1.6 ($-1$) & $<$  7.3 ($-2$) & $<$  1.8 ($-2$) & $<$  1.0 ($-2$)\\ 
L1448-C & $<$  5.5 ($-2$) &  1.3 ($-1$) &  6.5 ($-2$) &  2.4 ($-2$) &  2.0 ($-3$) &  3.1 ($-2$) & $<$  4.9 ($-3$) & $<$  2.2 ($-2$) & $<$  1.3 ($-2$) &  1.0 ($-2$) & $<$  1.0 ($-3$)\\ 
SerpS-MM18a &  7.8 ($-2$) &  1.2 ($-1$) &  1.4 ($-1$) &  2.8 ($-2$) &  5.3 ($-3$) &  5.9 ($-2$) &  8.8 ($-3$) & $<$  1.8 ($-2$) & $<$  1.2 ($-2$) &  3.2 ($-2$) & $<$  1.5 ($-3$)\\ 
L1157 & $<$  2.5 ($-1$) & $<$  1.2 ($-1$) &  7.3 ($-2$) & $<$  3.9 ($-2$) & $<$  5.8 ($-3$) &  3.5 ($-2$) & $<$  2.1 ($-2$) & $<$  9.9 ($-2$) & $<$  5.6 ($-2$) & $<$  7.8 ($-3$) & $<$  3.9 ($-3$)\\ 
\textbf{Group 3} & \textbf{ 7.8 ($-2$)} & \textbf{ 1.2 ($-1$)} & \textbf{ 9.1 ($-2$)} & \textbf{ 2.6 ($-2$)} & \textbf{ 3.7 ($-3$)} & \textbf{ 4.2 ($-2$)} & \textbf{ 8.8 ($-3$)} & --  & --  & \textbf{ 2.1 ($-2$)} & -- \smallskip \\ 
\hline 
\hline 
\multicolumn{1}{c}{Model} & \multicolumn{11}{l}{\citep{Bergner17}} \\ \hline 
1 $L_\odot$ & -- & 6($-$3) & 4($-$3) & 1.5($-$4) & -- & 1.3($-$4) & -- & -- & -- & 1.9($-5$) & -- \\ 10 $L_\odot$ & -- & 6($-$3) & 4($-$3) & 1.3($-$4) & -- & 1.3($-$4) & -- & -- & -- & 4($-5$) & -- \\ \hline 
 \end{tabular}
 \end{center}
 \vspace*{-2.5ex}
 \tablefoot{The values in bold face are mean values over the group of sources that precedes. $X$ ($Y$) means $X \times 10^Y$. Tentative detections are indicated with a star symbol. n.a. means not available, that is the column density of the molecule was not reported in the articles we compiled for L483 and IRAS16293B (see references in Sect.~\ref{ss:chemcomp}).
 }
 \end{sidewaystable}

\begin{figure*}[!ht]
 \centerline{\resizebox{1.0\hsize}{!}{\includegraphics[angle=270]{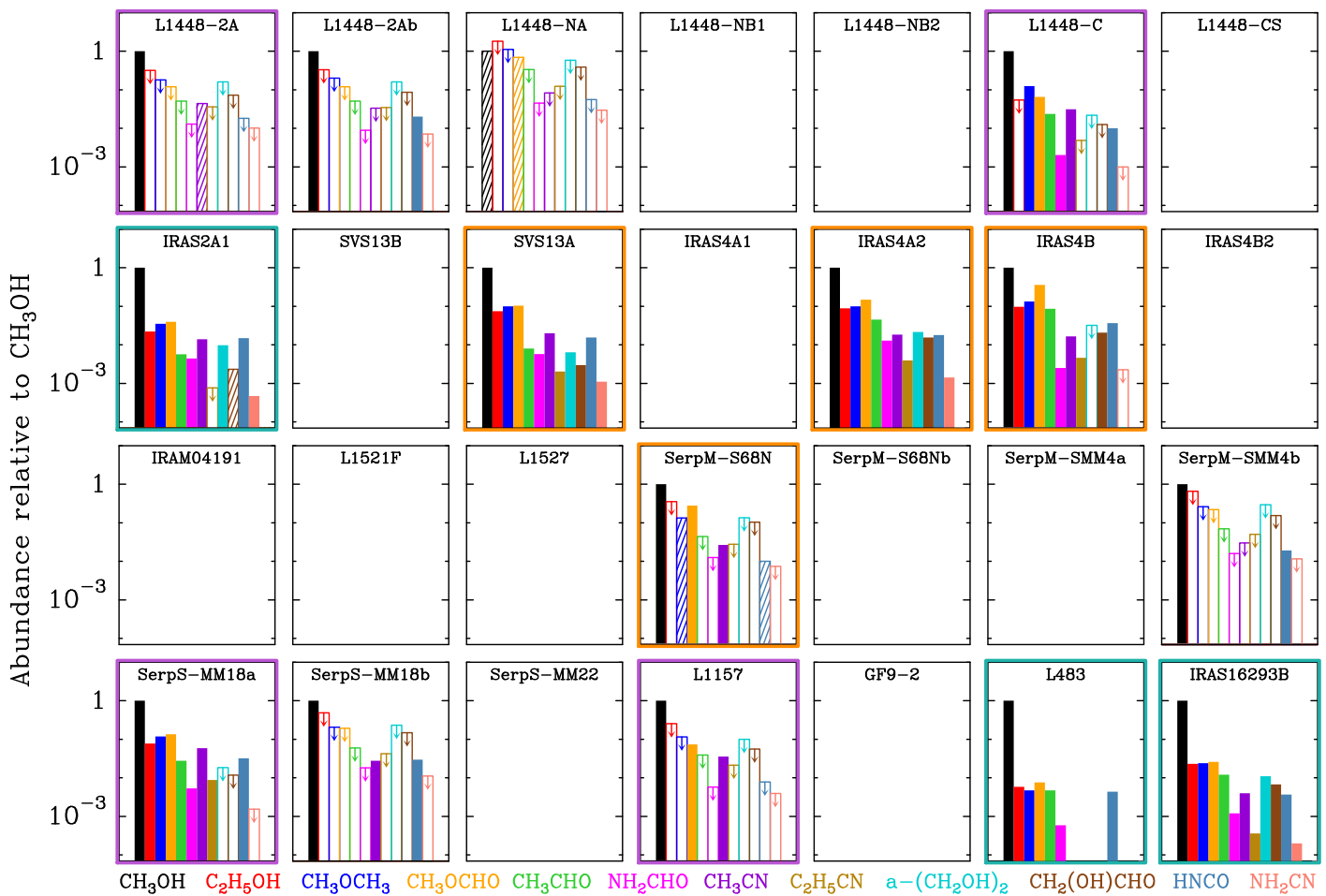}}}
 \caption{Same as Fig.~\ref{f:chemcomp} for the molecular abundance relative 
to methanol. Sources where methanol is not detected have an empty panel. 
The colored boxes highlight three groups of sources defined on the basis of 
similarities in their chemical composition: group 1 in blue, group 2 in 
orange, and group 3 in magenta.}
 \label{f:chemcompch3oh}
\end{figure*}

Because the previous normalizations may not represent robust proxies of the
molecular abundances relative to H$_2$, we plot in Fig.~\ref{f:chemcompch3oh}
the abundances relative to methanol. These relative abundances are summarized
in Table~\ref{t:chemcomp}. Three groups of sources stand out on the basis of
the abundances of oxygen-bearing molecules, cyanides, and CHO-bearing 
molecules relative to methanol. Group 1
(highlighted in blue) includes IRAS2A1, L483, and IRAS16293B. It is
characterized by low abundances of COMs relative to methanol, 
in particular C$_2$H$_5$OH, CH$_3$OCH$_3$, CH$_3$OCHO, and CH$_3$CHO ($<$0.02 on
average). Group 2 (in orange) includes SVS13A, IRAS4A2, IRAS4B, and SerpM-S68N.
It has abundances of these four oxygen-bearing molecules relative to methanol
higher by a factor $\sim$6 with respect to group 1. The other molecules are
enhanced by a smaller factor ($\sim$2--3 on average), except for
C$_2$H$_5$CN which is enhanced by a factor $\sim$10 with respect to group 1.
Finally, group 3 (in magenta) includes L1448-2A, L1448C, SerpS-MM18a, and
L1157. It is similar to group 2, but it has average abundances of CH$_3$CN and
C$_2$H$_5$CN relative to methanol enhanced by a further factor $\sim$2 and
average abundances of CH$_3$OCHO, CH$_3$CHO, and NH$_2$CHO relative to
methanol reduced by a factor $\sim$2 with respect to group 2.

We have assigned L1448-2A to group 3 because of its high CH$_3$CN abundance
relative to methanol. The CH$_3$CN detection being only tentative, the
assignment of L1448-2A to this group is also tentative. No other COM is
detected toward this source, but the upper limits are consistent with the
average abundances relative to methanol derived for this group.

Both HNCO and CH$_3$CN are detected toward SerpS-MM18b, with abundances
relative to methanol of 0.029 and 0.028, respectively. The abundance of HNCO
is similar to the average abundances of both groups 2 and 3, and the abundance
of CH$_3$CN lies in between these groups. Without other COM detections, it is
thus difficult to tell if this source belongs to group 2 or group 3. HNCO is
detected toward L1448-2Ab and SerpM-SMM4b with abundances relative to methanol
of 0.020 and 0.019, respectively, which are similar to the average abundances
of both groups 2 and 3 and a factor $\sim$3 higher than the one of group 1.
However, no COM apart from methanol is detected toward these sources, which
prevents us to distinguish between groups 2 and 3. The status of L1448-NA with
CH$_3$OCHO tentatively as abundant as methanol but no other COM detected is
unclear.

\section{Analysis: search for correlations}
\label{s:correlations}

\subsection{Correlations between COM abundances}
\label{ss:correl_coms}

\begin{figure*}
 \centerline{\resizebox{0.85\hsize}{!}{\includegraphics[angle=0]{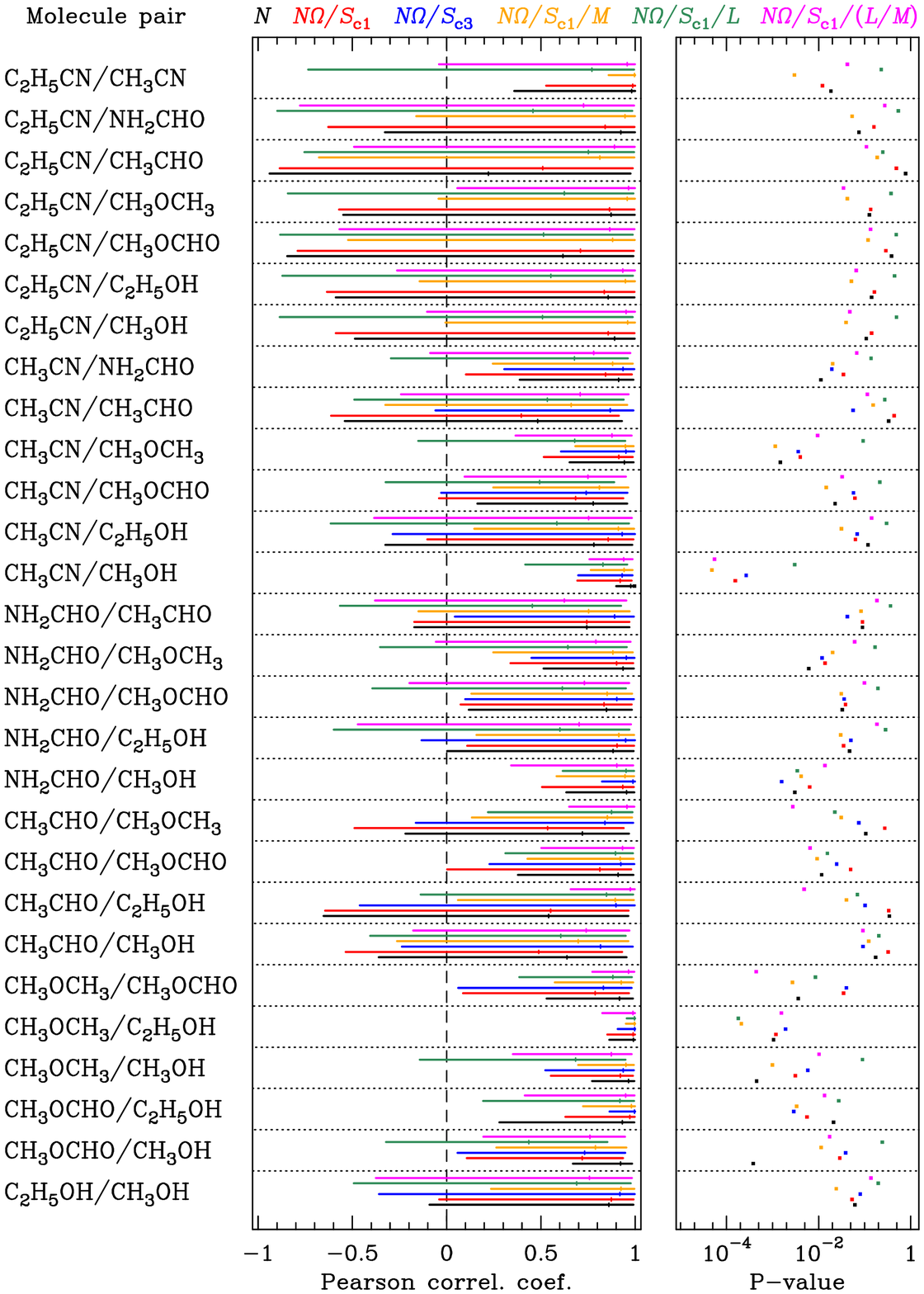}}}
 \caption{Pearson correlation coefficients with 95\% confidence interval 
(\textit{left}) and P-value (\textit{right}) for various pairs of COMs for
the CALYPSO source sample. The P-value is the probability of observing a 
correlation plot under the null hypothesis of no correlation. These parameters 
were derived from the plots shown 
in Figs.~\ref{f:correl_coldens}--\ref{f:correl_normsolangcontfluxmenvlint}.
The colors indicate the variables used to evaluate the degree of correlation
between molecules: column density (black), column density multiplied by the
solid angle $\Omega$ of the COM emission and divided by either the 1.3~mm
continuum peak flux density or the 1.3~mm continuum flux density integrated
over the size of the COM emission, $S_{\rm c1}$ (red), colum density multiplied
by the solid angle of the COM emission and divided by the 3~mm 
continuum peak flux density (blue), column density multiplied by
$\Omega/S_{\rm c1}$ and divided by the envelope mass (orange), column density
multiplied by $\Omega/S_{\rm c1}$ and divided by the internal luminosity
(green), column density multiplied by $\Omega/S_{\rm c1}$ and divided by the
ratio of internal luminosity to envelope mass (magenta).}
 \label{f:pearson}
\end{figure*}

To investigate the chemical composition of the CALYPSO sources one step further,
we show in
Figs.~\ref{f:correl_coldens}--\ref{f:correl_normsolangcontfluxmenvlint}
correlation plots for all pairs of COMs that have at least four detections,
except for CH$_2$(OH)CHO. For this analysis, we considered the 20 sources that 
have an envelope mass and an internal luminosity listed in 
Table~\ref{t:sources}. We normalized the column densities in a different way 
in each figure, as stated at the bottom of the figure. The Pearson 
correlation coefficients and their 95\% confidence intervals are indicated in 
each correlation plot and displayed in the lower left panel of each figure
\citep[see, e.g.,][for the calculation of confidence levels of correlation 
coefficients via a Fisher transformation]{Edwards76}. 
They are also summarized for all normalizations in Fig.~\ref{f:pearson} and 
in Table~\ref{t:correl}.

\begin{figure*}
 \centerline{\resizebox{0.8\hsize}{!}{\includegraphics[angle=270]{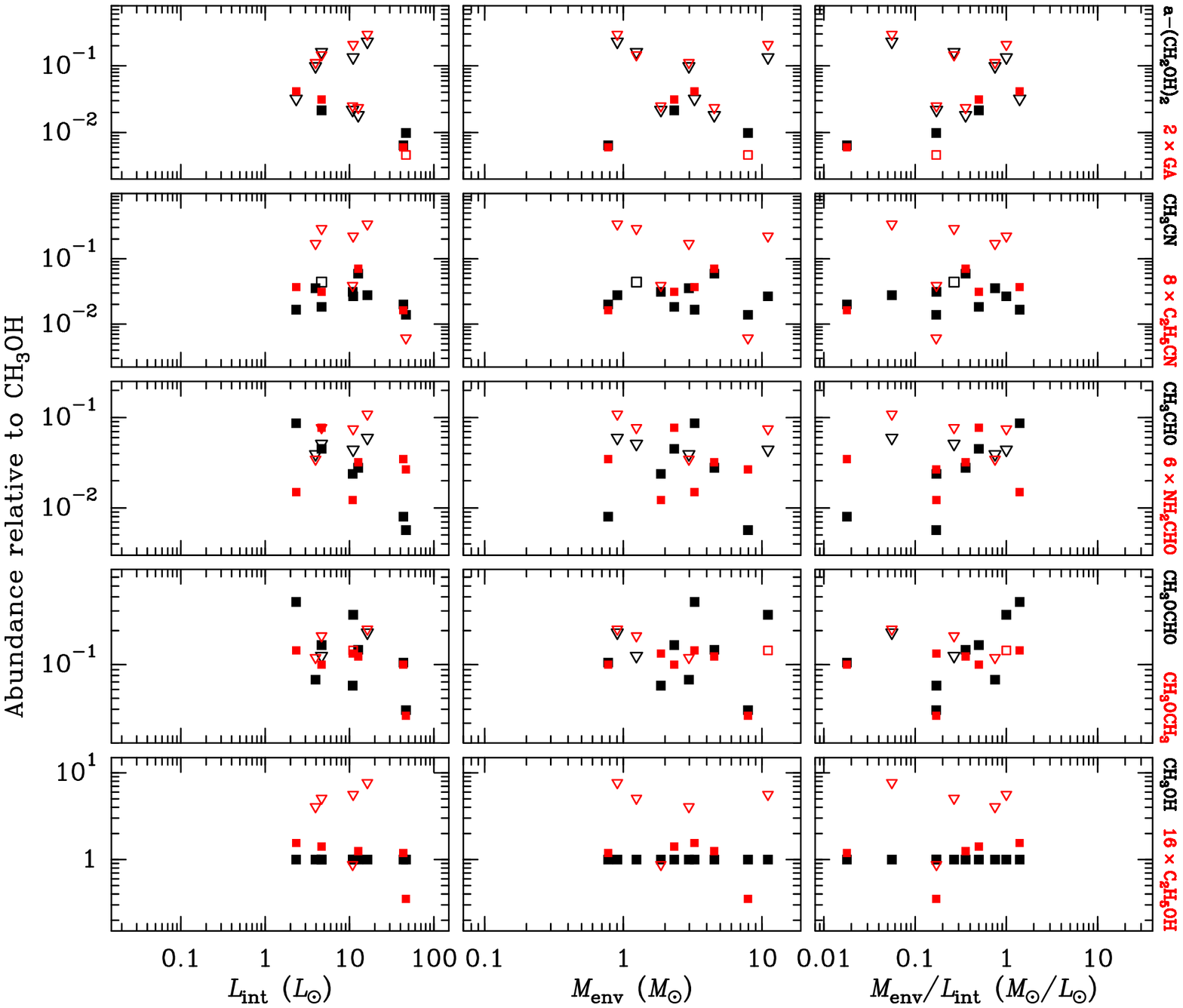}}}
 \caption{Abundances of ten complex organic molecules relative to methanol as
a function of internal luminosity, envelope mass, and their ratio. The
abundances of some molecules were multiplied by a scaling factor as indicated
on the right. Filled and open squares represent robust and tentative
detections, respectively, while open triangles show upper limits. GA stands
for glycolaldehyde, CH$_2$(OH)CHO.}
 \label{f:coldens_normch3oh}
\end{figure*}

Most pairs of COMs (25 out 28) have a Pearson correlation coefficient higher
than 0.6 when we consider their column densities, suggesting that they are to
some degree all correlated with each other. However, the low number of
detections implies a large 95\% confidence interval in many cases, meaning
that an absence of correlation is not ruled out in these cases. A more robust
correlation is found for 11 pairs out of
28, with a 95\% confidence interval that does not extend below 0.3 (numbers in
bold face in Table~\ref{t:correl}). The results are similar when we normalize
the column densities by the 1.3~mm or 3~mm proxies for H$_2$ introduced in
Sect.~\ref{ss:chemcomp}: 23 out of 28 pairs and 21 out of 21 pairs have a
Pearson correlation coefficient higher than 0.6 for the 1.3~mm and 3~mm
normalizations, respectively, and 8 out of 28 pairs and 8 out of 21 pairs,
respectively, have a 95\% confidence interval that does not extend below 0.3.
The degree of correlation is overall poorer when we normalize the
``abundances'' with the internal luminosity while it is higher when the
normalization is done with the envelope mass or the ratio of internal
luminosity to envelope mass: the previous numbers (23/8 out of 28) fall to
19/5 in the first case but increase to 28/9 and 28/10 in the latter cases.
Given that both normalization factors, envelope mass and ratio of internal
luminosity to envelope mass, have a dynamical range of two orders of
magnitude, they may introduce a bias that increases the degree of correlation
of the normalized abundances.

The best correlations with a narrow 95\% confidence interval (that is an
interval ending not lower than $\sim$0.6), whatever
the type of normalization, occur for the following pairs of COMs:
CH$_3$CN/CH$_3$OH, NH$_2$CHO/CH$_3$OH, and CH$_3$OCH$_3$/C$_2$H$_5$OH,
followed with a lower degree of confidence by CH$_3$CN/CH$_3$OCH$_3$, 
CH$_3$CHO/CH$_3$OCHO, CH$_3$OCH$_3$/CH$_3$OCHO, CH$_3$OCH$_3$/CH$_3$OH, and
CH$_3$OCHO/C$_2$H$_5$OH.

Coming back now to the individual correlation plots, for instance in 
Fig.~\ref{f:correl_normsolangcontflux}, we can pay attention to the 
non-detections that were not taken into account in the correlation analysis.
None of the non-detections in the correlation plots of CH$_3$CN/CH$_3$OH and
NH$_2$CHO/CH$_3$OH is inconsistent with the correlations noted above. In the
correlation plot of CH$_3$OCH$_3$/C$_2$H$_5$OH, the data point corresponding
to L1448-C with a detection of CH$_3$OCH$_3$ but an upper limit for
C$_2$H$_5$OH is marginally consistent with the correlation. In the correlation
plot of CH$_3$CN/CH$_3$OCH$_3$, the data point corresponding to L1157 with a
detection of CH$_3$CN but an upper limit for CH$_3$OCH$_3$ is marginally
consistent with the correlation. In the correlation plot of
CH$_3$OCH$_3$/CH$_3$OH, the data point corresponding to L1448-CS with a
tentative detection of CH$_3$OCH$_3$ but an upper limit for CH$_3$OH is
largely inconsistent with the correlation. This casts some doubts on the
tentative detection of CH$_3$OCH$_3$ in this source, which relies on 
only one line just below the 3$\sigma$ level. For the three other pairs
mentioned in the previous paragraph, the upper limits are not inconsistent
with the correlations.

\subsection{Correlations between molecules and source properties}
\label{ss:correl_prop}

\begin{figure*}
 \centerline{\resizebox{0.6\hsize}{!}{\includegraphics[angle=0]{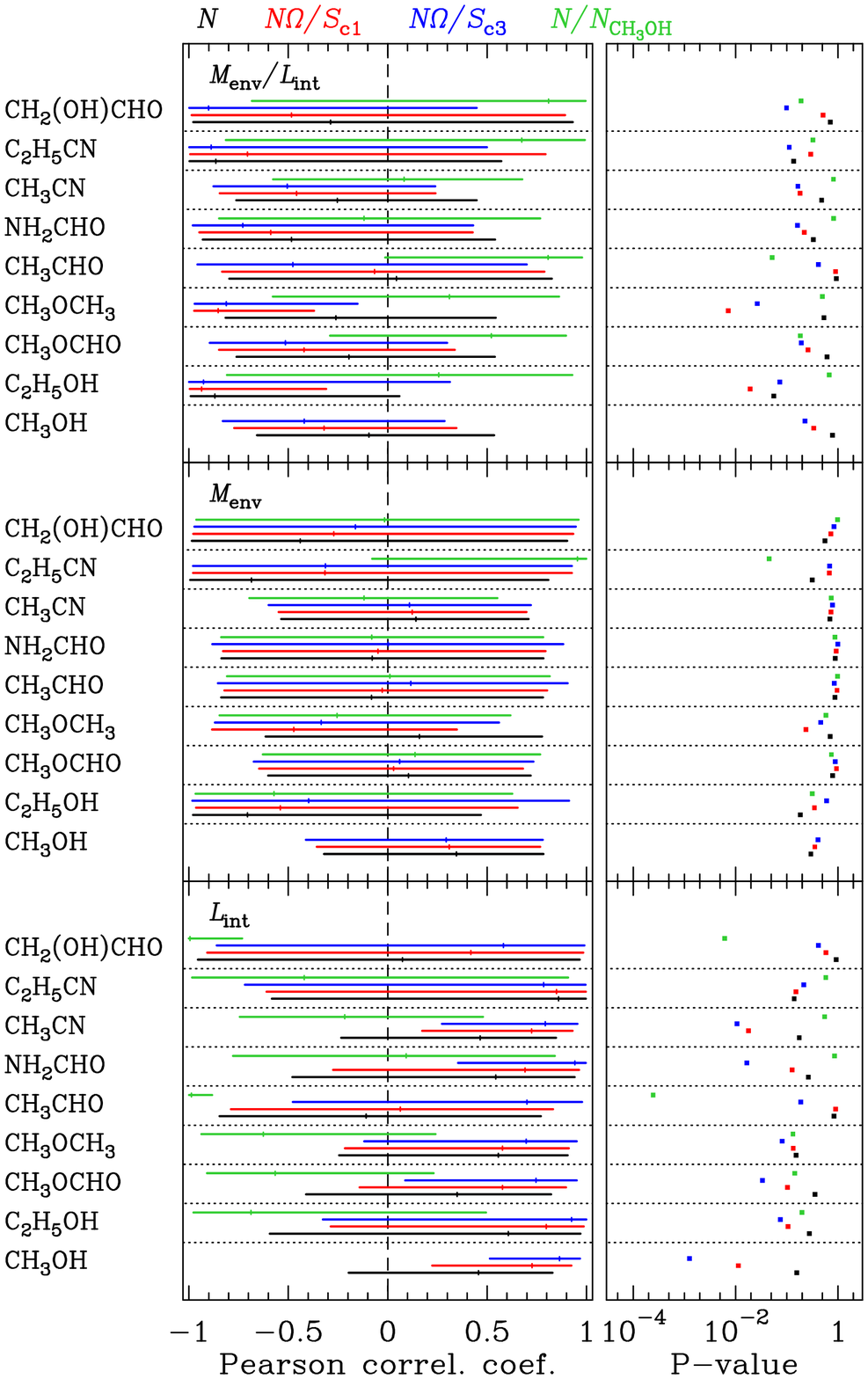}}}
 \caption{Pearson correlation coefficients with 95\% confidence interval 
(\textit{left}) and P-value (\textit{right}) between nine COMs and several 
source properties for the CALYPSO source sample. These parameters were derived 
from the plots shown in Figs.~\ref{f:coldens}--\ref{f:coldens_normsolangcont3} 
and \ref{f:coldens_normch3oh}. The colors indicate the variables used to 
evaluate the degree of correlation between the molecules and the source 
properties: column density (black), column density multiplied by the solid 
angle $\Omega$ of the COM emission and divided by either the 1.3~mm continuum
peak flux density or the 1.3~mm continuum flux density integrated over the size
of the COM emission, $S_{\rm c1}$ (red), column density multiplied by the solid 
angle $\Omega$ of the COM emission and divided by the 3~mm continuum peak flux 
density, $S_{\rm c3}$ (blue), column density normalized to that of CH$_3$OH 
(green). The source properties investigated here are the internal luminosity 
(\textit{bottom}), the envelope mass (\textit{middle}), and the ratio envelope 
mass over internal luminosity (\textit{top}).}
 \label{f:pearson_coldens}
\end{figure*}

\begin{table*}
 \begin{center}
 \caption{
 Correlations between column densities of ten COMs and properties of the CALYPSO sources.
}
 \label{t:correl_coldens}
 \vspace*{-1.2ex}
 \begin{tabular}{llccccccccc}
 \hline\hline
 \multicolumn{1}{c}{Prop.\tablefootmark{a}} & \multicolumn{1}{c}{Molecule} & \multicolumn{1}{c}{$N_{\rm pts}$\tablefootmark{b}} & \multicolumn{2}{c}{$N$\tablefootmark{c}} & \multicolumn{2}{c}{$N\Omega/S_{\rm c1}$\tablefootmark{d}} & \multicolumn{2}{c}{$N\Omega/S_{\rm c3}$\tablefootmark{e}} & \multicolumn{2}{c}{$N/N_{\rm CH3OH}$\tablefootmark{f}} \\ 
 \multicolumn{1}{c}{} & & & \multicolumn{1}{c}{$100\times\rho$} & \multicolumn{1}{c}{$P$} & \multicolumn{1}{c}{$100\times\rho$} & \multicolumn{1}{c}{$P$} & \multicolumn{1}{c}{$100\times\rho$} & \multicolumn{1}{c}{$P$} & \multicolumn{1}{c}{$100\times\rho$} & \multicolumn{1}{c}{$P$} \\ 
 \hline
\noalign{\smallskip} 
$L_{\rm int}$ & CH$_3$OH & 11 & $ 46^{+ 37}_{- 65}$ & 2(-1) & $ 73^{+ 20}_{- 50}$ & 1(-2) & $\mathbf{ 86}^{+ 10}_{- 35}$ & 1(-3) & -- & -- \\ 
\noalign{\smallskip} 
 & C$_2$H$_5$OH & 5 & $ 61^{+ 36}_{-120}$ & 3(-1) & $ 80^{+ 19}_{-108}$ & 1(-1) & $ 92^{+  7}_{-125}$ & 8(-2) & $-69^{+118}_{- 29}$ & 2(-1) \\ 
\noalign{\smallskip} 
 & CH$_3$OCHO & 9 & $ 35^{+ 47}_{- 76}$ & 4(-1) & $ 58^{+ 32}_{- 72}$ & 1(-1) & $ 75^{+ 20}_{- 66}$ & 3(-2) & $-57^{+ 80}_{- 34}$ & 1(-1) \\ 
\noalign{\smallskip} 
 & CH$_3$OCH$_3$ & 8 & $ 56^{+ 35}_{- 80}$ & 2(-1) & $ 58^{+ 33}_{- 79}$ & 1(-1) & $ 70^{+ 25}_{- 81}$ & 8(-2) & $-63^{+ 87}_{- 31}$ & 1(-1) \\ 
\noalign{\smallskip} 
 & CH$_3$CHO & 6 & $-11^{+ 88}_{- 74}$ & 8(-1) & $  6^{+ 77}_{- 85}$ & 9(-1) & $ 70^{+ 28}_{-118}$ & 2(-1) & $\mathbf{-99}^{+ 10}_{-  1}$ & 2(-4) \\ 
\noalign{\smallskip} 
 & NH$_2$CHO & 6 & $ 54^{+ 40}_{-102}$ & 3(-1) & $ 69^{+ 27}_{- 97}$ & 1(-1) & $\mathbf{ 94}^{+  5}_{- 59}$ & 2(-2) & $  9^{+ 75}_{- 87}$ & 9(-1) \\ 
\noalign{\smallskip} 
 & CH$_3$CN & 10 & $ 46^{+ 38}_{- 70}$ & 2(-1) & $ 72^{+ 21}_{- 55}$ & 2(-2) & $ 79^{+ 16}_{- 52}$ & 1(-2) & $-22^{+ 69}_{- 53}$ & 5(-1) \\ 
\noalign{\smallskip} 
 & C$_2$H$_5$CN & 4 & $ 86^{+ 14}_{-144}$ & 1(-1) & $ 85^{+ 15}_{-146}$ & 2(-1) & $ 78^{+ 21}_{-150}$ & 2(-1) & $-42^{+133}_{- 56}$ & 6(-1) \\ 
\noalign{\smallskip} 
 & a-(CH$_2$OH)$_2$ & 3 & -- & -- & -- & -- & -- & -- & -- & -- \\ 
\noalign{\smallskip} 
 & CH$_2$(OH)CHO & 4 & $  7^{+ 89}_{-103}$ & 9(-1) & $ 42^{+ 56}_{-133}$ & 6(-1) & $ 58^{+ 41}_{-144}$ & 4(-1) & $\mathbf{-99}^{+ 26}_{-  1}$ & 6(-3) \\ 
\noalign{\smallskip} 
\hline 
\noalign{\smallskip} 
$M_{\rm env}$ & CH$_3$OH & 11 & $ 35^{+ 44}_{- 67}$ & 3(-1) & $ 31^{+ 46}_{- 67}$ & 4(-1) & $ 29^{+ 48}_{- 71}$ & 4(-1) & -- & -- \\ 
\noalign{\smallskip} 
 & C$_2$H$_5$OH & 5 & $-70^{+117}_{- 27}$ & 2(-1) & $-54^{+119}_{- 42}$ & 3(-1) & $-40^{+131}_{- 59}$ & 6(-1) & $-57^{+120}_{- 39}$ & 3(-1) \\ 
\noalign{\smallskip} 
 & CH$_3$OCHO & 9 & $ 10^{+ 61}_{- 71}$ & 8(-1) & $  3^{+ 65}_{- 68}$ & 9(-1) & $  6^{+ 67}_{- 73}$ & 9(-1) & $ 14^{+ 63}_{- 77}$ & 7(-1) \\ 
\noalign{\smallskip} 
 & CH$_3$OCH$_3$ & 8 & $ 16^{+ 62}_{- 77}$ & 7(-1) & $-47^{+ 82}_{- 41}$ & 2(-1) & $-33^{+ 89}_{- 53}$ & 5(-1) & $-25^{+ 87}_{- 59}$ & 6(-1) \\ 
\noalign{\smallskip} 
 & CH$_3$CHO & 6 & $ -8^{+ 86}_{- 76}$ & 9(-1) & $ -3^{+ 83}_{- 79}$ & 1(0) & $ 12^{+ 79}_{- 97}$ & 9(-1) & $  1^{+ 80}_{- 82}$ & 1(0) \\ 
\noalign{\smallskip} 
 & NH$_2$CHO & 6 & $ -8^{+ 86}_{- 76}$ & 9(-1) & $ -5^{+ 84}_{- 78}$ & 9(-1) & $  0^{+ 88}_{- 88}$ & 1(0) & $ -8^{+ 86}_{- 76}$ & 9(-1) \\ 
\noalign{\smallskip} 
 & CH$_3$CN & 10 & $ 14^{+ 57}_{- 68}$ & 7(-1) & $ 12^{+ 58}_{- 67}$ & 7(-1) & $ 11^{+ 61}_{- 71}$ & 8(-1) & $-12^{+ 67}_{- 58}$ & 7(-1) \\ 
\noalign{\smallskip} 
 & C$_2$H$_5$CN & 4 & $-69^{+149}_{- 31}$ & 3(-1) & $-32^{+124}_{- 66}$ & 7(-1) & $-31^{+124}_{- 67}$ & 7(-1) & $ 95^{+  4}_{-103}$ & 5(-2) \\ 
\noalign{\smallskip} 
 & a-(CH$_2$OH)$_2$ & 3 & -- & -- & -- & -- & -- & -- & -- & -- \\ 
\noalign{\smallskip} 
 & CH$_2$(OH)CHO & 4 & $-44^{+134}_{- 54}$ & 6(-1) & $-27^{+120}_{- 71}$ & 7(-1) & $-16^{+111}_{- 81}$ & 8(-1) & $ -2^{+ 98}_{- 95}$ & 1(0) \\ 
\noalign{\smallskip} 
\hline 
\noalign{\smallskip} 
$M_{\rm env}$/$L_{\rm int}$ & CH$_3$OH & 11 & $ -9^{+ 63}_{- 56}$ & 8(-1) & $-32^{+ 67}_{- 45}$ & 3(-1) & $-42^{+ 71}_{- 41}$ & 2(-1) & -- & -- \\ 
\noalign{\smallskip} 
 & C$_2$H$_5$OH & 5 & $-87^{+ 93}_{- 12}$ & 6(-2) & $\mathbf{-94}^{+ 63}_{-  6}$ & 2(-2) & $-93^{+124}_{-  7}$ & 7(-2) & $ 26^{+ 67}_{-107}$ & 7(-1) \\ 
\noalign{\smallskip} 
 & CH$_3$OCHO & 9 & $-20^{+ 73}_{- 57}$ & 6(-1) & $-42^{+ 76}_{- 43}$ & 3(-1) & $-51^{+ 81}_{- 38}$ & 2(-1) & $ 52^{+ 38}_{- 81}$ & 2(-1) \\ 
\noalign{\smallskip} 
 & CH$_3$OCH$_3$ & 8 & $-26^{+ 80}_{- 56}$ & 5(-1) & $\mathbf{-85}^{+ 48}_{- 12}$ & 7(-3) & $-81^{+ 66}_{- 16}$ & 3(-2) & $ 31^{+ 55}_{- 89}$ & 5(-1) \\ 
\noalign{\smallskip} 
 & CH$_3$CHO & 6 & $  4^{+ 78}_{- 84}$ & 9(-1) & $ -7^{+ 85}_{- 77}$ & 9(-1) & $-48^{+118}_{- 48}$ & 4(-1) & $ 81^{+ 17}_{- 82}$ & 5(-2) \\ 
\noalign{\smallskip} 
 & NH$_2$CHO & 6 & $-48^{+102}_{- 45}$ & 3(-1) & $-59^{+102}_{- 36}$ & 2(-1) & $-73^{+116}_{- 25}$ & 2(-1) & $-12^{+ 89}_{- 73}$ & 8(-1) \\ 
\noalign{\smallskip} 
 & CH$_3$CN & 10 & $-25^{+ 70}_{- 51}$ & 5(-1) & $-46^{+ 70}_{- 39}$ & 2(-1) & $-51^{+ 74}_{- 37}$ & 2(-1) & $  8^{+ 59}_{- 66}$ & 8(-1) \\ 
\noalign{\smallskip} 
 & C$_2$H$_5$CN & 4 & $-86^{+144}_{- 13}$ & 1(-1) & $-71^{+150}_{- 29}$ & 3(-1) & $-89^{+139}_{- 11}$ & 1(-1) & $ 68^{+ 32}_{-149}$ & 3(-1) \\ 
\noalign{\smallskip} 
 & a-(CH$_2$OH)$_2$ & 3 & -- & -- & -- & -- & -- & -- & -- & -- \\ 
\noalign{\smallskip} 
 & CH$_2$(OH)CHO & 4 & $-29^{+122}_{- 69}$ & 7(-1) & $-48^{+138}_{- 50}$ & 5(-1) & $-90^{+135}_{- 10}$ & 1(-1) & $ 81^{+ 19}_{-149}$ & 2(-1) \\ 
\noalign{\smallskip} 
\hline 
 \end{tabular}
 \end{center}
 \vspace*{-2.5ex}
 \tablefoot{
 \tablefoottext{a}{Source property: internal luminosity ($L_{\rm int}$), envelope mass ($M_{\rm env}$), and ratio envelope mass over internal luminosity.}
 \tablefoottext{b}{Number of CALYPSO sources detected or tentatively detected. Pearson correlations are evaluated for the following quantities, only when at least four sources are detected ($N_{\rm pts}$ > 3):}
 \tablefoottext{c}{column density;}
 \tablefoottext{d}{column density times solid angle of COM emission divided by continuum flux density at 1.3~mm;}
 \tablefoottext{e}{column density times solid angle of COM emission divided by continuum peak flux density at 3~mm;}
 \tablefoottext{f}{column density normalized to that of CH$_3$OH;}
 For each type of correlation, $\rho$ is the Pearson correlation coefficient with its 95\% confidence interval, and $P$ is the P-value. $X$($Y$) means $X \times 10^Y$. Pearson coefficients with a confidence interval outside [$-$0.3, 0.3] are highlighted in bold face.}
 \end{table*}

We now search for correlations between the COM column densities and the
following source properties: internal luminosity, envelope mass, and ratio of
envelope mass to internal luminosity that has been proposed as an evolutionary
indicator decreasing with time \citep[][]{Andre00}. We test four different 
normalizations
of the column densities: column densities (Fig.~\ref{f:coldens}), column
densities multiplied by the solid angle of the COM emission and divided by the
continuum flux density at 1.3~mm (Fig.~\ref{f:coldens_normsolangcont}) and
3~mm (Fig.~\ref{f:coldens_normsolangcont3}), and abundances relative to
methanol (Fig.~\ref{f:coldens_normch3oh}). The Pearson correlation
coefficients of all individual plots and their 95\% confidence level are shown
in Fig.~\ref{f:pearson_coldens} and listed in Table~\ref{t:correl_coldens}.

Figure~\ref{f:coldens} reveals that no COMs are detected for sources with an 
internal luminosity lower than 2~$L_\odot$, while there is no such obvious
threshold in envelope mass for the range explored with the CALYPSO survey 
(0.2--20~$M_\odot$). We notice also that no COMs are detected for sources with
$M_{\rm env}/L_{\rm int}$ higher than 1.5~$M_\odot/L_\odot$. However the opposite is
not true: there are sources with internal luminosities higher than 
2~$L_\odot$ or with $M_{\rm env}/L_{\rm int}$ lower than 1.5~$M_\odot/L_\odot$ that
have no COM detection.

\begin{figure*}
 \centerline{\resizebox{0.9\hsize}{!}{\includegraphics[angle=270]{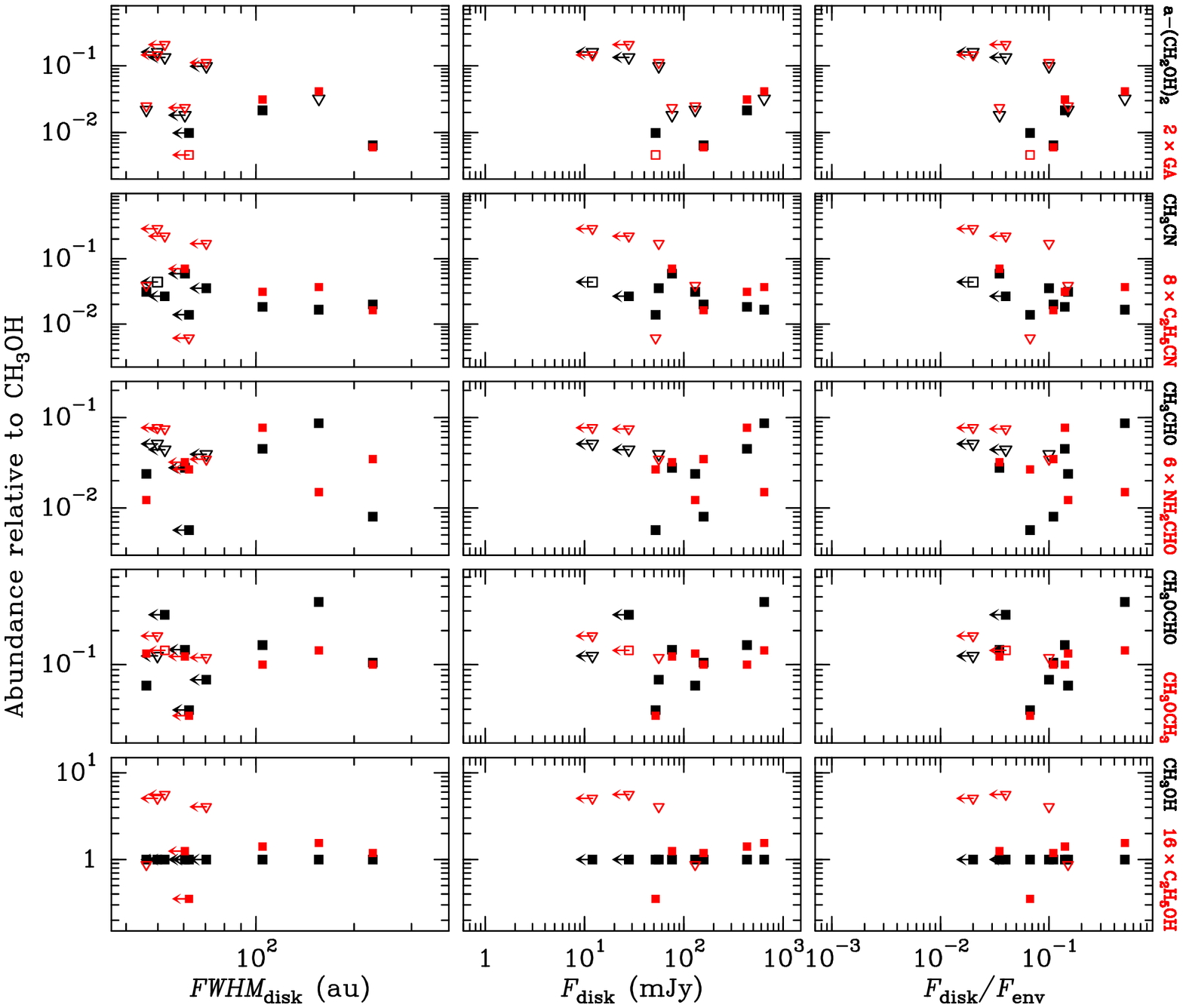}}}
 \caption{Column densities of ten complex organic molecules normalized to 
the column density of methanol as a function of disk size ($FWHM$), disk flux 
density, and ratio of disk to envelope flux densities taken from 
\citet{Maury19}. The column densities of some molecules were multiplied by a 
scaling factor as indicated on the right. Filled and open squares represent
robust and tentative detections, respectively, while open triangles show upper 
limits. Arrows indicate upper limits along the horizontal axis. GA stands for 
glycolaldehyde, CH$_2$(OH)CHO.}
 \label{f:coldens_disk_normch3oh}
\end{figure*}

Because of the low number of detections, most plots in 
Figs.~\ref{f:coldens}--\ref{f:coldens_normsolangcont3} and
\ref{f:coldens_normch3oh} are consistent with the absence of a correlation
between (normalized) column densities and source properties 
(Fig.~\ref{f:pearson_coldens} and Table~\ref{t:correl_coldens}). However, 
there is a clear anticorrelation between the abundances of CH$_3$CHO and 
CH$_2$(OH)CHO relative to methanol and the internal luminosities of the 
sources (green bars in the bottom panel of Fig.~\ref{f:pearson_coldens}). 
Although it is less significant, there also seems to be a correlation between 
the internal luminosities of the sources and the column densities of CH$_3$OH 
normalized with the ratio of the continuum flux density at 3~mm (and, to a 
lesser degree, 1.3~mm as well) to the COM solid angle (blue and red bars in 
bottom panel of Fig.~\ref{f:pearson_coldens}). The same applies to CH$_3$CN 
and NH$_2$CHO, though with even less significance. Finally, there seems to be 
an anticorrelation between the ratio of envelope mass to internal luminosity
and the column densities of C$_2$H$_5$OH and CH$_3$OCH$_3$ normalized with the 
ratio of the continuum flux density at 1.3~mm to the COM solid angle (red bars 
in top panel of Fig.~\ref{f:pearson_coldens}), but the significance of this
anticorrelation is low. The internal luminosity thus appears to be the 
parameter most impacting the COM chemical composition of the sources, followed 
by the ratio 
of envelope mass to internal luminosity, while the envelope mass itself does 
not imprint any obvious signature in the COM chemical composition.

However, we should also pay attention to the non-detections, which were not
taken into account in the correlation analysis.
Figure~\ref{f:coldens_normch3oh} indicates that the CH$_3$CHO non-detections
are consistent with the anticorrelation mentioned above, except for 
L1157 which is marginally inconsistent. The CH$_2$(OH)CHO non-detections are
all consistent with the anticorrelation.

For CH$_3$OH, two non-detections are inconsistent with the correlation
mentioned above (see Fig.~\ref{f:coldens_normsolangcont3}). They 
correspond to SVS13B and SerpS-MM4a. The other non-detections are consistent
or marginally consistent with the correlation. For CH$_3$CN, the upper limits
are consistent with the (loose) correlation, three of them only marginally. 
For NH$_2$CHO, all non-detections are
consistent with the (loose) correlation noted above.

One upper limit (L1448-C) is inconsistent with the loose correlation of 
C$_2$H$_5$OH with $M_{\rm env}$/$L_{\rm int}$ in 
Fig.~\ref{f:coldens_normsolangcont}, one is marginally consistent 
(SerpS-MM18b), and the other ones are consistent. For CH$_3$OCH$_3$, three 
upper limits are inconsistent with the loose correlation, five are marginally
consistent, and four are consistent.

\subsection{Correlations between molecules and disk properties}
\label{ss:correl_prop_disk}

\begin{figure*}
 \centerline{\resizebox{0.6\hsize}{!}{\includegraphics[angle=0]{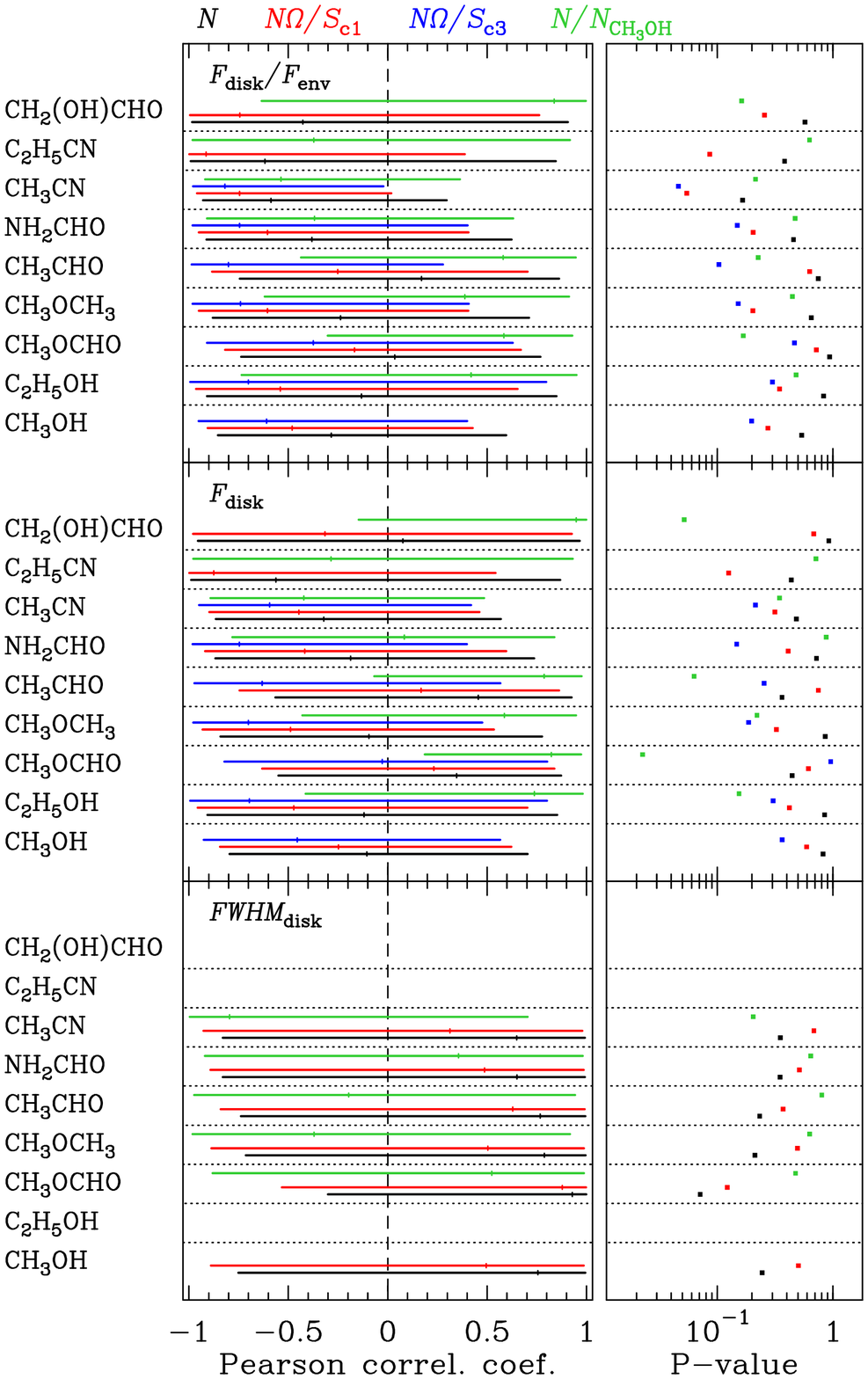}}}
 \caption{Pearson correlation coefficients with 95\% confidence interval 
(\textit{left}) and P-value (\textit{right}) between nine COMs and several 
disk properties for the CALYPSO source sample. These parameters were derived 
from the plots shown in 
Figs.~\ref{f:coldens_disk}--\ref{f:coldens_disk_normsolangcont3} and 
\ref{f:coldens_disk_normch3oh}. The colors indicate the variables used to 
evaluate the degree of correlation between the molecules and the disk 
properties, with the same meaning as in Fig.~\ref{f:pearson_coldens}. The disk 
properties investigated here are the disk radius (\textit{bottom}), the disk 
flux density (\textit{middle}), and the flux density ratio of the disk to the 
envelope (\textit{top}).}
 \label{f:pearson_coldens_disk}
\end{figure*}

\begin{table*}
 \begin{center}
 \caption{
 Correlations between column densities of ten COMs and disk properties of the CALYPSO sources.
}
 \label{t:correl_coldens_disk}
 \vspace*{-1.2ex}
 \begin{tabular}{llccccccccc}
 \hline\hline
 \multicolumn{1}{c}{Prop.\tablefootmark{a}} & \multicolumn{1}{c}{Molecule} & \multicolumn{1}{c}{$N_{\rm pts}$\tablefootmark{b}} & \multicolumn{2}{c}{$N$\tablefootmark{c}} & \multicolumn{2}{c}{$N\Omega/S_{\rm c1}$\tablefootmark{d}} & \multicolumn{2}{c}{$N\Omega/S_{\rm c3}$\tablefootmark{e}} & \multicolumn{2}{c}{$N/N_{\rm CH3OH}$\tablefootmark{f}} \\ 
 \multicolumn{1}{c}{} & & & \multicolumn{1}{c}{$100\times\rho$} & \multicolumn{1}{c}{$P$} & \multicolumn{1}{c}{$100\times\rho$} & \multicolumn{1}{c}{$P$} & \multicolumn{1}{c}{$100\times\rho$} & \multicolumn{1}{c}{$P$} & \multicolumn{1}{c}{$100\times\rho$} & \multicolumn{1}{c}{$P$} \\ 
 \hline
\noalign{\smallskip} 
$R_{\rm disk}$ & CH$_3$OH & 4 & $ 76^{+ 24}_{-151}$ & 2(-1) & $ 50^{+ 49}_{-138}$ & 5(-1) & -- & -- & -- & -- \\ 
\noalign{\smallskip} 
 & C$_2$H$_5$OH & 3 & -- & -- & -- & -- & -- & -- & -- & -- \\ 
\noalign{\smallskip} 
 & CH$_3$OCHO & 4 & $ 93^{+  7}_{-123}$ & 7(-2) & $ 88^{+ 12}_{-141}$ & 1(-1) & -- & -- & $ 52^{+ 46}_{-140}$ & 5(-1) \\ 
\noalign{\smallskip} 
 & CH$_3$OCH$_3$ & 4 & $ 79^{+ 21}_{-150}$ & 2(-1) & $ 50^{+ 48}_{-139}$ & 5(-1) & -- & -- & $-37^{+129}_{- 61}$ & 6(-1) \\ 
\noalign{\smallskip} 
 & CH$_3$CHO & 4 & $ 77^{+ 23}_{-151}$ & 2(-1) & $ 63^{+ 36}_{-147}$ & 4(-1) & -- & -- & $-20^{+114}_{- 78}$ & 8(-1) \\ 
\noalign{\smallskip} 
 & NH$_2$CHO & 4 & $ 65^{+ 34}_{-148}$ & 3(-1) & $ 49^{+ 50}_{-138}$ & 5(-1) & -- & -- & $ 36^{+ 63}_{-128}$ & 6(-1) \\ 
\noalign{\smallskip} 
 & CH$_3$CN & 4 & $ 65^{+ 34}_{-148}$ & 4(-1) & $ 31^{+ 67}_{-124}$ & 7(-1) & -- & -- & $-80^{+150}_{- 20}$ & 2(-1) \\ 
\noalign{\smallskip} 
 & C$_2$H$_5$CN & 3 & -- & -- & -- & -- & -- & -- & -- & -- \\ 
\noalign{\smallskip} 
 & a-(CH$_2$OH)$_2$ & 2 & -- & -- & -- & -- & -- & -- & -- & -- \\ 
\noalign{\smallskip} 
 & CH$_2$(OH)CHO & 3 & -- & -- & -- & -- & -- & -- & -- & -- \\ 
\noalign{\smallskip} 
\hline 
\noalign{\smallskip} 
$F_{\rm disk}$ & CH$_3$OH & 7 & $-10^{+ 81}_{- 69}$ & 8(-1) & $-25^{+ 87}_{- 60}$ & 6(-1) & $-45^{+102}_{- 47}$ & 4(-1) & -- & -- \\ 
\noalign{\smallskip} 
 & C$_2$H$_5$OH & 5 & $-12^{+ 97}_{- 79}$ & 8(-1) & $-47^{+118}_{- 48}$ & 4(-1) & $-70^{+150}_{- 30}$ & 3(-1) & $ 74^{+ 24}_{-115}$ & 2(-1) \\ 
\noalign{\smallskip} 
 & CH$_3$OCHO & 7 & $ 35^{+ 52}_{- 90}$ & 4(-1) & $ 23^{+ 61}_{- 86}$ & 6(-1) & $ -3^{+ 83}_{- 79}$ & 1(0) & $ 82^{+ 15}_{- 64}$ & 2(-2) \\ 
\noalign{\smallskip} 
 & CH$_3$OCH$_3$ & 6 & $ -9^{+ 87}_{- 75}$ & 9(-1) & $-49^{+102}_{- 44}$ & 3(-1) & $-70^{+118}_{- 28}$ & 2(-1) & $ 59^{+ 36}_{-102}$ & 2(-1) \\ 
\noalign{\smallskip} 
 & CH$_3$CHO & 6 & $ 46^{+ 47}_{-102}$ & 4(-1) & $ 17^{+ 69}_{- 91}$ & 8(-1) & $-63^{+120}_{- 34}$ & 3(-1) & $ 79^{+ 19}_{- 85}$ & 6(-2) \\ 
\noalign{\smallskip} 
 & NH$_2$CHO & 6 & $-19^{+ 92}_{- 68}$ & 7(-1) & $-42^{+101}_{- 50}$ & 4(-1) & $-75^{+114}_{- 24}$ & 1(-1) & $  8^{+ 75}_{- 86}$ & 9(-1) \\ 
\noalign{\smallskip} 
 & CH$_3$CN & 7 & $-32^{+ 89}_{- 54}$ & 5(-1) & $-45^{+ 91}_{- 45}$ & 3(-1) & $-59^{+101}_{- 35}$ & 2(-1) & $-42^{+ 91}_{- 47}$ & 3(-1) \\ 
\noalign{\smallskip} 
 & C$_2$H$_5$CN & 4 & $-56^{+143}_{- 43}$ & 4(-1) & $-87^{+142}_{- 12}$ & 1(-1) & -- & -- & $-29^{+122}_{- 69}$ & 7(-1) \\ 
\noalign{\smallskip} 
 & a-(CH$_2$OH)$_2$ & 3 & -- & -- & -- & -- & -- & -- & -- & -- \\ 
\noalign{\smallskip} 
 & CH$_2$(OH)CHO & 4 & $  8^{+ 89}_{-103}$ & 9(-1) & $-32^{+124}_{- 66}$ & 7(-1) & -- & -- & $ 95^{+  5}_{-109}$ & 5(-2) \\ 
\noalign{\smallskip} 
\hline 
\noalign{\smallskip} 
$F_{\rm disk}$/$F_{\rm env}$ & CH$_3$OH & 7 & $-28^{+ 88}_{- 57}$ & 5(-1) & $-48^{+ 91}_{- 43}$ & 3(-1) & $-61^{+101}_{- 34}$ & 2(-1) & -- & -- \\ 
\noalign{\smallskip} 
 & C$_2$H$_5$OH & 5 & $-13^{+ 98}_{- 78}$ & 8(-1) & $-54^{+119}_{- 42}$ & 3(-1) & $-70^{+150}_{- 29}$ & 3(-1) & $ 42^{+ 53}_{-115}$ & 5(-1) \\ 
\noalign{\smallskip} 
 & CH$_3$OCHO & 7 & $  4^{+ 73}_{- 77}$ & 9(-1) & $-17^{+ 84}_{- 65}$ & 7(-1) & $-37^{+100}_{- 54}$ & 5(-1) & $ 58^{+ 34}_{- 89}$ & 2(-1) \\ 
\noalign{\smallskip} 
 & CH$_3$OCH$_3$ & 6 & $-24^{+ 95}_{- 64}$ & 7(-1) & $-61^{+101}_{- 34}$ & 2(-1) & $-74^{+115}_{- 24}$ & 2(-1) & $ 39^{+ 52}_{-101}$ & 4(-1) \\ 
\noalign{\smallskip} 
 & CH$_3$CHO & 6 & $ 17^{+ 69}_{- 91}$ & 7(-1) & $-25^{+ 96}_{- 63}$ & 6(-1) & $-80^{+108}_{- 19}$ & 1(-1) & $ 58^{+ 37}_{-102}$ & 2(-1) \\ 
\noalign{\smallskip} 
 & NH$_2$CHO & 6 & $-38^{+100}_{- 53}$ & 5(-1) & $-60^{+101}_{- 35}$ & 2(-1) & $-74^{+115}_{- 24}$ & 1(-1) & $-37^{+100}_{- 54}$ & 5(-1) \\ 
\noalign{\smallskip} 
 & CH$_3$CN & 7 & $-59^{+ 88}_{- 34}$ & 2(-1) & $-75^{+ 76}_{- 21}$ & 5(-2) & $-82^{+ 80}_{- 16}$ & 5(-2) & $-54^{+ 90}_{- 38}$ & 2(-1) \\ 
\noalign{\smallskip} 
 & C$_2$H$_5$CN & 4 & $-62^{+146}_{- 37}$ & 4(-1) & $-91^{+130}_{-  8}$ & 9(-2) & -- & -- & $-37^{+129}_{- 61}$ & 6(-1) \\ 
\noalign{\smallskip} 
 & a-(CH$_2$OH)$_2$ & 3 & -- & -- & -- & -- & -- & -- & -- & -- \\ 
\noalign{\smallskip} 
 & CH$_2$(OH)CHO & 4 & $-43^{+133}_{- 56}$ & 6(-1) & $-74^{+151}_{- 25}$ & 3(-1) & -- & -- & $ 84^{+ 16}_{-147}$ & 2(-1) \\ 
\noalign{\smallskip} 
\hline 
 \end{tabular}
 \end{center}
 \vspace*{-2.5ex}
 \tablefoot{
 Same as Table~\ref{t:correl_coldens} for the following disk properties: radius ($R_{\rm disk}$), flux density ($F_{\rm disk}$), and flux density ratio of disk and envelope ($F_{\rm disk}/F_{\rm env}$).}
 \end{table*}

We also search for correlations between the COM column densities and the
following properties of the candidate disk-like structures  detected in 
continuum emission in the CALYPSO sample by 
\citet{Maury19}: disk size ($FWHM$), disk flux density, and ratio of disk
to envelope flux densities\footnote{The kinematical analysis of the CALYPSO 
data on small scales has revealed Keplerian rotation in only two sources 
(L1448-C and L1527) while no sign of Keplerian rotation has been found for 
radii larger than 50 au in the other Class 0 sources of the sample 
\citep[][]{Maret20}.}. 
We test the same four normalizations as in
Sect.~\ref{ss:correl_prop}. The distributions are displayed in
Figs.~\ref{f:coldens_disk}--\ref{f:coldens_disk_normsolangcont3} and 
\ref{f:coldens_disk_normch3oh}. The Pearson correlation coefficients of all
individual plots and their 95\% confidence level are shown in 
Fig.~\ref{f:pearson_coldens_disk} and listed in 
Table~\ref{t:correl_coldens_disk}.
There are no obvious correlations between the COM column densities and the
disk properties, whatever the chosen normalization of the column densities. 
All distributions have a 95\% confidence level interval of their Pearson 
correlation coefficient that is so large that it includes 0, which means that 
the distributions are consistent with the absence of a correlation. 

Figure.~\ref{f:coldens_disk} shows that, out of four sources with a large 
($FWHM_{\rm disk} > 100$~au) candidate disk-like structure, three have COM 
detections (IRAS4A2, IRAS4B, SVS13A), but the source with the largest one does 
not (SerpM-SMM4a). However, no sign of Keplerian detection was found by
\citet{Maret20} in SerpM-SMM4a, and \citet{Maury19} argued in their 
Appendix~C.12 that the true nature of its candidate disk-like structure is 
unclear. One source with a small ($FWHM_{\rm disk} = 46$~au), resolved disk-like 
structure has COM detections (L1448-C) while both sources with an 
intermediate-size ($FWHM_{\rm disk} \sim 81$ and 54~au), resolved disk-like
structure do not (SerpS-MM22 and L1527). Among the 14 sources with a detected 
(resolved or unresolved) disk-like structure, the seven sources with a 
methanol column density higher than $8 \times 10^{16}$~cm$^{-2}$ 
all have disk flux densities higher than 50~mJy. However, three sources with 
similarly high disk flux densities do not have methanol detections (SVS13B, 
L1527\footnote{Methanol has been detected at small scales toward this source 
with ALMA \citep[][]{Sakai14b}.}, SerpM-SMM4a). Two sources detected in 
methanol with a column density lower than $8 \times 10^{16}$~cm$^{-2}$ have no 
disk detection (SerpM-S68N, L1448-2A).

\begin{figure*}[!t]
 \centerline{\resizebox{0.9\hsize}{!}{\includegraphics[angle=0]{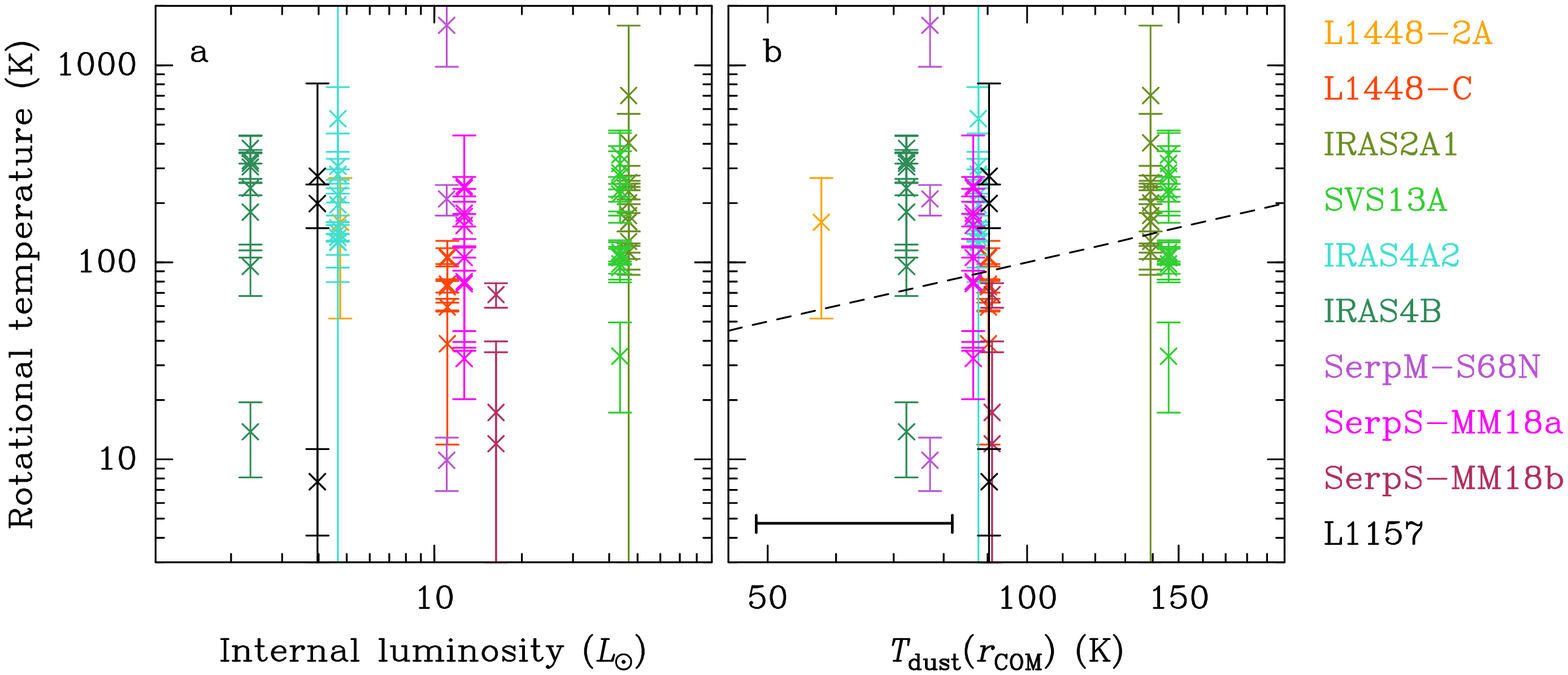}}}
 \caption{Rotational temperatures derived from population diagrams of various
COMs versus (\textbf{a}) internal luminosity and (\textbf{b}) dust temperature 
expected at the radius of the COM emitting region. The sources are 
color-coded following the color scheme shown on the right. The rotational 
temperatures and their uncertainties are listed in Table~\ref{t:popfit}. The
radii, $r_{\rm COM}$, correspond to half the sizes listed in 
Table~\ref{t:sizes}. The internal luminosities are given in
Table~\ref{t:sources}. The data points corresponding to L1448-2A and 
L1448-C in panel \textbf{a} were slightly shifted to the right (internal 
luminosity multiplied by a factor 1.02) for clarity. The dashed line shows
$T_{\rm rot} = T_{\rm dust}$ for comparison purposes. The horizonthal error bar
represents the likely uncertainty on $T_{\rm dust}$($r_{\rm COM}$).}
 \label{f:trot-lint}
\end{figure*}

Among the nine CALYPSO sources that we classified into groups (1 to 3) on the 
basis of their COM composition relative to methanol (Sect.~\ref{ss:chemcomp}), 
only two have no detection of a disk-like structure: SerpM-S68N (group 2) and 
L1448-2A (group 3), with
$F_{\rm disk}/F_{\rm env} <$ 4\% and 2\%, respectively, while the other seven
sources have a detection with $F_{\rm disk}/F_{\rm env}$ ranging from 3.5\%
to 50\%. We do not see any obvious differences in terms of disk detection or
disk properties between the three groups. One slight difference is that, while
75\% of the sources have a disk detection for both groups 2 and 3, all three 
disk-like structures in group 2 are resolved while only one disk-like structure
out of three is resolved in group 3: the disk-like structures in group 2, when 
detected, are larger than the disk-like structures in group 3. The single 
CALYPSO source in group 1 has a detected disk-like structure that is 
unresolved, like two sources of group 3.

\subsection{Correlations between rotational temperatures and source properties}
\label{ss:trot-lint}

We use Fig.~\ref{f:trot-lint} to investigate possible correlations between 
some properties of the sources and the rotational temperatures derived from 
the COM population diagrams shown in 
Figs.~\ref{f:popdiag_l1448-2a_p1_ch3oh}--\ref{f:popdiag_l1157_p1_ch3cn}. 
The rotational temperatures are summarized for each source in 
Figs.~\ref{f:trot_l1448-2a_p1}--\ref{f:trot_l1157_p1} and they are listed in 
Table~\ref{t:popfit}. Figure~\ref{f:trot-lint}a shows that there is no 
correlation between the rotational temperatures and the internal luminosities
of the sources. 

We then compare the rotational temperatures derived from the COMs to 
the gas kinetic temperature expected at the radius of the COM emission, 
$r_{\rm COM}$, equal to half of the size reported in Table~\ref{t:popfit}. 
Given the high densities expected at these small scales, we expect the dust 
and gas to be well coupled thermally, and the rotational temperatures of the 
COMs to trace the kinetic temperature of the gas. Thus we could expect a 
correlation between the rotational temperatures of the COMs and the dust 
temperature of the COM emitting region if the COMs trace the region where they 
desorb from the icy mantles of dust grains as a result of the heating by the 
central protostar. 

The method that we adopted to estimate the dust temperature at $r_{\rm COM}$ is
described in Appendix~\ref{a:tdust}. We estimate the uncertainty on this dust
temperature to be at least a factor 1.3.
Figure~\ref{f:trot-lint}b does not show any obvious correlation between the 
rotational temperatures and the expected dust temperatures at the scale of the 
COM emission. Given the large dispersion (and uncertainties) of rotational 
temperatures and the large uncertainty on the dust temperatures, it is 
possible that the rotation temperatures do trace the dust temperatures at the
radius of the COM emission, which would be consistent with the fact that
the presence of COMs in these sources is directly related to the heating by 
the nascent protostar. However, the large uncertainties 
that affect Figure~\ref{f:trot-lint}b may mask deviations that would point to 
other processes (e.g., accretion shocks, shocks in outflows) as the source of 
the COM emission in these sources. An accurate modeling of the dust radiative
transfer of the CALYPSO sources at subarcsecond scale would be necessary to
make progress.

\subsection{Comparison of COM, disk, and thermal heating sizes}
\label{ss:rcom_rdisk_rsub}

\begin{figure}[!t]
 \centerline{\resizebox{1.0\hsize}{!}{\includegraphics[angle=0]{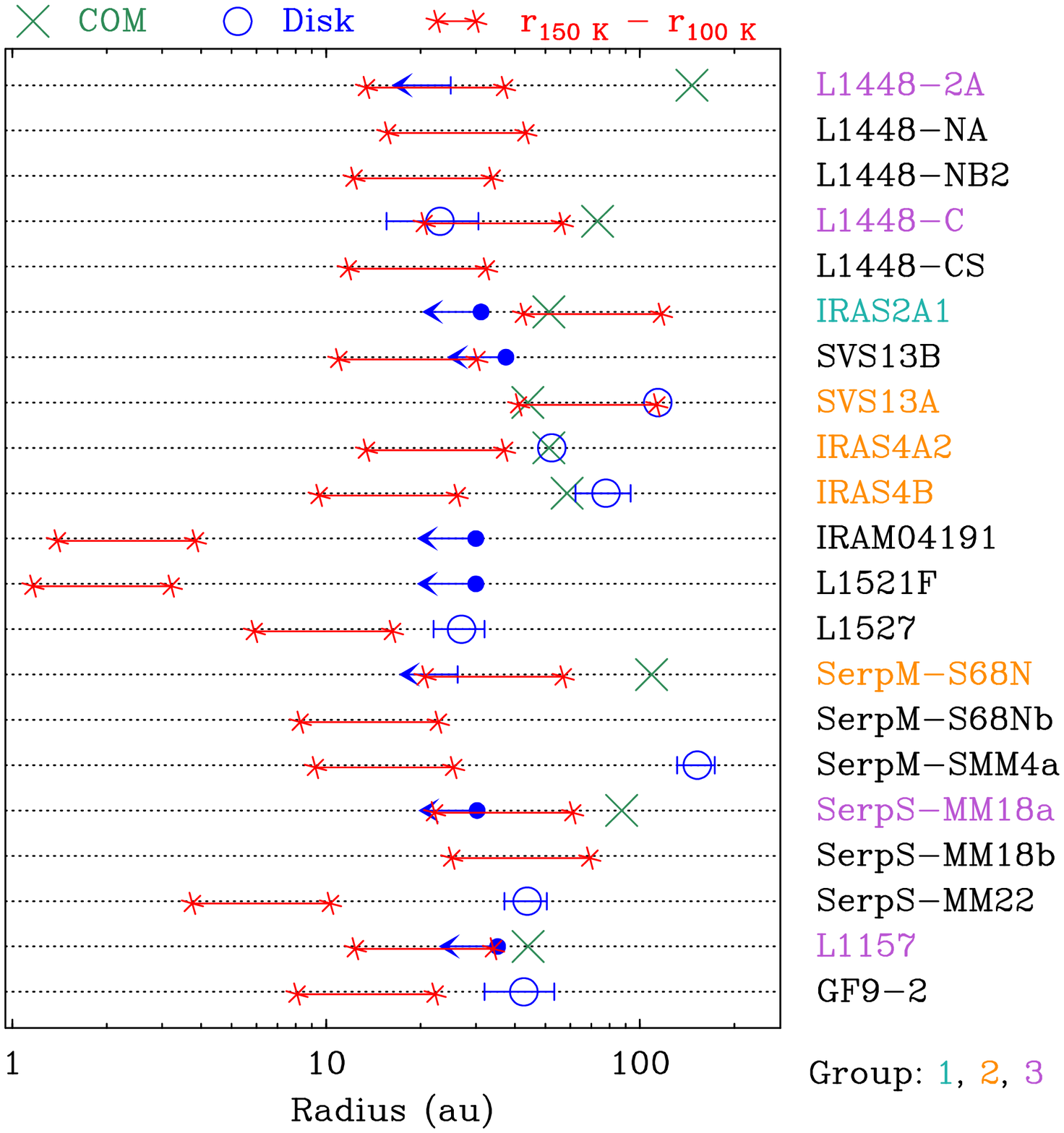}}}
 \caption{Comparison of the COM emission radius (green cross) to the disk 
radius (blue circle, with error bar when available) and the range of radii 
over which ices are expected to sublimate from the grain mantles (red bar 
terminated by stars). The COM emission radius corresponds to half the FWHM 
size reported in Table~\ref{t:sizes}, when COMs are detected. The disk radius 
corresponds to half the FWHM size derived by \citet{Maury19} when they detect 
a candidate disk-like structure, corrected for the new distance when needed. 
The radii of the candidate disk-like structures for SVS13A 
(circumbinary in this case) and IRAS4A2 come from a similar analysis 
(A.~Maury, priv. comm.). A blue dot with an arrow pointing to the left 
indicates the upper limit on the radius of a detected but unresolved disk. A 
blue bar with an arrow pointing to the left represents a disk non-detection 
with an upper limit on the radius that could still be estimated by 
\citet{Maury19}. All disk parameters were rescaled to the new distances when 
necessary. For each source, the two red stars mark the temperature range 
100--150~K (from right to left) expected from the luminosity given in 
Table~\ref{t:sources}. The names of the sources belonging to groups 1, 2, and 
3 are color-coded with the color scheme indicated in the bottom right corner, 
as the frames in Fig.~\ref{f:chemcompch3oh} and as described in 
Sect.~\ref{ss:chemcomp}.}
\label{f:rcom_rdisk_rsub}
\end{figure}

Figure~\ref{f:rcom_rdisk_rsub} compares, for the 21 sources that have an 
internal luminosity listed in Table~\ref{t:sources}, the radius of the COM 
emission (green), when COMs are detected, to the disk radius (blue), when a 
candidate disk-like structure is detected in continuum emission by 
\citet{Maury19} in the CALYPSO survey, and the range of 
radii over which the thermal heating by the protostar is expected to produce 
temperatures of 100--150~K (red). This range of radii was computed from the 
luminosities listed in Table~\ref{t:sources} using Eq.~\ref{e:tdust} and 
the correction factor defined in Appendix~\ref{a:tdust}. We want to 
investigate whether the presence of COMs is due to accretion shocks at the 
edge of the disk, or simply reveals a classical hot corino picture where the 
COMs become detectable in the gas phase once they have thermally desorbed from 
the dust grain mantles under the influence of the protostellar luminosity. 

There are several systematic uncertainties that are difficult to evaluate 
properly in Fig.~\ref{f:rcom_rdisk_rsub}: a) the relationship between the size 
of the disk-like structure derived by \citet{Maury19} and the location of the 
centrifugal barrier where accretion shocks may occur or the radial extent of a 
disk atmosphere such as the one detected in HH212 \citep[][]{Lee19}, b) the
uncertainty on the correction factor used to compute the dust temperature (see 
Appendix~\ref{a:tdust}), c) the uncertainties in the derivation of the 
protostellar luminosities especially in the case of binaries unresolved by 
\textit{Herschel}, d) the uncertainty on the radius of the COM emission, which 
is most of the time barely resolved in the CALYPSO survey, e) the complexity 
of the small-scale structure, for instance the binarity of SVS13A and 
L1448-2A. Still, this figure is interesting in a statistical sense. It 
reveals a diversity of configurations in the CALYPSO sample: 
\begin{itemize}
\item two sources have a COM radius similar to the radius of the disk-like
structure and larger than the 100--150~K region (IRAS4A2, IRAS4B).
\item two sources have a COM radius falling in the 100--150~K region with a 
smaller (IRAS2A1) or larger disk-like structure (SVS13A).
\item five sources have COM radii larger than both the 100--150~K radius range
and the radius of the disk-like structure or its upper limit (L1448-2A, 
L1448-C, SerpM-S68N, SerpS-MM18a, L1157). In the case of L1157, the COM radius 
is in fact only slightly larger than the upper limit on the radius of the 
disk-like structure and the radius at 100~K.
\item four sources have a detected and resolved disk-like structure but no COM 
emission detected with CALYPSO (L1527, SerpM-SMM4, SerpS-MM22, GF9-2).
\item three sources have a detected but unresolved disk-like structure and no
COM emission (IRAM04191, L1521F, SVS13B)
\item for the remaining five sources, we have no information about their disk, 
and they have either no COM emission (L1448-NB2, SerpM-S68Nb) or they have 
only tentative COM detections (L1448-NA, L1448-CS) or COM emission but no
good estimate of its size (SerpS-MM18b).
\end{itemize}
We conclude from this analysis that the detection of a disk-like structure does 
not imply the detection of COM emission, and vice versa, and that the size of 
the COM emission, when detected, is not systematically related to the 
size of the disk-like structure nor to the extent of the hot (100--150~K) 
inner envelope. Figure~\ref{f:rcom_rdisk_rsub} shows, however, that all four 
sources belonging to group 3 have a COM radius larger than both the radius of 
the disk-like structure and the extent of the hot inner envelope. No 
systematic pattern is visible in this figure for group 2, though.

\section{Discussion}
\label{s:discussion}

\subsection{Spatial origin of COMs in protostars}
\label{ss:origin}

This study that relies on the CALYPSO survey presents the largest sample of 
solar-type Class~0 protostars investigated for COM emission at sub-arcsecond
scale. Compact emission of at least one COM (methanol) has been detected for
12 sources out of 26. Even though the angular resolution of the survey is 
barely sufficient to resolve the COM emission in the sources where COMs are 
detected, we attempt to draw conclusions about the spatial origin of COMs 
around Class 0 (and I) protostars. Table~\ref{t:summary} summarizes the COM 
results that we obtained in the previous sections.

\begin{table*}[!ht]
 \begin{center}
 \caption{
 Summary of COM properties of CALYPSO sources.
}
 \label{t:summary}
 \vspace*{-1.2ex}
 \begin{tabular}{lccccccccccc}
 \hline\hline
 \multicolumn{1}{c}{Source} & \multicolumn{1}{c}{Class\tablefootmark{a}} & \multicolumn{1}{c}{$L_{\rm int}$\tablefootmark{b}} & \multicolumn{1}{c}{$M_{\rm env}/L_{\rm int}$\tablefootmark{b}} & \multicolumn{1}{c}{Cand.} & \multicolumn{1}{c}{CH$_3$OH\tablefootmark{d}} &  \multicolumn{1}{c}{$\geqslant$3} & \multicolumn{1}{c}{Chem.} & \multicolumn{1}{c}{COM} & \multicolumn{1}{c}{$r_{\rm COM}$} & \multicolumn{1}{c}{$r_{\rm COM}$} & \multicolumn{1}{c}{COM} \\ 
  & & \multicolumn{1}{c}{($L_\odot$)} & \multicolumn{1}{c}{($M_\odot/L_\odot$)} & \multicolumn{1}{c}{disk\tablefootmark{c}} & & \multicolumn{1}{c}{COMs\tablefootmark{e}} & \multicolumn{1}{c}{group\tablefootmark{f}} & \multicolumn{1}{c}{$\parallel$ jet\tablefootmark{g}} & \multicolumn{1}{c}{$\sim$$r_{\rm disk}$\tablefootmark{h}} &  \multicolumn{1}{c}{$\sim$$r_{\rm100-150K}$\tablefootmark{i}} & \multicolumn{1}{c}{orig.\tablefootmark{j}} \\ 
  \multicolumn{1}{c}{(1)} & \multicolumn{1}{c}{(2)} & \multicolumn{1}{c}{(3)} & \multicolumn{1}{c}{(4)} & \multicolumn{1}{c}{(5)} & \multicolumn{1}{c}{(6)} & \multicolumn{1}{c}{(7)} & \multicolumn{1}{c}{(8)} & \multicolumn{1}{c}{(9)} & \multicolumn{1}{c}{(10)} & \multicolumn{1}{c}{(11)} & \multicolumn{1}{c}{(12)} \\ 
 \hline
 L1448-2A & 0  & 4.7 & 0.27 &  n &  y & n &  3 & -- &  n & n &  ?  \\ 
 L1448-2Ab & 0?  & $<$ 4.7 & $>$ 0.13 &  -- &  y & n &  -- & -- &  -- & -- &  --  \\ 
 L1448-NA & I  & 6.4 & 0.12 &  -- &  t & n &  -- & -- &  -- & -- &  --  \\ 
 L1448-NB1 & 0  & $<$ 3.9 & $>$ 0.8 &  n &  n & n &  -- & -- &  -- & -- &  --  \\ 
 L1448-NB2 & 0?  & 3.9 & 0.40 &  -- &  n & n &  -- & -- &  -- & -- &  --  \\ 
 L1448-C & 0  & 11 & 0.17 &  y &  y & y &  3 & n &  n & t &  hc?  \\ 
 L1448-CS & I  & 3.6 & 0.043 &  -- &  n & n &  -- & -- &  -- & -- &  --  \\ 
 IRAS2A1 & 0  & 47 & 0.17 &  $<$y &  y & y &  1 & y &  n & y &  hc, o  \\ 
 SVS13B & 0  & 3.1 & 0.9 &  $<$y &  n & n &  -- & -- &  -- & -- &  --  \\ 
 SVS13A & I  & 44 & 0.018 &  y &  y & y &  2 & n &  n & y &  hc  \\ 
 IRAS4A1 & 0  & $<$ 4.7 & $>$ 2.1 &  n &  n & n &  -- & -- &  -- & -- &  --  \\ 
 IRAS4A2 & I?  & 4.7 & 0.5 &  y &  y & y &  2 & n &  y & n &  a  \\ 
 IRAS4B & 0  & 2.3 & 1.4 &  y &  y & y &  2 & y &  y & n &  o, a  \\ 
 IRAS4B2 & 0?  & $<$ 0.16 & $>$ 9.0 &  -- &  n & n &  -- & -- &  -- & -- &  --  \\ 
 IRAM04191 & 0  & 0.05 & 10 &  $<$y &  n & n &  -- & -- &  -- & -- &  --  \\ 
 L1521F & 0  & 0.035 & 20 &  $<$y &  n & n &  -- & -- &  -- & -- &  --  \\ 
 L1527 & 0  & 0.9 & 1.3 &  y &  n & n &  -- & -- &  -- & -- &  --  \\ 
 SerpM-S68N & 0  & 11 & 1.0 &  n &  y & y &  2 & -- &  -- & n &  ?  \\ 
 SerpM-S68Nb & 0  & 1.8 & -- &  -- &  n & n &  -- & -- &  -- & -- &  --  \\ 
 SerpM-SMM4a & 0  & 2.2 & 3.0 &  y &  n & n &  -- & -- &  -- & -- &  --  \\ 
 SerpM-SMM4b & 0?  & -- & -- &  -- &  y & n &  -- & -- &  -- & -- &  --  \\ 
 SerpS-MM18a & 0  & 13 & 0.36 &  $<$y &  y & y &  3 & y &  n & n &  hc, o  \\ 
 SerpS-MM18b & 0?  & 16 & 0.06 &  -- &  y & n &  -- & -- &  -- & -- &  --  \\ 
 SerpS-MM22 & 0  & 0.36 & 2.5 &  y &  n & n &  -- & -- &  -- & -- &  --  \\ 
 L1157 & 0  & 4.0 & 0.7 &  $<$y &  y & y &  3 & -- &  ? & t &  hc?, a?  \\ 
 GF9-2 & 0  & 1.7 & 1.7 &  y &  n & n &  -- & -- &  -- & -- &  --  \\ 
 \hline
 \end{tabular}
 \end{center}
 \vspace*{-2.5ex}
 \tablefoot{
 \tablefoottext{a}{Class according to Table~3 of \citet{Maury19}}.
 \tablefoottext{b}{Internal luminosity and ratio of envelope mass to internal luminosity (see Table~\ref{t:sources} for references and uncertainties).}
 \tablefoottext{c}{Candidate disk-like structure detected and resolved (y), detected but unresolved ($<$y), or not detected (n) in CALYPSO continuum data \citep[][]{Maury19}.}
 \tablefoottext{d}{Methanol detected (y), tentatively detected (t), or not detected (n).}
 \tablefoottext{e}{y stands for at least three COMs detected, n for less than three.}
 \tablefoottext{f}{Chemical group (see Sect.~\ref{ss:chemcomp}, Fig.~\ref{f:chemcompch3oh}, and Table~\ref{t:chemcomp}).}
 \tablefoottext{g}{Elongation of COM emission parallel to jet/outflow (y) or not (n) (see Sect.~\ref{ss:elongations}).}
 \tablefoottext{h}{y if radius of COM emission similar to radius of candidate disk reported by \citet{Maury19}, n if not (see Fig.~\ref{f:rcom_rdisk_rsub}).}
 \tablefoottext{i}{Dust temperature at radius of COM emission within  100--150~K (y: yes, t: tentative) or not (n) (see Fig.~\ref{f:rcom_rdisk_rsub}).}
 \tablefoottext{j}{COM possible origin: hot corino (hc), outflow (o), accretion shock/disk atmosphere (a) (see Sect.~\ref{ss:origin}).}
 }
 \end{table*}

An enhanced desorption of COMs in accretion shocks at the centrifugal
barrier has been proposed by \citet{Oya16} and \citet{Csengeri18} to explain
the results of observations performed with ALMA toward IRAS16293A and the 
high-mass protostar G328.2551-0.5321, respectively. Theoretical calculations
show that this is indeed a possible mechanism if the grain population carrying
the adsorbed molecules is dominated by small grains of size $\sim$0.01~$\mu$m 
\citep[][]{Miura17}.
If we assume that each protostellar disk has an accretion shock at 
the radius of its centrifugal barrier, and that this shock should lead to the 
desorption of the COMs formed in the ice mantles of dust grains
then we would expect a strong correlation between the detection of a disk and 
the detection of COM emission, as well as a good match between the disk radius 
and the radius of the COM emission. The analysis presented in 
Sect.~\ref{ss:rcom_rdisk_rsub} shows that this is not the case: although this 
picture appears to work well for IRAS4A2 and IRAS4B, the detection of COM 
emission is not systematically correlated with the detection of a disk-like
structure in the full CALYPSO sample. 

One possibility for this lack of correlation is that the shock
parameters (pre-shock density and velocity) or the size distribution of dust
grains are not the same in all sources. \citet{Miura17} showed that the 
desorption efficiency significantly depends on these parameters.
Another possibility could be that sources with a disk
radius (or upper limit thereof) smaller than the radius at which COMs would 
thermally desorb in the envelope under the influence of the protostar 
luminosity (at temperatures higher than 100--150 K) could have COM emission 
dominated by this thermal process. This hot corino scenario appears to work 
for IRAS2A1 (and, marginally, L1448-C and L1157), as well as for L483 for 
which \citet{Jacobsen19} found a COM emission radius of $\sim$40~au with ALMA, 
consistent with their estimate of the sublimation radius in the envelope 
($\sim$50~au), while they did not detect a disk and obtained an upper limit to 
its radius of 15~au. However, this hot corino scenario does not explain the 
presence of COMs in L1448-2A, SerpM-S68N, and SerpS-MM18a which all have a COM 
emission radius $\sim$1.5--4 times larger than the expected radius of the hot 
($T>$100~K) inner envelope. SVS13A has a COM radius consistent with the radius 
of the hot inner envelope, but its disk-like structure is larger and thus the 
accretion shock scenario does not work for this source.

As mentioned in Sect.~\ref{s:intro}, the presence of COMs in the gas phase 
of a protostar could also be related to its jet/outflow. We found in 
Sect.~\ref{ss:elongations} that the COM emission tends to be elongated along
a direction close to the outflow axis in three sources of the CALYPSO sample:
IRAS2A1, IRAS4B, and SerpS-MM18a. This will need to be confirmed with
observations at higher angular resolution, for instance with ALMA.

If we merge the previous results, we find possible COM origin scenarii for 
seven sources out of the nine sources with a measured COM emission size: (1)
hot corino origin for IRAS2A1, SVS13A, and, marginally, L1448-C and L1157; (2) 
accretion shock or disk atmosphere origin for IRAS4A2 and IRAS4B; (3) outflow 
origin for IRAS2A1, IRAS4B, and SerpS-MM18a. L1157 could also fit into the 
accretion shock or disk atmosphere scenario if its actual disk size is close 
to the CALYPSO upper limit. The two remaining sources, L1448-2A and 
SerpM-S68N, do not fit into any of the three scenarii.

The lack of COM emission in sources with detected and resolved disk-like 
structures (L1527, SerpM-SMM4, SerpS-MM22, GF9-2) 
questions the accretion shock scenario as a general mechanism for the release 
of COMs in the gas phase around protostars. Specific shock parameters may be 
required for this process to be efficient 
\citep[see, e.g.,][]{Miura17}. However, methanol emission was 
detected with a size (FWHM) of about 1$\arcsec$ by \citet{Sakai14b} toward 
L1527 with ALMA, which roughly corresponds to the radius of the centrifugal 
barrier ($\sim$100~au) determined by \citet{Sakai14a} in this source. 
\citet{Sakai14b} derived a methanol column density more than one order of 
magnitude lower than the upper limit we obtained from the CALYPSO data, 
even after accounting for the different temperatures assumed in both studies.
Therefore, we cannot exclude that the CALYPSO non-detection of COM 
emission in SerpM-SMM4, SerpS-MM22, and GF9-2 is simply due to a lack of 
sensitivity like in L1527. 

Given that L1527 has a luminosity of 0.9~$L_\odot$ only, the 
ALMA detection of methanol toward this source also means that the upper limit 
in internal luminosity of 2~$L_\odot$ below which COMs are not detected with 
CALYPSO (see Sect.~\ref{ss:correl_prop}) may simply be due to a lack of 
sensitivity and not to an intrinsic property of sources fainter than 
2~$L_\odot$. Several COMs were also detected with ALMA by 
\citet{Imai16,Imai19} toward the isolated Class~0 protostar B335 which is even 
fainter than L1527 \citep[0.7~$L_\odot$,][]{Evans15}. The radius of the 
methanol emission, $\sim$15~au \citep[][]{Imai19,Bjerkeli19}, is consistent 
with the hot corino scenario for this source given its luminosity. In 
addition, the radius of the centrifugal barrier inferred by \citet{Imai19} for 
B335 is more than three times smaller than the radius of the methanol emission, 
suggesting that the accretion shock or disk atmosphere scenario does not work 
for this source.

The lack of COM emission in CALYPSO sources with a detected but 
unresolved disk-like structure (SVS13B, IRAM04191, L1521F) may be due to a 
lack of sensitivity if the COM emission is unresolved as well. Similarly, the 
hot ($T>100$~K) region in IRAM04191 and L1521F is expected to be tiny 
($R<4$~au, Fig.~\ref{f:rcom_rdisk_rsub}) so the CALYPSO survey certainly lacks 
sensitivity to detect COM emission from a hot corino in both sources. However, 
the hot region in SVS13B is expected to be as extended as in IRAS4A2 so the 
lack of COM emission in SVS13B suggests that an accretion shock origin for the 
COM emission in IRAS4A2 may indeed be a better scenario than a pure hot corino 
origin.

\subsection{COM composition as an evolutionary tracer?}
\label{ss:evolution}

The CALYPSO sample is not homogeneous in terms of COM detection and, when
COMs are detected, in terms of chemical composition. Slightly less than half
of the sources have COM detections. Our analysis of the COM composition of
the nine CALYPSO sources with a sufficient number of detections, plus L483 and
IRAS16293B, revealed the existence of three groups (Sect.~\ref{ss:chemcomp}).
Group 1 has low abundances of COMs (especially the oxygen-bearing ones) 
relative to methanol compared to groups 2 and 3, and group 3 has abundances of
cyanides and CHO-bearing molecules relative to methanol enhanced and reduced,
respectively, by a factor $\sim$2 with respect to group 2.

The first question to address is whether these chemical groups reveal an 
evolutionary sequence. The sources in groups 1, 2, and 3 have a ratio of
envelope mass to internal luminosity, $M_{\rm env}/L_{\rm int}$, ranging from 
0.14 to 0.26, 0.018 to 1.43, and 0.17 to 0.75, respectively. For L483 and 
IRAS16293 (taken as system) in group 1, we assumed a luminosity of 
13~$L_\odot$ and an envelope mass of 1.8~$M_\odot$ \citep[][]{Shirley00}, and 
21~$L_\odot$ and 5.4~$M_\odot$ \citep[][]{Jorgensen05,Schoeier02}, 
respectively. If $M_{\rm env}/L_{\rm int}$ is an evolutionary tracer, then 
sources in groups 1 and 3 are in similar evolutionary stages, but yet with
different chemical composition, while group 2 contains sources that span a 
broad range of evolutionary stages, from Class~0 (IRAS4B, SerpM-S68N) to 
Class~I (SVS13A and, maybe, IRAS4A2), but yet share a similar
chemical composition. This absence of correlation between the chemical groups 
and $M_{\rm env}/L_{\rm int}$ is consistent with the lack of correlation between 
this ratio and the COM column densities (Sect.~\ref{ss:correl_prop}). Both 
results tend to imply that the COM chemical composition of protostars is not 
an evolutionary tracer, if $M_{\rm env}/L_{\rm int}$ is such a tracer. 

Episodic accretion, which is known to occur during the protostellar evolution 
\citep[][]{Safron15} and is thereby thought to solve the ``luminosity 
problem'' \citep[][]{Kenyon90,Dunham10}, may however change the picture 
as it suggests that $M_{\rm env}/L_{\rm int}$ can only be used as a robust
evolutionary indicator in a statistical sense and not necessarily in all
individual cases.
In turn, the presence and extent of COM emission and the 
COM composition of the sources might be in some way related to (present or 
past) bursts of accretion \citep{Taquet16}. Evidence for past accretion 
bursts has been claimed in a few low-mass protostars and very low luminosity 
objects on the basis of simple molecules like CO and N$_2$H$^+$
\citep[]{Jorgensen15,Hsieh18,Hsieh19}.
However, none of the six CALYPSO sources (L1448-C, IRAS2A, SVS13A, IRAS4A, 
IRAS4B, and L1527) included in the sample studied by \citet{Jorgensen15} show 
a clear sign of a past accretion burst according to their analysis. No 
evidence for a past accretion burst was found for IRAS4A, IRAS4B, L1448-C, and 
L1157 by \citet{Anderl16} either, on the basis of the C$^{18}$O and N$_2$H$^+$ 
data of the CALYPSO survey itself. Therefore, it is unlikely that the sample 
of CALYPSO sources with COM emission is as a whole strongly affected by 
episodic accretion. We note, however, that signs of episodic accretion has
been inferred for SerpS-MM18 by \citet{Plunkett15} from their detection with
ALMA of 22 knots in the outflow that they interpret as being the result of 
episodic ejection events. Given their high luminosities, it could also be that
IRAS2A1 and SVS13A are currently experiencing an accretion burst 
\citep[][]{Hsieh19}. However, they are not classified in the same COM 
chemical group (group 1 versus group 2, respectively), and SerpS-MM18 belongs
yet to the third group, so episodic accretion may not be the determining 
factor in the COM chemical composition.

Interestingly, L1448-2A, which does not fit in any of the three COM scenarii 
described in Sect.~\ref{ss:origin}, may have experienced an accretion burst 
less than 10$^3$ yr ago \citep[][]{Hsieh19}. This could explain why its 
COM radius is a factor 4--10 times larger than the size of the current 
100--150~K region of its envelope. \citet{Hsieh19} deduced an accretion 
luminosity of 8--36~$L_\odot$ during the past burst from their snowline 
analysis. Such a burst would have pushed the 100-150~K region away by a factor 
2.8-6, in rough agreement with the size of the current COM emission. It would
be interesting to investigate if a past accretion burst could also explain the
size of the COM emission in SerpM-S68N, the other source that we could not
assign to any of the three COM scenarii in Sect.~\ref{ss:origin}.

The chemical groups do not seem to correlate with the COM origin scenarii 
tentatively assigned to the CALYPSO sources in Sect.~\ref{ss:origin}. Sources 
with a possible hot corino origin or a possible outflow origin for their COM 
emission are found in all three groups. However, those with a possible 
accretion shock origin of their COM emission are found in group 2 only (and
possibly one in group 3). The enhanced abundance of cyanides relative to 
methanol in group 2 looks inconsistent with the chemical differentiation found 
by \citet{Csengeri18,Csengeri19} in G328.2551-0.5321 where, 
according to their interpretation, the accretion shocks at the centrifugal
barrier are traced by oxygen-bearing molecules such as methanol while the 
cyanides trace a more compact region closer to the protostar. It is therefore
unclear whether the accretion shock scenario is the correct interpretation for
these group 2 sources.

\citet{Lee19} reported the column densities of several COMs associated with the
disk atmosphere of the Class 0 protostar HH212. These column densities turn
into abundances relative to methanol of 0.021 for C$_2$H$_5$OH, 0.025 for 
CH$_3$OCHO, 0.012 for CH$_3$CHO, and 0.0012 for NH$_2$CHO, with uncertainties 
of a factor of a few. Among the three chemical groups identified in the 
CALYPSO survey, HH212 seems to fit best into group 1. Given that a two-sided 
disk atmosphere corresponds to an elongation along the outflow axis, it may
not be surprising that we found a possible outflow origin for IRAS2A1 in 
group 1. Higher angular resolution observations would be needed to verify
if the COM emission in IRAS2A1 traces a disk atmosphere like in HH212.

\subsection{Relation between chemical composition of protostars and cloud 
environment}
\label{ss:clouds}

The three chemical groups identified in this study are not related, on a 
one-to-one basis, to the molecular clouds which the sources belong to. Group 3 
contains sources in L1448 (L1448-2A, L1448-C), Serpens South (SerpS-MM18a), 
and Cepheus (L1157), group 2 contains sources in NGC1333 (SVS13A, IRAS4A2, 
IRAS4B) and Serpens Main (SerpM-S68N), while another source in NGC1333 
(IRAS2A1) belongs to group 1 that also contains sources in Ophiuchus 
(IRAS16293B) and the Aquila rift (L483). 

However, one cloud of our survey stands out: no COM was detected with CALYPSO 
toward the three protostars located in Taurus (IRAM04191, L1521F, L1527).
IRAM04191 and L1521F are the sources with the lowest luminosity in
the CALYPSO sample, and the size of their hot (100-150~K), inner envelope is 
expected to be more than one order of magnitude smaller than the CALYPSO beam 
(see Fig.~\ref{f:rcom_rdisk_rsub}), so it is maybe not surprising that we did 
not detect any COM toward these sources. Methanol has been detected toward 
L1527 with ALMA \citep[][]{Sakai14b} and CALYPSO was not sensitive enough to 
see it. However, the upper limit that we obtained for its column density 
normalized by the dust continuum emission is about one order of magnitude 
lower than the normalized column density of methanol in the sources where 
CALYPSO detected it (see Fig.~\ref{f:chemcompsolang1mmcont}). This suggests 
that the methanol abundance in L1527 is significantly lower than in these 
sources. However, sources such as IRAS4A1, IRAS4B2, or SerpM-SMM4a have 
similarly low or even lower upper limits for methanol in 
Fig.~\ref{f:chemcompsolang1mmcont}, so the low abundances of COMs in the 
Taurus protostars observed with CALYPSO do not necessarily result from the 
Taurus environment itself.

We conclude from all this that the differences in COM chemical composition in 
the  CALYPSO sample do not reflect the global environment in which these 
sources are embedded. Source-to-source variations in chemical composition 
within a given cloud rather suggest an evolutionary effect or the influence of 
local conditions (episodic accretion?).

\subsection{Correlation between column densities does not imply chemical link}
\label{ss:chemical_link}

We reported in Sect.~\ref{ss:correl_coms} strong correlations for three pairs 
of molecules: CH$_3$CN/CH$_3$OH, NH$_2$CHO/CH$_3$OH, and 
CH$_3$OCH$_3$/C$_2$H$_5$OH. It has been argued in the past that correlations
between the column densities or abundances of molecules reveal chemical links 
between them. For instance \citet{Jaber14} found a strong correlation over 
five orders of magnitude between methyl formate, CH$_3$OCHO, and dimethyl 
ether, CH$_3$OCH$_3$. They suggested that this correlation results from the two 
species having the same precursor, as previously advocated by 
\citet{Brouillet13} on the basis of similar arguments, or from one being the 
precursor of the other. They drew the same conclusion for methyl formate and 
formamide, which showed a similar correlation. The results we obtain from
the CALYPSO survey shed a new light on this matter. To our knowledge, there
is no obvious chemical link between methyl cyanide and methanol: the formation
of methyl cyanide is dominated by the reaction of CH$_3$$^+$ with HCN in the
gas phase in R. Garrod's chemical models for instance 
\citep[e.g.,][]{Belloche16}, while methanol is known to form efficiently only 
on the grain surfaces by hydrogenation of CO. Yet we find a strong correlation 
between methanol and methyl cyanide. We conclude from this that the
existence of a strong correlation between two molecules in a sample of sources
does not imply that these molecules are related chemically. 
A similar conclusion was drawn by \citet{Quenard18} for the correlation
between HNCO and NH$_2$CHO reported in earlier studies. They deduced from
their chemical modelling that this correlation comes from the same response of 
the two molecules to the temperature rather than from a direct chemical link.

\subsection{Comparison to water surveys}
\label{ss:water}

\citet{Mottram14,Mottram17} analyzed water emission observed with 
\textit{Herschel} in two samples of Class~0 and I protostars (the WISH and 
WILL samples). These samples have the following sources in common with 
CALYPSO: L1448-C (L1448-MM), IRAS2A, SVS13 (IRAS~3A), IRAS4A, IRAS4B, L1527, 
Serp-SMM4 (Ser-SMM4), L1157, L1448-2 (PER01), L1448-N (PER02), and SerpS-MM18 
(SER02). L1448-C, IRAS2A1, IRAS4A2, IRASB, and SerpS-MM18a, which all have 
many COMs detected with CALYPSO, have relatively strong detections of water 
components corresponding to cavity shocks or spot shocks, whereas L1527, with 
no COM detected with CALYPSO, has much weaker water emission. This could 
suggest a link between the detection of COMs and the interaction of the 
jet/outflow with the envelope. However, L1448-N and SerpM-SMM4, which have no 
firm COM detection or only a methanol detection, have strong water detections 
like the former sources, and L1157, which has three COMs detected with 
CALYPSO, has a much weaker water emission. Furthermore, the \textit{Herschel} 
surveys did not have the angular resolution needed to separate the individual 
components of multiple systems, which show very different COM properties in 
the CALYPSO sample. Overall, it is thus unclear whether the COM emission in 
Class~0 protostars is related to their jet/outflow activity. SVS13-A, a 
Class~I protostar with many COMs detected with CALYPSO, was observed in only 
one low-energy water line, so we cannot compare it to the other sources in a 
meaningful way.

\subsection{Comparison to other COM surveys}
\label{ss:othersurveys}

\citet{Taquet15} reported results of their observations of IRAS2A and 
IRAS4A with PdBI at 2~mm with an angular resolution of $\sim$2$\arcsec$. The 
column densities they derived for C$_2$H$_5$OH, CH$_3$OCH$_3$, CH$_3$OCHO, 
NH$_2$CHO, CH$_3$CN, C$_2$H$_5$CN, and CH$_2$(OH)CHO all agree within a factor
of two with the ones we obtained with CALYPSO, after rescaling the column
densities of IRAS2A to account for the different source size that they assumed
(0.2$\arcsec$ versus 0.35$\arcsec$ here). However, their methanol column 
densities are a factor 3 and 9 times higher than the ones we derived for these 
sources. This may be due to the larger beam of their observations which may be 
more contaminated by the outflow. This would then imply that the abundances
relative to methanol of the COMs listed above are much lower in the outflow
than in the inner region traced with CALYPSO. While the agreement between the
CALYPSO column densities and the ones derived by \citet{Taquet15} is good, we
find discrepancies larger than a factor of 2 for most COMs detected by 
\citet{LopezSepulcre17} toward IRAS4A2 with ALMA and PdBI at $\sim$0.5$\arcsec$
and $\sim$1.2$\arcsec$ angular resolution, respectively, after rescaling their 
column densities from a source size of 0.3$\arcsec$ to 0.35$\arcsec$. The 
largest discrepancy, a factor 7, occurs for NH$_2$CHO. The reasons for these 
discrepancies is unclear. Like \citet{LopezSepulcre17}, we do not detect any
COM toward IRAS4A1.

\citet{Graninger16} and \citet{Bergner17} reported the results of a survey of 
16 embedded protostars with the IRAM 30\,m telescope. Most of their targets are 
Class~I objects and the COM emission they detected is characterized by low
rotational temperatures, suggesting that their survey probes the cold and 
low-density part of the protostellar envelopes rather than the hot, 
high-density, inner regions traced with CALYPSO. Their sources were all 
detected in methanol, while CH$_3$CHO and CH$_3$CN were detected toward 6 and 
7 sources, respectively. Only two sources were detected in CH$_3$OCH$_3$ and 
CH$_3$OCHO. All sources but two were detected in HNCO. The HNCO abundances 
relative to methanol range from $\sim$0.06 to $\sim$0.5, with most values 
around 0.2. This is more than one order order of magnitude higher than what we 
obtained for the CALYPSO sample, which is dominated by Class~0 protostars. 
This difference unlikely results from an evolutionary effect, because the 
Class~I protostar SVS13A in the CALYPSO sample has a HNCO abundance relative 
to methanol also one order of magnitude lower than the IRAM 30\,m sample. It 
probably rather reflects a difference in chemical composition between the 
(cold) envelope traced with the IRAM 30\,m telescope and the small-scale, hot 
emission traced with PdBI. 

The two sources of \citet{Bergner17} with detections of CH$_3$OCH$_3$ and 
CH$_3$OCHO have abundances relative to methanol of $\sim$0.09 and $\sim$0.06 
for these two molecules, respectively, which is similar to what we obtain for 
group 3 for both molecules and group 2 for CH$_3$OCH$_3$, but a factor 3--5 
higher than group 1 (Table~\ref{t:chemcomp}). In addition, \citet{Bergner17} 
report a median value of CH$_3$CHO abundance relative to methanol similar to 
the values of groups 2 and 3 but a factor $\sim$5 higher than group 1. Group 1 
may thus represent sources that have undergone a different type of chemical 
processing of O-bearing COMs on small scale compared to the dominant processing 
in the cold envelope. The situation is less clear for the N-bearing 
COMs: the CH$_3$CN abundances relative to methanol derived by 
\citet{Bergner17} range from 0.005 to 0.05, which encompasses the values we 
derive for the three groups of the CALYPSO sample.

\citet{Bergner17} found that methanol is well correlated with CH$_3$CHO and
CH$_3$CN in their source sample. We find a good correlation between methanol 
and CH$_3$CN for the CALYPSO sources as well, but the correlation is much 
weaker between methanol and CH$_3$CHO. They also claim finding a correlation 
between the envelope mass and the column densities of all molecules but this 
does not seem to be really the case for the COMs shown in their Fig. 8 
(CH$_3$CN, CH$_3$OH, and CH$_3$CHO). Their claim of a positive correlation
with bolometric luminosity is not convincing either (see their Fig. 9). Their 
CH$_3$CHO abundances relative to methanol do not correlate with the luminosity 
either, while we find an anti-correlation between the internal luminosities 
and the abundances of CH$_3$CHO and CH$_2$(OH)CHO relative to methanol 
(Sect.~\ref{ss:correl_prop}). 

On the basis of a compilation of literature results on \hbox{low-,} 
intermediate-, and high-mass hot cores combined with their 
measurements on IRAS4A and IRAS2A, \citet{Taquet15} found a correlation 
between the luminosity and the abundances of C$_2$H$_5$CN and CH$_2$(OH)CHO 
relative to methanol (see their Fig. 8). However, the sample of sources with 
CH$_2$(OH)CHO is small in both their and our studies (four sources), and their 
correlation does not hold for the (three) sources with luminosities below 
100~$L_\odot$. \citet{OspinaZamudio18} also found a correlation between the
abundance of C$_2$H$_5$CN relative to methyl formate and the luminosity for
a compilation of eight low-, \hbox{intermediate-,} and high-mass hot cores, 
but they 
did not find any systematic variations of the abundances of O-bearing COMs
with luminosity between the low- and intermediate-mass sources which cover 
nearly two orders of magnitude in luminosity. However, their sample does not 
cover luminosities below 9~$L_\odot$ while our sample of sources with COM 
detection extends down to 2 $L_\odot$. The luminosity may thus have a bigger 
impact on the O-bearing COM chemical composition over the range 
1--10~$L_\odot$ than over the range 10--500~$L_\odot$.

The ASAI survey performed with the IRAM 30~m telescope has revealed the 
chemical composition at large scale of a sample of low-mass
star forming regions, from prestellar cores to protostars. \citet{Lefloch18}
used the ratio of the number of detected O-bearing species to hydrocarbons to
classify the sources into WCCC sources or hot corinos. Four sources are
in common with CALYPSO: L1157 and L1527 were classified as WCCC sources, and
IRAS4A and SVS13A as hot corinos on the basis of the ASAI results. The former 
two sources have a low channel count peak in the CALYPSO survey, while the 
latter two have a high channel count peak. The chemical richness of 
protostars probed at small scale with PdBI thus seems to be related to the
chemical composition probed at large scale with the single-dish telescope.
However, L1157 has several COMs detected on small scale with CALYPSO, so 
its WCCC nature at large scale does not prevent the existence of a hot corino
at small scale. This is similar to the candidate WCCC source L483 which
also harbors a hot corino \citep[][]{Oya17}. No COMs are detected at small
scale toward L1527 with CALYPSO, but methanol was detected with ALMA 
\citep[][]{Sakai14b}. Given that we classified L1157 in the same COM chemical 
group as SerpS-MM18a, L1448-C, and L1448-2A (group 3), it would be interesting 
to check if the latter three sources also present a WCCC nature at large scale.

\subsection{Comparison to COM chemical models}
\label{ss:models}

\citet{Bergner17} presented results of chemical simulations computed with the
three-phase chemical kinetics code MAGICKAL \citep{Garrod13}. They used a grid
of simulations to map the computed abundances to the physical structure of a 
low-mass protostellar envelope. They considered two cases, a 1~$L_\odot$ and a 
10~$L_\odot$ protostar. They compared the results of these 
simulations to their single-dish observations of a sample of Class 0 and I 
protostars, which were also sensitive to the large-scale COM emission of 
the envelopes. The CALYPSO interferometric data that we have used here
probe only the inner parts of the envelopes because of the spatial filtering 
by the interferometer. Therefore, in order to compare the chemical composition
derived from the CALYPSO survey to the one predicted by the models, we
considered only the inner regions of the models of \citet{Bergner17}. In 
practice, we integrated the modeled column densities of the molecules over the 
inner region where the abundance of methanol is the highest and forms a plateau
\citep[see Fig.~12 of][]{Bergner17}. The radius of this region is about 6~au 
and 20~au for the 1~$L_\odot$ and 10~$L_\odot$ simulations, respectively. The
resulting average abundances of several COMs relative to methanol are listed at
the bottom of Table~\ref{t:chemcomp}. 

Table~\ref{t:chemcomp} shows that the modeled COM abundances relative to
methanol are relatively insensitive to the adopted luminosity of the 
protostar. Both models underestimate the abundances derived from the CALYPSO 
survey by at least a factor three and up to two orders of magnitude for the
COMs, and even three orders of magnitude for HNCO. Nevertheless, we notice 
that, among the three chemical groups of the CALYPSO sample, group 1 comes the 
closest to the model predictions, in particular with respect to CH$_3$OCH$_3$ 
and CH$_3$OCHO that agree within a factor of three and six with the model, 
respectively, which can be considered as satisfactory given the uncertainties 
on the reaction rates in the chemical network used for the simulations. We
could then speculate that group 1 is the most consistent with a hot-corino
origin of the COM emission, but this is not in agreement with our conclusion
stated in Sect.~\ref{ss:evolution} that sources matching a possible hot-corino
scenario are found in all three groups. In addition, the discrepancy for the 
other molecules is about two orders of magnitude for group 1.

The models of \citet{Bergner17} underpredict by typically one order of 
magnitude the abundances of most COMs in their sample of protostars. Only 
CH$_3$CHO showed a good match. They argued that the discrepancy may result 
from cold (large-scale) methanol being overabundant in the model, 
due to an overactive chemical desorption. We find, however, that this 
discrepancy holds also at the small scales probed by CALYPSO, where thermal 
desorption is most likely dominant. The reason for the discrepancy between
models and observations must lie somewhere else. Higher abundances relative
to methanol of COMs such as CH$_3$OCHO and CH$_3$OCH$_3$ were obtained in models
that include proton-transfer reactions with ammonia in the gas phase and/or
luminosity outbursts \citep[][]{Taquet16}. In the latter case, the abundance
of these two molecules relative to methanol are enhanced after the burst 
because they recondense more slowly than methanol. However, the timescale for
this recondensation is short according to \citeauthor{Taquet16}'s simulations 
($< 1000$~yr), and we have concluded in Sect.~\ref{ss:evolution} that the 
sample of CALYPSO sources with COM detections does not seem to be, as a whole, 
affected by episodic accretion so the solution for the discrepancy is likely
not this one. We do not know the abundance of ammonia in the 
CALYPSO sources, so it is difficult to conclude whether the proton-transfer 
mechanism proposed by \citet{Taquet16} can explain the discrepancy between the 
COM abundances relative to methanol obtained with CALYPSO and the ones 
predicted by the models of \citet{Bergner17}.

\subsection{Implications for the formation of COMs}

\citet{Chuang16} investigated the cold surface formation of 
glycolaldehyde, ethylene glycol, and methyl formate with laboratory 
experiments that involve the recombination of free radicals formed via H-atom
addition and abstraction reactions, starting from ice mixtures of CO, H$_2$CO,
and CH$_3$OH. In most of their experiments, glycolaldehyde is found to have a 
similar abundance as methyl formate, or to be more abundant, while only two 
experiments (CO+H$_2$CO+H and H$_2$CO+CH$_3$OH+H) produce less glycolaldehyde 
than methyl formate, with an abundance ratio of about 1:3. In their more 
recent work where they compared the production of these three molecules under
H-atom addition and/or UV irradiation, glycolaldehyde and ethylene glycol 
are both always more abundant than methyl formate \citep[][]{Chuang17}. The 
numerical simulations of \citet{Garrod13} also produce glycolaldehyde 3--6 
times more abundant than methyl formate in the gas phase after sublimation 
from the ice mantles of dust grains where they are formed. 

These experimental 
and numerical results are much different from the abundance ratios 
[CH$_2$(OH)CHO]/[CH$_3$OCHO] that we obtained for the CALYPSO sample: the 
derived ratios range from 3\% (SVS13A) to 11\% (IRAS4A2), and the upper limits 
in sources where methyl formate is detected but not glycolaldehyde range from 
9\% (SerpS-MM18a) to 80\% (L1157). \citet{Taquet17} reported a similarly low 
upper limit ($<$6\%) in the cold dark cloud Barnard~5 and concluded that the 
discrepancy with the experimental results of \citet{Chuang16} suggests that 
surface chemistry is not the dominant mechanism for the formation of methyl 
formate. Comparing their experimental results to a compilation of 
observational results including IRAS2A and IRAS4A from \citet{Taquet15}, 
\citet{Chuang17} also conclude that the overabundance of glycolaldehyde over
methyl formate suggests that gas-phase chemistry plays a significant role,
either through the destruction of glycolaldehyde or an enhanced production of 
methyl formate. A similar conclusion could be drawn now from the CALYPSO 
sample. \citet{Skouteris18} have, however, argued that glycolaldehyde could be
formed in the gas phase. It would thus be interesting to investigate
if pure gas phase chemistry can produce a [CH$_2$(OH)CHO]/[CH$_3$OCHO] ratio
consistent with the observed ones.

\subsection{Comparison to comet 67P/Churyumov-Gerasimenko}
\label{ss:comets}

\citet{Drozdovskaya19} reported a correlation between the chemical composition
of IRAS16293B and comet 67P/Churyumov-Gerasimenko, concluding that the volatile
composition of cometesimals and planetesimals is partially inherited from the
prestellar and protostellar phases. Our analysis of the CALYPSO
sample shows that IRAS16293B shares a similar COM chemical composition as 
IRAS2A1 and L483 (chemical group 1, see Sect.~\ref{ss:chemcomp}), suggesting
that comet 67P shares a similar composition as group 1. However, groups 2 and 
3 are characterized by abundances relative to methanol of O-bearing molecules 
higher by a factor of $\sim$6 compared to group 1. This factor is similar to 
the dispersion of the correlation found by \citet{Drozdovskaya19}. 
Therefore, on the basis of the limited sample of COMs analyzed here, it is 
still too premature to conclude if one of the three chemical groups is more 
correlated to the chemical composition of comet 67P than the others.

\section{Conclusions}
\label{s:conclusions}

We have taken advantage of the CALYPSO survey to explore the presence of COMs
in a large sample of 22 Class~0 and four Class~I protostars at high angular 
resolution. Methanol is detected in 12 sources and tentatively detected in one 
source, which represents half of the sample. Eight sources (30\%) have 
detections of at least three COMs. We derived the column
densities of the detected COMs and searched for correlations with various
source properties, either collected from the literature or derived from 
the CALYPSO survey. The main conclusions of this analysis are the following:

\begin{enumerate}
 \item The high angular resolution of the CALYPSO survey has revealed a strong
 COM chemical differentiation in multiple systems: five systems have at least
 three COMs detected in one component while the other component is devoid of 
 COM emission. This is markedly different from the prototypical hot-corino 
 source IRAS16293 where many COMs have been reported towards both components 
 of the binary in the literature. This also raises the question whether all 
 protostars go through a phase showing COM emission.
 \item All CALYPSO sources with an internal luminosity higher than 4~$L_\odot$ 
 have at least one detected COM (methanol). On the contrary, no COM emission 
 is detected in sources with an internal luminosity lower than 2~$L_\odot$. 
 This seems to be due to a lack of sensitivity rather than an intrinsic 
 property of low-luminosity sources.
 \item The internal luminosity is the source parameter impacting the most the
 COM chemical composition. The abundances of CH$_3$CHO and CH$_2$(OH)CHO 
 relative to methanol are anti-correlated with the internal luminosity. There 
 seems to be a correlation between the internal luminosity and the column 
 density of CH$_3$OH normalized to the continuum emission.
 \item The detection of a disk-like structure in continuum emission does not 
 imply the detection of COM emission, and vice versa. The size of the COM 
 emission, when detected, is not systematically related to the size of the
 disk-like structure nor to the extent of the hot inner envelope.
 \item No single scenario can explain the origin of COMs in all
 the CALYPSO sources with COM detections. For seven sources out of the
 nine sources with a measured COM emission size, we find that a canonical
 hot-corino origin may explain the COM emission in four sources, an 
 accretion-shock origin in two or possibly three sources, and an outflow 
 origin in
 three sources, whereby three of these sources fit into two of these three 
 scenarii. One of the two remaining sources may fit into a hot-corino scenario 
 coupled to a recent accretion burst.
 \item The CALYPSO sources with COM detections show different chemical 
 compositions. We identified three groups on the basis of the abundances of 
 oxygen-bearing molecules, cyanides, and CHO-bearing molecules relative to 
 methanol. These chemical groups do not correlate with the three scenarii 
 mentioned above. They do not seem to correlate either with the evolutionary
 status of the sources if we take the ratio of envelope mass to internal
 luminosity as an evolutionary tracer. However, the chemical groups do not 
 correlate either with the cloud environment in which the sources are 
 embedded. The source-to-source variations in COM chemical composition may thus
 rather reflect an evolutionary effect or the influence of local conditions 
 such as episodic accretion. 
 \item The column densities of several pairs of COMs correlate well with each
 other although some of these pairs, such as CH$_3$OH and CH$_3$CN, are not 
 linked chemically. Therefore, the existence of a strong correlation between 
 two molecules does not imply that these molecules are related chemically.
\end{enumerate}

While the CALYPSO survey was initially not designed for an extensive study of 
the COM emission in young protostars, its high angular resolution and 
sensitivity has allowed us to start shedding light on the presence of COMs in 
a more statistical way than has been done before. However, no single scenario 
that can explain the origin of COMs in the CALYPSO sample emerges from our 
analysis. Future imaging spectral line surveys of a larger sample of young 
protostars at even higher angular resolution sufficient to resolve the expected
disk scales (a few tens of au) will be necessary to make further
progress. The determination of individual internal luminosities in close 
binaries and multiple systems will also be necessary to search for correlations
in a more robust way. Searching for correlations between the COM emission and
the jet/outflow properties of the sources may also be promising.

\begin{acknowledgements}
We thank Rob Garrod for sending us the abundance profiles of the models of 
\citet{Bergner17} in electronic format. We are grateful to Bilal Ladjelate for
sending us his estimates of the internal luminosities prior to publication
and Beno\^it Tabone for discussions about the relation between disks and COMs. 
We thank Sandrine Bottinelli, Beno\^it Commer{\c c}on, Cornelis Dullemond, 
Patrick Hennebelle, Ralf Klessen, and Ralf Launhardt for their participation 
to the preparation of the CALYPSO project. Ph.A.\ and A.M.\ thank the European 
Research Council for funding under the European Union's Horizon 2020 research 
and innovation programme (ORISTARS grant agreement No. 291294 and MagneticYSOS 
grant agreement No. 679937). S.M.\ acknowledges support from the French Agence 
Nationale de la Recherche (ANR), under reference ANR-12-JS05-0005. S.C.\
gratefully acknowledges support by the Programme National ``Physique et Chimie 
du Milieu Interstellaire'' (PCMI) of CNRS/INSU with INC/INP co-funded by CEA 
and CNES, and by the Conseil Scientifique of Observatoire de Paris 
(AF ALMA-NOEMA).
L.T.\ thanks the following institutions for funding: the Italian Ministero 
dell' Istruzione, Universita e Ricerca through the grant Progetti Premiali 
2012 - iALMA (CUP C52I13000140001); the European Union's Horizon 2020 research 
and innovation programme under the Marie Sklodowska-Curie grant agreement No.
823823; the Deutsche Forschungsgemeinschaft (DFG, German Research Foundation) 
- Ref. No. FOR 2634/1ER685/11-1; and the DFG cluster of excellence ORIGINS 
(www.origins-cluster.de). 
\end{acknowledgements}

%
%

\begin{appendix}
\label{Appendix}

\section{Beam sizes and noise levels}
\label{a:beam_rms}

Table~\ref{t:beam_rms} lists the beam sizes, positions angles, and the noise
levels of the data cubes used to analyze the COM emission of the CALYPSO 
sources.

\begin{table*}[!ht]
 \begin{center}
 \caption{
 Beam sizes and noise levels.
}
 \label{t:beam_rms}
 \vspace*{-1.2ex}
 \begin{tabular}{lccccccccccc}
 \hline\hline
 \multicolumn{1}{c}{Field} & \multicolumn{1}{c}{Setup} & & \multicolumn{4}{c}{LSB} & & \multicolumn{4}{c}{USB} \\ 
 \cline{4-7} \cline{9-12}
 \noalign{\smallskip}
  & & & \multicolumn{1}{c}{Maj\tablefootmark{a}} & \multicolumn{1}{c}{Min\tablefootmark{a}} & \multicolumn{1}{c}{PA\tablefootmark{b}} & \multicolumn{1}{c}{rms\tablefootmark{c}} & & \multicolumn{1}{c}{Maj\tablefootmark{a}} & \multicolumn{1}{c}{Min\tablefootmark{a}} & \multicolumn{1}{c}{PA\tablefootmark{b}} & \multicolumn{1}{c}{rms\tablefootmark{c}} \\ 
  & & & \multicolumn{1}{c}{$\arcsec$} & \multicolumn{1}{c}{$\arcsec$} & \multicolumn{1}{c}{$^\circ$} & \multicolumn{1}{c}{mJy~beam$^{-1}$} & & \multicolumn{1}{c}{$\arcsec$} & \multicolumn{1}{c}{$\arcsec$} & \multicolumn{1}{c}{$^\circ$} & \multicolumn{1}{c}{mJy~beam$^{-1}$} \\ 
 \hline
 L1448-2A & S1 & & 1.34 & 1.05 & $ 24$ &  3.2  & & 1.33 & 1.05 & $ 23$ &  3.5  \\ 
 & S2 & & 1.00 & 0.71 & $ 26$ &  2.6  & & 1.01 & 0.72 & $ 25$ &  2.7  \\ 
 & S3 & & 1.79 & 1.42 & $ 72$ &  1.0  & & 1.76 & 1.39 & $ 68$ &  1.1  \\ 
 L1448-N & S1 & & 1.31 & 1.09 & $ 37$ &  7.0  & & 1.32 & 1.09 & $ 38$ &  7.6  \\ 
 & S2 & & 1.05 & 0.86 & $ 21$ &  3.6  & & 1.05 & 0.86 & $ 20$ &  3.8  \\ 
 & S3 & & 1.75 & 1.33 & $ 56$ &  1.1  & & 1.73 & 1.30 & $ 53$ &  1.1  \\ 
 L1448-C & S1 & & 1.22 & 0.95 & $ 29$ &  3.1  & & 1.22 & 0.95 & $ 29$ &  3.4  \\ 
 & S2 & & 1.02 & 0.73 & $ 27$ &  2.3  & & 1.03 & 0.73 & $ 26$ &  2.4  \\ 
 & S3 & & 1.82 & 1.45 & $ 70$ &  1.1  & & 1.79 & 1.42 & $ 66$ &  1.1  \\ 
 IRAS2A & S1 & & 1.18 & 0.97 & $ 56$ &  6.4  & & 1.17 & 0.96 & $ 56$ &  6.7  \\ 
 & S2 & & 1.02 & 0.85 & $ 23$ &  3.2  & & 1.03 & 0.85 & $ 23$ &  3.2  \\ 
 & S3 & & 1.74 & 1.33 & $ 57$ &  1.1  & & 1.72 & 1.30 & $ 54$ &  1.1  \\ 
 SVS13B & S1 & & 0.57 & 0.33 & $ 28$ &  2.7  & & 0.56 & 0.33 & $ 28$ &  3.0  \\ 
 & S2 & & 0.67 & 0.50 & $ 37$ &  2.3  & & 0.67 & 0.50 & $ 37$ &  2.3  \\ 
 & S3 & & 1.73 & 1.25 & $ 42$ &  0.8  & & 1.76 & 1.25 & $ 39$ &  0.8  \\ 
 IRAS4A & S1 & & 0.59 & 0.35 & $ 24$ &  2.2  & & 0.61 & 0.36 & $ 25$ &  2.7  \\ 
 & S2 & & 0.69 & 0.55 & $ 34$ &  2.7  & & 0.68 & 0.55 & $ 34$ &  2.4  \\ 
 & S3 & & 1.54 & 0.99 & $ 25$ &  1.1  & & 1.54 & 0.99 & $ 25$ &  1.2  \\ 
 IRAS4B & S1 & & 0.68 & 0.42 & $ 24$ &  3.0  & & 0.71 & 0.46 & $ 23$ &  2.9  \\ 
 & S2 & & 0.69 & 0.56 & $ 32$ &  3.0  & & 0.68 & 0.56 & $ 32$ &  2.9  \\ 
 & S3 & & 1.71 & 1.24 & $ 43$ &  1.1  & & 1.70 & 1.20 & $ 36$ &  1.1  \\ 
 IRAM04191 & S1 & & 1.38 & 1.20 & $ 25$ &  4.7  & & 1.36 & 1.19 & $ 16$ &  5.0  \\ 
 & S2 & & 1.12 & 0.92 & $ 15$ &  2.5  & & 1.13 & 0.93 & $ 15$ &  2.6  \\ 
 & S3 & & 1.88 & 1.61 & $ 56$ &  0.7  & & 1.91 & 1.57 & $ 48$ &  0.7  \\ 
 L1521F & S1 & & 1.35 & 1.05 & $ 23$ &  2.0  & & 1.35 & 1.04 & $ 22$ &  2.2  \\ 
 & S2 & & 1.04 & 0.74 & $ 27$ &  1.6  & & 1.05 & 0.74 & $ 26$ &  1.6  \\ 
 & S3 & & 1.88 & 1.50 & $ 83$ &  0.6  & & 1.84 & 1.48 & $ 79$ &  0.7  \\ 
 L1527 & S1 & & 0.57 & 0.35 & $ 41$ &  2.5  & & 0.57 & 0.35 & $ 40$ &  2.8  \\ 
 & S2 & & 0.64 & 0.51 & $ 51$ &  2.3  & & 0.64 & 0.51 & $ 52$ &  2.3  \\ 
 & S3 & & 1.64 & 1.24 & $ 46$ &  0.8  & & 1.65 & 1.24 & $ 44$ &  0.9  \\ 
 SerpM-S68N & S1 & & 0.92 & 0.50 & $ 24$ &  4.1  & & 0.91 & 0.45 & $ 23$ &  4.6  \\ 
 & S2 & & 1.03 & 0.56 & $ 26$ &  4.5  & & 1.02 & 0.56 & $ 26$ &  4.6  \\ 
 & S3 & & 2.25 & 1.44 & $ 28$ &  1.0  & & 2.22 & 1.34 & $ 27$ &  1.1  \\ 
 SerpM-SMM4 & S1 & & 0.91 & 0.52 & $ 28$ &  4.4  & & 0.91 & 0.51 & $ 30$ &  5.0  \\ 
 & S2 & & 1.12 & 0.69 & $ 33$ &  4.6  & & 1.13 & 0.68 & $ 34$ &  4.8  \\ 
 & S3 & & 2.36 & 1.43 & $ 27$ &  1.0  & & 2.33 & 1.33 & $ 26$ &  1.1  \\ 
 SerpS-MM18 & S1 & & 0.84 & 0.49 & $ 27$ &  3.2  & & 0.83 & 0.49 & $ 27$ &  4.0  \\ 
 & S2 & & 1.05 & 0.60 & $ 23$ &  4.7  & & 1.04 & 0.60 & $ 23$ &  4.8  \\ 
 & S3 & & 2.39 & 1.48 & $ 25$ &  1.1  & & 2.36 & 1.38 & $ 24$ &  1.1  \\ 
 SerpS-MM22 & S1 & & 0.90 & 0.50 & $ 23$ &  5.0  & & 0.89 & 0.49 & $ 23$ &  5.3  \\ 
 & S2 & & 1.15 & 0.58 & $ 20$ &  5.0  & & 1.13 & 0.58 & $ 20$ &  5.1  \\ 
 & S3 & & 2.33 & 1.35 & $ 25$ &  1.1  & & 2.29 & 1.29 & $ 23$ &  1.1  \\ 
 L1157 & S1 & & 0.61 & 0.50 & $ 19$ &  4.2  & & 0.61 & 0.50 & $ 19$ &  4.7  \\ 
 & S2 & & 0.59 & 0.46 & $ -1$ &  3.2  & & 0.59 & 0.46 & $ -1$ &  3.2  \\ 
 & S3 & & 1.43 & 1.04 & $ 62$ &  1.4  & & 1.43 & 1.06 & $ 60$ &  1.4  \\ 
 GF9-2 & S1 & & 0.92 & 0.61 & $ 18$ &  2.7  & & 0.93 & 0.62 & $ 18$ &  3.0  \\ 
 & S2 & & 0.87 & 0.73 & $ 20$ &  2.1  & & 0.82 & 0.72 & $ 11$ &  2.1  \\ 
 & S3 & & 1.44 & 1.02 & $ 66$ &  0.8  & & 1.43 & 1.03 & $ 61$ &  0.8  \\ 
 \hline
 \end{tabular}
 \end{center}
 \vspace*{-2.5ex}
 \tablefoot{
 \tablefoottext{a}{Beam sizes along major and minor axes (HPBW).}
 \tablefoottext{b}{Beam position angle measured east from north.}
 \tablefoottext{c}{Median noise level in channel maps.}
 }
 \end{table*}

\section{Spectra}
\label{a:spectra}

\begin{figure*}
 \centerline{\resizebox{1.0\hsize}{!}{\includegraphics[angle=0]{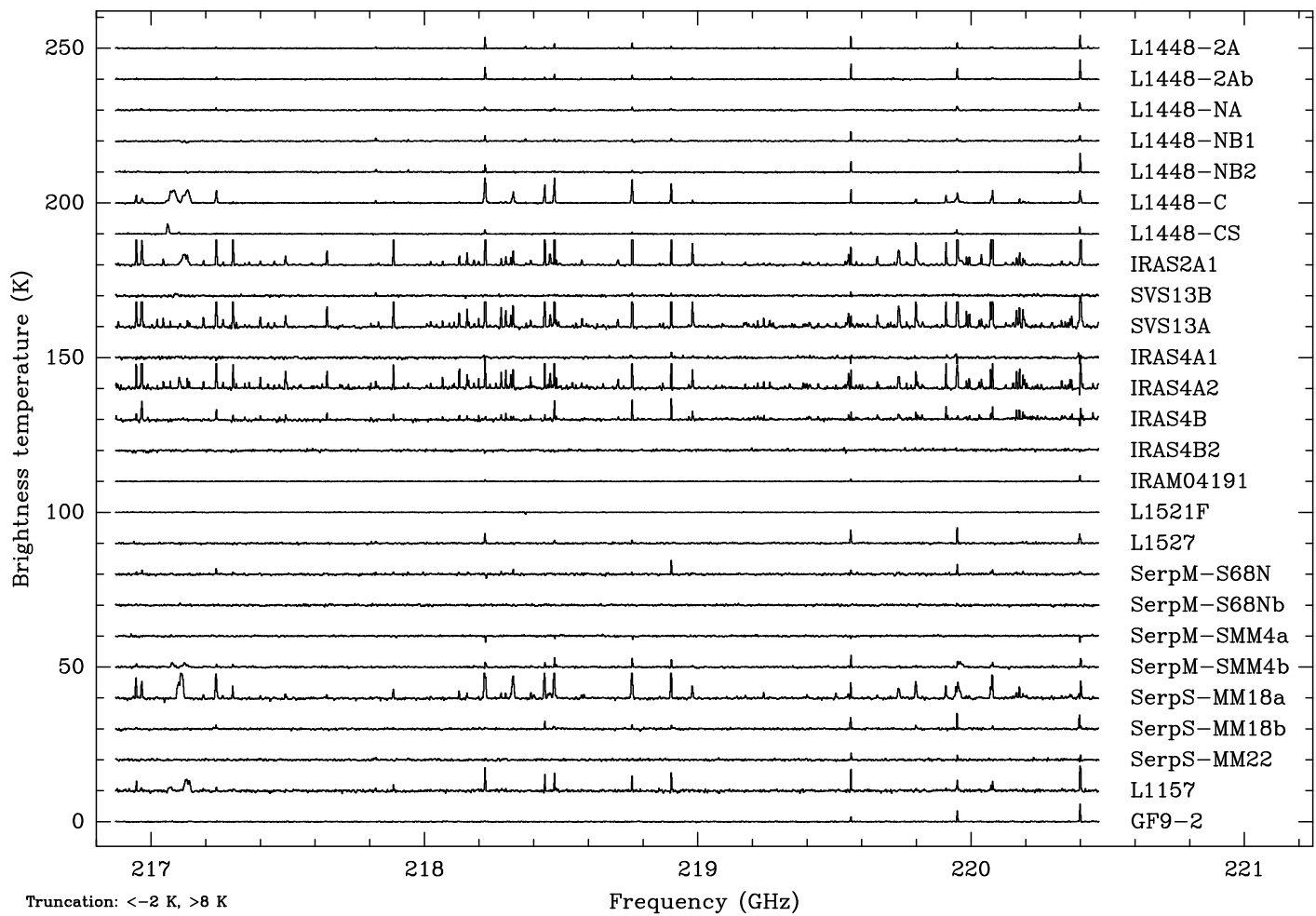}}}
 \caption{Continuum-subtracted WideX spectra at 1.4~mm toward the continuum 
emission peaks of the CALYPSO sources. For display purposes, the spectra were 
truncated to the range [-5, 8.5] K and shifted by multiples of 10~K vertically.}
 \label{f:spectraS2}
\end{figure*}

\begin{figure*}
 \centerline{\resizebox{1.0\hsize}{!}{\includegraphics[angle=0]{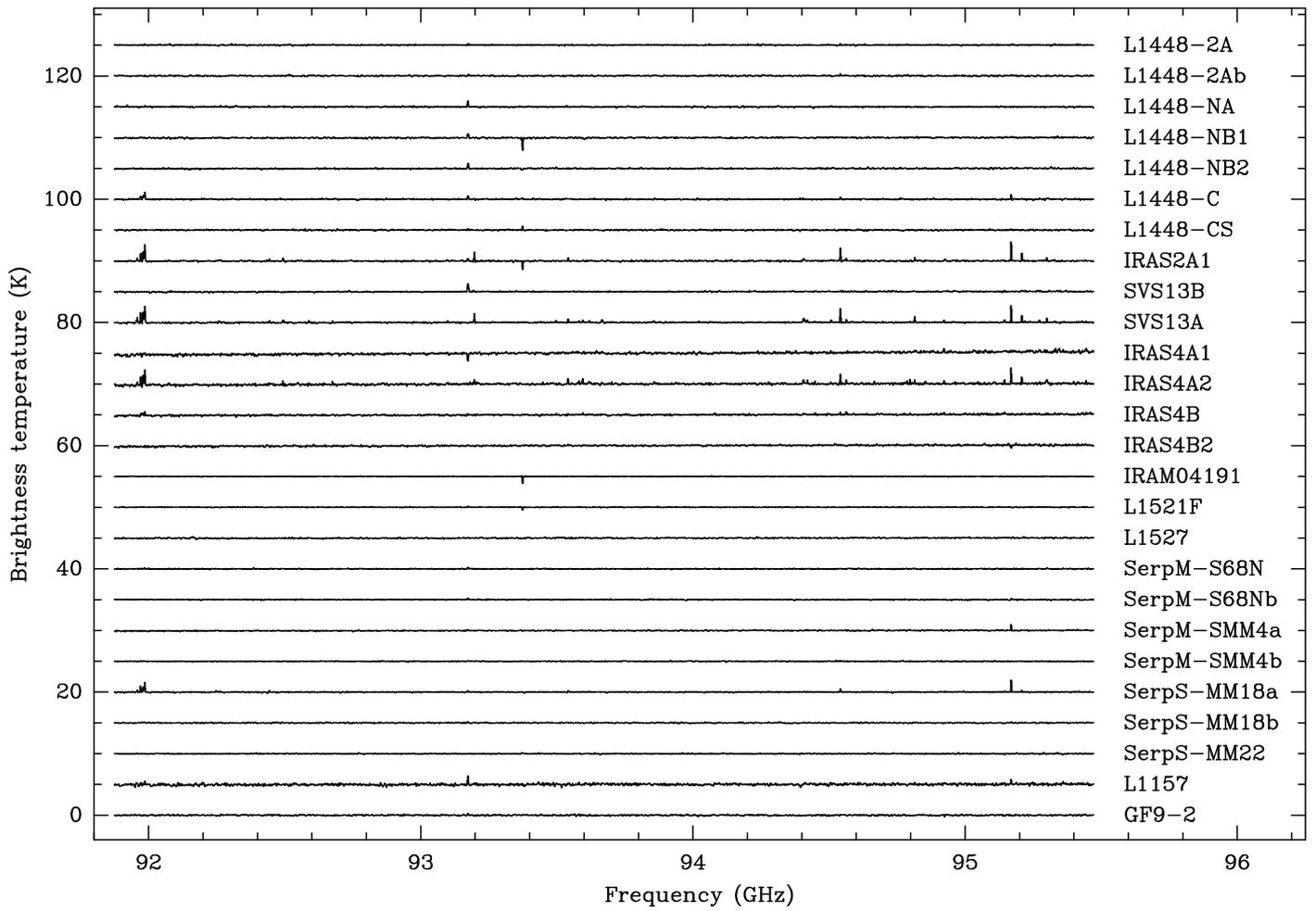}}}
 \caption{Continuum-subtracted WideX spectra at 3~mm toward the continuum 
emission peaks of the CALYPSO sources. For display purposes, the spectra were
shifted by multiples of 10~K vertically.}
 \label{f:spectraS3}
\end{figure*}

Figures~\ref{f:spectraS2} and \ref{f:spectraS3} show the S2 and S3 spectra 
obtained toward some of the continuum peaks of the CALYPSO sources like in 
Fig.~\ref{f:spectraS1} for setup S1.

\clearpage
\newpage

\section{Additional maps and coordinates}
\label{a:coordinates}

\begin{figure*}
 \centerline{\resizebox{0.24\hsize}{!}{\includegraphics[angle=0]{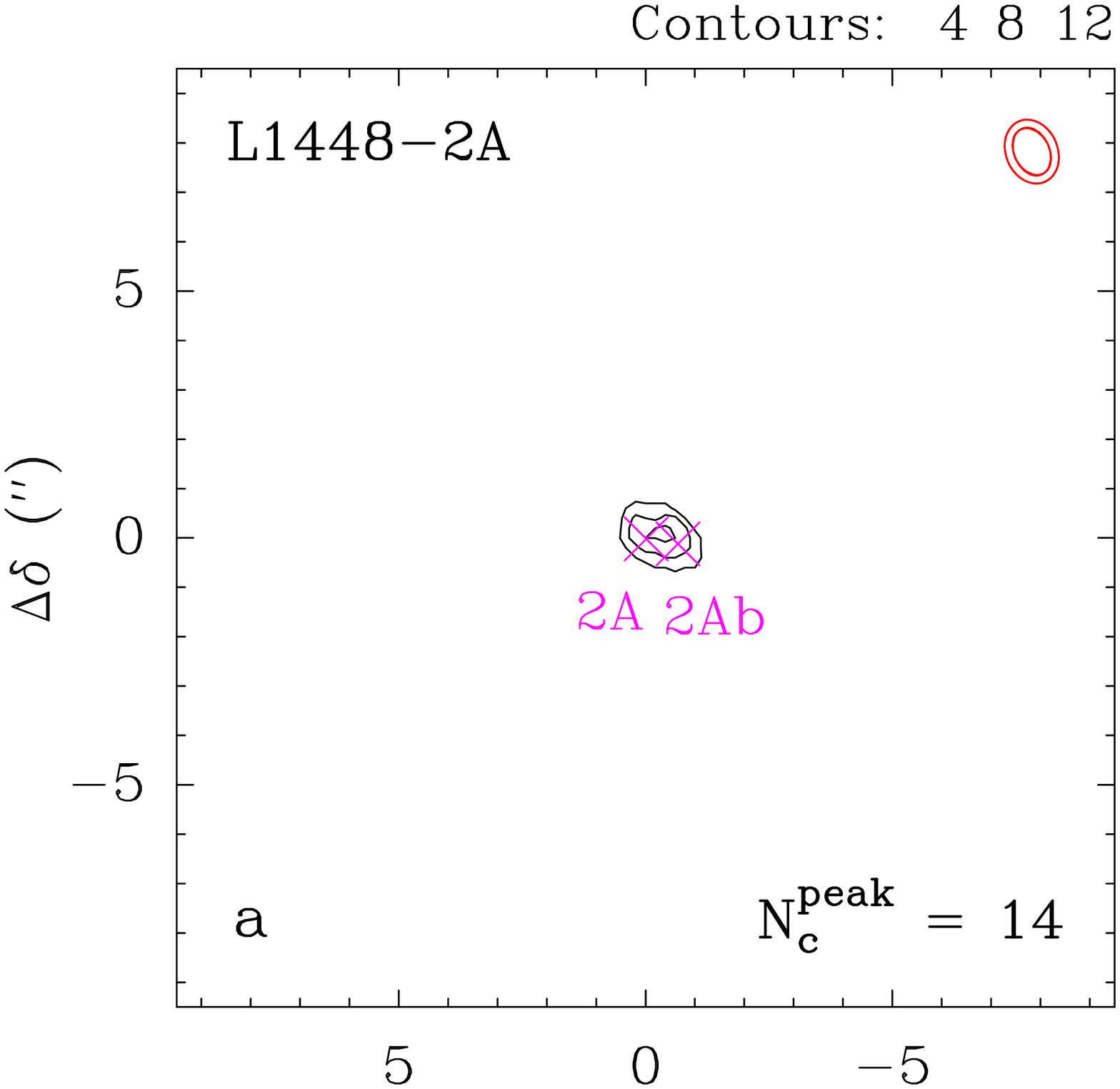}}\resizebox{0.24\hsize}{!}{\includegraphics[angle=0]{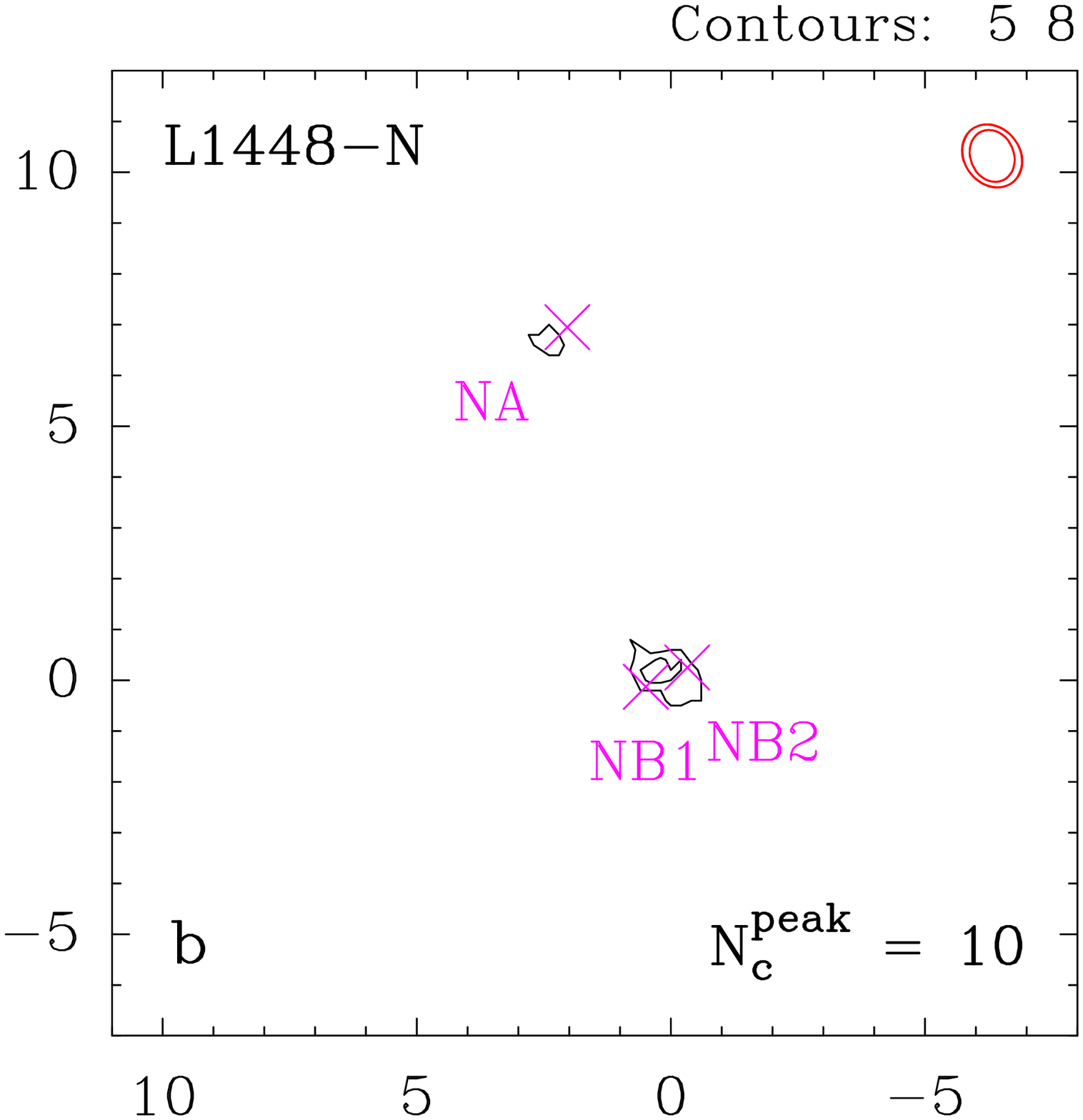}}\resizebox{0.24\hsize}{!}{\includegraphics[angle=0]{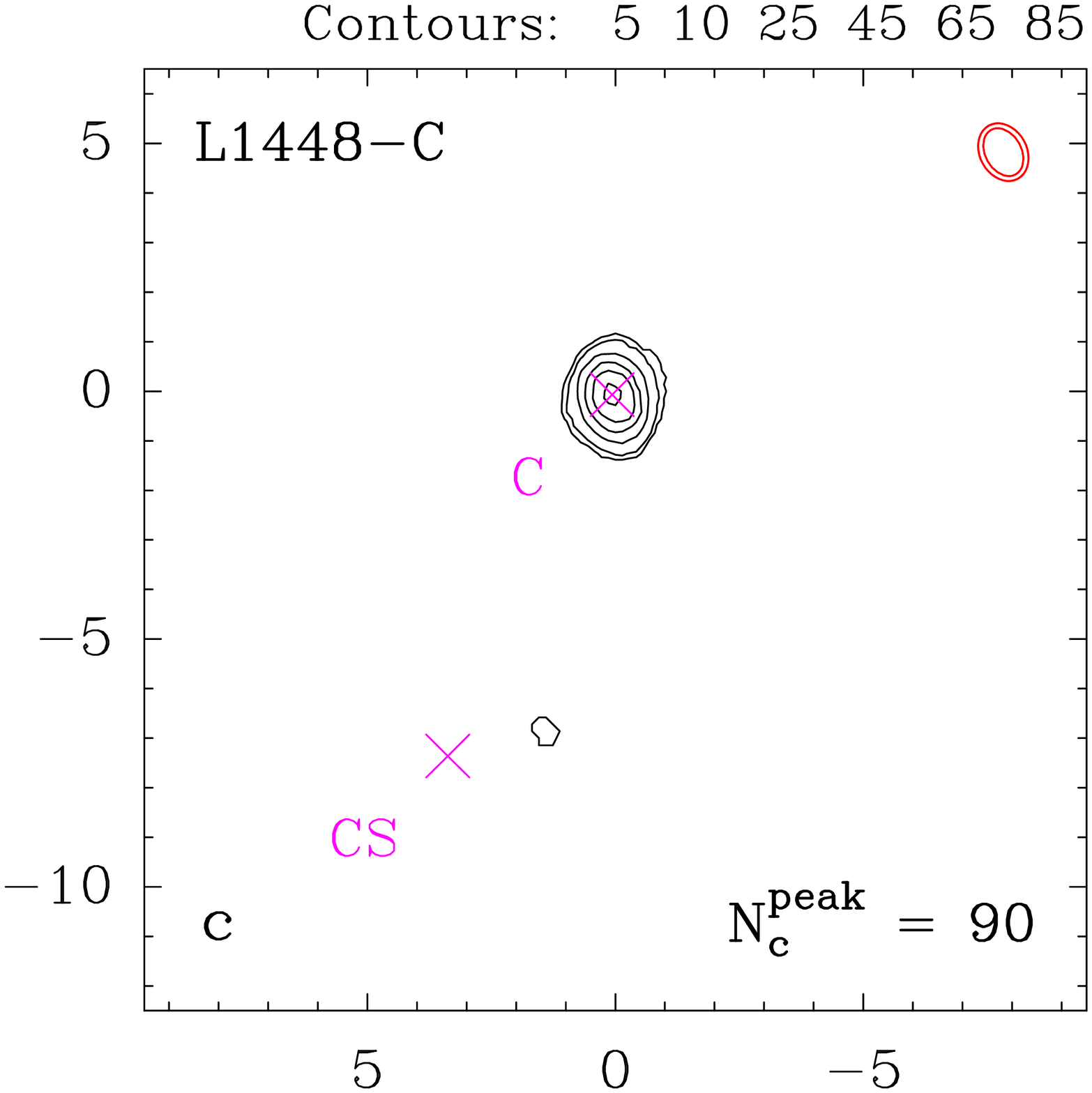}}\resizebox{0.24\hsize}{!}{\includegraphics[angle=0]{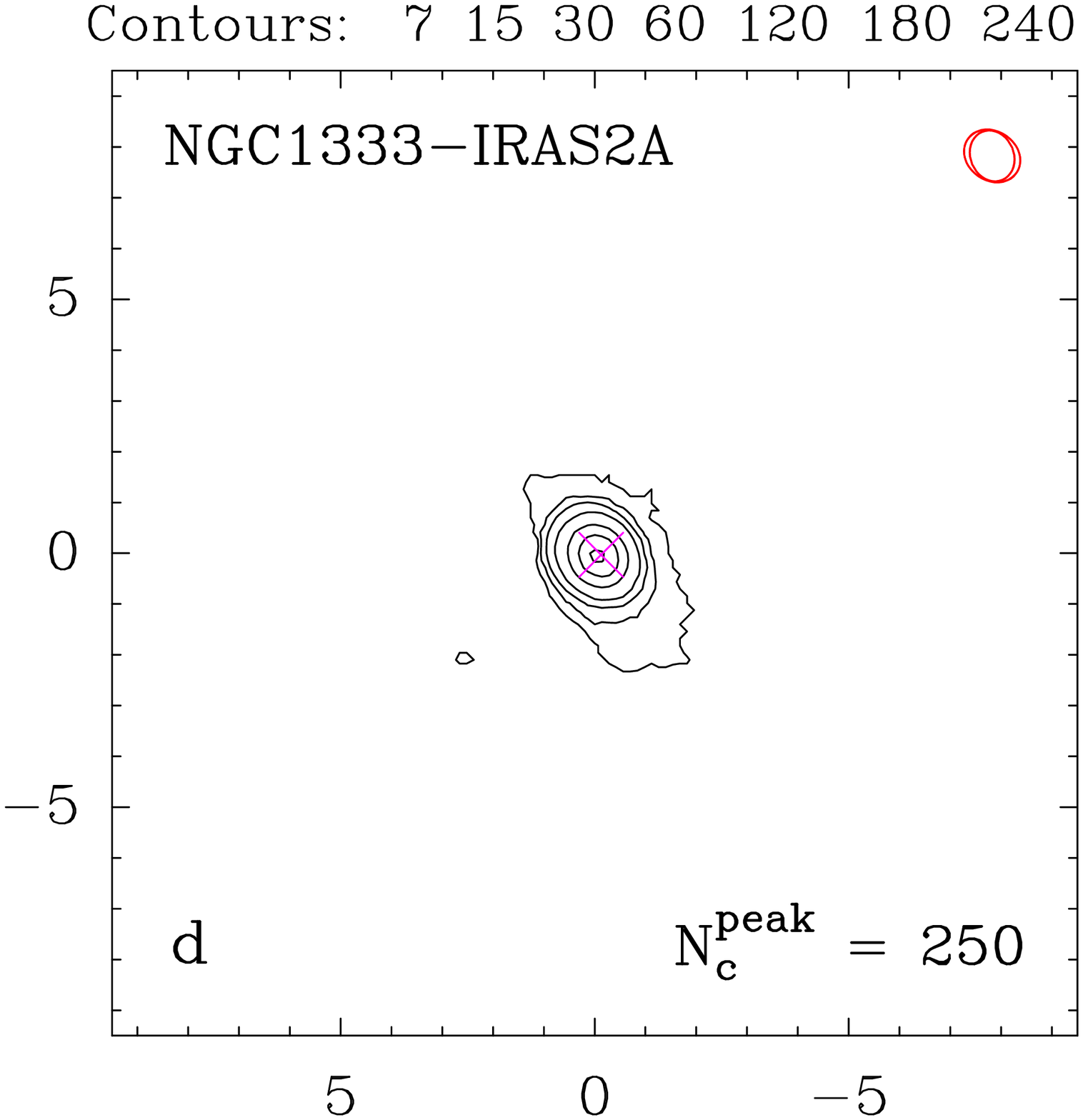}}}
 \centerline{\resizebox{0.24\hsize}{!}{\includegraphics[angle=0]{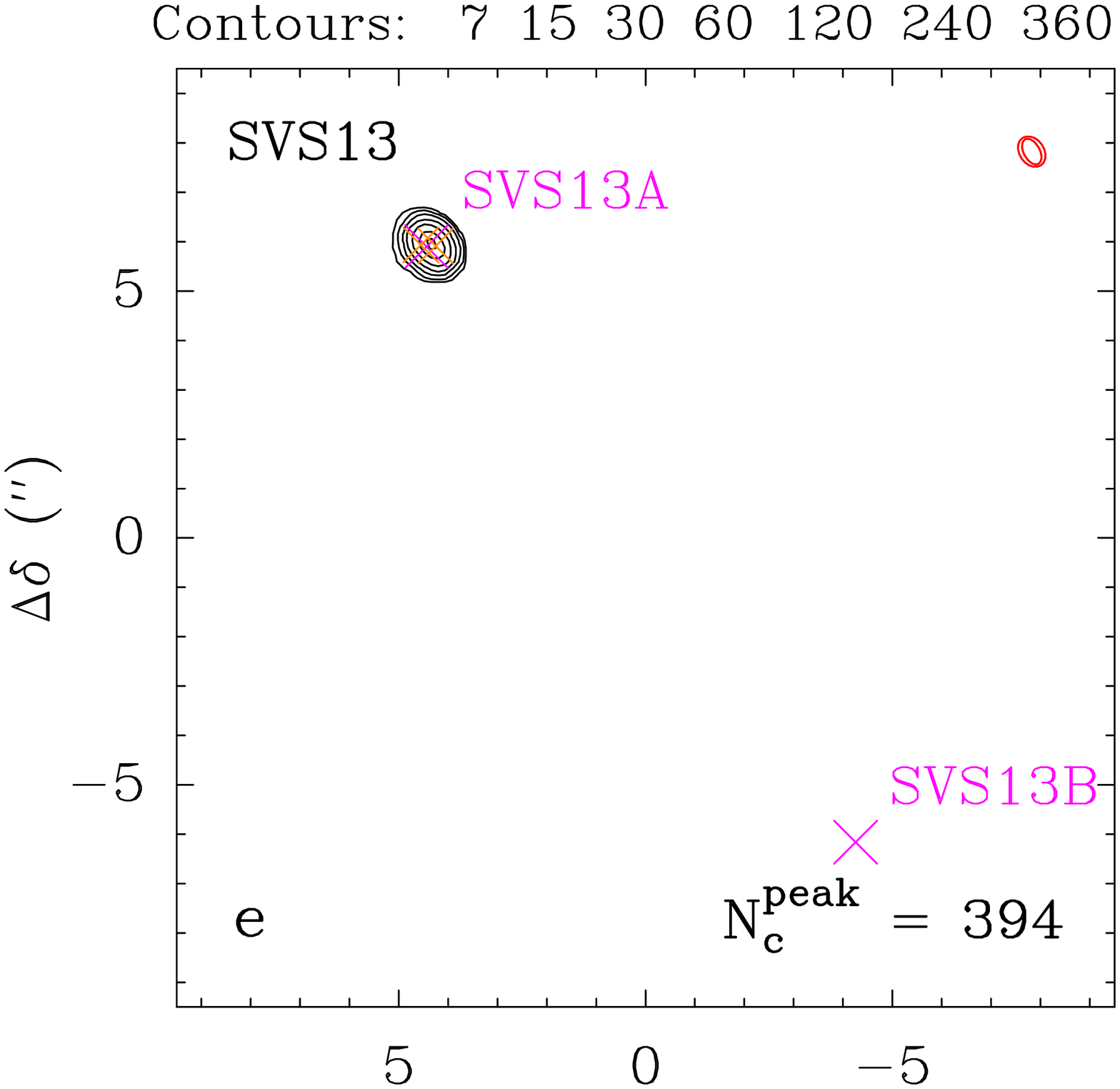}}\resizebox{0.24\hsize}{!}{\includegraphics[angle=0]{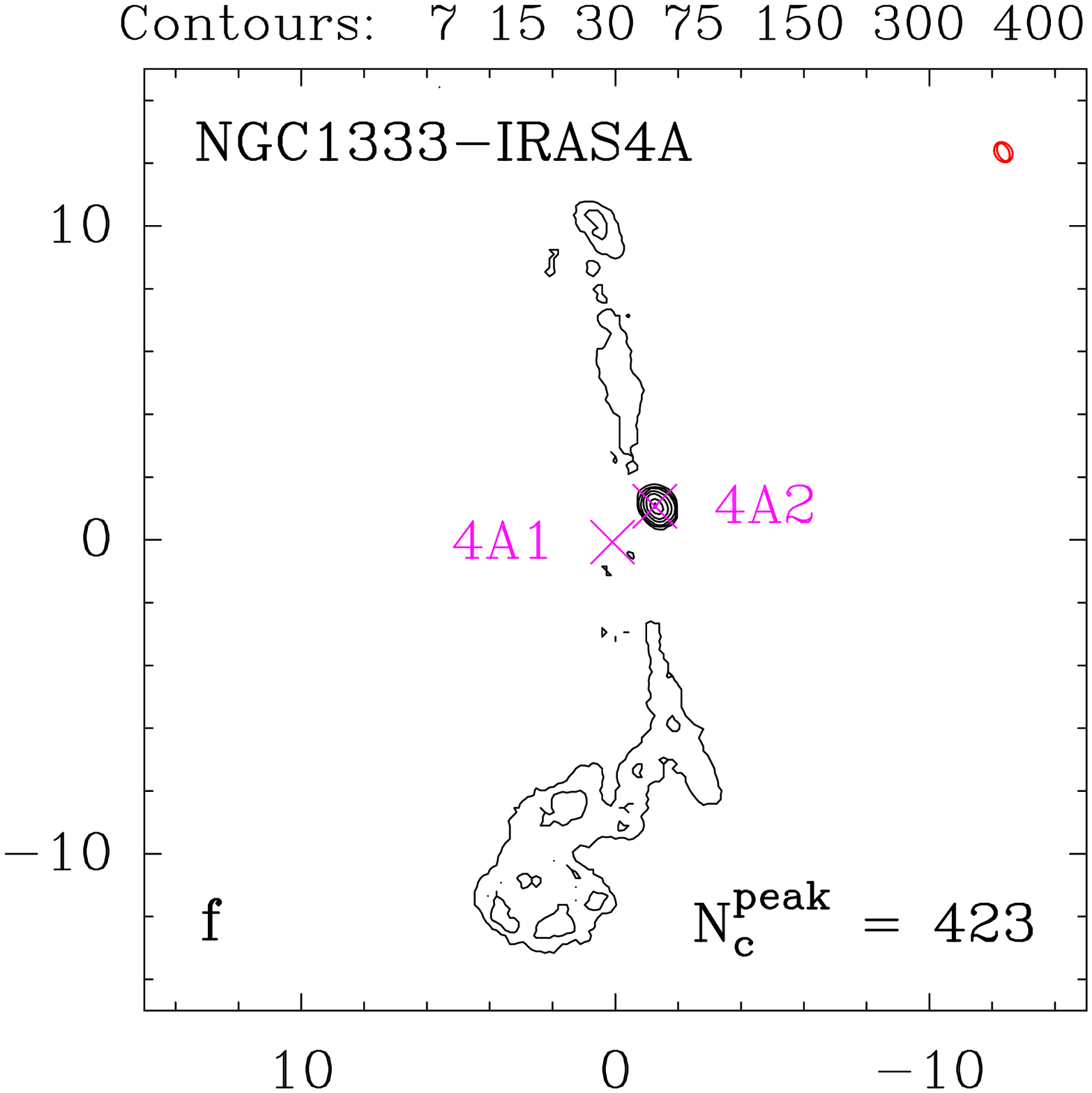}}\resizebox{0.24\hsize}{!}{\includegraphics[angle=0]{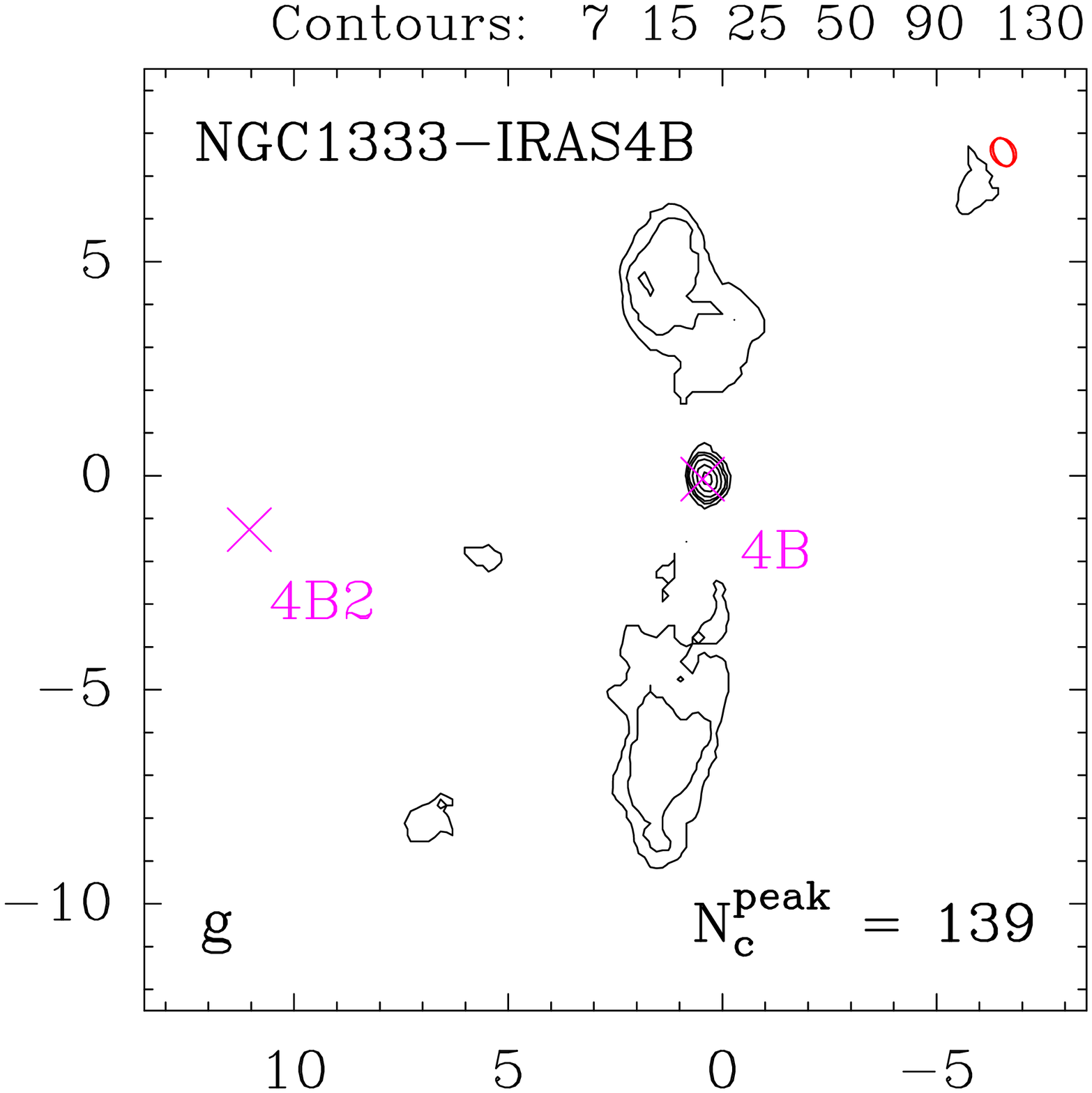}}\resizebox{0.24\hsize}{!}{\includegraphics[angle=0]{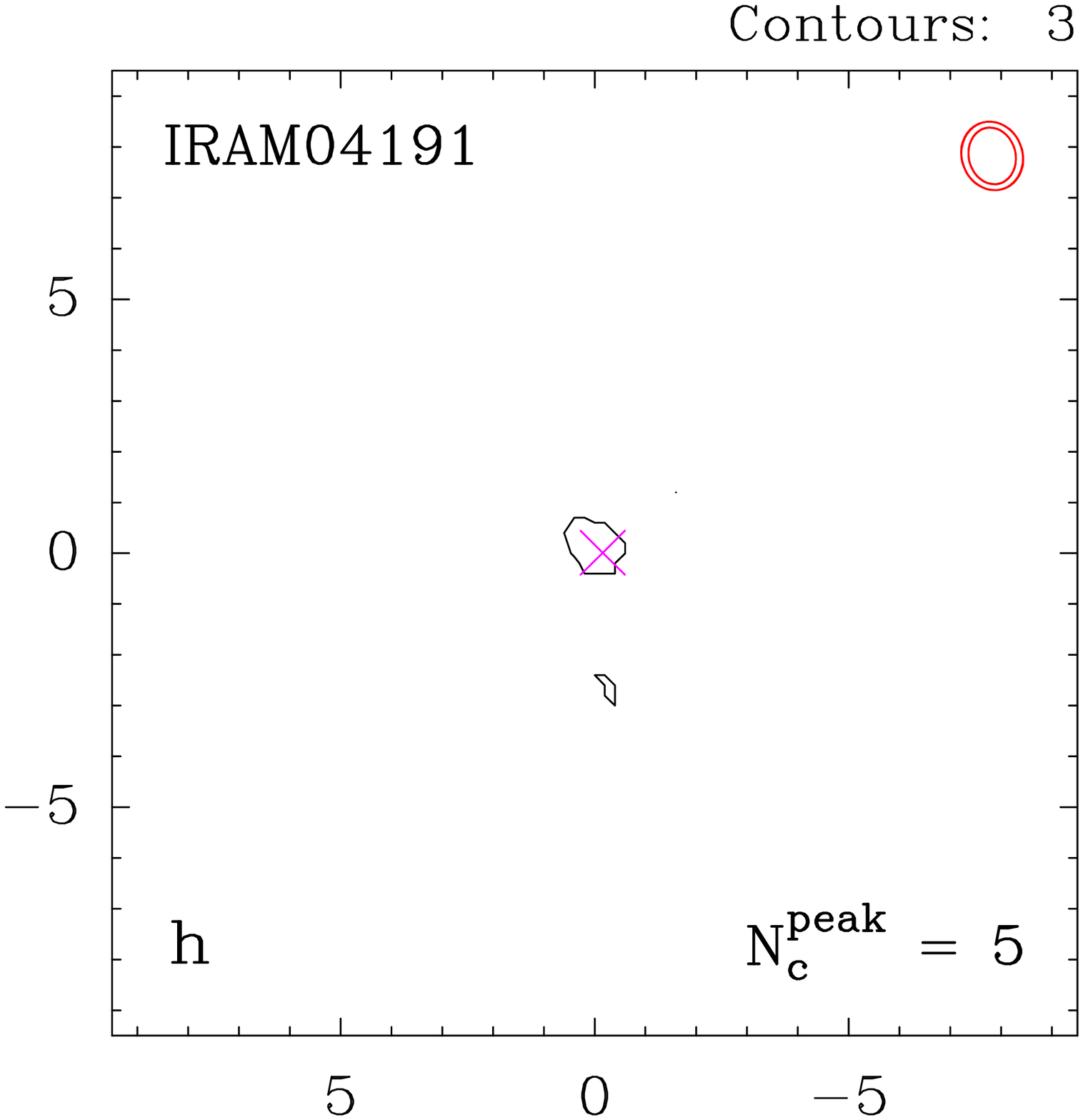}}}
 \centerline{\resizebox{0.24\hsize}{!}{\includegraphics[angle=0]{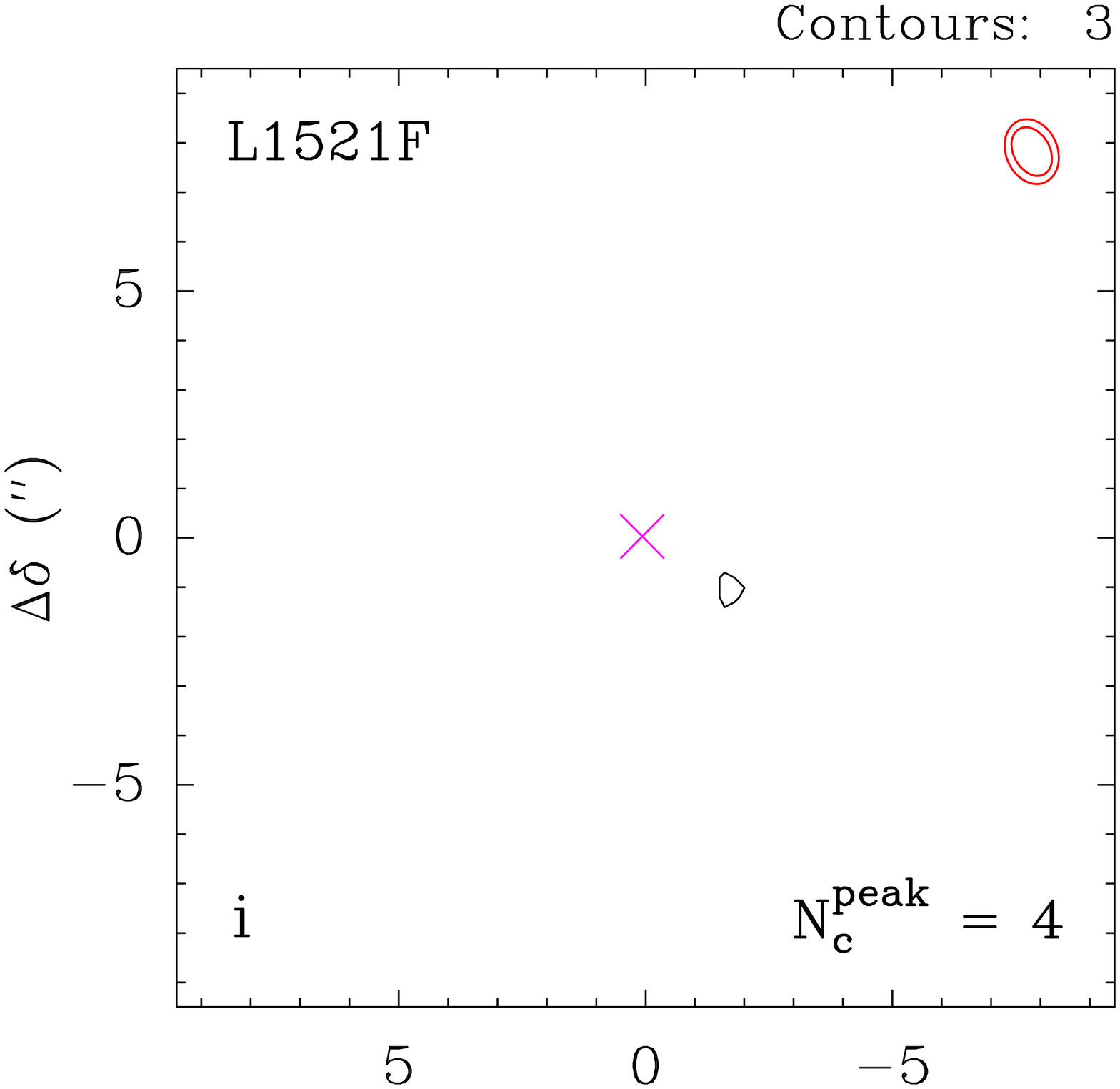}}\resizebox{0.24\hsize}{!}{\includegraphics[angle=0]{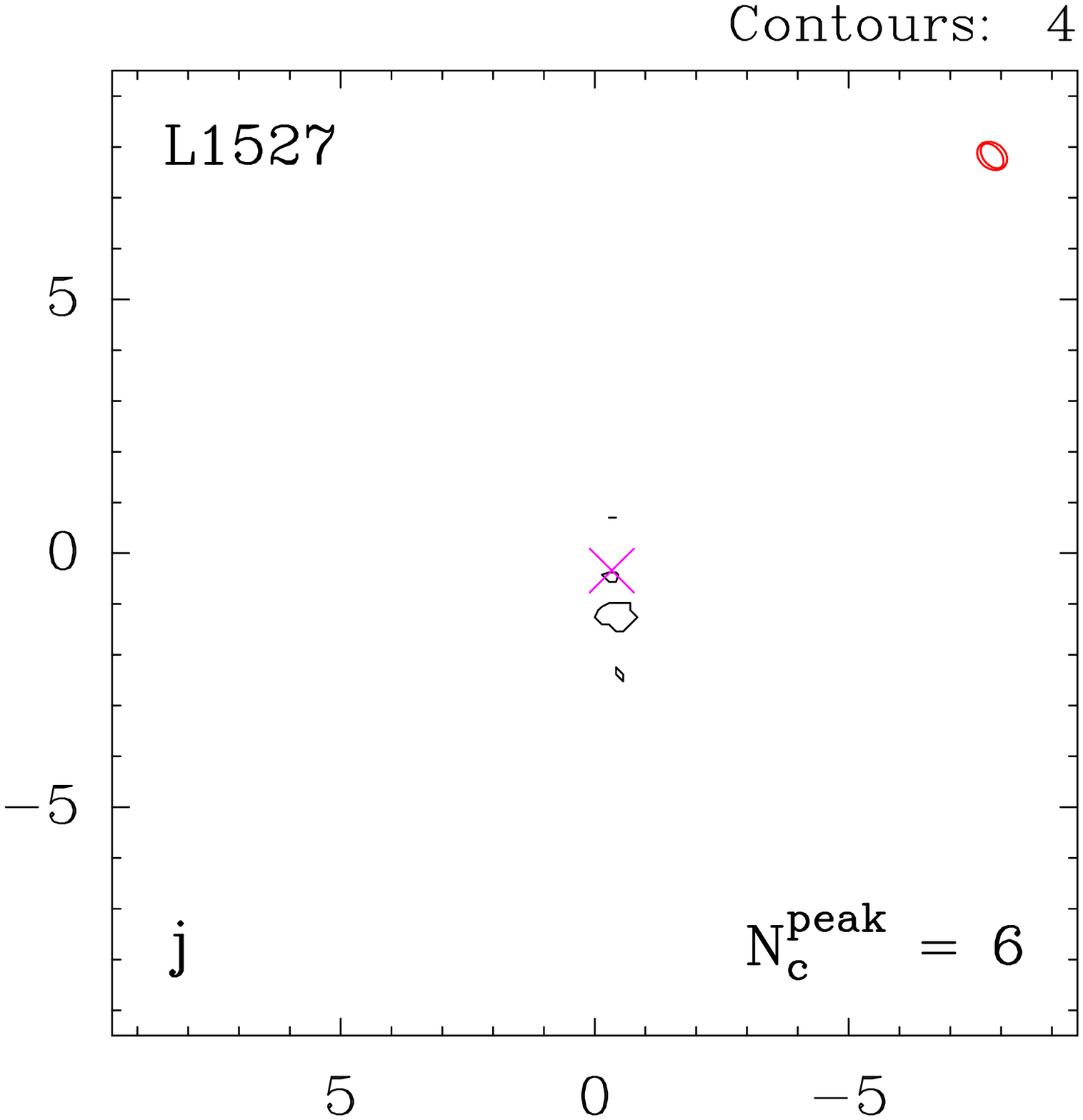}}\resizebox{0.24\hsize}{!}{\includegraphics[angle=0]{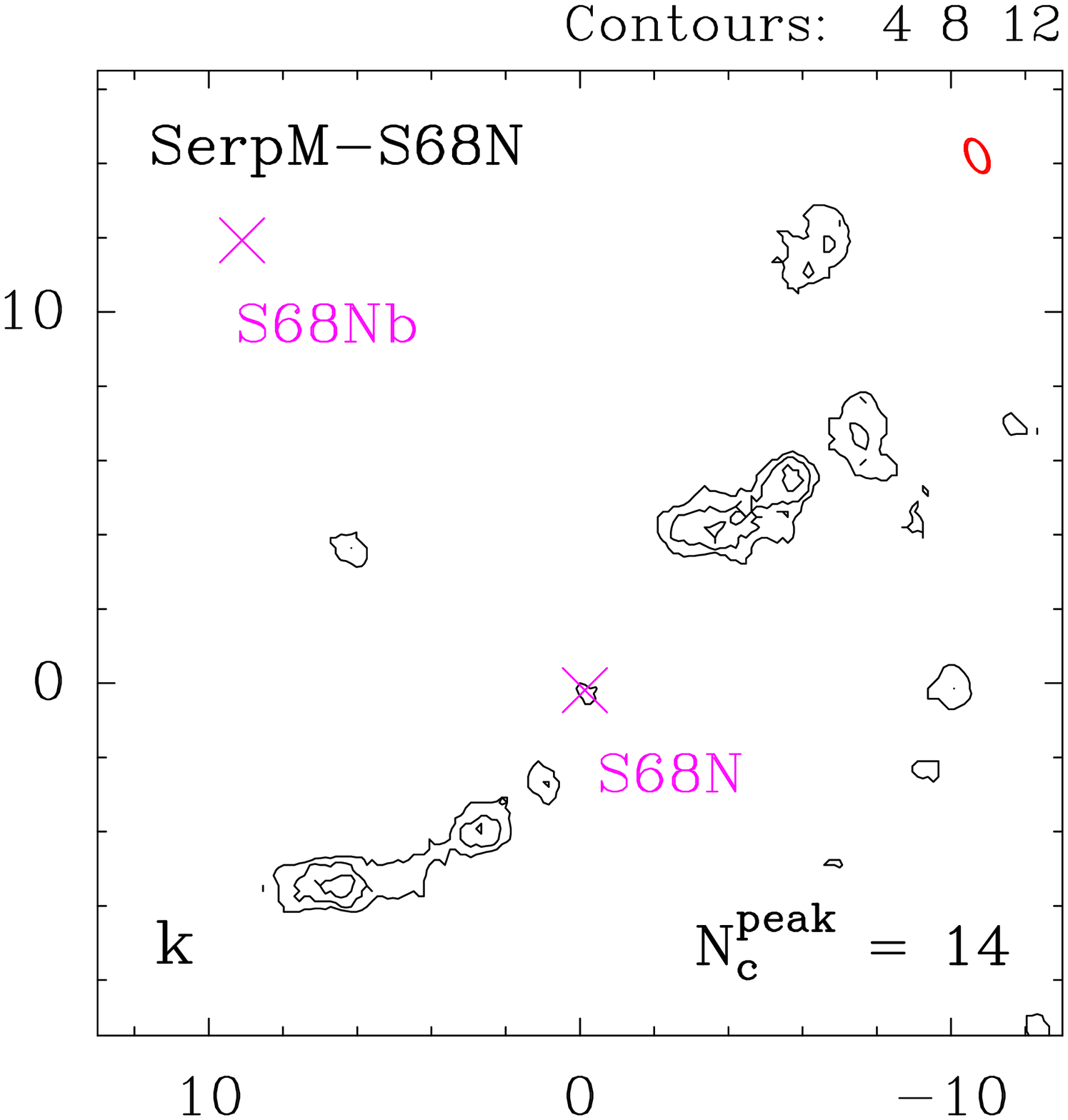}}\resizebox{0.24\hsize}{!}{\includegraphics[angle=0]{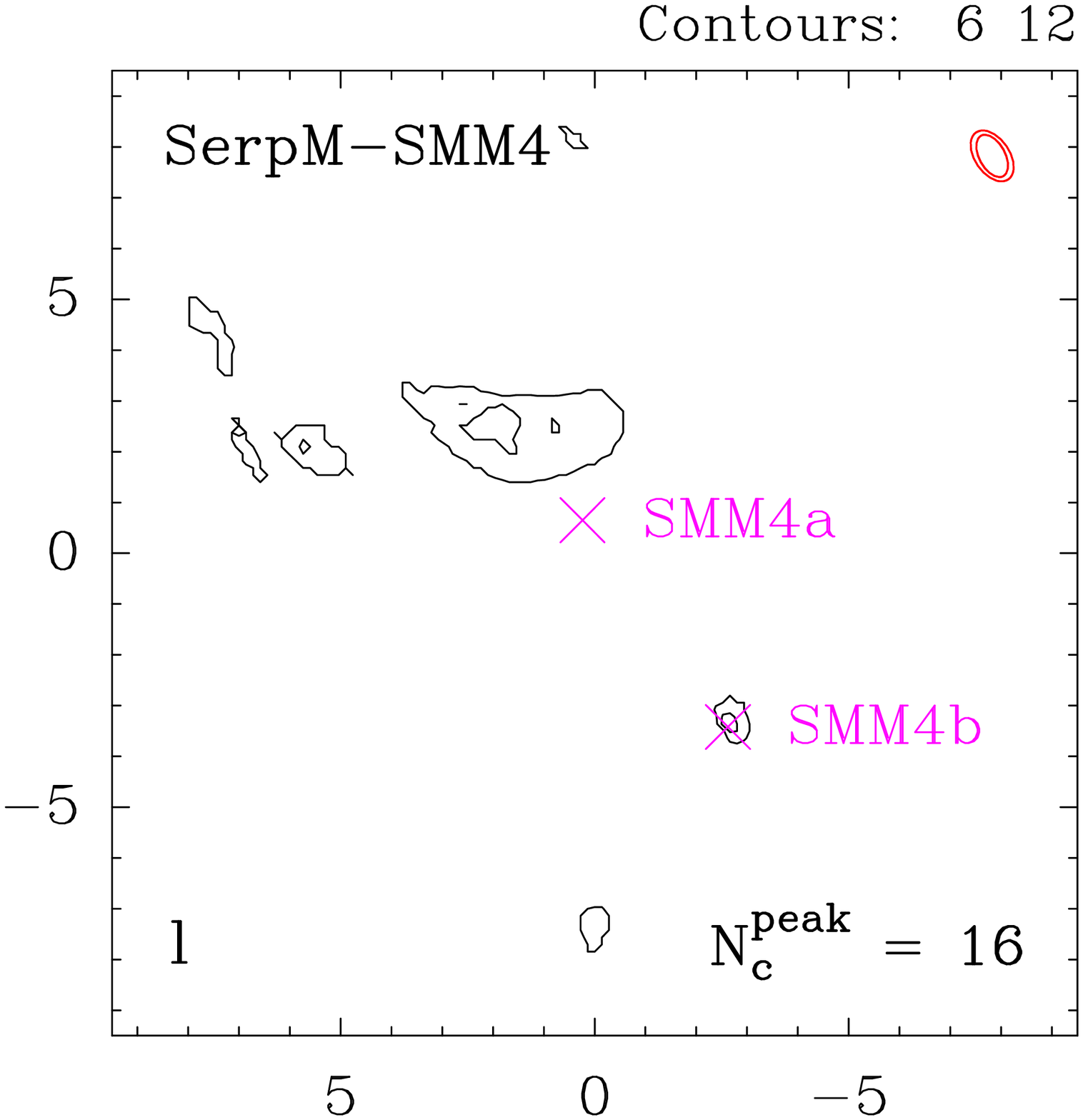}}}
 \centerline{\resizebox{0.24\hsize}{!}{\includegraphics[angle=0]{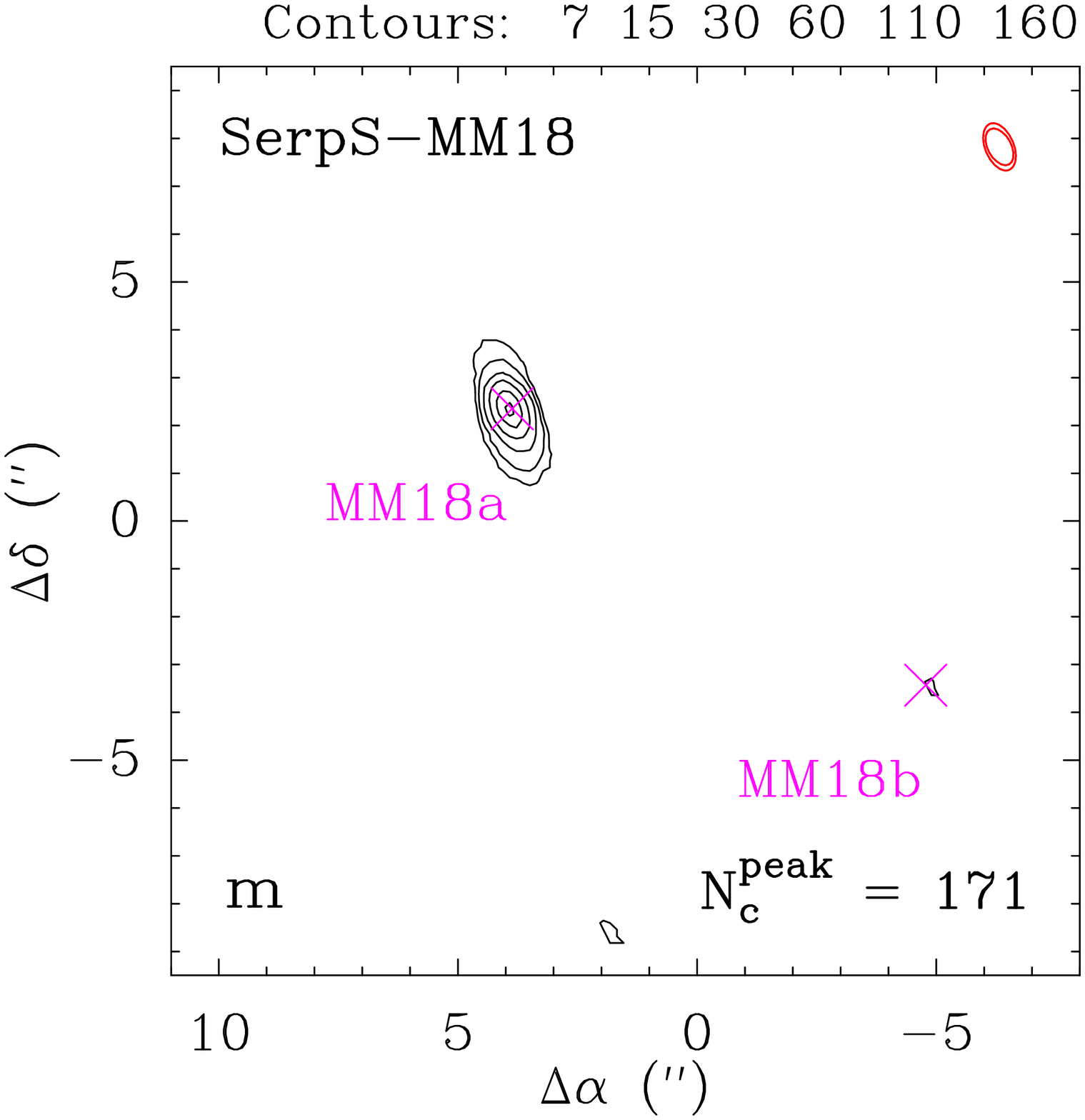}}\resizebox{0.24\hsize}{!}{\includegraphics[angle=0]{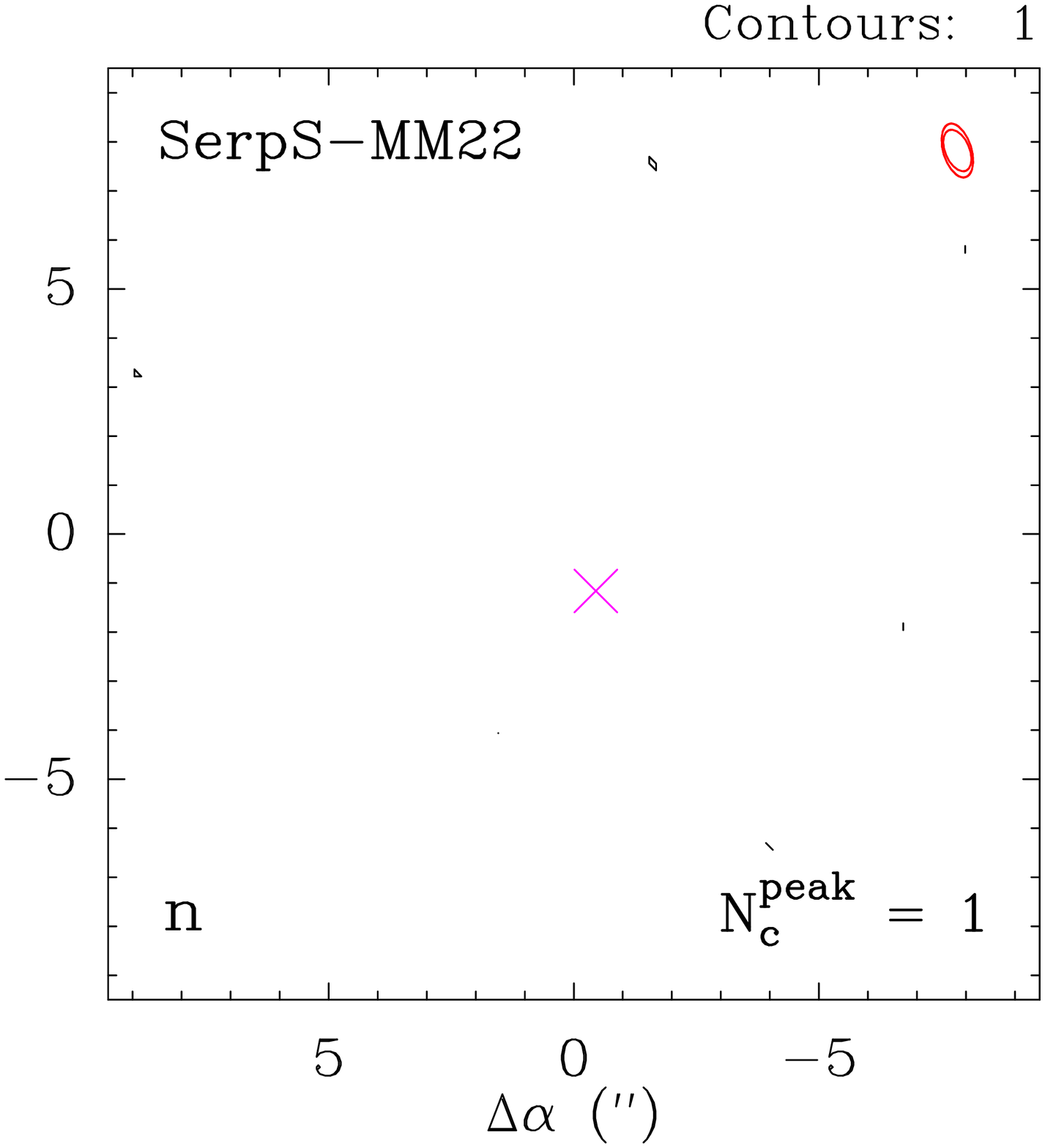}}\resizebox{0.24\hsize}{!}{\includegraphics[angle=0]{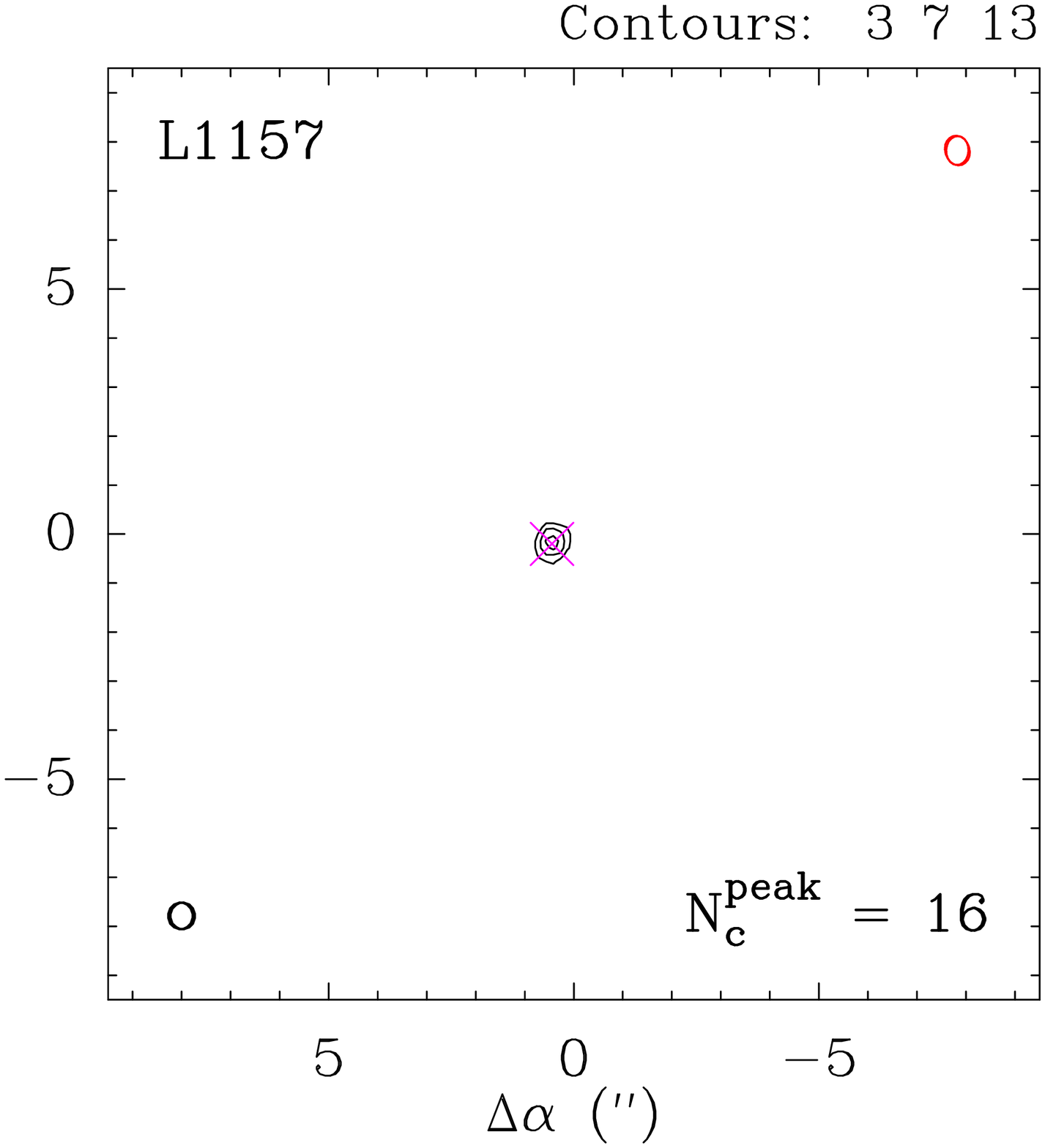}}\resizebox{0.24\hsize}{!}{\includegraphics[angle=0]{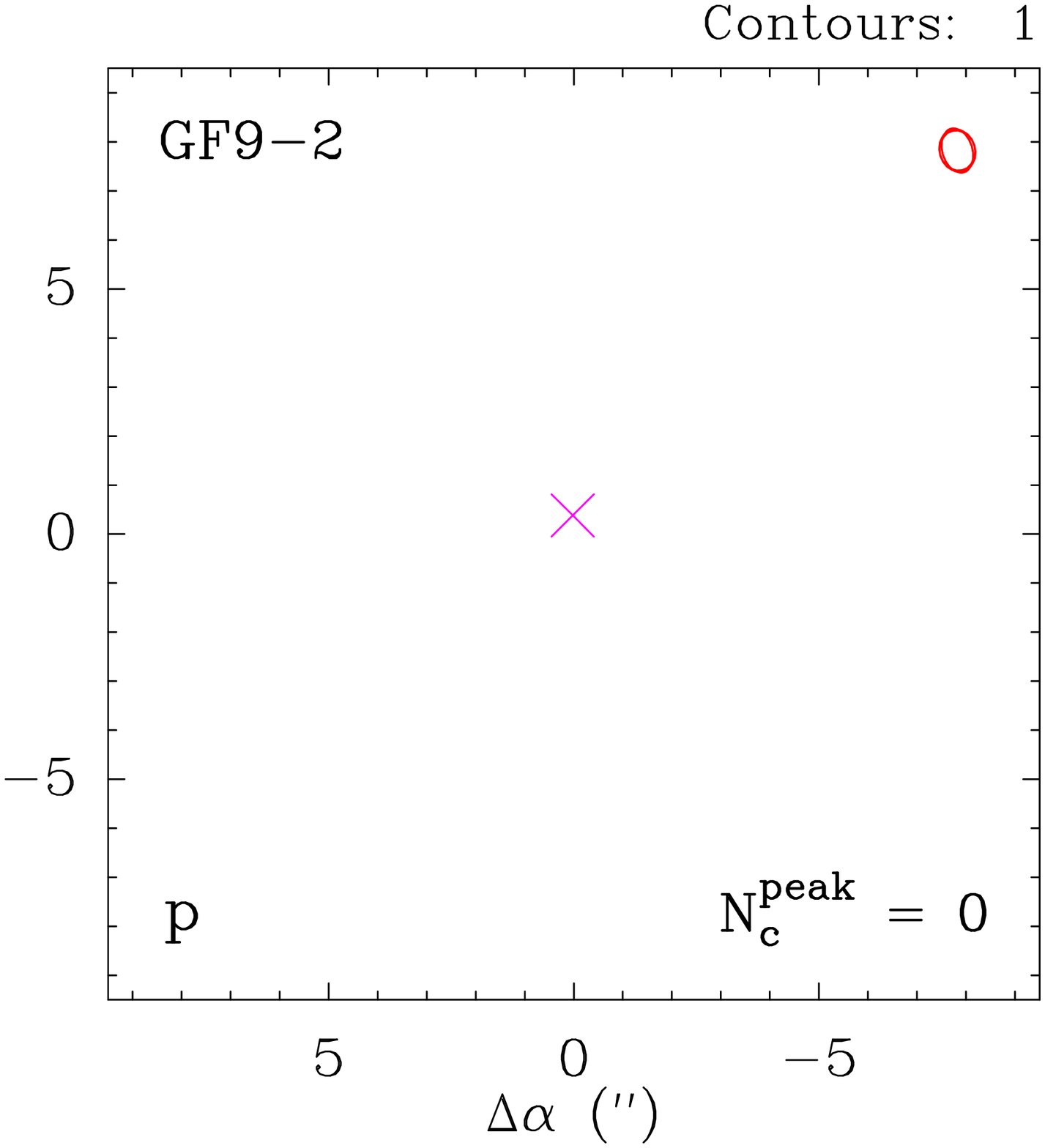}}}
 \caption{Same as Fig.~\ref{f:channelcounts} but for a larger field of
view.}
 \label{f:channelcounts_large}
\end{figure*}

Figure~\ref{f:channelcounts_large} shows the same maps as in 
Fig.~\ref{f:channelcounts} but over a larger field of view.
Table~\ref{t:maporigins} lists the equatorial coordinates of the origins of 
the maps shown in Figs.~\ref{f:channelcounts} and \ref{f:channelcounts_large},
as well as the systemic velocity assumed for the sources located in each map.

\begin{table}[!ht]
 \begin{center}
 \caption{
 Coordinates of map centers and assumed systemic velocities.
}
 \label{t:maporigins}
 \vspace*{-1.2ex}
 \begin{tabular}{lccc}
 \hline\hline
 \multicolumn{1}{c}{Field} & \multicolumn{2}{c}{Coordinates (J2000)} & \multicolumn{1}{c}{$V_{\rm lsr}$} \\ 
  & \multicolumn{1}{c}{$\alpha$ (hh:mm:ss)} & \multicolumn{1}{c}{$\delta$ (dd:mm:ss)} & \multicolumn{1}{c}{km~s$^{-1}$} \\ 
 \hline
 L1448-2A & 03:25:22.406 & 30:45:13.28 & $4.2$ \\ 
 L1448-N & 03:25:36.340 & 30:45:14.90 & $4.7$ \\ 
 L1448-C & 03:25:38.870 & 30:44:05.40 & $5.7$ \\ 
 IRAS2A & 03:28:55.580 & 31:14:37.10 & $7.3$ \\ 
 SVS13B & 03:29:03.410 & 31:15:57.90 & $8.4$ \\ 
 IRAS4A & 03:29:10.530 & 31:13:31.05 & $7.0$ \\ 
 IRAS4B & 03:29:11.980 & 31:13:08.10 & $7.0$ \\ 
 IRAM04191 & 04:21:56.910 & 15:29:46.10 & $6.7$ \\ 
 L1521F & 04:28:38.936 & 26:51:35.11 & $6.5$ \\ 
 L1527 & 04:39:53.900 & 26:03:10.00 & $5.9$ \\ 
 SerpM-S68N & 18:29:48.100 & 01:16:43.60 & $8.8$ \\ 
 SerpM-SMM4 & 18:29:56.700 & 01:13:15.00 & $8.1$ \\ 
 SerpS-MM18 & 18:30:03.860 & --02:03:04.90 & $7.4$ \\ 
 SerpS-MM22 & 18:30:12.340 & --02:06:52.40 & $7.4$ \\ 
 L1157 & 20:39:06.190 & 68:02:15.90 & $2.6$ \\ 
 GF9-2 & 20:51:29.820 & 60:18:38.06 & $-2.5$ \\ 
 \hline
 \end{tabular}
 \end{center}
 \vspace*{-2.5ex}
 \tablefoot{The coordinates correspond to the origins of the maps shown in Figs.~\ref{f:channelcounts} and \ref{f:channelcounts_large}. The last column gives the systemic velocity assumed for the sources located in each map.
 }
 \end{table}

\clearpage
\newpage

\section{Source sizes}
\label{a:sizes}

\begin{figure*}
\centerline{\resizebox{0.95\hsize}{!}{\includegraphics[angle=270]{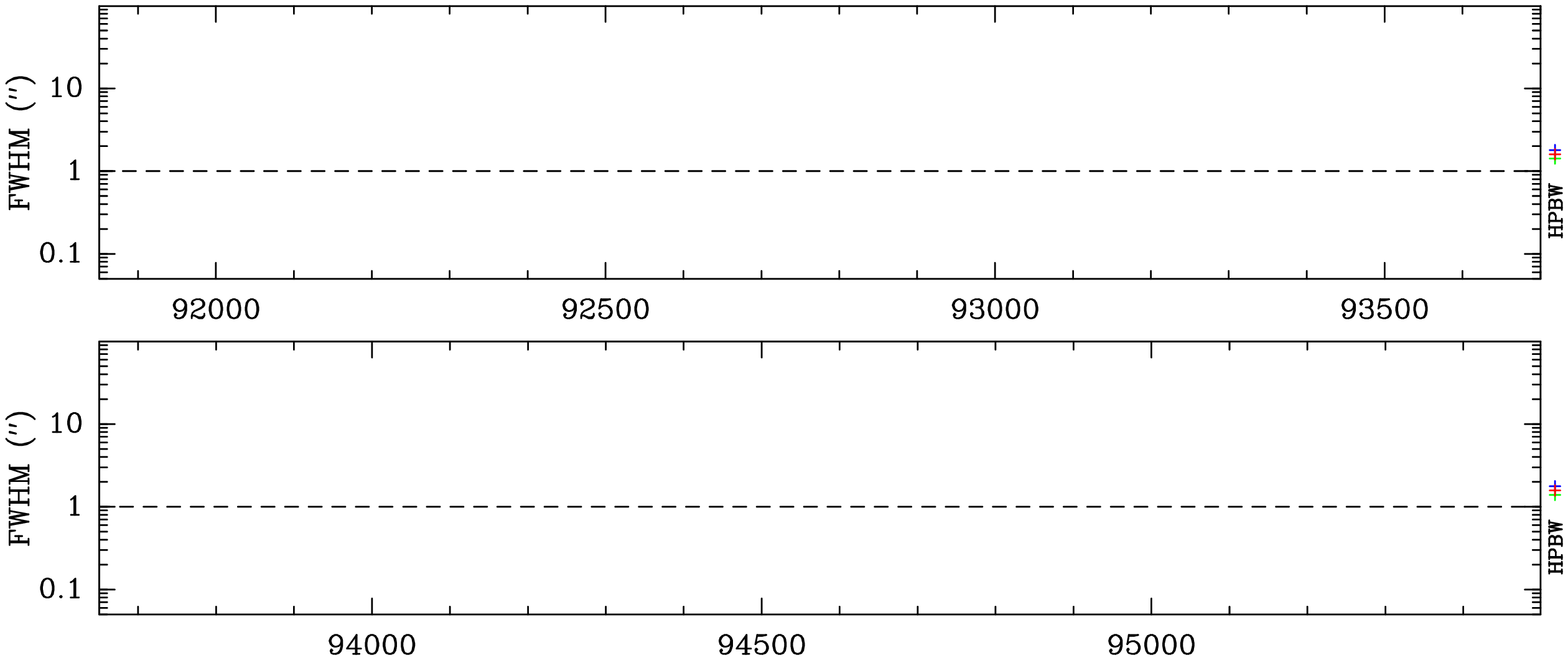}}}
\vspace*{1.5ex}
\centerline{\resizebox{0.95\hsize}{!}{\includegraphics[angle=270]{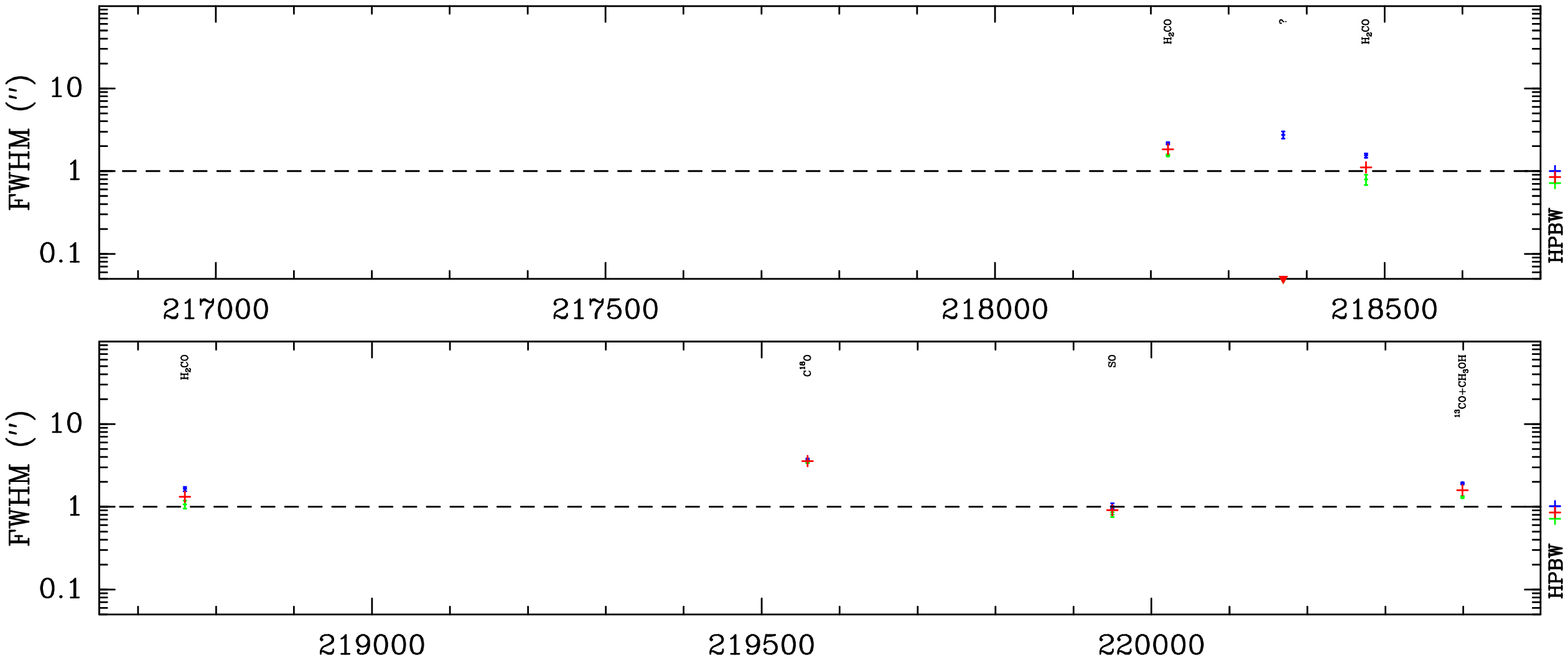}}}
\vspace*{1.5ex}
\centerline{\resizebox{0.95\hsize}{!}{\includegraphics[angle=270]{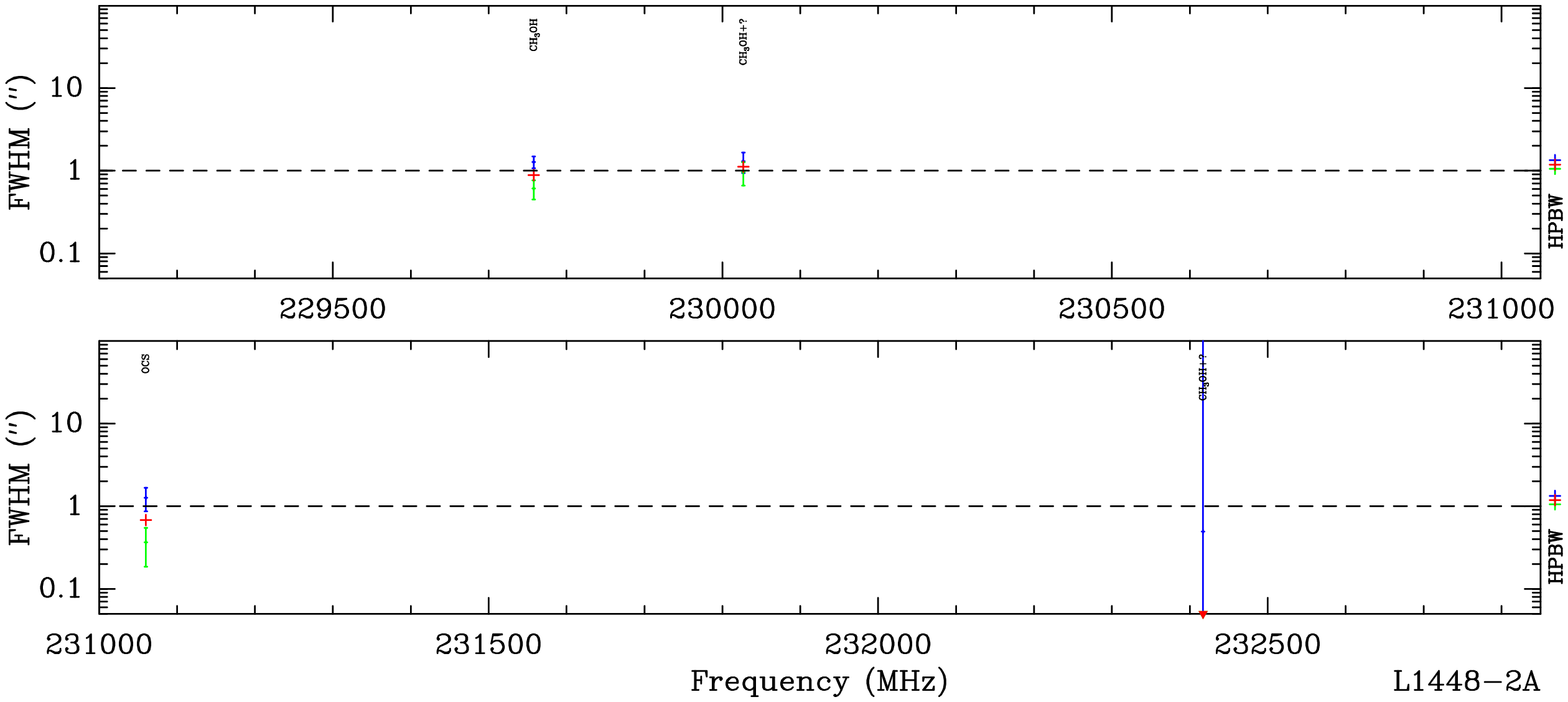}}}
\caption{Deconvolved size of spectral line emission as a function of frequency 
for L1448-2A. For each spectral line, the blue and green symbols represent 
the major and minor axes with their associated uncertainties, respectively, and 
the red cross is their geometrical mean. The molecule(s) assigned to each 
detected line is (are) indicated at the top of each panel. A question mark 
means either that the line is unidentified or that less than $\sim$70\%
of the measured peak flux density is 
accounted for by our model. The synthesized beam size ($HPBW$) is 
shown to the right of each panel. The dashed line shows the size assumed for 
the spectral line modeling in Sect.~\ref{ss:modeling}, and is reported in
Table~\ref{t:sizes}.}
\label{f:sizes_l1448-2a}
\end{figure*}

\begin{figure*}
\centerline{\resizebox{0.95\hsize}{!}{\includegraphics[angle=270]{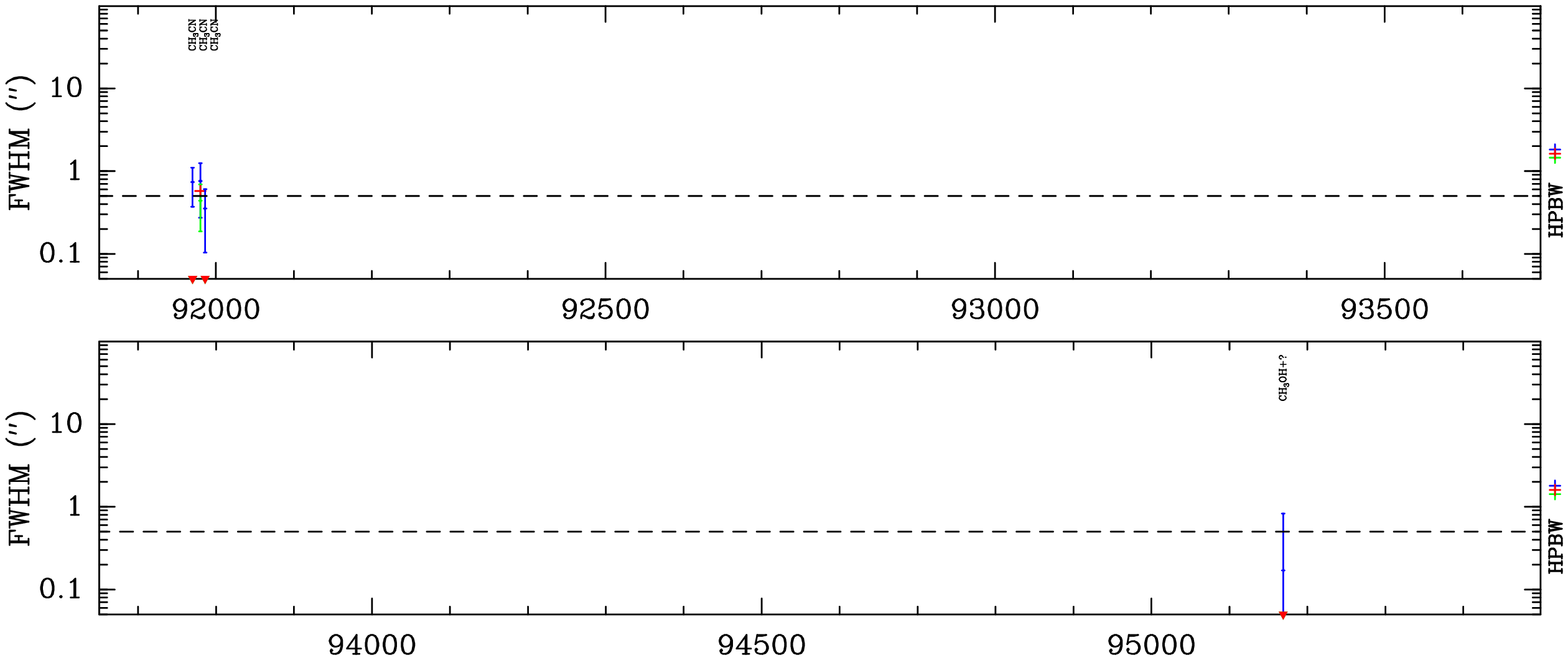}}}
\vspace*{1.5ex}
\centerline{\resizebox{0.95\hsize}{!}{\includegraphics[angle=270]{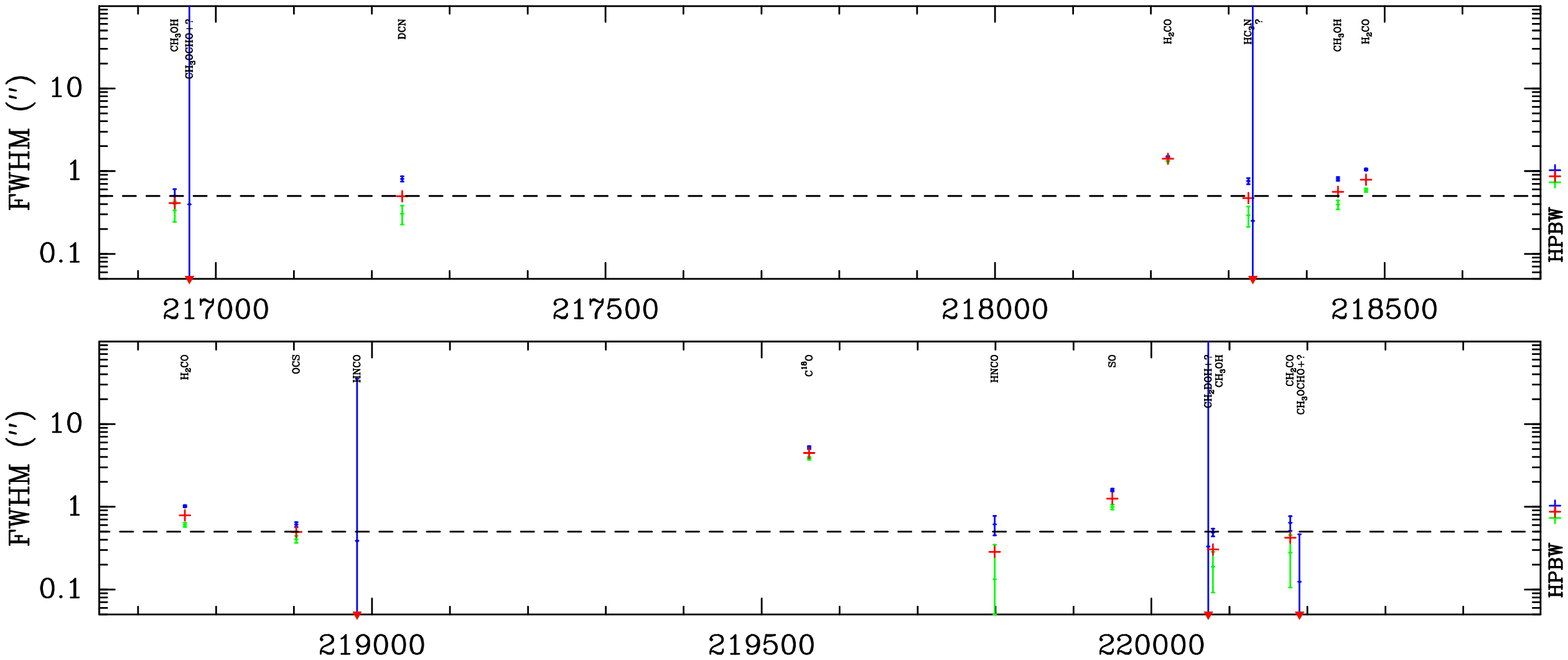}}}
\vspace*{1.5ex}
\centerline{\resizebox{0.95\hsize}{!}{\includegraphics[angle=270]{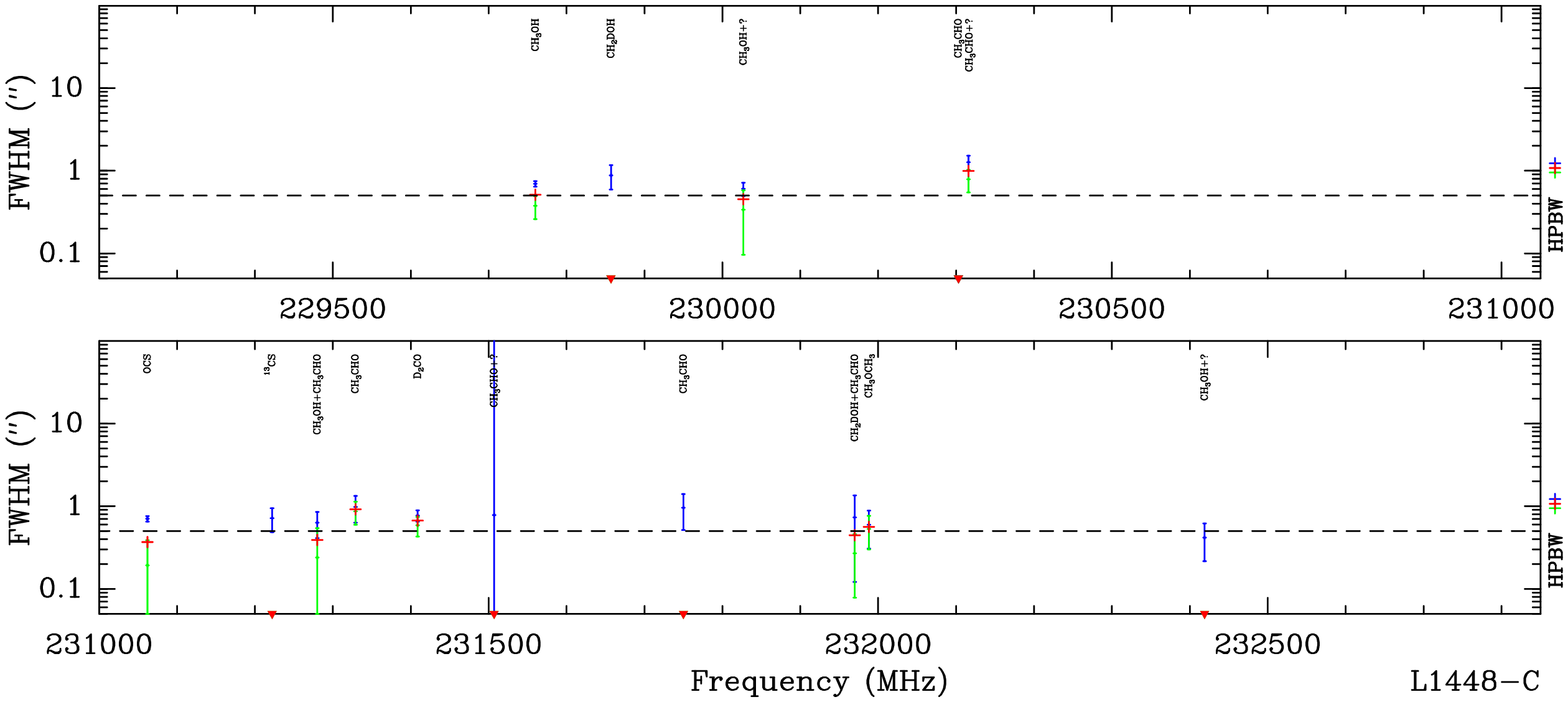}}}
\caption{Same as Fig.~\ref{f:sizes_l1448-2a} for L1448-C.}
\label{f:sizes_l1448-c}
\end{figure*}

\begin{figure*}
\centerline{\resizebox{0.95\hsize}{!}{\includegraphics[angle=270]{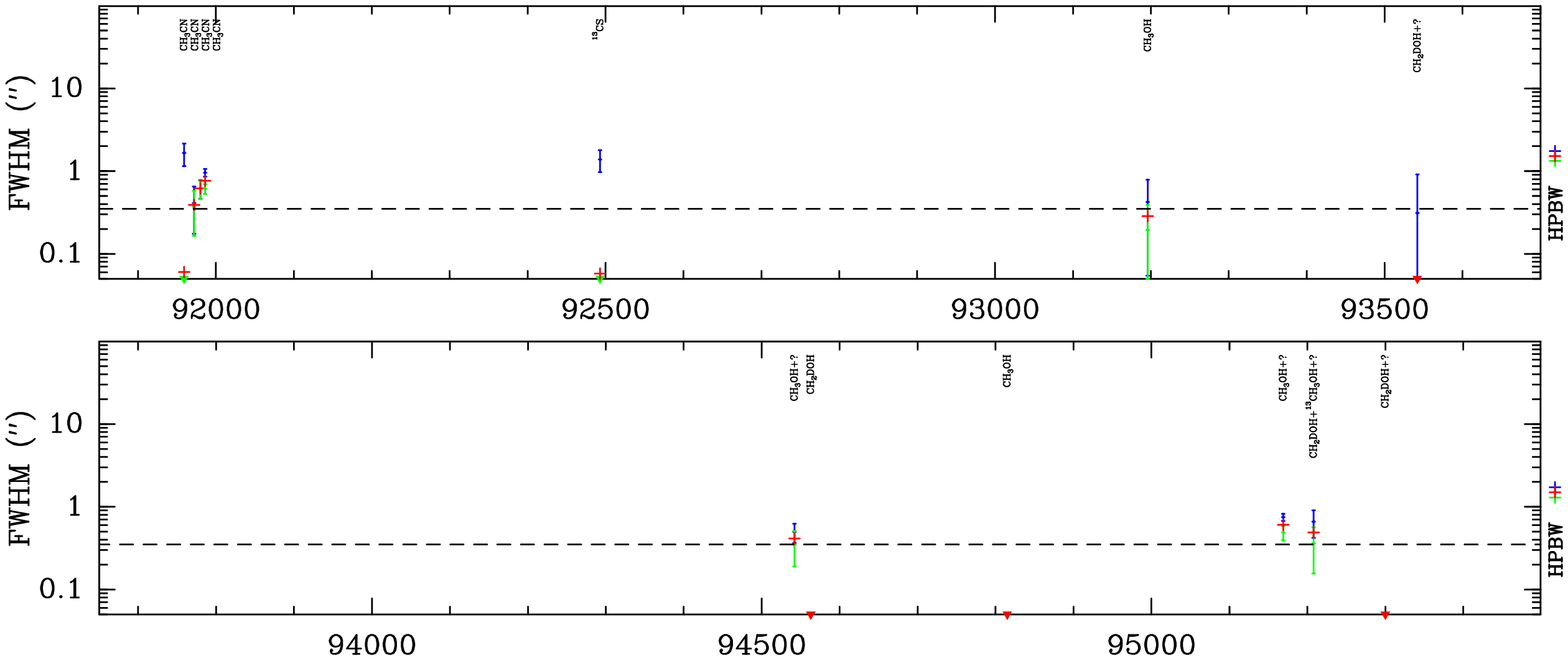}}}
\vspace*{1.5ex}
\centerline{\resizebox{0.95\hsize}{!}{\includegraphics[angle=270]{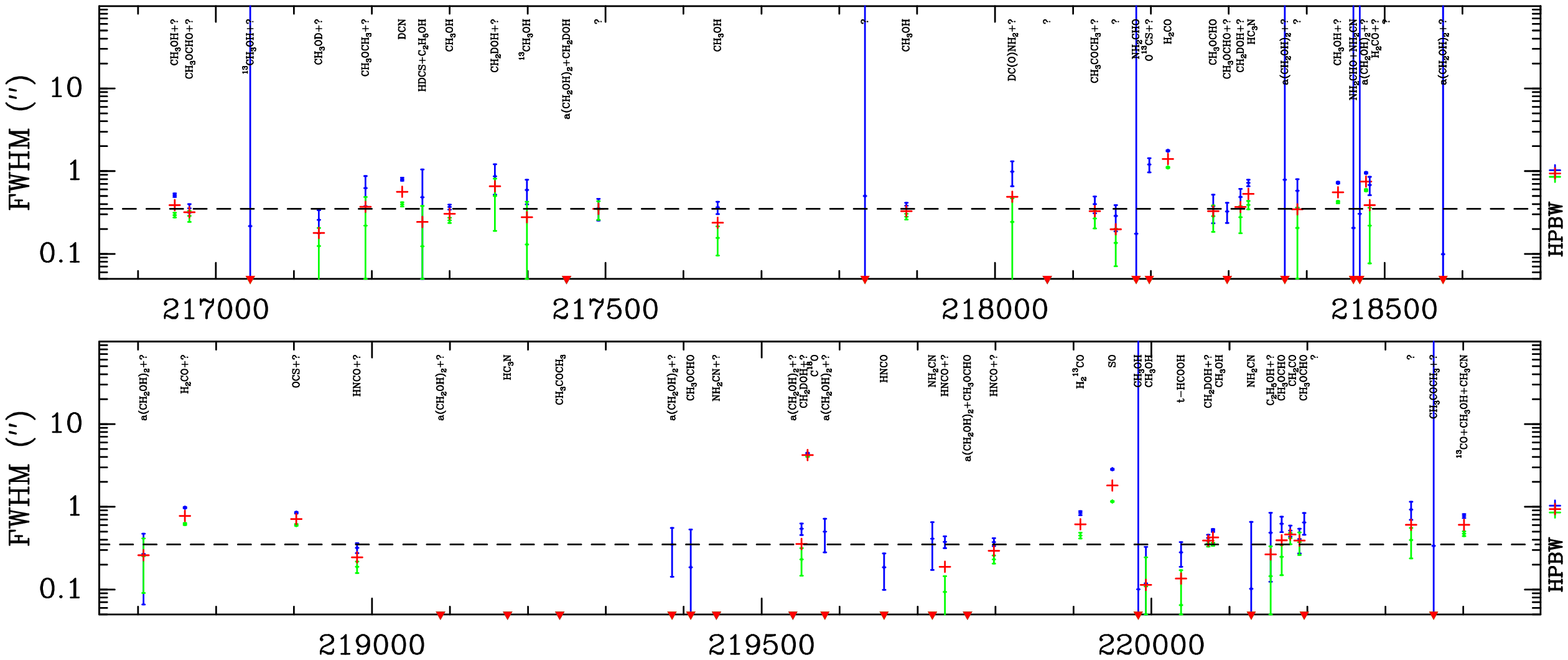}}}
\vspace*{1.5ex}
\centerline{\resizebox{0.95\hsize}{!}{\includegraphics[angle=270]{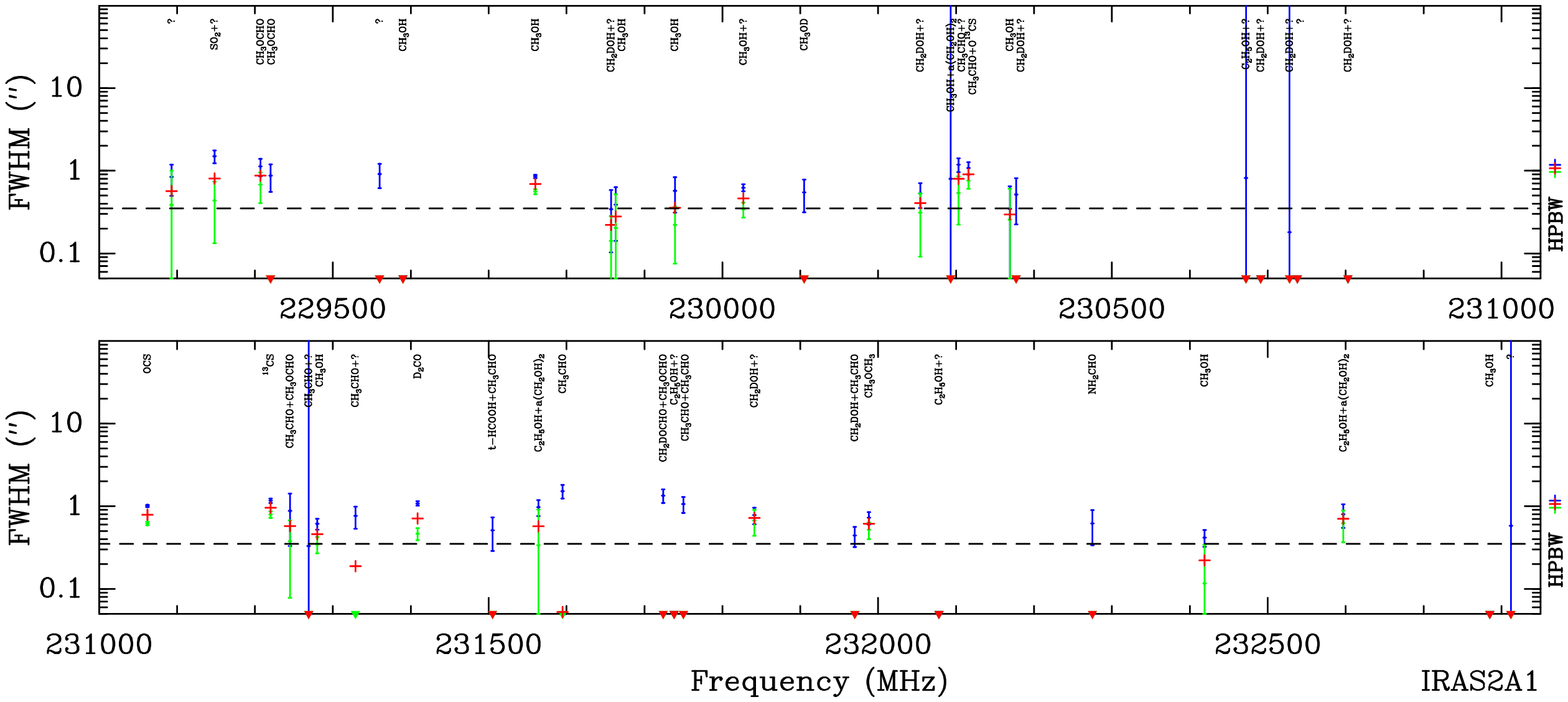}}}
\caption{Same as Fig.~\ref{f:sizes_l1448-2a} for IRAS2A1.}
\label{f:sizes_iras2a1}
\end{figure*}

\begin{figure*}
\centerline{\resizebox{0.95\hsize}{!}{\includegraphics[angle=270]{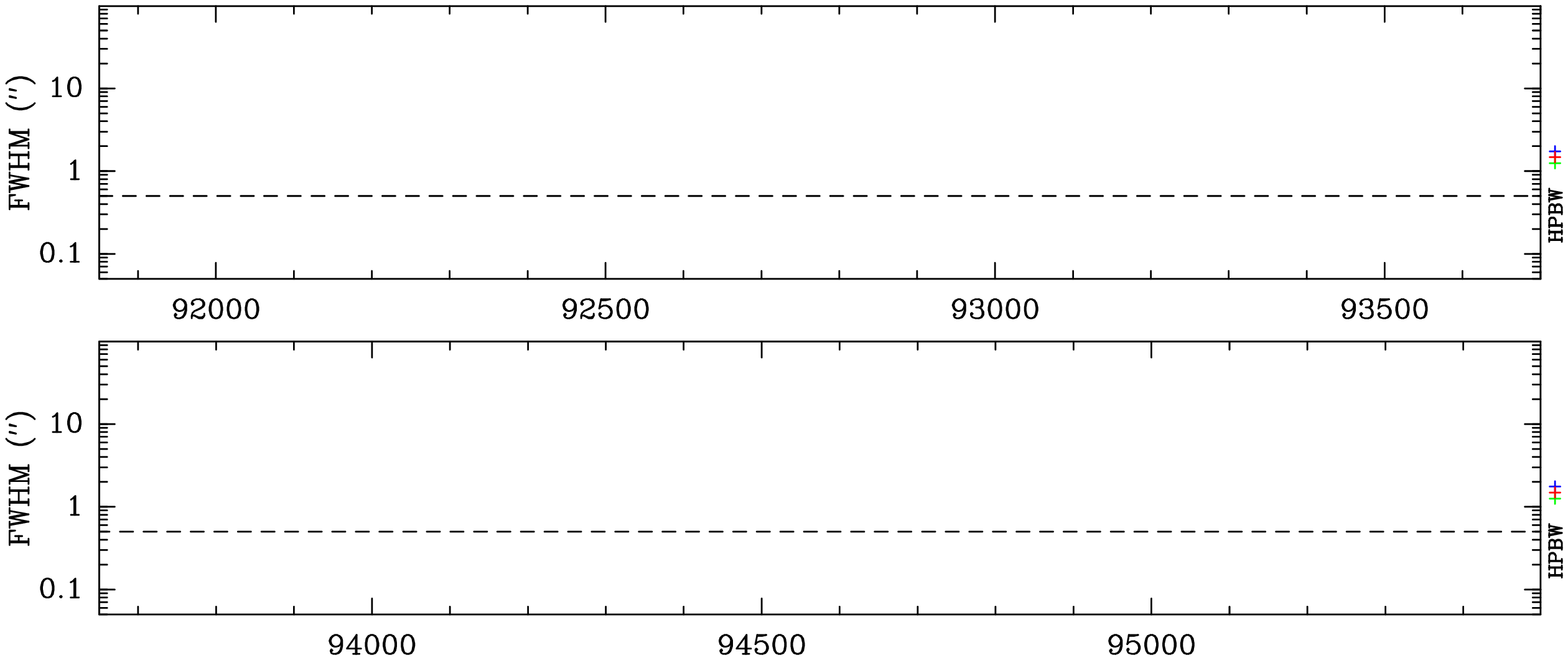}}}
\vspace*{1.5ex}
\centerline{\resizebox{0.95\hsize}{!}{\includegraphics[angle=270]{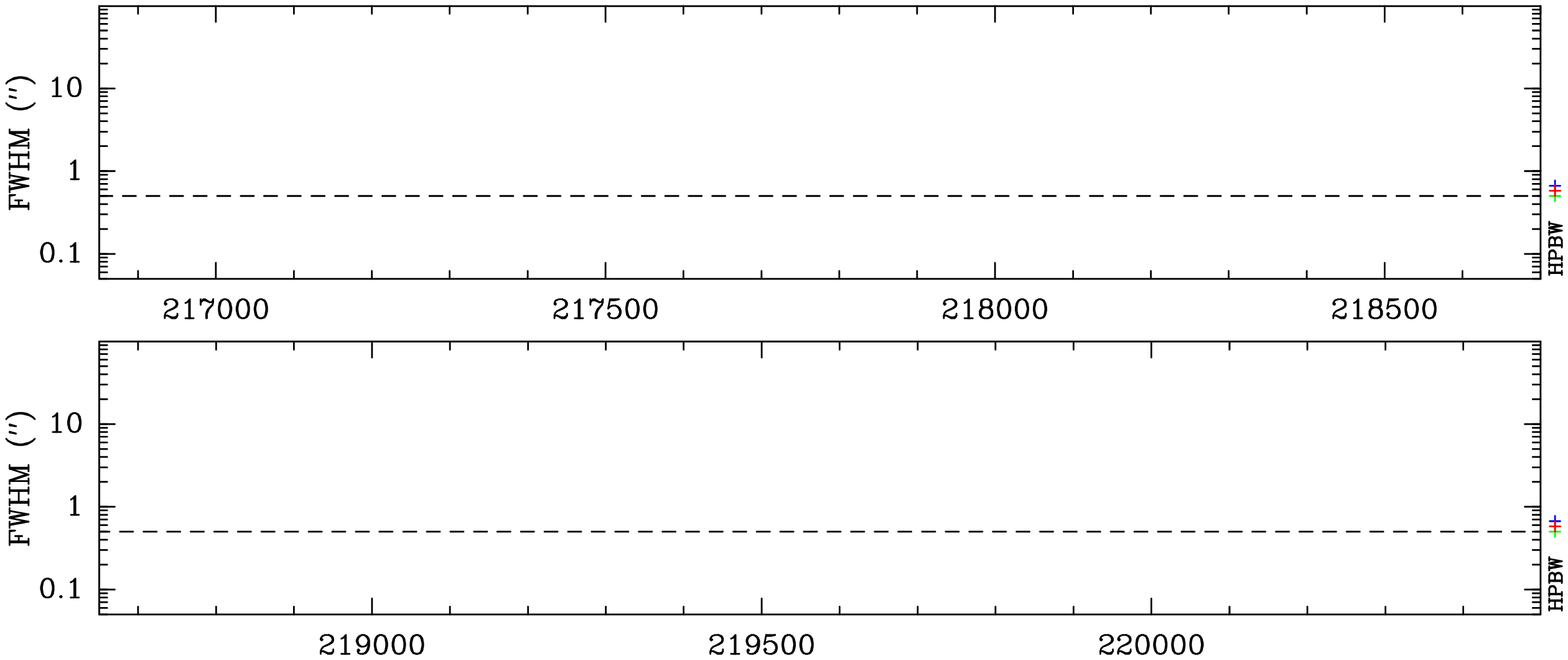}}}
\vspace*{1.5ex}
\centerline{\resizebox{0.95\hsize}{!}{\includegraphics[angle=270]{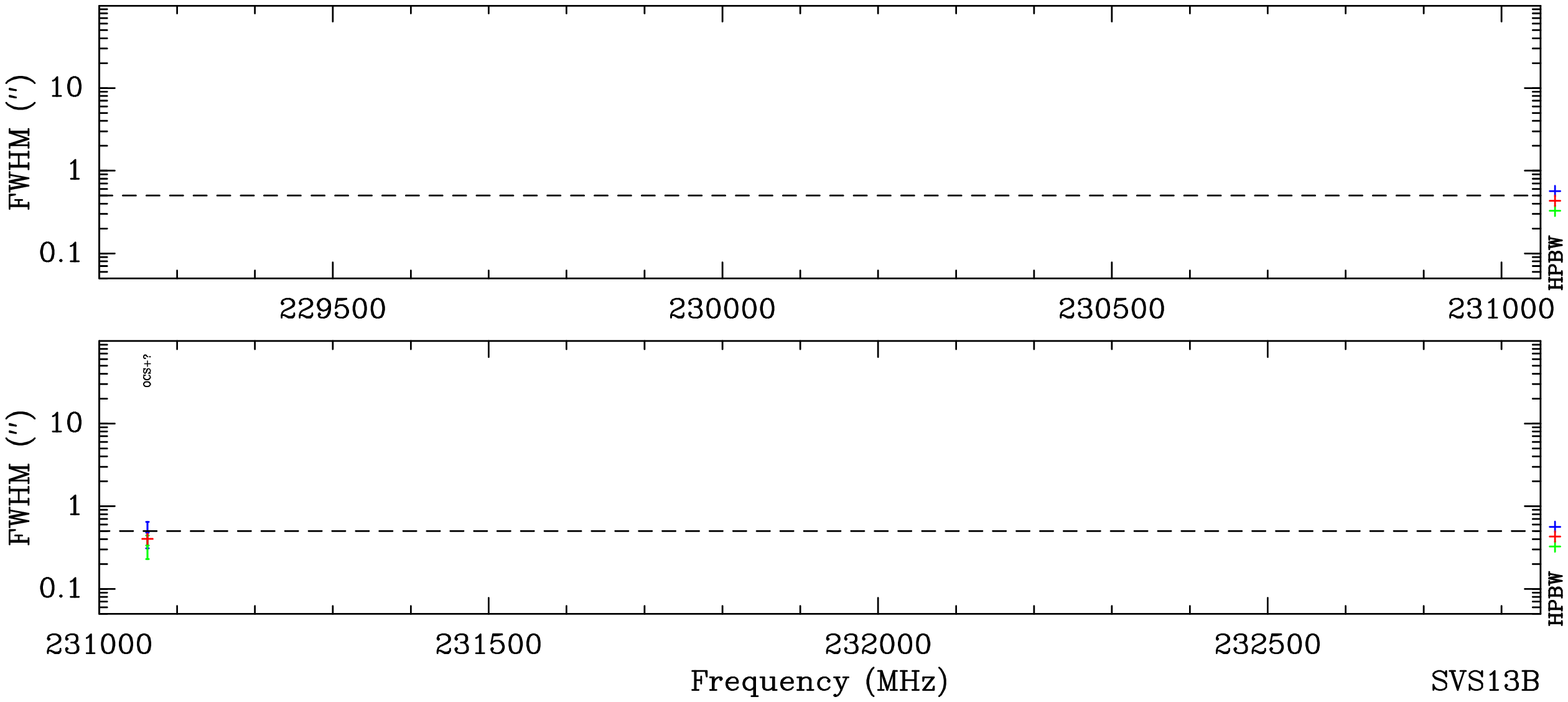}}}
\caption{Same as Fig.~\ref{f:sizes_l1448-2a} for SVS13B.}
\label{f:sizes_svs13-b}
\end{figure*}

\begin{figure*}
\centerline{\resizebox{0.95\hsize}{!}{\includegraphics[angle=270]{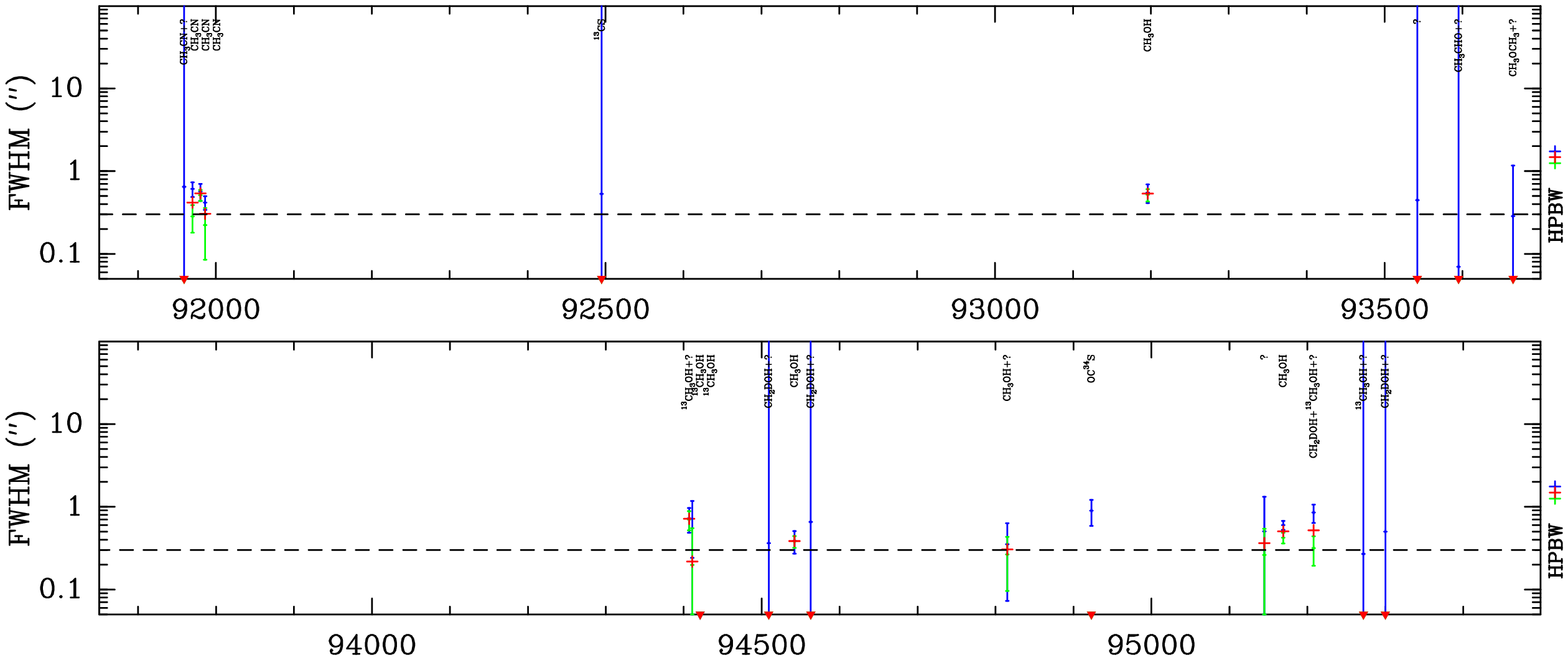}}}
\vspace*{1.5ex}
\centerline{\resizebox{0.95\hsize}{!}{\includegraphics[angle=270]{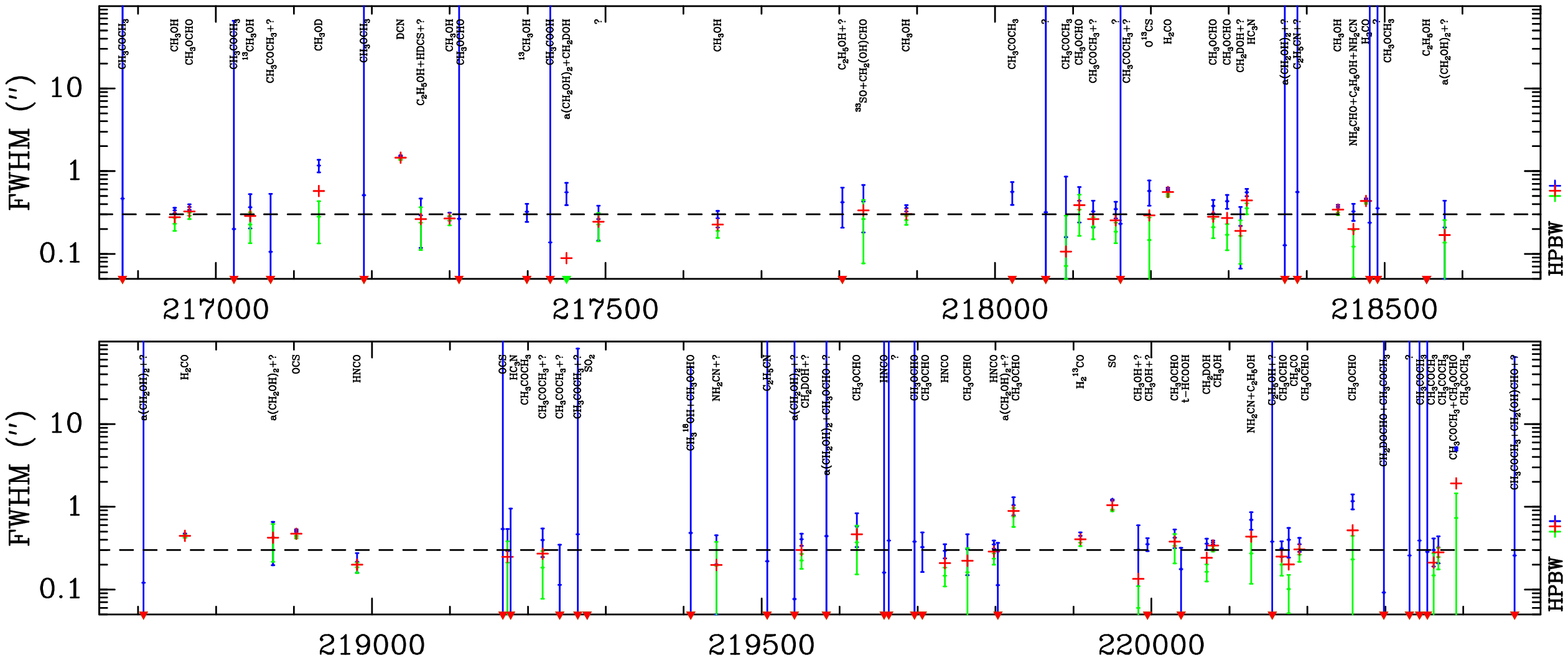}}}
\vspace*{1.5ex}
\centerline{\resizebox{0.95\hsize}{!}{\includegraphics[angle=270]{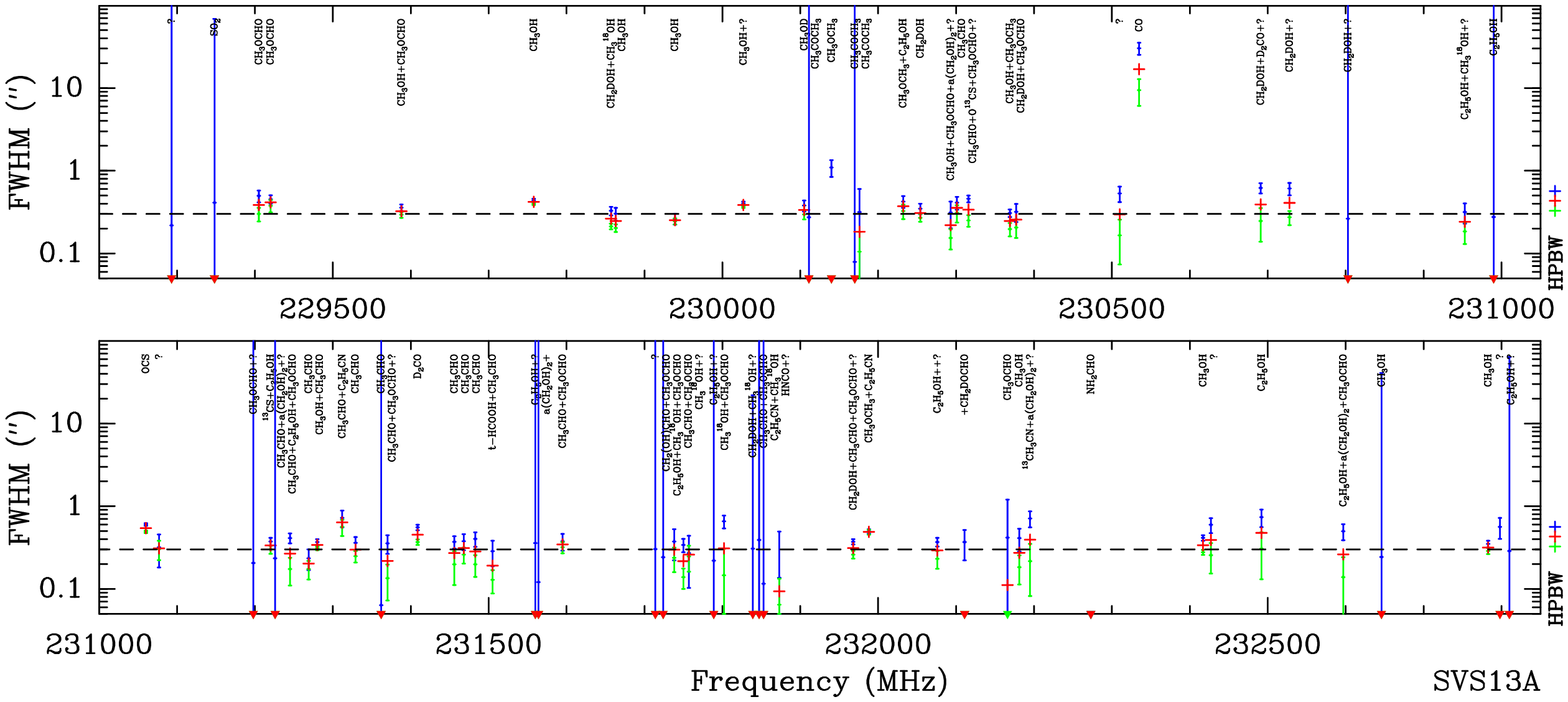}}}
\caption{Same as Fig.~\ref{f:sizes_l1448-2a} for SVS13A.}
\label{f:sizes_svs13-a}
\end{figure*}

\begin{figure*}
\centerline{\resizebox{0.95\hsize}{!}{\includegraphics[angle=270]{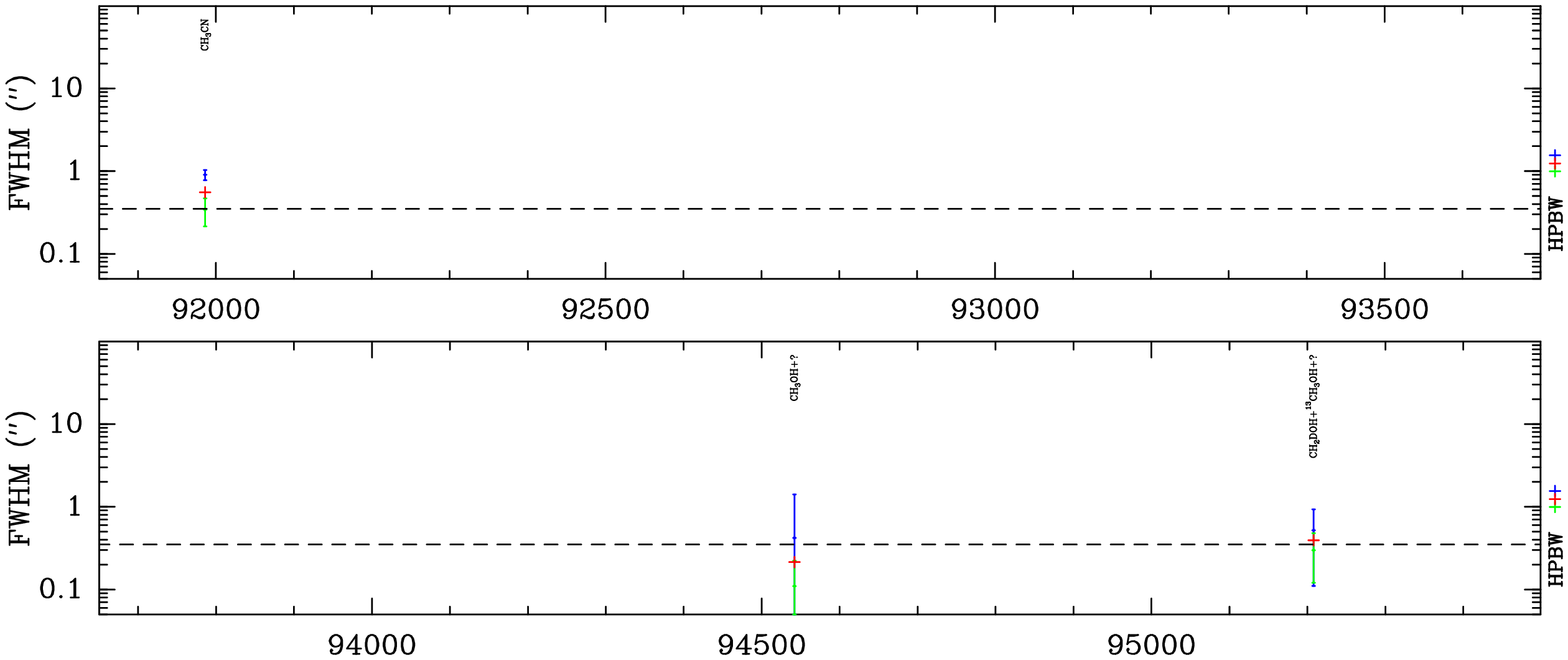}}}
\vspace*{1.5ex}
\centerline{\resizebox{0.95\hsize}{!}{\includegraphics[angle=270]{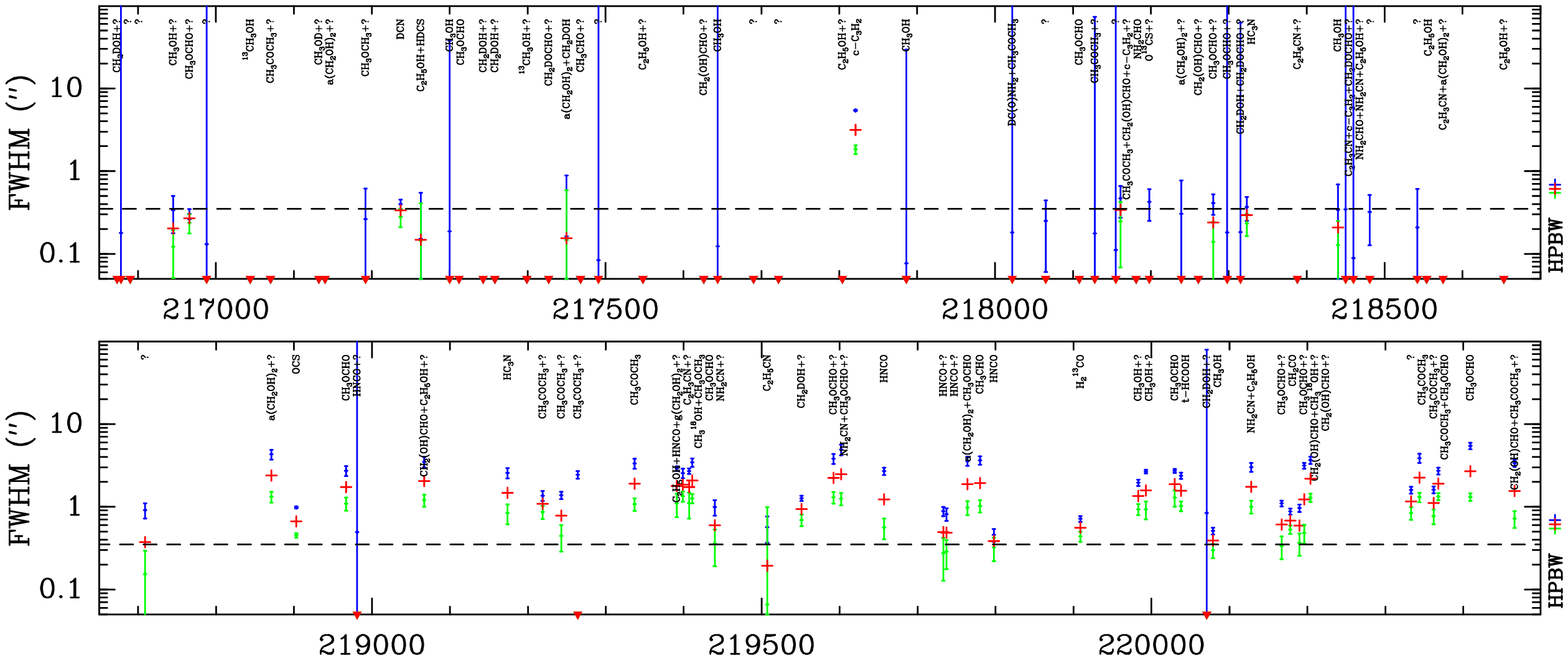}}}
\vspace*{1.5ex}
\centerline{\resizebox{0.95\hsize}{!}{\includegraphics[angle=270]{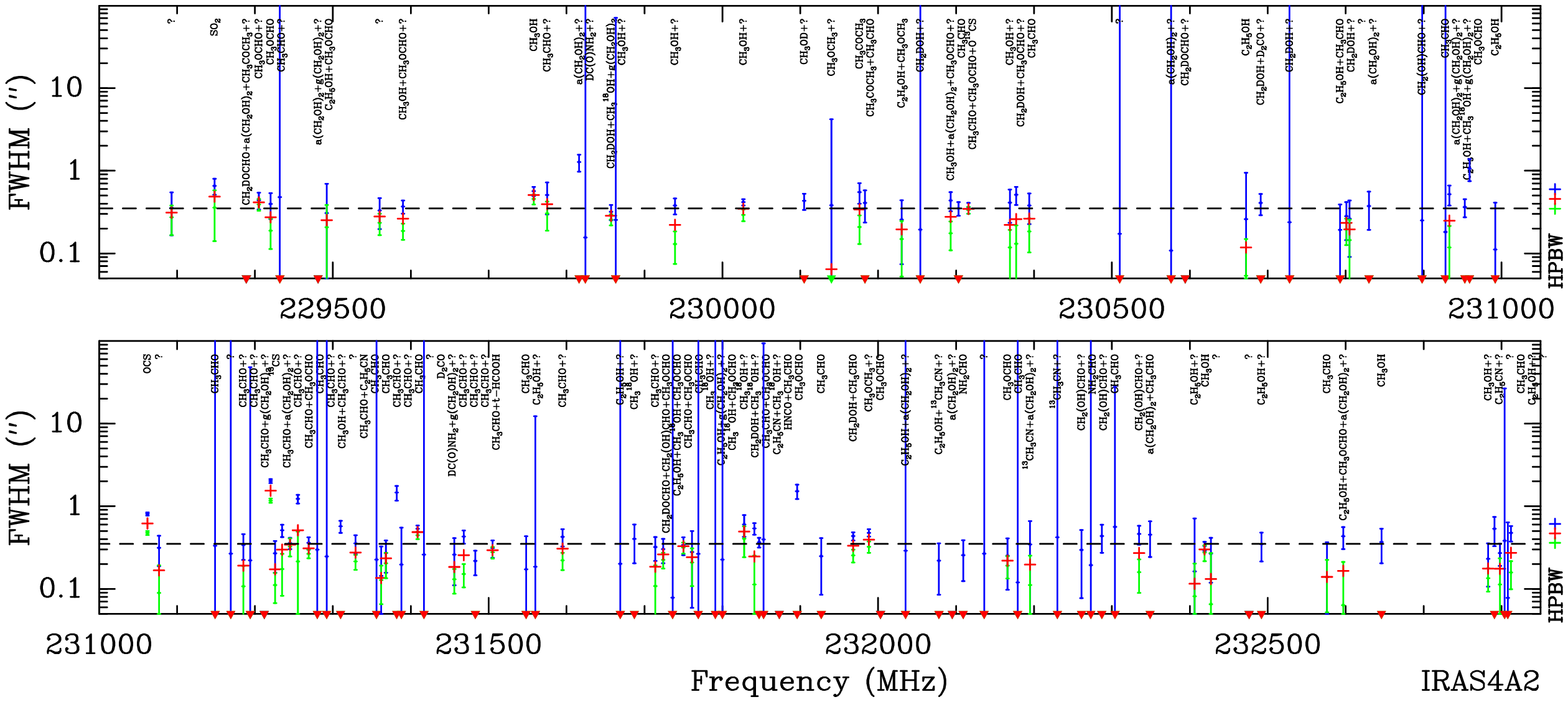}}}
\caption{Same as Fig.~\ref{f:sizes_l1448-2a} for IRAS4A2.}
\label{f:sizes_iras4a2}
\end{figure*}

\begin{figure*}
\centerline{\resizebox{0.95\hsize}{!}{\includegraphics[angle=270]{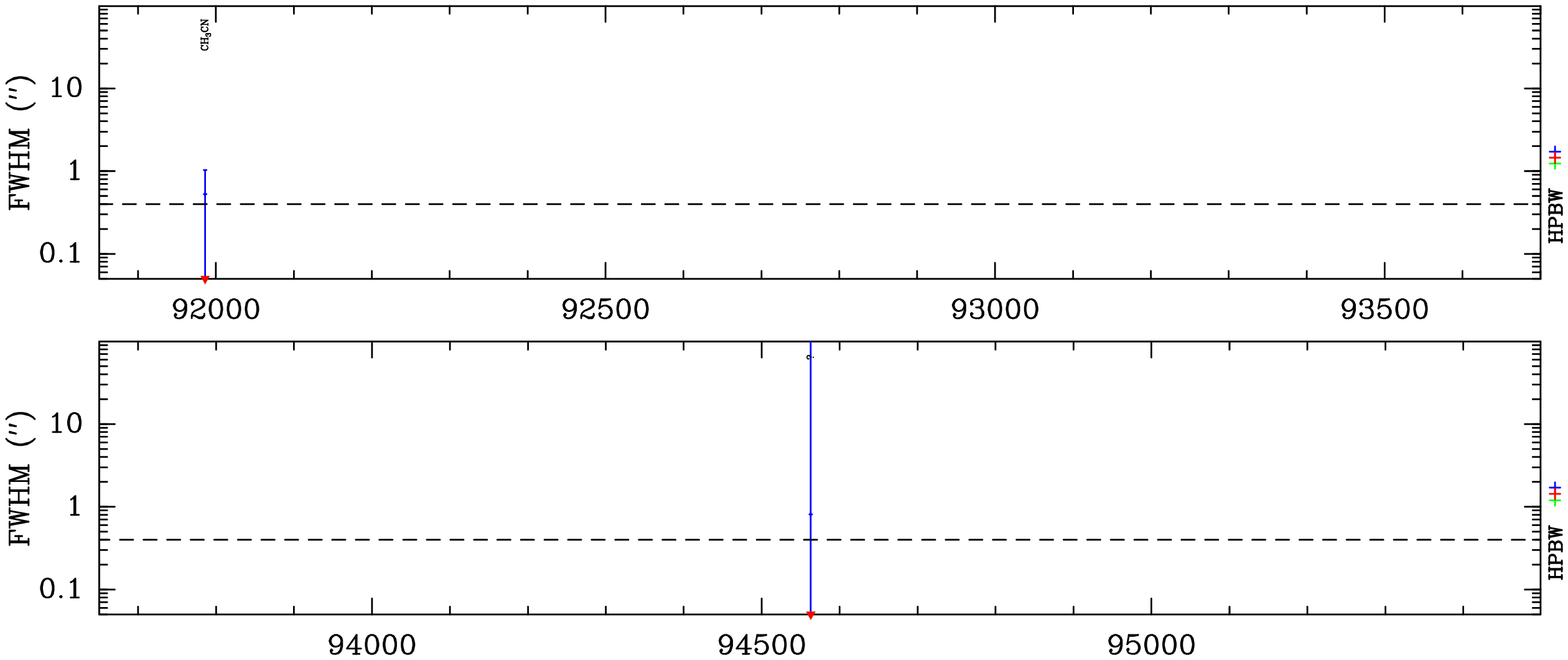}}}
\vspace*{1.5ex}
\centerline{\resizebox{0.95\hsize}{!}{\includegraphics[angle=270]{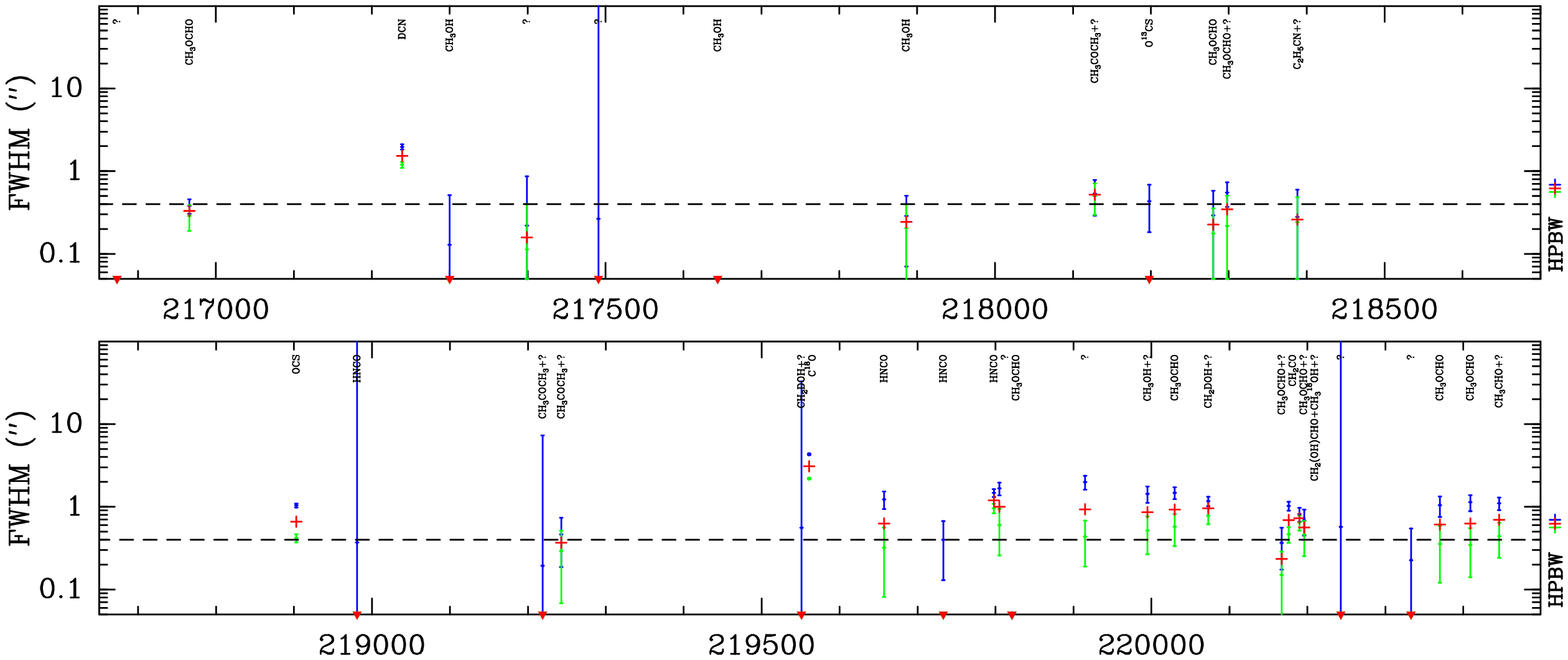}}}
\vspace*{1.5ex}
\centerline{\resizebox{0.95\hsize}{!}{\includegraphics[angle=270]{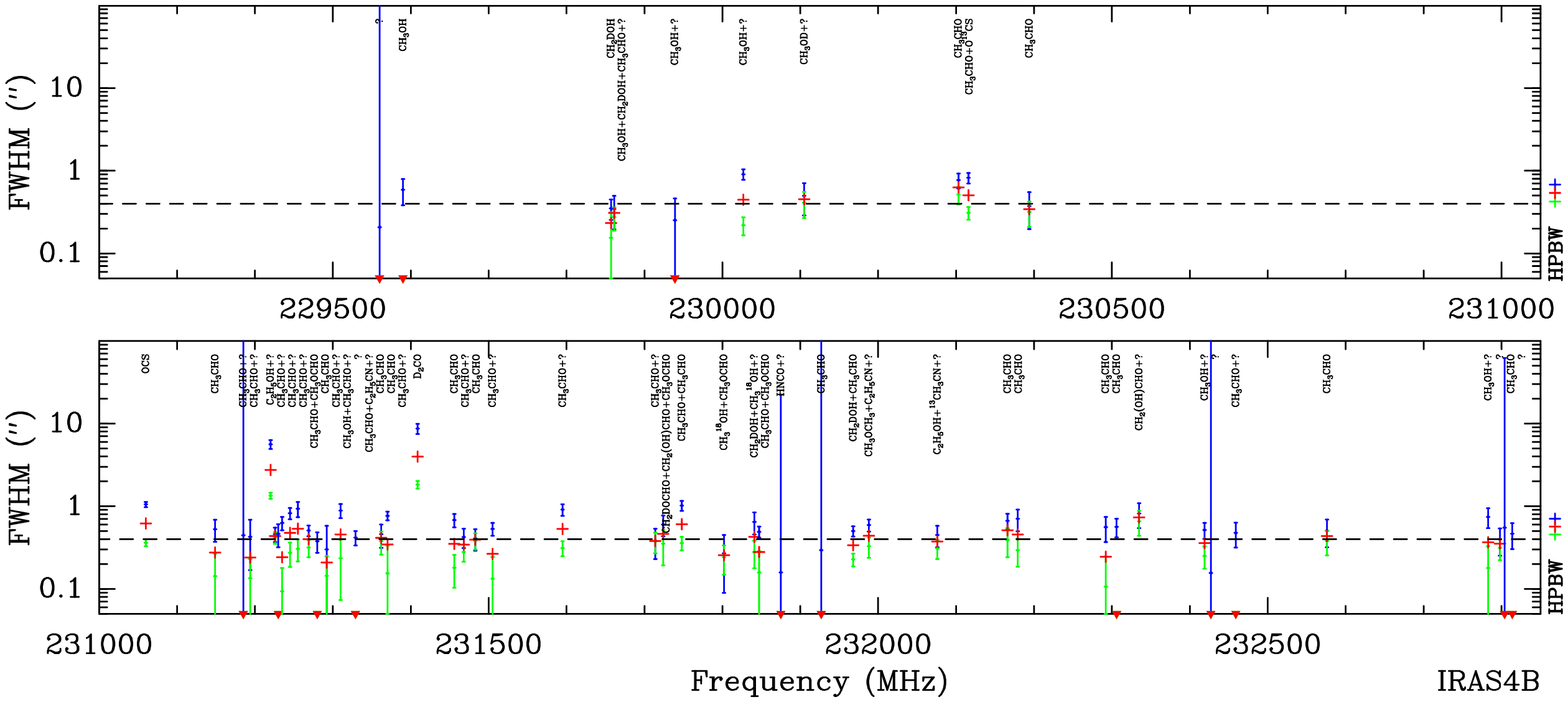}}}
\caption{Same as Fig.~\ref{f:sizes_l1448-2a} for IRAS4B.}
\label{f:sizes_iras4b}
\end{figure*}

\begin{figure*}
\centerline{\resizebox{0.95\hsize}{!}{\includegraphics[angle=270]{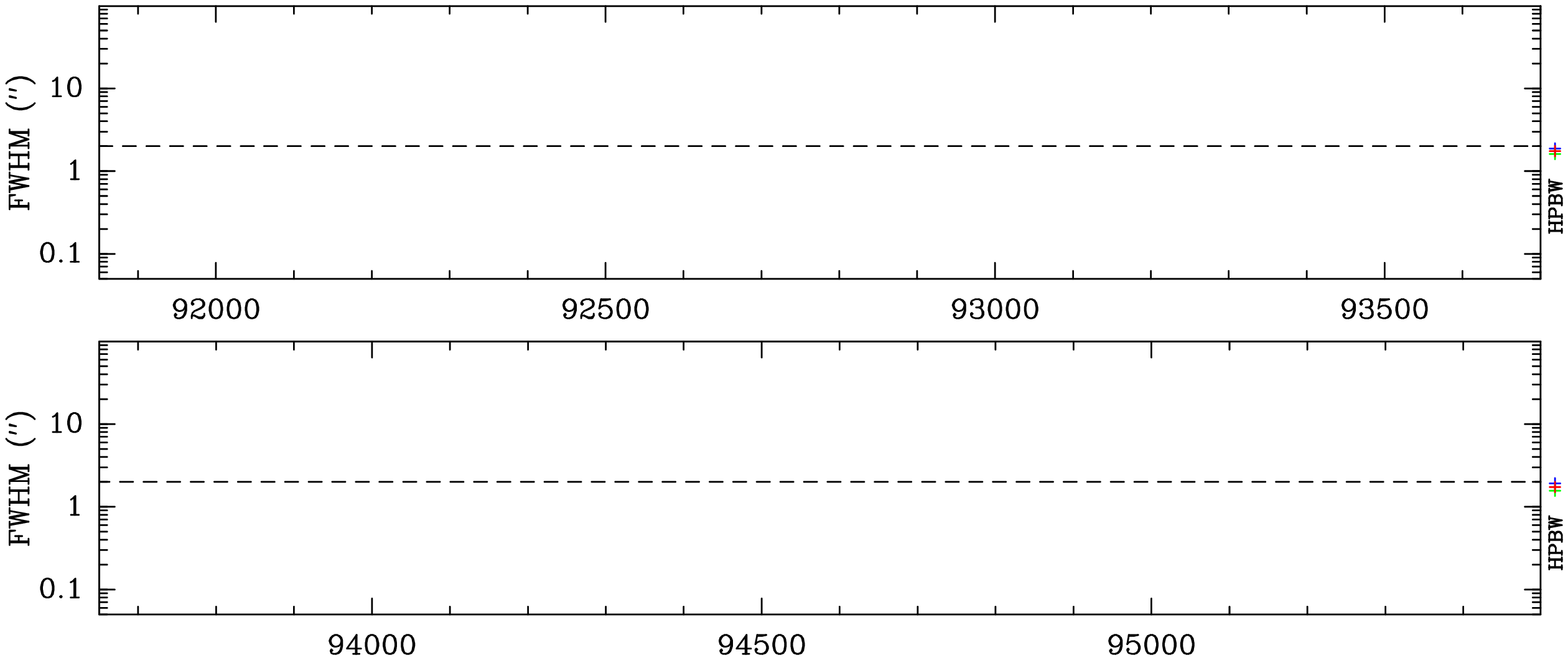}}}
\vspace*{1.5ex}
\centerline{\resizebox{0.95\hsize}{!}{\includegraphics[angle=270]{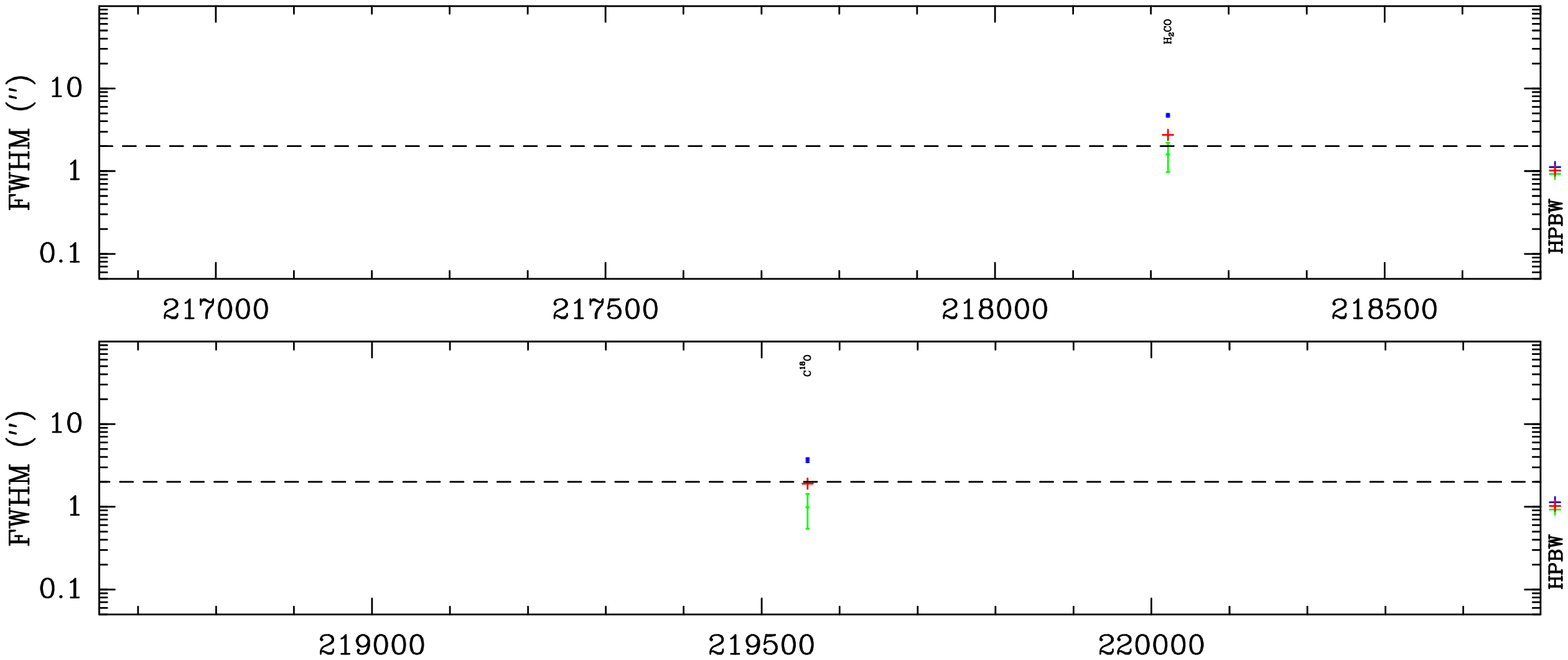}}}
\vspace*{1.5ex}
\centerline{\resizebox{0.95\hsize}{!}{\includegraphics[angle=270]{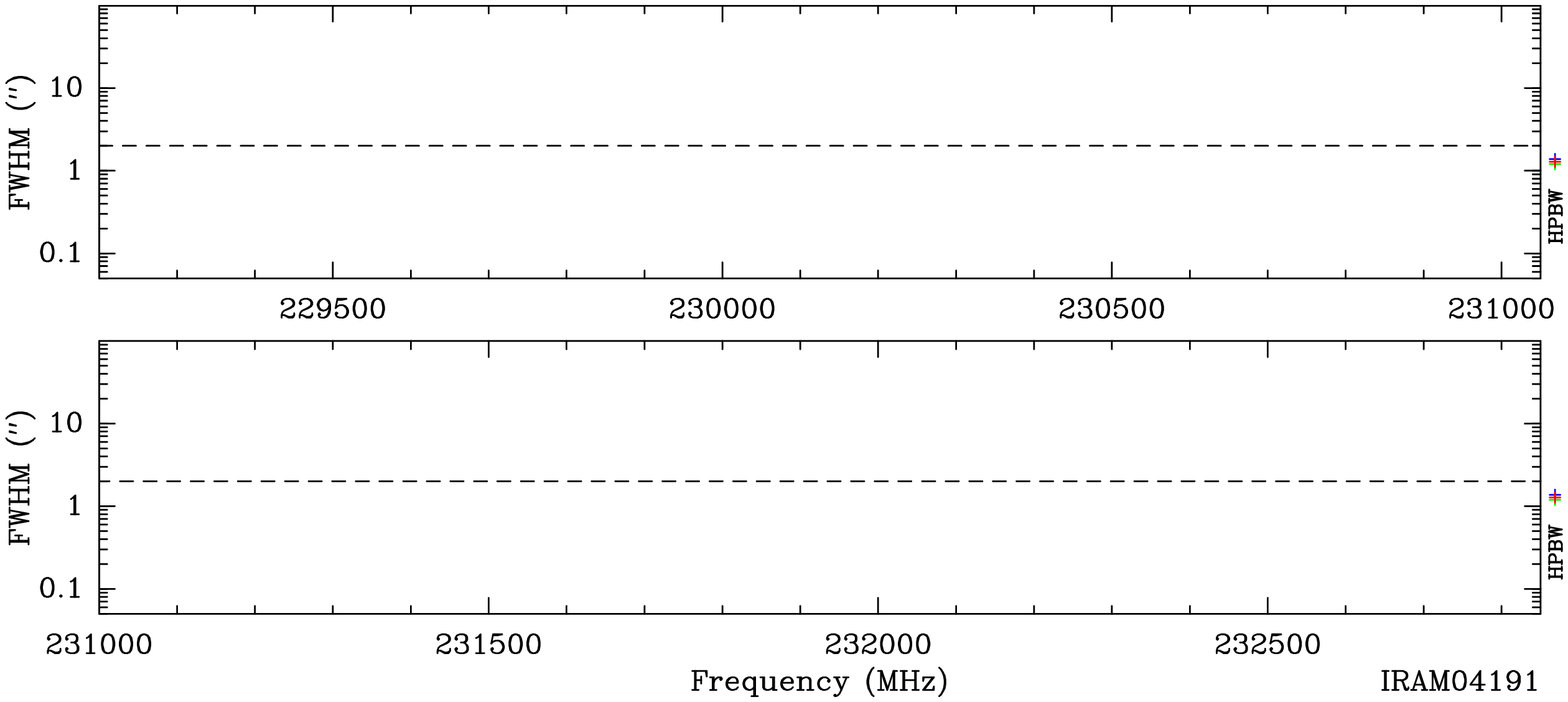}}}
\caption{Same as Fig.~\ref{f:sizes_l1448-2a} for IRAM04191.}
\label{f:sizes_iram04191}
\end{figure*}

\begin{figure*}
\centerline{\resizebox{0.95\hsize}{!}{\includegraphics[angle=270]{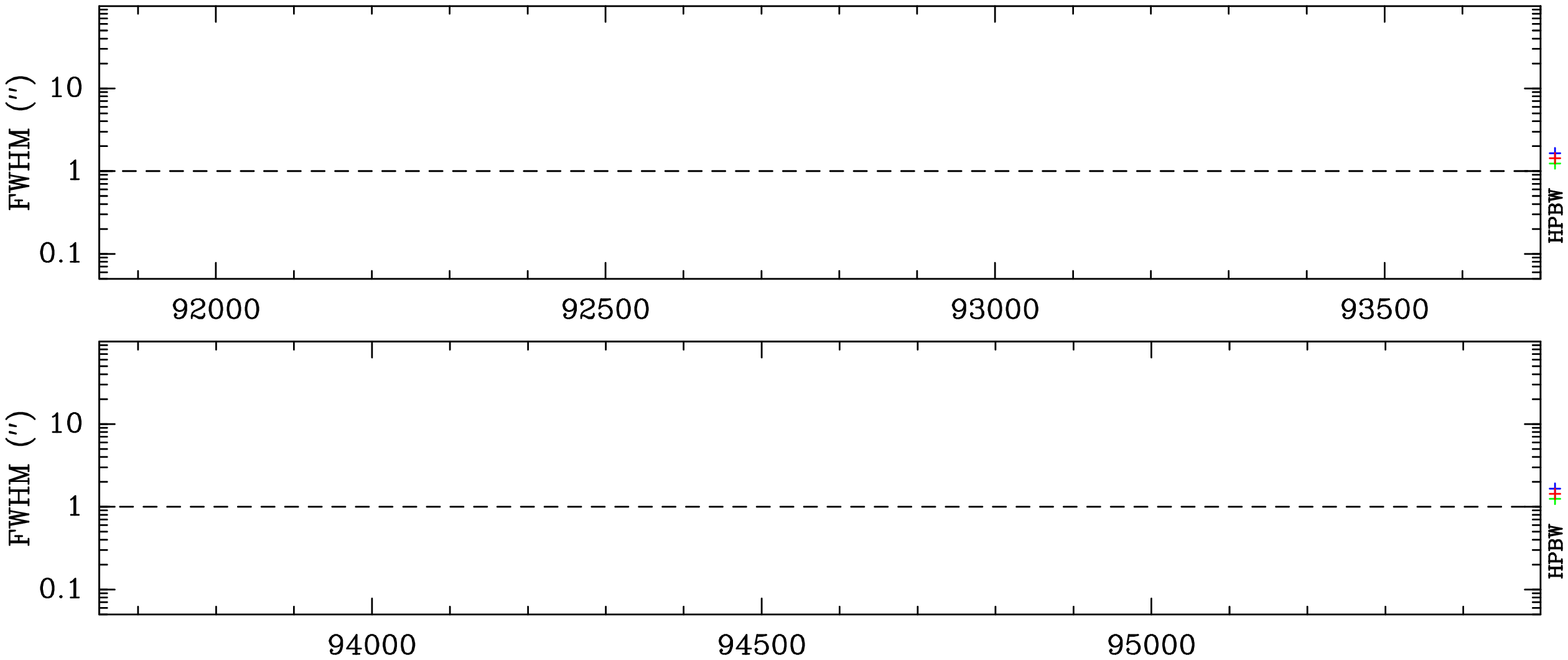}}}
\vspace*{1.5ex}
\centerline{\resizebox{0.95\hsize}{!}{\includegraphics[angle=270]{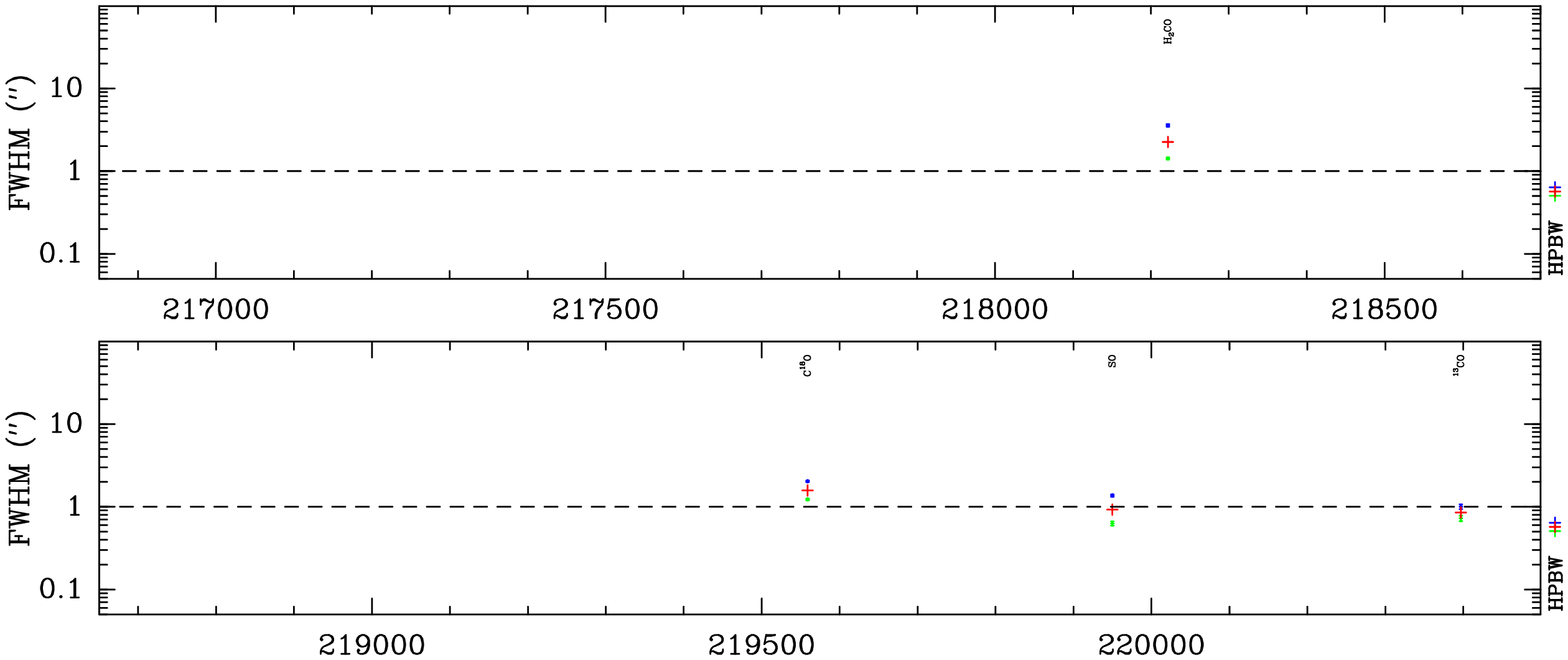}}}
\vspace*{1.5ex}
\centerline{\resizebox{0.95\hsize}{!}{\includegraphics[angle=270]{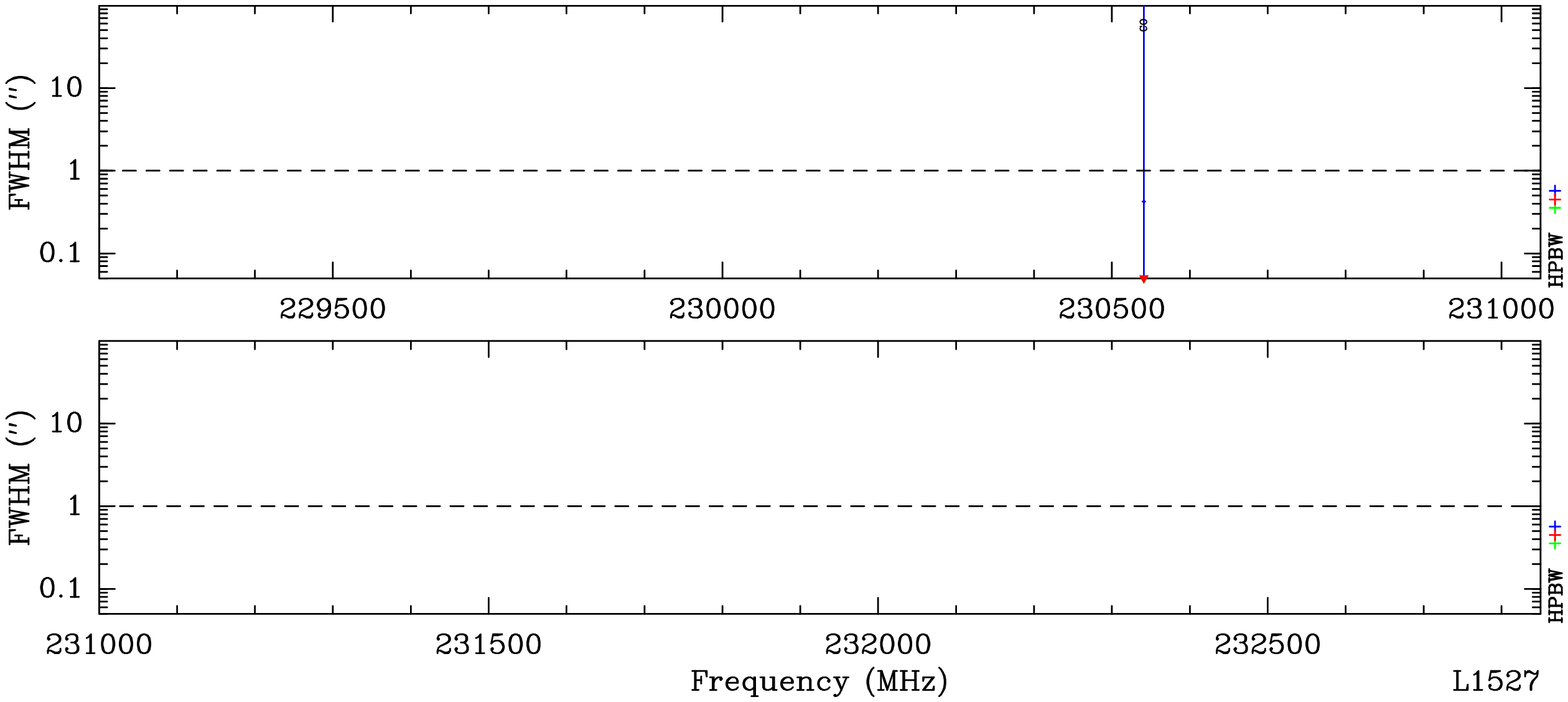}}}
\caption{Same as Fig.~\ref{f:sizes_l1448-2a} for L1527.}
\label{f:sizes_l1527}
\end{figure*}

\begin{figure*}
\centerline{\resizebox{0.95\hsize}{!}{\includegraphics[angle=270]{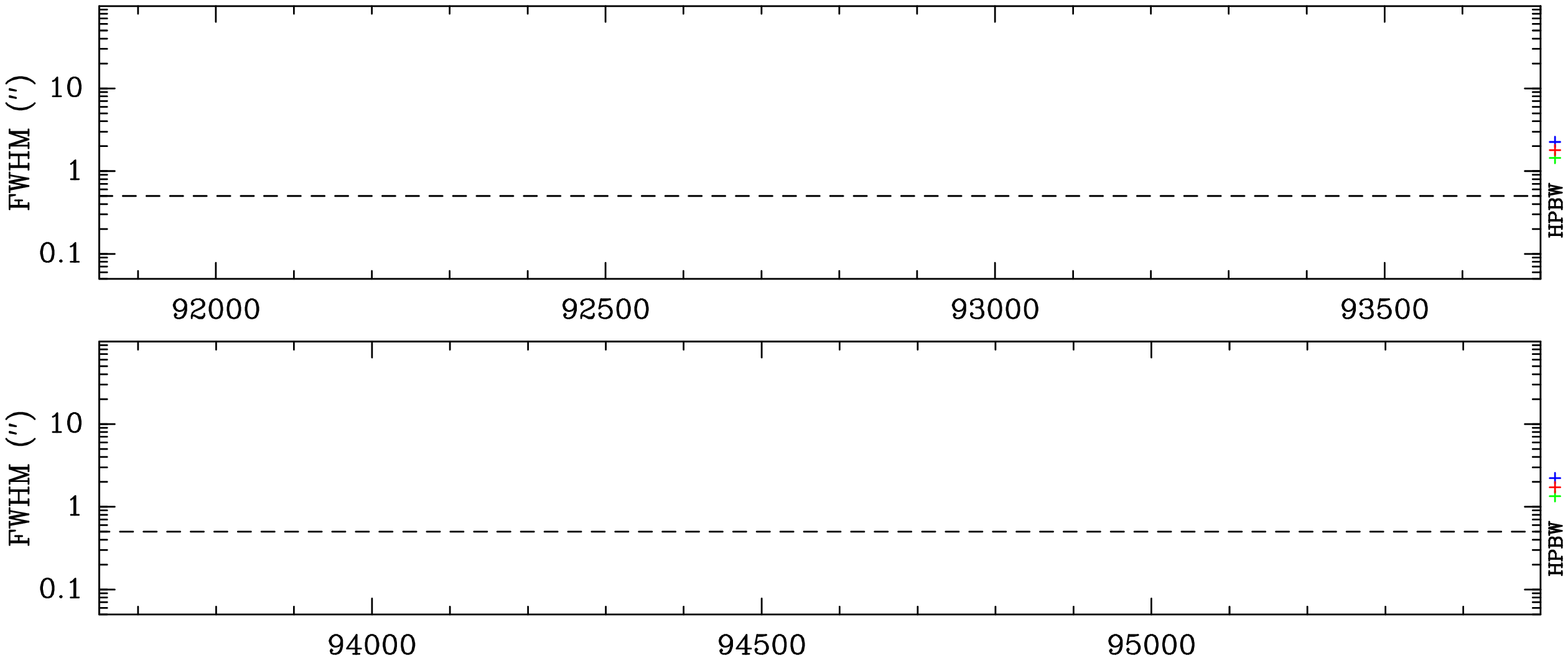}}}
\vspace*{1.5ex}
\centerline{\resizebox{0.95\hsize}{!}{\includegraphics[angle=270]{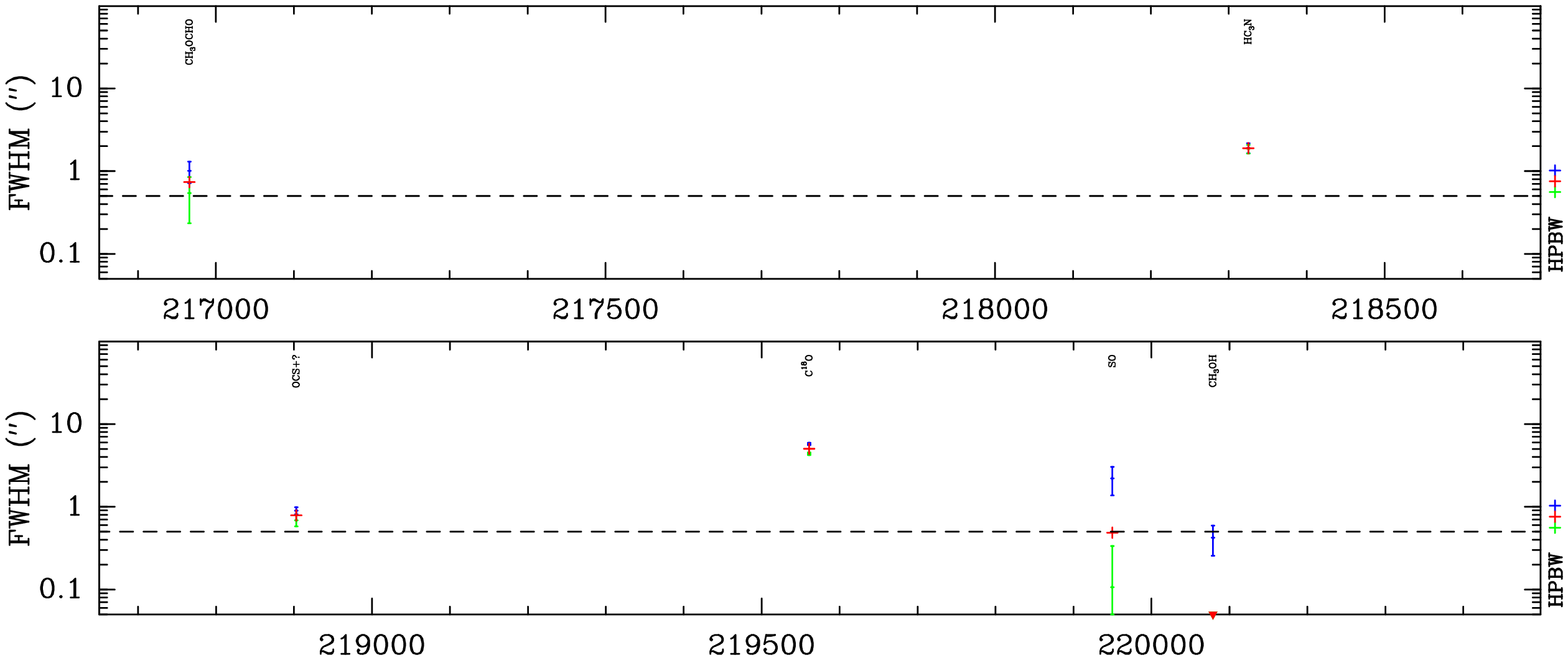}}}
\vspace*{1.5ex}
\centerline{\resizebox{0.95\hsize}{!}{\includegraphics[angle=270]{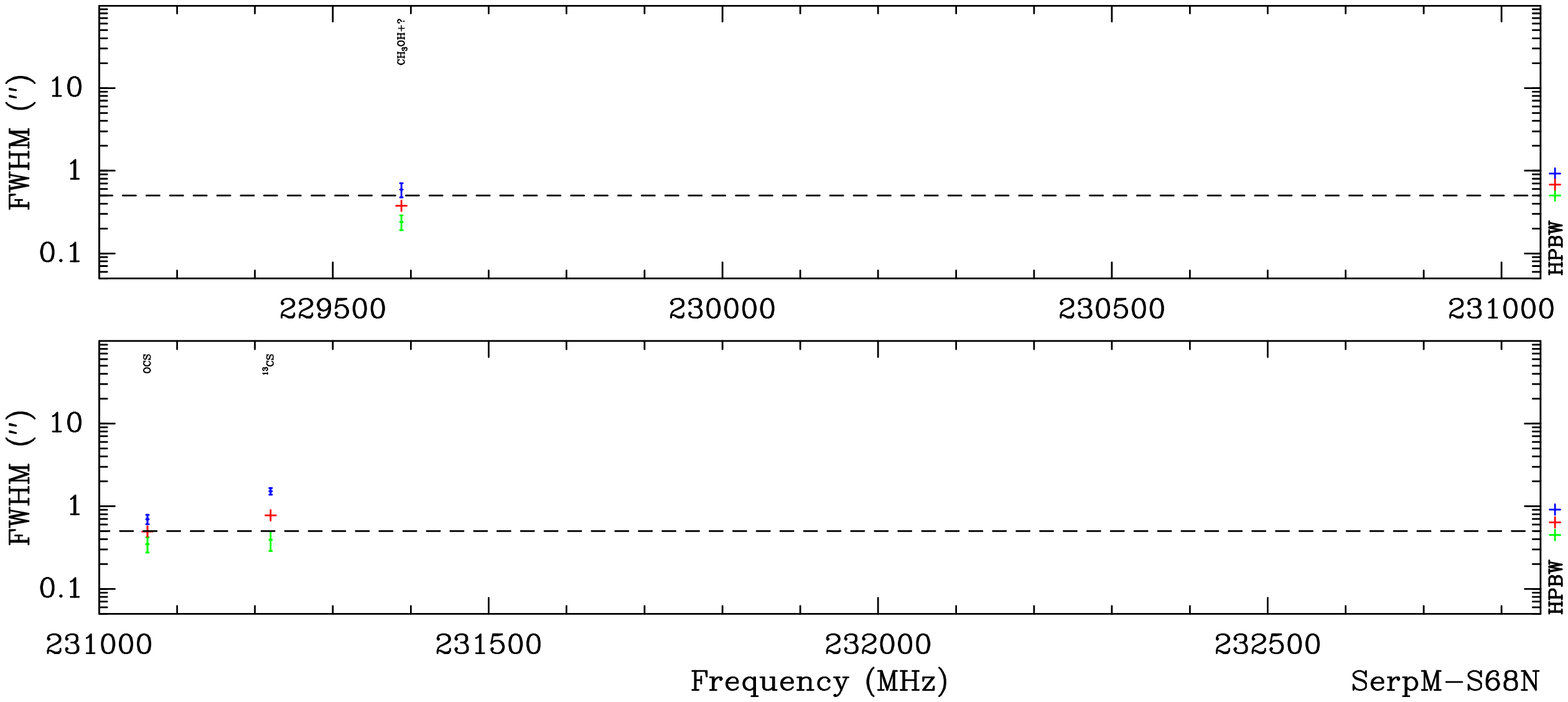}}}
\caption{Same as Fig.~\ref{f:sizes_l1448-2a} for SerpM-S68N.}
\label{f:sizes_serp-s68n}
\end{figure*}

\begin{figure*}
\centerline{\resizebox{0.95\hsize}{!}{\includegraphics[angle=270]{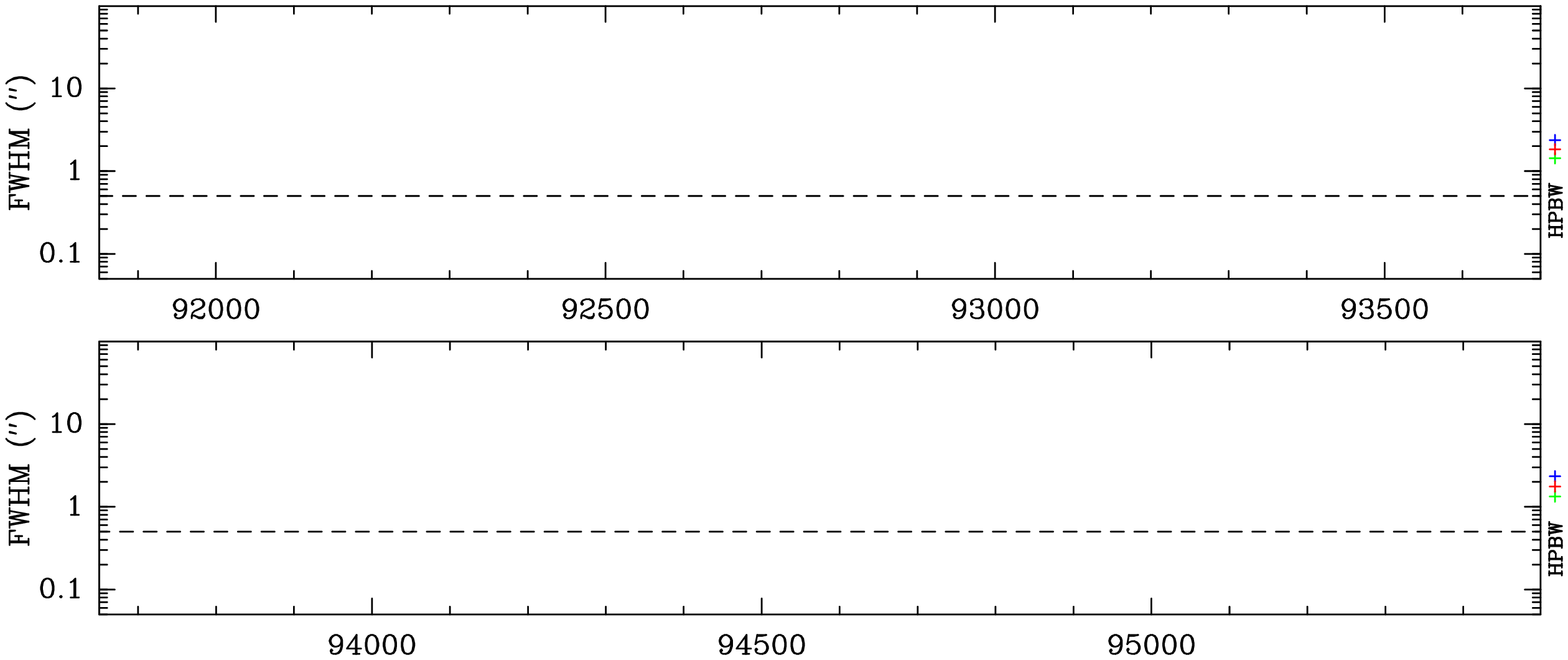}}}
\vspace*{1.5ex}
\centerline{\resizebox{0.95\hsize}{!}{\includegraphics[angle=270]{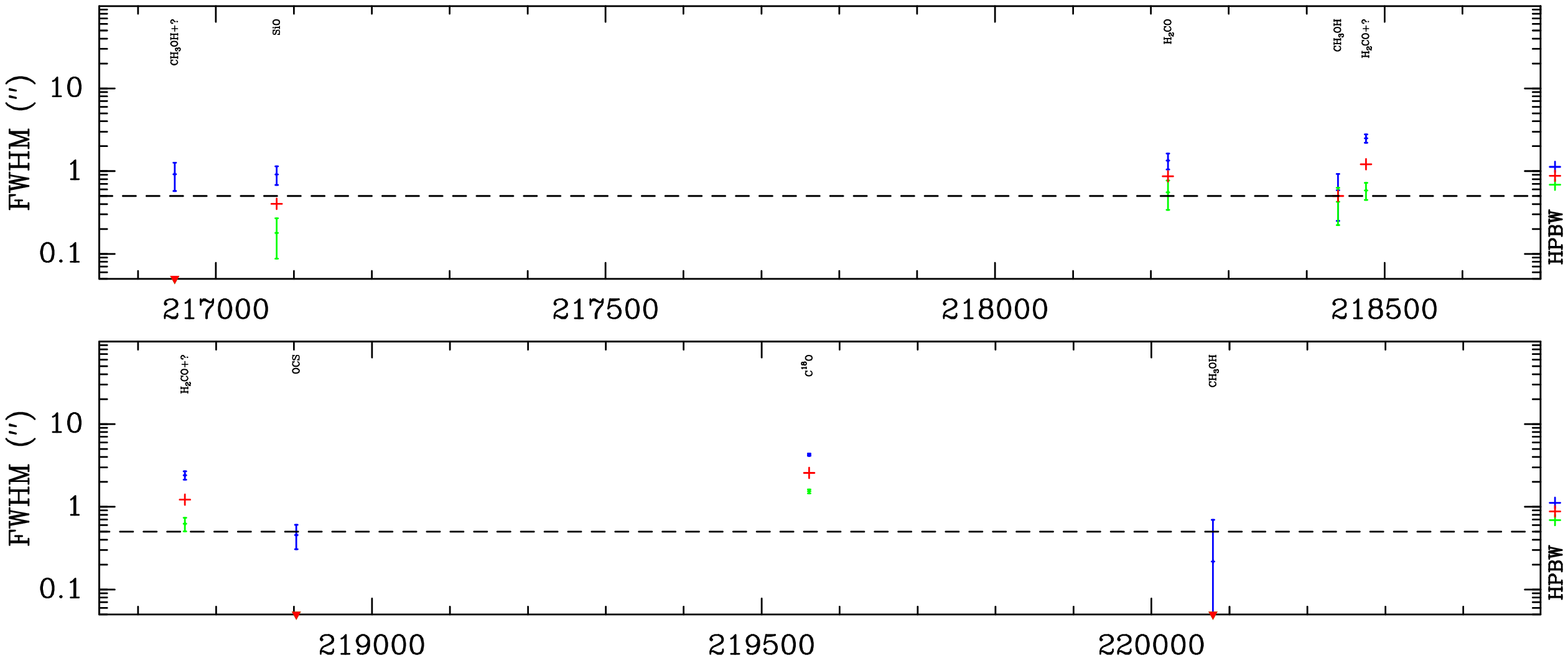}}}
\vspace*{1.5ex}
\centerline{\resizebox{0.95\hsize}{!}{\includegraphics[angle=270]{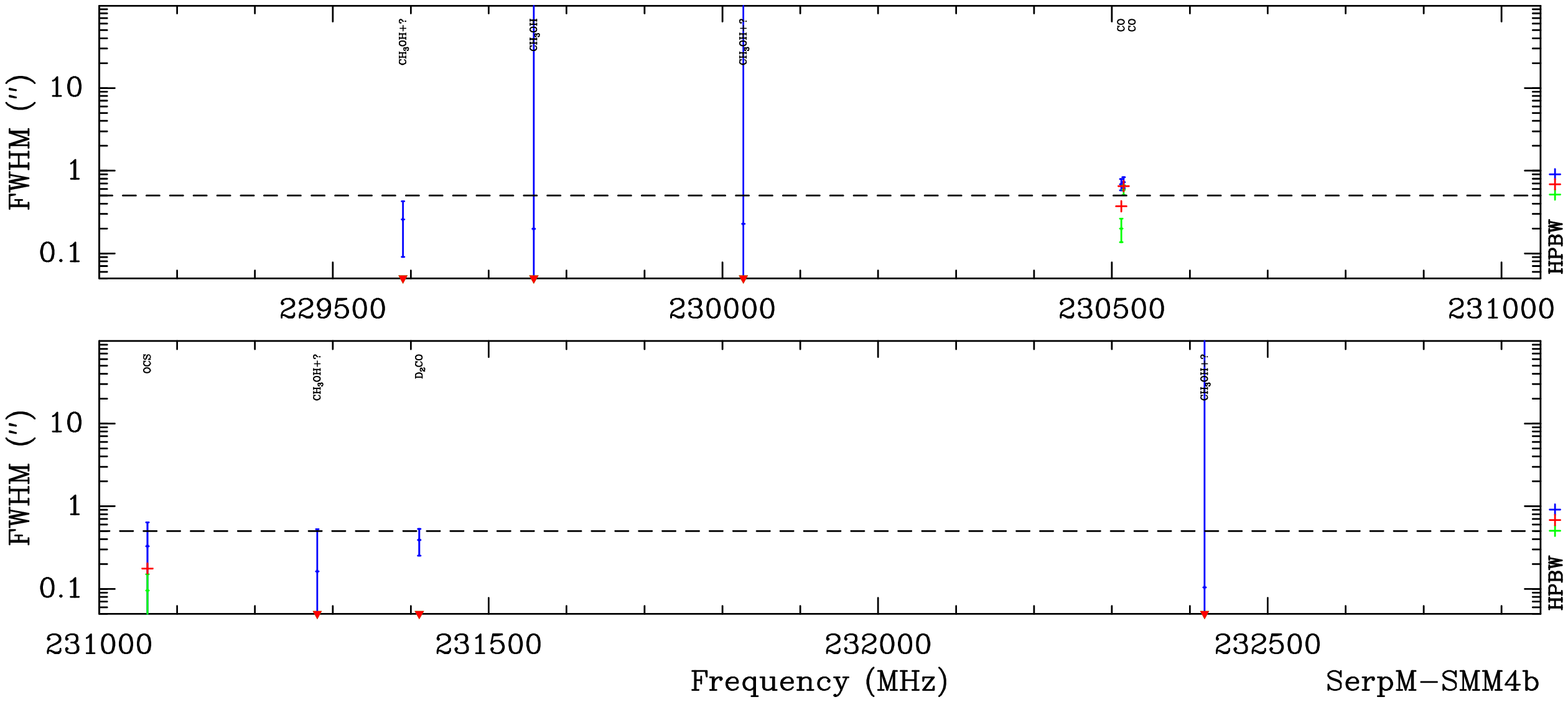}}}
\caption{Same as Fig.~\ref{f:sizes_l1448-2a} for SerpM-SMM4b.}
\label{f:sizes_serp-smm4}
\end{figure*}

\begin{figure*}
\centerline{\resizebox{0.95\hsize}{!}{\includegraphics[angle=270]{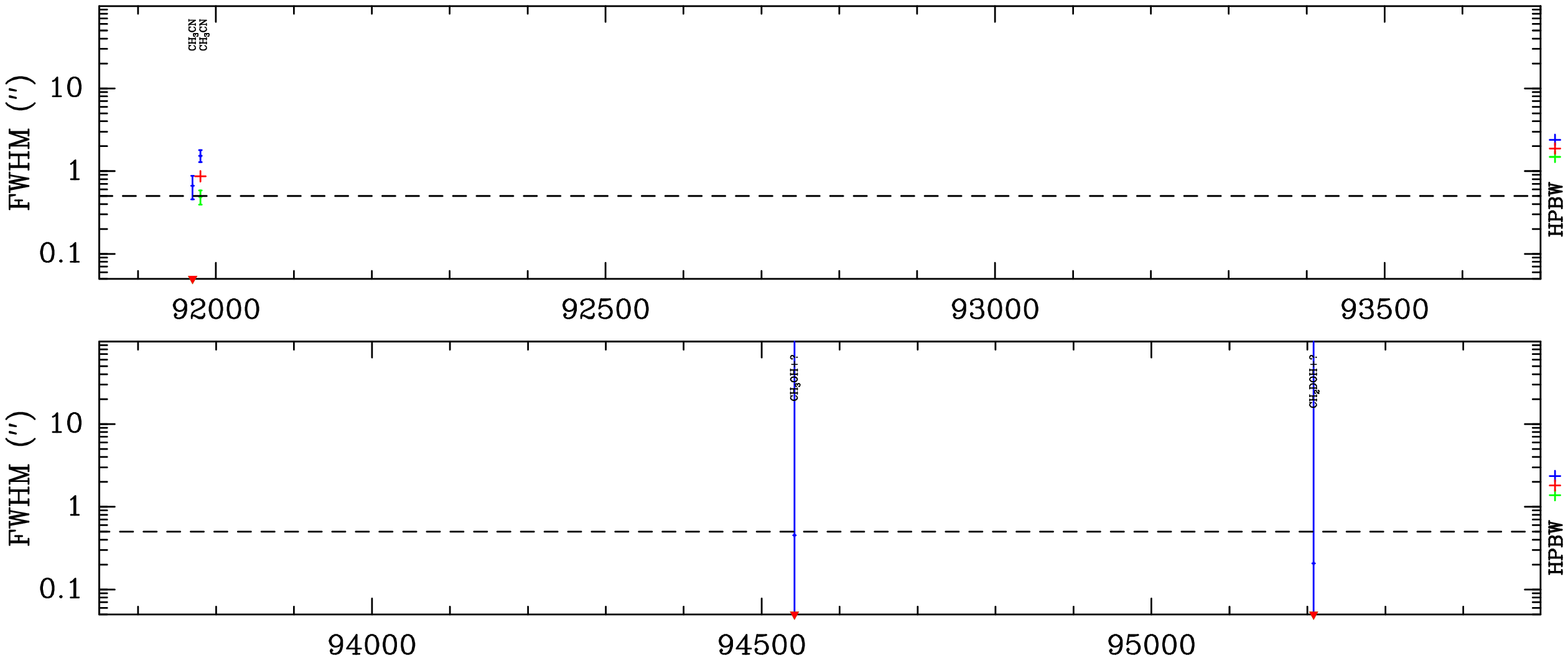}}}
\vspace*{1.5ex}
\centerline{\resizebox{0.95\hsize}{!}{\includegraphics[angle=270]{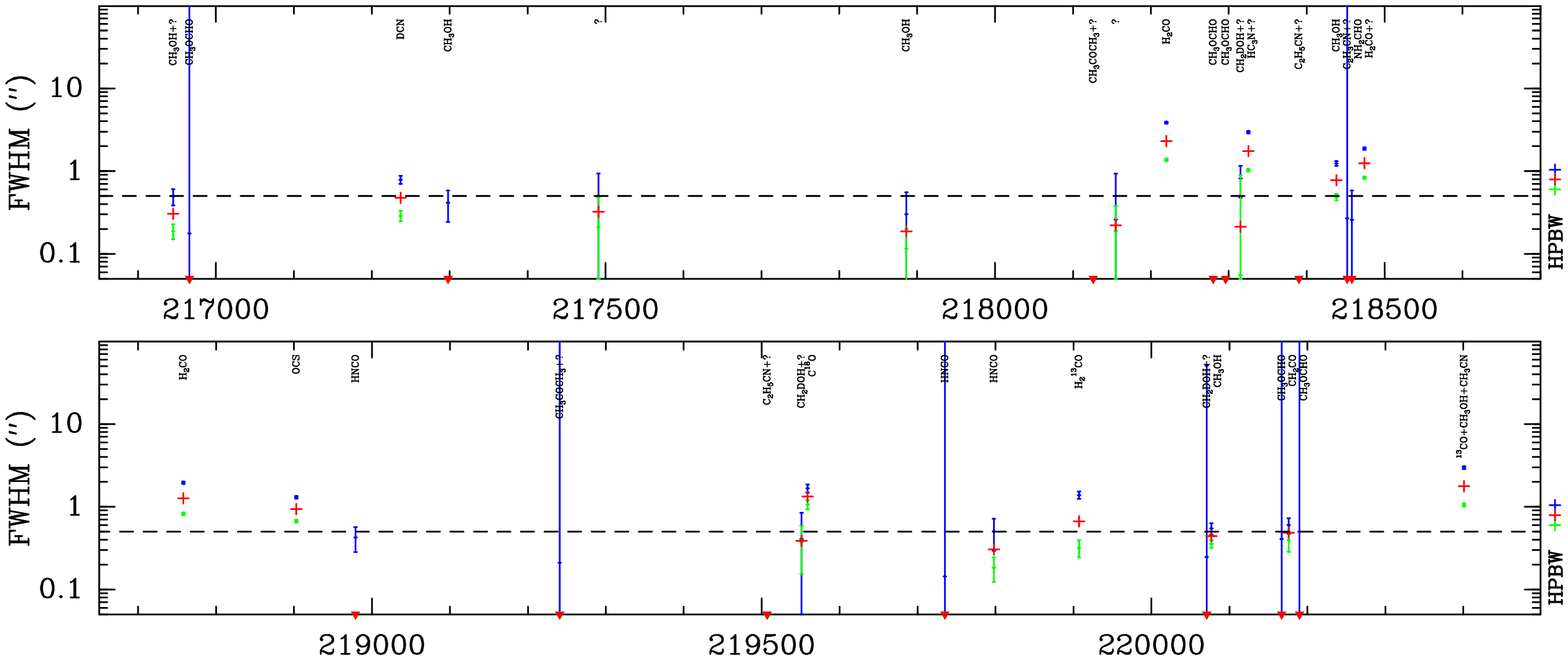}}}
\vspace*{1.5ex}
\centerline{\resizebox{0.95\hsize}{!}{\includegraphics[angle=270]{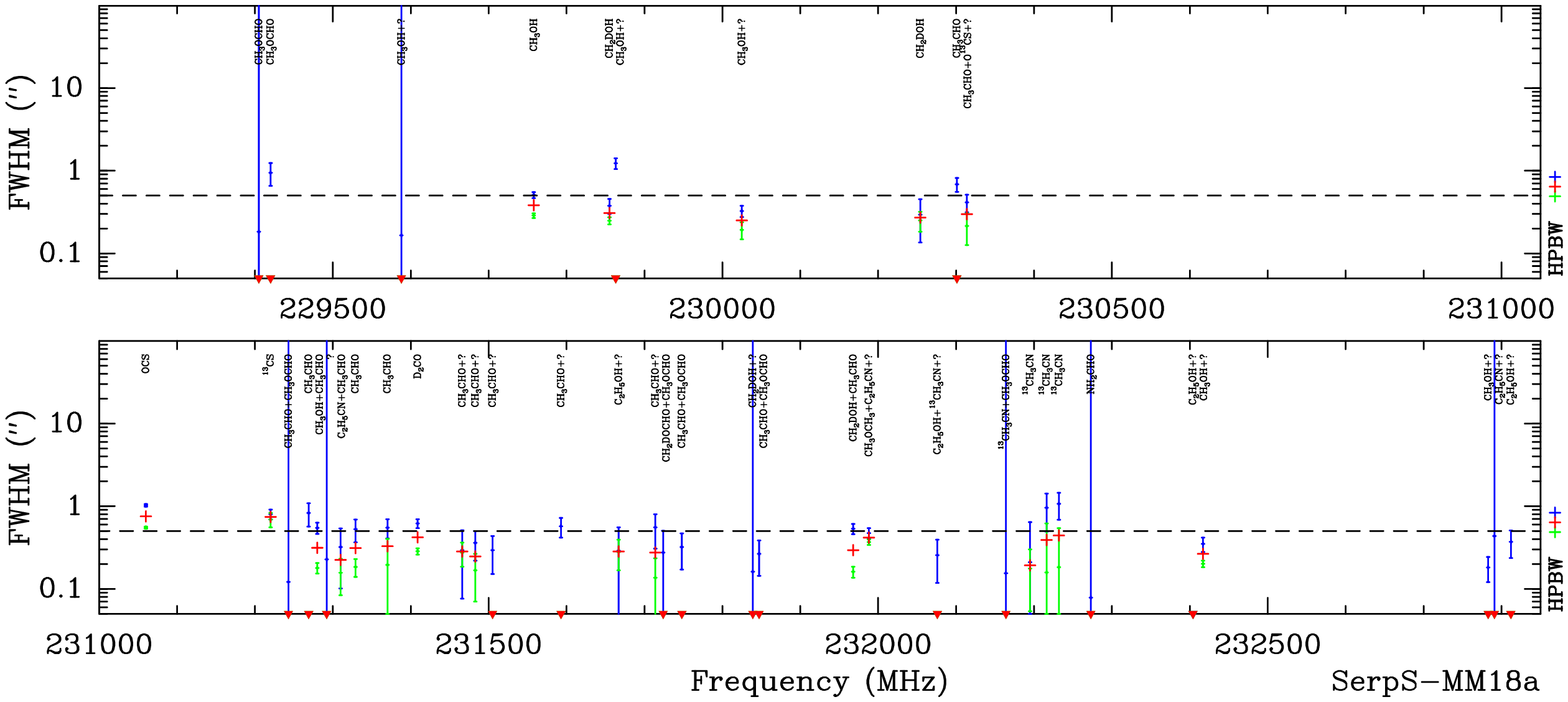}}}
\caption{Same as Fig.~\ref{f:sizes_l1448-2a} for SerpS-MM18a.}
\label{f:sizes_aqu-mms1}
\end{figure*}

\begin{figure*}
\centerline{\resizebox{0.95\hsize}{!}{\includegraphics[angle=270]{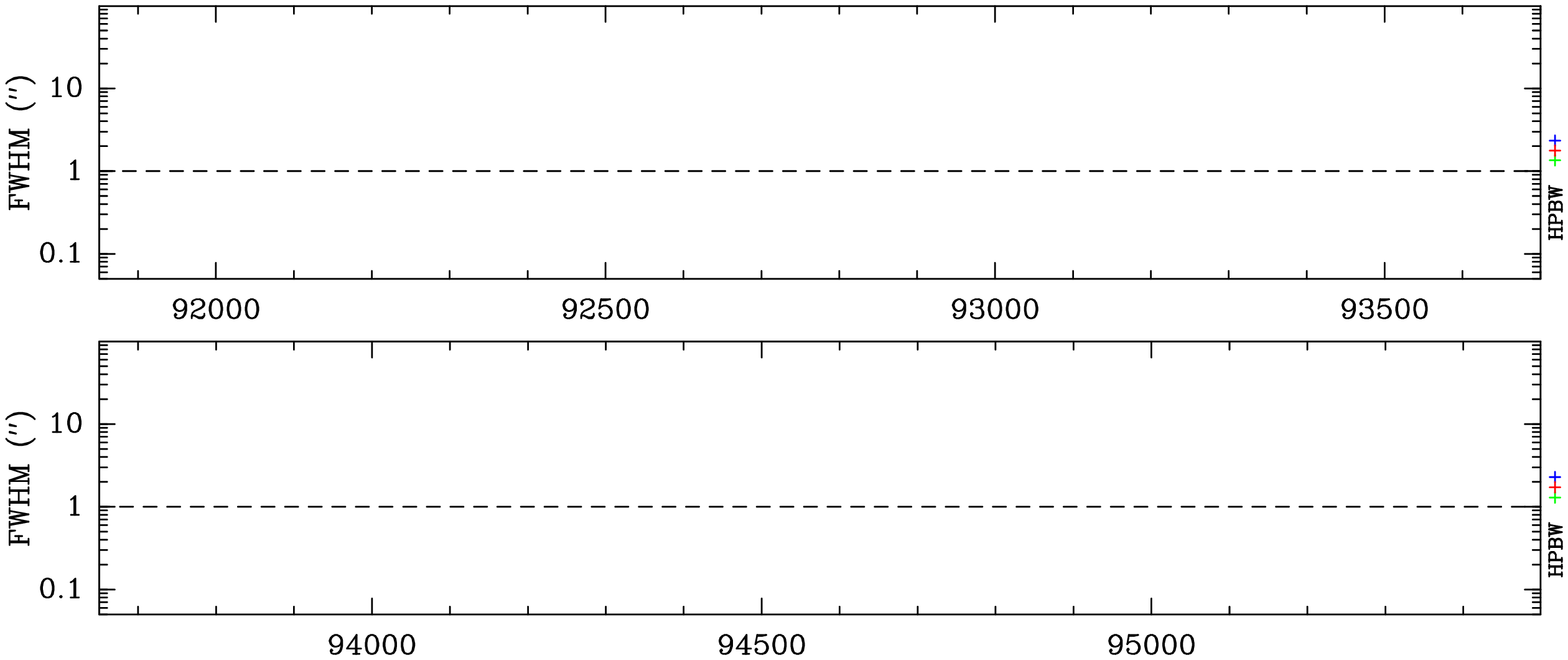}}}
\vspace*{1.5ex}
\centerline{\resizebox{0.95\hsize}{!}{\includegraphics[angle=270]{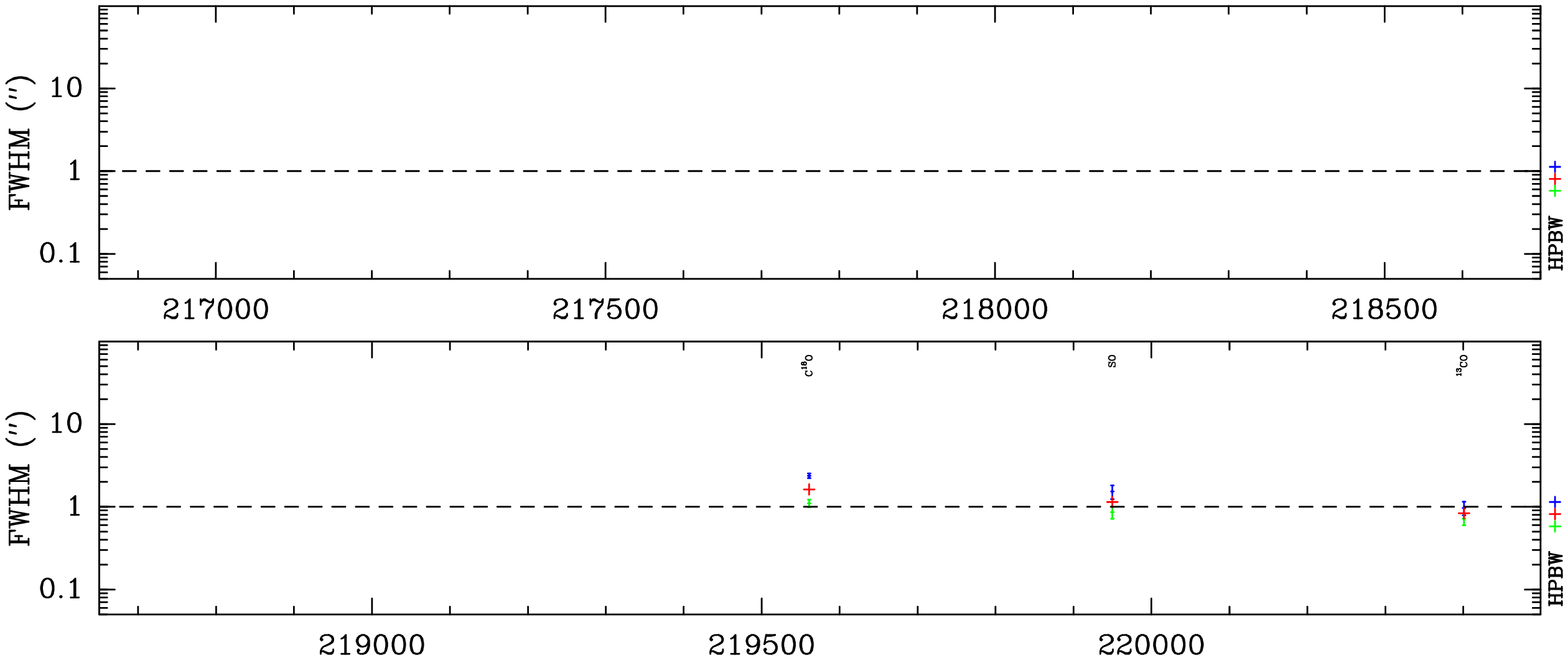}}}
\vspace*{1.5ex}
\centerline{\resizebox{0.95\hsize}{!}{\includegraphics[angle=270]{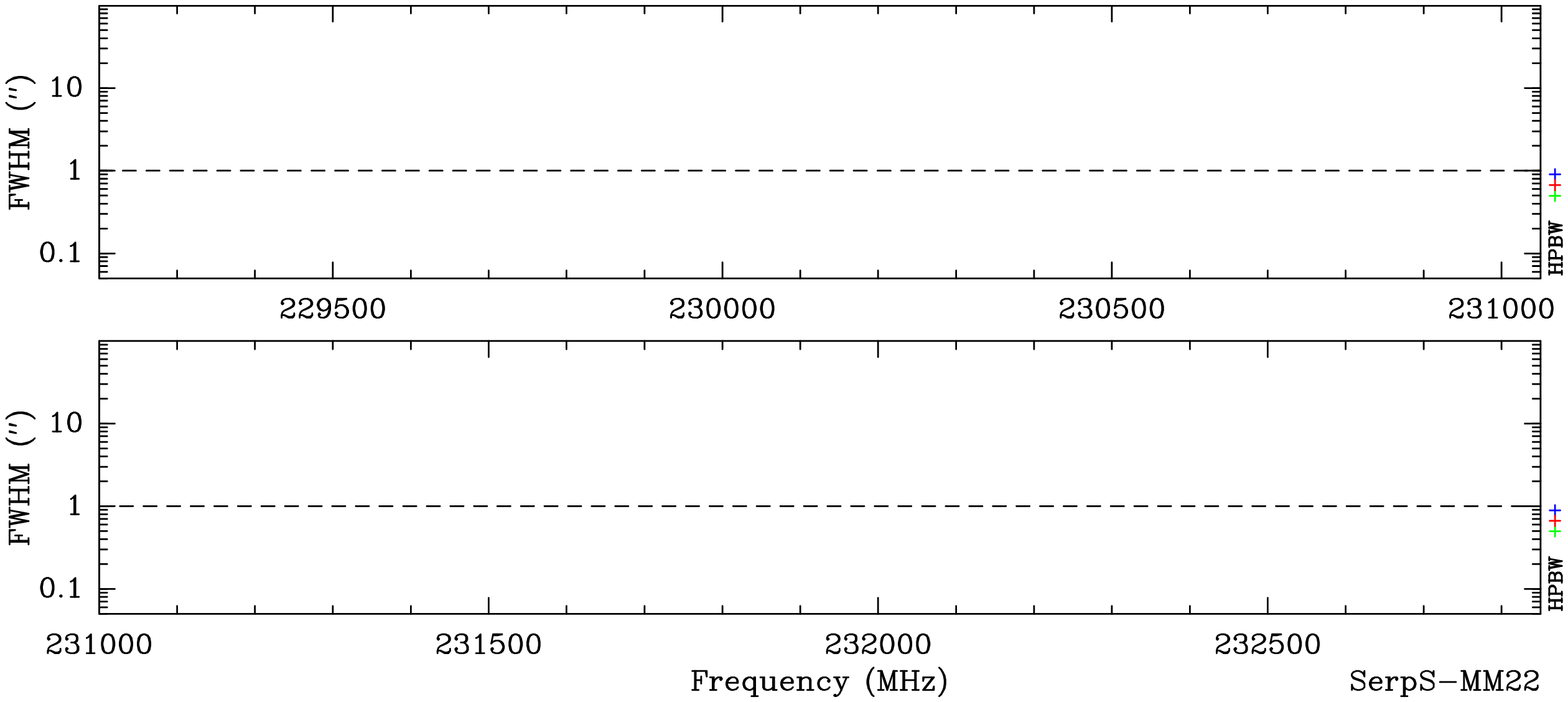}}}
\caption{Same as Fig.~\ref{f:sizes_l1448-2a} for SerpS-MM22.}
\label{f:sizes_aqu-mms2}
\end{figure*}

\begin{figure*}
\centerline{\resizebox{0.95\hsize}{!}{\includegraphics[angle=270]{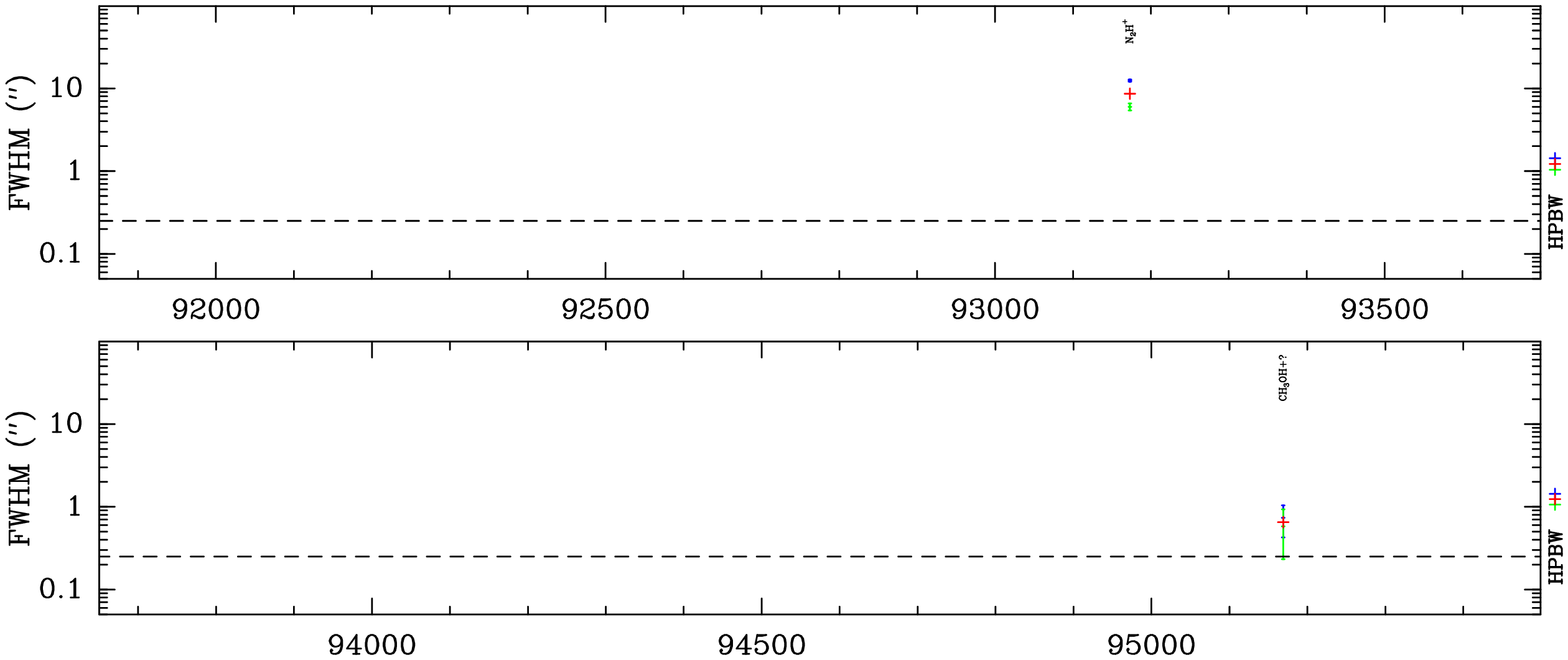}}}
\vspace*{1.5ex}
\centerline{\resizebox{0.95\hsize}{!}{\includegraphics[angle=270]{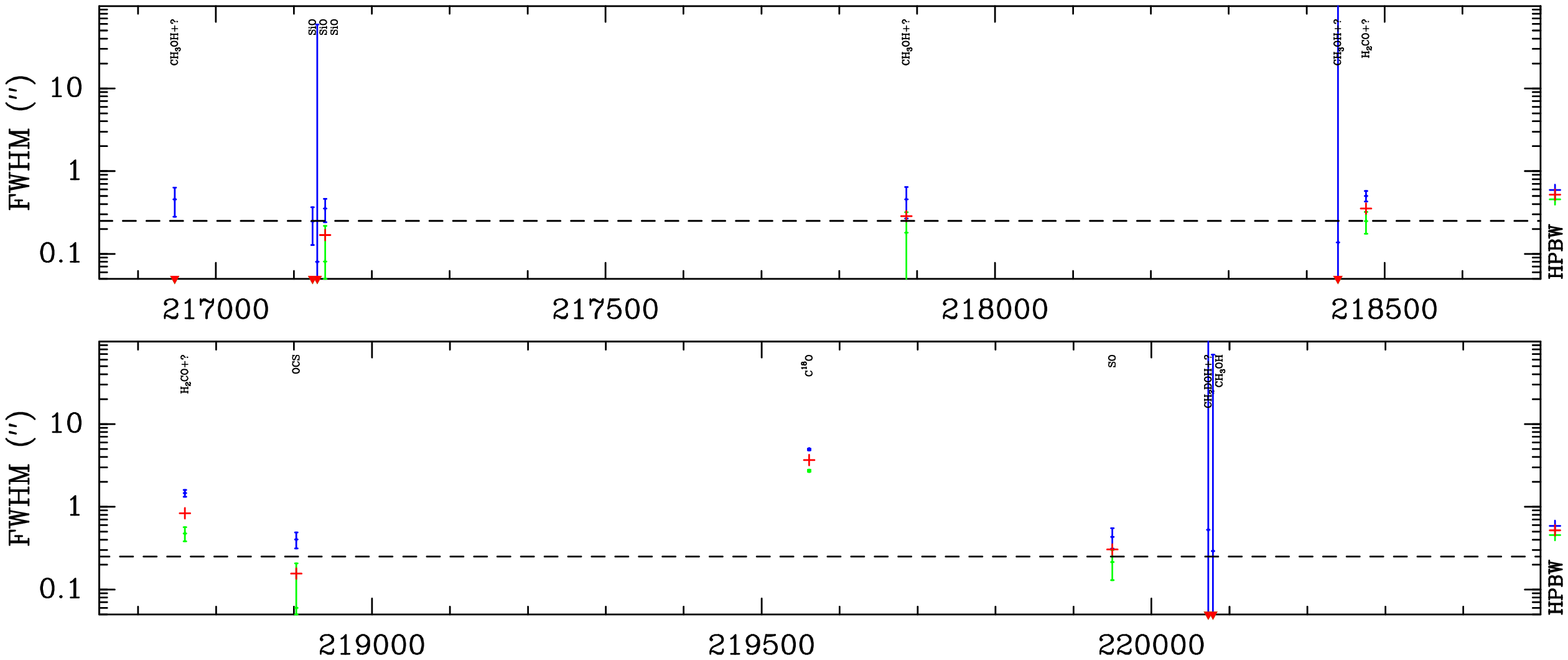}}}
\vspace*{1.5ex}
\centerline{\resizebox{0.95\hsize}{!}{\includegraphics[angle=270]{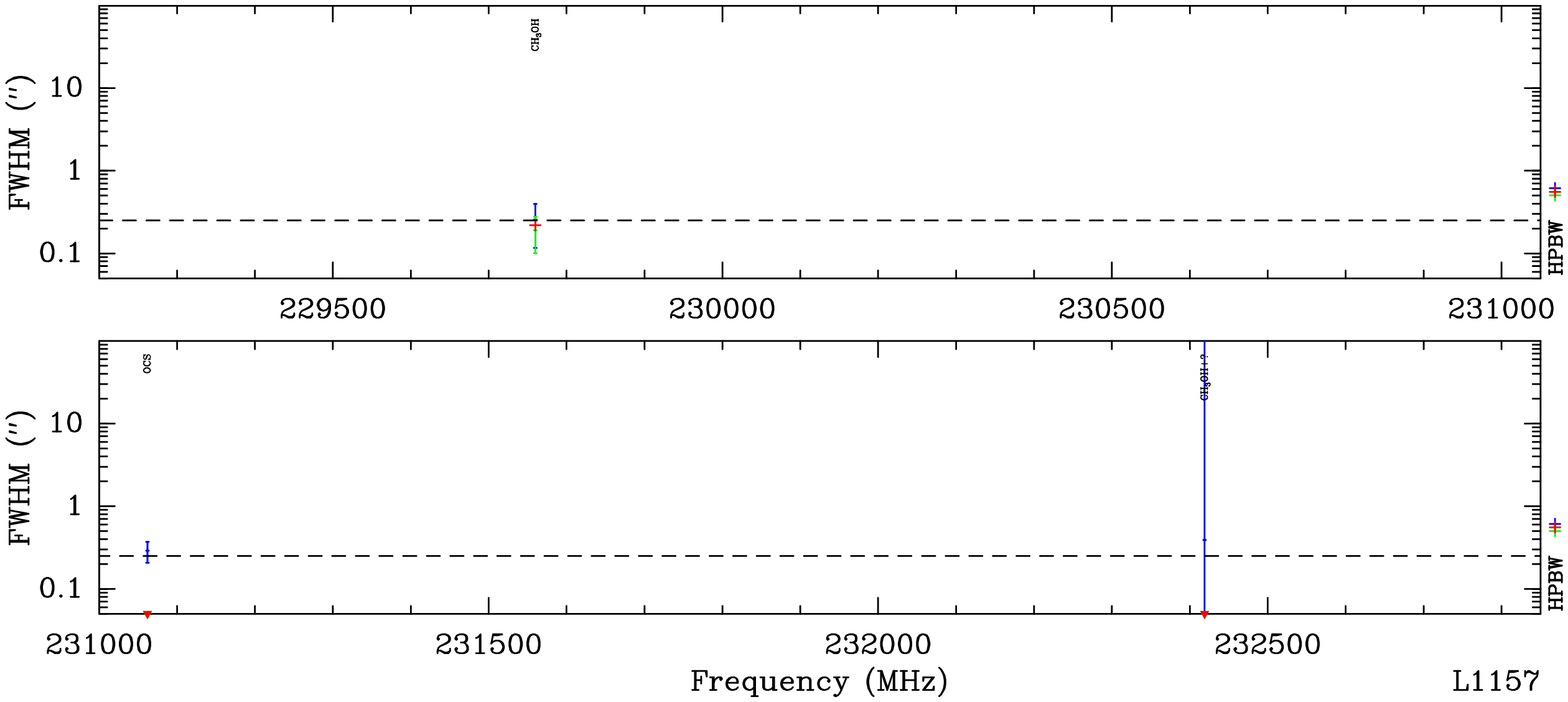}}}
\caption{Same as Fig.~\ref{f:sizes_l1448-2a} for L1157.}
\label{f:sizes_l1157}
\end{figure*}

\begin{figure*}
\centerline{\resizebox{0.95\hsize}{!}{\includegraphics[angle=270]{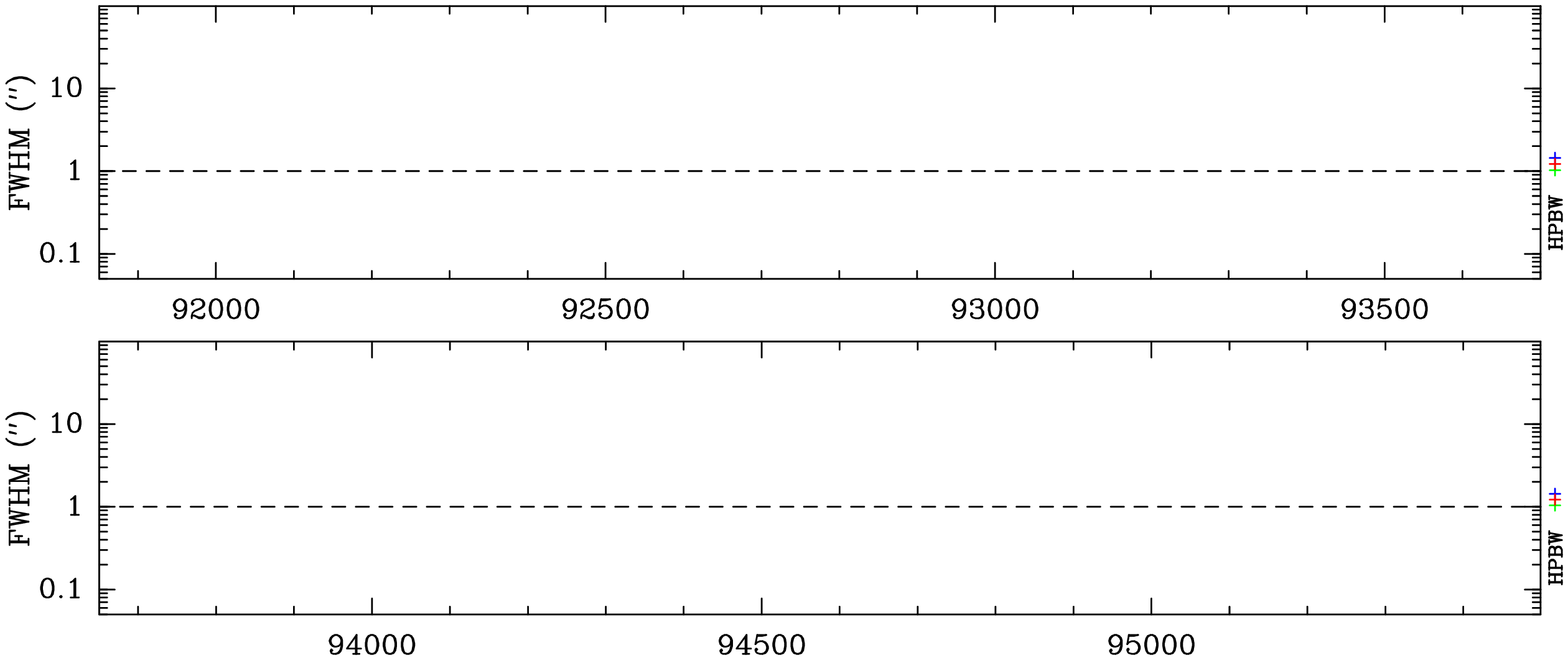}}}
\vspace*{1.5ex}
\centerline{\resizebox{0.95\hsize}{!}{\includegraphics[angle=270]{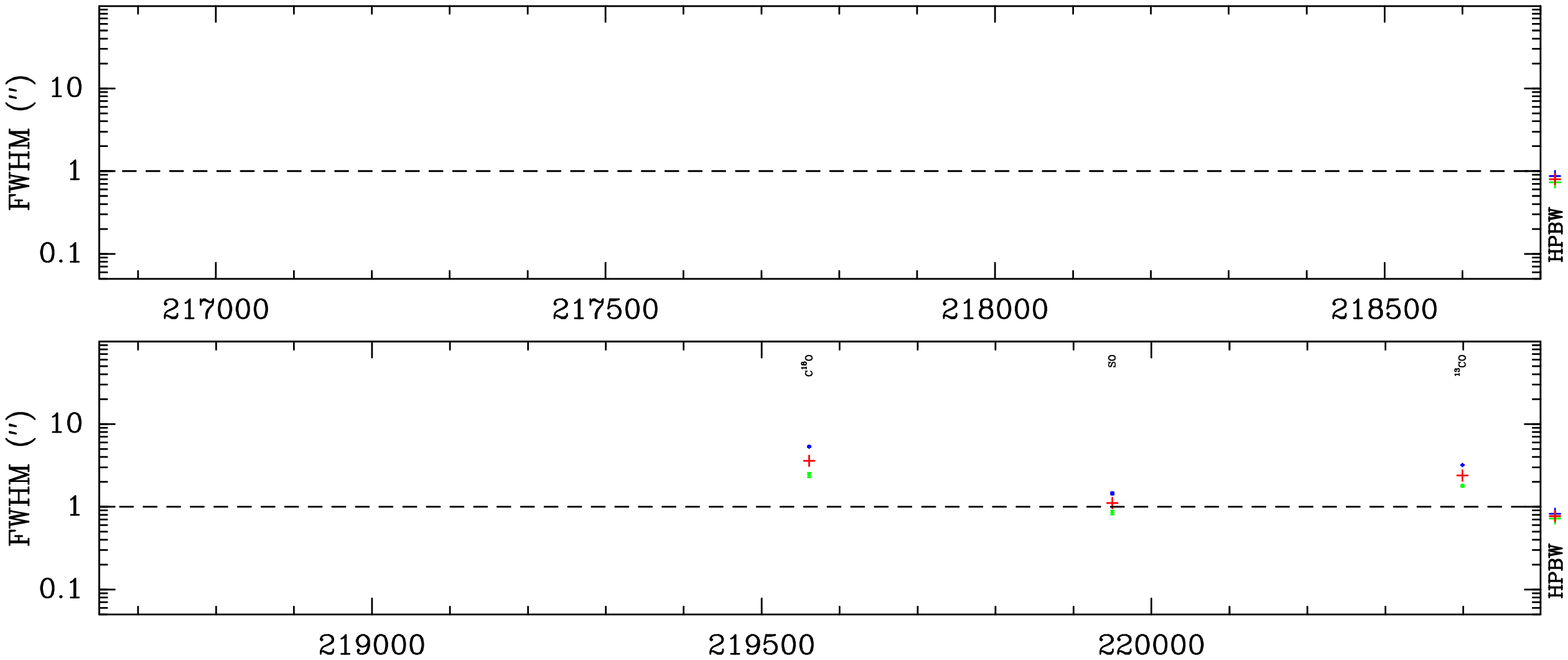}}}
\vspace*{1.5ex}
\centerline{\resizebox{0.95\hsize}{!}{\includegraphics[angle=270]{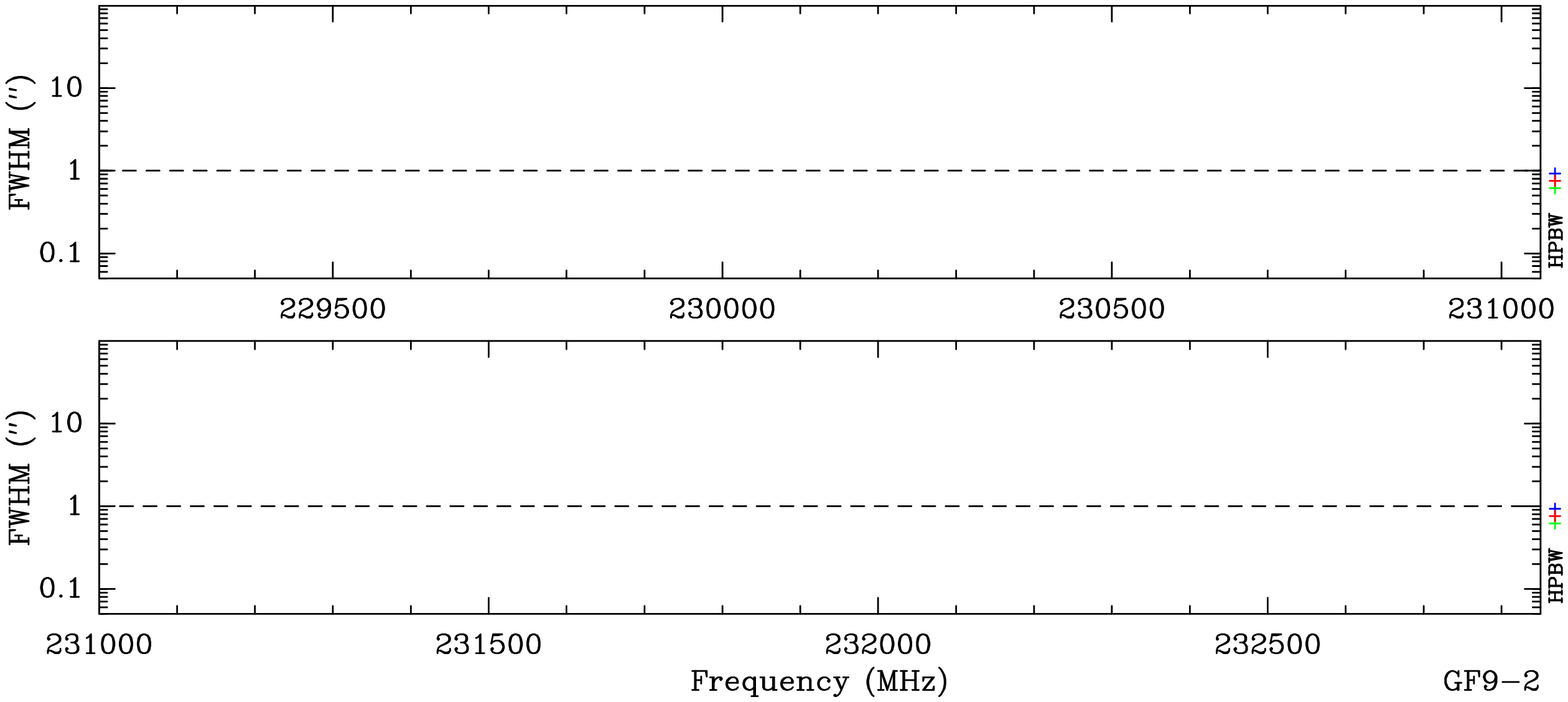}}}
\caption{Same as Fig.~\ref{f:sizes_l1448-2a} for GF9-2.}
\label{f:sizes_gf9-2}
\end{figure*}

Figs.~\ref{f:sizes_l1448-2a}--\ref{f:sizes_gf9-2} show the deconvolved sizes 
of the line emission as a function of frequency, as measured from Gaussian 
fits performed in the $uv$ plane.

\clearpage

\clearpage

\clearpage

\clearpage

\clearpage

\clearpage

\clearpage

\clearpage

\clearpage

\clearpage

\clearpage

\clearpage

\clearpage

\clearpage

\clearpage
\newpage

\section{Rotational temperatures}
\label{a:trot}

\begin{figure}[!htbp]
\centerline{\resizebox{0.83\hsize}{!}{\includegraphics[angle=0]{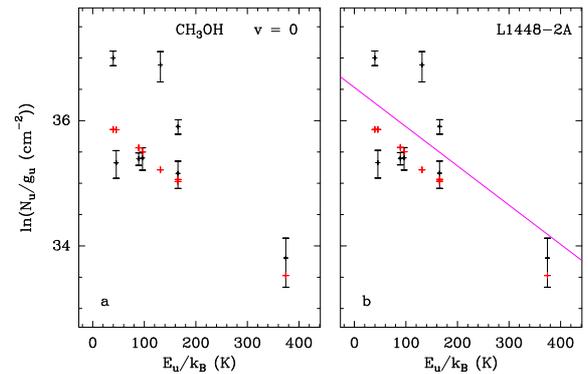}}}
\caption{Population diagram of CH$_3$OH in L1448-2A. The observed data points are shown in various colors (but not red) as indicated in the upper right corner of panel \textbf{a)} while the synthetic populations are shown in red. No correction is applied in panel \textbf{a)}. In panel \textbf{b)}, the optical depth correction has been applied to both the observed and synthetic populations and the contamination by all other species included in the full model has been removed from the observed data points. The purple line is a linear fit to the observed populations (in linear-logarithmic space).}
\label{f:popdiag_l1448-2a_p1_ch3oh}
\end{figure}

\begin{figure}[!htbp]
\centerline{\resizebox{0.83\hsize}{!}{\includegraphics[angle=0]{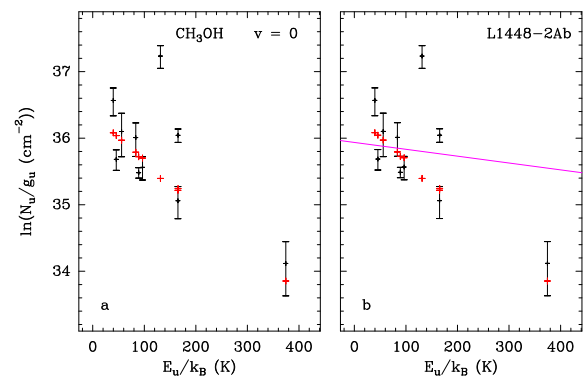}}}
\caption{Same as Fig.~\ref{f:popdiag_l1448-2a_p1_ch3oh} for CH$_3$OH in L1448-2Ab.}
\label{f:popdiag_l1448-2a_p2_ch3oh}
\end{figure}

\begin{figure}[!htbp]
\centerline{\resizebox{0.83\hsize}{!}{\includegraphics[angle=0]{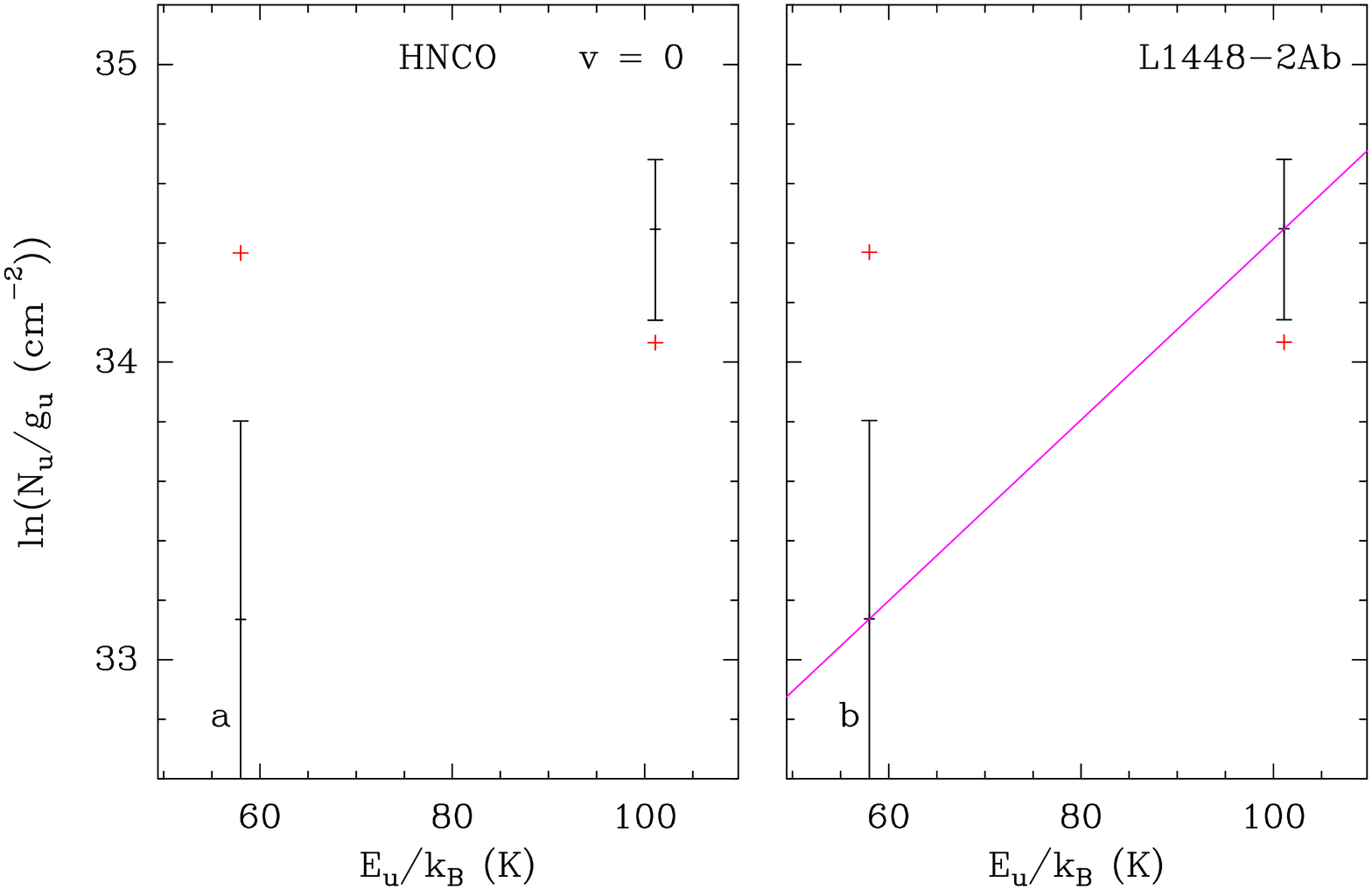}}}
\caption{Same as Fig.~\ref{f:popdiag_l1448-2a_p1_ch3oh} for HNCO in L1448-2Ab.}
\label{f:popdiag_l1448-2a_p2_hnco}
\end{figure}

\begin{figure}[!htbp]
\centerline{\resizebox{0.83\hsize}{!}{\includegraphics[angle=0]{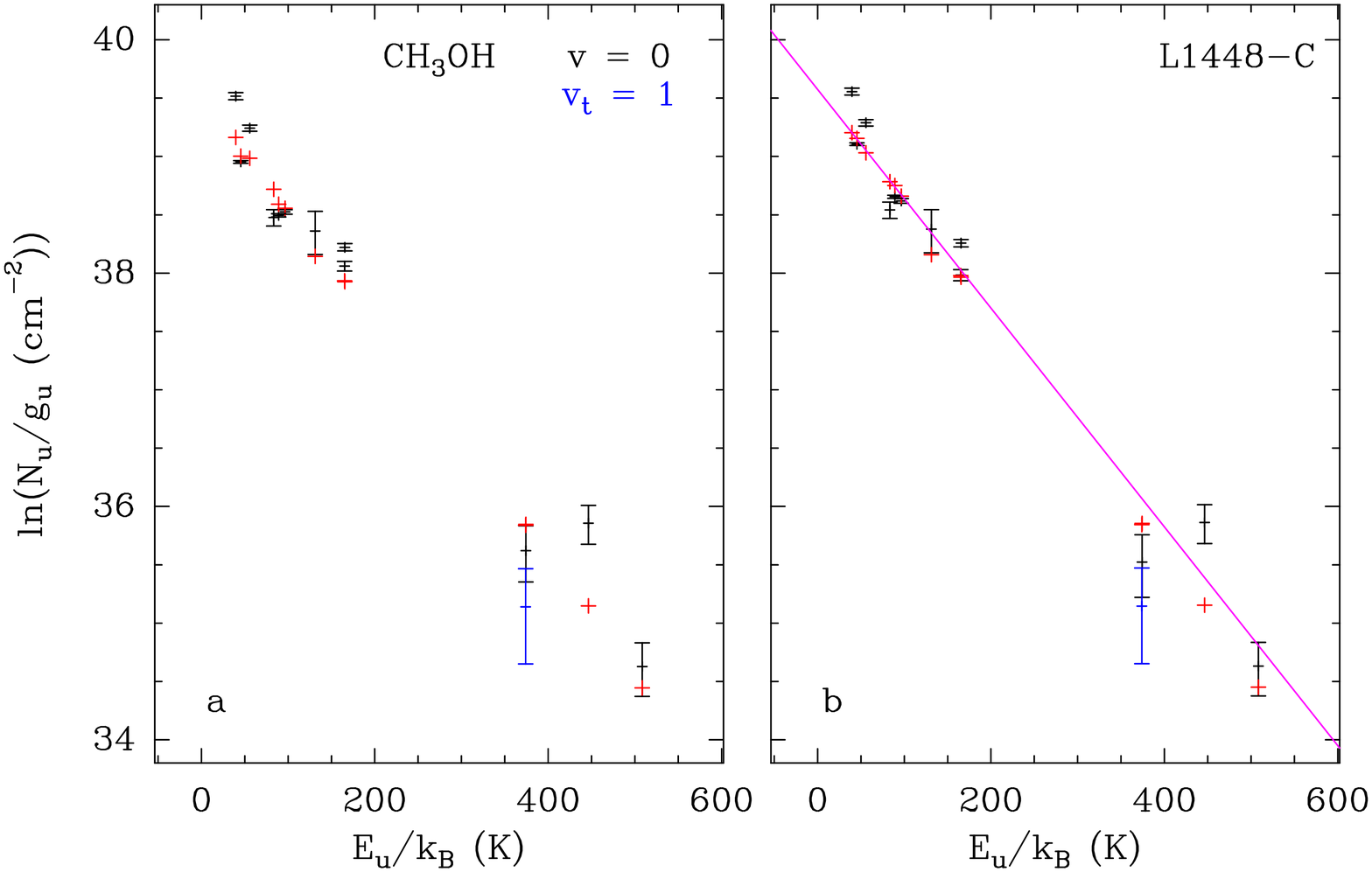}}}
\caption{Same as Fig.~\ref{f:popdiag_l1448-2a_p1_ch3oh} for CH$_3$OH in L1448-C.}
\label{f:popdiag_l1448-c_p1_ch3oh}
\end{figure}

\begin{figure}[!htbp]
\centerline{\resizebox{0.83\hsize}{!}{\includegraphics[angle=0]{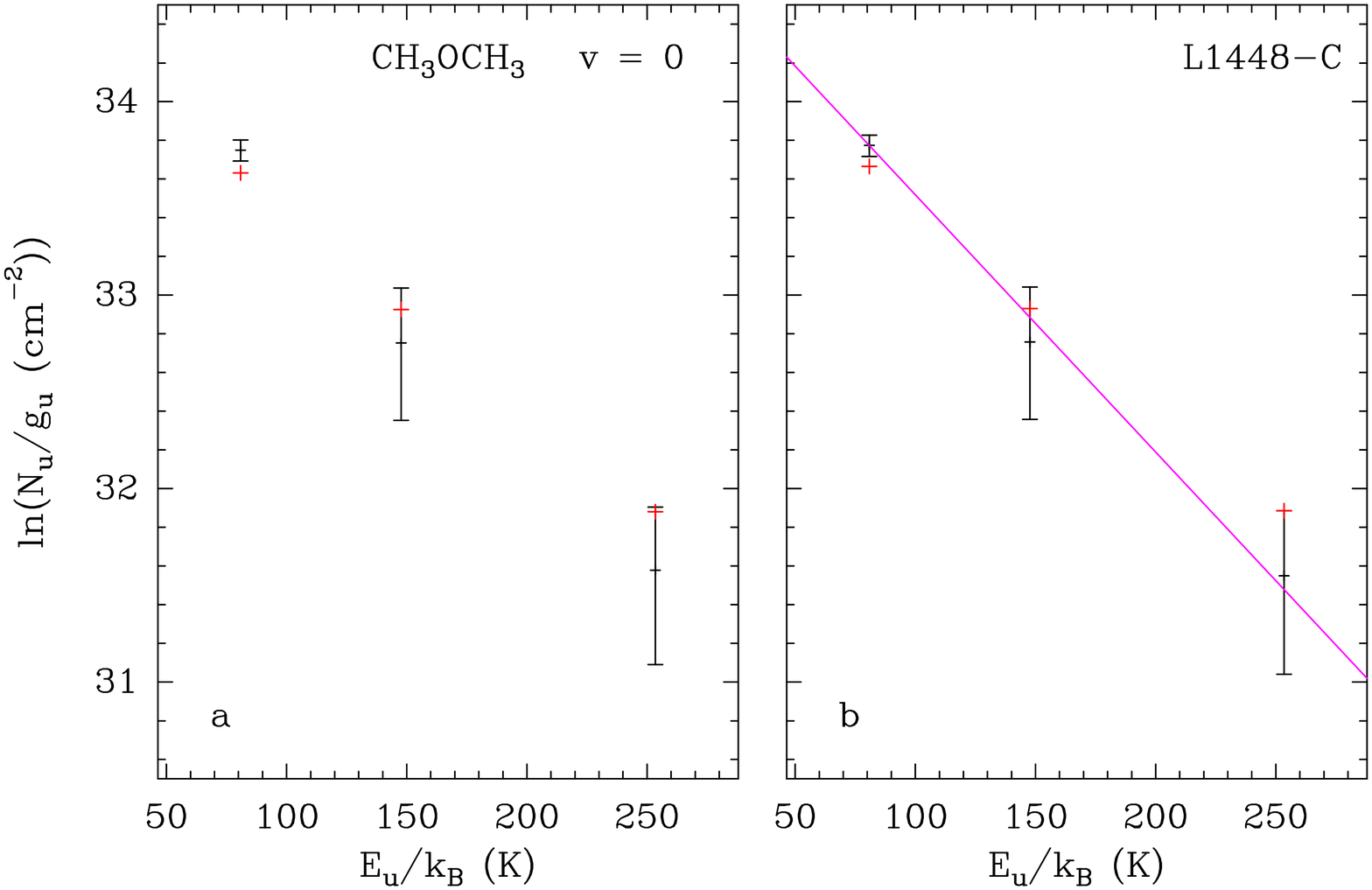}}}
\caption{Same as Fig.~\ref{f:popdiag_l1448-2a_p1_ch3oh} for CH$_3$OCH$_3$ in L1448-C.}
\label{f:popdiag_l1448-c_p1_ch3och3}
\end{figure}

\begin{figure}[!htbp]
\centerline{\resizebox{0.83\hsize}{!}{\includegraphics[angle=0]{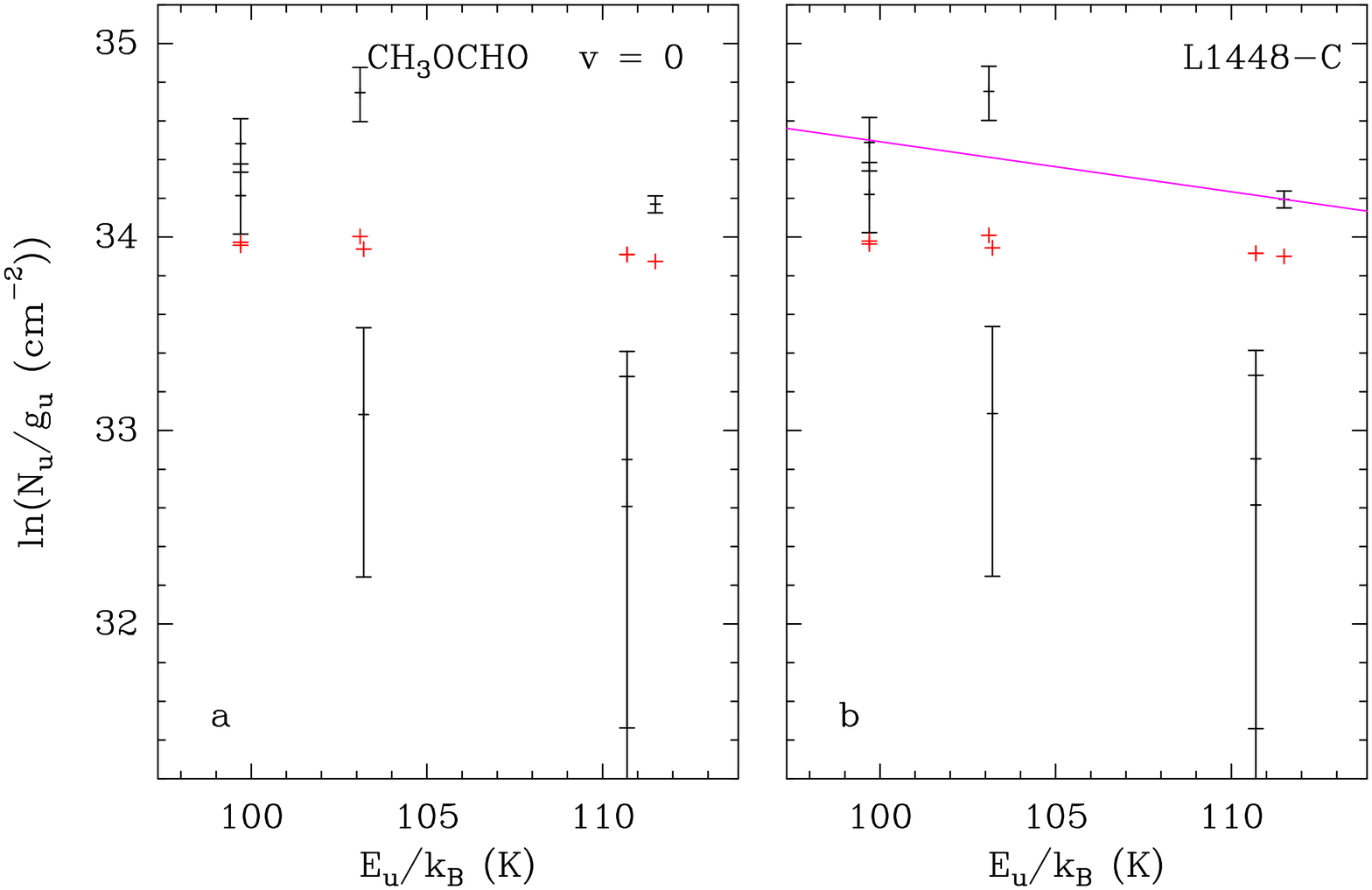}}}
\caption{Same as Fig.~\ref{f:popdiag_l1448-2a_p1_ch3oh} for CH$_3$OCHO in L1448-C.}
\label{f:popdiag_l1448-c_p1_ch3ocho}
\end{figure}

\begin{figure}[!htbp]
\centerline{\resizebox{0.83\hsize}{!}{\includegraphics[angle=0]{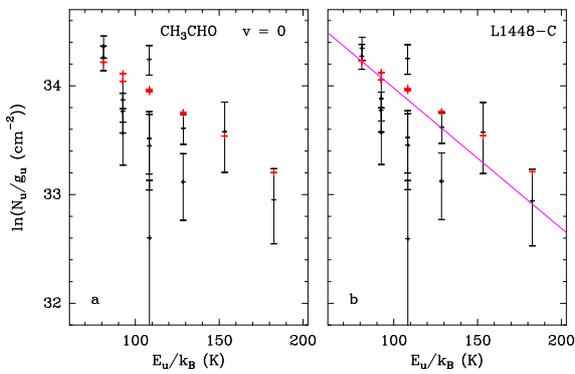}}}
\caption{Same as Fig.~\ref{f:popdiag_l1448-2a_p1_ch3oh} for CH$_3$CHO in L1448-C.}
\label{f:popdiag_l1448-c_p1_ch3cho}
\end{figure}

\begin{figure}[!htbp]
\centerline{\resizebox{0.83\hsize}{!}{\includegraphics[angle=0]{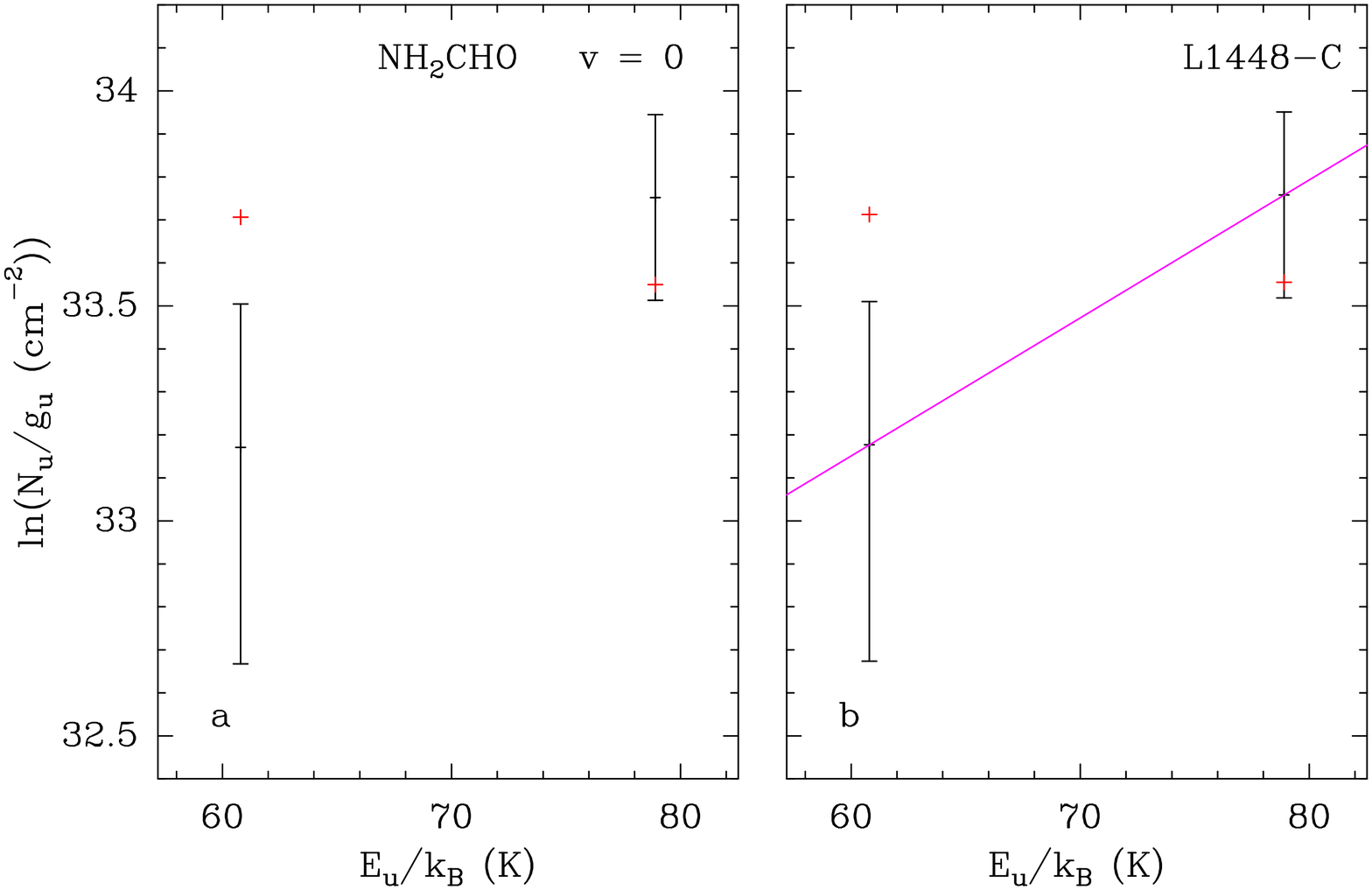}}}
\caption{Same as Fig.~\ref{f:popdiag_l1448-2a_p1_ch3oh} for NH$_2$CHO in L1448-C.}
\label{f:popdiag_l1448-c_p1_nh2cho}
\end{figure}

\begin{figure}[!htbp]
\centerline{\resizebox{0.83\hsize}{!}{\includegraphics[angle=0]{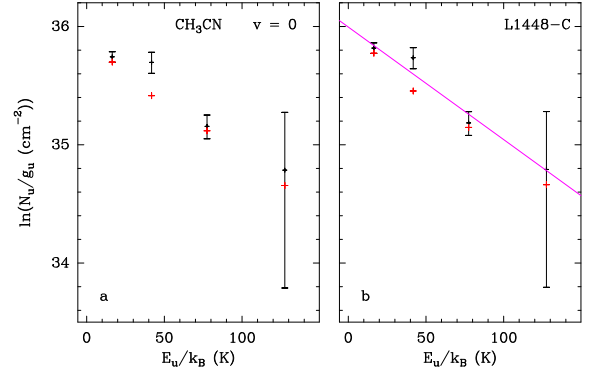}}}
\caption{Same as Fig.~\ref{f:popdiag_l1448-2a_p1_ch3oh} for CH$_3$CN in L1448-C.}
\label{f:popdiag_l1448-c_p1_ch3cn}
\end{figure}

\begin{figure}[!htbp]
\centerline{\resizebox{0.83\hsize}{!}{\includegraphics[angle=0]{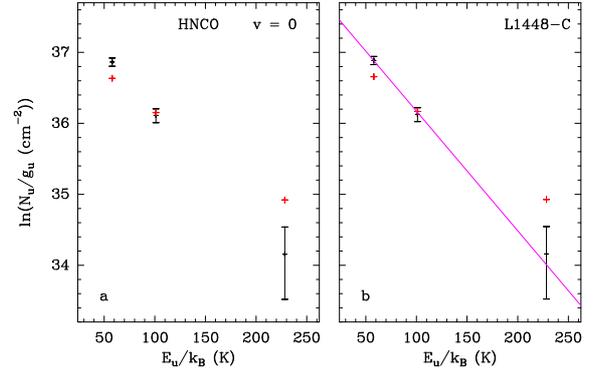}}}
\caption{Same as Fig.~\ref{f:popdiag_l1448-2a_p1_ch3oh} for HNCO in L1448-C.}
\label{f:popdiag_l1448-c_p1_hnco}
\end{figure}

\begin{figure}[!htbp]
\centerline{\resizebox{0.83\hsize}{!}{\includegraphics[angle=0]{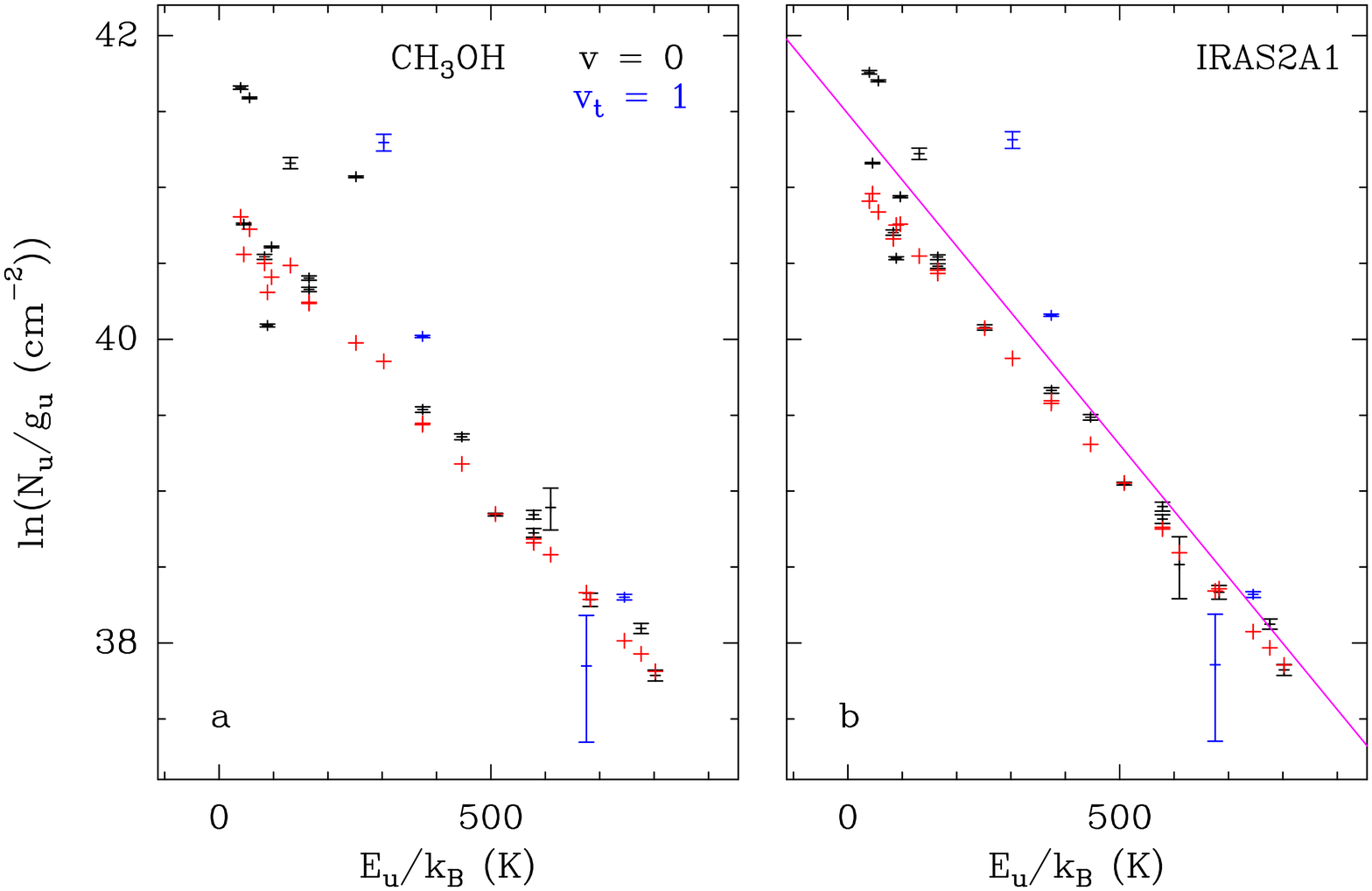}}}
\caption{Same as Fig.~\ref{f:popdiag_l1448-2a_p1_ch3oh} for CH$_3$OH in IRAS2A1.}
\label{f:popdiag_n1333-irs2a_p1_ch3oh}
\end{figure}

\begin{figure}[!htbp]
\centerline{\resizebox{0.83\hsize}{!}{\includegraphics[angle=0]{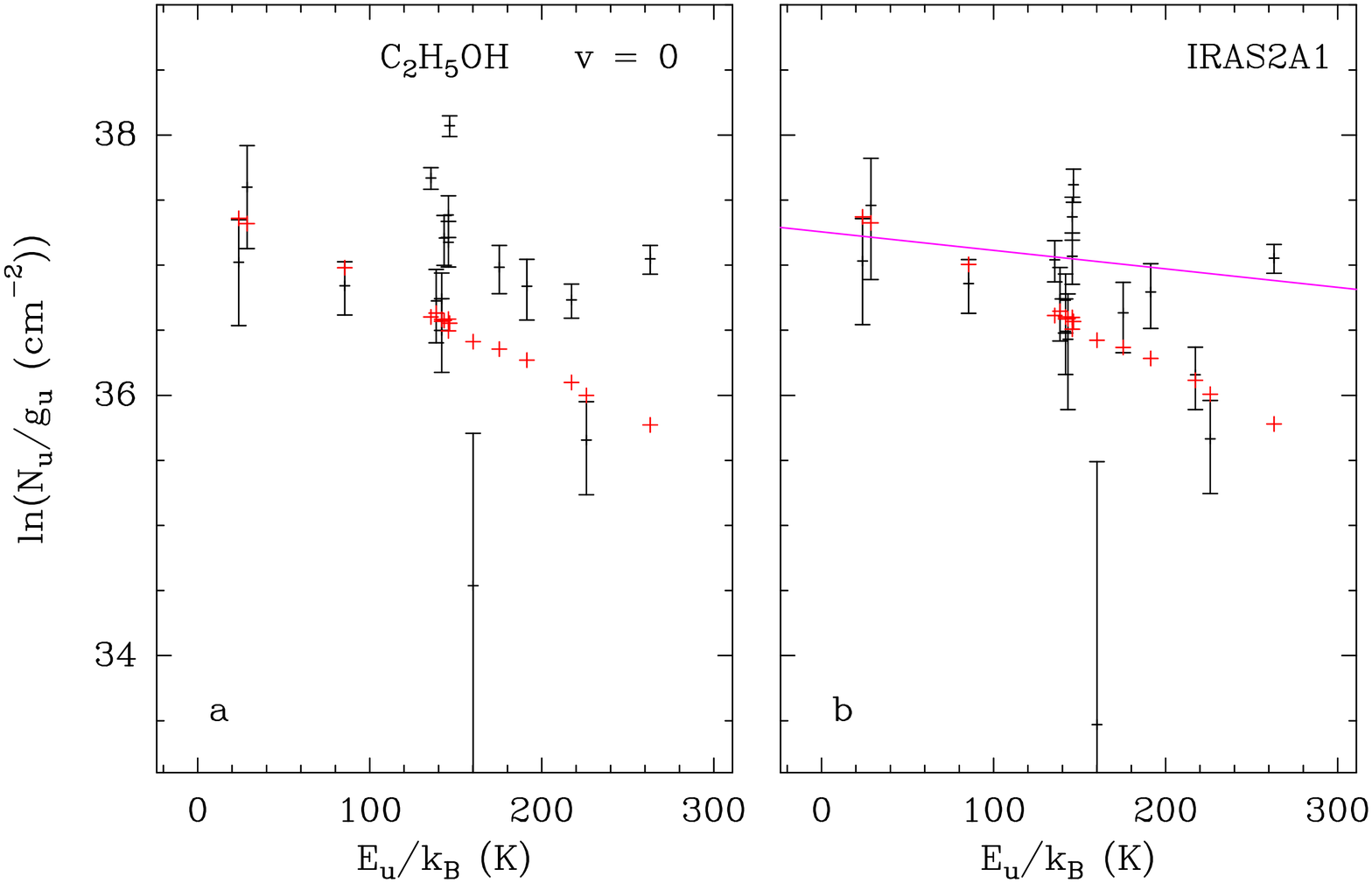}}}
\caption{Same as Fig.~\ref{f:popdiag_l1448-2a_p1_ch3oh} for C$_2$H$_5$OH in IRAS2A1.}
\label{f:popdiag_n1333-irs2a_p1_c2h5oh}
\end{figure}

\begin{figure}[!htbp]
\centerline{\resizebox{0.83\hsize}{!}{\includegraphics[angle=0]{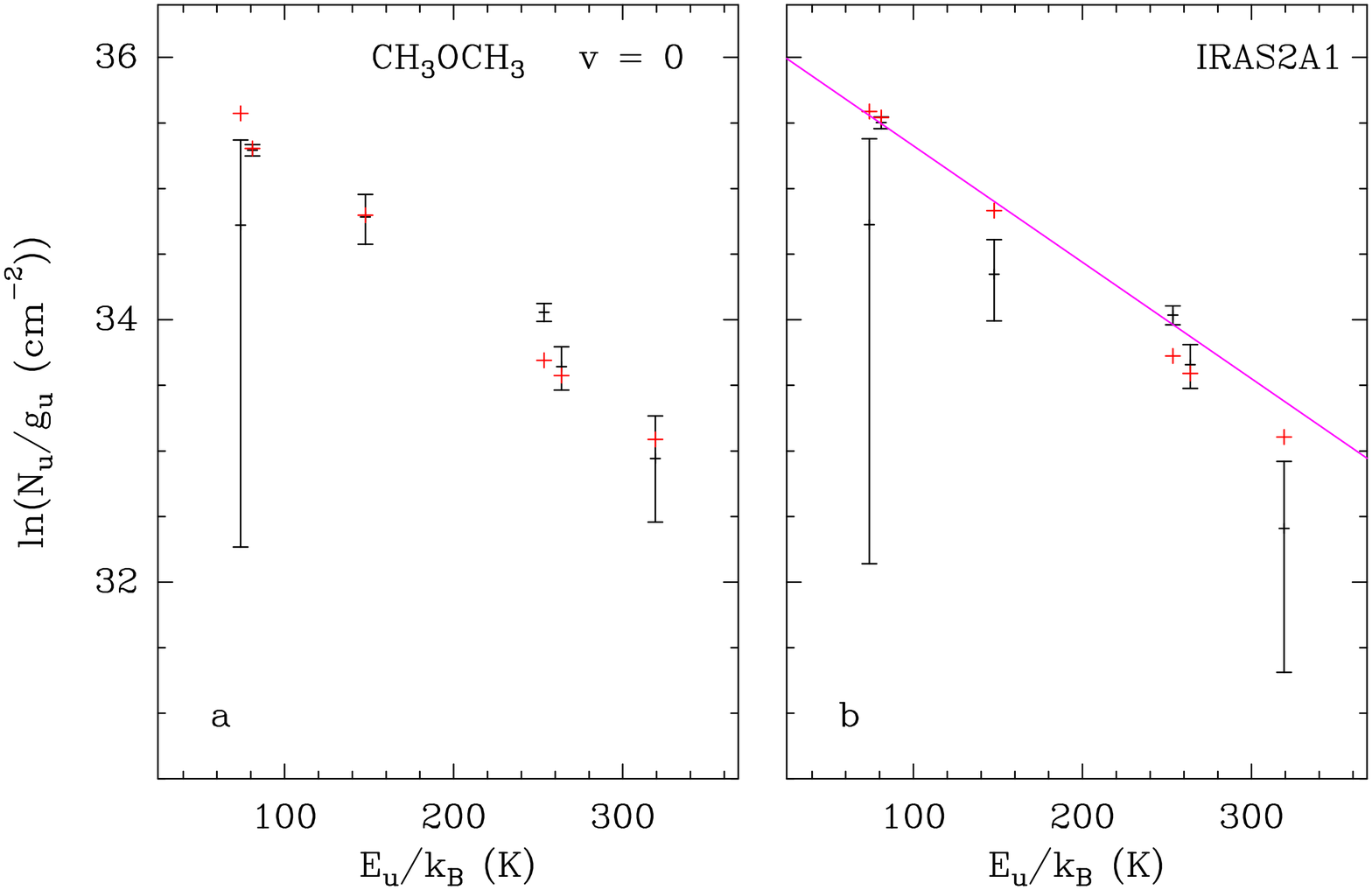}}}
\caption{Same as Fig.~\ref{f:popdiag_l1448-2a_p1_ch3oh} for CH$_3$OCH$_3$ in IRAS2A1.}
\label{f:popdiag_n1333-irs2a_p1_ch3och3}
\end{figure}

\begin{figure}[!htbp]
\centerline{\resizebox{0.83\hsize}{!}{\includegraphics[angle=0]{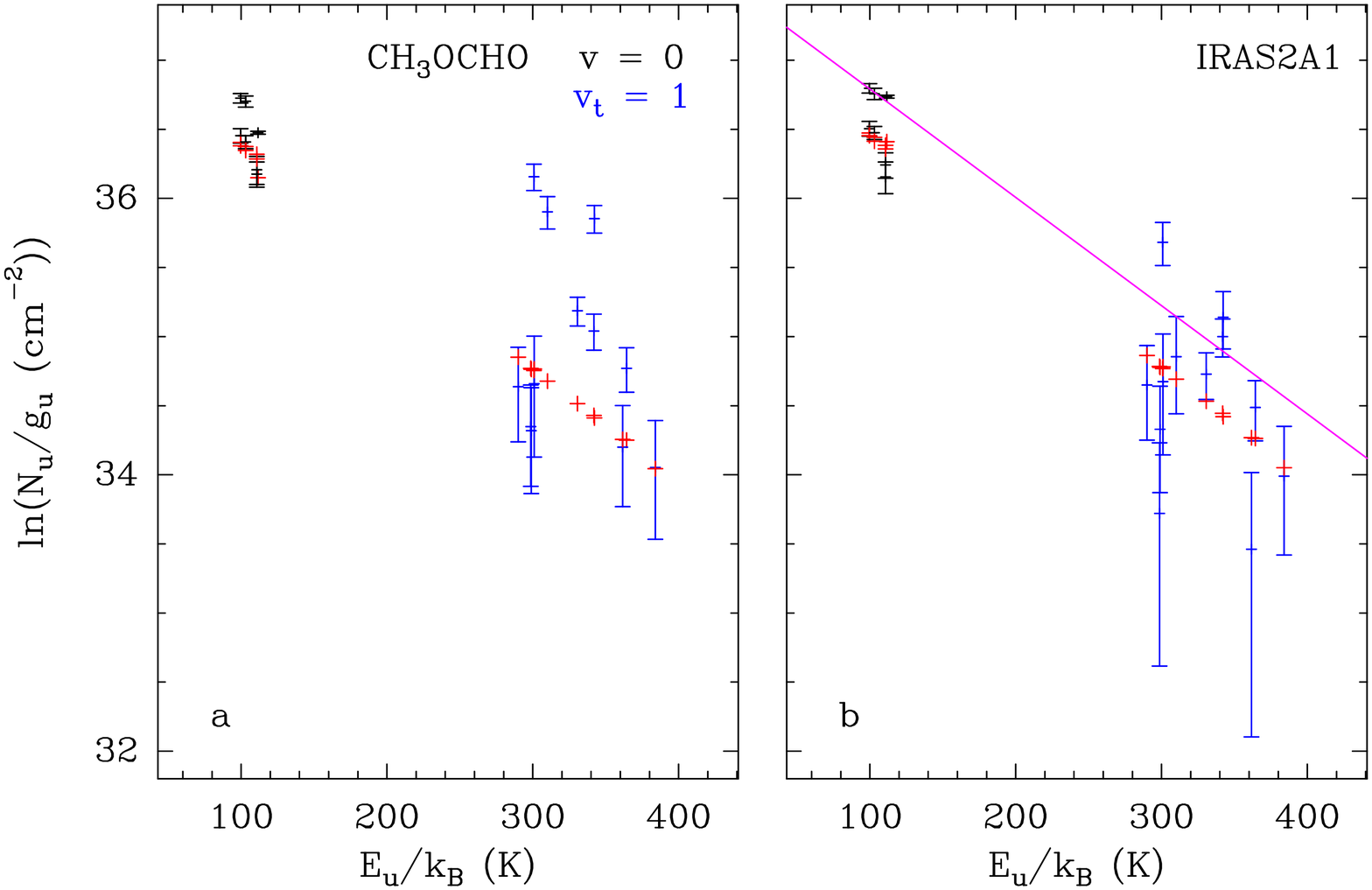}}}
\caption{Same as Fig.~\ref{f:popdiag_l1448-2a_p1_ch3oh} for CH$_3$OCHO in IRAS2A1.}
\label{f:popdiag_n1333-irs2a_p1_ch3ocho}
\end{figure}

\begin{figure}[!htbp]
\centerline{\resizebox{0.83\hsize}{!}{\includegraphics[angle=0]{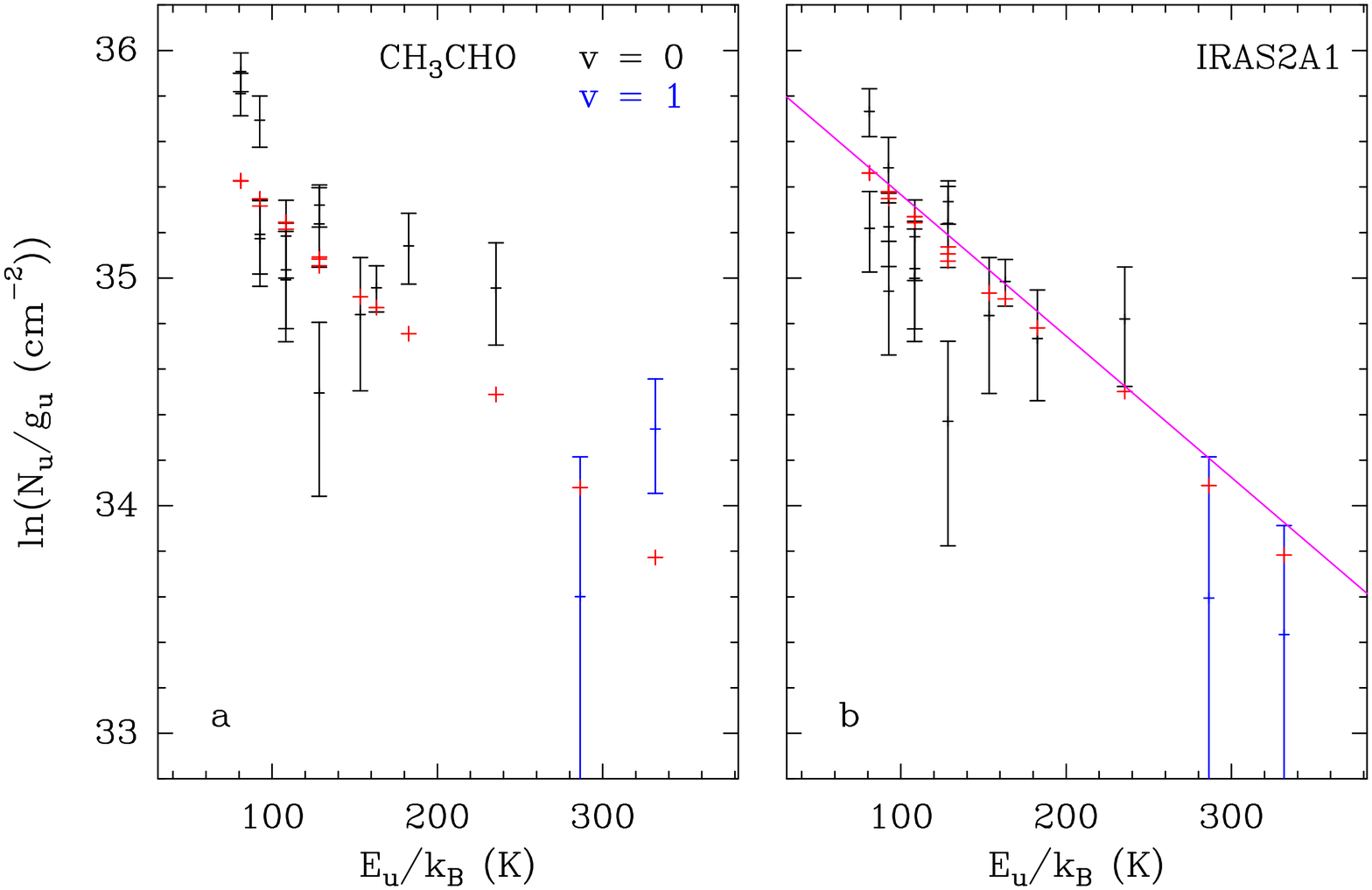}}}
\caption{Same as Fig.~\ref{f:popdiag_l1448-2a_p1_ch3oh} for CH$_3$CHO in IRAS2A1.}
\label{f:popdiag_n1333-irs2a_p1_ch3cho}
\end{figure}

\begin{figure}[!htbp]
\centerline{\resizebox{0.83\hsize}{!}{\includegraphics[angle=0]{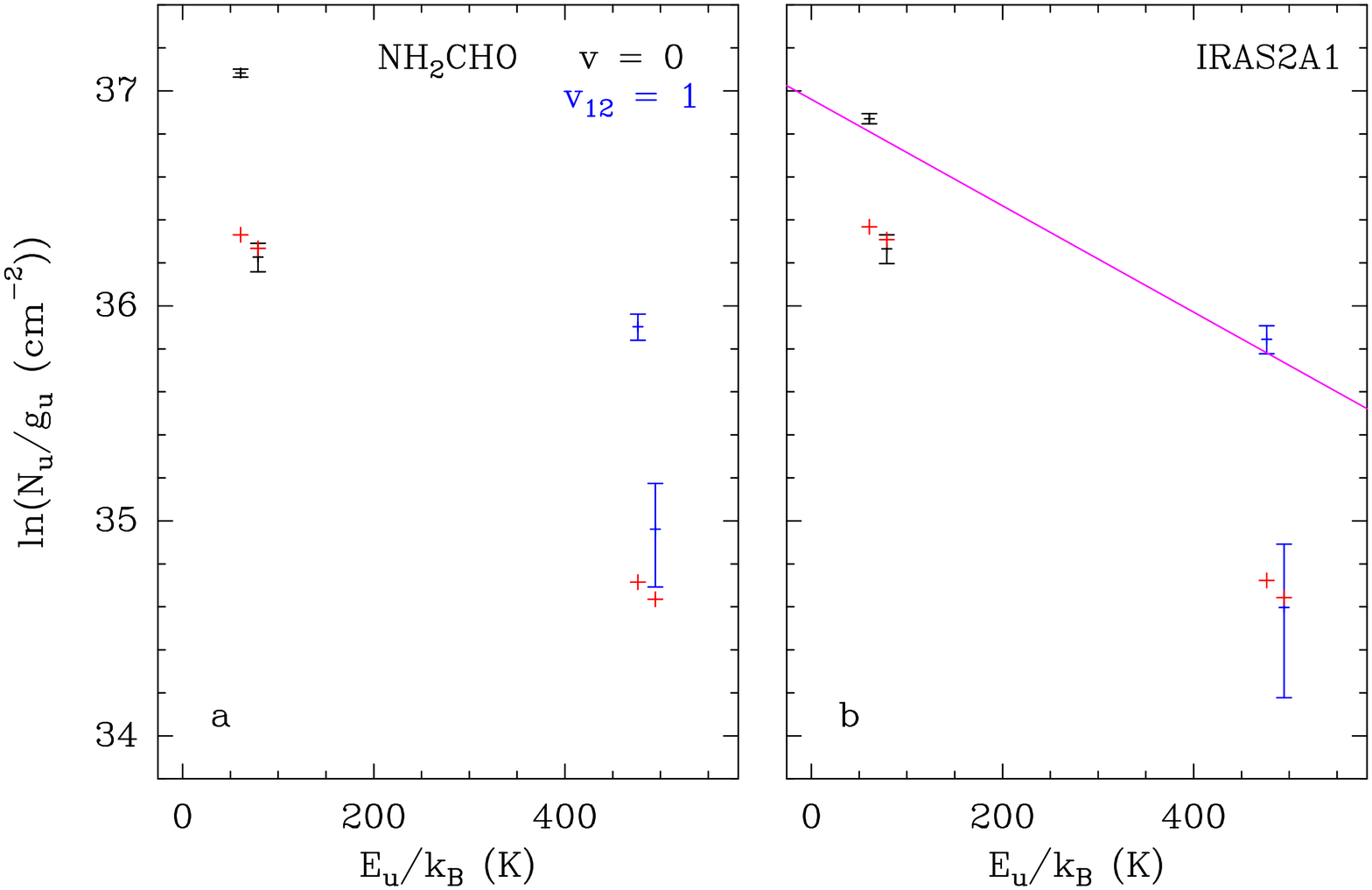}}}
\caption{Same as Fig.~\ref{f:popdiag_l1448-2a_p1_ch3oh} for NH$_2$CHO in IRAS2A1.}
\label{f:popdiag_n1333-irs2a_p1_nh2cho}
\end{figure}

\begin{figure}[!htbp]
\centerline{\resizebox{0.83\hsize}{!}{\includegraphics[angle=0]{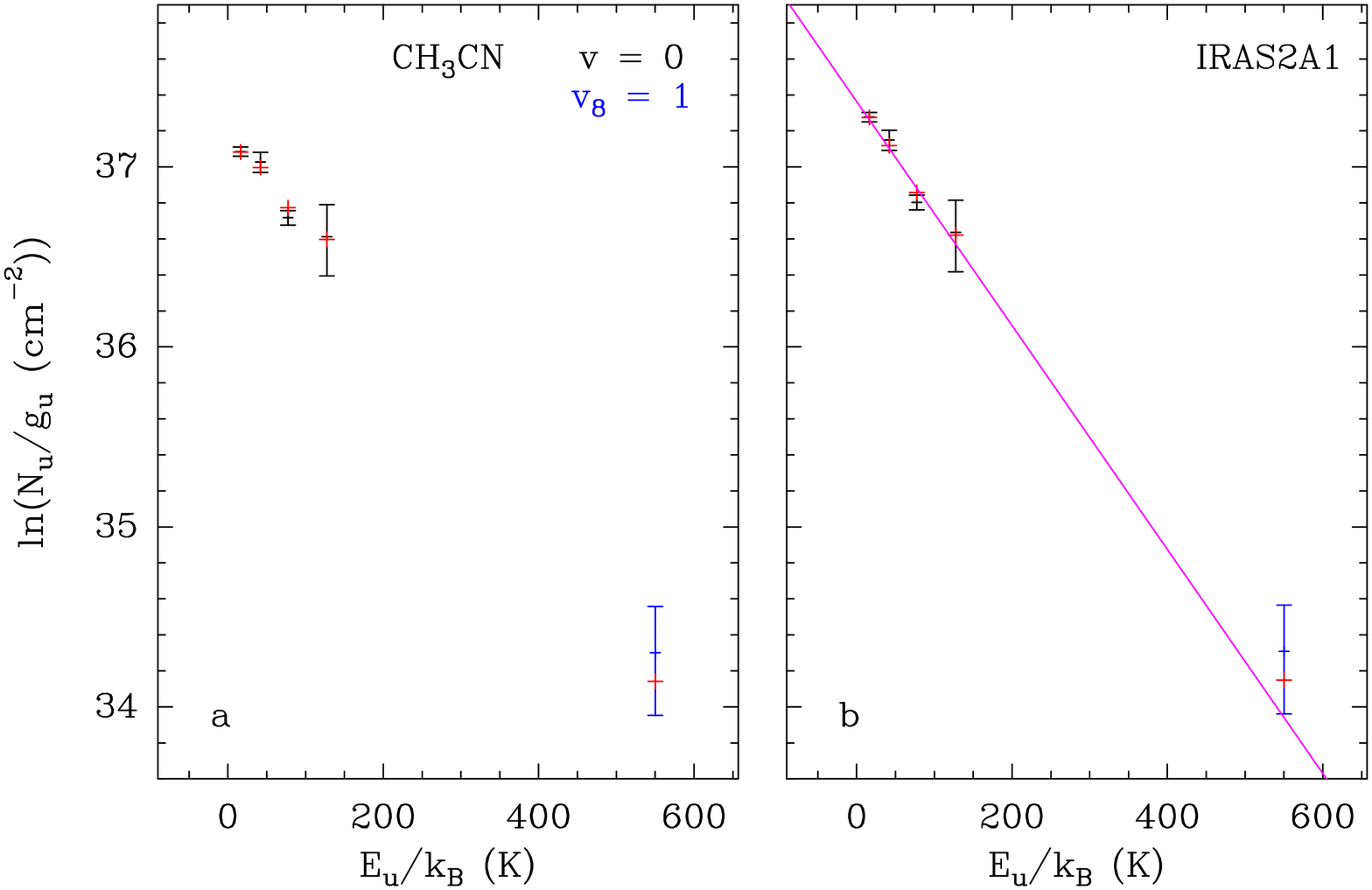}}}
\caption{Same as Fig.~\ref{f:popdiag_l1448-2a_p1_ch3oh} for CH$_3$CN in IRAS2A1.}
\label{f:popdiag_n1333-irs2a_p1_ch3cn}
\end{figure}

\begin{figure}[!htbp]
\centerline{\resizebox{0.83\hsize}{!}{\includegraphics[angle=0]{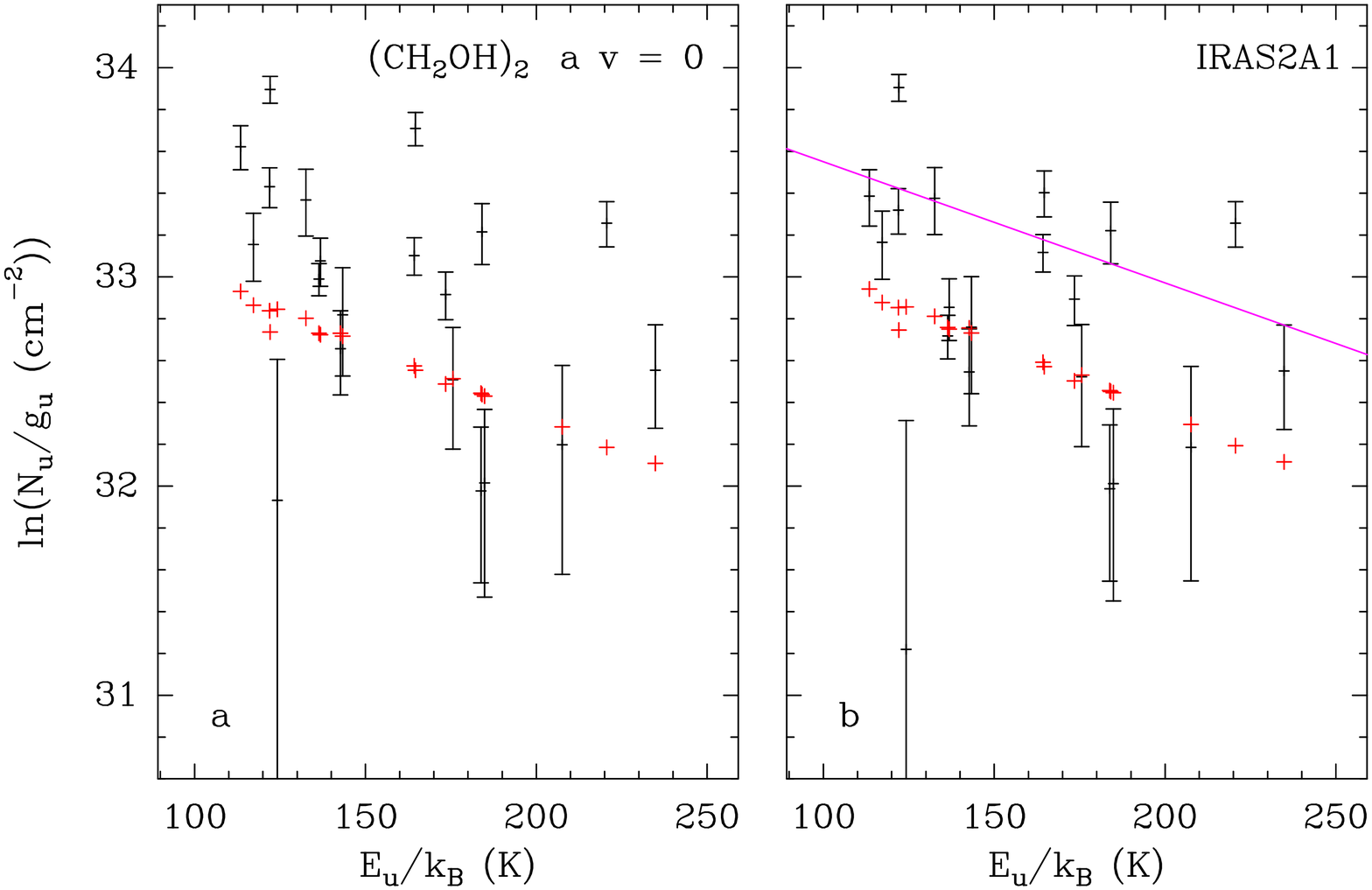}}}
\caption{Same as Fig.~\ref{f:popdiag_l1448-2a_p1_ch3oh} for (CH$_2$OH)$_2$ in IRAS2A1.}
\label{f:popdiag_n1333-irs2a_p1_ch2oh-2}
\end{figure}

\begin{figure}[!htbp]
\centerline{\resizebox{0.83\hsize}{!}{\includegraphics[angle=0]{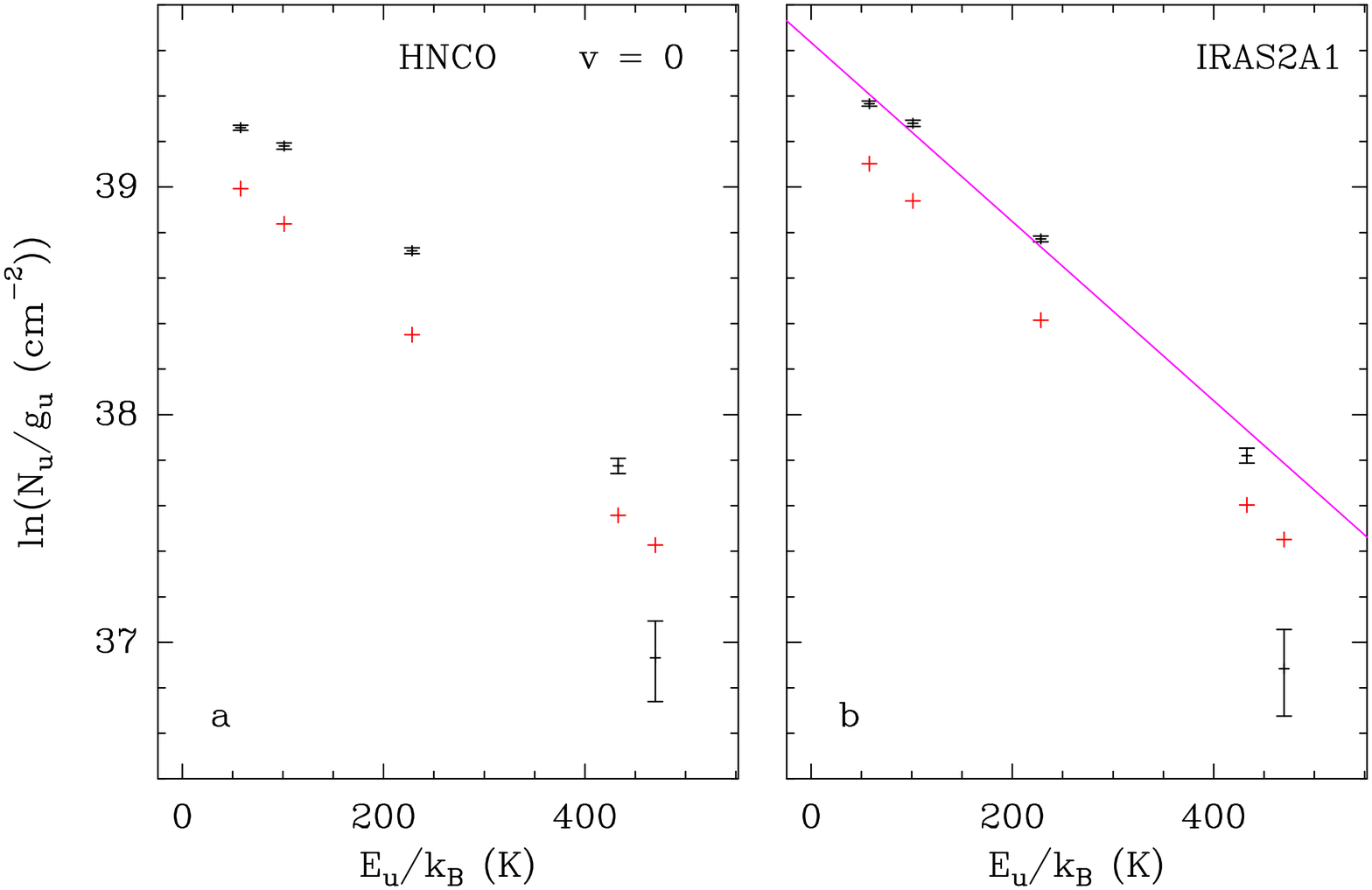}}}
\caption{Same as Fig.~\ref{f:popdiag_l1448-2a_p1_ch3oh} for HNCO in IRAS2A1.}
\label{f:popdiag_n1333-irs2a_p1_hnco}
\end{figure}

\begin{figure}[!htbp]
\centerline{\resizebox{0.83\hsize}{!}{\includegraphics[angle=0]{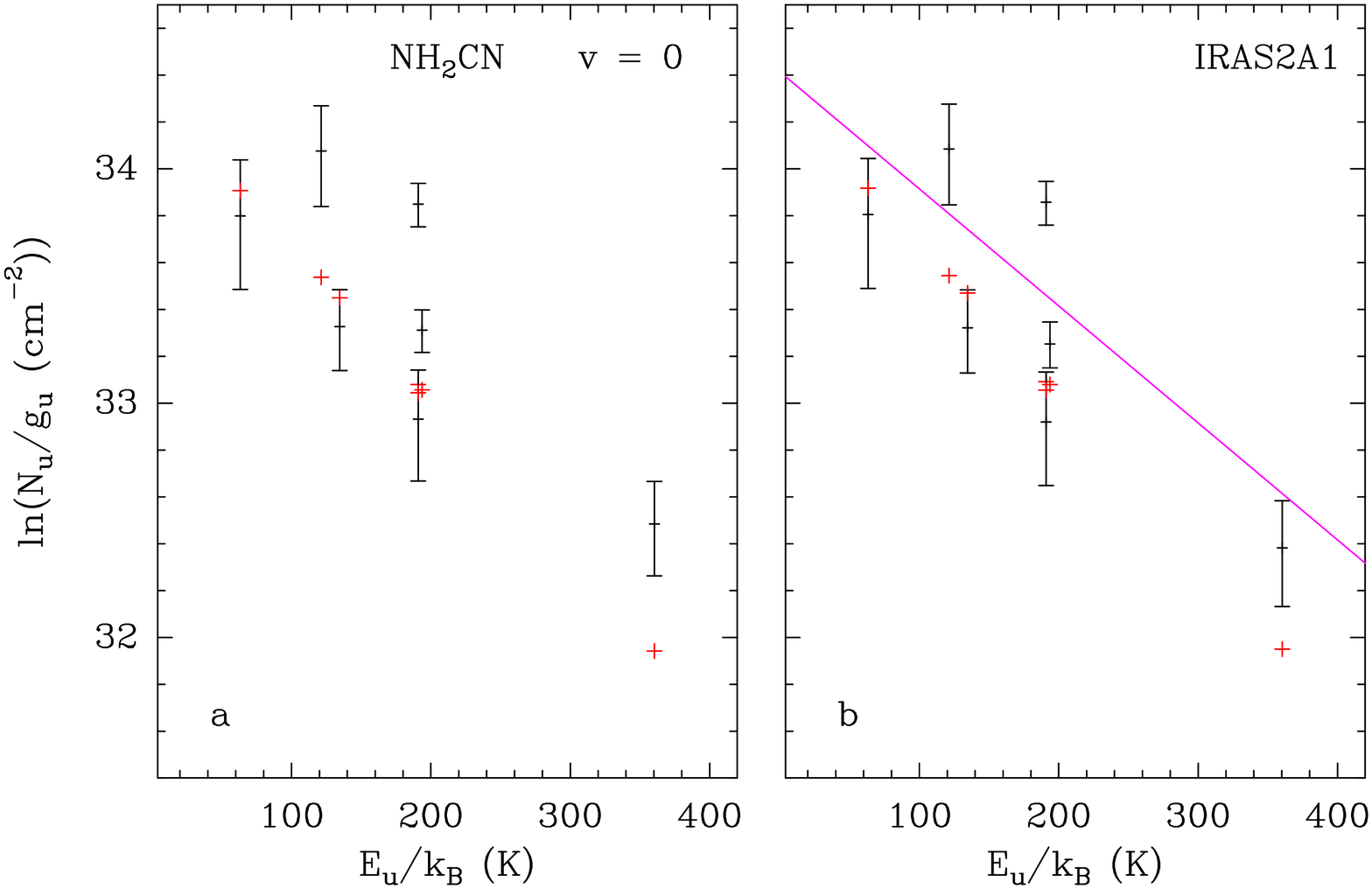}}}
\caption{Same as Fig.~\ref{f:popdiag_l1448-2a_p1_ch3oh} for NH$_2$CN in IRAS2A1.}
\label{f:popdiag_n1333-irs2a_p1_nh2cn}
\end{figure}

\begin{figure}[!htbp]
\centerline{\resizebox{0.83\hsize}{!}{\includegraphics[angle=0]{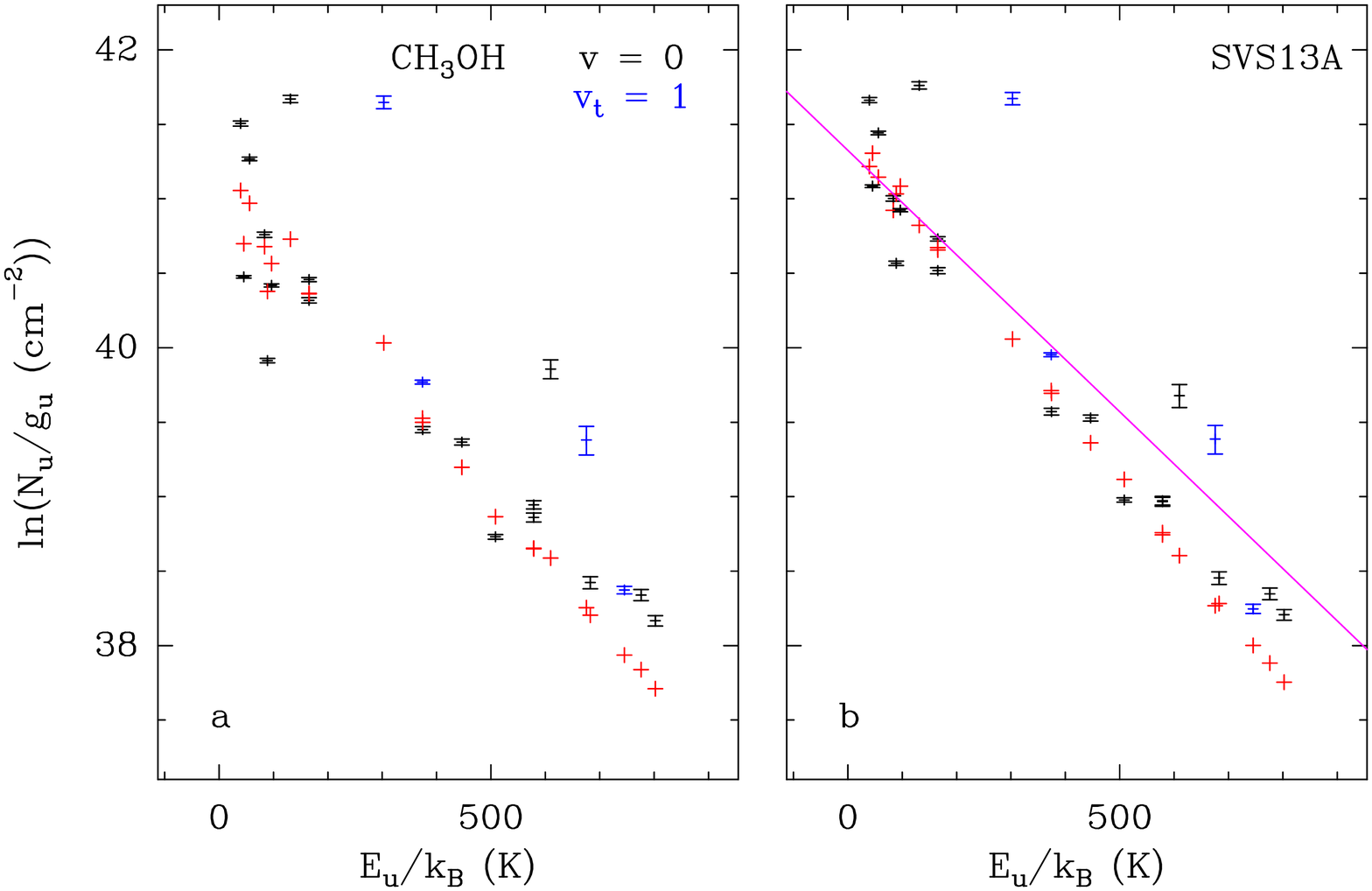}}}
\caption{Same as Fig.~\ref{f:popdiag_l1448-2a_p1_ch3oh} for CH$_3$OH in SVS13A.}
\label{f:popdiag_svs13-b_p1_ch3oh}
\end{figure}

\begin{figure}[!htbp]
\centerline{\resizebox{0.83\hsize}{!}{\includegraphics[angle=0]{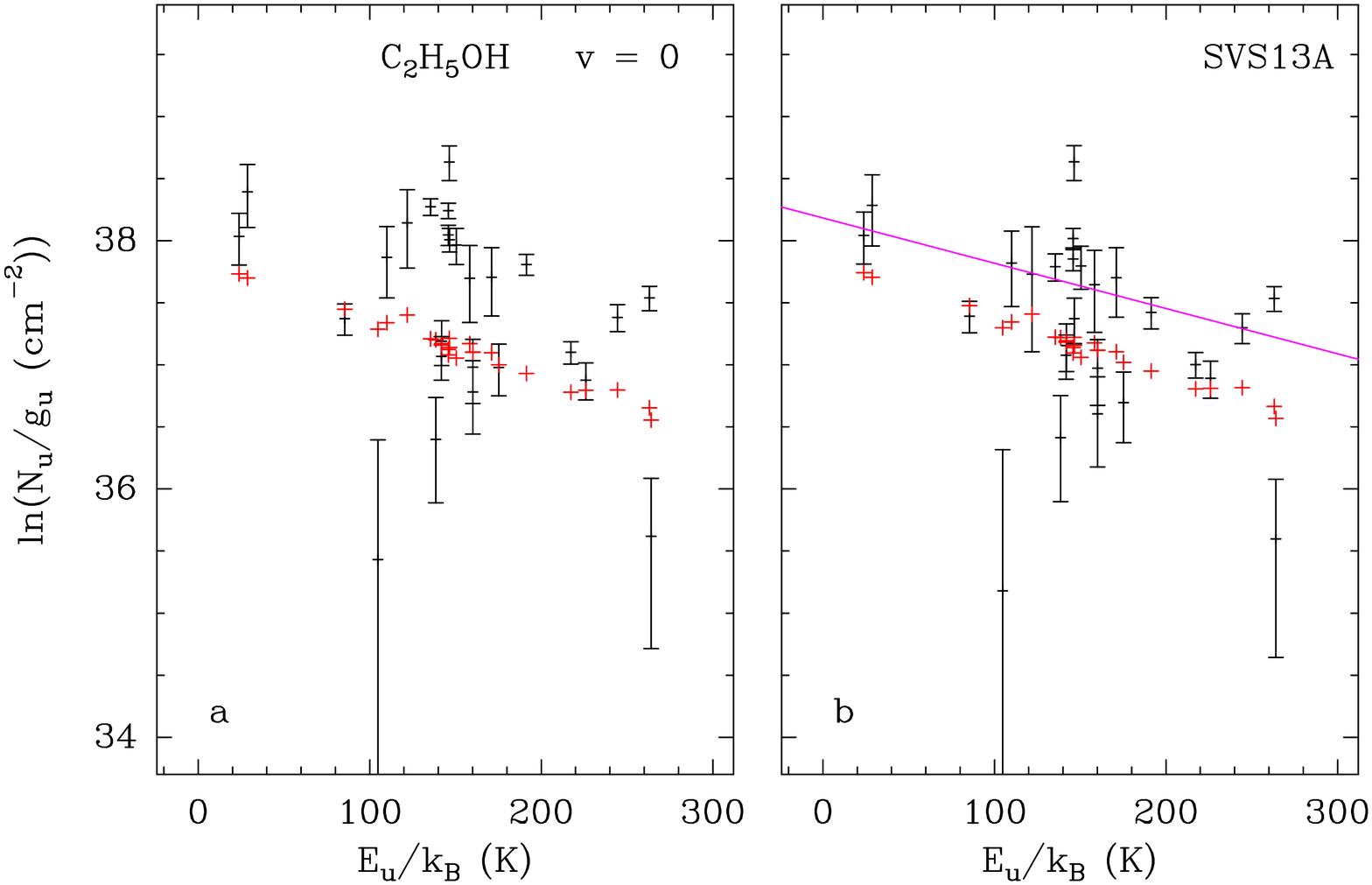}}}
\caption{Same as Fig.~\ref{f:popdiag_l1448-2a_p1_ch3oh} for C$_2$H$_5$OH in SVS13A.}
\label{f:popdiag_svs13-b_p1_c2h5oh}
\end{figure}

\begin{figure}[!htbp]
\centerline{\resizebox{0.83\hsize}{!}{\includegraphics[angle=0]{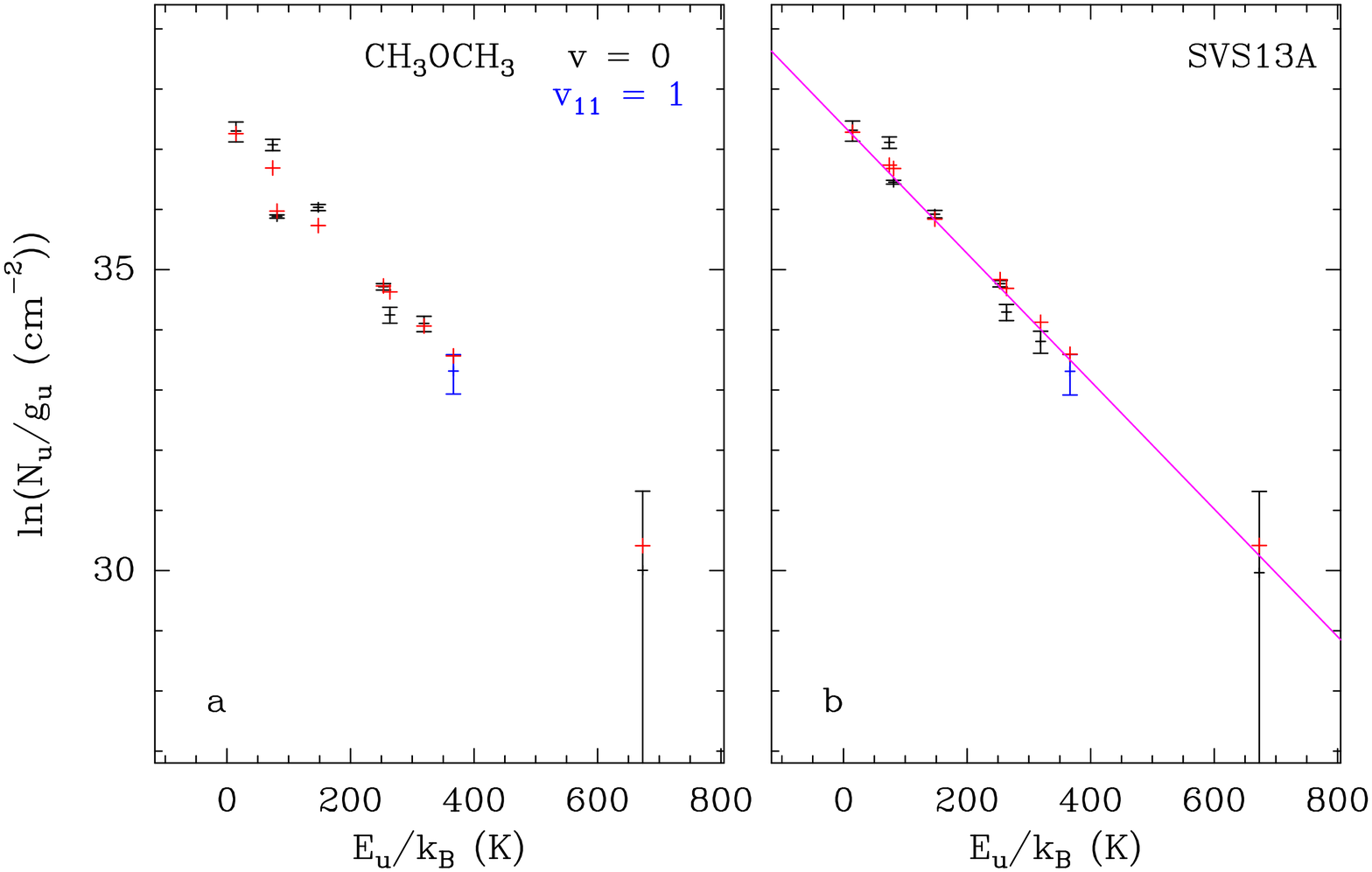}}}
\caption{Same as Fig.~\ref{f:popdiag_l1448-2a_p1_ch3oh} for CH$_3$OCH$_3$ in SVS13A.}
\label{f:popdiag_svs13-b_p1_ch3och3}
\end{figure}

\begin{figure}[!htbp]
\centerline{\resizebox{0.83\hsize}{!}{\includegraphics[angle=0]{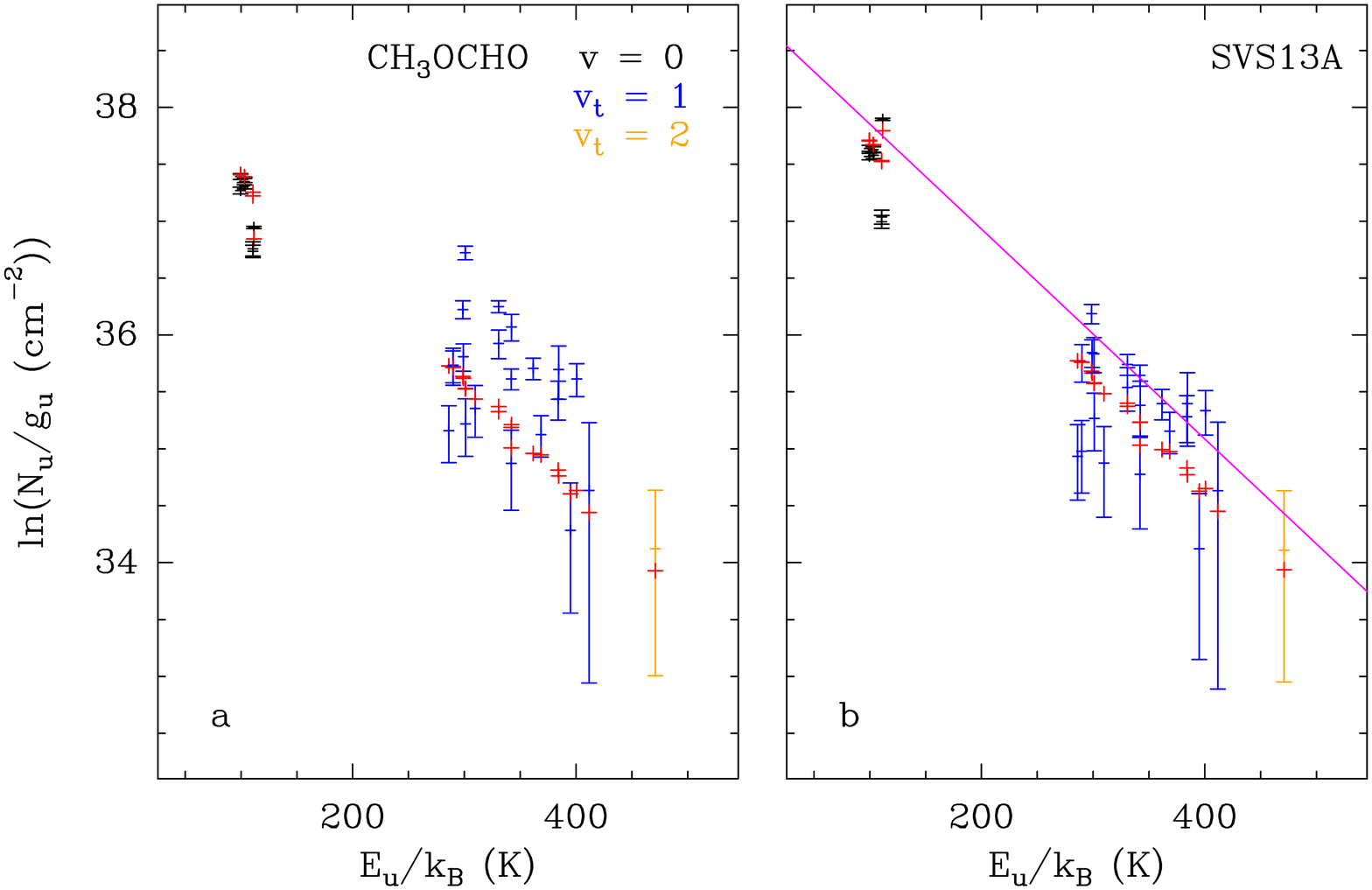}}}
\caption{Same as Fig.~\ref{f:popdiag_l1448-2a_p1_ch3oh} for CH$_3$OCHO in SVS13A.}
\label{f:popdiag_svs13-b_p1_ch3ocho}
\end{figure}

\begin{figure}[!htbp]
\centerline{\resizebox{0.83\hsize}{!}{\includegraphics[angle=0]{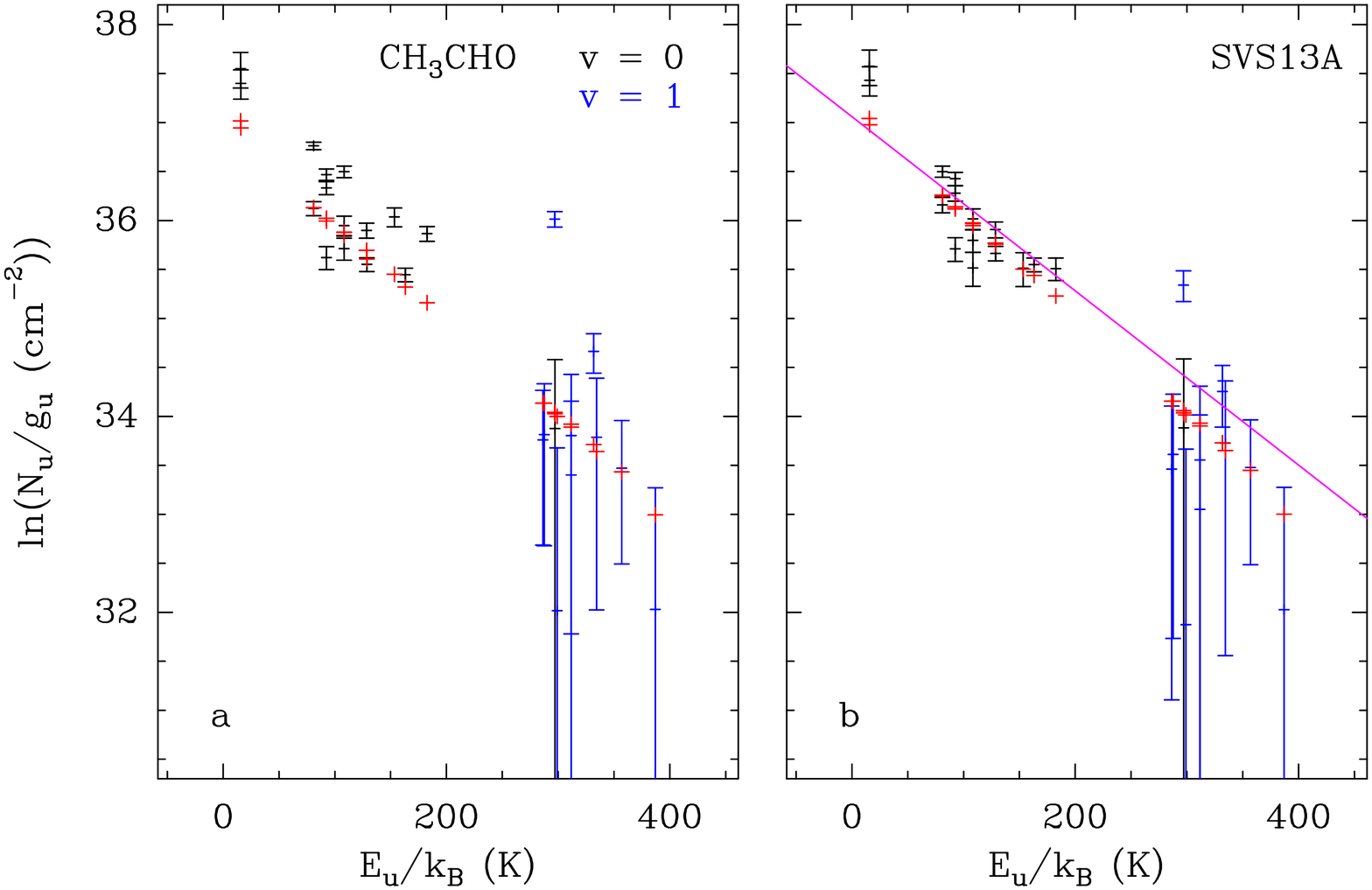}}}
\caption{Same as Fig.~\ref{f:popdiag_l1448-2a_p1_ch3oh} for CH$_3$CHO in SVS13A.}
\label{f:popdiag_svs13-b_p1_ch3cho}
\end{figure}

\begin{figure}[!htbp]
\centerline{\resizebox{0.83\hsize}{!}{\includegraphics[angle=0]{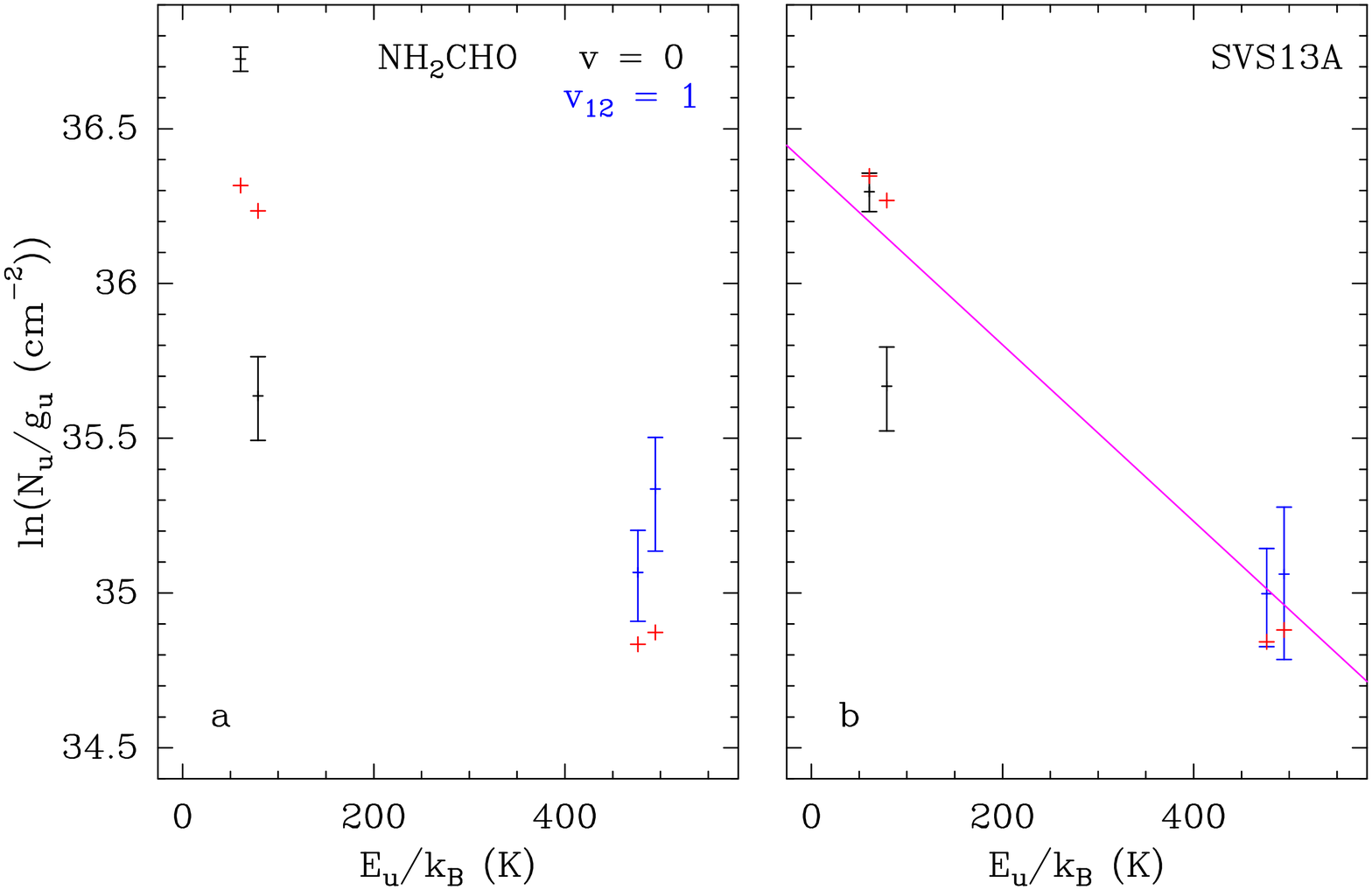}}}
\caption{Same as Fig.~\ref{f:popdiag_l1448-2a_p1_ch3oh} for NH$_2$CHO in SVS13A.}
\label{f:popdiag_svs13-b_p1_nh2cho}
\end{figure}

\begin{figure}[!htbp]
\centerline{\resizebox{0.83\hsize}{!}{\includegraphics[angle=0]{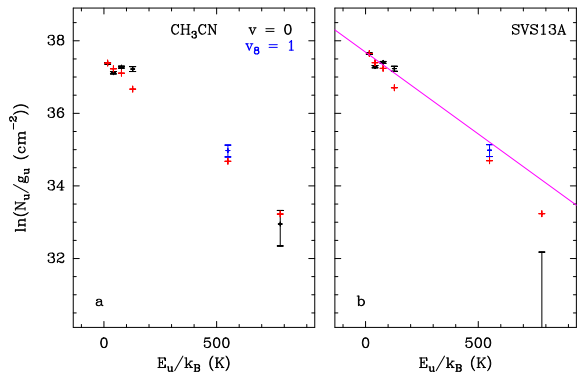}}}
\caption{Same as Fig.~\ref{f:popdiag_l1448-2a_p1_ch3oh} for CH$_3$CN in SVS13A.}
\label{f:popdiag_svs13-b_p1_ch3cn}
\end{figure}

\begin{figure}[!htbp]
\centerline{\resizebox{0.83\hsize}{!}{\includegraphics[angle=0]{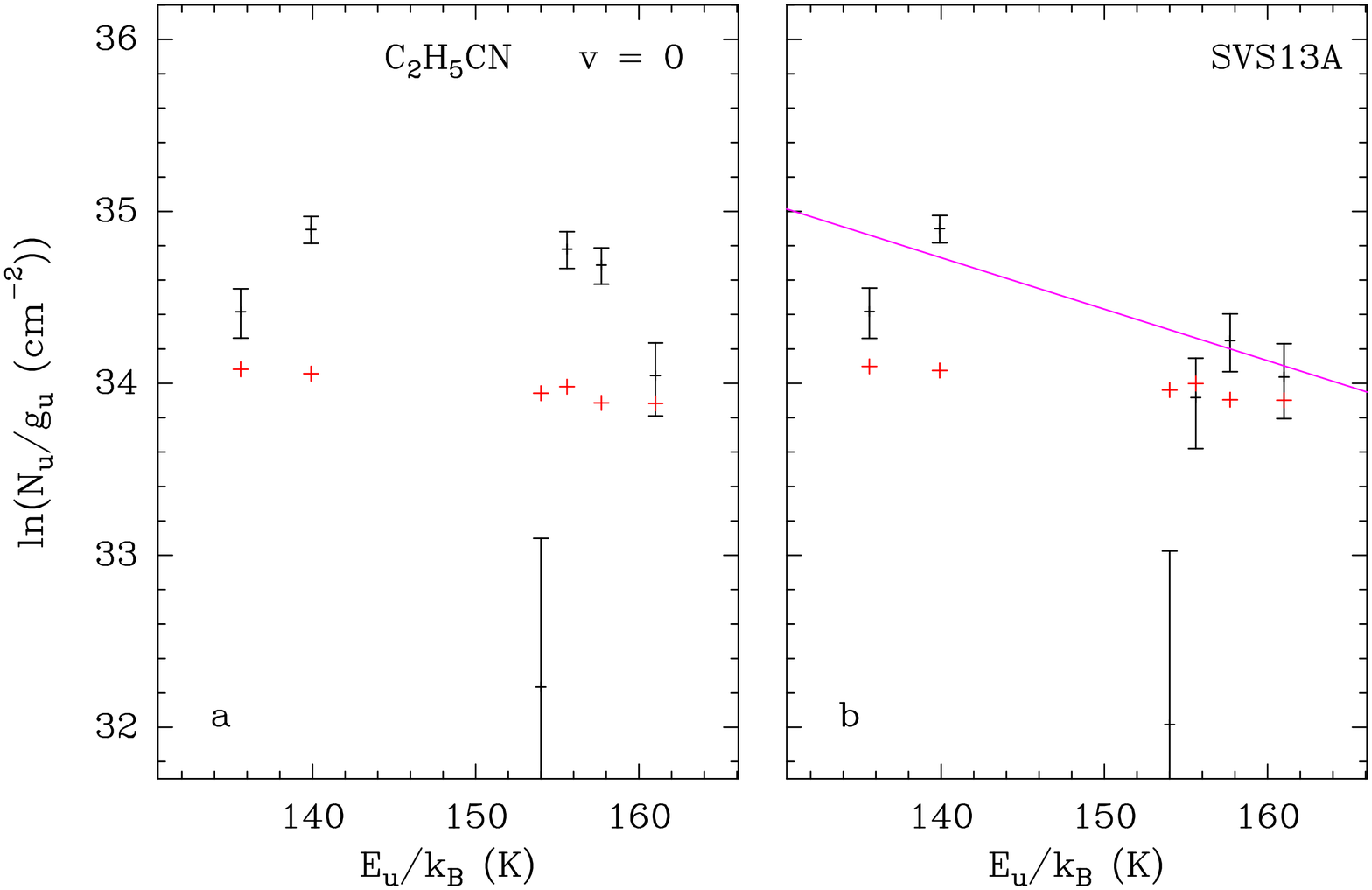}}}
\caption{Same as Fig.~\ref{f:popdiag_l1448-2a_p1_ch3oh} for C$_2$H$_5$CN in SVS13A.}
\label{f:popdiag_svs13-b_p1_c2h5cn}
\end{figure}

\begin{figure}[!htbp]
\centerline{\resizebox{0.83\hsize}{!}{\includegraphics[angle=0]{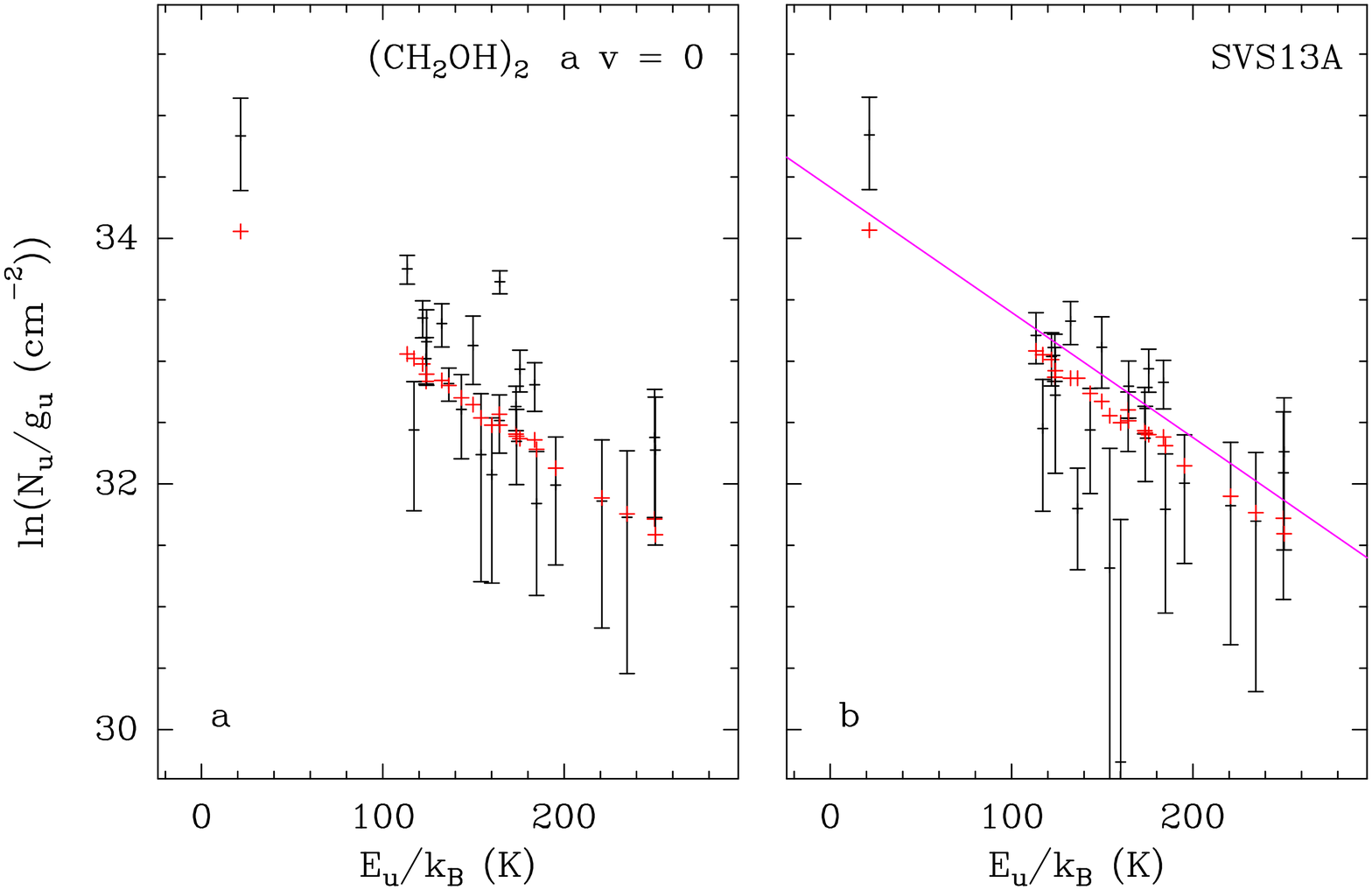}}}
\caption{Same as Fig.~\ref{f:popdiag_l1448-2a_p1_ch3oh} for (CH$_2$OH)$_2$ in SVS13A.}
\label{f:popdiag_svs13-b_p1_ch2oh-2}
\end{figure}

\begin{figure}[!htbp]
\centerline{\resizebox{0.83\hsize}{!}{\includegraphics[angle=0]{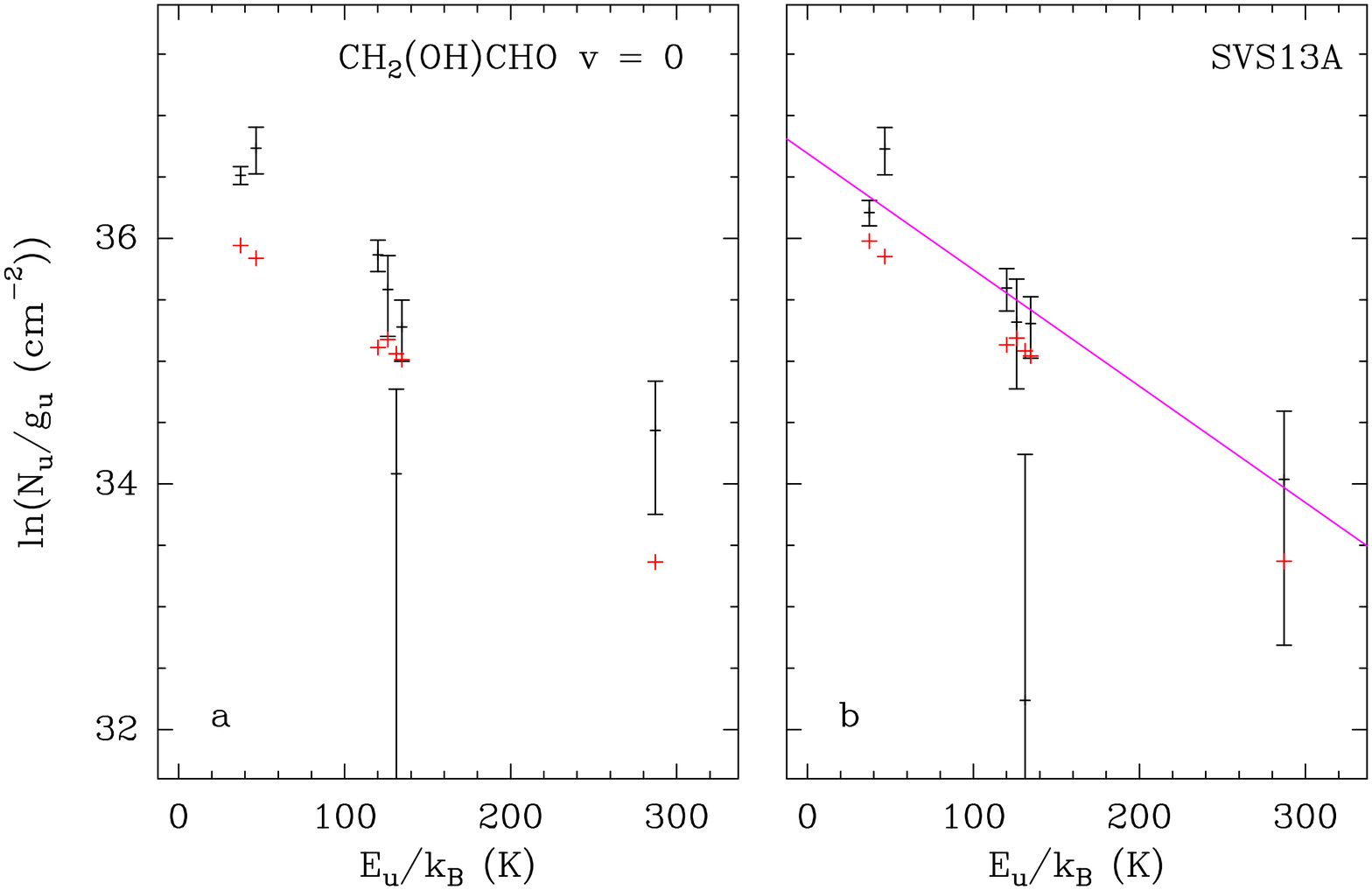}}}
\caption{Same as Fig.~\ref{f:popdiag_l1448-2a_p1_ch3oh} for CH$_2$(OH)CHO in SVS13A.}
\label{f:popdiag_svs13-b_p1_ch2ohcho}
\end{figure}

\begin{figure}[!htbp]
\centerline{\resizebox{0.83\hsize}{!}{\includegraphics[angle=0]{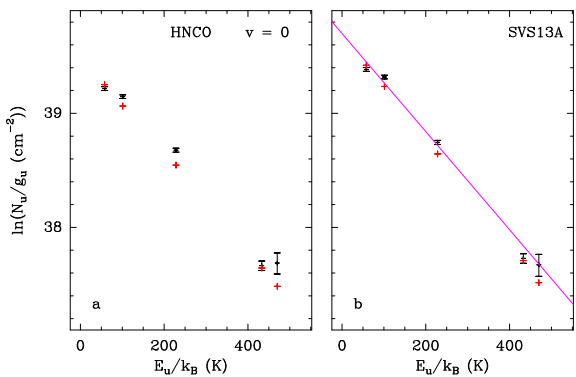}}}
\caption{Same as Fig.~\ref{f:popdiag_l1448-2a_p1_ch3oh} for HNCO in SVS13A.}
\label{f:popdiag_svs13-b_p1_hnco}
\end{figure}

\begin{figure}[!htbp]
\centerline{\resizebox{0.83\hsize}{!}{\includegraphics[angle=0]{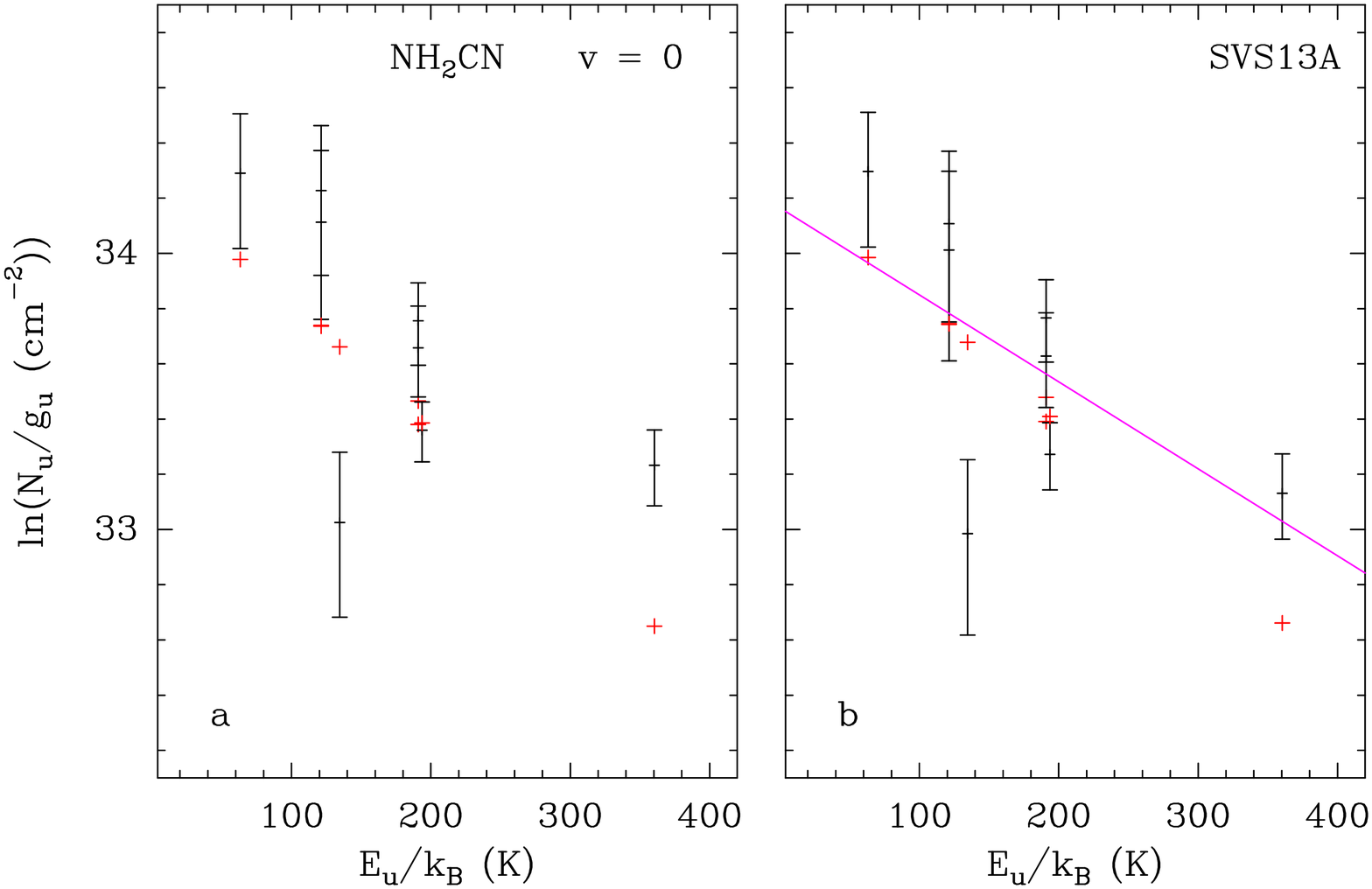}}}
\caption{Same as Fig.~\ref{f:popdiag_l1448-2a_p1_ch3oh} for NH$_2$CN in SVS13A.}
\label{f:popdiag_svs13-b_p1_nh2cn}
\end{figure}

\begin{figure}[!htbp]
\centerline{\resizebox{0.83\hsize}{!}{\includegraphics[angle=0]{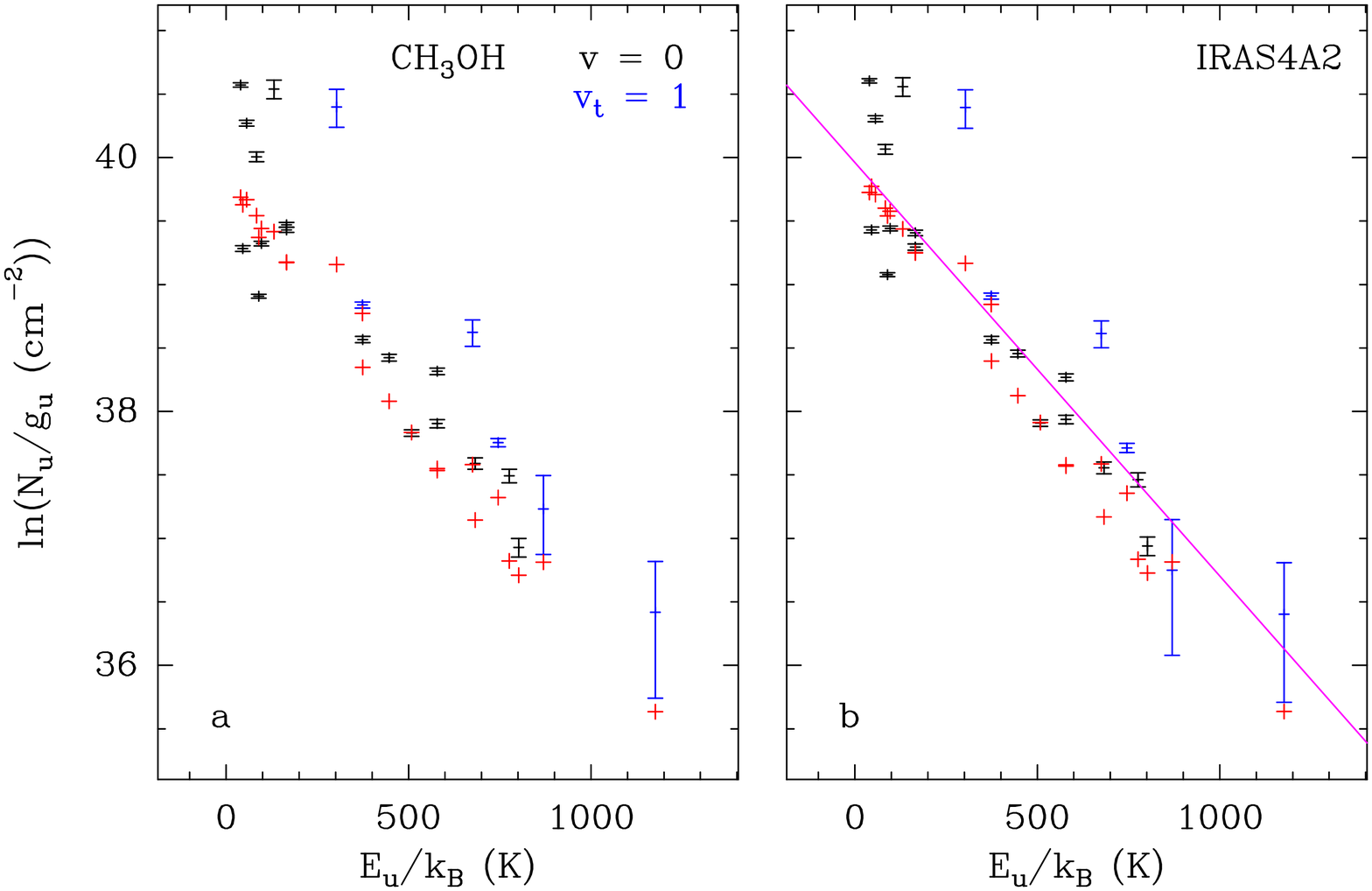}}}
\caption{Same as Fig.~\ref{f:popdiag_l1448-2a_p1_ch3oh} for CH$_3$OH in IRAS4A2.}
\label{f:popdiag_n1333-irs4a_p2_ch3oh}
\end{figure}

\begin{figure}[!htbp]
\centerline{\resizebox{0.83\hsize}{!}{\includegraphics[angle=0]{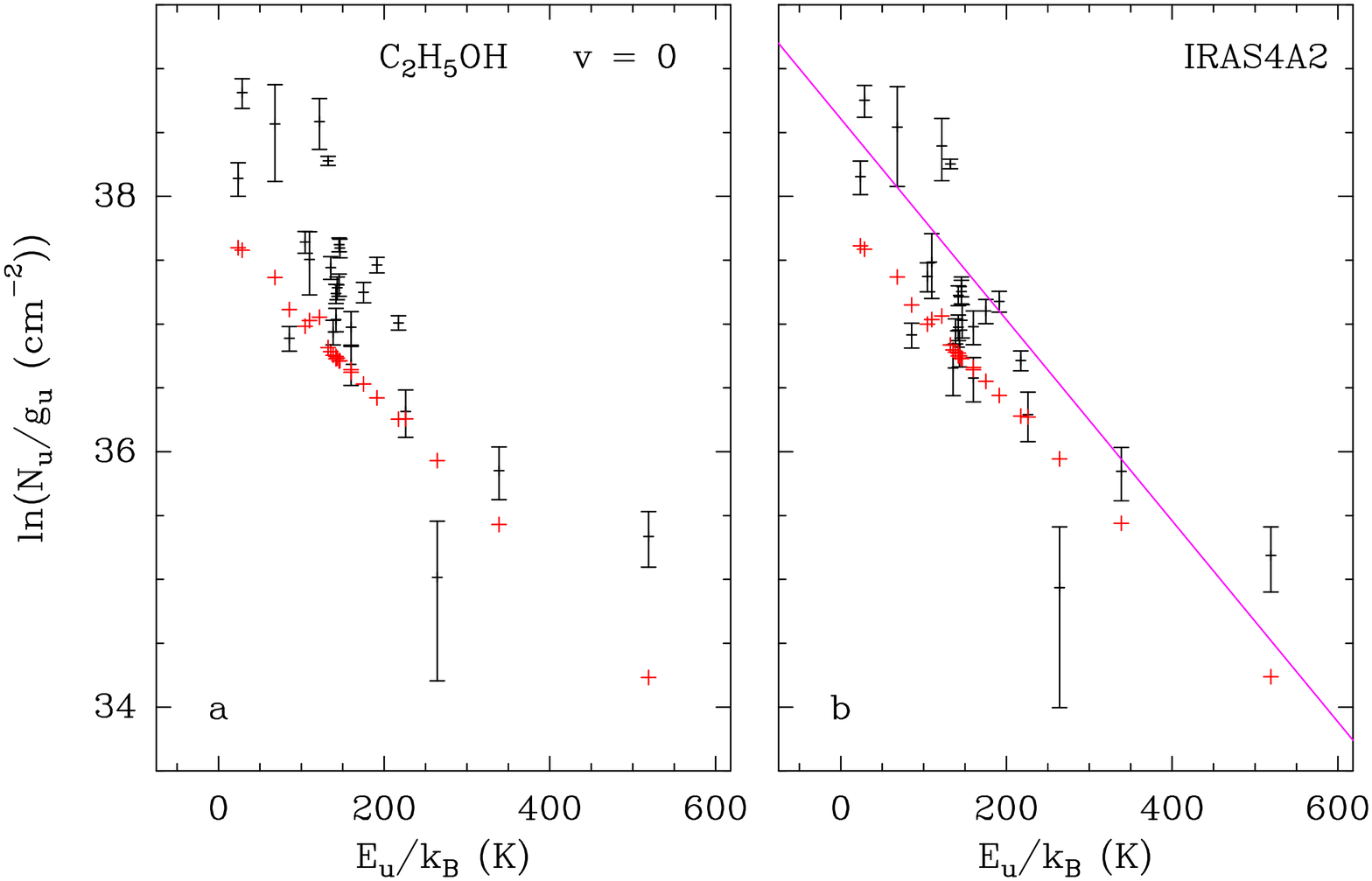}}}
\caption{Same as Fig.~\ref{f:popdiag_l1448-2a_p1_ch3oh} for C$_2$H$_5$OH in IRAS4A2.}
\label{f:popdiag_n1333-irs4a_p2_c2h5oh}
\end{figure}

\begin{figure}[!htbp]
\centerline{\resizebox{0.83\hsize}{!}{\includegraphics[angle=0]{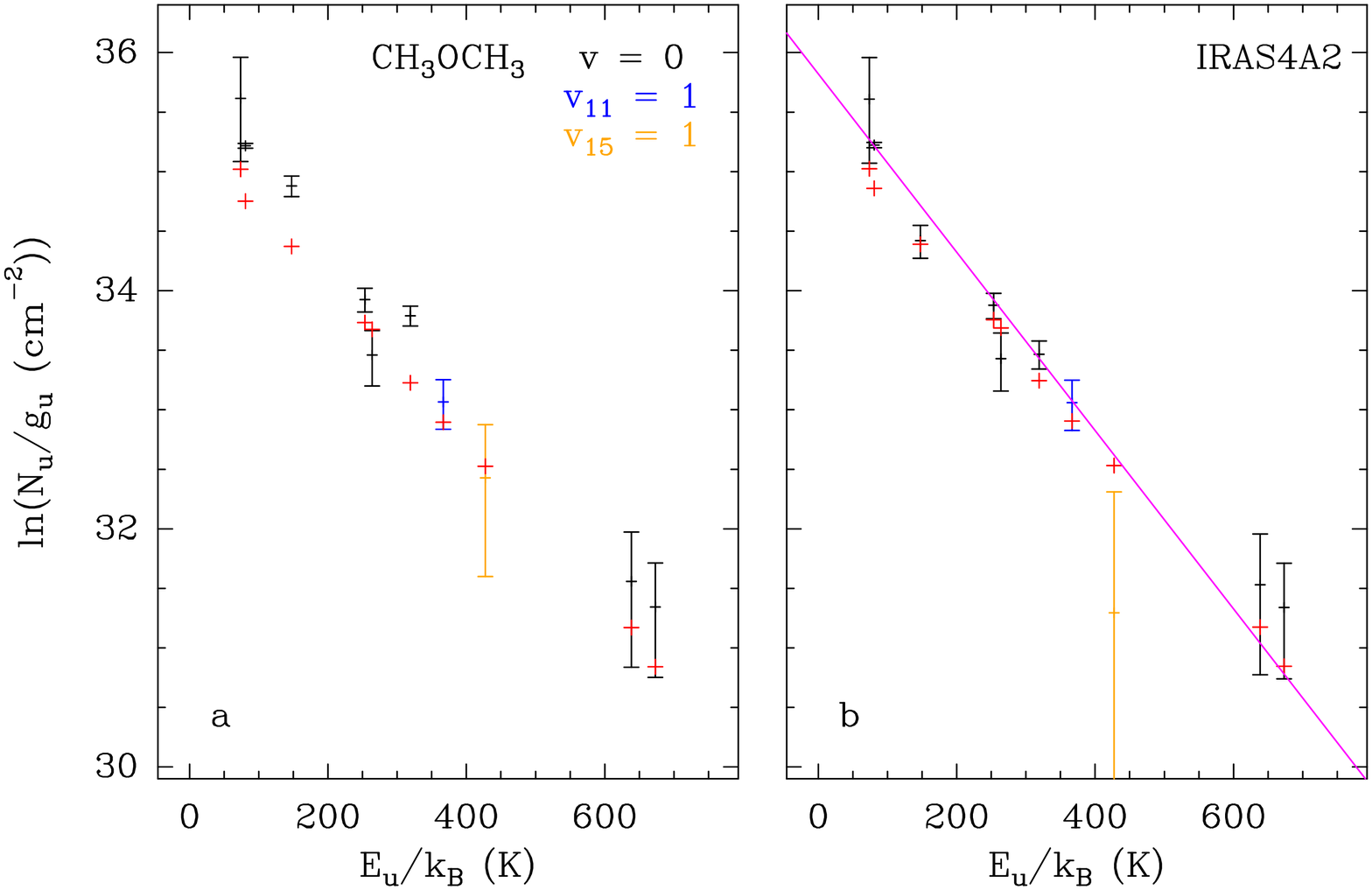}}}
\caption{Same as Fig.~\ref{f:popdiag_l1448-2a_p1_ch3oh} for CH$_3$OCH$_3$ in IRAS4A2.}
\label{f:popdiag_n1333-irs4a_p2_ch3och3}
\end{figure}

\begin{figure}[!htbp]
\centerline{\resizebox{0.83\hsize}{!}{\includegraphics[angle=0]{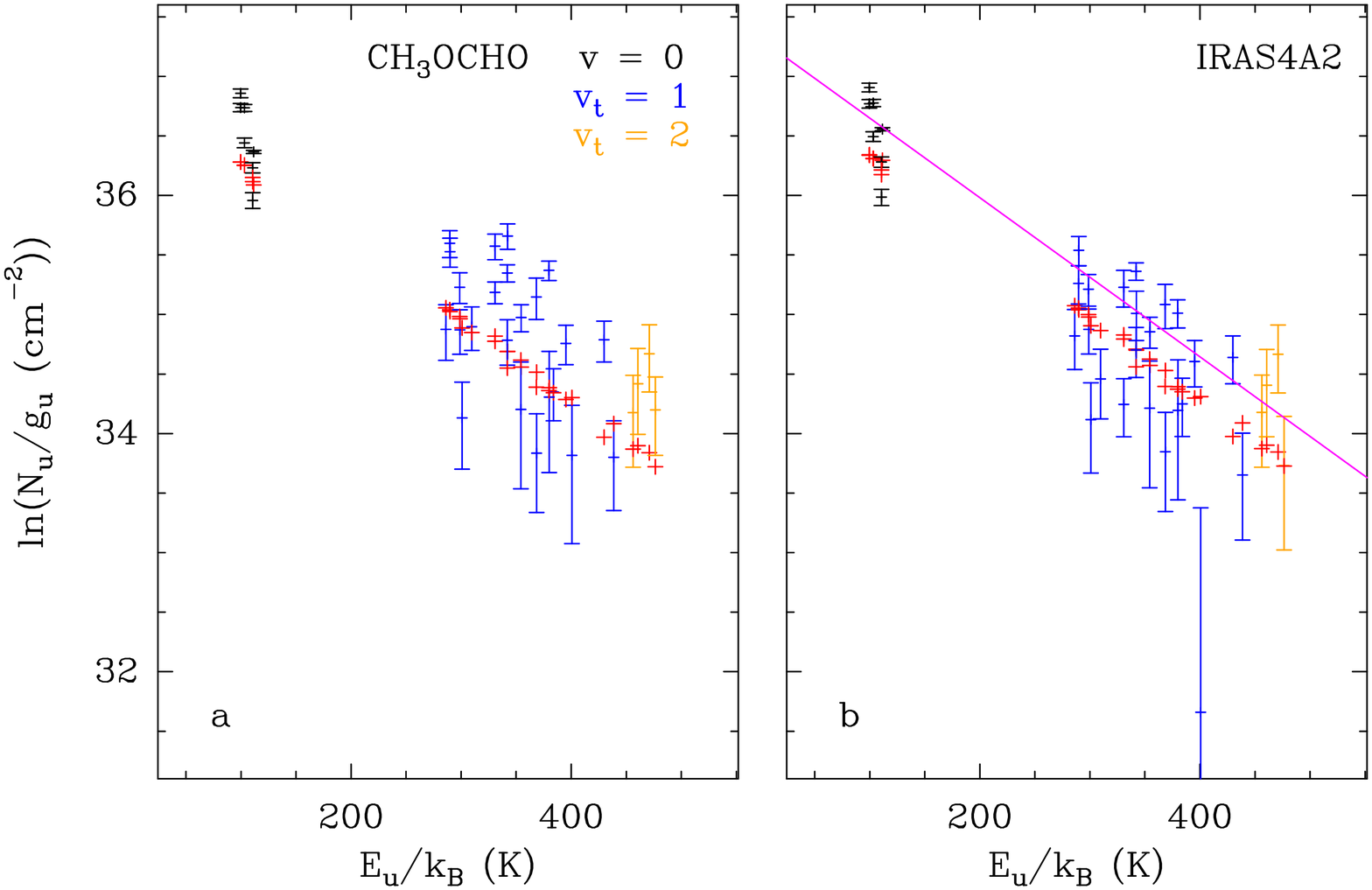}}}
\caption{Same as Fig.~\ref{f:popdiag_l1448-2a_p1_ch3oh} for CH$_3$OCHO in IRAS4A2.}
\label{f:popdiag_n1333-irs4a_p2_ch3ocho}
\end{figure}

\begin{figure}[!htbp]
\centerline{\resizebox{0.83\hsize}{!}{\includegraphics[angle=0]{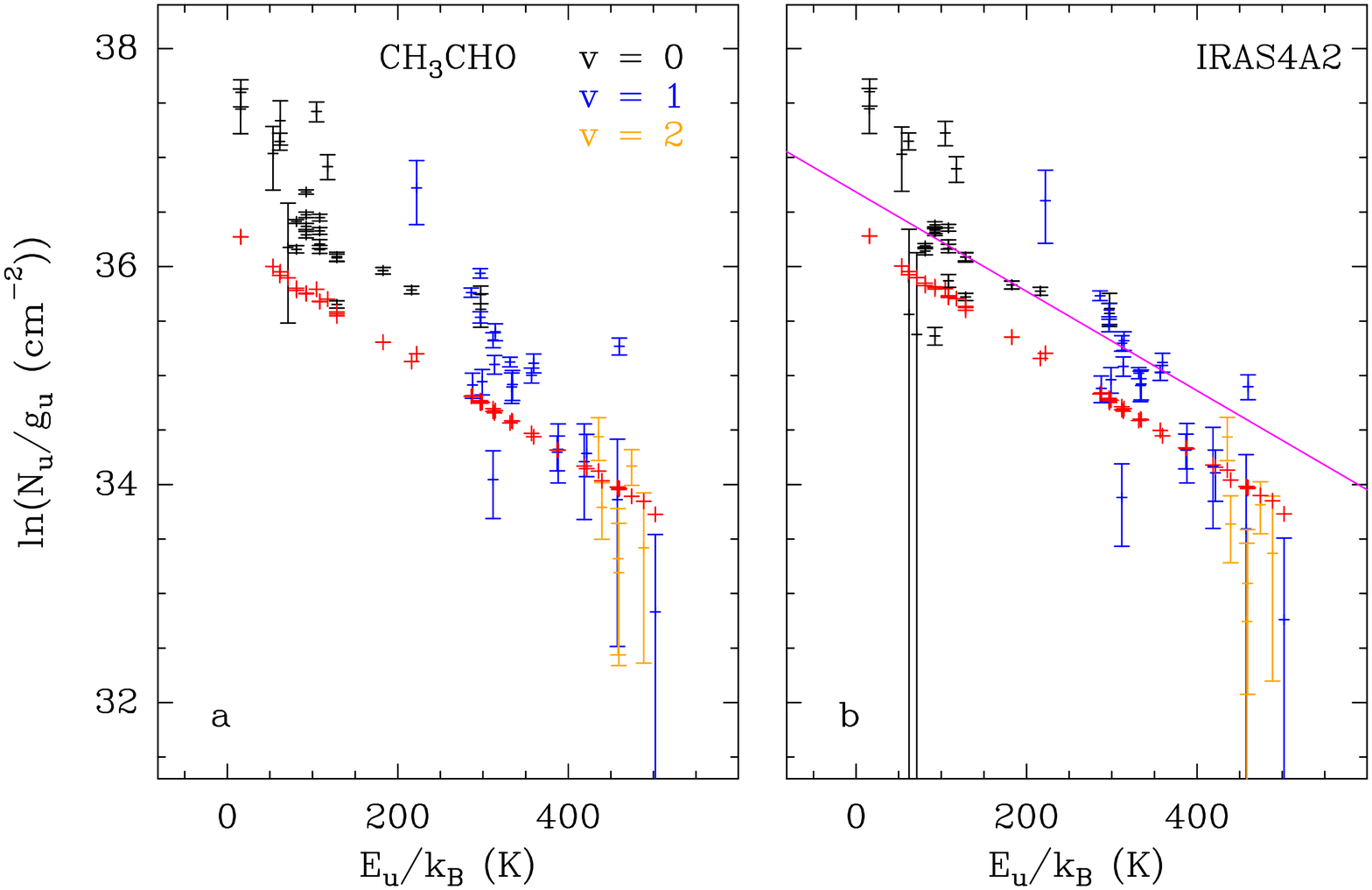}}}
\caption{Same as Fig.~\ref{f:popdiag_l1448-2a_p1_ch3oh} for CH$_3$CHO in IRAS4A2.}
\label{f:popdiag_n1333-irs4a_p2_ch3cho}
\end{figure}

\begin{figure}[!htbp]
\centerline{\resizebox{0.83\hsize}{!}{\includegraphics[angle=0]{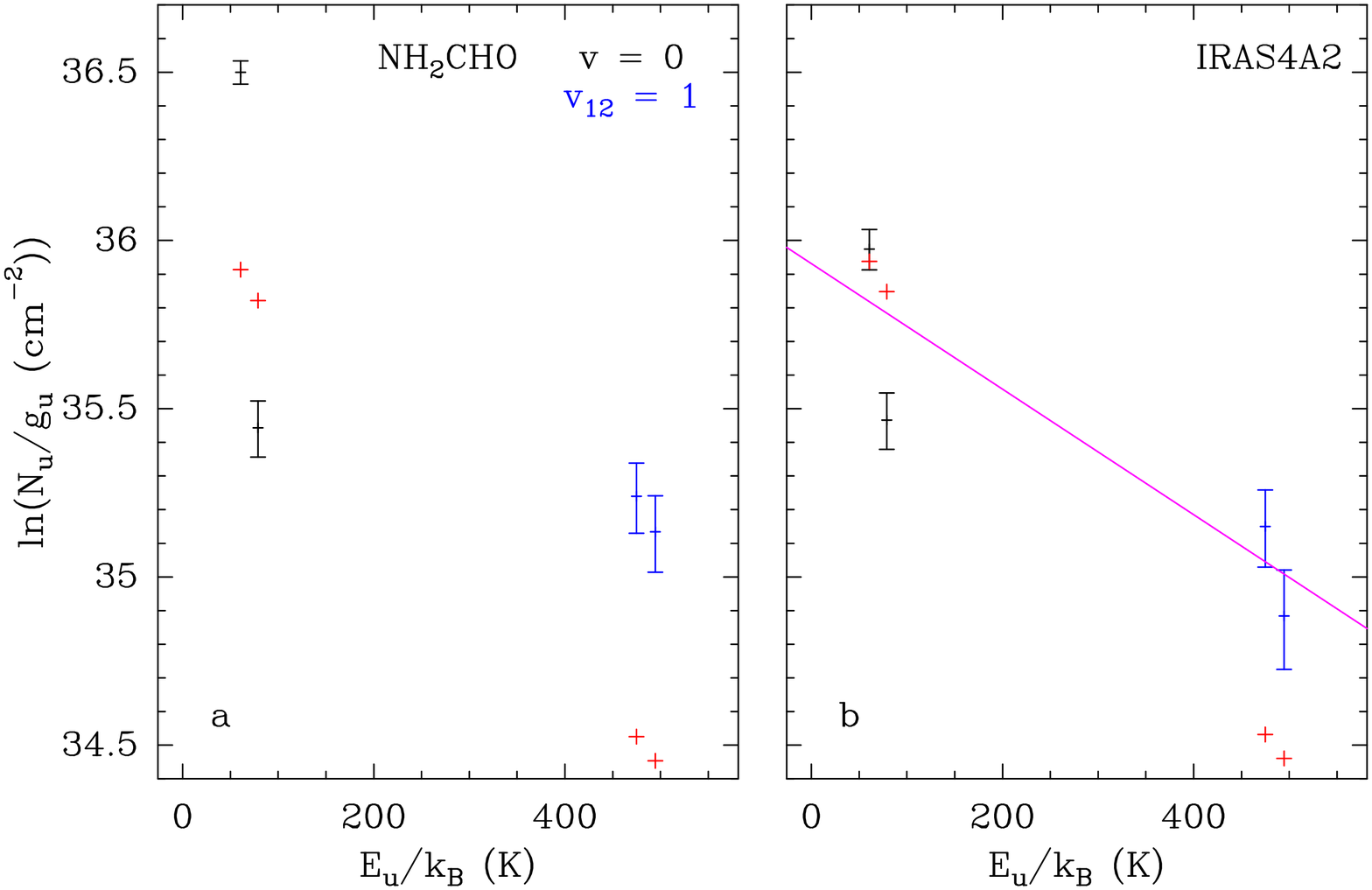}}}
\caption{Same as Fig.~\ref{f:popdiag_l1448-2a_p1_ch3oh} for NH$_2$CHO in IRAS4A2.}
\label{f:popdiag_n1333-irs4a_p2_nh2cho}
\end{figure}

\begin{figure}[!htbp]
\centerline{\resizebox{0.83\hsize}{!}{\includegraphics[angle=0]{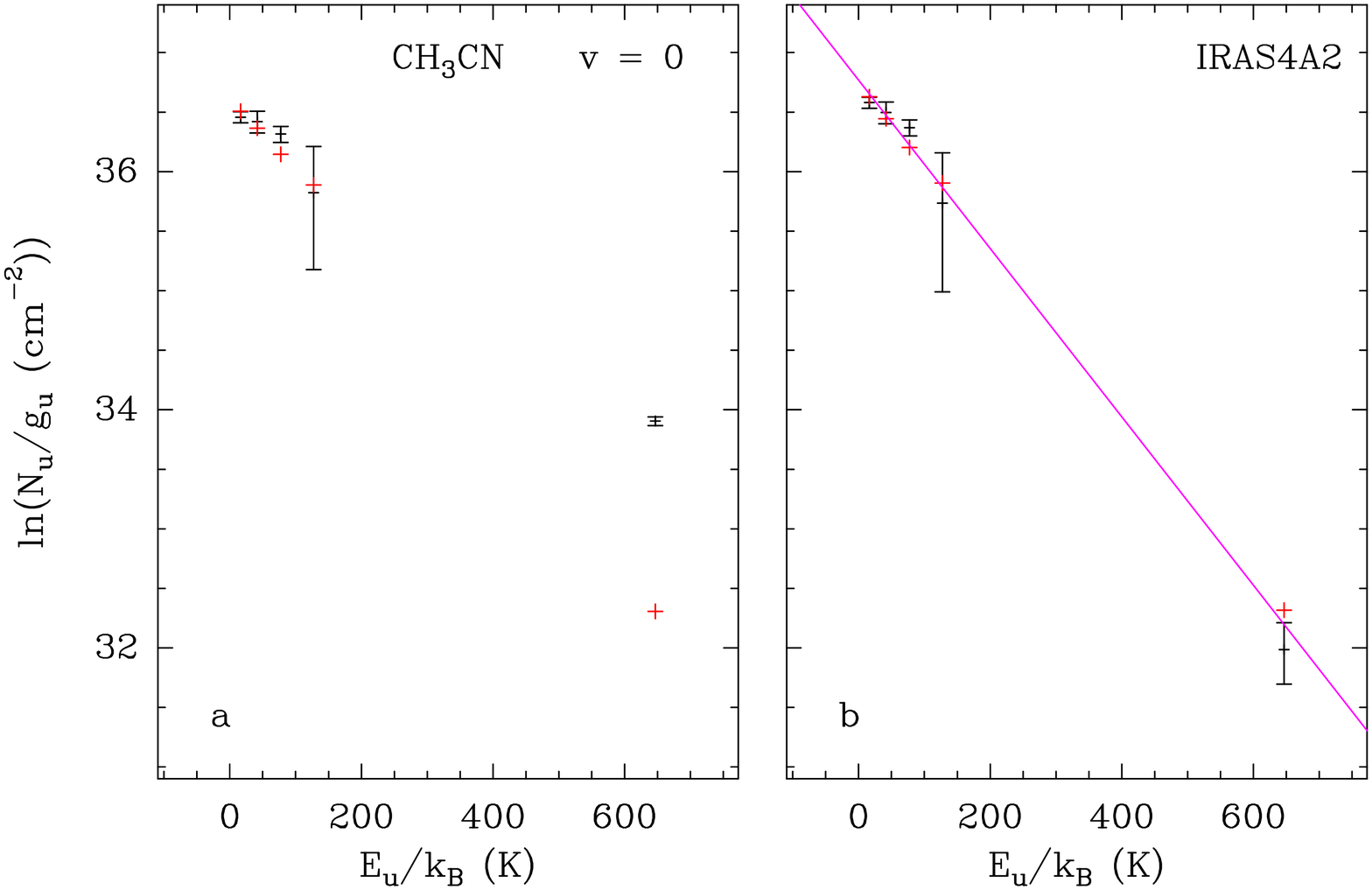}}}
\caption{Same as Fig.~\ref{f:popdiag_l1448-2a_p1_ch3oh} for CH$_3$CN in IRAS4A2.}
\label{f:popdiag_n1333-irs4a_p2_ch3cn}
\end{figure}

\begin{figure}[!htbp]
\centerline{\resizebox{0.83\hsize}{!}{\includegraphics[angle=0]{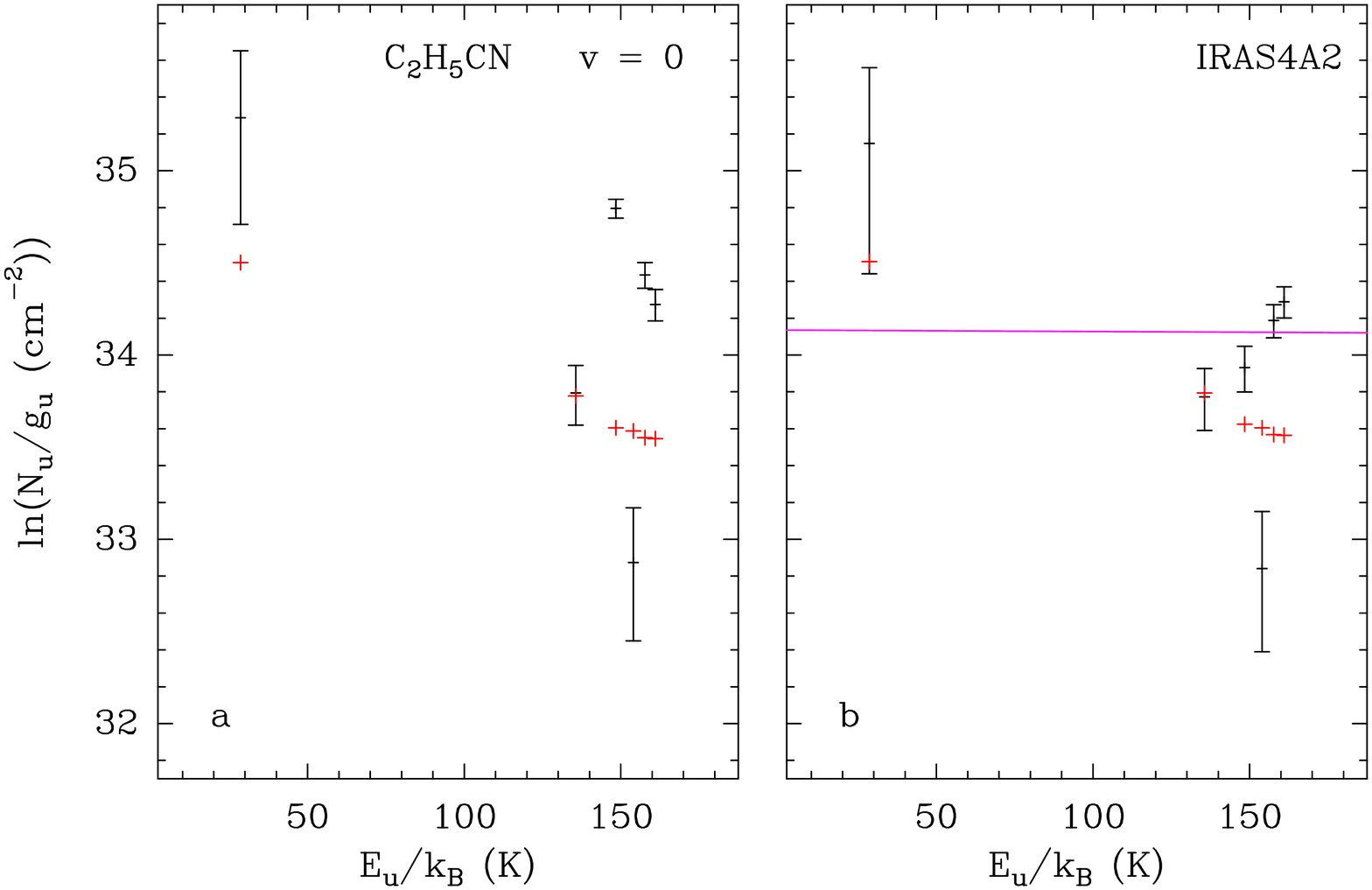}}}
\caption{Same as Fig.~\ref{f:popdiag_l1448-2a_p1_ch3oh} for C$_2$H$_5$CN in IRAS4A2.}
\label{f:popdiag_n1333-irs4a_p2_c2h5cn}
\end{figure}

\begin{figure}[!htbp]
\centerline{\resizebox{0.83\hsize}{!}{\includegraphics[angle=0]{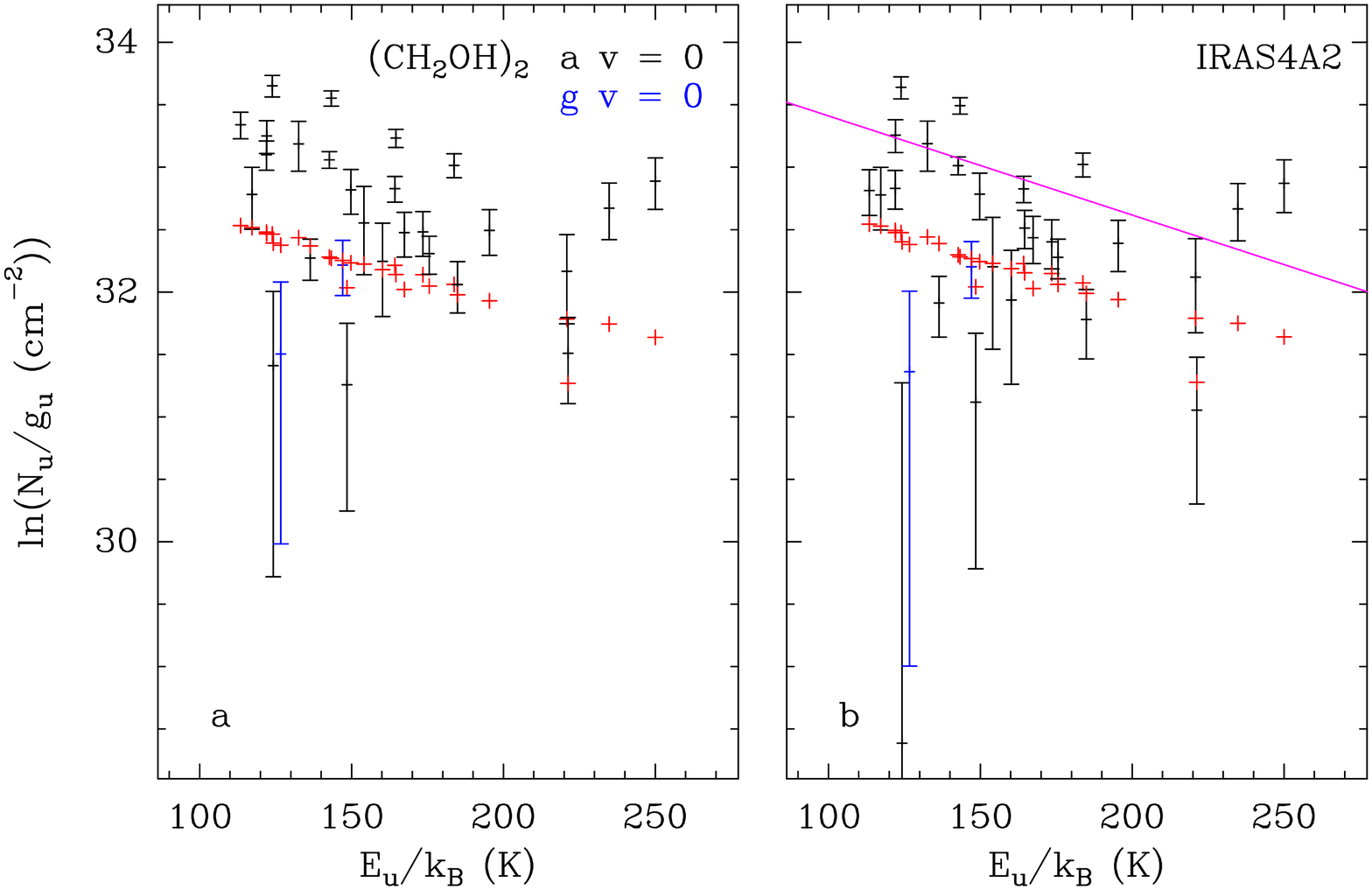}}}
\caption{Same as Fig.~\ref{f:popdiag_l1448-2a_p1_ch3oh} for (CH$_2$OH)$_2$ in IRAS4A2.}
\label{f:popdiag_n1333-irs4a_p2_ch2oh-2}
\end{figure}

\begin{figure}[!htbp]
\centerline{\resizebox{0.83\hsize}{!}{\includegraphics[angle=0]{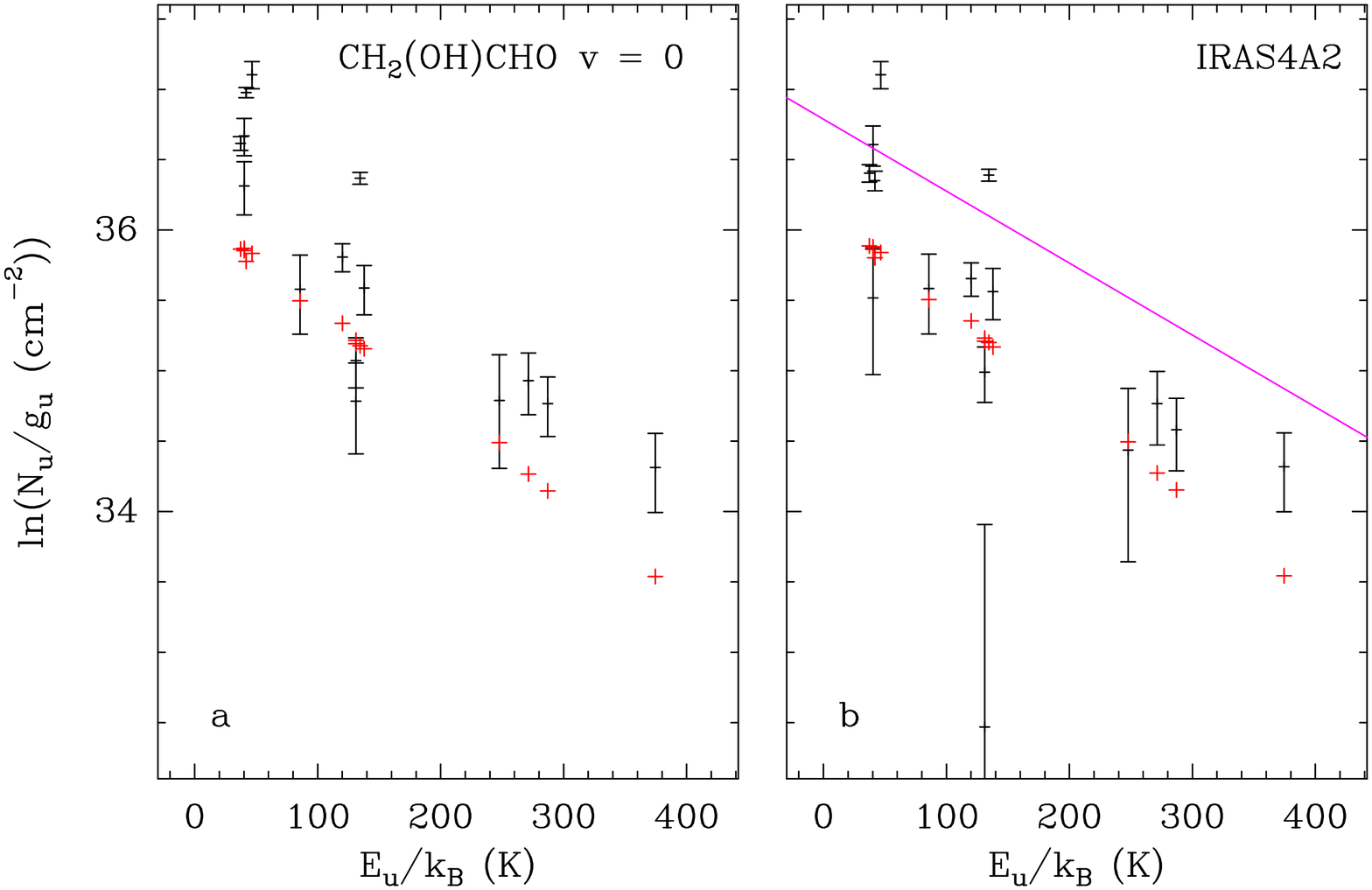}}}
\caption{Same as Fig.~\ref{f:popdiag_l1448-2a_p1_ch3oh} for CH$_2$(OH)CHO in IRAS4A2.}
\label{f:popdiag_n1333-irs4a_p2_ch2ohcho}
\end{figure}

\begin{figure}[!htbp]
\centerline{\resizebox{0.83\hsize}{!}{\includegraphics[angle=0]{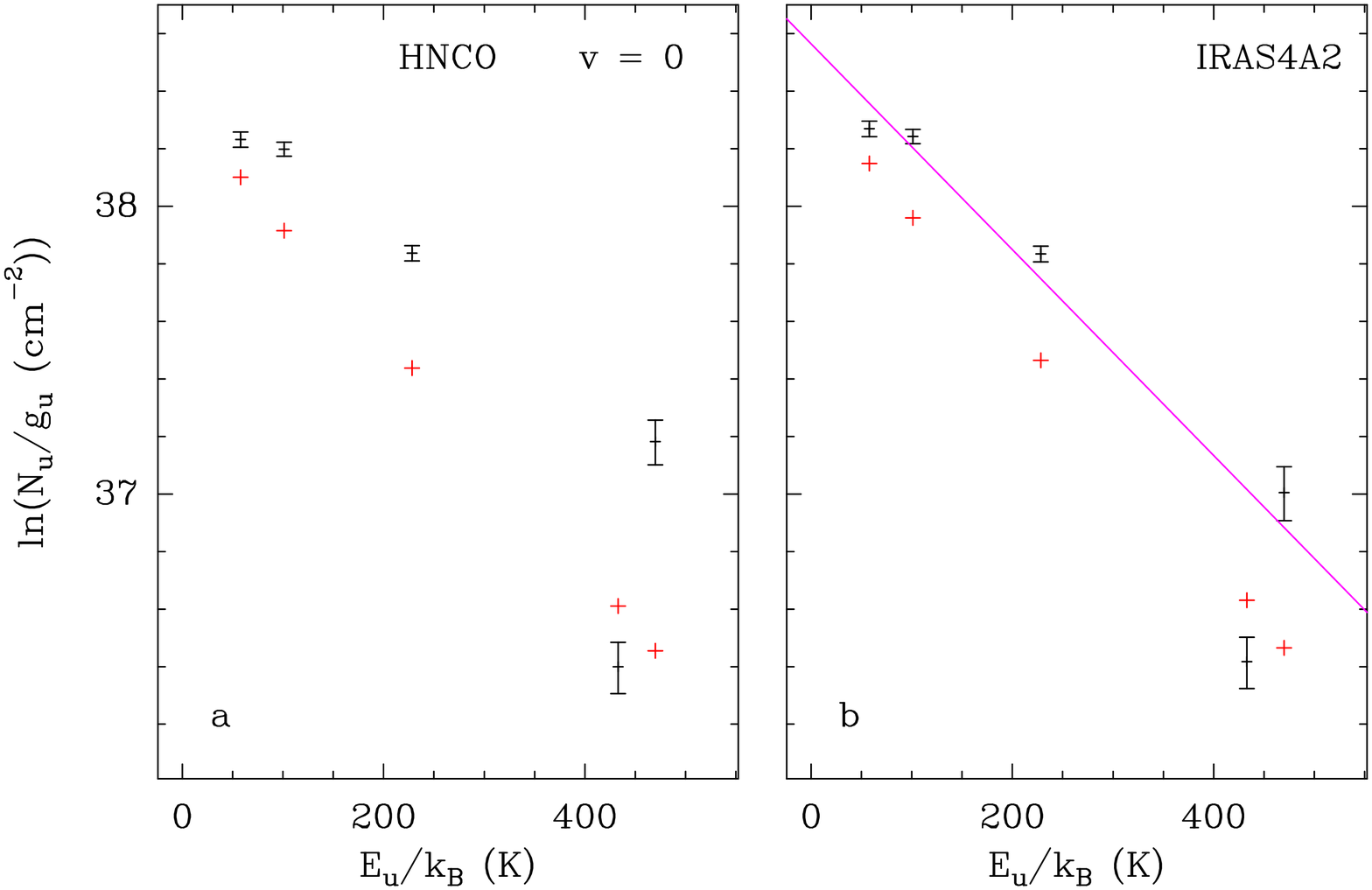}}}
\caption{Same as Fig.~\ref{f:popdiag_l1448-2a_p1_ch3oh} for HNCO in IRAS4A2.}
\label{f:popdiag_n1333-irs4a_p2_hnco}
\end{figure}

\begin{figure}[!htbp]
\centerline{\resizebox{0.83\hsize}{!}{\includegraphics[angle=0]{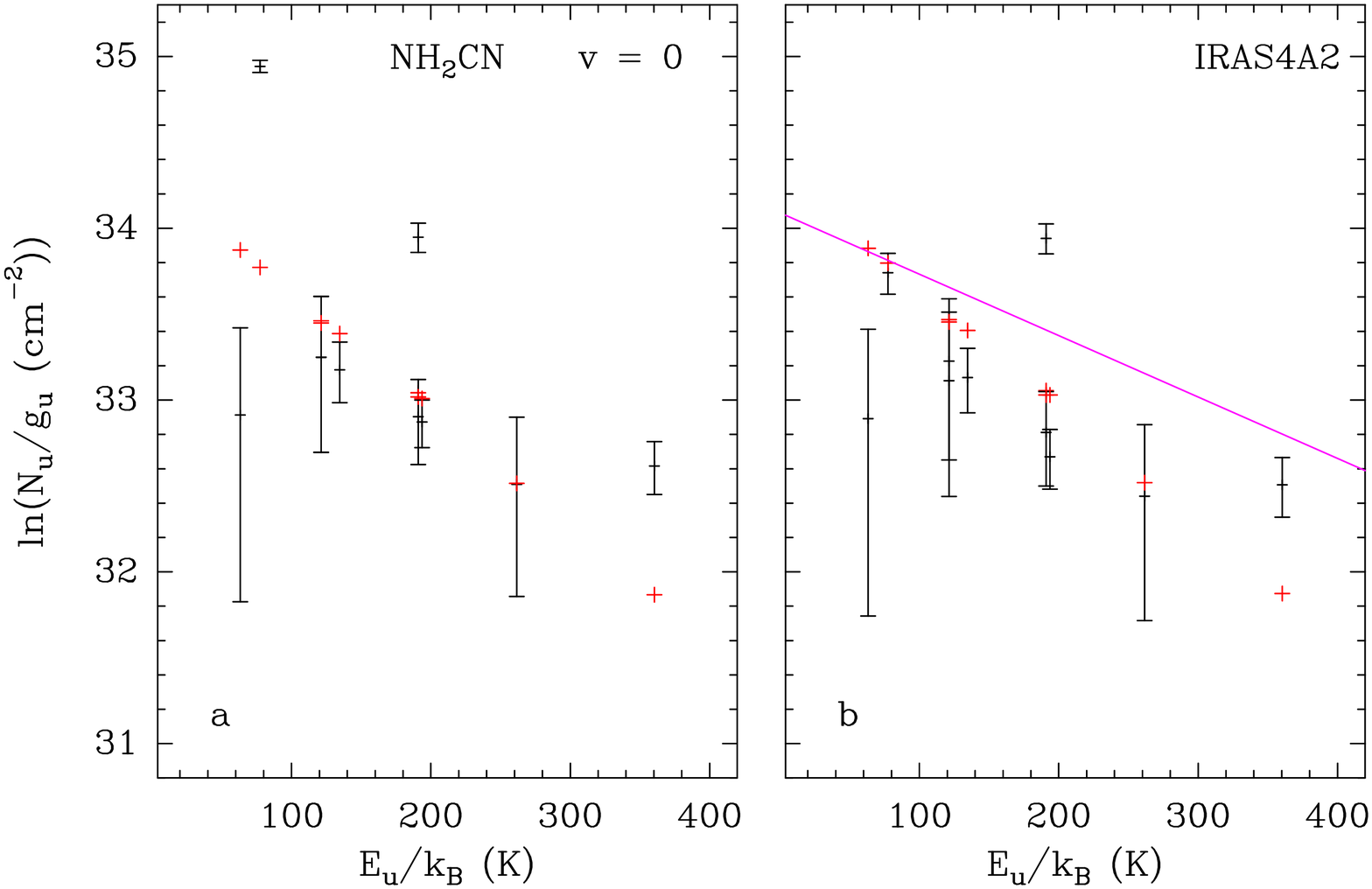}}}
\caption{Same as Fig.~\ref{f:popdiag_l1448-2a_p1_ch3oh} for NH$_2$CN in IRAS4A2.}
\label{f:popdiag_n1333-irs4a_p2_nh2cn}
\end{figure}

\begin{figure}[!htbp]
\centerline{\resizebox{0.83\hsize}{!}{\includegraphics[angle=0]{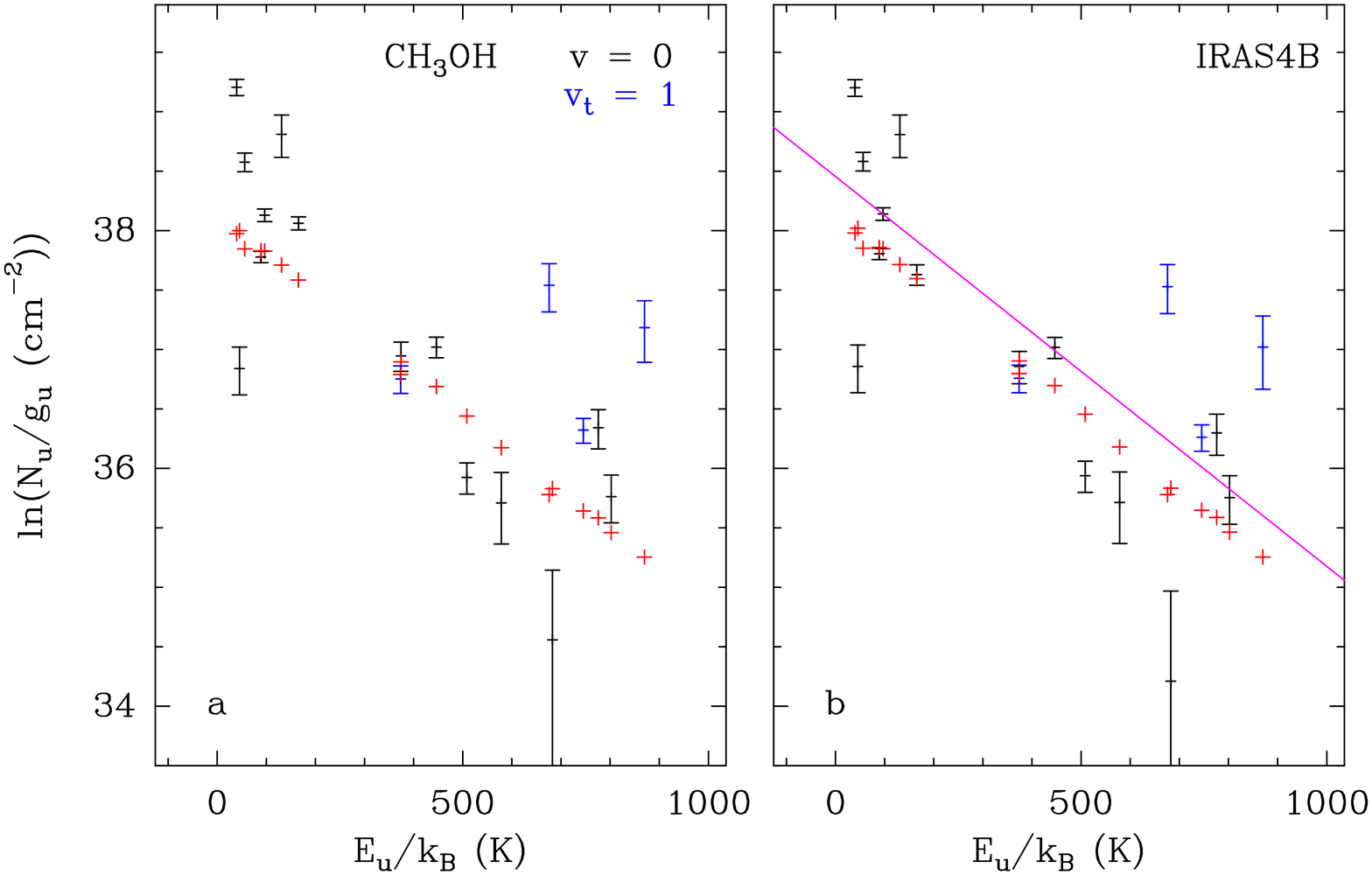}}}
\caption{Same as Fig.~\ref{f:popdiag_l1448-2a_p1_ch3oh} for CH$_3$OH in IRAS4B.}
\label{f:popdiag_n1333-irs4b_p1_ch3oh}
\end{figure}

\begin{figure}[!htbp]
\centerline{\resizebox{0.83\hsize}{!}{\includegraphics[angle=0]{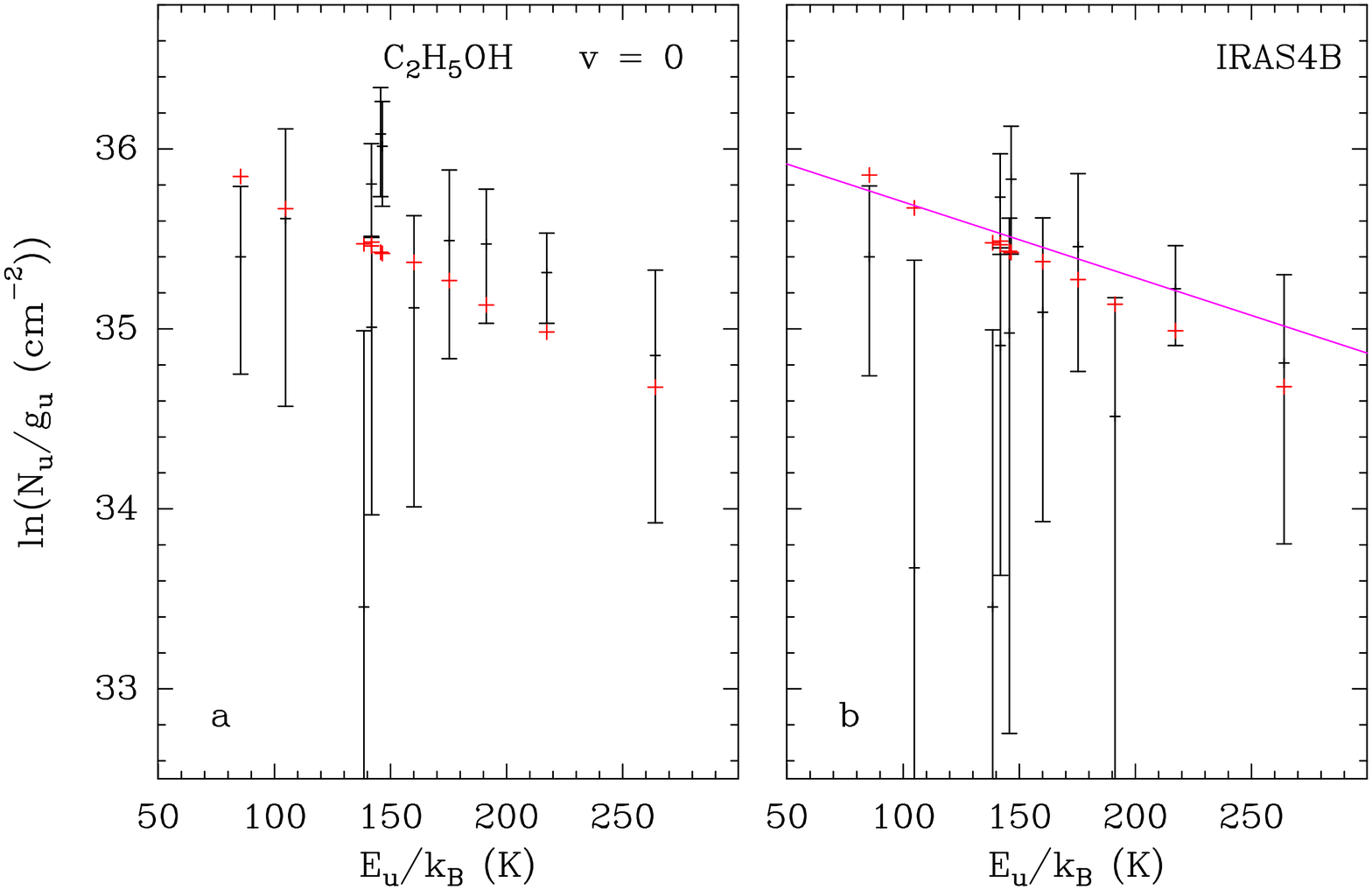}}}
\caption{Same as Fig.~\ref{f:popdiag_l1448-2a_p1_ch3oh} for C$_2$H$_5$OH in IRAS4B.}
\label{f:popdiag_n1333-irs4b_p1_c2h5oh}
\end{figure}

\begin{figure}[!htbp]
\centerline{\resizebox{0.83\hsize}{!}{\includegraphics[angle=0]{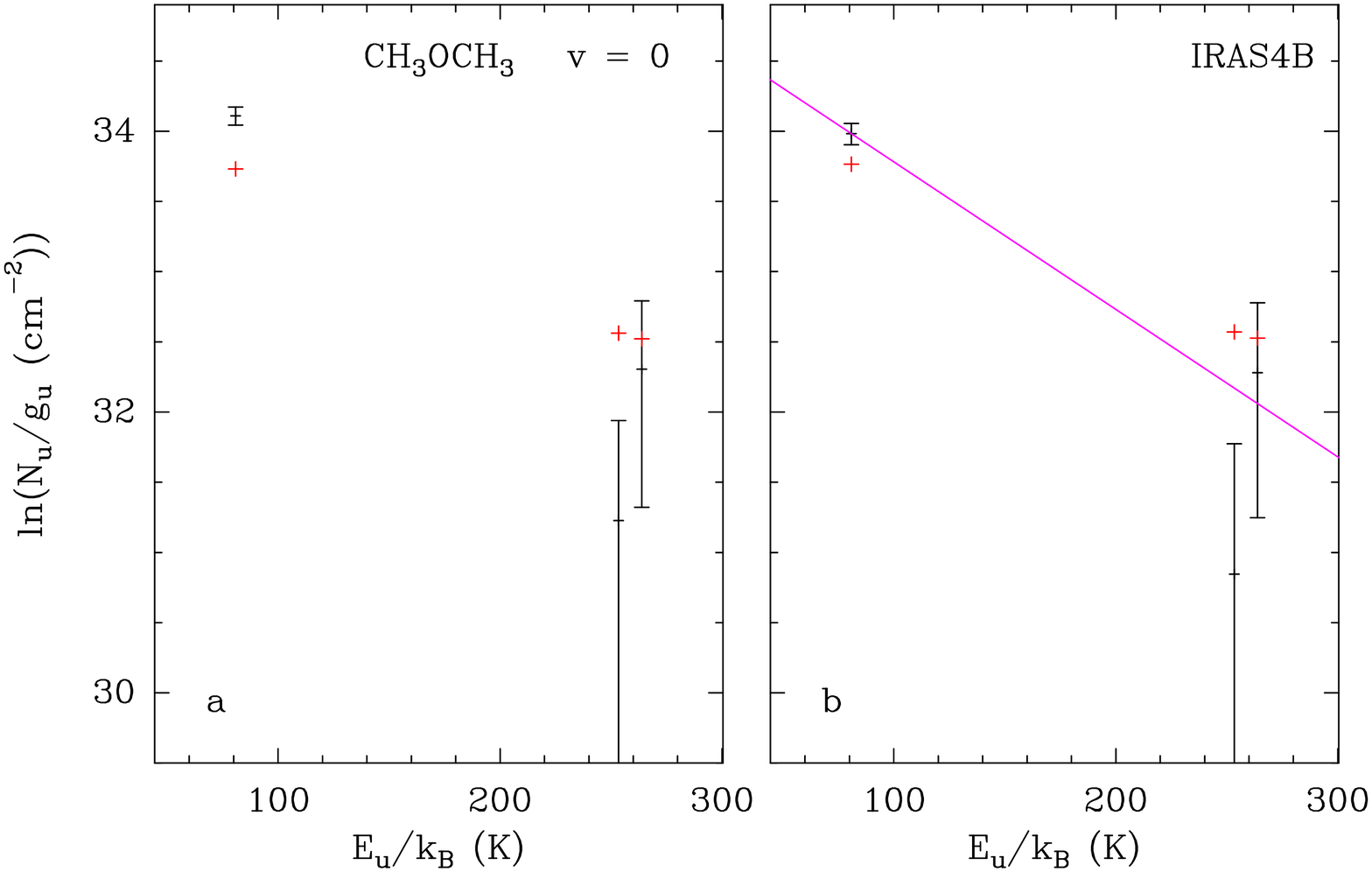}}}
\caption{Same as Fig.~\ref{f:popdiag_l1448-2a_p1_ch3oh} for CH$_3$OCH$_3$ in IRAS4B.}
\label{f:popdiag_n1333-irs4b_p1_ch3och3}
\end{figure}

\begin{figure}[!htbp]
\centerline{\resizebox{0.83\hsize}{!}{\includegraphics[angle=0]{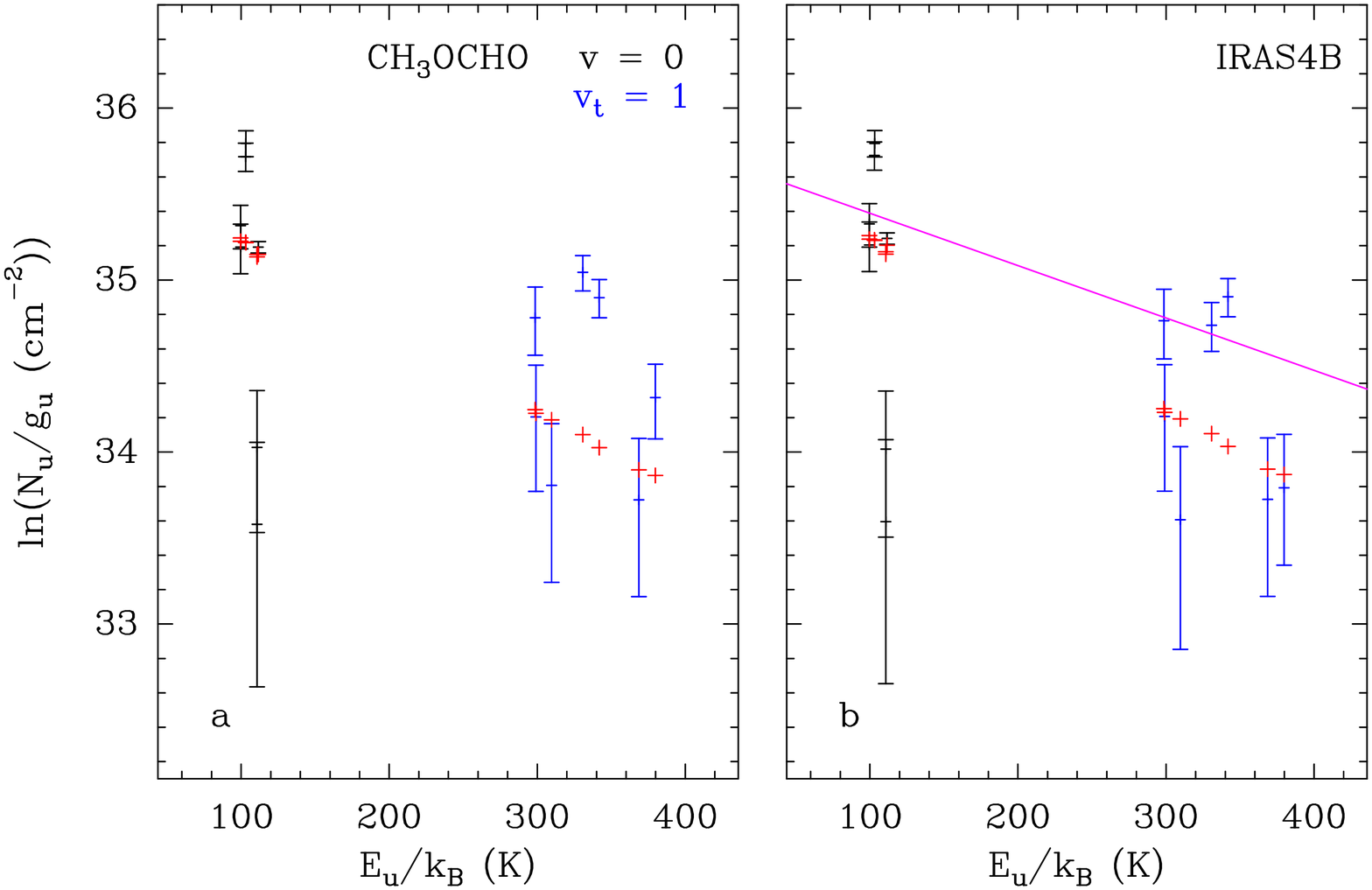}}}
\caption{Same as Fig.~\ref{f:popdiag_l1448-2a_p1_ch3oh} for CH$_3$OCHO in IRAS4B.}
\label{f:popdiag_n1333-irs4b_p1_ch3ocho}
\end{figure}

\begin{figure}[!htbp]
\centerline{\resizebox{0.83\hsize}{!}{\includegraphics[angle=0]{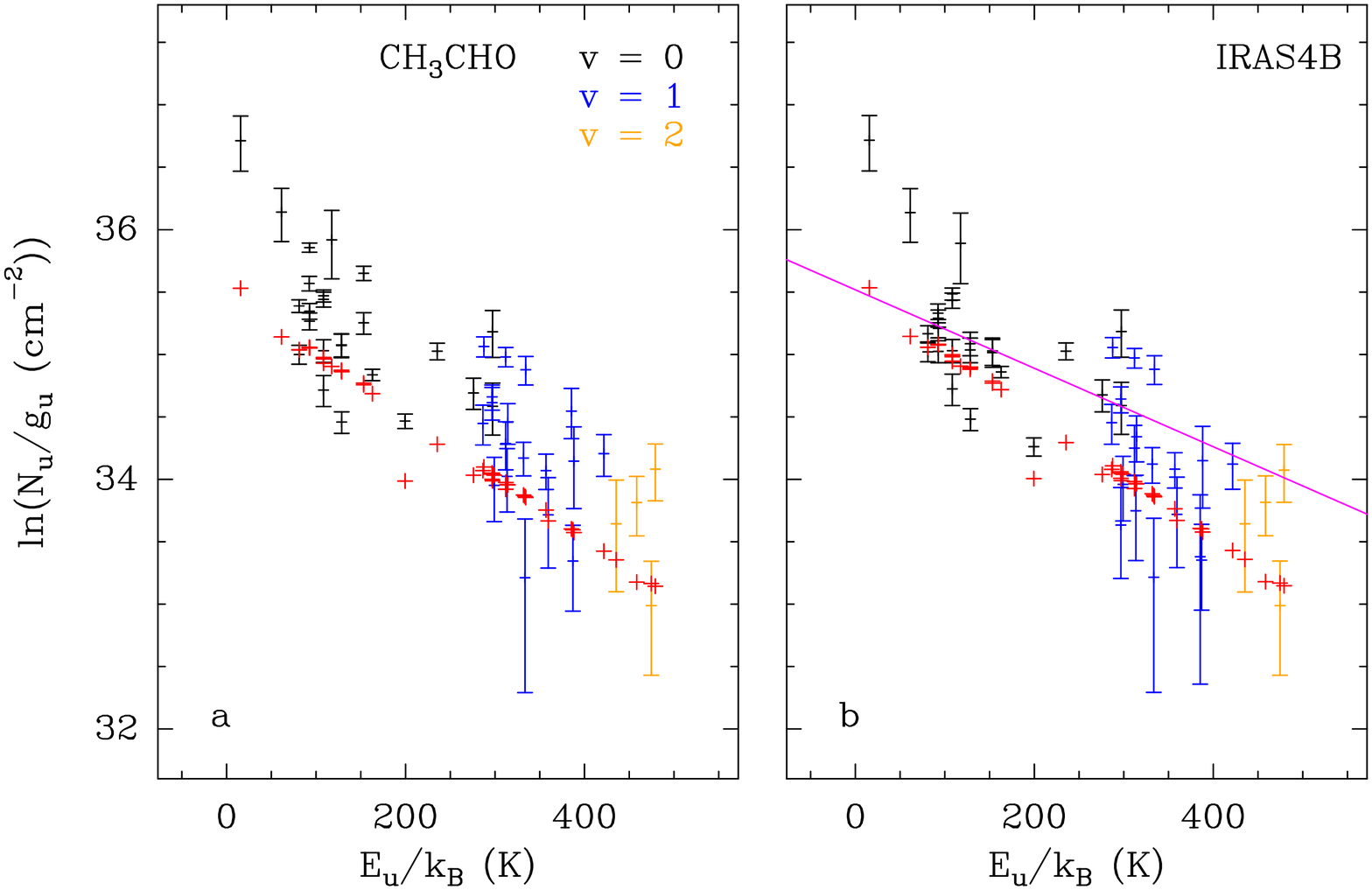}}}
\caption{Same as Fig.~\ref{f:popdiag_l1448-2a_p1_ch3oh} for CH$_3$CHO in IRAS4B.}
\label{f:popdiag_n1333-irs4b_p1_ch3cho}
\end{figure}

\begin{figure}[!htbp]
\centerline{\resizebox{0.83\hsize}{!}{\includegraphics[angle=0]{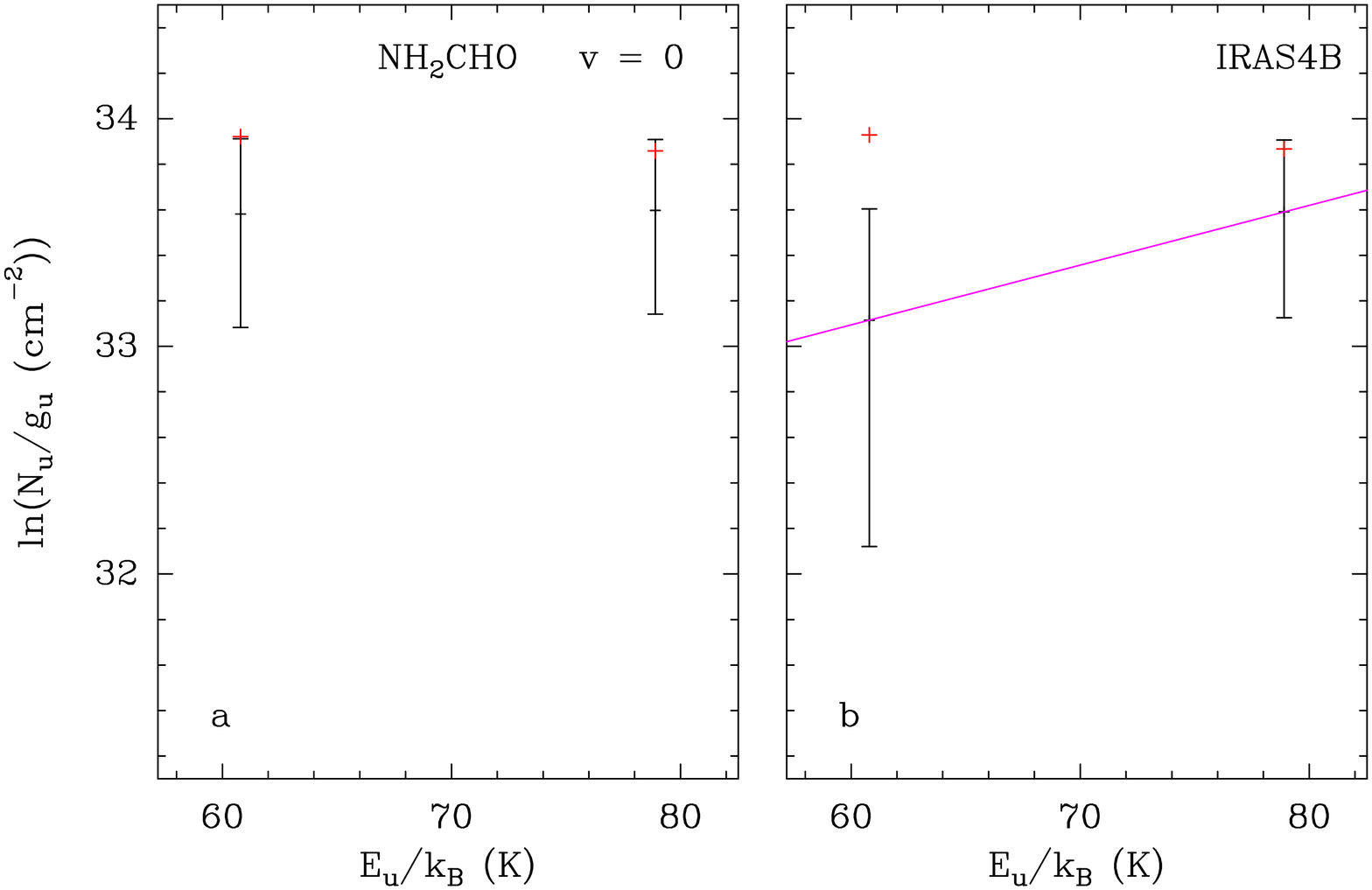}}}
\caption{Same as Fig.~\ref{f:popdiag_l1448-2a_p1_ch3oh} for NH$_2$CHO in IRAS4B.}
\label{f:popdiag_n1333-irs4b_p1_nh2cho}
\end{figure}

\begin{figure}[!htbp]
\centerline{\resizebox{0.83\hsize}{!}{\includegraphics[angle=0]{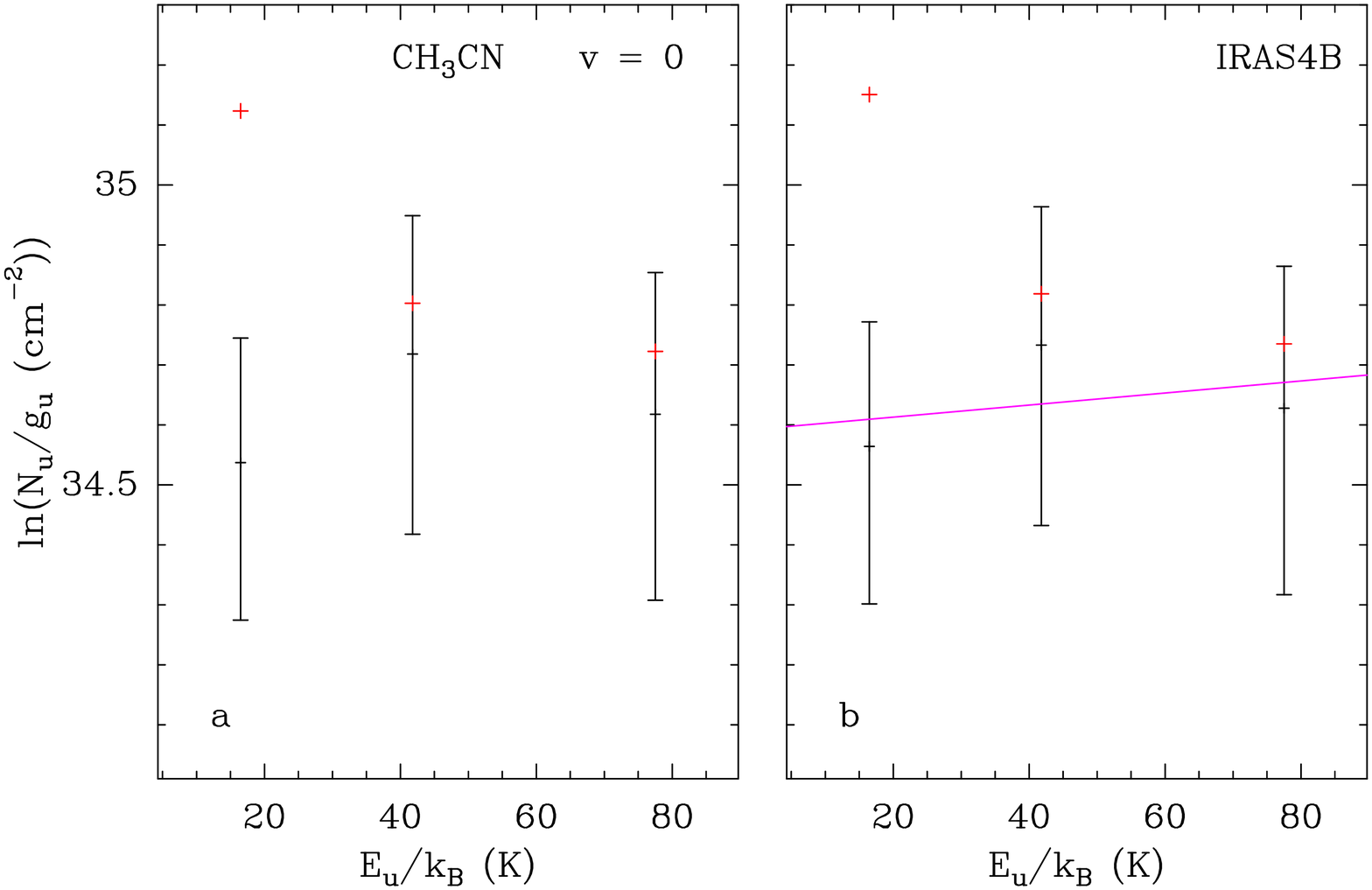}}}
\caption{Same as Fig.~\ref{f:popdiag_l1448-2a_p1_ch3oh} for CH$_3$CN in IRAS4B.}
\label{f:popdiag_n1333-irs4b_p1_ch3cn}
\end{figure}

Population diagrams of (complex) organic molecules detected toward the 
CALYPSO sources are shown in 
Figs.~\ref{f:popdiag_l1448-2a_p1_ch3oh}--\ref{f:popdiag_l1157_p1_ch3cn}.
Figures~\ref{f:trot_l1448-2a_p1}--\ref{f:trot_l1157_p1} display the derived
rotational temperatures and their uncertainties for each source separately,
which are all listed in Table~\ref{t:popfit}. 

\paragraph{L1448-2A:} The population diagram of methanol yields a 
rotational temperature that is consistent with 150~K albeit with a large
uncertainty. We adopted this temperature to derive the column density (or upper 
limit) of all selected organic molecules.

\paragraph{L1448-2Ab:} Only two population diagrams could be constructed.
Given the large dispersion of the measured integrated intensities of 
transitions with $E_{\rm u}/k_{\rm B} <$ 200~K, the rotational temperature of 
methanol is not constrained. However, the low peak intensities of one 
transition with $E_{\rm u}/k_{\rm B} >$ 200 K and the non-detection of 
transitions from within the first torsionally excited state suggest a 
rotational temperature below 200~K. We assumed a value of 150~K, which we also 
used for the other complex molecules. The population diagram of HNCO contains 
only two data points and the uncertainties are too large to constrain the 
rotational temperature of this molecule.

\paragraph{L1448-NA:} We assumed a temperature of 100~K to derive upper 
limits to the column densities of the selected organic molecules.

\paragraph{L1448-NB1:} The non-detection of the $8_{-1}$--$7_0$ transition of 
methanol at 229758.756 MHz suggests a temperature lower than 100~K, but this
is uncertain. We assumed a temperature of 100 K to derive column density upper 
limits for the other selected organic molecules.

\paragraph{L1448-NB2:} We assumed a temperature of 100~K to derive upper 
limits to the column densities of the selected organic molecules.

\paragraph{L1448-C:} The fitted rotational temperatures of CH$_3$OH, CH$_3$CHO, 
and CH$_3$CN are consistent with a temperature of 100~K within $\sim$$1\sigma$. 
CH$_3$OCH$_3$ and HNCO have lower temperatures of 75~K and 59~K, 
respectively, but these values are based on three data points only. The lower 
value derived for CH$_3$OCHO is uncertain due to the narrow range of upper 
level energies. It is consistent within $3\sigma$ with a temperature of 
100~K. The temperature is not constrained for NH$_2$CHO due to the low 
signal-to-noise ratios and the small number of data points. We assumed a 
temperature of 100~K for all selected organic species.

\paragraph{L1448-CS:} We assumed the same temperature as for L1448-C.

\paragraph{IRAS2A1:} For all species but NH$_2$CHO, we assumed a 
rotational temperature that is within $1\sigma$ of the value derived from the 
fit to the population diagram. The fit of NH$_2$CHO relies on only four 
transitions. The high fitted temperature is strongly biased by the 
$\varv_{12}=1$ group of transitions at 218.18 GHz that is most likely blended 
with a transition of an unidentified species. We did not trust
this high value and assumed a temperature of 250~K like methanol instead.

\paragraph{SVS13B:} We assumed a temperature of 150~K to derive upper 
limits to the column densities of the selected organic molecules.

\paragraph{SVS13A:} For all species but C$_2$H$_5$CN, we assumed a rotational 
temperature that is within $1\sigma$ of the value derived from the fit to the 
population diagram. The transition of C$_2$H$_5$CN with 
$E_{\rm u}/k_{\rm B} \sim 140$~K at $\sim$218390~MHz is most likely severely 
contaminated by emission from an unidentified species. This strongly biases 
the population diagram and prevents a reliable estimate of its rotational 
temperature. We assumed the same temperature as for CH$_3$CN.

\paragraph{IRAS4A1:} We assumed a temperature of 150~K to derive upper 
limits to the column densities of the selected organic molecules.

\paragraph{IRAS4A2:} For all species but two, we used a temperature that is 
consistent within $1\sigma$ with the value derived from the population 
diagram. The temperature used for CH$_3$OCH$_3$ is consistent with the fitted 
rotational temperature within $2.5\sigma$. The fit to the population diagram 
of C$_2$H$_5$CN is unconstrained. We assumed the same temperature as for 
CH$_3$CN. 

\paragraph{IRAS4B:} The population diagrams of CH$_3$OH and HNCO indicate 
temperatures on the order of 300~K, which we adopted for the modeling. The 
population diagrams of CH$_3$OCHO and CH$_3$CHO
also suggest temperatures on the order of 300~K, but such high temperatures
would overestimate some transitions which are not detected. As a compromise,
we adopted a temperature of 200~K for both species. The other population 
diagrams do not constrain the rotational temperatures well and we assumed a 
temperature of 150~K for all other complex species. The low rotational 
temperature derived from the population diagram of C$_2$H$_5$CN is not 
reliable due to the narrow range of $E_{\rm u}/k_{\rm B}$.

\paragraph{IRAS4B2:} We assumed a temperature of 150~K to derive upper 
limits to the column densities of the selected organic molecules.

\paragraph{IRAM04191:} We assumed a temperature of 100~K to derive upper 
limits to the column densities of the selected organic molecules.

\paragraph{L1521F:} We assumed a temperature of 100~K to derive upper 
limits to the column densities of the selected organic molecules.

\paragraph{L1527:} We assumed a temperature of 100~K to derive upper 
limits to the column densities of the selected organic molecules.

\paragraph{SerpM-S68N:} Only three population diagrams could be constructed.
The fit to the population diagram of CH$_3$OH indicates a temperature of 
about 200~K, which we adopted. The population diagram of CH$_3$OCHO does not
constrain the temperature because of the too narrow range of 
$E_{\rm u}/k_{\rm B}$. The one of CH$_3$CN is too noisy to constrain the 
temperature. We assumed a temperature of 150~K for both molecules as well as 
all other selected organic molecules.

\paragraph{SerpM-S68Nb:} We assumed a temperature of 150~K to derive upper 
limits to the column densities of the selected organic molecules.

\paragraph{SerpM-SMM4a:} We assumed a temperature of 150~K to derive upper 
limits to the column densities of the selected organic molecules.

\paragraph{SerpM-SMM4b:} Only two population diagrams could be constructed.
The fit to the population diagram of CH$_3$OH indicates a temperature of 
about 220~K. We adopted a value of 250~K, consistent with the latter within 
$1\sigma$. The population diagram of HNCO does not constrain the temperature 
well. For this molecules and all the other selected organic molecules, we 
assumed a temperature of 250~K.

\paragraph{SerpS-MM18a:} 
The population diagrams of nine molecules could be constructed and fitted. 
The rotational temperature of CH$_3$OH is well constrained, around 150~K. 
This temperature is consistent within 2$\sigma$ with the rotational 
temperatures derived for most of the other molecules. There are three 
exceptions, CH$_3$OCH$_3$, CH$_2$CN, and C$_2$H$_5$CN. The fit of C$_2$H$_5$CN 
is highly uncertain due to the narrow energy range of the covered transitions. 
Therefore, we also assumed a temperature of 150~K for this molecule. The 
population diagram of CH$_3$CN suggest a somewhat higher temperature and we 
adopted 200~K, consistent within $2\sigma$ with the fit result. The population 
diagram of CH$_3$OCH$_3$ suggests a lower temperature, and we used 110~K, 
consistent within $1\sigma$ with the fit result.

\paragraph{SerpS-MM18b:} The populations diagrams of CH$_3$CN and HNCO have
only two points and are unconstrained. The fit to the population diagram of 
methanol yields a low temperature of $\sim 70$~K. However, fitting the spectum 
with such a low temperature predicts that a transition with 
$E_{\rm u}/k_{\rm B} = 40$~K at 230027~MHz should be detected while it is not. As 
a compromise we used a temperature of 120~K for methanol and all other selected 
organic molecules.

\paragraph{SerpS-MM22:} We assumed a temperature of 150~K to derive upper 
limits to the column densities of the selected organic molecules.

\paragraph{L1157:} The population diagram of methanol indicates a temperature
on the order of 200~K, which we adopted for this molecule. The other two 
population diagrams (CH$_3$OCHO and CH$_3$CN) poorly constrain the rotational
temperature of these molecules. However, the non-detection of the 12--11 
transition of CH$_3$CN at 220.4 GHz is not consistent with a temperature of 
200~K or higher. We thus adopted a temperature of 150~K for this molecule as
well as for CH$_3$OCHO. The column density upper limits for the other complex
organic molecules were derived with this temperature.

\paragraph{GF9-2:} We assumed a temperature of 100~K to derive upper 
limits to the column densities of the selected organic molecules.

\clearpage 

\clearpage 

\clearpage 

\clearpage 

\clearpage 

\clearpage 

\clearpage 
\begin{figure}[!htbp]
\centerline{\resizebox{0.83\hsize}{!}{\includegraphics[angle=0]{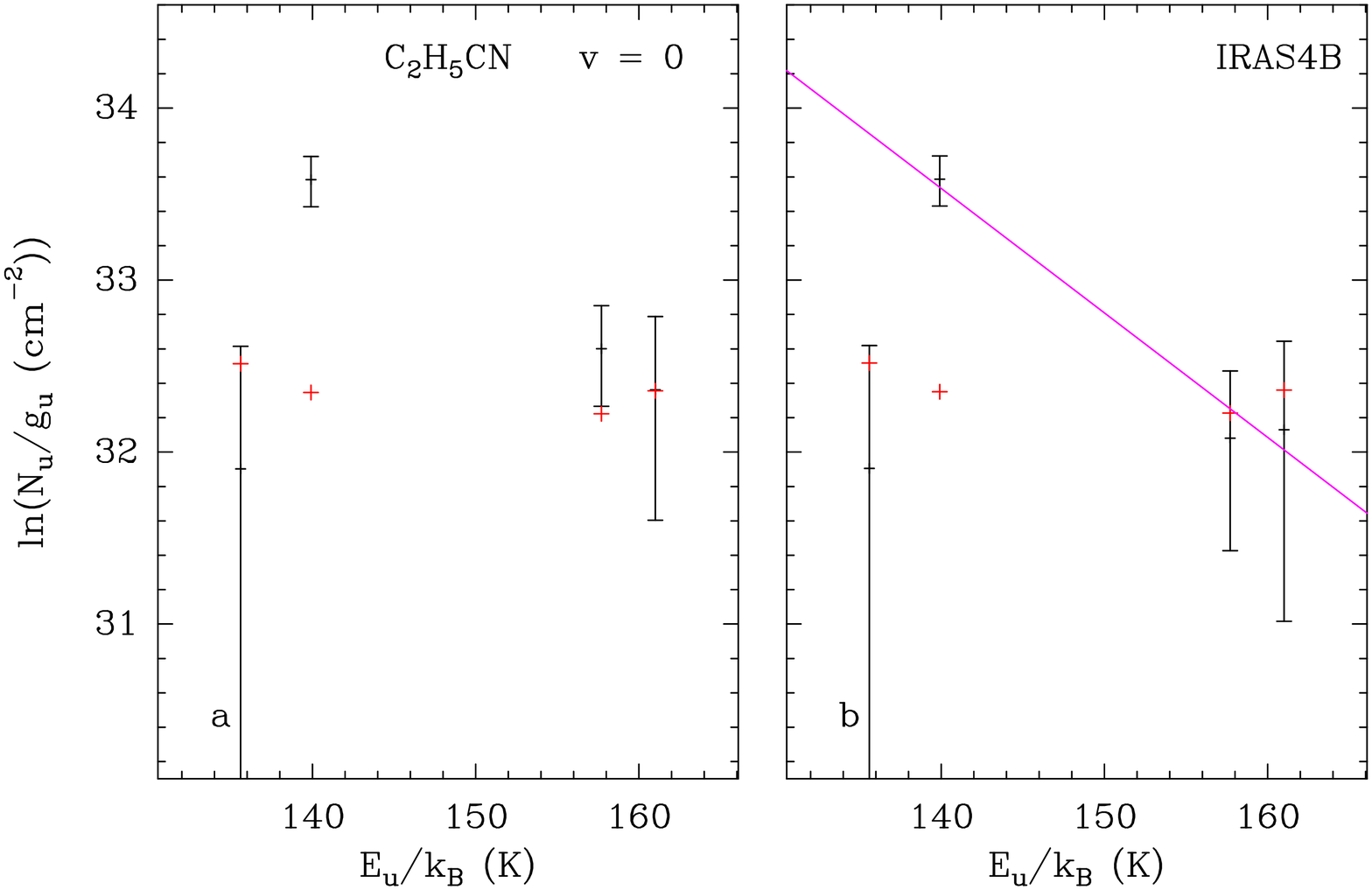}}}
\caption{Same as Fig.~\ref{f:popdiag_l1448-2a_p1_ch3oh} for C$_2$H$_5$CN in IRAS4B.}
\label{f:popdiag_n1333-irs4b_p1_c2h5cn}
\end{figure}

\begin{figure}[!htbp]
\centerline{\resizebox{0.83\hsize}{!}{\includegraphics[angle=0]{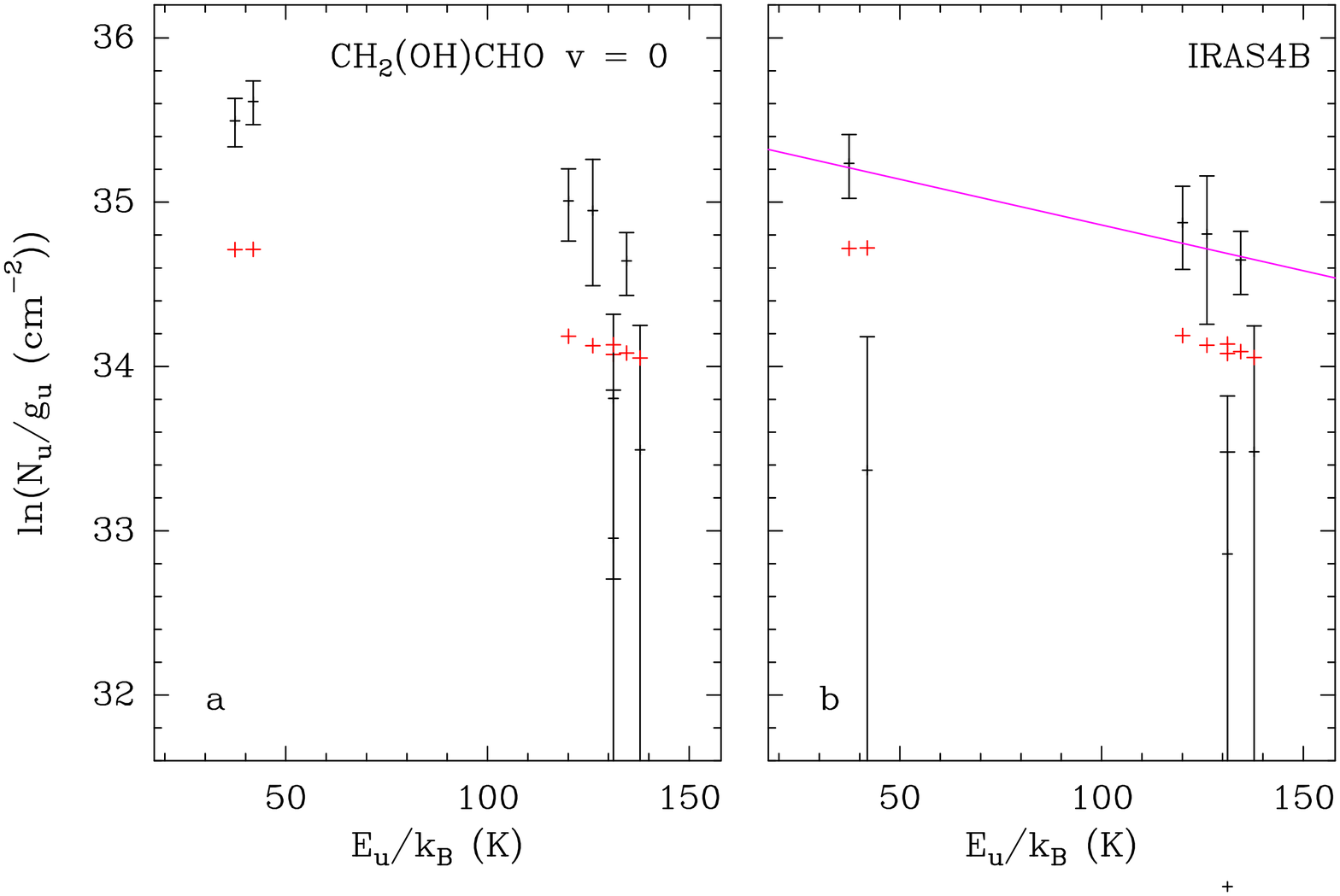}}}
\caption{Same as Fig.~\ref{f:popdiag_l1448-2a_p1_ch3oh} for CH$_2$(OH)CHO in IRAS4B.}
\label{f:popdiag_n1333-irs4b_p1_ch2ohcho}
\end{figure}

\begin{figure}[!htbp]
\centerline{\resizebox{0.83\hsize}{!}{\includegraphics[angle=0]{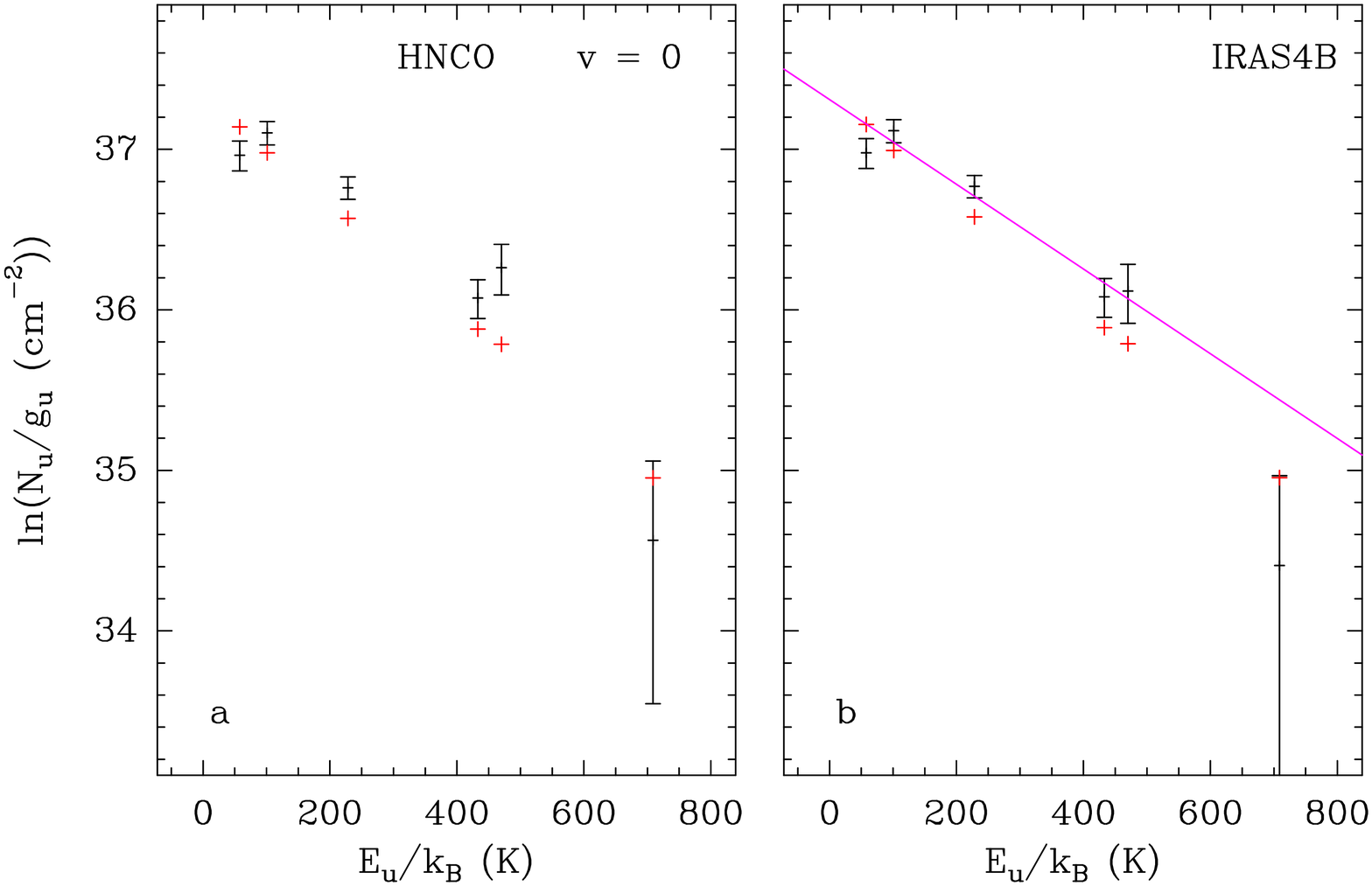}}}
\caption{Same as Fig.~\ref{f:popdiag_l1448-2a_p1_ch3oh} for HNCO in IRAS4B.}
\label{f:popdiag_n1333-irs4b_p1_hnco}
\end{figure}

\begin{figure}[!htbp]
\centerline{\resizebox{0.83\hsize}{!}{\includegraphics[angle=0]{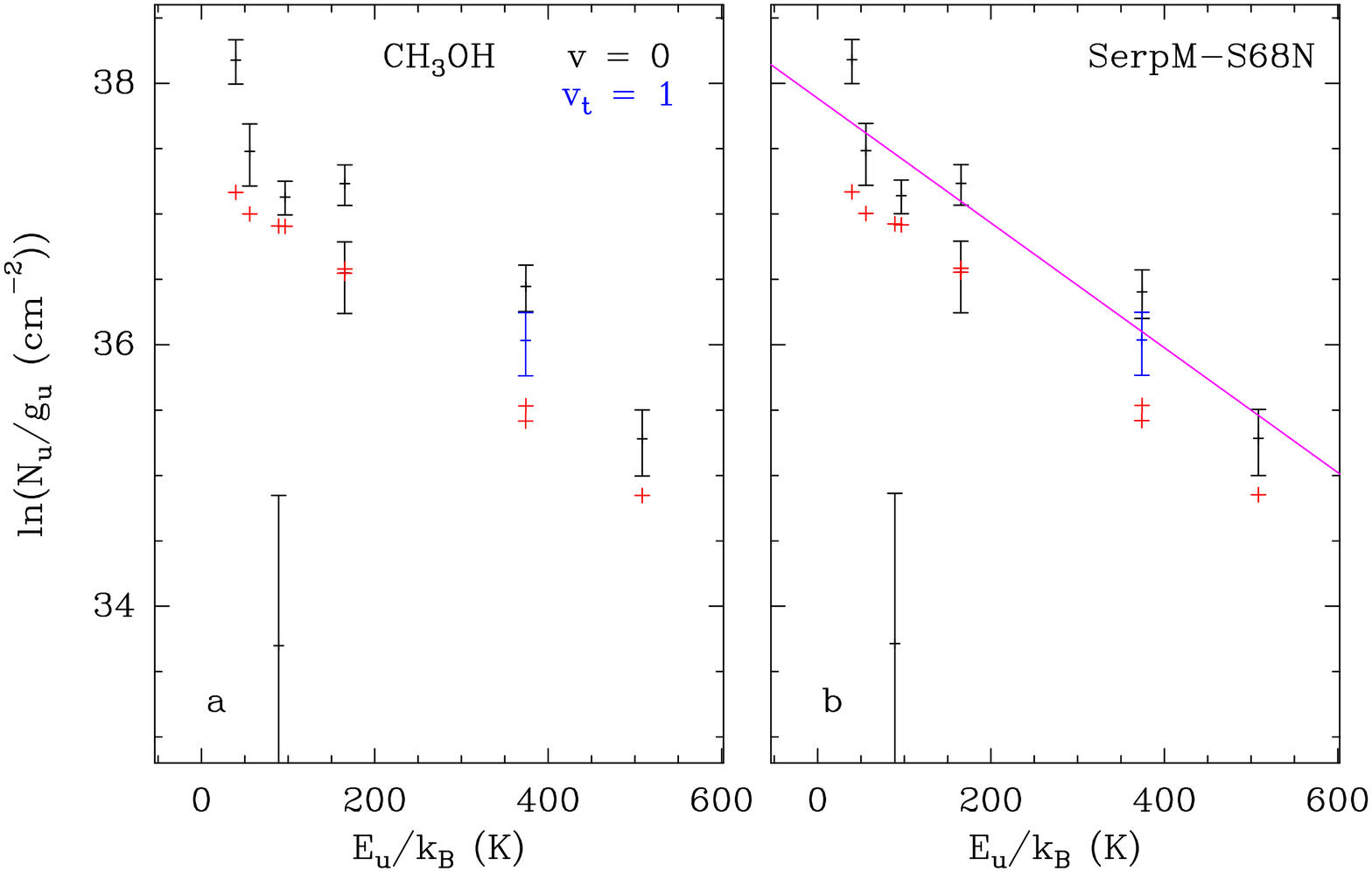}}}
\caption{Same as Fig.~\ref{f:popdiag_l1448-2a_p1_ch3oh} for CH$_3$OH in SerpM-S68N.}
\label{f:popdiag_serp-s68n_p1_ch3oh}
\end{figure}

\begin{figure}[!htbp]
\centerline{\resizebox{0.83\hsize}{!}{\includegraphics[angle=0]{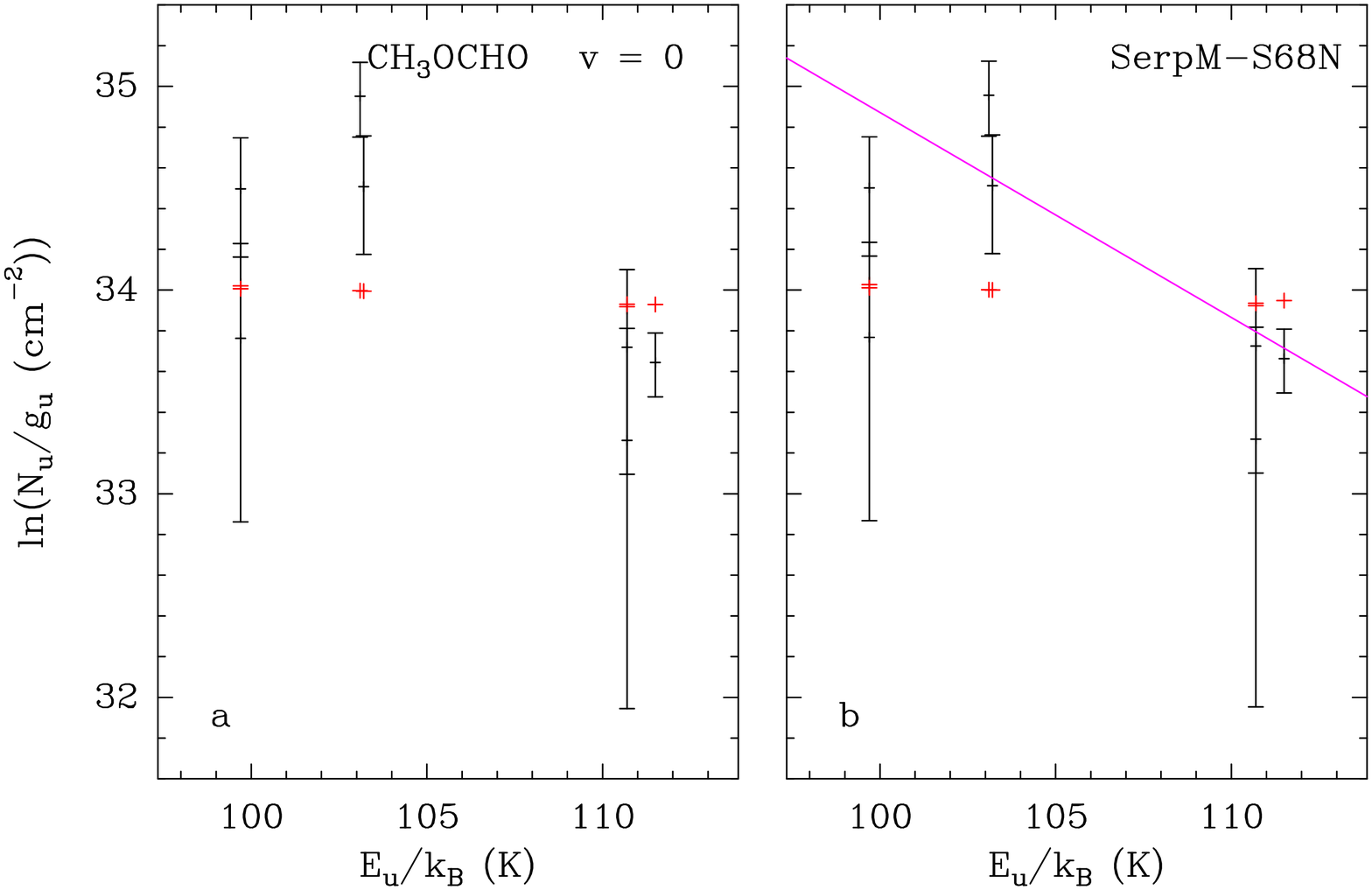}}}
\caption{Same as Fig.~\ref{f:popdiag_l1448-2a_p1_ch3oh} for CH$_3$OCHO in SerpM-S68N.}
\label{f:popdiag_serp-s68n_p1_ch3ocho}
\end{figure}

\begin{figure}[!htbp]
\centerline{\resizebox{0.83\hsize}{!}{\includegraphics[angle=0]{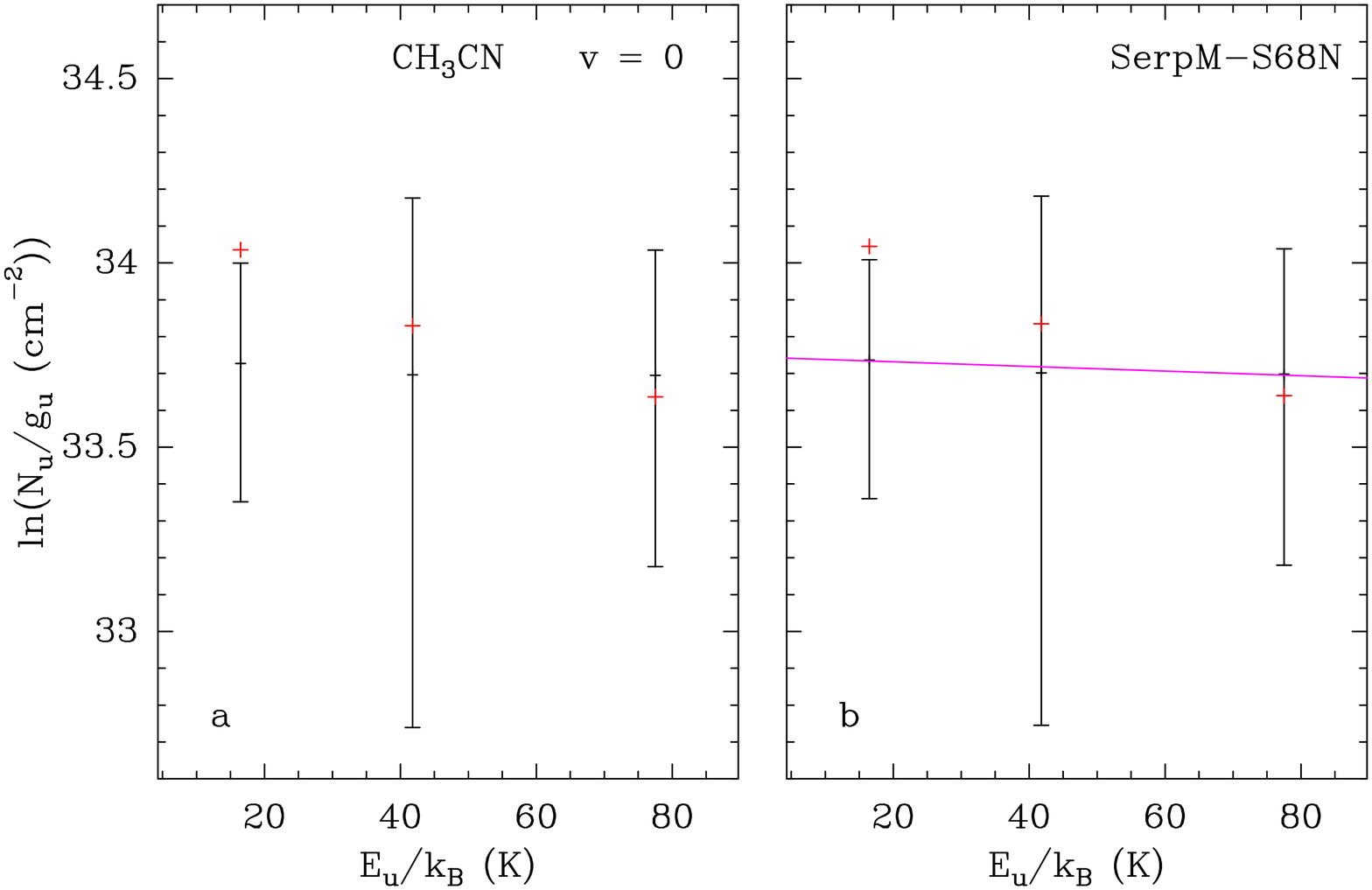}}}
\caption{Same as Fig.~\ref{f:popdiag_l1448-2a_p1_ch3oh} for CH$_3$CN in SerpM-S68N.}
\label{f:popdiag_serp-s68n_p1_ch3cn}
\end{figure}

\begin{figure}[!htbp]
\centerline{\resizebox{0.83\hsize}{!}{\includegraphics[angle=0]{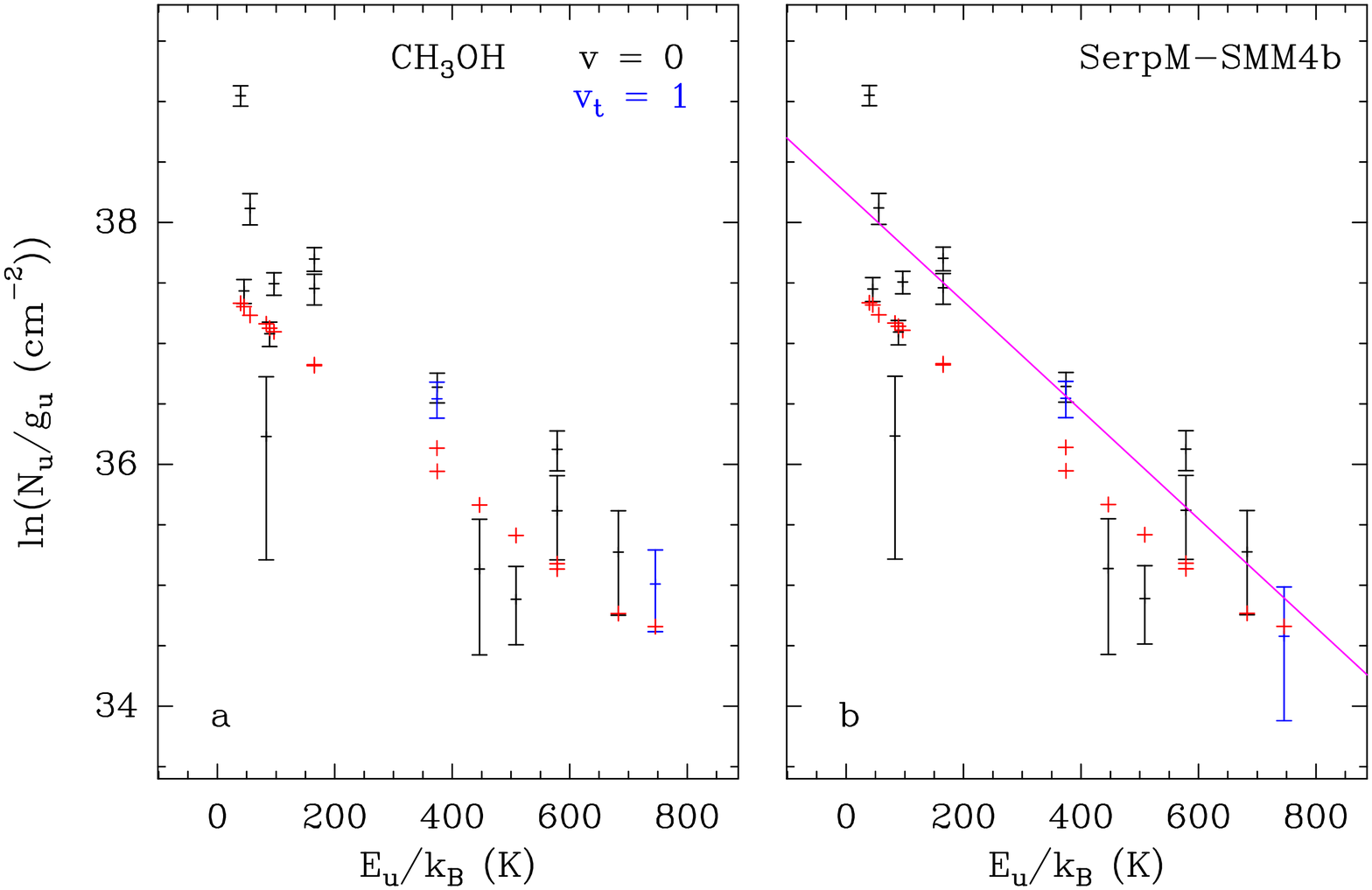}}}
\caption{Same as Fig.~\ref{f:popdiag_l1448-2a_p1_ch3oh} for CH$_3$OH in SerpM-SMM4b.}
\label{f:popdiag_serp-smm4_p2_ch3oh}
\end{figure}

\begin{figure}[!htbp]
\centerline{\resizebox{0.83\hsize}{!}{\includegraphics[angle=0]{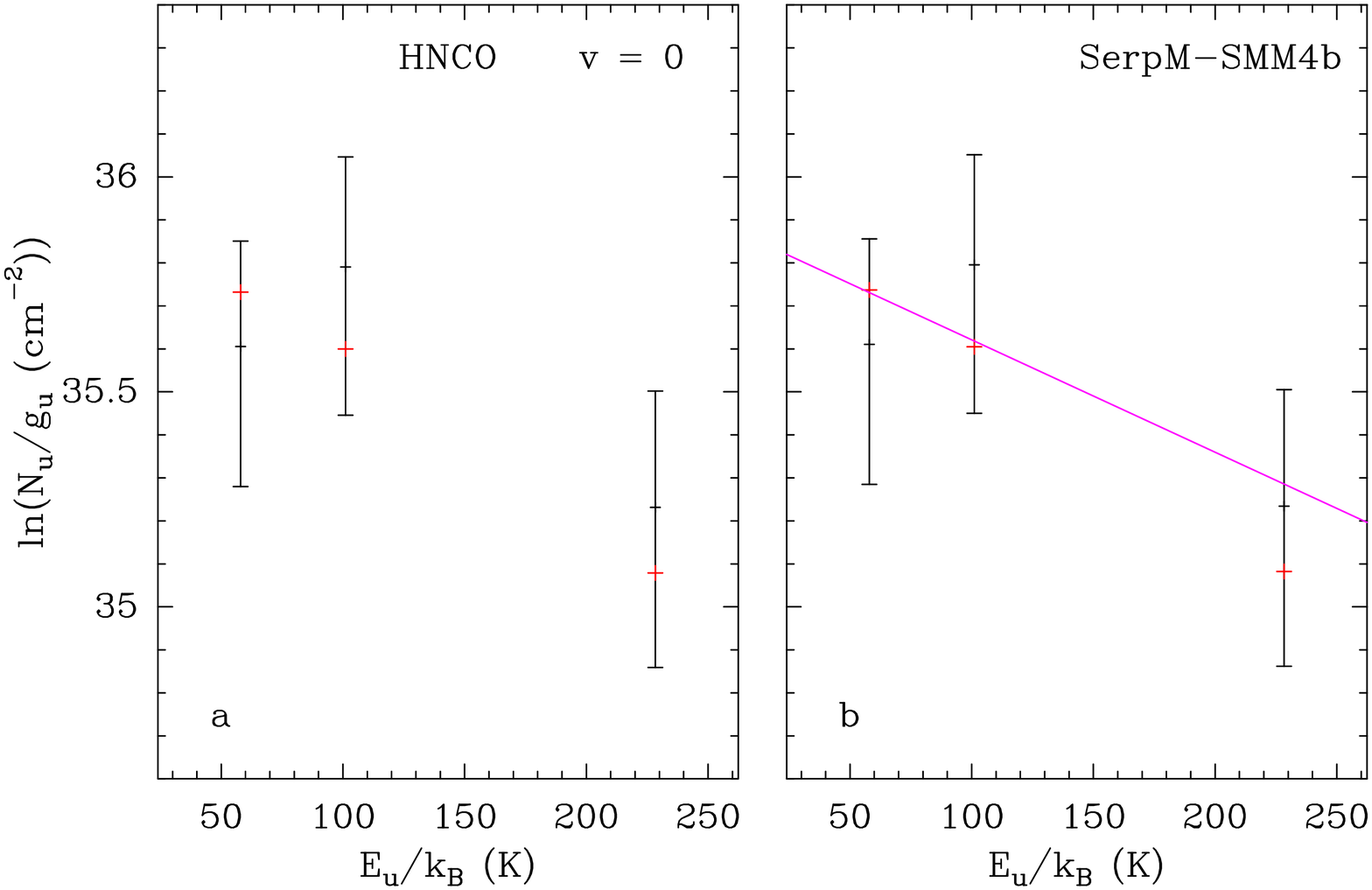}}}
\caption{Same as Fig.~\ref{f:popdiag_l1448-2a_p1_ch3oh} for HNCO in SerpM-SMM4b.}
\label{f:popdiag_serp-smm4_p2_hnco}
\end{figure}

\clearpage 
\begin{figure}[!htbp]
\centerline{\resizebox{0.83\hsize}{!}{\includegraphics[angle=0]{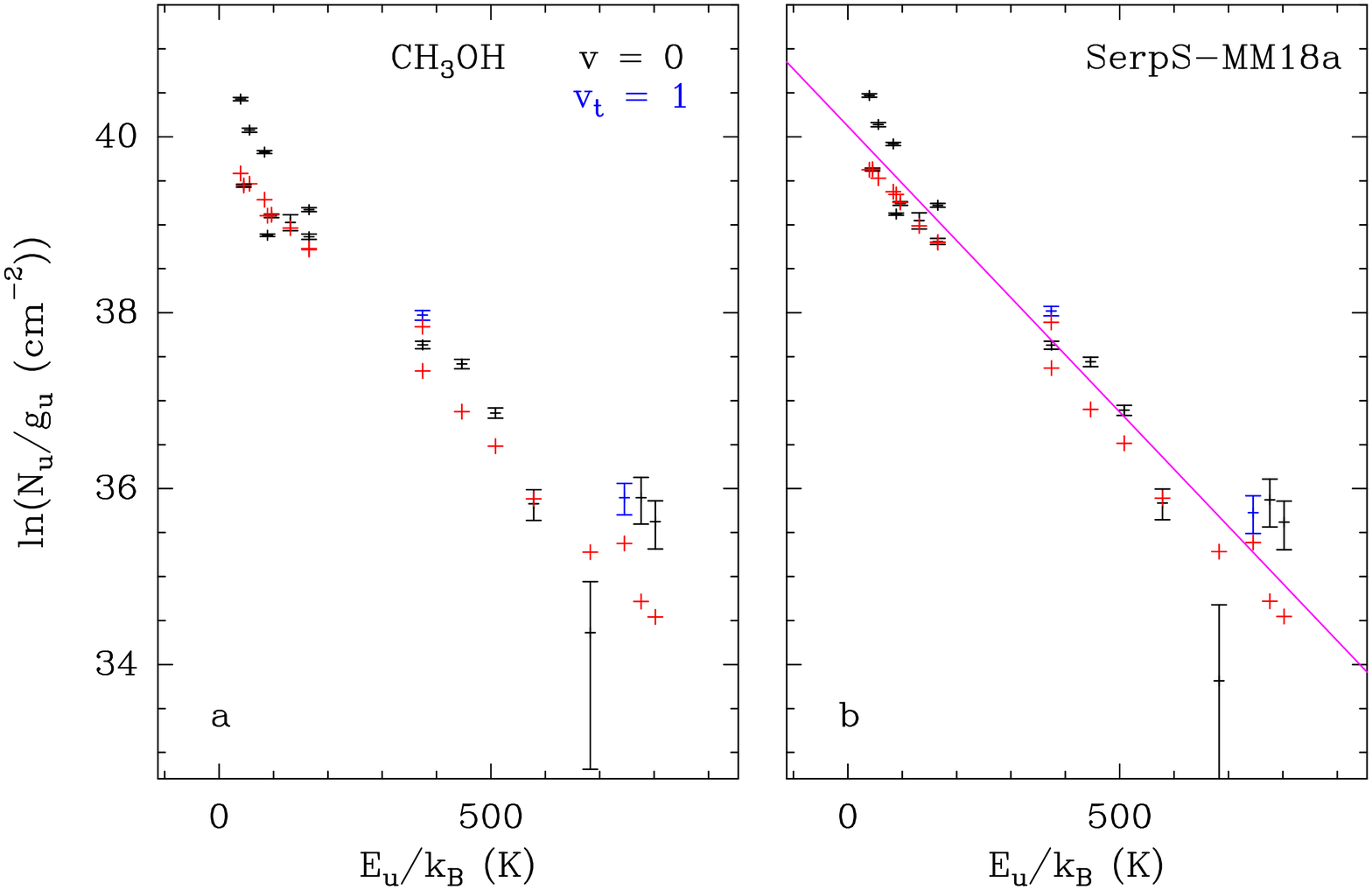}}}
\caption{Same as Fig.~\ref{f:popdiag_l1448-2a_p1_ch3oh} for CH$_3$OH in SerpS-MM18a.}
\label{f:popdiag_aqu-mms1_p1_ch3oh}
\end{figure}

\begin{figure}[!htbp]
\centerline{\resizebox{0.83\hsize}{!}{\includegraphics[angle=0]{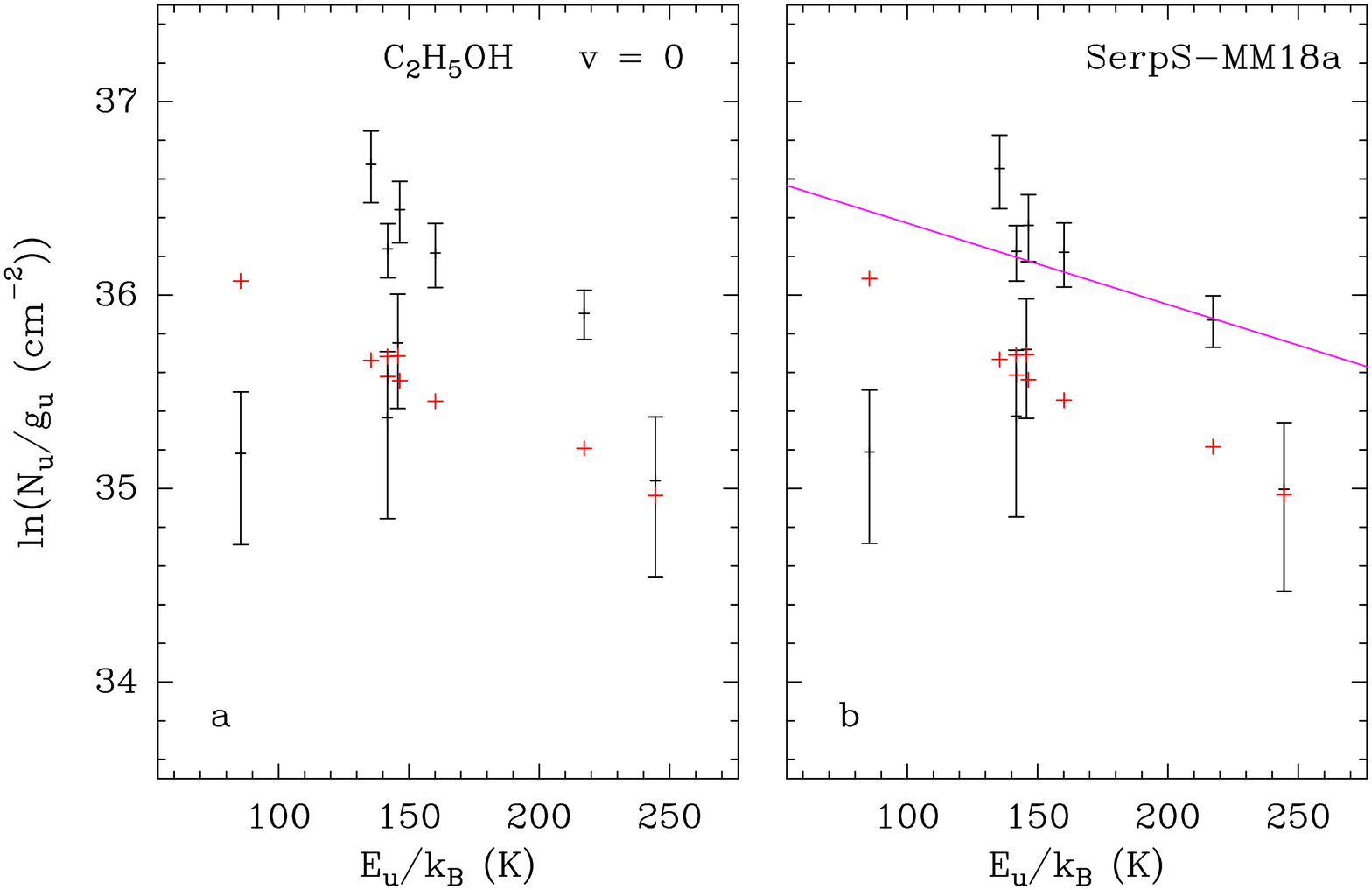}}}
\caption{Same as Fig.~\ref{f:popdiag_l1448-2a_p1_ch3oh} for C$_2$H$_5$OH in SerpS-MM18a.}
\label{f:popdiag_aqu-mms1_p1_c2h5oh}
\end{figure}

\begin{figure}[!htbp]
\centerline{\resizebox{0.83\hsize}{!}{\includegraphics[angle=0]{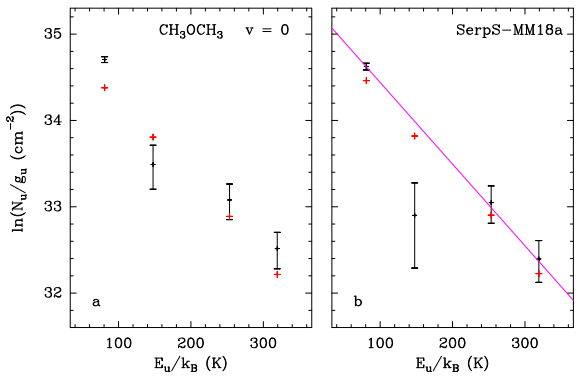}}}
\caption{Same as Fig.~\ref{f:popdiag_l1448-2a_p1_ch3oh} for CH$_3$OCH$_3$ in SerpS-MM18a.}
\label{f:popdiag_aqu-mms1_p1_ch3och3}
\end{figure}

\begin{figure}[!htbp]
\centerline{\resizebox{0.83\hsize}{!}{\includegraphics[angle=0]{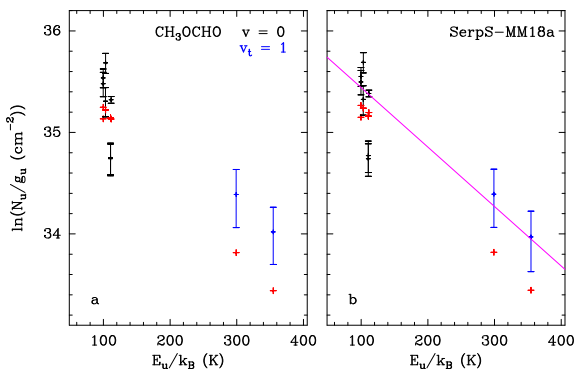}}}
\caption{Same as Fig.~\ref{f:popdiag_l1448-2a_p1_ch3oh} for CH$_3$OCHO in SerpS-MM18a.}
\label{f:popdiag_aqu-mms1_p1_ch3ocho}
\end{figure}

\begin{figure}[!htbp]
\centerline{\resizebox{0.83\hsize}{!}{\includegraphics[angle=0]{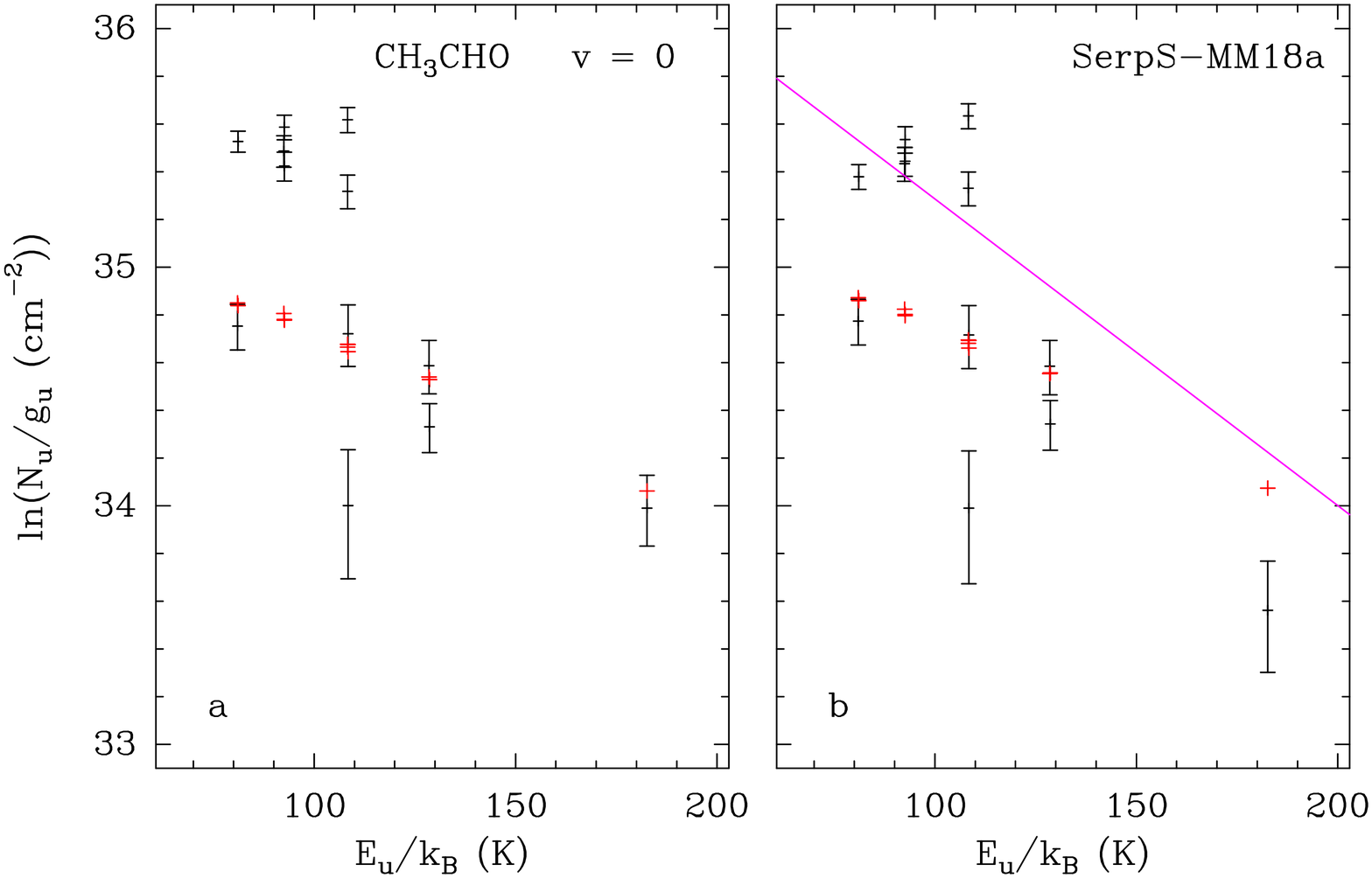}}}
\caption{Same as Fig.~\ref{f:popdiag_l1448-2a_p1_ch3oh} for CH$_3$CHO in SerpS-MM18a.}
\label{f:popdiag_aqu-mms1_p1_ch3cho}
\end{figure}

\begin{figure}[!htbp]
\centerline{\resizebox{0.83\hsize}{!}{\includegraphics[angle=0]{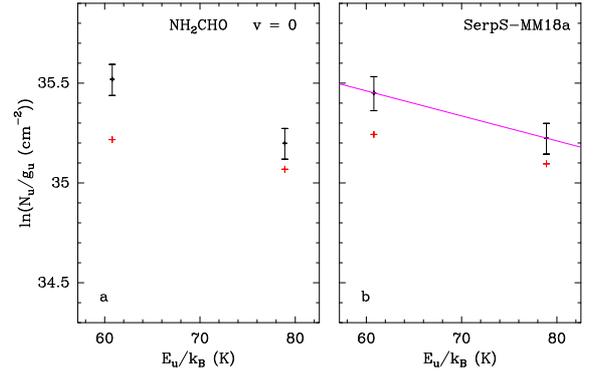}}}
\caption{Same as Fig.~\ref{f:popdiag_l1448-2a_p1_ch3oh} for NH$_2$CHO in SerpS-MM18a.}
\label{f:popdiag_aqu-mms1_p1_nh2cho}
\end{figure}

\begin{figure}[!htbp]
\centerline{\resizebox{0.83\hsize}{!}{\includegraphics[angle=0]{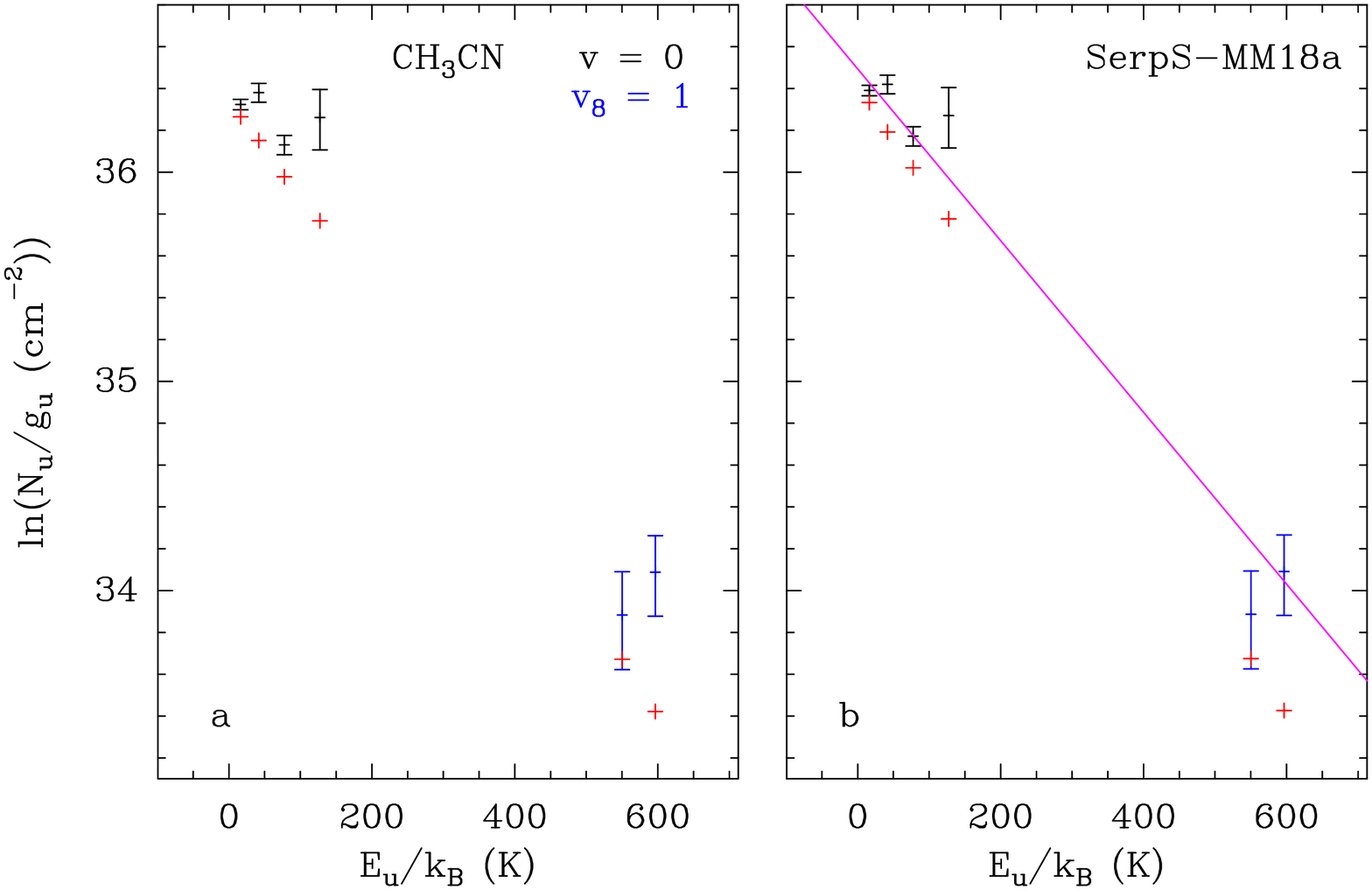}}}
\caption{Same as Fig.~\ref{f:popdiag_l1448-2a_p1_ch3oh} for CH$_3$CN in SerpS-MM18a.}
\label{f:popdiag_aqu-mms1_p1_ch3cn}
\end{figure}

\begin{figure}[!htbp]
\centerline{\resizebox{0.83\hsize}{!}{\includegraphics[angle=0]{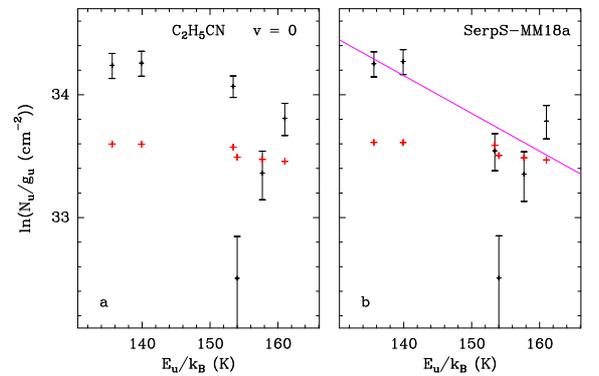}}}
\caption{Same as Fig.~\ref{f:popdiag_l1448-2a_p1_ch3oh} for C$_2$H$_5$CN in SerpS-MM18a.}
\label{f:popdiag_aqu-mms1_p1_c2h5cn}
\end{figure}

\clearpage 
\begin{figure}[!htbp]
\centerline{\resizebox{0.83\hsize}{!}{\includegraphics[angle=0]{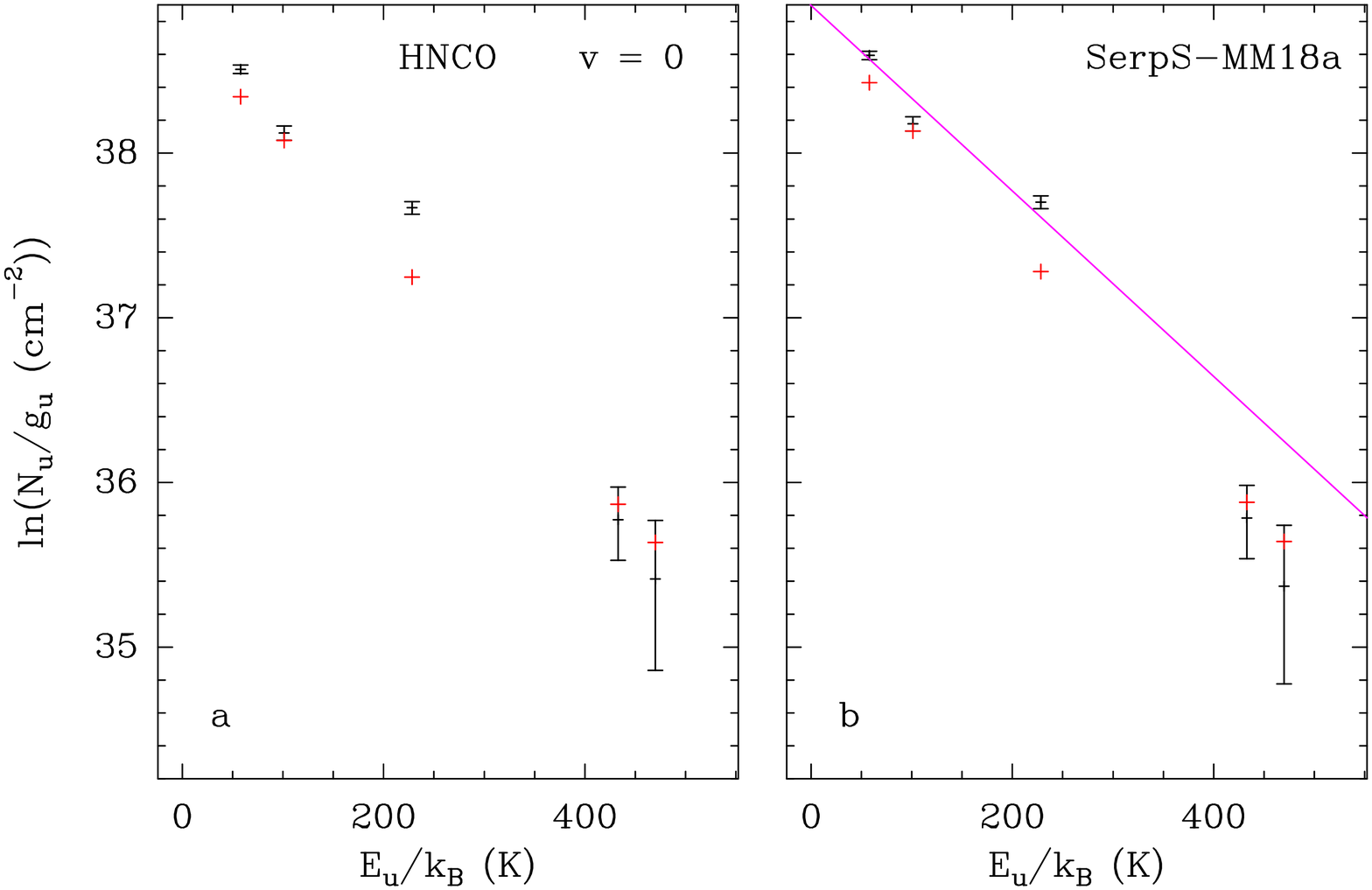}}}
\caption{Same as Fig.~\ref{f:popdiag_l1448-2a_p1_ch3oh} for HNCO in SerpS-MM18a.}
\label{f:popdiag_aqu-mms1_p1_hnco}
\end{figure}

\begin{figure}[!htbp]
\centerline{\resizebox{0.83\hsize}{!}{\includegraphics[angle=0]{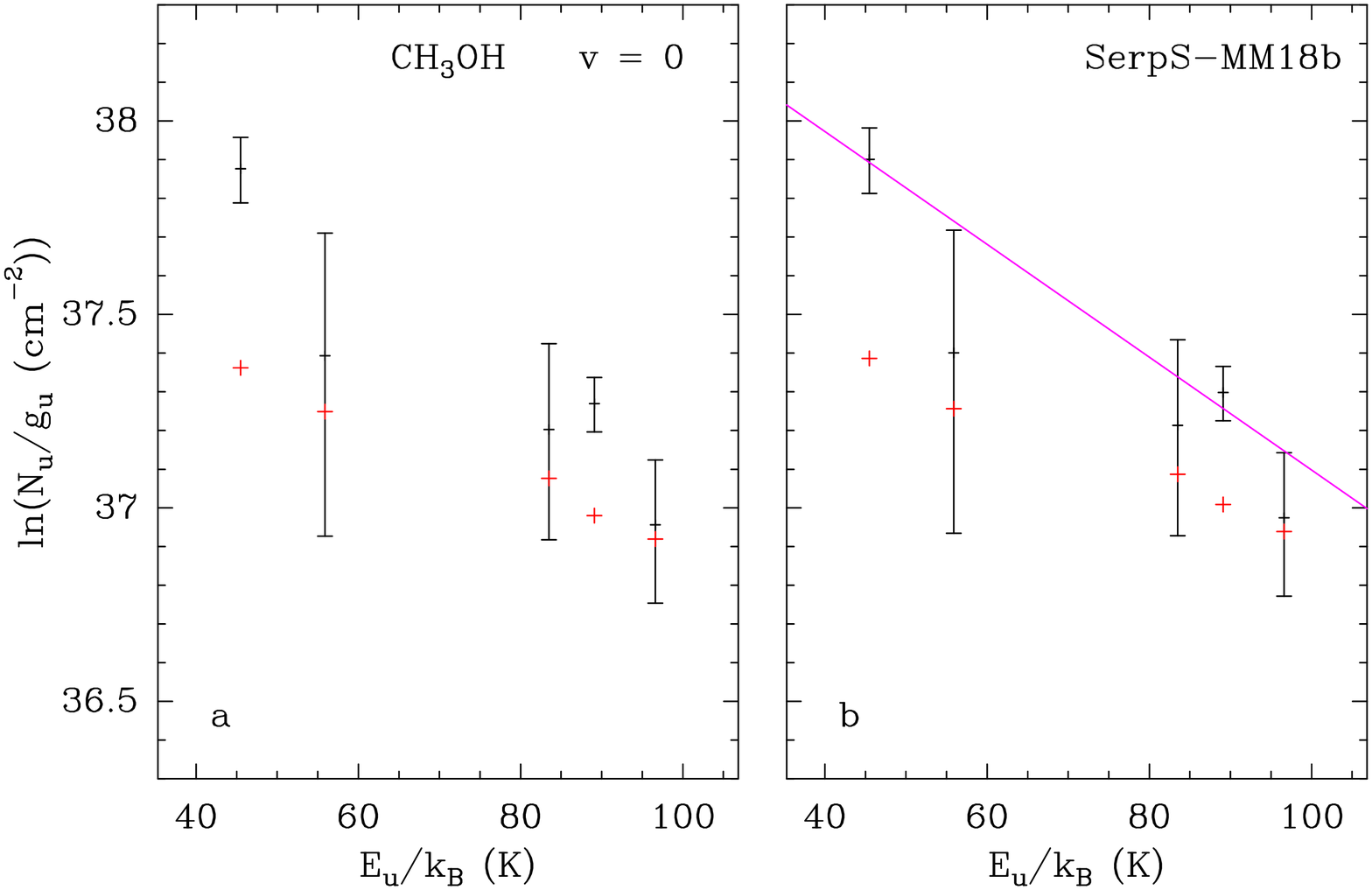}}}
\caption{Same as Fig.~\ref{f:popdiag_l1448-2a_p1_ch3oh} for CH$_3$OH in SerpS-MM18b.}
\label{f:popdiag_aqu-mms1_p2_ch3oh}
\end{figure}

\begin{figure}[!htbp]
\centerline{\resizebox{0.83\hsize}{!}{\includegraphics[angle=0]{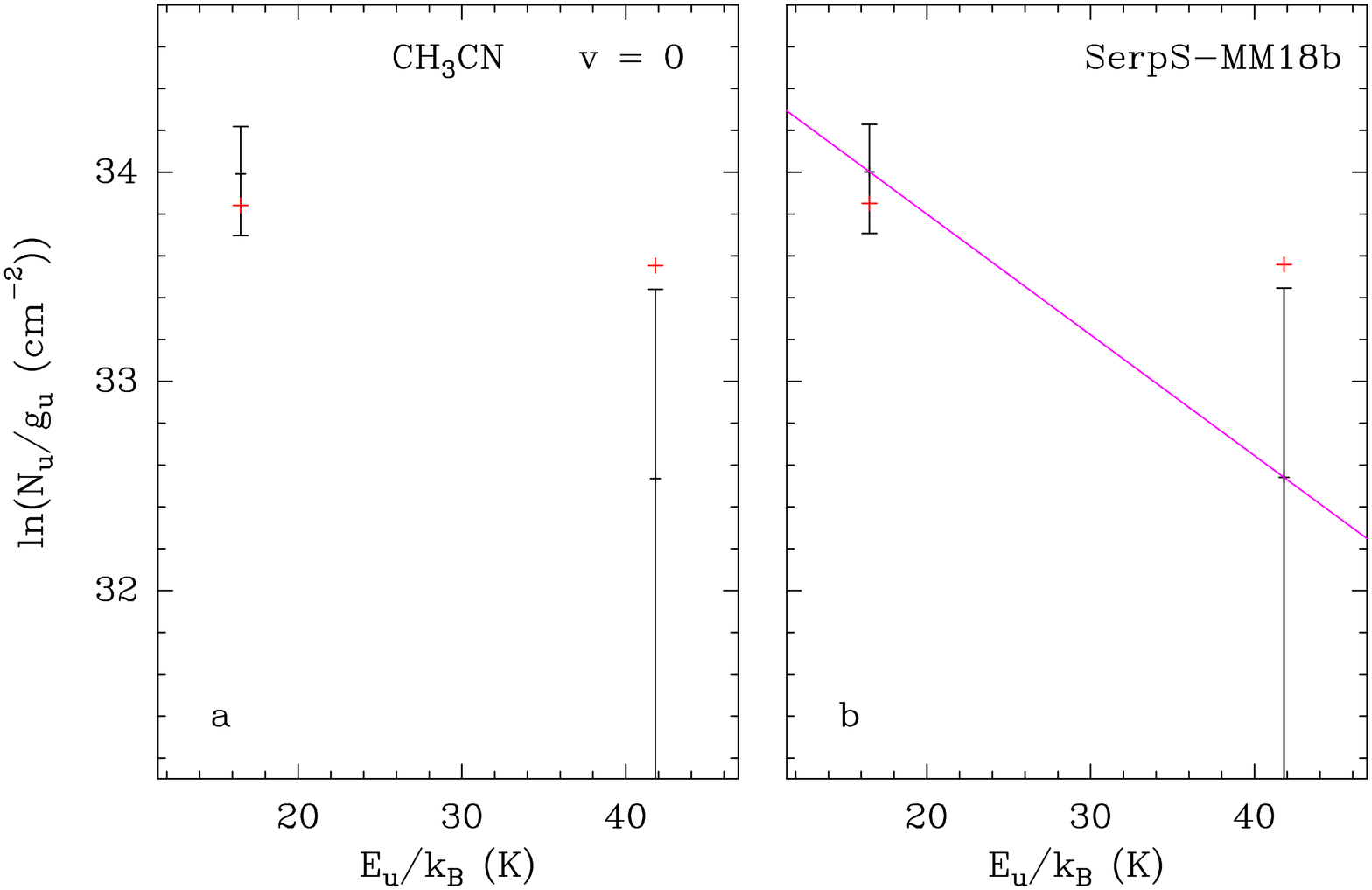}}}
\caption{Same as Fig.~\ref{f:popdiag_l1448-2a_p1_ch3oh} for CH$_3$CN in SerpS-MM18b.}
\label{f:popdiag_aqu-mms1_p2_ch3cn}
\end{figure}

\begin{figure}[!htbp]
\centerline{\resizebox{0.83\hsize}{!}{\includegraphics[angle=0]{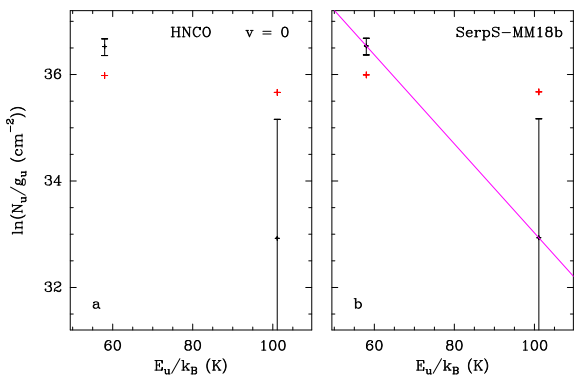}}}
\caption{Same as Fig.~\ref{f:popdiag_l1448-2a_p1_ch3oh} for HNCO in SerpS-MM18b.}
\label{f:popdiag_aqu-mms1_p2_hnco}
\end{figure}

\begin{figure}[!htbp]
\centerline{\resizebox{0.83\hsize}{!}{\includegraphics[angle=0]{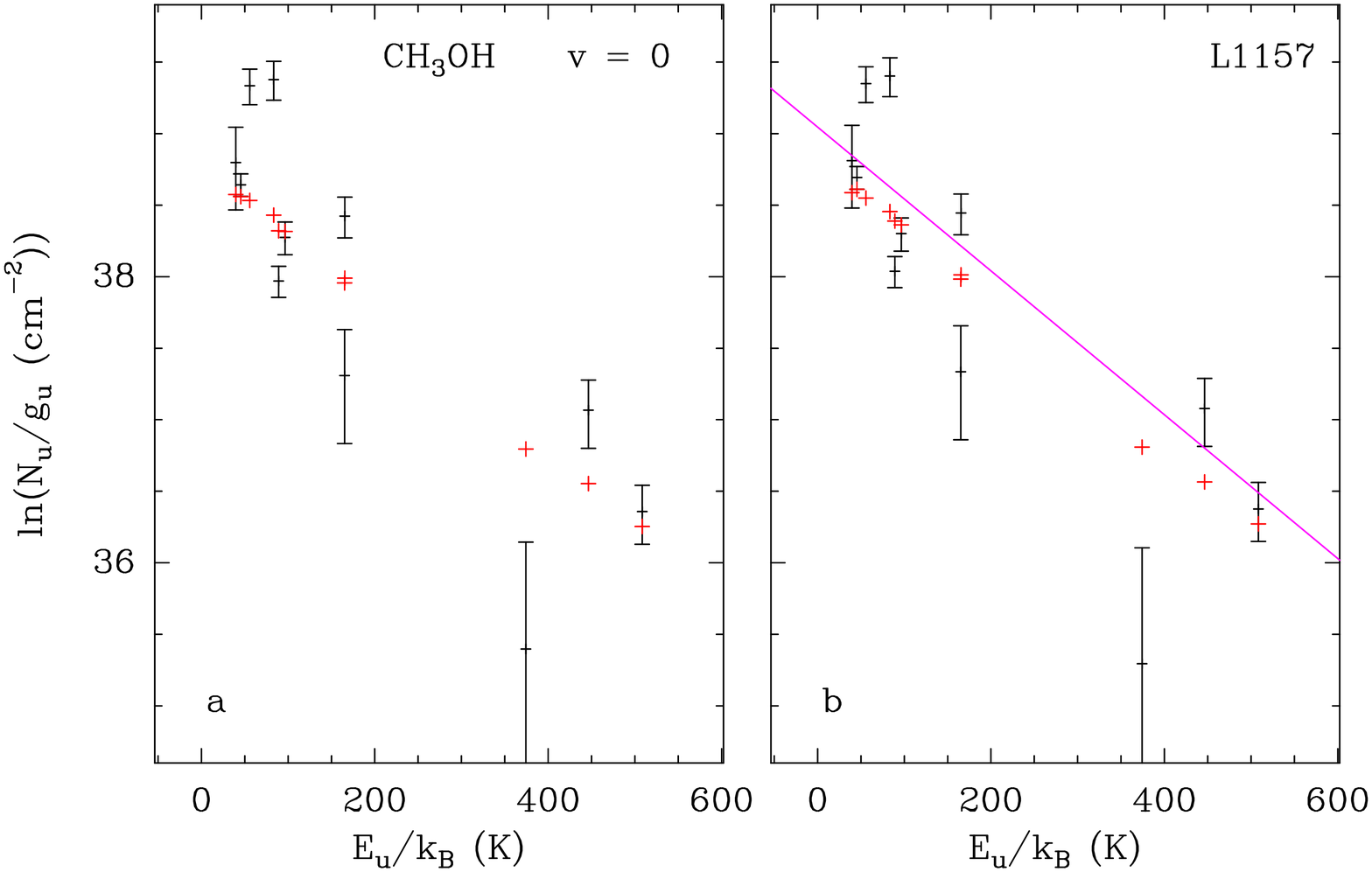}}}
\caption{Same as Fig.~\ref{f:popdiag_l1448-2a_p1_ch3oh} for CH$_3$OH in L1157.}
\label{f:popdiag_l1157_p1_ch3oh}
\end{figure}

\begin{figure}[!htbp]
\centerline{\resizebox{0.83\hsize}{!}{\includegraphics[angle=0]{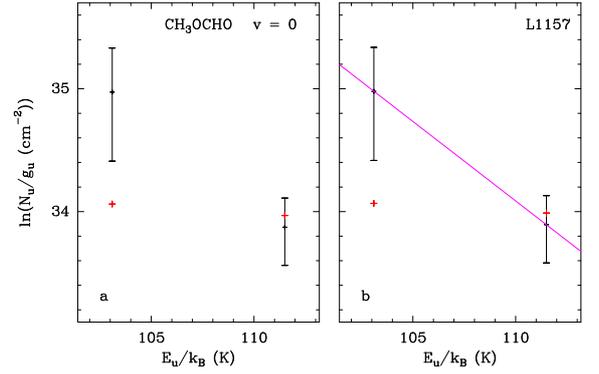}}}
\caption{Same as Fig.~\ref{f:popdiag_l1448-2a_p1_ch3oh} for CH$_3$OCHO in L1157.}
\label{f:popdiag_l1157_p1_ch3ocho}
\end{figure}

\begin{figure}[!htbp]
\centerline{\resizebox{0.83\hsize}{!}{\includegraphics[angle=0]{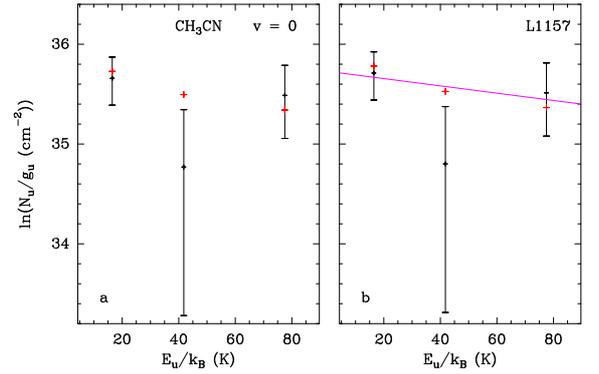}}}
\caption{Same as Fig.~\ref{f:popdiag_l1448-2a_p1_ch3oh} for CH$_3$CN in L1157.}
\label{f:popdiag_l1157_p1_ch3cn}
\end{figure}

\begin{figure}[!htbp]
\centerline{\resizebox{0.75\hsize}{!}{\includegraphics[angle=0]{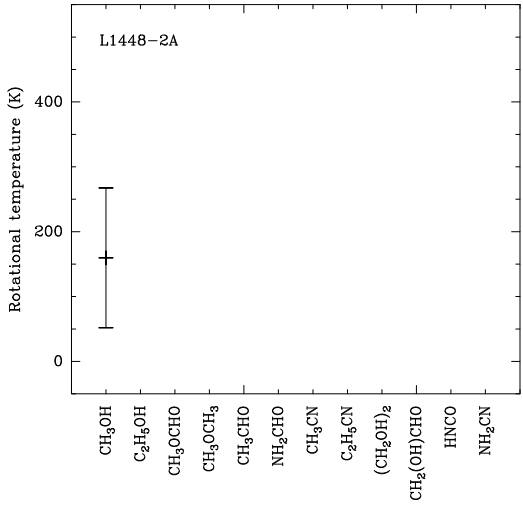}}}
\caption{Rotational temperatures of (complex) organic molecules derived toward L1448-2A. The uncertainties represent the standard deviation of the rotational temperature fit to each population diagram.}
\label{f:trot_l1448-2a_p1}
\end{figure}

\begin{figure}[!htbp]
\centerline{\resizebox{0.75\hsize}{!}{\includegraphics[angle=0]{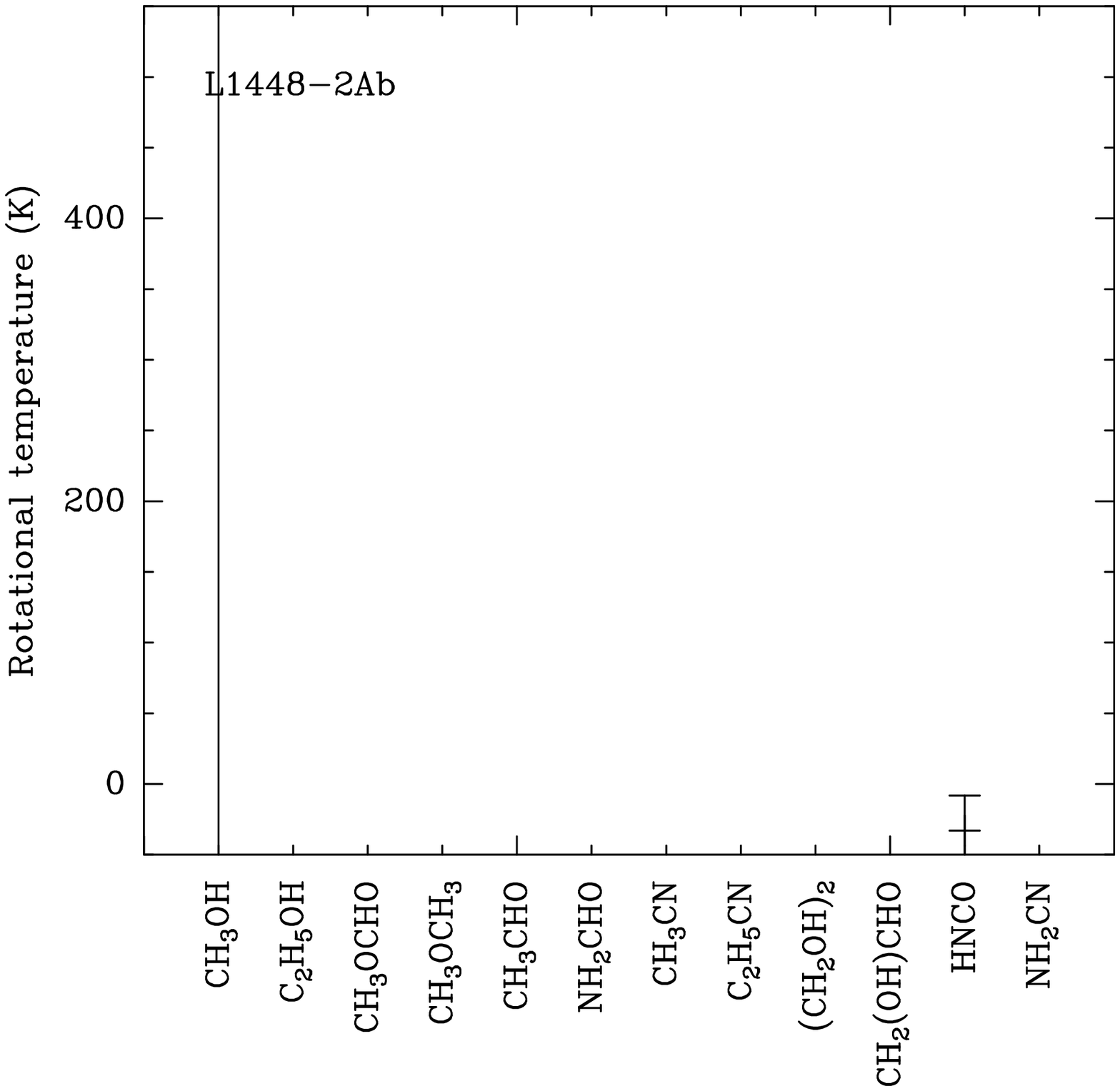}}}
\caption{Same as Fig.~\ref{f:trot_l1448-2a_p1} for L1448-2Ab.}
\label{f:trot_l1448-2a_p2}
\end{figure}

\begin{figure}[!htbp]
\centerline{\resizebox{0.75\hsize}{!}{\includegraphics[angle=0]{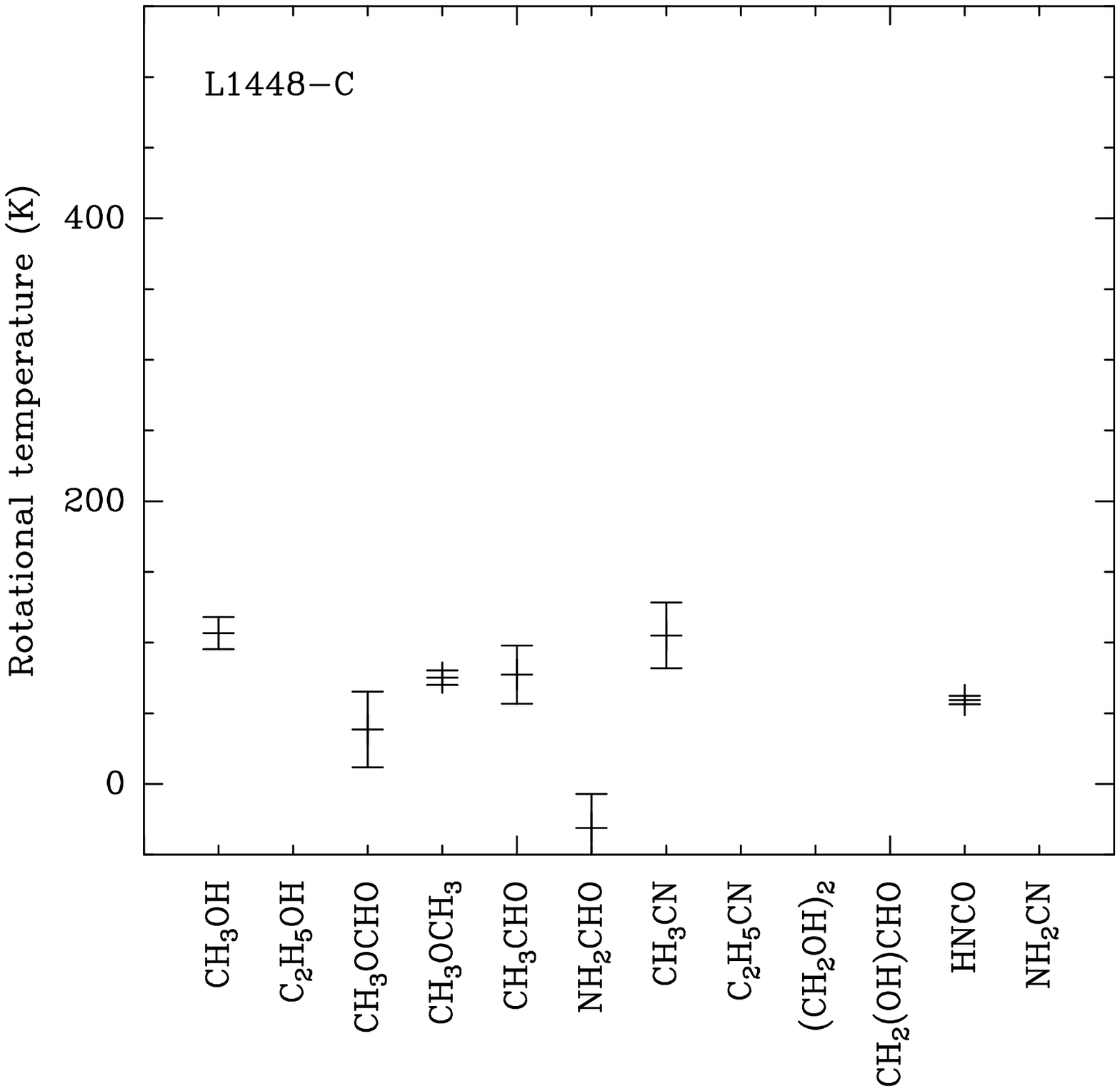}}}
\caption{Same as Fig.~\ref{f:trot_l1448-2a_p1} for L1448-C.}
\label{f:trot_l1448-c_p1}
\end{figure}

\begin{figure}[!htbp]
\centerline{\resizebox{0.75\hsize}{!}{\includegraphics[angle=0]{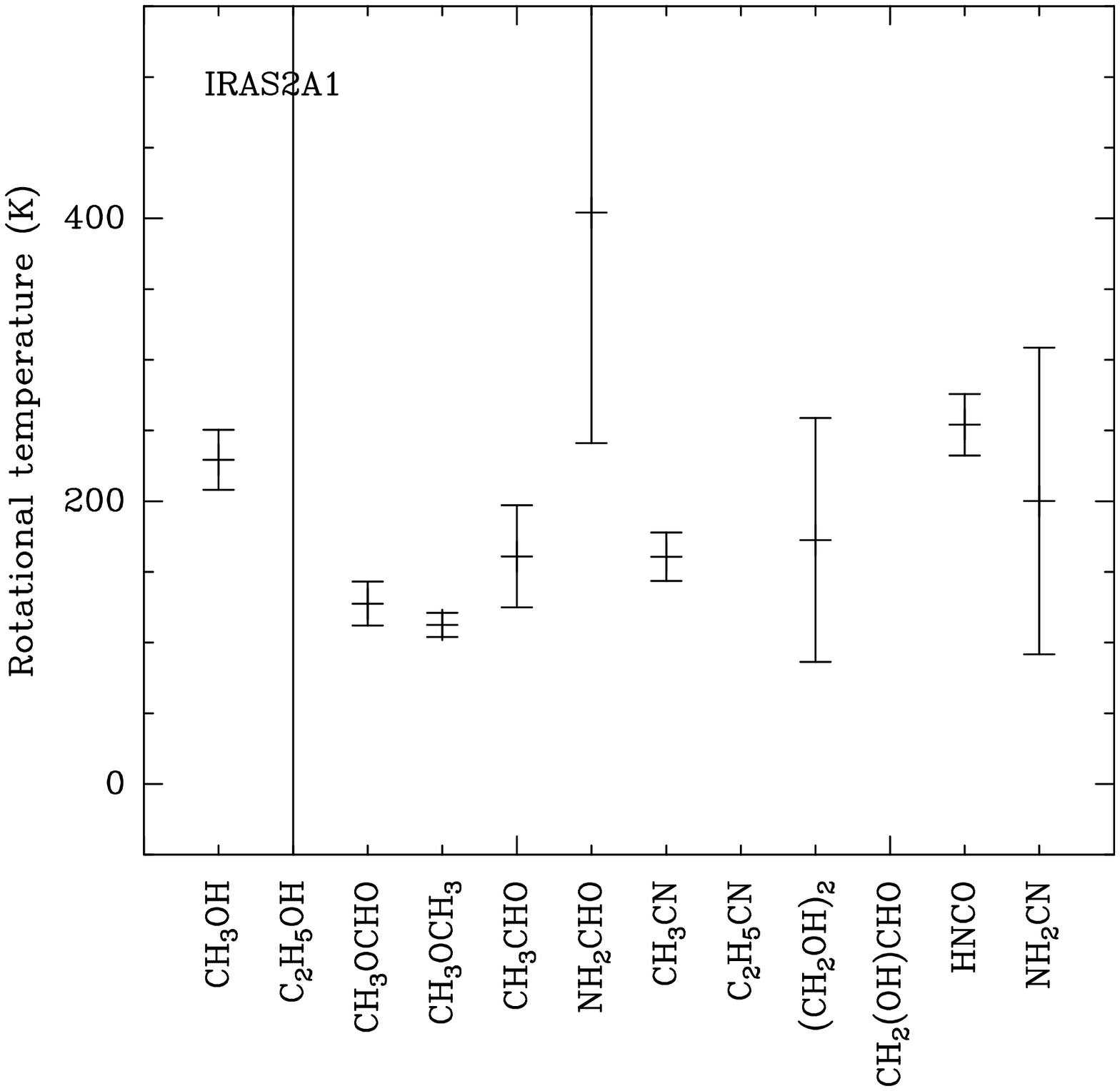}}}
\caption{Same as Fig.~\ref{f:trot_l1448-2a_p1} for IRAS2A1.}
\label{f:trot_n1333-irs2a_p1}
\end{figure}

\begin{figure}[!htbp]
\centerline{\resizebox{0.75\hsize}{!}{\includegraphics[angle=0]{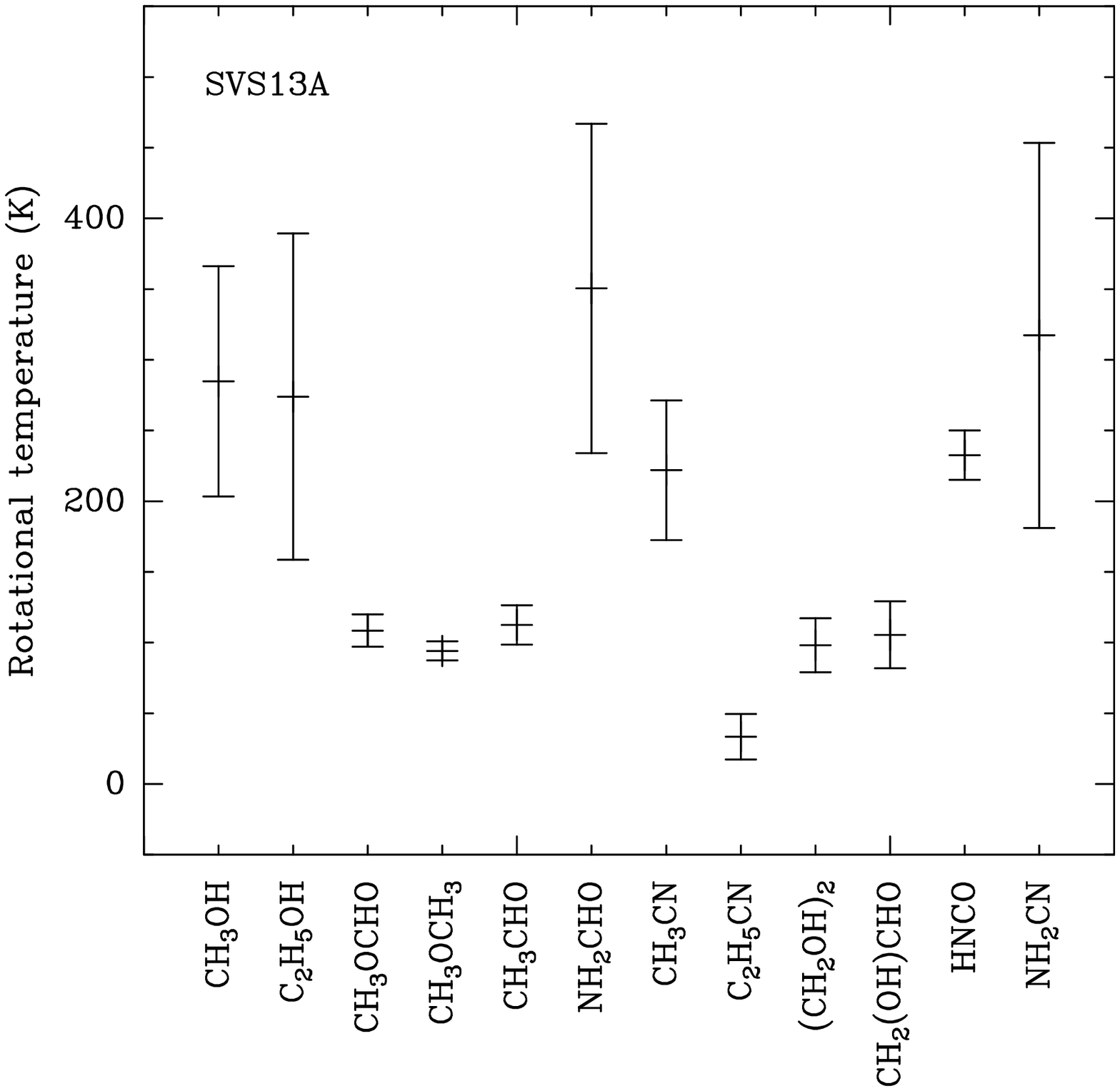}}}
\caption{Same as Fig.~\ref{f:trot_l1448-2a_p1} for SVS13A.}
\label{f:trot_svs13-b_p1}
\end{figure}

\begin{figure}[!htbp]
\centerline{\resizebox{0.75\hsize}{!}{\includegraphics[angle=0]{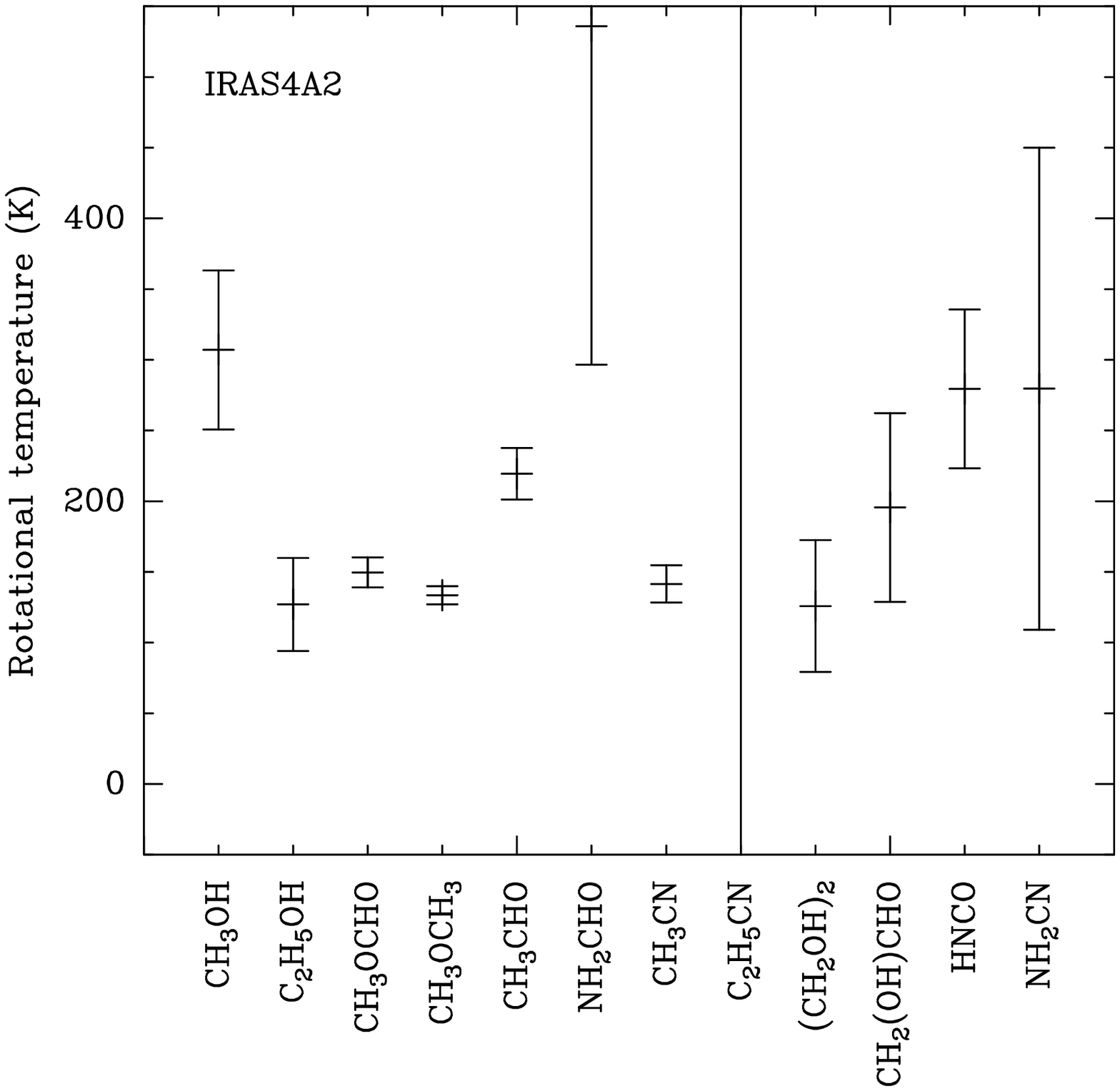}}}
\caption{Same as Fig.~\ref{f:trot_l1448-2a_p1} for IRAS4A2.}
\label{f:trot_n1333-irs4a_p2}
\end{figure}

\clearpage 
\begin{figure}[!htbp]
\centerline{\resizebox{0.75\hsize}{!}{\includegraphics[angle=0]{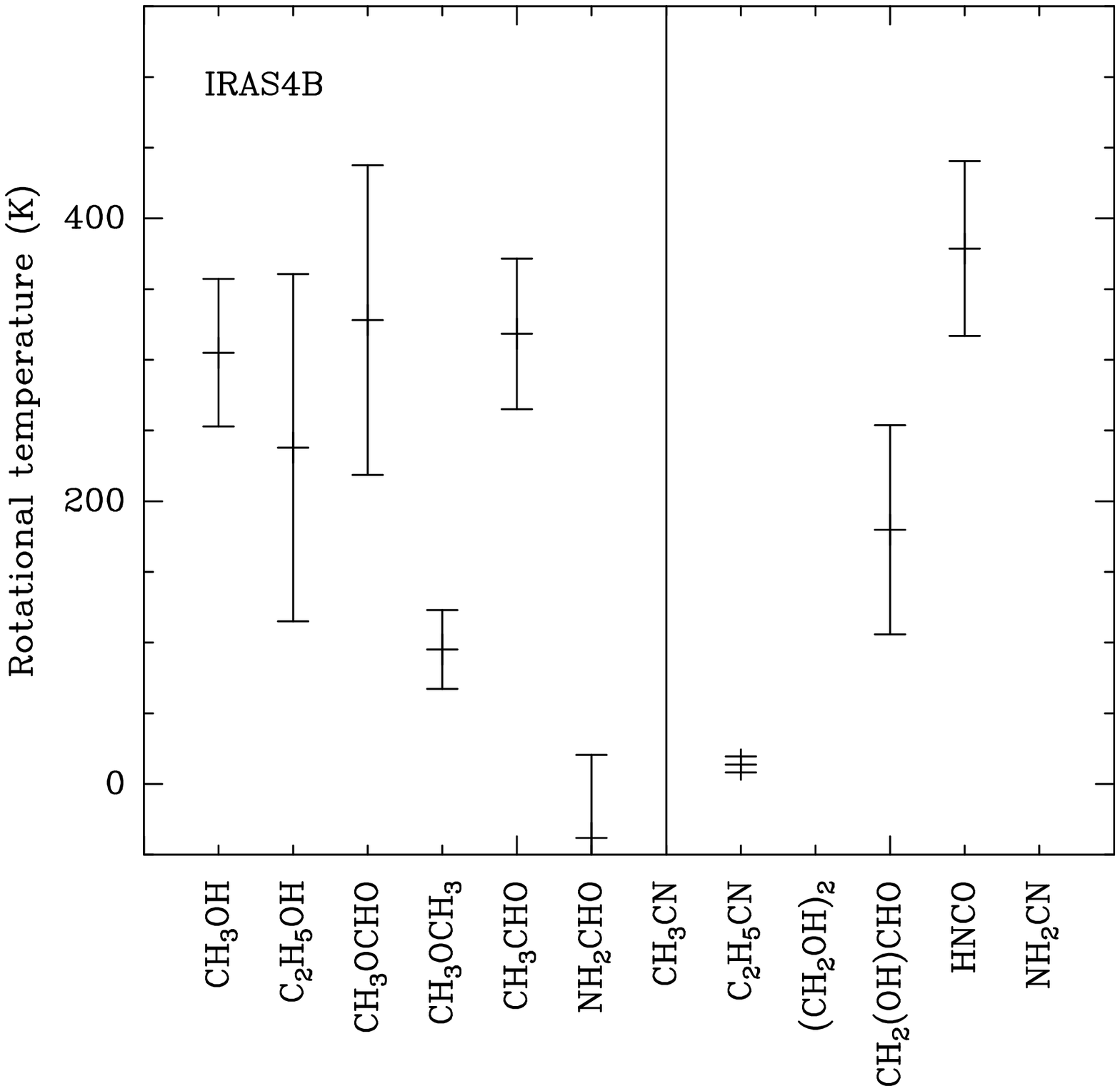}}}
\caption{Same as Fig.~\ref{f:trot_l1448-2a_p1} for IRAS4B.}
\label{f:trot_n1333-irs4b_p1}
\end{figure}

\begin{figure}[!htbp]
\centerline{\resizebox{0.75\hsize}{!}{\includegraphics[angle=0]{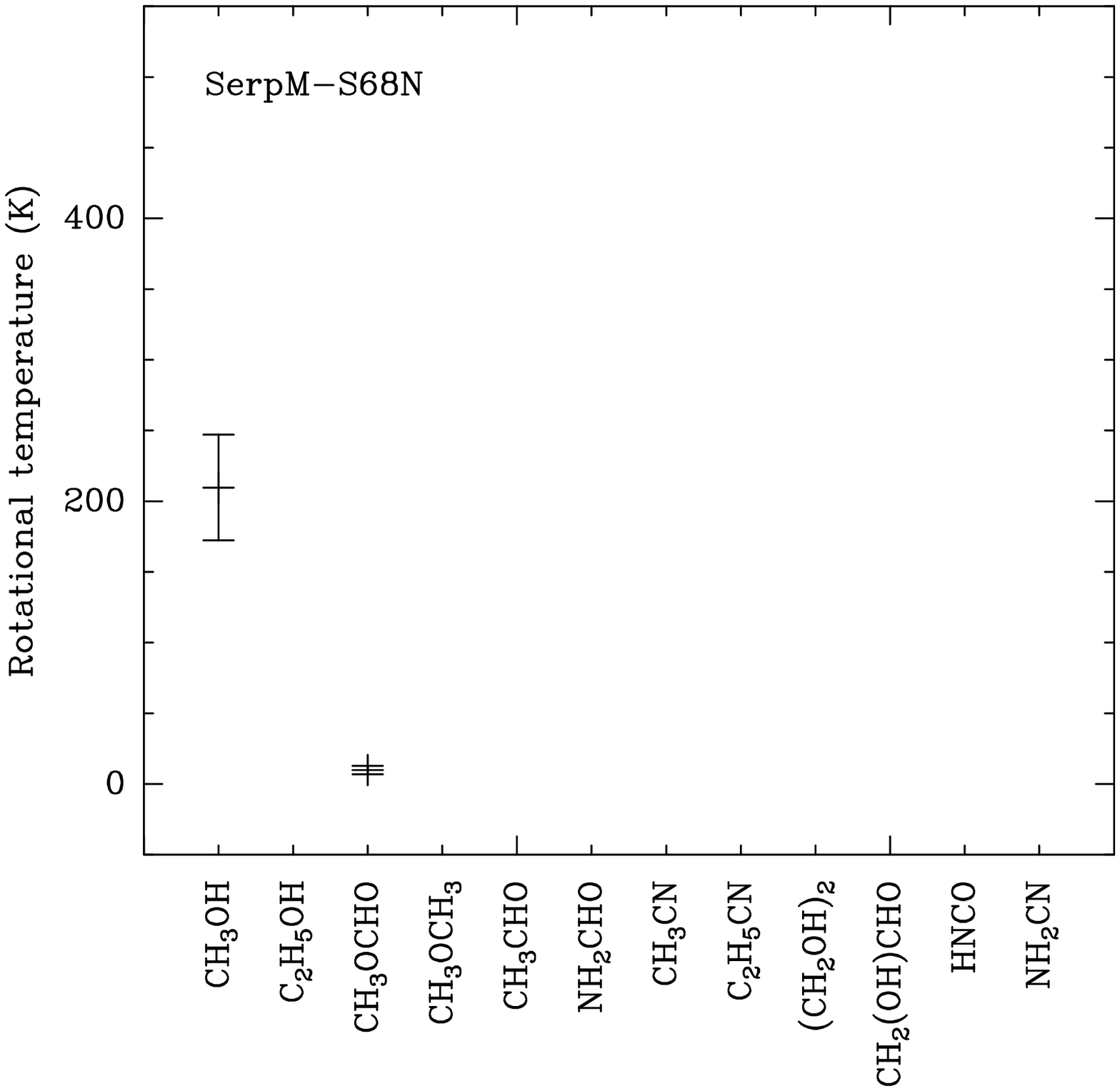}}}
\caption{Same as Fig.~\ref{f:trot_l1448-2a_p1} for SerpM-S68N.}
\label{f:trot_serp-s68n_p1}
\end{figure}

\begin{figure}[!htbp]
\centerline{\resizebox{0.75\hsize}{!}{\includegraphics[angle=0]{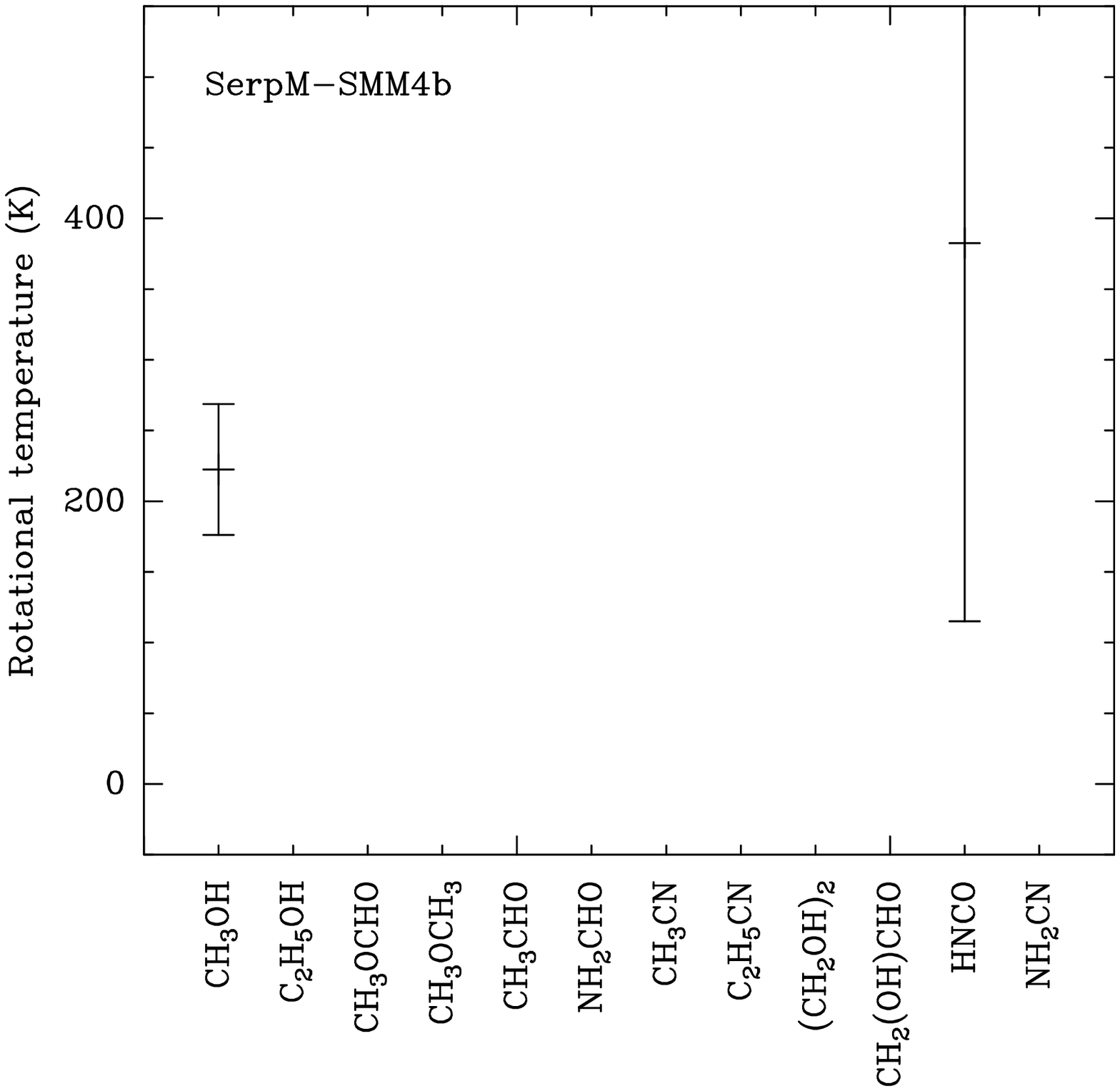}}}
\caption{Same as Fig.~\ref{f:trot_l1448-2a_p1} for SerpM-SMM4b.}
\label{f:trot_serp-smm4_p2}
\end{figure}

\begin{figure}[!htbp]
\centerline{\resizebox{0.75\hsize}{!}{\includegraphics[angle=0]{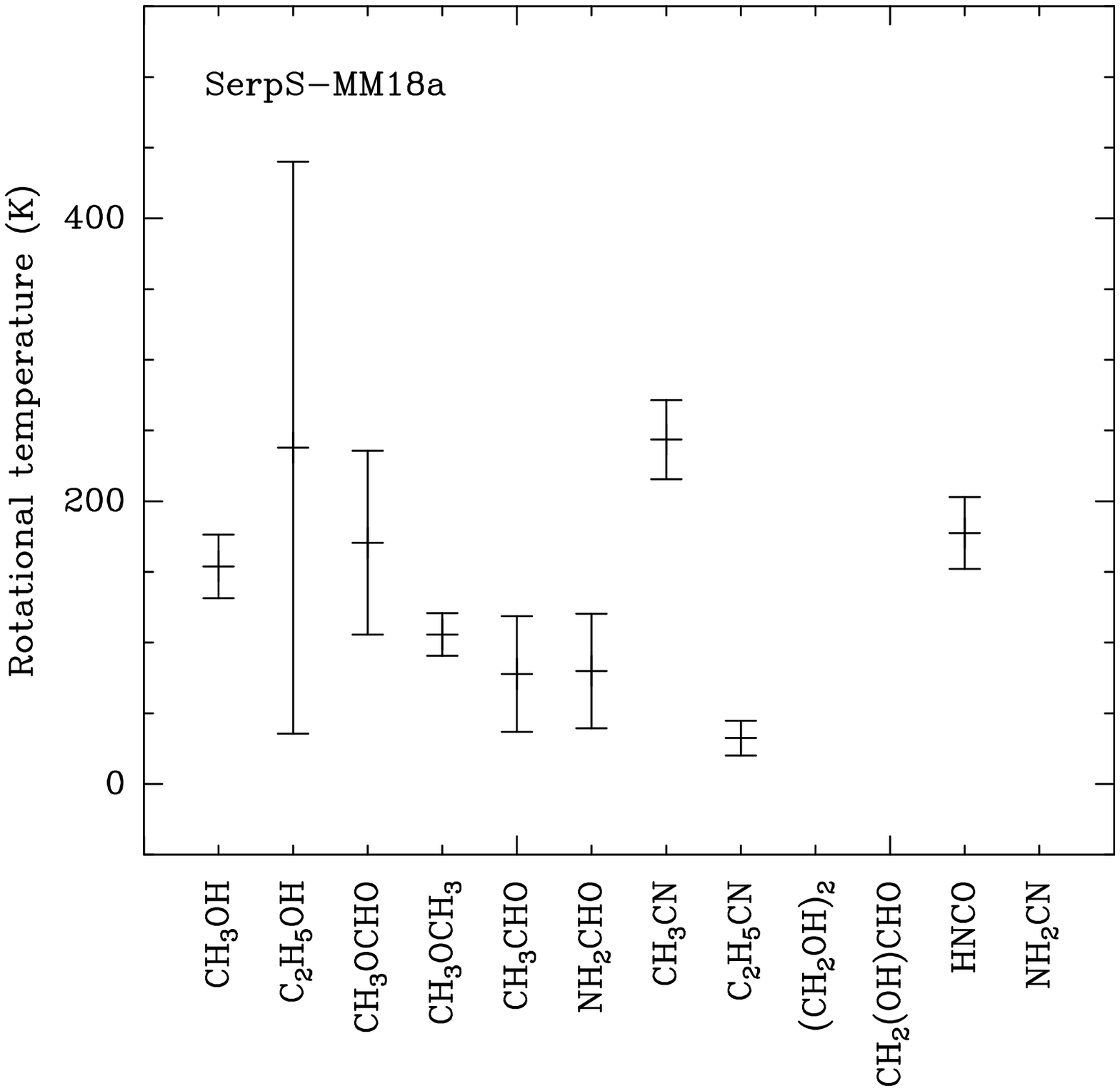}}}
\caption{Same as Fig.~\ref{f:trot_l1448-2a_p1} for SerpS-MM18a.}
\label{f:trot_aqu-mms1_p1}
\end{figure}

\begin{figure}[!htbp]
\centerline{\resizebox{0.75\hsize}{!}{\includegraphics[angle=0]{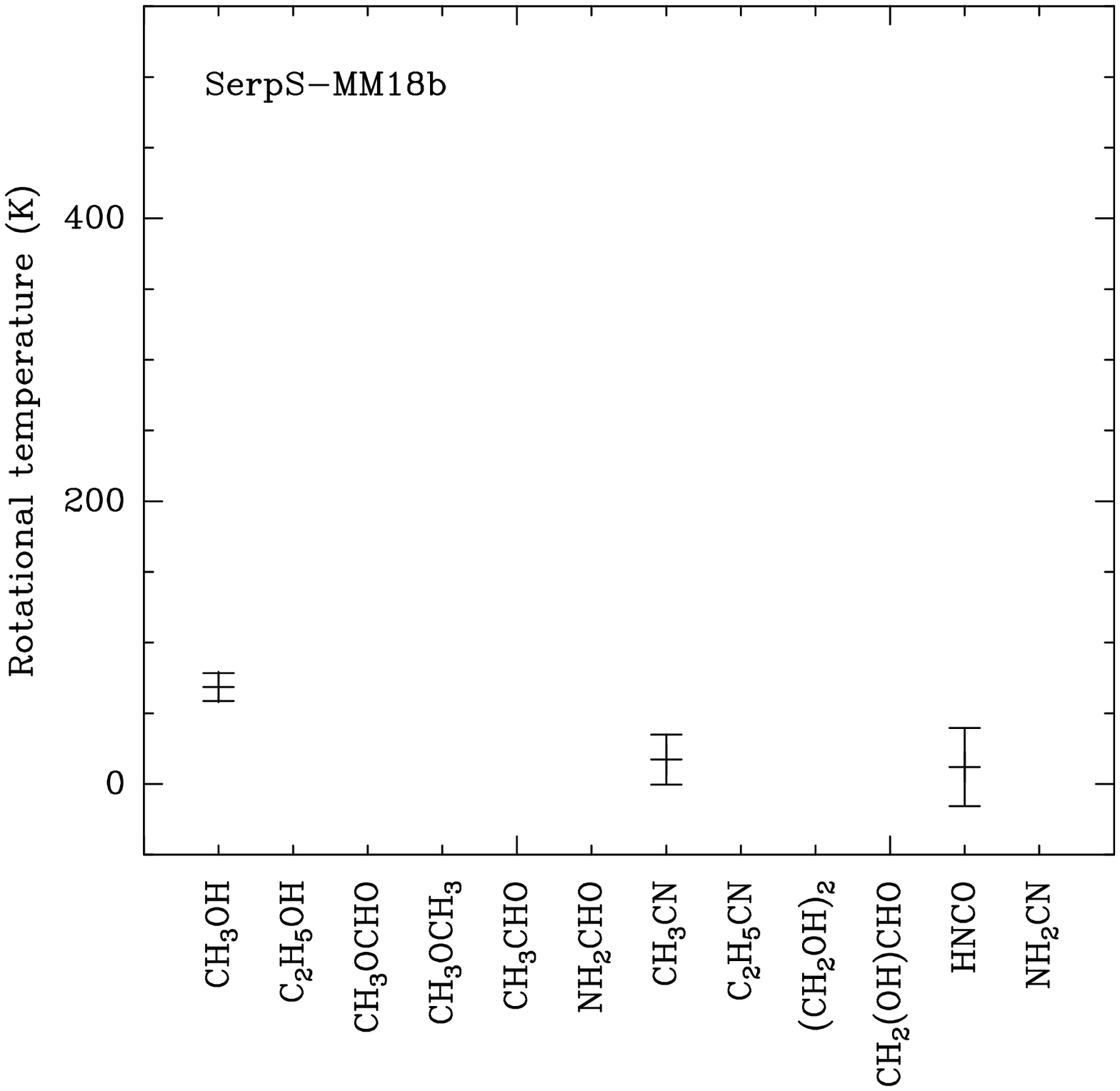}}}
\caption{Same as Fig.~\ref{f:trot_l1448-2a_p1} for SerpS-MM18b.}
\label{f:trot_aqu-mms1_p2}
\end{figure}

\begin{figure}[!htbp]
\centerline{\resizebox{0.75\hsize}{!}{\includegraphics[angle=0]{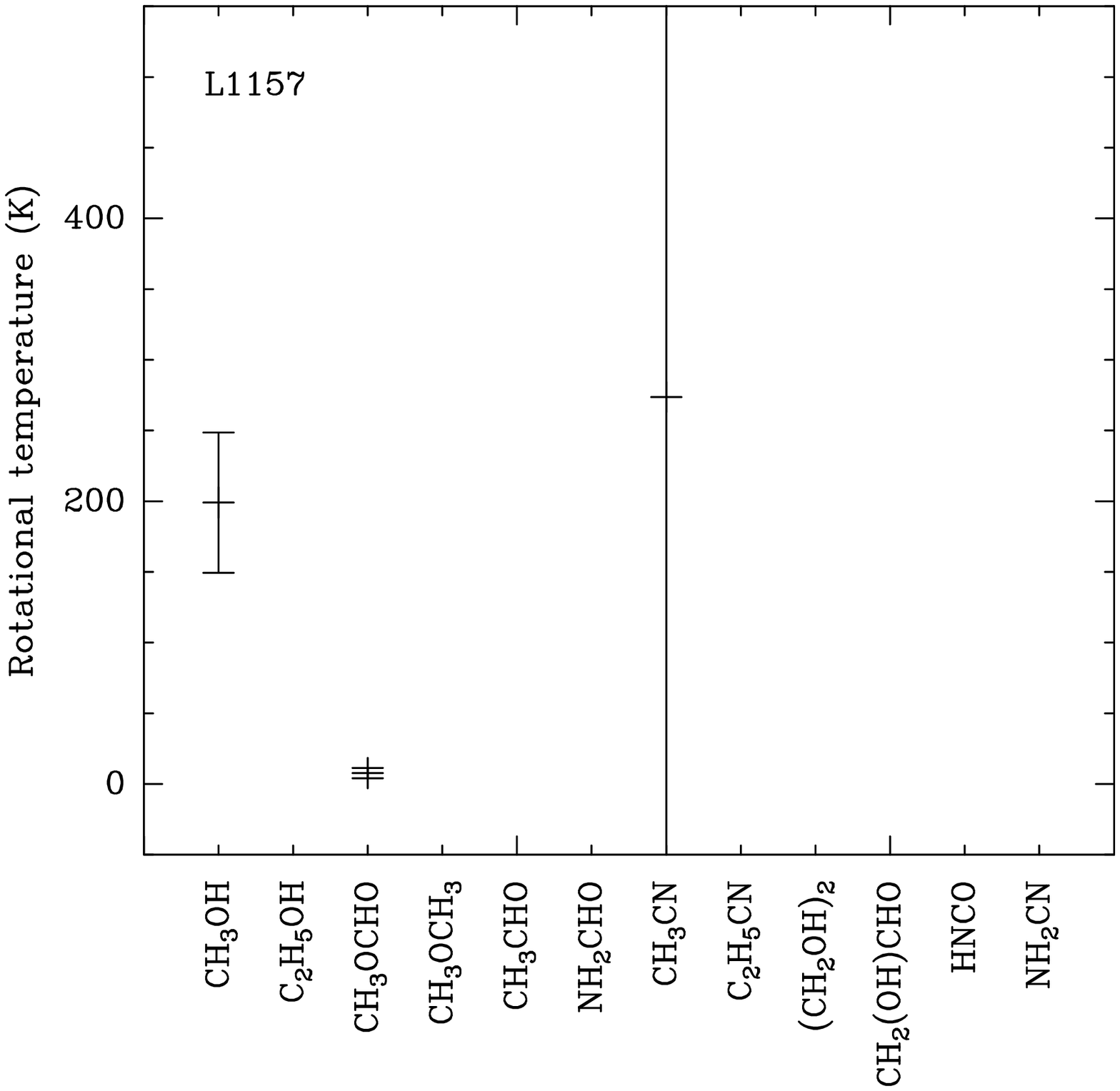}}}
\caption{Same as Fig.~\ref{f:trot_l1448-2a_p1} for L1157.}
\label{f:trot_l1157_p1}
\end{figure}

\clearpage

\clearpage
\newpage

\section{Parameters of radiative transfer models}
\label{a:weedsmodels}

Table~\ref{t:coldens} lists the parameters of the best-fit radiative transfer 
models computed with Weeds. 

\clearpage
\newpage

\section{Chemical composition}
\label{a:chemcomp}

\begin{figure*}[!ht]
 \centerline{\resizebox{1.0\hsize}{!}{\includegraphics[angle=270]{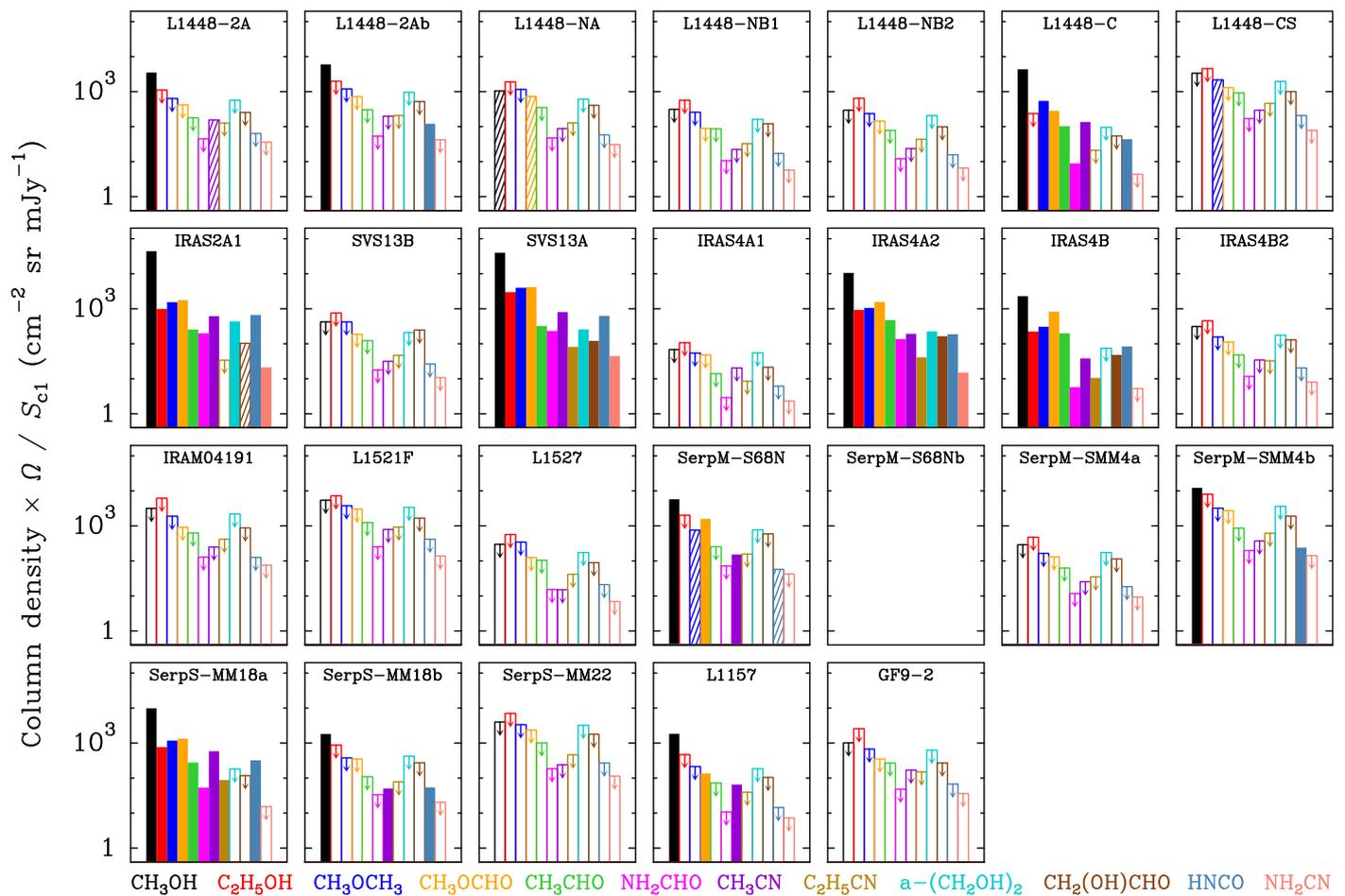}}}
 \caption{Same as Fig.~\ref{f:chemcomp} for the column density multiplied by 
the solid angle of the COM emission and divided by either the 1.3~mm continuum 
peak flux density or the 1.3~mm continuum flux density integrated over the 
size of the COM emission. SerpM-S68Nb is located outside the primary beam at 
1.3~mm so we did not attempt any normalization by the continuum emission.}
 \label{f:chemcompsolang1mmcont}
\end{figure*}

\begin{figure*}[!ht]
 \centerline{\resizebox{1.0\hsize}{!}{\includegraphics[angle=270]{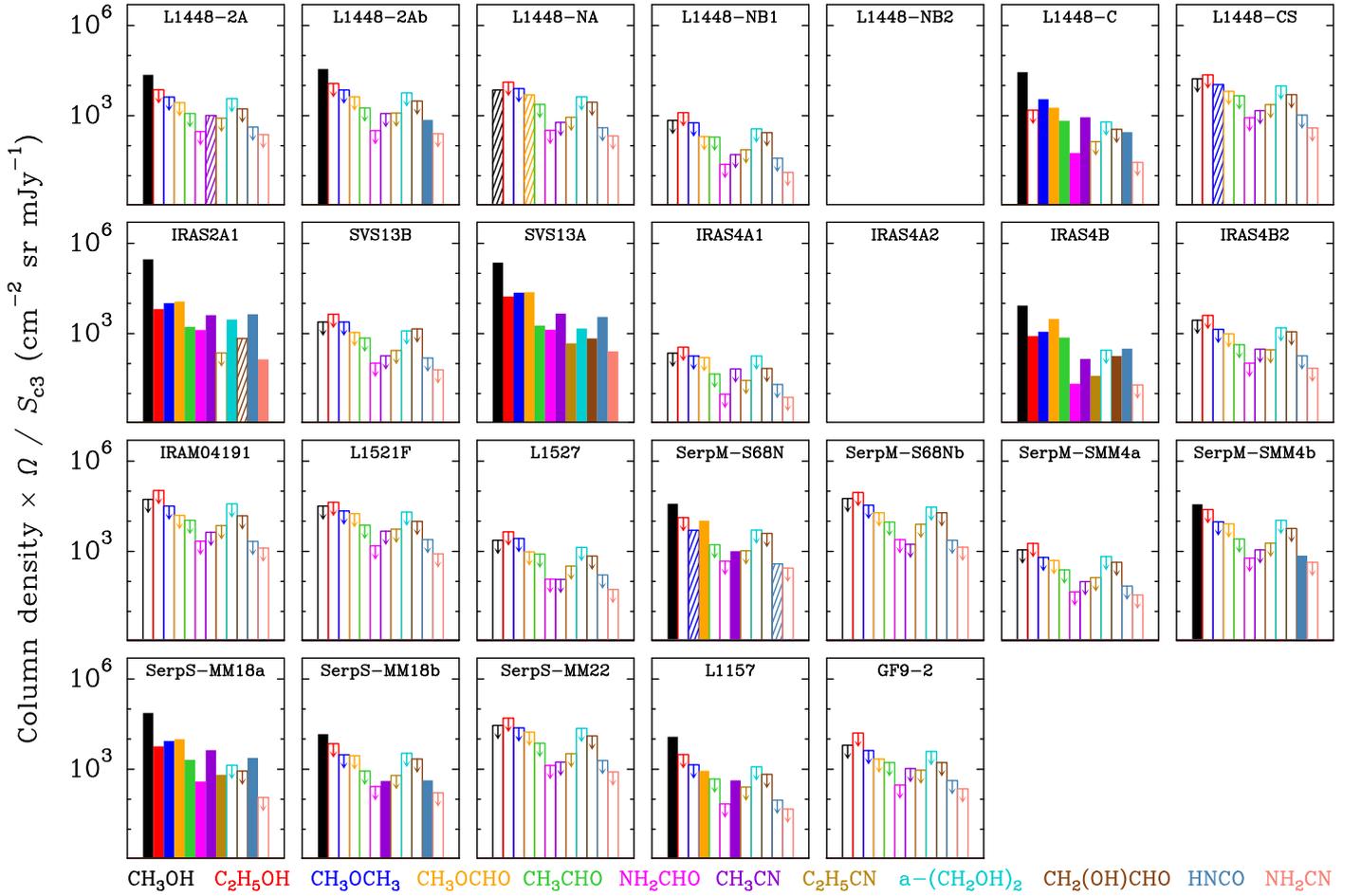}}}
 \caption{Same as Fig.~\ref{f:chemcomp} for the column density multiplied by 
the solid angle of the COM emission and divided by the 3~mm continuum peak 
flux density. No 3~mm continuum flux density is available for L1448-NB2 and 
IRAS4A2 because these sources are unresolved from their close primary 
protostar at this wavelength.}
 \label{f:chemcompsolang3mmcont}
\end{figure*}

Figures~\ref{f:chemcompsolang1mmcont} and \ref{f:chemcompsolang3mmcont} show
the chemical composition of the \hbox{CALYPSO} sources normalized by the 
continuum flux density measured at 1.3~mm and 3~mm, respectively, and the 
solid angle of the COM emission.

\clearpage
\newpage

\section{Column density correlations}
\label{a:correlations}

\begin{figure*}
\centerline{\resizebox{0.95\hsize}{!}{\includegraphics[angle=270]{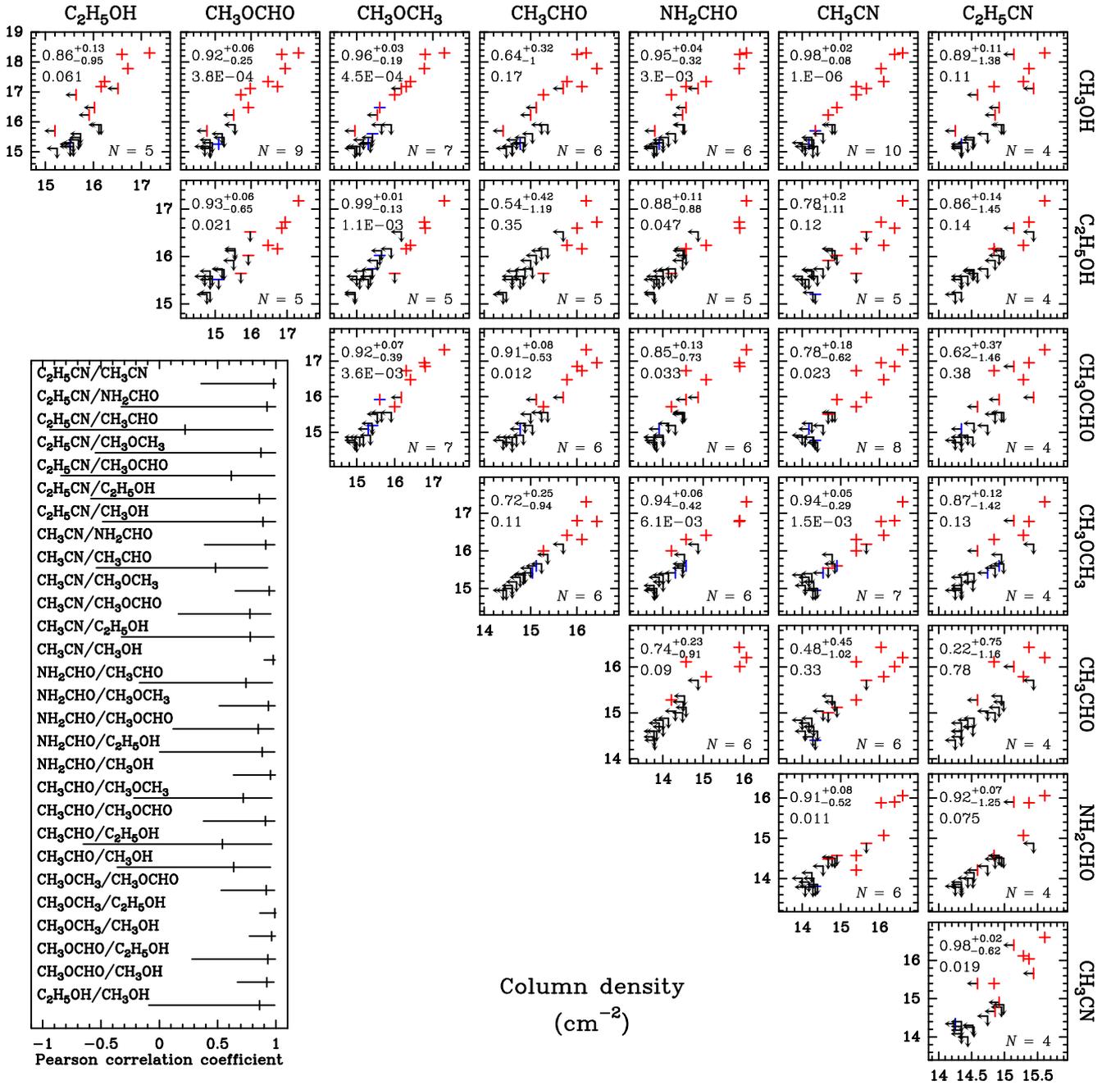}}}
\caption{Correlation plots of the column densities of various pairs of COMs. 
The labels of the axes indicate the decimal logarithm of the column densities.
Firm and tentative detections are indicated with red and blue crosses (or bars) 
respectively. Upper limits are indicated with black arrows. In each panel, the 
Pearson correlation coefficient and its 95\% confidence interval are displayed 
in the top left corner, followed by the P-value just below. The number of 
measurements used to compute the correlation coefficient is given in the 
bottom right corner of each panel. 
The plot in the lower left corner of the figure displays the Pearson 
correlation coefficients of all pairs of COMs with their 95\% confidence
intervals.}
\label{f:correl_coldens}
\end{figure*}

\begin{figure*}
\centerline{\resizebox{0.95\hsize}{!}{\includegraphics[angle=0]{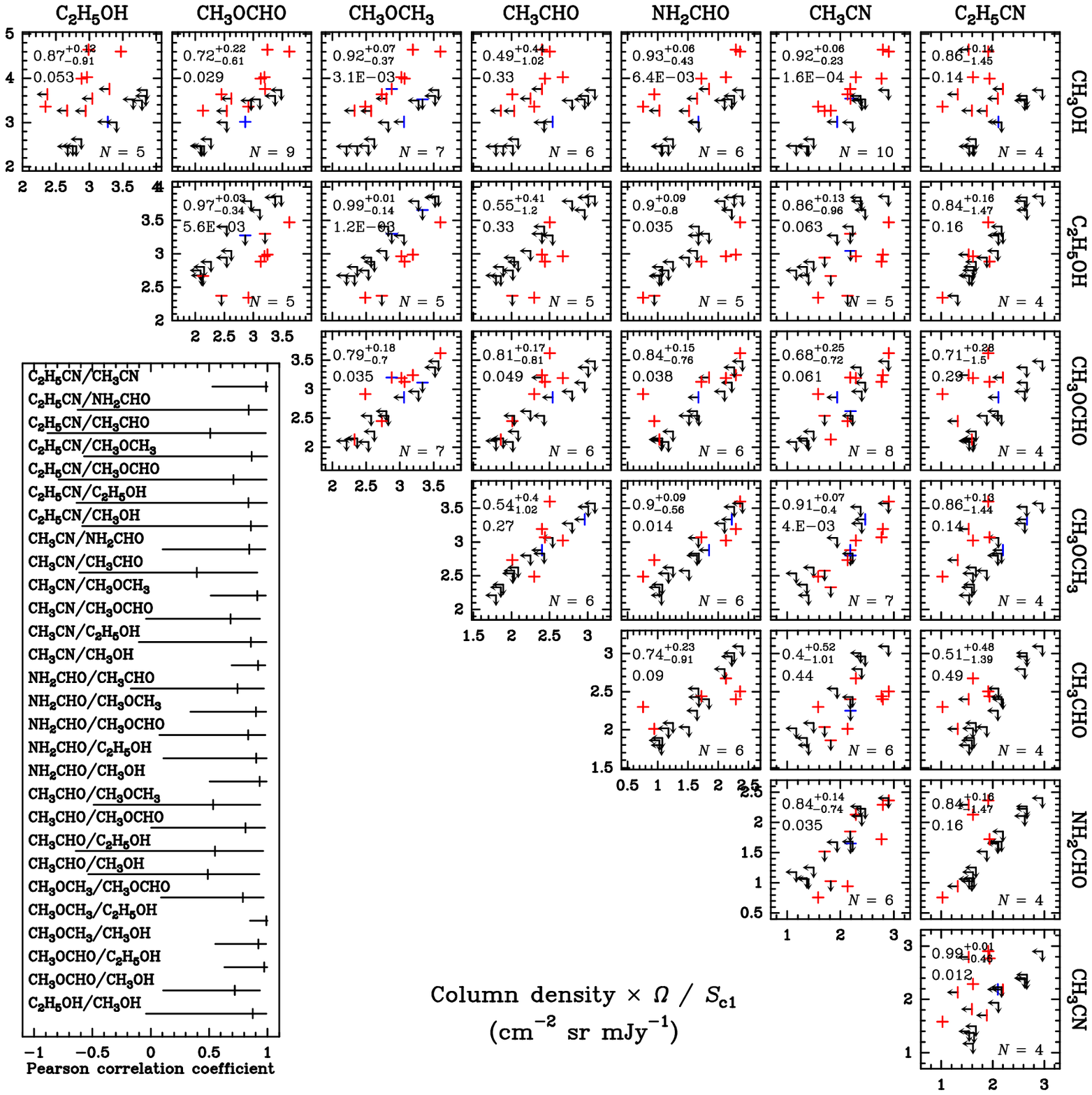}}}
\caption{Same as Fig.~\ref{f:correl_coldens} but using the column density 
multiplied by the solid angle of the COM emission and divided by either the
1.3~mm continuum peak flux density or the 1.3~mm continuum flux density
integrated over the size of the COM emission.}
\label{f:correl_normsolangcontflux}
\end{figure*}

\begin{figure*}
\centerline{\resizebox{0.95\hsize}{!}{\includegraphics[angle=0]{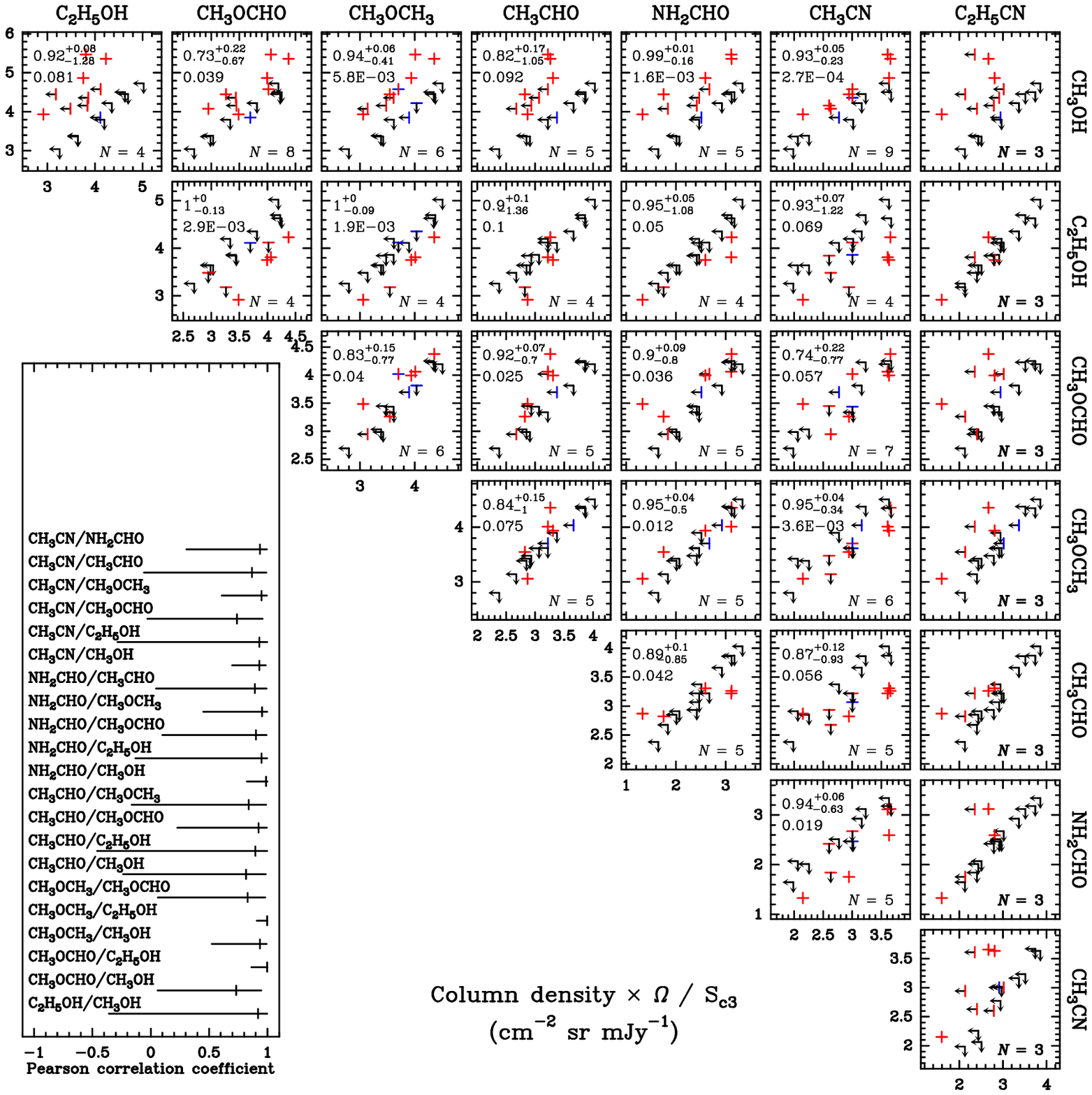}}}
\caption{Same as Fig.~\ref{f:correl_coldens} but using the column density 
multiplied by the solid angle of the COM emission and divided by the 3~mm
continuum peak flux density.}
\label{f:correl_normsolangcontflux3}
\end{figure*}

\begin{figure*}
\centerline{\resizebox{0.95\hsize}{!}{\includegraphics[angle=0]{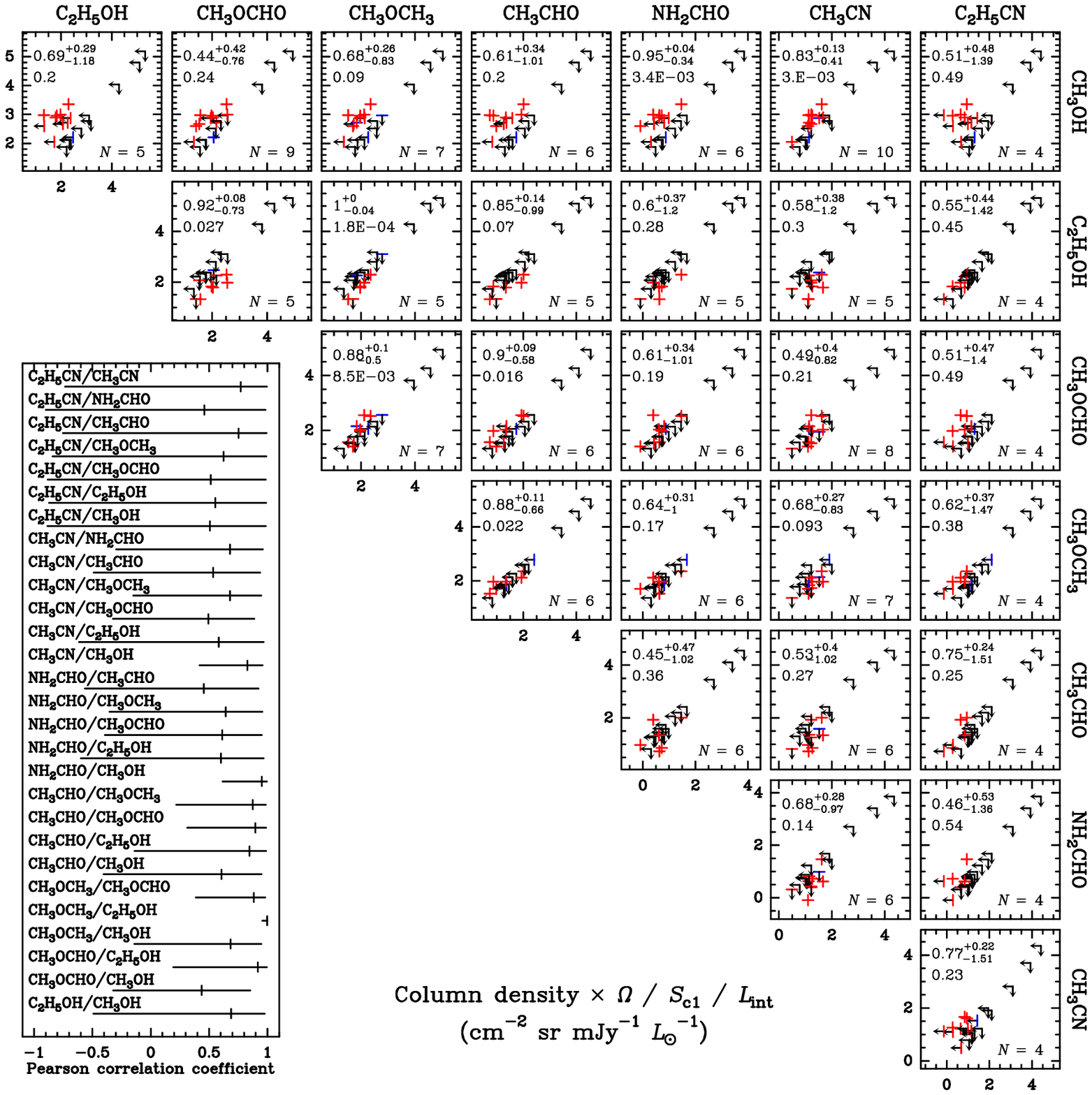}}}
\caption{Same as Fig.~\ref{f:correl_coldens} but using the column density 
multiplied by the solid angle of the COM emission, divided by either the
1.3~mm continuum peak flux density or the 1.3~mm continuum flux density
integrated over the size of the COM emission, and divided by the internal
luminosity.}
\label{f:correl_normsolangcontfluxlint}
\end{figure*}

\begin{figure*}
\centerline{\resizebox{0.95\hsize}{!}{\includegraphics[angle=0]{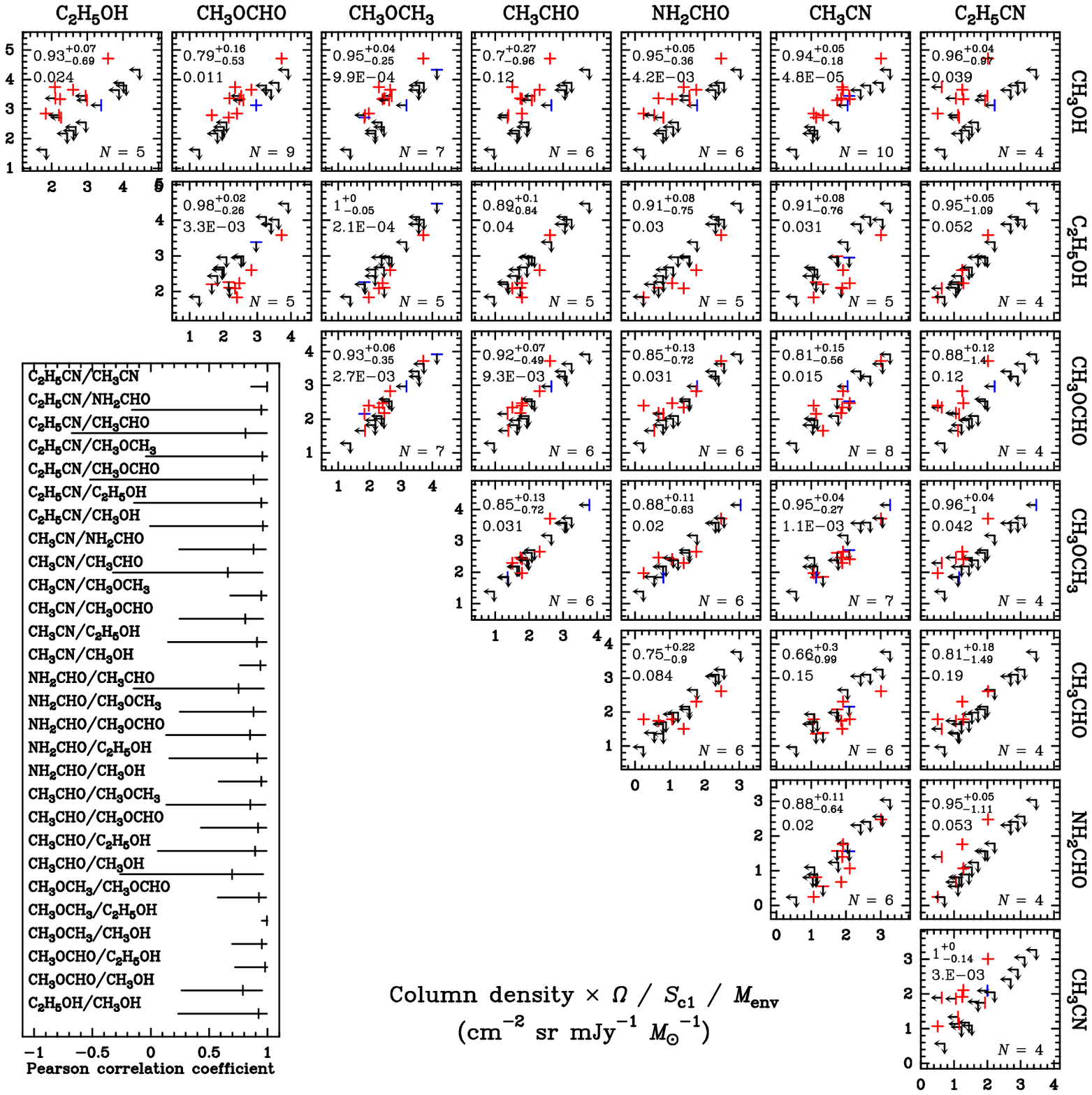}}}
\caption{Same as Fig.~\ref{f:correl_coldens} but using the column density 
multiplied by the solid angle of the COM emission, divided by either the
1.3~mm continuum peak flux density or the 1.3~mm continuum flux density
integrated over the size of the COM emission, and divided by the envelope mass.}
\label{f:correl_normsolangcontfluxmenv}
\end{figure*}

\begin{figure*}
\centerline{\resizebox{0.95\hsize}{!}{\includegraphics[angle=0]{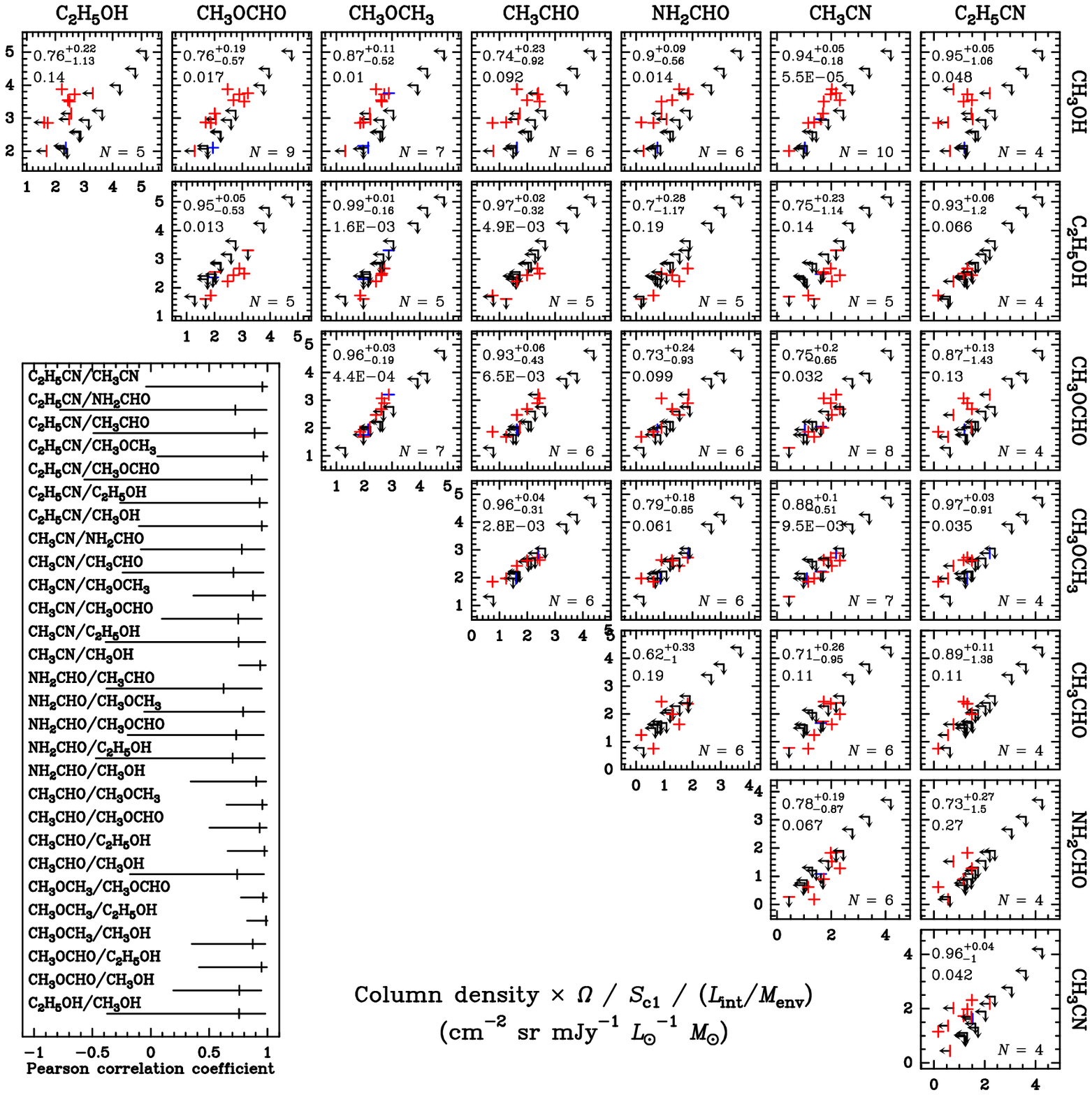}}}
\caption{Same as Fig.~\ref{f:correl_coldens} but using the column density 
multiplied by the solid angle of the COM emission, divided by either the
1.3~mm continuum peak flux density or the 1.3~mm continuum flux density
integrated over the size of the COM emission, and divided by the ratio of the
internal luminosity to the envelope mass.}
\label{f:correl_normsolangcontfluxmenvlint}
\end{figure*}

Figures~\ref{f:correl_coldens}--\ref{f:correl_normsolangcontfluxmenvlint} show
correlation plots of the column densities of various pairs of COMs, normalized
or not to other quantities. Table~\ref{t:correl} summarizes the correlation
parameters.

\begin{table*}
 \begin{center}
 \caption{
 Correlations between column densities normalized by various quantities.
}
 \label{t:correl}
 \vspace*{-1.2ex}
 \begin{tabular}{lcccccccccccccccc}
 \hline\hline
 \multicolumn{1}{c}{Molecule pair} & \multicolumn{1}{c}{\hspace*{-3ex} $N_{\rm pts}$\tablefootmark{a}} & \multicolumn{2}{c}{$N$\tablefootmark{b}} & \multicolumn{2}{c}{$N\Omega/S_{\rm c1}$\tablefootmark{c}} & \multicolumn{2}{c}{$N\Omega/S_{\rm c3}$\tablefootmark{d}} & \multicolumn{2}{c}{$N\Omega/S_{\rm c1}/L$\tablefootmark{e}} & \multicolumn{2}{c}{$N\Omega/S_{\rm c1}/M$\tablefootmark{f}} & \multicolumn{2}{c}{$N\Omega/S_{\rm c1}/(L/M)$\tablefootmark{g}} \\ 
  & & \multicolumn{1}{c}{\hspace*{-1ex} $100\times\rho$} & \multicolumn{1}{c}{\hspace*{-2ex} $P$} & \multicolumn{1}{c}{\hspace*{-1ex} $100\times\rho$} & \multicolumn{1}{c}{\hspace*{-2ex} $P$} & \multicolumn{1}{c}{\hspace*{-1ex} $100\times\rho$} & \multicolumn{1}{c}{\hspace*{-2ex} $P$} & \multicolumn{1}{c}{\hspace*{-1ex} $100\times\rho$} & \multicolumn{1}{c}{\hspace*{-2ex} $P$} & \multicolumn{1}{c}{\hspace*{-1ex} $100\times\rho$} & \multicolumn{1}{c}{\hspace*{-2ex} $P$} & \multicolumn{1}{c}{\hspace*{-1ex} $100\times\rho$} & \multicolumn{1}{c}{\hspace*{-2ex} $P$} \\ 
 \hline
C$_2$H$_5$OH -- CH$_3$OH & \hspace*{-3ex} 5 & \hspace*{-1ex} $ 86^{+ 13}_{- 95}$ & \hspace*{-2ex} 6(-2) & \hspace*{-1ex} $ 87^{+ 12}_{- 91}$ & \hspace*{-2ex} 5(-2) & \hspace*{-1ex} $ 92^{+  8}_{-128}$ & \hspace*{-2ex} 8(-2) & \hspace*{-1ex} $ 69^{+ 29}_{-118}$ & \hspace*{-2ex} 2(-1) & \hspace*{-1ex} $ 93^{+  7}_{- 69}$ & \hspace*{-2ex} 2(-2) & \hspace*{-1ex} $ 76^{+ 22}_{-113}$ & \hspace*{-2ex} 1(-1) \\ 
\noalign{\smallskip} 
CH$_3$OCHO -- CH$_3$OH & \hspace*{-3ex} 9 & \hspace*{-1ex} \textbf{ 92}$^{+  6}_{- 25}$ & \hspace*{-2ex} 4(-4) & \hspace*{-1ex} $ 72^{+ 22}_{- 61}$ & \hspace*{-2ex} 3(-2) & \hspace*{-1ex} $ 73^{+ 22}_{- 67}$ & \hspace*{-2ex} 4(-2) & \hspace*{-1ex} $ 44^{+ 42}_{- 76}$ & \hspace*{-2ex} 2(-1) & \hspace*{-1ex} $ 79^{+ 16}_{- 53}$ & \hspace*{-2ex} 1(-2) & \hspace*{-1ex} $ 76^{+ 19}_{- 57}$ & \hspace*{-2ex} 2(-2) \\ 
\noalign{\smallskip} 
CH$_3$OCHO -- C$_2$H$_5$OH & \hspace*{-3ex} 5 & \hspace*{-1ex} $ 93^{+  6}_{- 65}$ & \hspace*{-2ex} 2(-2) & \hspace*{-1ex} \textbf{ 97}$^{+  3}_{- 34}$ & \hspace*{-2ex} 6(-3) & \hspace*{-1ex} \textbf{100}$^{+  0}_{- 13}$ & \hspace*{-2ex} 3(-3) & \hspace*{-1ex} $ 92^{+  8}_{- 73}$ & \hspace*{-2ex} 3(-2) & \hspace*{-1ex} \textbf{ 98}$^{+  2}_{- 26}$ & \hspace*{-2ex} 3(-3) & \hspace*{-1ex} \textbf{ 95}$^{+  5}_{- 53}$ & \hspace*{-2ex} 1(-2) \\ 
\noalign{\smallskip} 
CH$_3$OCH$_3$ -- CH$_3$OH & \hspace*{-3ex} 7 & \hspace*{-1ex} \textbf{ 96}$^{+  3}_{- 19}$ & \hspace*{-2ex} 4(-4) & \hspace*{-1ex} \textbf{ 92}$^{+  7}_{- 37}$ & \hspace*{-2ex} 3(-3) & \hspace*{-1ex} \textbf{ 94}$^{+  6}_{- 41}$ & \hspace*{-2ex} 6(-3) & \hspace*{-1ex} $ 68^{+ 26}_{- 83}$ & \hspace*{-2ex} 9(-2) & \hspace*{-1ex} \textbf{ 95}$^{+  4}_{- 25}$ & \hspace*{-2ex} 1(-3) & \hspace*{-1ex} \textbf{ 87}$^{+ 11}_{- 52}$ & \hspace*{-2ex} 1(-2) \\ 
\noalign{\smallskip} 
CH$_3$OCH$_3$ -- C$_2$H$_5$OH & \hspace*{-3ex} 5 & \hspace*{-1ex} \textbf{ 99}$^{+  1}_{- 13}$ & \hspace*{-2ex} 1(-3) & \hspace*{-1ex} \textbf{ 99}$^{+  1}_{- 14}$ & \hspace*{-2ex} 1(-3) & \hspace*{-1ex} \textbf{100}$^{+  0}_{-  9}$ & \hspace*{-2ex} 2(-3) & \hspace*{-1ex} \textbf{100}$^{+  0}_{-  4}$ & \hspace*{-2ex} 2(-4) & \hspace*{-1ex} \textbf{100}$^{+  0}_{-  5}$ & \hspace*{-2ex} 2(-4) & \hspace*{-1ex} \textbf{ 99}$^{+  1}_{- 16}$ & \hspace*{-2ex} 2(-3) \\ 
\noalign{\smallskip} 
CH$_3$OCH$_3$ -- CH$_3$OCHO & \hspace*{-3ex} 7 & \hspace*{-1ex} \textbf{ 92}$^{+  7}_{- 39}$ & \hspace*{-2ex} 4(-3) & \hspace*{-1ex} $ 79^{+ 18}_{- 70}$ & \hspace*{-2ex} 4(-2) & \hspace*{-1ex} $ 83^{+ 15}_{- 77}$ & \hspace*{-2ex} 4(-2) & \hspace*{-1ex} \textbf{ 88}$^{+ 10}_{- 50}$ & \hspace*{-2ex} 9(-3) & \hspace*{-1ex} \textbf{ 93}$^{+  6}_{- 35}$ & \hspace*{-2ex} 3(-3) & \hspace*{-1ex} \textbf{ 96}$^{+  3}_{- 19}$ & \hspace*{-2ex} 4(-4) \\ 
\noalign{\smallskip} 
CH$_3$CHO -- CH$_3$OH & \hspace*{-3ex} 6 & \hspace*{-1ex} $ 64^{+ 32}_{-100}$ & \hspace*{-2ex} 2(-1) & \hspace*{-1ex} $ 49^{+ 44}_{-102}$ & \hspace*{-2ex} 3(-1) & \hspace*{-1ex} $ 82^{+ 17}_{-105}$ & \hspace*{-2ex} 9(-2) & \hspace*{-1ex} $ 61^{+ 34}_{-101}$ & \hspace*{-2ex} 2(-1) & \hspace*{-1ex} $ 70^{+ 27}_{- 96}$ & \hspace*{-2ex} 1(-1) & \hspace*{-1ex} $ 74^{+ 23}_{- 92}$ & \hspace*{-2ex} 9(-2) \\ 
\noalign{\smallskip} 
CH$_3$CHO -- C$_2$H$_5$OH & \hspace*{-3ex} 5 & \hspace*{-1ex} $ 54^{+ 42}_{-119}$ & \hspace*{-2ex} 3(-1) & \hspace*{-1ex} $ 55^{+ 41}_{-120}$ & \hspace*{-2ex} 3(-1) & \hspace*{-1ex} $ 90^{+ 10}_{-136}$ & \hspace*{-2ex} 1(-1) & \hspace*{-1ex} $ 85^{+ 14}_{- 99}$ & \hspace*{-2ex} 7(-2) & \hspace*{-1ex} $ 89^{+ 10}_{- 84}$ & \hspace*{-2ex} 4(-2) & \hspace*{-1ex} \textbf{ 97}$^{+  2}_{- 32}$ & \hspace*{-2ex} 5(-3) \\ 
\noalign{\smallskip} 
CH$_3$CHO -- CH$_3$OCHO & \hspace*{-3ex} 6 & \hspace*{-1ex} \textbf{ 91}$^{+  8}_{- 53}$ & \hspace*{-2ex} 1(-2) & \hspace*{-1ex} $ 81^{+ 17}_{- 81}$ & \hspace*{-2ex} 5(-2) & \hspace*{-1ex} $ 92^{+  7}_{- 70}$ & \hspace*{-2ex} 2(-2) & \hspace*{-1ex} \textbf{ 90}$^{+  9}_{- 58}$ & \hspace*{-2ex} 2(-2) & \hspace*{-1ex} \textbf{ 92}$^{+  7}_{- 49}$ & \hspace*{-2ex} 9(-3) & \hspace*{-1ex} \textbf{ 93}$^{+  6}_{- 43}$ & \hspace*{-2ex} 7(-3) \\ 
\noalign{\smallskip} 
CH$_3$CHO -- CH$_3$OCH$_3$ & \hspace*{-3ex} 6 & \hspace*{-1ex} $ 72^{+ 25}_{- 94}$ & \hspace*{-2ex} 1(-1) & \hspace*{-1ex} $ 54^{+ 40}_{-102}$ & \hspace*{-2ex} 3(-1) & \hspace*{-1ex} $ 84^{+ 15}_{-100}$ & \hspace*{-2ex} 8(-2) & \hspace*{-1ex} $ 88^{+ 11}_{- 66}$ & \hspace*{-2ex} 2(-2) & \hspace*{-1ex} $ 85^{+ 13}_{- 72}$ & \hspace*{-2ex} 3(-2) & \hspace*{-1ex} \textbf{ 96}$^{+  4}_{- 31}$ & \hspace*{-2ex} 3(-3) \\ 
\noalign{\smallskip} 
NH$_2$CHO -- CH$_3$OH & \hspace*{-3ex} 6 & \hspace*{-1ex} \textbf{ 95}$^{+  4}_{- 32}$ & \hspace*{-2ex} 3(-3) & \hspace*{-1ex} \textbf{ 93}$^{+  6}_{- 43}$ & \hspace*{-2ex} 6(-3) & \hspace*{-1ex} \textbf{ 99}$^{+  1}_{- 16}$ & \hspace*{-2ex} 2(-3) & \hspace*{-1ex} \textbf{ 95}$^{+  4}_{- 34}$ & \hspace*{-2ex} 3(-3) & \hspace*{-1ex} \textbf{ 95}$^{+  5}_{- 36}$ & \hspace*{-2ex} 4(-3) & \hspace*{-1ex} \textbf{ 90}$^{+  9}_{- 56}$ & \hspace*{-2ex} 1(-2) \\ 
\noalign{\smallskip} 
NH$_2$CHO -- C$_2$H$_5$OH & \hspace*{-3ex} 5 & \hspace*{-1ex} $ 88^{+ 11}_{- 88}$ & \hspace*{-2ex} 5(-2) & \hspace*{-1ex} $ 90^{+  9}_{- 80}$ & \hspace*{-2ex} 4(-2) & \hspace*{-1ex} $ 95^{+  5}_{-108}$ & \hspace*{-2ex} 5(-2) & \hspace*{-1ex} $ 60^{+ 37}_{-120}$ & \hspace*{-2ex} 3(-1) & \hspace*{-1ex} $ 91^{+  8}_{- 75}$ & \hspace*{-2ex} 3(-2) & \hspace*{-1ex} $ 70^{+ 28}_{-117}$ & \hspace*{-2ex} 2(-1) \\ 
\noalign{\smallskip} 
NH$_2$CHO -- CH$_3$OCHO & \hspace*{-3ex} 6 & \hspace*{-1ex} $ 85^{+ 13}_{- 73}$ & \hspace*{-2ex} 3(-2) & \hspace*{-1ex} $ 84^{+ 15}_{- 76}$ & \hspace*{-2ex} 4(-2) & \hspace*{-1ex} $ 90^{+  9}_{- 80}$ & \hspace*{-2ex} 4(-2) & \hspace*{-1ex} $ 61^{+ 34}_{-101}$ & \hspace*{-2ex} 2(-1) & \hspace*{-1ex} $ 85^{+ 13}_{- 72}$ & \hspace*{-2ex} 3(-2) & \hspace*{-1ex} $ 73^{+ 24}_{- 93}$ & \hspace*{-2ex} 1(-1) \\ 
\noalign{\smallskip} 
NH$_2$CHO -- CH$_3$OCH$_3$ & \hspace*{-3ex} 6 & \hspace*{-1ex} \textbf{ 94}$^{+  6}_{- 42}$ & \hspace*{-2ex} 6(-3) & \hspace*{-1ex} \textbf{ 90}$^{+  9}_{- 56}$ & \hspace*{-2ex} 1(-2) & \hspace*{-1ex} \textbf{ 95}$^{+  4}_{- 50}$ & \hspace*{-2ex} 1(-2) & \hspace*{-1ex} $ 64^{+ 31}_{-100}$ & \hspace*{-2ex} 2(-1) & \hspace*{-1ex} $ 88^{+ 11}_{- 63}$ & \hspace*{-2ex} 2(-2) & \hspace*{-1ex} $ 79^{+ 18}_{- 85}$ & \hspace*{-2ex} 6(-2) \\ 
\noalign{\smallskip} 
NH$_2$CHO -- CH$_3$CHO & \hspace*{-3ex} 6 & \hspace*{-1ex} $ 74^{+ 23}_{- 91}$ & \hspace*{-2ex} 9(-2) & \hspace*{-1ex} $ 74^{+ 23}_{- 91}$ & \hspace*{-2ex} 9(-2) & \hspace*{-1ex} $ 89^{+ 10}_{- 85}$ & \hspace*{-2ex} 4(-2) & \hspace*{-1ex} $ 45^{+ 47}_{-102}$ & \hspace*{-2ex} 4(-1) & \hspace*{-1ex} $ 75^{+ 22}_{- 90}$ & \hspace*{-2ex} 8(-2) & \hspace*{-1ex} $ 62^{+ 33}_{-100}$ & \hspace*{-2ex} 2(-1) \\ 
\noalign{\smallskip} 
CH$_3$CN -- CH$_3$OH & \hspace*{-3ex} 10 & \hspace*{-1ex} \textbf{ 98}$^{+  2}_{-  8}$ & \hspace*{-2ex} 1(-6) & \hspace*{-1ex} \textbf{ 92}$^{+  6}_{- 23}$ & \hspace*{-2ex} 2(-4) & \hspace*{-1ex} \textbf{ 93}$^{+  5}_{- 23}$ & \hspace*{-2ex} 3(-4) & \hspace*{-1ex} \textbf{ 83}$^{+ 13}_{- 41}$ & \hspace*{-2ex} 3(-3) & \hspace*{-1ex} \textbf{ 94}$^{+  5}_{- 18}$ & \hspace*{-2ex} 5(-5) & \hspace*{-1ex} \textbf{ 94}$^{+  5}_{- 18}$ & \hspace*{-2ex} 6(-5) \\ 
\noalign{\smallskip} 
CH$_3$CN -- C$_2$H$_5$OH & \hspace*{-3ex} 5 & \hspace*{-1ex} $ 78^{+ 20}_{-111}$ & \hspace*{-2ex} 1(-1) & \hspace*{-1ex} $ 86^{+ 13}_{- 96}$ & \hspace*{-2ex} 6(-2) & \hspace*{-1ex} $ 93^{+  7}_{-122}$ & \hspace*{-2ex} 7(-2) & \hspace*{-1ex} $ 58^{+ 38}_{-120}$ & \hspace*{-2ex} 3(-1) & \hspace*{-1ex} $ 91^{+  8}_{- 76}$ & \hspace*{-2ex} 3(-2) & \hspace*{-1ex} $ 75^{+ 23}_{-114}$ & \hspace*{-2ex} 1(-1) \\ 
\noalign{\smallskip} 
CH$_3$CN -- CH$_3$OCHO & \hspace*{-3ex} 8 & \hspace*{-1ex} $ 78^{+ 18}_{- 62}$ & \hspace*{-2ex} 2(-2) & \hspace*{-1ex} $ 68^{+ 25}_{- 72}$ & \hspace*{-2ex} 6(-2) & \hspace*{-1ex} $ 74^{+ 22}_{- 77}$ & \hspace*{-2ex} 6(-2) & \hspace*{-1ex} $ 49^{+ 40}_{- 82}$ & \hspace*{-2ex} 2(-1) & \hspace*{-1ex} $ 81^{+ 15}_{- 56}$ & \hspace*{-2ex} 1(-2) & \hspace*{-1ex} $ 75^{+ 20}_{- 65}$ & \hspace*{-2ex} 3(-2) \\ 
\noalign{\smallskip} 
CH$_3$CN -- CH$_3$OCH$_3$ & \hspace*{-3ex} 7 & \hspace*{-1ex} \textbf{ 94}$^{+  5}_{- 29}$ & \hspace*{-2ex} 1(-3) & \hspace*{-1ex} \textbf{ 91}$^{+  7}_{- 40}$ & \hspace*{-2ex} 4(-3) & \hspace*{-1ex} \textbf{ 95}$^{+  4}_{- 34}$ & \hspace*{-2ex} 4(-3) & \hspace*{-1ex} $ 68^{+ 27}_{- 83}$ & \hspace*{-2ex} 9(-2) & \hspace*{-1ex} \textbf{ 95}$^{+  4}_{- 27}$ & \hspace*{-2ex} 1(-3) & \hspace*{-1ex} \textbf{ 88}$^{+ 10}_{- 51}$ & \hspace*{-2ex} 9(-3) \\ 
\noalign{\smallskip} 
CH$_3$CN -- CH$_3$CHO & \hspace*{-3ex} 6 & \hspace*{-1ex} $ 48^{+ 45}_{-102}$ & \hspace*{-2ex} 3(-1) & \hspace*{-1ex} $ 40^{+ 52}_{-101}$ & \hspace*{-2ex} 4(-1) & \hspace*{-1ex} $ 87^{+ 12}_{- 93}$ & \hspace*{-2ex} 6(-2) & \hspace*{-1ex} $ 53^{+ 40}_{-102}$ & \hspace*{-2ex} 3(-1) & \hspace*{-1ex} $ 66^{+ 30}_{- 99}$ & \hspace*{-2ex} 2(-1) & \hspace*{-1ex} $ 71^{+ 26}_{- 95}$ & \hspace*{-2ex} 1(-1) \\ 
\noalign{\smallskip} 
CH$_3$CN -- NH$_2$CHO & \hspace*{-3ex} 6 & \hspace*{-1ex} \textbf{ 91}$^{+  8}_{- 52}$ & \hspace*{-2ex} 1(-2) & \hspace*{-1ex} $ 84^{+ 14}_{- 74}$ & \hspace*{-2ex} 3(-2) & \hspace*{-1ex} \textbf{ 94}$^{+  6}_{- 63}$ & \hspace*{-2ex} 2(-2) & \hspace*{-1ex} $ 68^{+ 28}_{- 97}$ & \hspace*{-2ex} 1(-1) & \hspace*{-1ex} $ 88^{+ 11}_{- 64}$ & \hspace*{-2ex} 2(-2) & \hspace*{-1ex} $ 78^{+ 19}_{- 87}$ & \hspace*{-2ex} 7(-2) \\ 
\noalign{\smallskip} 
C$_2$H$_5$CN -- CH$_3$OH & \hspace*{-3ex} 4 & \hspace*{-1ex} $ 89^{+ 11}_{-138}$ & \hspace*{-2ex} 1(-1) & \hspace*{-1ex} $ 86^{+ 14}_{-145}$ & \hspace*{-2ex} 1(-1) & \hspace*{-1ex} -- & \hspace*{-2ex} -- & \hspace*{-1ex} $ 51^{+ 48}_{-139}$ & \hspace*{-2ex} 5(-1) & \hspace*{-1ex} $ 96^{+  4}_{- 97}$ & \hspace*{-2ex} 4(-2) & \hspace*{-1ex} $ 95^{+  5}_{-106}$ & \hspace*{-2ex} 5(-2) \\ 
\noalign{\smallskip} 
C$_2$H$_5$CN -- C$_2$H$_5$OH & \hspace*{-3ex} 4 & \hspace*{-1ex} $ 86^{+ 14}_{-145}$ & \hspace*{-2ex} 1(-1) & \hspace*{-1ex} $ 84^{+ 16}_{-147}$ & \hspace*{-2ex} 2(-1) & \hspace*{-1ex} -- & \hspace*{-2ex} -- & \hspace*{-1ex} $ 55^{+ 44}_{-142}$ & \hspace*{-2ex} 4(-1) & \hspace*{-1ex} $ 95^{+  5}_{-109}$ & \hspace*{-2ex} 5(-2) & \hspace*{-1ex} $ 93^{+  6}_{-120}$ & \hspace*{-2ex} 7(-2) \\ 
\noalign{\smallskip} 
C$_2$H$_5$CN -- CH$_3$OCHO & \hspace*{-3ex} 4 & \hspace*{-1ex} $ 62^{+ 37}_{-146}$ & \hspace*{-2ex} 4(-1) & \hspace*{-1ex} $ 71^{+ 28}_{-150}$ & \hspace*{-2ex} 3(-1) & \hspace*{-1ex} -- & \hspace*{-2ex} -- & \hspace*{-1ex} $ 51^{+ 47}_{-140}$ & \hspace*{-2ex} 5(-1) & \hspace*{-1ex} $ 88^{+ 12}_{-140}$ & \hspace*{-2ex} 1(-1) & \hspace*{-1ex} $ 87^{+ 13}_{-143}$ & \hspace*{-2ex} 1(-1) \\ 
\noalign{\smallskip} 
C$_2$H$_5$CN -- CH$_3$OCH$_3$ & \hspace*{-3ex} 4 & \hspace*{-1ex} $ 87^{+ 12}_{-142}$ & \hspace*{-2ex} 1(-1) & \hspace*{-1ex} $ 86^{+ 13}_{-144}$ & \hspace*{-2ex} 1(-1) & \hspace*{-1ex} -- & \hspace*{-2ex} -- & \hspace*{-1ex} $ 62^{+ 37}_{-147}$ & \hspace*{-2ex} 4(-1) & \hspace*{-1ex} $ 96^{+  4}_{-100}$ & \hspace*{-2ex} 4(-2) & \hspace*{-1ex} $ 97^{+  3}_{- 91}$ & \hspace*{-2ex} 3(-2) \\ 
\noalign{\smallskip} 
C$_2$H$_5$CN -- CH$_3$CHO & \hspace*{-3ex} 4 & \hspace*{-1ex} $ 22^{+ 75}_{-116}$ & \hspace*{-2ex} 8(-1) & \hspace*{-1ex} $ 51^{+ 48}_{-139}$ & \hspace*{-2ex} 5(-1) & \hspace*{-1ex} -- & \hspace*{-2ex} -- & \hspace*{-1ex} $ 75^{+ 24}_{-151}$ & \hspace*{-2ex} 2(-1) & \hspace*{-1ex} $ 81^{+ 18}_{-149}$ & \hspace*{-2ex} 2(-1) & \hspace*{-1ex} $ 89^{+ 11}_{-138}$ & \hspace*{-2ex} 1(-1) \\ 
\noalign{\smallskip} 
C$_2$H$_5$CN -- NH$_2$CHO & \hspace*{-3ex} 4 & \hspace*{-1ex} $ 92^{+  7}_{-125}$ & \hspace*{-2ex} 8(-2) & \hspace*{-1ex} $ 84^{+ 16}_{-147}$ & \hspace*{-2ex} 2(-1) & \hspace*{-1ex} -- & \hspace*{-2ex} -- & \hspace*{-1ex} $ 46^{+ 53}_{-136}$ & \hspace*{-2ex} 5(-1) & \hspace*{-1ex} $ 95^{+  5}_{-111}$ & \hspace*{-2ex} 5(-2) & \hspace*{-1ex} $ 73^{+ 27}_{-150}$ & \hspace*{-2ex} 3(-1) \\ 
\noalign{\smallskip} 
C$_2$H$_5$CN -- CH$_3$CN & \hspace*{-3ex} 4 & \hspace*{-1ex} \textbf{ 98}$^{+  2}_{- 62}$ & \hspace*{-2ex} 2(-2) & \hspace*{-1ex} \textbf{ 99}$^{+  1}_{- 46}$ & \hspace*{-2ex} 1(-2) & \hspace*{-1ex} -- & \hspace*{-2ex} -- & \hspace*{-1ex} $ 77^{+ 22}_{-151}$ & \hspace*{-2ex} 2(-1) & \hspace*{-1ex} \textbf{100}$^{+  0}_{- 14}$ & \hspace*{-2ex} 3(-3) & \hspace*{-1ex} $ 96^{+  4}_{-100}$ & \hspace*{-2ex} 4(-2) \\ 
\noalign{\smallskip} 
 \hline
 \end{tabular}
 \end{center}
 \vspace*{-2.5ex}
 \tablefoot{
 \tablefoottext{a}{Number of CALYPSO sources detected in both tracers. Pearson correlations are evaluated for the following quantities:}
 \tablefoottext{b}{column density;}
 \tablefoottext{c}{column density multiplied by the solid angle of the COM emission and divided by either the 1.3~mm continuum peak flux density or the 1.3~mm continuum flux density integrated over the size of the COM emission;}
 \tablefoottext{d}{column density multiplied by the solid angle of the COM emission and divided by the 3~mm continuum peak flux density;}
 \tablefoottext{e}{column density multiplied by the solid angle of the COM emission, divided by either the 1.3~mm continuum peak flux density or the 1.3~mm continuum flux density integrated over the size of the COM emission, and divided by the internal luminosity;}
 \tablefoottext{f}{column density multiplied by the solid angle of the COM emission, divided by either the 1.3~mm continuum peak flux density or the 1.3~mm continuum flux density integrated over the size of the COM emission, and divided by the envelope mass;}
 \tablefoottext{g}{column density multiplied by the solid angle of the COM emission, divided by either the 1.3~mm continuum peak flux density or the 1.3~mm continuum flux density integrated over the size of the COM emission, and divided by the ratio of internal luminosity to the envelope mass.}
 For each type of correlation, $\rho$ is the Pearson correlation coefficient with its 95\% confidence interval, and $P$ is the P-value. $X$($Y$) means $X \times 10^Y$. Pearson coefficients with a confidence interval outside [$-$0.3, 0.3] are highlighted in bold face.}
 \end{table*}

\clearpage
\newpage

\section{Correlations of column densities with source properties}
\label{a:correl_prop}

\begin{figure*}
 \centerline{\resizebox{0.8\hsize}{!}{\includegraphics[angle=270]{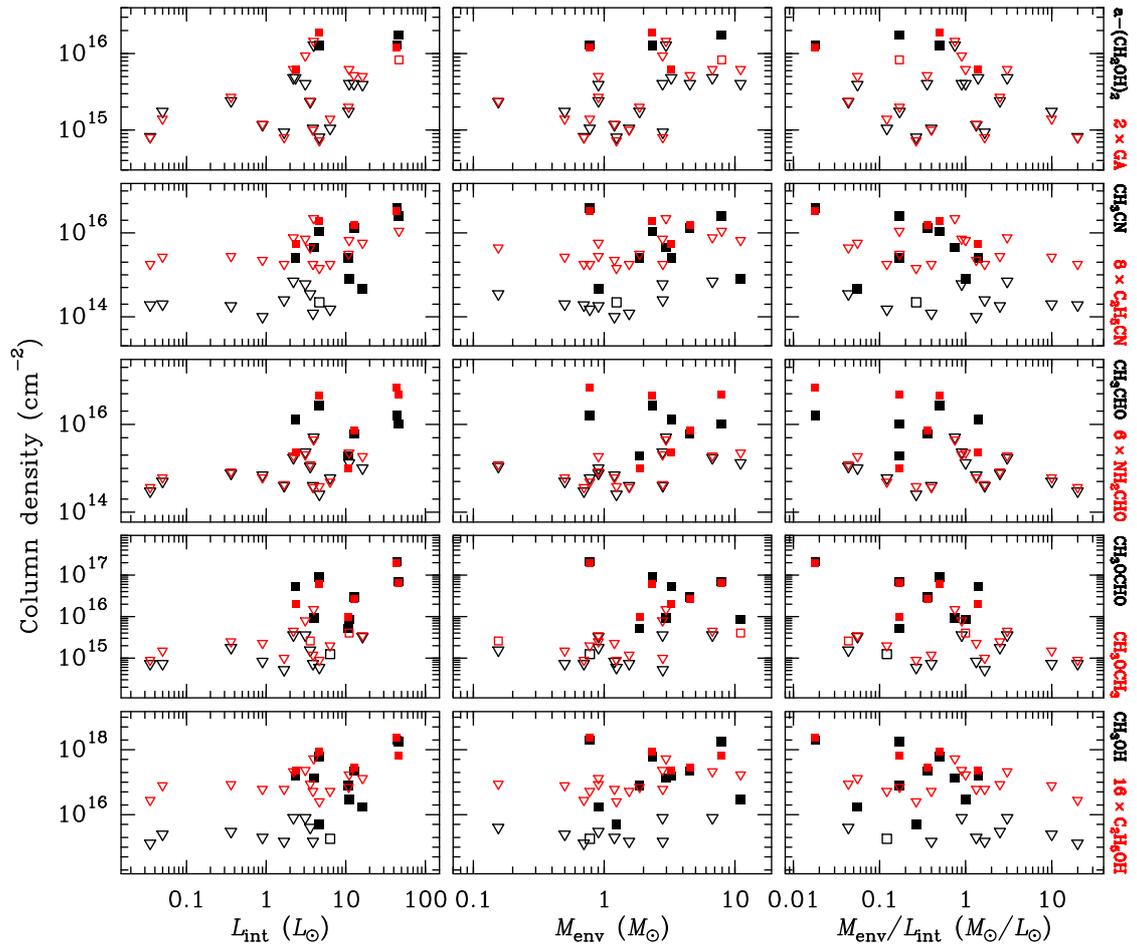}}}
 \caption{Column densities of ten complex organic molecules as a function
of internal luminosity, envelope mass, and ratio of envelope mass to internal
luminosity. The column densities of some molecules were multiplied by a scaling 
factor as indicated on the right. Filled and open squares represent robust and
tentative detections, respectively, while open triangles show upper limits. GA 
stands for glycolaldehyde, CH$_2$(OH)CHO.}
 \label{f:coldens}
\end{figure*}

\begin{figure*}
 \centerline{\resizebox{0.8\hsize}{!}{\includegraphics[angle=270]{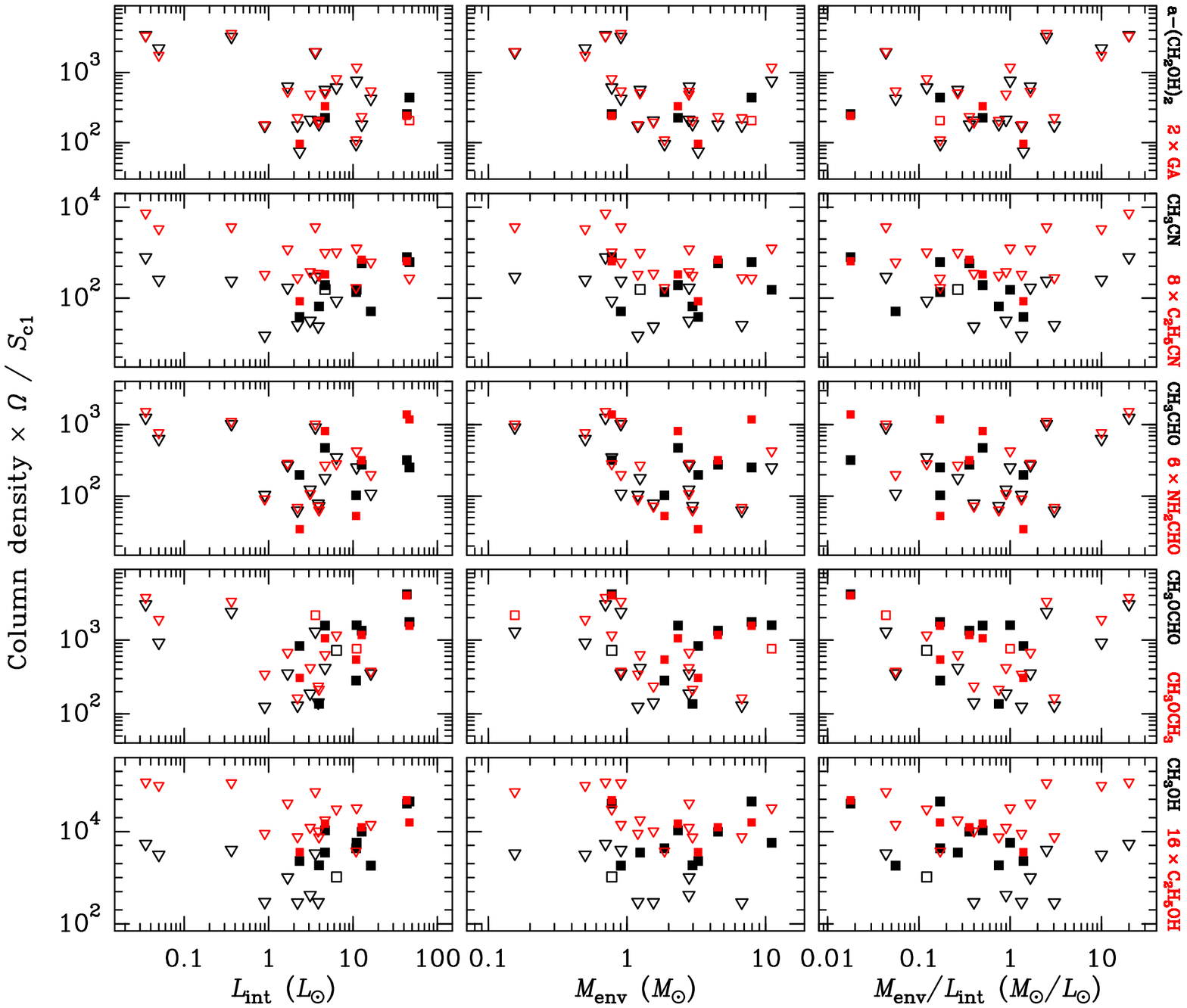}}}
 \caption{Same as Fig.~\ref{f:coldens} but for the column densities multiplied
by the solid angle of the COM emission and divided by either the 1.3~mm
continuum peak flux density or the 1.3~mm continuum flux density integrated
over the size of the COM emission when it is larger than the beam.}
 \label{f:coldens_normsolangcont}
\end{figure*}

\begin{figure*}
 \centerline{\resizebox{0.8\hsize}{!}{\includegraphics[angle=270]{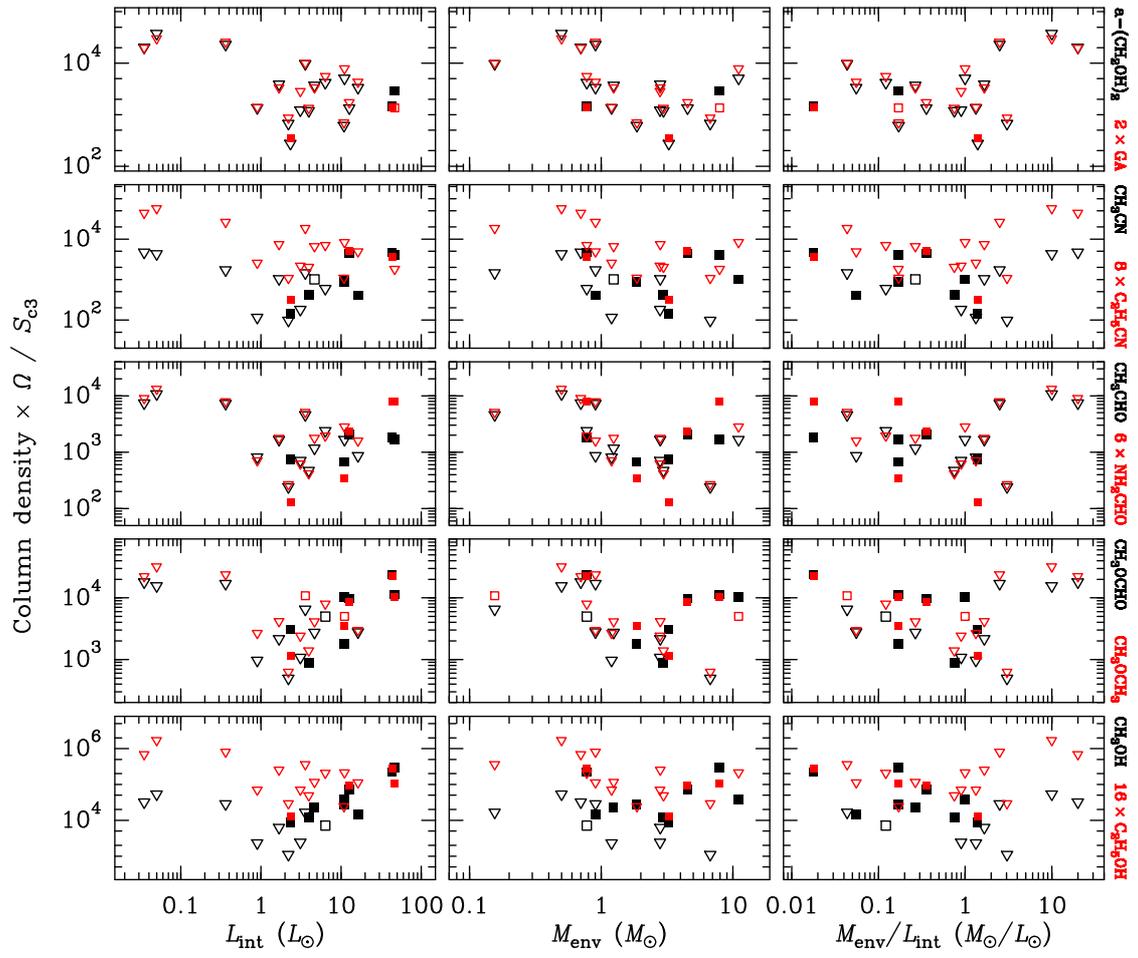}}}
 \caption{Same as Fig.~\ref{f:coldens} but for the column densities multiplied
by the solid angle of the COM emission and divided by the 3~mm continuum peak
flux density.}
 \label{f:coldens_normsolangcont3}
\end{figure*}

Figures~\ref{f:coldens}--\ref{f:coldens_normsolangcont3} show correlation plots
between COM column densities and source properties, with different 
normalizations.

\clearpage
\newpage

\section{Correlations of column densities with disk properties}
\label{a:correl_prop_disk}

\begin{figure*}
 \centerline{\resizebox{0.8\hsize}{!}{\includegraphics[angle=270]{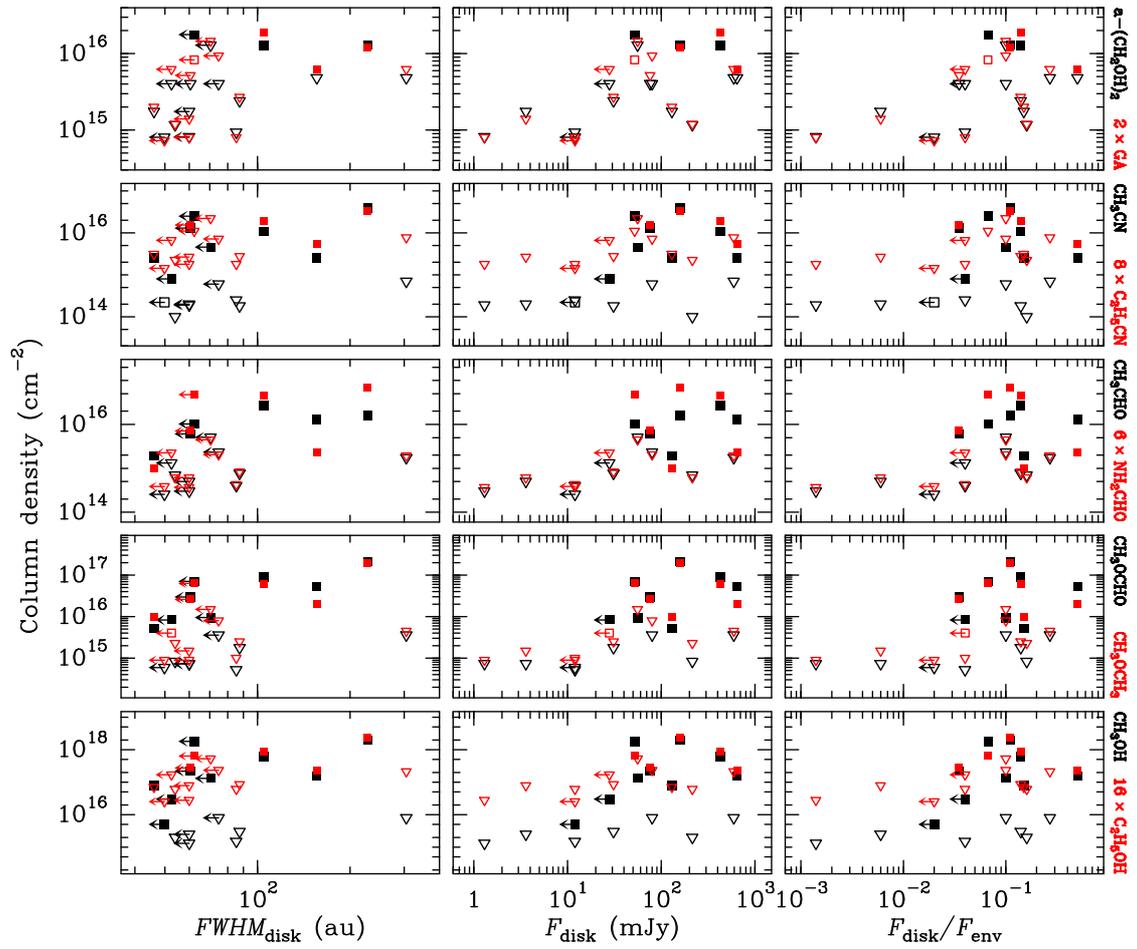}}}
 \caption{Column densities of ten complex organic molecules as a function
of disk size ($FWHM$), disk flux density, and ratio of disk to envelope flux 
densities. The column densities of some molecules were multiplied by a scaling 
factor as indicated on the right. Filled and open squares represent robust and
tentative detections, respectively, while open triangles show upper limits. 
Arrows indicate upper limits along the horizontal axis. GA stands for 
glycolaldehyde, CH$_2$(OH)CHO.}
 \label{f:coldens_disk}
\end{figure*}

\begin{figure*}
 \centerline{\resizebox{0.8\hsize}{!}{\includegraphics[angle=270]{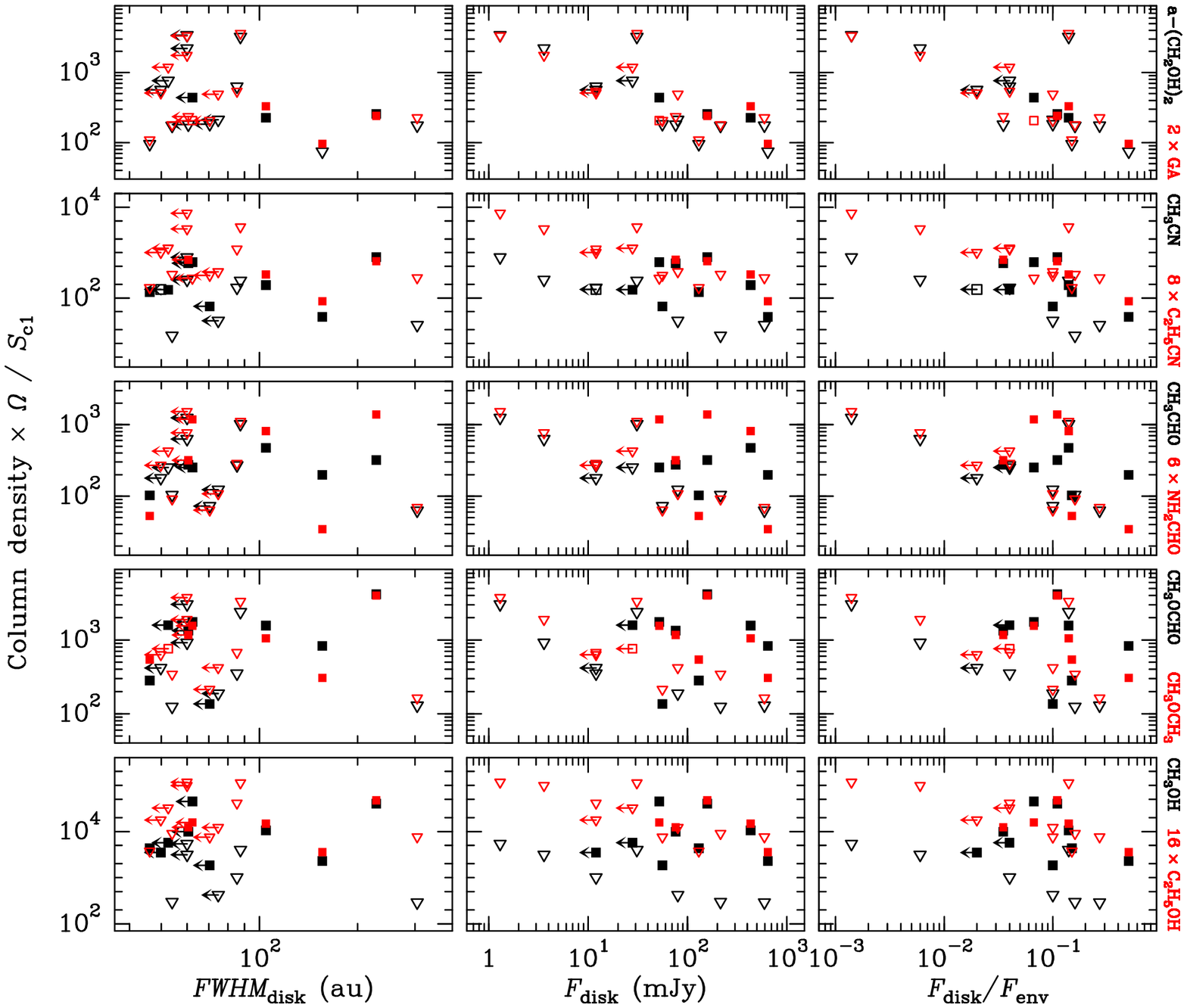}}}
 \caption{Same as Fig.~\ref{f:coldens_disk} but for the column densities
multiplied by the solid angle of the COM emission and divided by either the
1.3~mm continuum peak flux density or the 1.3~mm continuum flux density
integrated over the size of the COM emission when it is larger than the beam.}
 \label{f:coldens_disk_normsolangcont}
\end{figure*}

\begin{figure*}
 \centerline{\resizebox{0.8\hsize}{!}{\includegraphics[angle=270]{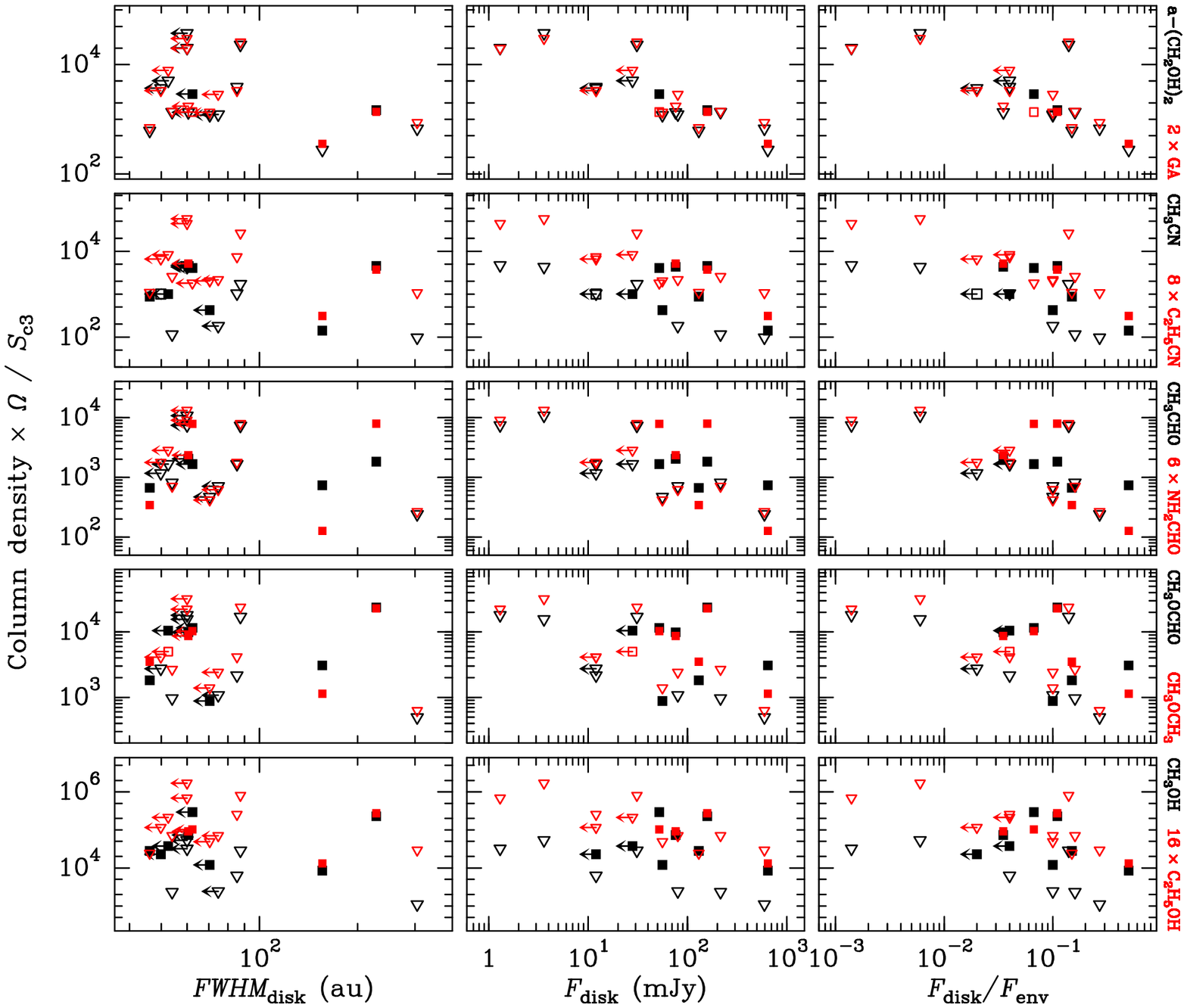}}}
 \caption{Same as Fig.~\ref{f:coldens_disk} but for the column densities
multiplied by the solid angle of the COM emission and divided by the 3~mm
continuum peak flux density.}
 \label{f:coldens_disk_normsolangcont3}
\end{figure*}

Figures~\ref{f:coldens_disk}--\ref{f:coldens_disk_normsolangcont3} show 
correlation plots between COM column densities and the properties of the
disk-like structures of the sources when detected, with different 
normalizations.

\clearpage
\newpage

\section{Estimation of dust temperatures}
\label{a:tdust}

The dust temperature can be computed using Eq.~2 of \citet{Motte01}: 
\begin{equation}
\label{e:tdust}
T_{\rm dust} = 38\,{\rm K} \times 
\left(\frac{r_{COM}}{100\,{\rm au}}\right)^{-0.4} 
\left(\frac{L_{\rm int}}{1\,L_\odot}\right)^{0.2},
\end{equation}
which is derived from Eq.~3 of \citet{Terebey93}. 
This equation assumes optically thin emission of dust heated by a central 
protostar in spherical symmetry and a dust opacity index, $\beta$, of 1.0. 
$L_{\rm int}$ is the internal luminosity given in Table~\ref{t:sources}. 
Given that the assumption of optically thin emission may not be valid on the
small scales where the COMs emit, we verify the reliability of 
Eq.~\ref{e:tdust} by comparing it to the temperature profiles derived by 
\citet{Kristensen12} with the one-dimensional radiative transfert code DUSTY.
They modeled five of the sources shown in Fig.~\ref{f:trot-lint} (L1448-C, 
IRAS2A, IRAS4A, IRAS4B, and L1157). For this comparison, we assume the same 
distances and luminosities as \citet{Kristensen12}. We find that the temperature
profiles agree relatively well with each other at large radii (beyond an
angular radius of 0.5$\arcsec$, 0.7$\arcsec$, 1.0$\arcsec$, 0.3$\arcsec$, and 
0.23$\arcsec$, respectively), while the dust temperature profile obtained with 
DUSTY is much steeper in the inner part, likely because of the dust optical
depth\footnote{\citet{Maury19} find a ratio of the continuuum effective 
radiation temperature to the dust temperature of 0.16 for L1448-C, 0.13 for 
IRAS2A, 0.5 for IRAS4B, and 0.20 for L1157, which correspond to optical depths 
of 0.18, 0.14, 0.7, and 0.22, respectively, meaning that the optically thin 
assumption does not hold anymore for the emission on scales smaller than the 
beam.}. Because $r_{\rm COM}$ is in all five 
sources smaller than this radius, Eq.~\ref{e:tdust} underestimates the
dust temperature at $r_{\rm COM}$ in these cases. The correction factors to apply
to Eq.~\ref{e:tdust} are 1.3, 1.7, 2.5, 1.3, and 1.3, respectively. 

For a given angular radius, the temperature computed with 
Eq.~\ref{e:tdust} is independent of the distance. We thus assume that the 
correction factors derived above are also applicable at the revised distances
listed in Table~\ref{t:sources}. As a first caveat, we mention that, while the 
luminosities used by \citet{Kristensen12} for L1157, L1448-C, and IRAS2A are 
similar to ours (once rescaled to the same distance), the luminosities we use 
for IRAS4A and IRAS4B are a factor $\sim3$ lower. This may have a small impact 
on the correction factor to apply to Eq.~\ref{e:tdust} for these sources, 
but we neglect this additional correction because it is not straightforward to 
evaluate. The second caveat concerns the density profile, which is the key 
parameter controlling where the dust emission becomes optically thick. This
is critical, in particular, in the case of binaries where the individual 
density profiles of the components are not well known. For instance, the
large correction factor obtained for IRAS4A does most likely not apply to 
IRAS4A2 which likely dominates the luminosity of the system but not the mass.
A dedicated radiative transfer simulation of IRAS4A2 with a reduced mass would
yield a smaller correction factor. Given these uncertainties on the density
profiles of the sources, we decide to use a single correction factor of 1.3
for all sources shown in Fig.~\ref{f:trot-lint}. On the basis of the discussion
above, we think that the dust temperatures derived at $r_{\rm COM}$ with 
Eq.~\ref{e:tdust} and this correction factor are uncertain by at least a 
factor 1.3.

\end{appendix}

\end{document}